\documentclass{MMws-book975x65}
 \usepackage[utf8]{inputenc}
 \usepackage{amsmath}

\usepackage{wrapfig}
\usepackage{wrapft}
\usepackage[utf8]{inputenc}
\usepackage{makeidx}

\makeindex

\usepackage{graphicx}

\newcommand{\be}{\begin{eqnarray}}
\newcommand{\ee}{\end{eqnarray}}
\newcommand{\bea}{\begin{eqnarray}}
\newcommand{\eea}{\end{eqnarray}}
\newcommand{\ba}{\begin{eqnarray}}
\newcommand{\ea}{\end{eqnarray}}
\newcommand{\Dslash}{D\hspace{-1.6ex}/\hspace{0.6ex} }
\newcommand{\Wslash}{W\hspace{-1.6ex}/\hspace{0.6ex} }
\newcommand{\pslash}{p\hspace{-1.6ex}/\hspace{0.6ex} }
\newcommand{\partslash}{\partial\hspace{-1.6ex}/\hspace{0.6ex} }

\newcommand{\fslash}{\hspace{-1.4ex}/\hspace{0.6ex} }
\newcommand{\nn}{\nonumber}
\newcommand{\Tr}{\mbox{Tr}\;}

\newcommand{\ket}[1]{\left|#1\right\rangle}
\newcommand{\bra}[1]{\left\langle#1\right|}
\newcommand{\braket}[3]{\langle#1|#2|#3\rangle}

\newcommand{\A}{\mbox{$A_\mu^a$}}

\newcommand{\pa}{\partial}

\newcommand{\la}{\lambda}
\newcommand{\om}{\omega}

\newcommand{\rar}{\rightarrow}

\newcommand{\pt}{\partial_\tau}
\newcommand{\po}{\partial_\omega}

\begin{document}
 \begin{center}
 {\bf \Large
 Nonperturbative Topological Phenomena \\ in QCD and Related Theories\\
 \vspace{2cm}
 Edward Shuryak }
 \end{center}
 \newpage
 


 {\bf Preface} \\ \\
The book is a summary of various lectures on nonperturbative QCD which I has given during the 
last three decades at Stony Brook. 
It is about  the {\em topological objects} present in gauge theories,  their semiclassical theory and applications to multiple physical phenomena.  
The overall content of the book can be grasped from the list of chapters and sections below,
so let me mention here, at the start, what is different in this book relative to others. 
There are good classic  reviews, books
and lecture notes
on {\em magnetic monopoles} and {\em instantons}.  Yet those about {\em instanton-dyons }, {\em sphalerons} and even such traditional object as {\em QCD flux tubes}
are yet to be found. 
Even for subjects on which there is extensive pedagogic-style literature,
they all have rather different focus. Usually these objects --
topological solitons, as one can call them collectively-- are 
 treated $individually$. (It can be compared to a visit to a zoo: here is a $lion$, and here is a $gazelle$, etc. )

Of course, we will have similar individual discussion of all these objects
 below as well, but the focus will be on their $ensembles$, and $phenomena$ which
 such objects generate  collectively. (Think of it as an actual trip to African savannas.) For example, at some settings such solitons can exist  in finite cluster or groups, and in another 
 in infinite (scaled as volume) ``condensates". Under certain conditions,
 $monopoles$  undergo Bose-Einstein condensation, related to ``confinement-deconfinement" 
 phase transitions. 
$Instantons$ lead to 
quark pairing and  condensation, breaking  spontaneously the chiral symmetry. This leads
to effective quark mass (and therefore large fraction of the nucleon's mass -- as well as our own!). $Sphalerons$ lead to chiral imbalance in heavy ion collisions and in early Universe,
producing also sounds, gravity waves and perhaps even the Baryon Asymmetry. 

  The semiclassical models of all these phenomena have also strong roots 
  in the first-principle numerical approach to gauge theories, known as {\em lattice gauge theories}. For all of these objects, observations of them ``on the lattice", and verification
  of their collective effects, would also be the important part of many chapters below.
  
   Crucial feature of modern field/string theories is a  notion of 
     $ ``dualities"$, namely existence of {\em different yet physically equivalent }
descriptions.   
 We will  discuss three of those:
 (i) the famed electric-magnetic duality; (ii) the ``Poisson duality",
e.g. between monopoles and instanton-dyons, as well as (iii)  the holographic gauge-string  or AdS/CFT duality. Discovery of each duality is always a surprise,
  complemented with original disbelief, then some confusion and finally
 multiple tests. This make internal logic of a book a bit complex, since some chapters are ``dual"
 to others. Let me illustrate the situation for ``Poisson duality", much less known than the others. It tells us that the partition function (and of course everything else, stemming from it)
 using $monopoles$ looks very different from that using $instanton-dyons$: but the results
 are the same. Therefore, one should either use one or the other formulation, not
 their sum or other combination:  they are just two different ways of organizing 
 evaluation of the same path inntegral!  

 \begin{tblofcontents}
 \end{tblofcontents}

\chapter{Introduction}

\section{What are the ``nonperturbative topological" phenomena?}

In {\bf quantum field theory (QFT)} courses  one first learns about {\em weakly coupled} field theories,
such as quantum electrodynamics and $perturbative$ QCD. 
They start with harmonic oscillator quantization, ``quanta" of the fields, and their weak
interaction, described by {\em Feynman
diagrams}\footnote{Of course, Feynman diagrams have now much wider than QFT range of applications. Readers
interested in the easiest introduction to Feynman diagrams can find it in my textbook "Manybody theory in the nutshell", Princeton University Press, where they are explained  using basic toy models from quantum statistical mechanics
. 
}.  Expanding the partition sum of a theory, represented by some
path integral, in  powers of the coupling one get results which  schematically looks like this \be Amplitude(p) = \sum_n (\alpha_s)^n C_n(p) \label{eqn_pert_series} \ee
where $\alpha_s\equiv g^2/4\pi$ is assumed to be small enough to make
series convergent in some ``practical sense"\footnote{While these series are known to be divergent or asymptotic,
several its terms approach certain limit at sufficiently small $\alpha_s$, before divergent away from it in higher order.
General issues related with perturbative series we will discuss in connection to the so called transseries later.
}. The coefficients $C_n$ 
are calculated functions of the kinematical parameters, e.g. products of 4-momenta of 
in and out-going quanta. This approach was created in 1950's in quantum electrodynamics,
QED.  
The corresponding coupling\footnote{We will use now standard high energy physics units $\hbar=c=1$,
showing them explicitly only in few cases such as this one. For more on units and notations, see Appendix.} $$e^2/\hbar c\approx 1/137 \ll 1$$ is small, serving as
 a natural small parameter of the perturbation theory. 

In non-Abelian gauge theories such as QCD the  coupling constant $\alpha_s(\mu)=g^2(\mu)/4\pi$ is
$running$, it depends on the momentum scale $\mu$ involved. 
The coupling is small at large momentum transfer (``hard") processes.
Perturbative QCD (pQCD) also uses Feynman diagrams: but we will not discuss any of that
in these lectures. (The most important perturbatively calculated function is the 
$beta$ function of the renormalization group, which is discussed in the Appendix. 
Another perturbatively calculated quantity we will need is
the {\em effective action}, created by quantum fluctuations for constant classical $\langle A_0 \rangle \neq 0$, also given in Appendix.)

These lectures are not about Feynman diagrams and perturbation theory, but about {\em nonperturbative phenomena}.
 Unfortunately,
 the word ``nonperturbative" (appearing in the book's title)  is  used in literature with several different meanings. The weakest of them is a situation in which the perturbative
  series  are re-summed. Sometimes the sum is simpler than each diagram by itself:
this for example happen if the answer is $non-integer$ power of $\alpha_s$. The most important re-summation will be that given by Renormalization Group (RG) 
  (Gell-Mann-Law beta function). 
  
    However, in this book we adopt  stronger meaning of the term, namely we will call ``nonperturbative" any  phenomena
  which are  {\em invisible\footnote{Strickly speaking, they are visible via  the so called Dyson phenomena explaining why the perturbative series are asymptotic (badly divergent) series.
  } 
  in the perturbative context}. What it means is exemplified by a 
  function of the coupling like this
  \be f(\alpha_s)\sim exp\big(-{const \over \alpha_s}\big) \big( \sum_{n=0}^\infty (\alpha_s)^n B_n \big) \ee 
  Attempting to expand it in powers of the coupling one finds that all coefficients of its Taylor expansion are zero.  The exponential term and the perturbative series build on it
are derived by the {\em semiclassical} methods, which we will study.  All exponential, perturbative and logarithmic (not shown) terms together form the so called $transseries$\footnote{A curious reader may wander if exponent, log and powers do form together 
a sufficiently complete set. The answer to it is affirmative.}. Relations between coefficients
$B_n$ and $C_n$ of (\ref{eqn_pert_series}), if found, are called {\em resurgence
relations}. 
  
    The {\em ``topological"} phenomena are related to  existence of {\em topological solitons}, made of so strong fields, that the interaction (non-quadratic) parts of Lagrangian  
    is as large as quadratic ones.  In the case of gauge theories 
    it means $\A_\mu=O( 1/g).$  Specifically,
    we will discuss
  :\\\\
    (i) {\bf  instantons} and their constituents, {\bf instanton-dyons}\\
    (ii) {\bf sphalerons}, unstable magnetic balls and their explosions \\
   (iii)   magnetic {\bf monopoles } \\
   (iv)  confining electric flux tubes, also known as the {\bf QCD strings}\\

As interesting as those objects are by themselves, 
as some mathematical curiosities, we will be mostly interested in ``what they can do to help us to understand the world around us", to mention a standard textbook-style sentence. 
Therefore, let me mention on the onset $why$ we are going to study them.

{\bf Instantons} are 4-dimensional solitons, historically important as
evidence of tunneling phenomena existing not only in quantum mechanical
systems (like nuclear $\alpha$ decays) but in QFT as well, in gauge theories especially.
They also were the first objects for  which semiclassical approximation 
was used in the QFT context\footnote{More generic semiclassical objects called {\bf fluctons}
were also developed, and will be discussed in Chapter \ref{sec_semi}. We did not list them
above since they are not generally topological, and their role in QFTs is not  worked out
in detail so far.
} .

Instantons also are very important in QCD because their {\em fermionic zero modes}
generate non-trivial interaction between light quarks (in QCD) or quarks and leptons (in
electroweak theory). If instantons are present
in large enough density, this effective multi-quark forces 
are strong enough to form the so called {\em quark condensate} 
$$ \langle \bar q q \rangle \neq 0 $$
This does happen in the QCD vacuum we leave in, making near-massless $u,d,s$ quarks look like
objects with an effective mass $\sim 400\, MeV$. So instantons are responsible for significant fraction
of the nucleon mass (and thus our mass as well!): this alone means that they deserve
to be studied. 	Quark condensate ``melts" at $T>T_\chi$, the so called temperature of the chiral symmetry restoration. In  QCD  with physical quarks both are close
$$T_\chi\approx T_c \approx 155 \, MeV $$
Another important phenomenon in QCD at finite temperatures is $deconfinement$:
both transitions seem to happen at the same temperature. 
This however seems to be occasional\footnote{ We know this because some deformations of  QCD does split them significantly,
and even make them of different transition order.}.

{\bf Monopoles} are 3-d solitons, made of the gauge and scalar fields, with
nonzero magnetic charges. They are bosons and their ensemble can 
undergo Bose-Einstein condensation. In QCD this happens  below some temperature $T_c$.
Their condensate expells the color-electric field into  confining electric flux tubes.
That is why $T_c$ is called the $deconfinement$ temperature: above it the matter is
quark-gluon plasma (QGP). 

{\bf Sphalerons}, like monopoles, are magnetic 3-d solutions to Yang-Mills equations. Unlike monopoles,
their magnetic field goes in circles, and so they lack magnetic charge. Although
they satisfy Yang-Mills equation, they are not
minima of the action but only saddle points. Therefore they are $unstable$ and, with small perturbation, can $explode$ along the so called {\em sphaleron path} configurations. The explosion itself allows for analytic solution and we will study it in sphaleron chapter. While sphalerons are important both in QCD and electroweak theory, 
we will discuss them mostly in the context of cosmological electroweak phase transitions (EWPT),
since they violate the baryon/lepton numbers conservation and are suspect to generate the observed baryon asymmetry of the Universe.

Certain properties of hadronic amplitudes and spectrum lead to the idea
that quarks  in the QCD vacuum are connected by the {\bf QCD strings}. Studies of effective string description
led to creation of the {\em String Theory}: but we will not go in this direction below. Instead
we will discuss their structure, interactions and in general their role in strong interaction physics. The main applications we will consider
are {\em inter-quark confining potentials}, closely related to {\em Pomerons and Reggeons},
the somewhat mysterious objects which emerged in 1960's from phenomenology of
hadronic scattering.  We will of course have a look at modern lattice and experimental data, as well as some modern derivation of the Pomeron amplitude.

Let un now return to ``perturbative" phenomena, for a moment.  
A  re-summation of certain sequence of diagrams
can lead to  properties of new quasiparticles, collective excitations etc. 
Even more advance theory can be based on the renormalization group:
the calculations may start in weak coupling, and lead to fixed points. They may also
indicate coupling flow to strong coupling, and explain existence of
qualitatively new phases of the theory. A classic example is 
the BCS theory of  superconductivity: weak phonon-induced attraction between electrons
get stronger near the Fermi surface, till finally it become strong enough to form Cooper pairs. 

%

 ``QCD-like gauge theories" we will discuss also 
posess the {\em asymptotic
freedom} phenomenon.  In  the UV (high momenta, small distances)
the coupling is 
 weak $g(p\rightarrow \infty )\rightarrow 0$.   
  To one-loop accuracy their coupling ``runs" according to 
\be {8\pi^2 \over g^2(p)}= ({11 N_c \over 3}- {2 N_f\over 3} ) log({p \over \Lambda_{QCD}}) \ee  
The he first  (one-loop)   coefficient in the r.h.s. $$b=({11 N_c \over 3}- {2 N_f\over 3} )$$ of the beta function give depends on the number of colors $N_c$ and light quark flavors $N_f$. It will always be positive in this course, so the number of fermions would be limited
 $11 N_c-2N_f>0$.

Substitution of this expression into perturbative series one gets series in inverse power of $log(p)$, small at large $p^2$. Since log is not a very strong function,
one would need to reach rather far in momentum scale,
compared to
the basic scale $\Lambda_{QCD}\approx 1/fm\sim 0.2\, GeV$
 to make coupling rather weak.
To give an idea, $\alpha_s(100\, GeV)\approx 0.11$ and $\alpha_s(1\, GeV)\approx 1/3$.

Now we return to non-perturbative phenomena.  
Substituting it into the exponential function mentioned above, one obtains certain $powers$ of 
the momentum scale 
\be exp\big[-C{8\pi^2 \over g^2(p)} \big] = \big[{ \Lambda_{QCD} \over p} \big]^{C\cdot b} \ee

 So, one may
reformulate our definition of the non-perturbative effects as those depending on the relevant momentum scale as its {\em inverse powers}.
Furthermore, one might expect that those powers would be some integers: and indeed, one expects
such effects to be present, they go under the term ``infrared renormalons" and have large literature
devoted to them. Nevertheless, the origin of those effects remains rather obscure, and will not be discussed.

Instead, we will  focus  on the {\em topological non-perturbative effects} induced by topological 
solitons of various kind. 
Unlike small perturbative fields -- photon or gluon waves -- they by no means
are small perturbations around ``classical vacuum", the zero fields. 
Their masses or actions  in the weak coupling are large
$O(1/g^2)$.   Account for such effects 
complement/generalize the
perturbative series to the so called {\em trans-series}. 

The main analytic method  to be used are $semiclassical$ approximation,
in various settings.
In this lectures we will compare the results with what is obtained 
in numerical 
simulations of the gauge theories, which is ``based on first principles of the theory"
and by definition include all effects, perturbative and non-perturbative. Needless to say,
the ultimate judge in physics is experiment, to which we will refer whenever possible.

\section{Brief history of Non-Abelian gauge theories and Quantum Chromodynamics} 
 While most readers can skip this section, some may still need some historic perspective.
Maxwellian electrodynamics is clearly the prototype field theory, on which
significant part of physics and its applications is firmly based. As shown by Lorentz and Poincare, electrodynamics require
 transforms from one moving frame into another by (then unusual) {\em Lorentz transformations}.
 Einstein in 1905 extended Lorentz invariance to all fields of physics,
including mechanics, explaining that previously used Galilean transformation
are but an approximation   at small velocities $v\ll c$. 

 After creation of Quantum Mechanics was complete, electrodynamics was also extended, from classical
 field theory to quantum one, known now as {\em Quantum Electrodynamics} (QED). Development
 of its perturbative formulation and working out multiple applications (splitting
 of atomic levels, anomalous magnetic moments of electron and muons, etc) was 
 quite a triumph. 
 
 QED is  an Abelian theory: this follows from the fact that photons, quanta of electromagnetic field $A_\mu$, are not themselves charged. Only other particles, say electrons, can 
 provide a source for electromagnetic fields.  It is a ``gauge theory": the wave function
 of charged objects allows transformation of its phase
 \be \psi(x) \rightarrow e^{i \phi(x)}  \psi(x) \ee
where $\phi(x)$ is an arbitrary function of the space-time coordinates. The invariance is
created by a ``compensation", subtracting gradient of $\phi$ from the gauge field.
The corresponding group of transformations is Abelian, which means that one can
multiply by these phase factors in any order.

In \cite{Yang:1954ek} QED was famously generalized to the case in which there are several types of charges to {\em non-Abelian gauge theories}. The local gauge symmetry
 is not a ``symmetry'' in the usual sense, because
it is indeed quite different from, say, translational, rotational
or flavor symmetries
we are used to. 
All physical observables  are singlets
under gauge rotations, and therefore are not really transformed. Thus
there cannot be any relations
 between them to be deduced
from this symmetry. Rather it tells us which degrees of freedom
of the fields are unphysical, or {\em redundant}, variables.

In order to explain its general  meaning  it
 is useful to trace an  analogy  with the  Einstein's  general
 relativity.
    Special  relativity  postulates  that  all
 inertial reference frames can equally be used.   General
 relativity goes further and suggests that one can make  arbitrary
 coordinate transformations, independently  {\em at any point}.  Similarly, in gauge theories
  "global  symmetries"  (rotations   in   the
  space of internal quantum numbers) the physicists  came  to
 local gauge invariance,  based  on  {\em independent}  rotations in space of charges,  at
 different space-time points.

Gauge transformation of the charged fields, now possessing some additional index, 
is generalized to unitary matrix\footnote{Like in other papers and books, we 
use here and below the Einstein's notations: any repeated index is summed.}  
 \be \psi_i(x) \rightarrow U_{ij} \psi_{j}(x) \ee
 reflecting the fact that in the space of charges a coordinate system  can be chosen arbitrarily and it can be done
 {\em independently} at each  space-time points. The corresponding groups of $U$ 
 are non-Abelian, since order of transformation is important: in general  unitary matrices do not commute with each other. This leads to many consequences, in particularly to 
 non-Maxwellian relation between the vector field $A_\mu$ and the field strength $G_{\mu\nu}$
 containing the commutator\footnote{The readers not familiar with this may need to
 consider more introductory texts.}.
 
 Yang and Mills first interpreted the index $i=1,2$ as $isospin$, the $SU(2)$ flavor symmetry
 due to (approximate) similarity of the two lightest quarks, $u$ and $d$.
 Therefore their  non-Abelian field was $SU(2)$
  adjoint  vector field identified with the  $\rho$ mesons. Subsequent natural application of such $SU(2)$ group 
  was related to the
 ``weak isospin" of weak interactions, with
  non-Abelian fields being   vector $W,Z$ bosons of weak interactions. 
 In both of these cases great  original difficulty was that gauge symmetry requires 
  the vector field quanta to be $massless$. Needless to say, neither $\rho$ mesons, nor  $W,Z$ bosons are massless.
 This puzzle was eventually solved, by ``soft" spontaneous
 breaking of the  $SU(2)$ gauge symmetry of weak interactions in Weinberg-Salam theory.
 
 Quantum chromodynamics (QCD) is based on the  $SU(3)$ group of rotations
 in space of quark's ``color" degree of freedom. This gauge group is $not$ broken,
 so gluons are massless. Yet they are strongly interacting and do not propagate
 individually due to {\em color confinement} phenomenon which we will discuss in the next subsection.
 
 Historically, QCD became standard theory of strong interaction after the theoretical discovery
 of ``asymptotic freedom" \cite{Gross:1973ju,Politzer:1973fx}
 (``running" of the coupling toward zero at small distances), nicely 
 correlated in time with the experimental discovery of weakly interacting pointlike quarks inside the nucleons (and other hadrons, of course). We will not discuss perturbative
 QCD as such, and thus  discussion and the relevant expressions  can be found in Appendix \ref{sec_as_freedom}. 
 Derivation of the two-loop beta function for supersymmetric relatives of QCD, based on instantons, will be given in
 section \ref{sec_NSVZ}.
 
 While the main playground of QED is in the atomic physics, that of  QCD 
 is the physics of strong interactions. Instead of atoms one has particles collectively called  ``hadrons". Some authors emphasize the fact that while atoms are nonrelativistic, 
 hadrons show large variety of regimes including very light and very heavy quarks. 
 We would emphasize even more important distinction between them: in QED the vacuum is
 ``empty", in QCD it is very nontrivial. It is the understanding of the ``vacuum structure"
 which is necessary in order to understand structure of various hadrons.
 
 We will not go into large field of hadronic models here, and just present  brief overview of few main directions:\\
 
 i){\em  Traditional quark models} (too many to mention here) are aimed at calculation of 
static properties (e.g. masses, radii, magnetic moments etc). Normally all calculations are done
in hadron's rest frame, using certain model Hamiltonians.  Typically, chiral symmetry breaking is included via
effective ``constituent quark" masses, the Coulomb-like and confinement  forces are included via some 
potentials. In some models also  a ``residual" interaction is also included, via some 4-quark terms of the Nambu-Jona-Lasinio type.\\

(ii) Numerical calculation of  {\em Euclidean-time} two- and three-point  correlation functions is another general approach, 
with a source and sink operators creating a state with needed quantum numbers, and the
third one in between, representing the observable. Originally started from small-distance
OPE and the QCD sum rule method, it moved to intermediate distances (see review
e.g. in  \cite{Shuryak:1993kg}), and is now mostly used at large time separations $\Delta \tau$ 
(compared to inverse mass gaps in the problem $1/\Delta M$) by the lattice gauge theory (LGT) simulations.  This condition ensure ``relaxation" of the correlators 
to the lowest mass hadron in a given channel. We will discuss Euclidean 
correlation functions  and lattice gauge theories in the corresponding chapters below.\\

(iii) Light-front quantization using also certain model Hamiltonians, aimed at
 the set of quantities, available from experiment. 
Deep inelastic scattering (DIS), as well as many other hard processes,
use factorization theorems of perturbative QCD and
 nonperturbative {\em parton distribution functions} (PDFs). Hard exclusive processes
(such as e.g. formfactors) 
are described in terms of  nonperturbative  hadron on-light-front wave functions (LCWFs), for reviews see \cite{Brodsky:1989pv,Chernyak:1983ej}. We will discuss it in chapter \ref{sec_light_front}.\\

(iv) Relatively new approach is ``holographic QCD", describing hadrons as quantum fields 
propagating in the ``bulk" space with extra dimensions. It originally supposed to 
be a dual description to some strong-coupling regime of QCD, and therefore was mostly
used for description of Quark-Gluon Plasma (QGP) phase at high temperatures. 
Nevertheless, its versions including confinement (via dilaton background with certain ``walls"
\cite{Erlich:2005qh} ) and quark-related fields (especially in the so called Veneziano limit in which both the number of flavors and colors are large $N_f,N_c\rightarrow \infty, N_f/N_c=fixed$ \cite{Jarvinen:2011qe} )
  do reproduce hadronic spectroscopy, with nice Regge trajectories. 
  The holographic models also led to interesting revival of baryons-as-solitons type models, generalizing skyrmions and including also vector meson clouds.  
  Brodsky and de Teramond  \cite{Brodsky:2006uqa} proposed to relate the wave functions in extra dimension $z$ to those on the light cone,
identifying  $z$ with certain combination of the light cone variables $\zeta$. 
 Needless to say, all of these are  models constructed ``bottom-up", but with well defined Lagrangians and some economic set of parameters, from
 which a lot of (mutually consistent) predictions can be worked out.

While (i) and (iv) remain basically in realm of model building,
  (ii) remains the most fundamental and
consistent approach. Lattice studies, starting  from the first principles
of QCD had convincingly demonstrated that they correctly include all nonperturbative phenomena.  They do display chiral symmetry breaking and confinement, and reproduce
accurately
hadronic masses. Yet its contact with PDFs and light-front wave functions remains difficult. The light front direction, on the other hand, for decades relied on perturbative QCD,
in denial of most of nonperturbative physics.

\section{introduction to color confinement}

This phenomenon, more precisely called {\em color-electric} confinement,
is perhaps the most complicated non-perturbative phenomenon.
In this section we introduce several ways in which its presence, 
or absence (called $deconfinement$) at high temperature/densities,
can be defined and studied. 

In a qualitative form, its definition can be that no object with color-electric charge
(such as quarks or gluons, or any combination of them with nonzero charge)  
appears in physical spectrum. However, to prove it in theory or in practice is
rather hard, so many other definitions are used.

In pure gauge $SU(N_c)$ theories deconfinement transition is related with
breaking of certain well defined symmetry, called center symmetry.
Therefore the transition is a phase transition with a well defined order parameter.
So we will start with the corresponding explanations and lattice data in
subsection \ref{sec_holonomy_intro}.

Another manifestation of confinement is formation of flux tubes between quarks,
and related linear confining potential $V(r)=\sigma r$. 
In \cite{Wilson:1974sk} this idea was reformulated 
for lattice studies, requiring that the so called  Wilson loop
have expectation value decreasing as an exponent of its $area$ if the area is large
\be  \langle W(large\, C) \rangle \sim exp[-\sigma_T Area(C)]
\ee
Discussion of this statement will be done in subsection \ref{sec_wilson_loop}.

Another interesting definition of confinement is related with Bose-Einstein
condensation (BEC) of magnetic monopoles. The introduction of this idea is given  below, with much more detailed discussion
 of it in chapter  devoted to magnetic monopoles. Let me here only
 mention that BEC can be detected in a number of ways, with certain 
 specially constructed order parameters. However, those parameters are 
 $nonlocal$ ones, and it is not clear where they do or do not show up as a singularity of the free energy.
 
 Confinement can be associated with existence of the electric flux tubes with a nonzero tension. We will discuss those in chapter  devoted to QCD flux tubes. This 
 ``operational definition" is not in fact correct: it has been theoretically argued 
and recently observed on the lattice  that the electric flux tubes can in fact exist even in the
deconfined phase.

Finally, the so called Hagedorn phenomenon -- a divergence of the partition funciton
of hadronic matter -- can also serve as an (approximate) indicator of the deconfinement
transition. 

\subsection{Polyakov lines} \label{sec_holonomy_intro}
\begin{figure}[h!]
\begin{center}
\includegraphics[width=10cm]{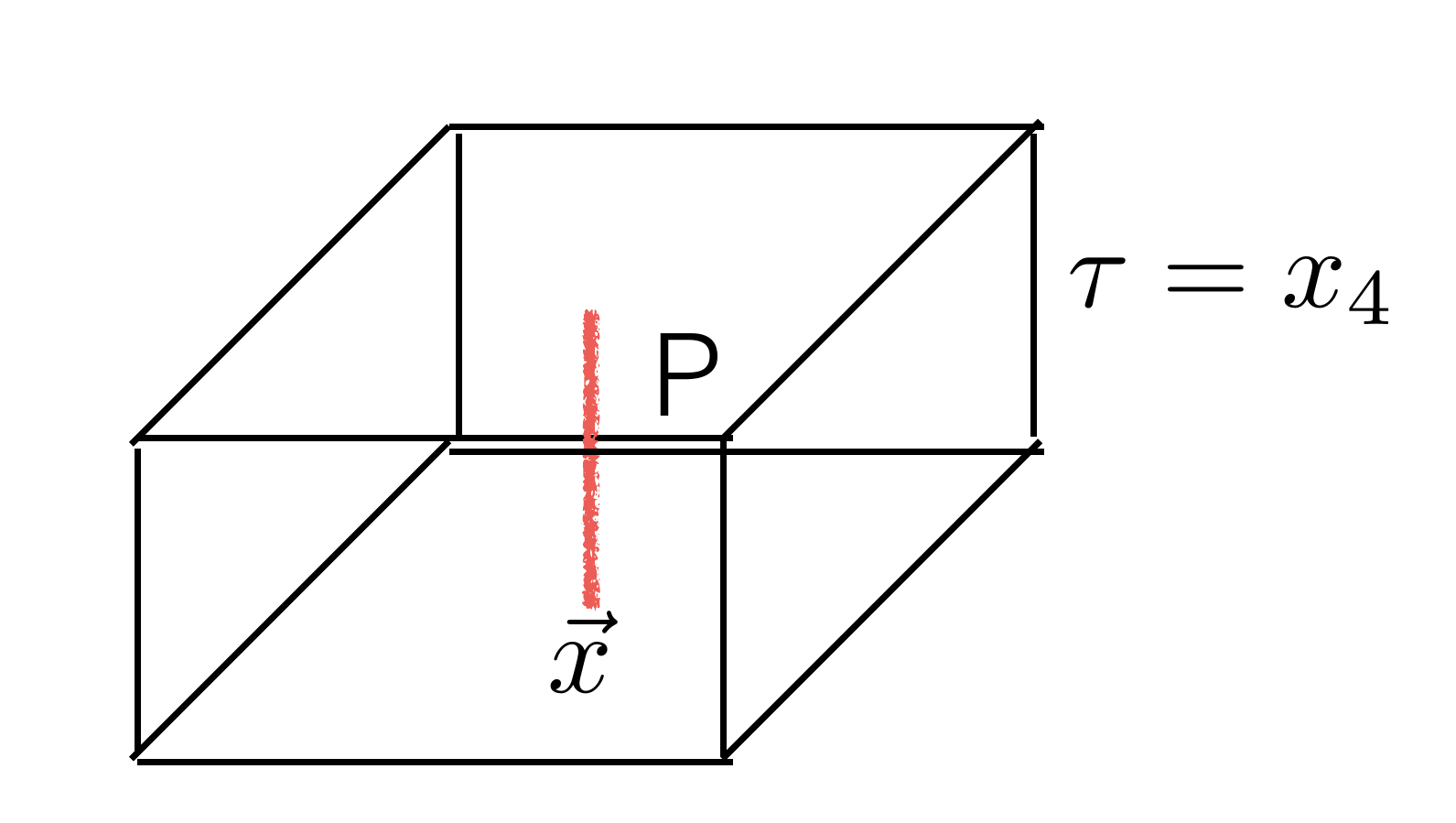}
\caption{The Polyakov line $P(\vec x)$ on the lattice. The vertical direction is that 
of Euclidean time $x_4=\tau$. Periodicity in it means that the fields on upper and the lower planes are identified.
}
\label{fig_P2}
\end{center}
\end{figure}

The Polyakov line \cite{Polyakov:1976fu} is defined as a similar integral as in $W$, but over the Euclidean time $\tau$ defined on a {\em closed circle} $C$ around Matsubara Euclidean time, see, Fig.\ref{fig_P2},
with circumference $\beta=\hbar/T$
\be \hat{P}=Pexp(i \int_C dx_4 \hat A_4), \,\,\,\,P = {1\over N_c}Tr\big( \hat P\big) \ee
The hats (here and below) remind us that it is still a matrix in color space.  Since $\hat A$ is hermitian, $ \hat{P}$ is unitary matrix $\in U(N_c)$. 

$Pexp$ means path-order exponent, defined as a limit of a product of matricies
describing small steps in time. 
On the lattice $ \hat{P}$ can be seen as a product of  link variables
 $\Pi_C(U) $ and the product should be done over the same circular contour $C$.
The corresponding integrals in mathematics are known as ``holonomies". Unlike gauge potential itself, $\hat P$ is {\em gauge invariant} due to
close contour $C$.
 
 The example of the simplest non-Abelian group $SU(2)$  will often be discussed. Assuming $ \hat{P}
$ is diagonal, it has the form $diag(e^{i\phi},e^{-i\phi})$ where the phase is
$$ \phi=(1/T)A_4^3 (1/2)= \pi \nu $$
where 1/2 is from the fact that color generator is $\hat \tau^a/2$ and we introduced 
a new parameter $2\pi T \nu=A_4^3$ for the average field strength. Then the trace is
$$ P= cos(\pi \nu) $$
 
 The physical meaning of vacuum expectation value (VEV) of $P$ is the effective quark free energy,
 $$ <P>=exp(-F_q/T) $$
In the confining phase of pure gauge theories $ <P>=0$ (which corresponds to infinitely heavy
point quark), while deconfinement means that it is finite $ <P>\neq 0$. So, for $SU(2)$ case the so called {\em trivial holonomy} corresponds to values $P=1,\nu=0$, while the confining value are $P=0, \nu=1/2$ .

\begin{figure}[h!]
\begin{center}
\includegraphics[width=10cm]{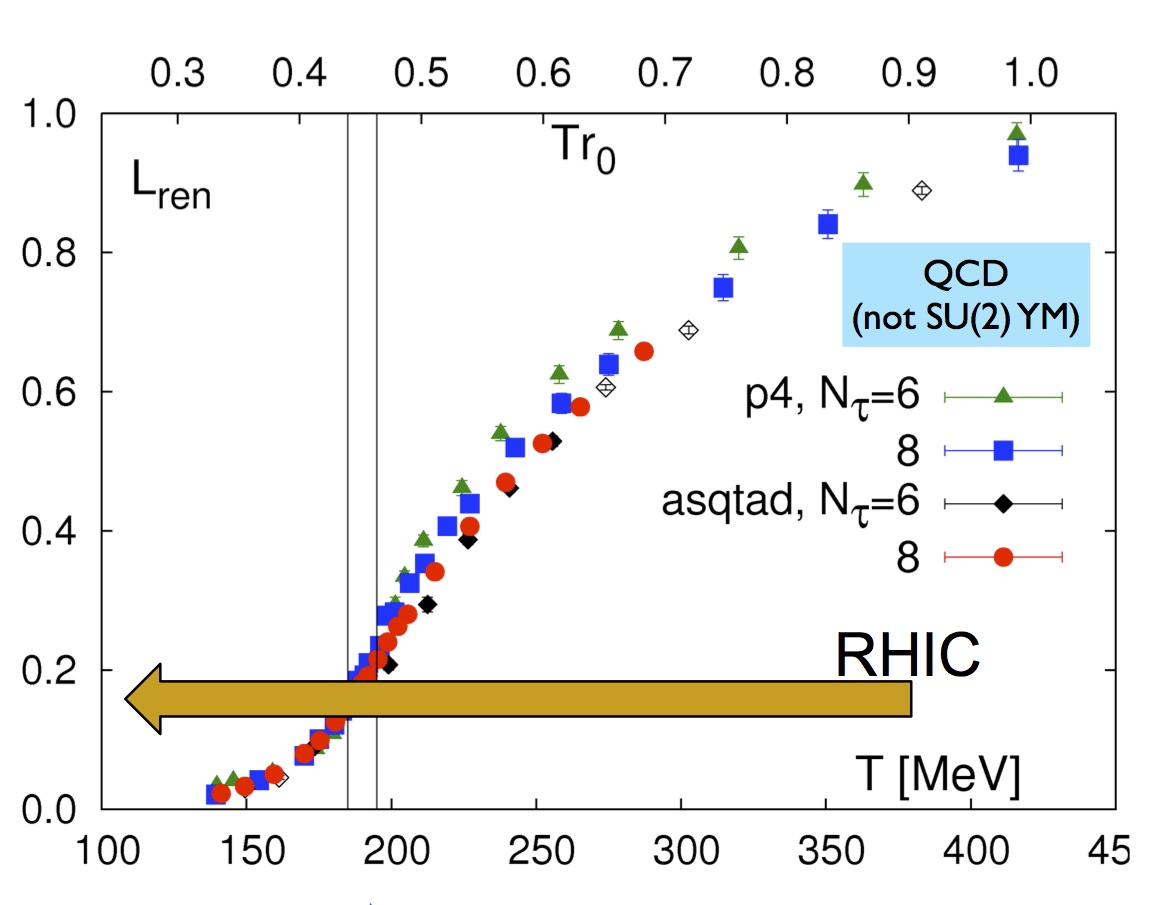}
\caption{Lattice data on the average value of the renormalized Polyakov line, as a function of the temperature $T$ in QCD. Different points correspond to different lattice actions. Two vertical lines indicate 
location of the critical point, following from studies of the thermodynamical observables.
}
\label{fig_PofT}
\end{center}
\end{figure}

 QCD is of course based on $SU(3)$ gauge theory and thus $\hat P, \hat A$ are $3\times 3$ matrices: we will discuss those when needed.  
Let us give an example of the
numerical simulations of lattice QCD  shown in Fig.\ref{fig_PofT}. In 
QCD transition is smooth crossover, while in pure gauge $SU(3)$ theory
without quarks (not shown) the VEV of $P$ at $T<T_c$ is strictly zero, with a jump at $T_c$.
Small $\langle P \rangle$ means large effective quark free energy and strong suppression
of their contribution to physical processes at low temperatures. 

\subsection{ Wilson lines and vortices}  \label{sec_wilson_loop}

\begin{figure}[h!]
\begin{center}
\includegraphics[width=10cm]{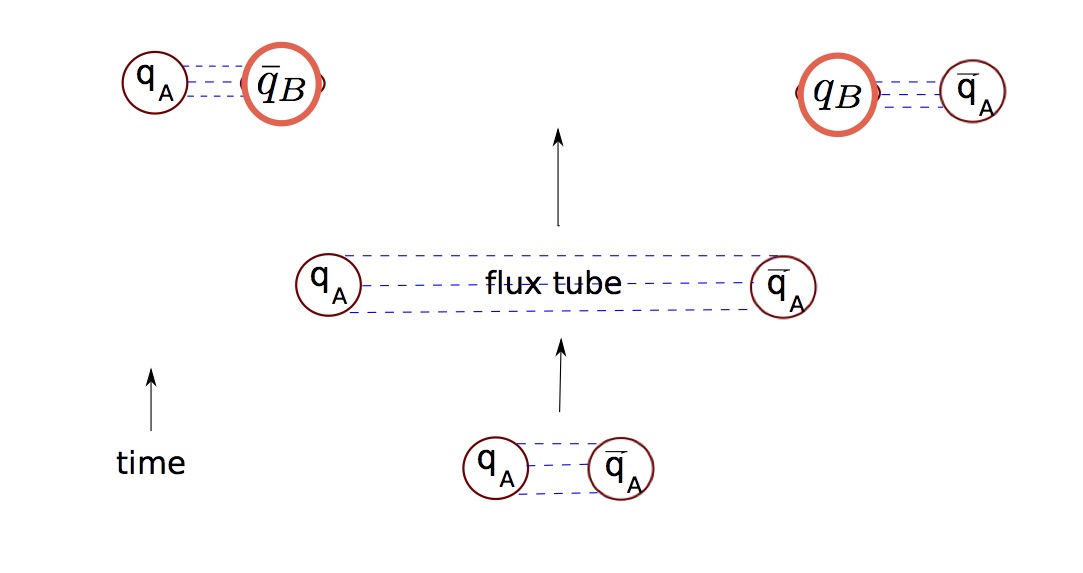}
\caption{A sketch of a heavy-heavy evolution, from  small distances (bottom) to broken flux tube and formation of
two heavy-light meson excitations (top)}
\label{fig_fluxtube_breaking}
\end{center}
\end{figure}

Using the language of color charges -- in particular very heavy external quarks -- 
one explains absence of colored states by the existence of  color-electric flux tubes. As shown in  Fig.\ref{fig_fluxtube_breaking}, it has a non-zero tension -- energy per length-- and thus creates a linear potential 
between charges. In QCD with light quarks, those string can be broken if its length is sufficient for
production of two heavy-light mesons.

Ken Wilson in mid-1970's played key role in formulation of the non-perturbative definition of QCD-like theories
on the lattice. He also promoted the statement about a linear potential to a more abstract mathematical form:
the VEV of the Wilson line 
$$ W=Tr Pexp(i \int_C dx_\mu A_\mu) $$
over some contour $C$ of sufficiently large size  with (the matrix valued) color gauge field.
$Pexp$ means product of exponents along a given contour $C$. Its VEV
should behave as follows
$$ <W>= e^{-\sigma*Area} $$
the Area means to be of a surface inclosed by the contour $C$. If it is a rectangular contour $T*L$ in 0-1 plane,
it is indeed corresponds to the $area=T*L$ and $\sigma $ is identified with the string tension.
This statement is known as the so called Wilson's
confinement criterium.
The main achievement of the very first numerical lattice studies, by M.Creutz in 1980, 
were demonstration of the area law, both in the strong and weak coupling settings.

What kind of gauge field configurations may lead to such ``area law"? This question 
lead to attention to the configurations with a nontrivial topology, known as $vortices$.
Quantized vortices in liquid helium and superconductors are well known, and
they are characterized by the fact that integral $\int_C dx_\mu A_\mu$ over any contour 
keep the same value. 

The center $Z_n\in SU(N)$ is defined as the set of elements with commute with each
group member: those are
$$ z_n=e^{i 2\pi {n\over N}} $$
Note that $(z_n)^N=1$.
In the simplest non-Abelian group with $N=2$ , there are two elements $z_0=1,z_1=-1$
with squares equal to 1.

In the gauge theory people looked at the so called
{\em center vortices} for which the circulation integral around them is $Z_n$. Let us in particular focus on  
$z_1=-1$ element in the $N=2$ case: each time such vortex pierces the Wilson line, there is a sign change.
Note that in 4 dimension linkage of the 2-d Wilson line with the 2-d vortex line history is a topological concept.
Note also, that since one thinks about $W$ in 0-1 plane, the  2-d vortex should be extended in the dual  2-3 plane.

Now, if there is a certain ensemble of center vortices, the area law follows. Suppose their locations is
random and $n$ of them are linked with smaller Wilson line with area $A$. The probability to have $n$ piercing it is
$$ P(n)= C_N^n ({A \over L^2})^n (1- {A \over L^2})^{N-n} $$
and 
$$W=\sum_n (-1)^n P_n=(1- {2A \over L^2})^N \rightarrow_{(N,L\rightarrow\infty,\rho=N/L^2=fixed)} =e^{-2\rho A} $$
The argument in this form is due to Engelhardt, Reinhardt et al. (1998).)

If it is sufficiently dense,
as lattice studies had shown, one obtains nearly all  experimental value of the string tension $\sigma$.
Removing center vortices from lattice gauge field configurations leads to zero string tension: thus the so called
``dominance" of center vortices claimed.

More details about the center vortices as the origin of confinement can be found in ``The confinement problem in lattice gauge theory", J.Greensite, Prog. Part. Nucl. Phys. 51 (2003) 1, hep-lat/0301023.


\subsection{Hadronic matter at $T<T_c$ and the Hagedorn phenomenon}
\label{sec_Hahedorn}

Thermodynamics  is normally derived from a statistical sum over physical excited states of the system
$$ Z=e^{-F(T)/T}=\sum_{n} exp(-E_n/T)=\int {d^3pV \over (2\pi)^3}dM \rho(M) exp(-\sqrt{p^2+M^2}/T)$$
where we introduced the spectral density of hadronic masses. 
Confinement in QCD-like theories is often stated as {\em  the absence of all colored states form the physical spectrum}.
All excited states are colorless hadrons: mesons $\bar q  q$ , baryons $qqq$, recently found $\bar q\bar q q q$ tetraquarks (with heavy quarks), etc.

The chiral symmetry is the property of the theories with massless quarks. Crudely speaking,
it means that the left and right-handed polarizations of the quark fields are independent of each other
and can be rotated separately. 
One may wander how the
small quark masses in QCD Lagrangian can be seen in a hadronic framework. In fact they can:
via $massless$ pions, the Goldstone modes of spontaneously broken chiral symmetry .

 One may also wander if the phenomenon of the $deconfinement$ can be really expressed in hadronic framework.
 Yes it can: via the so called Hagedorn phenomenon. Hagedorn noticed that  the spectral density of hadronic masses
grows very rapidly, approximately exponentially
 $$\rho(M)\sim e^{M/T_H} $$
and as a result when $T$ approaches $T_H$ the statistical  sum $Z$ diverges due to proliferation of many states. 

\begin{figure}[htbp]
\begin{center}
\includegraphics[width=10cm]{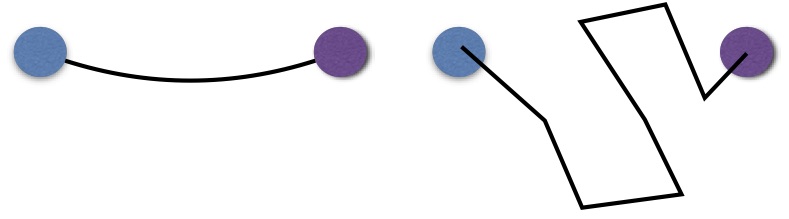}
\caption{A sketch of a meson structure at low $T$ (left) and high $T$ (right),
with a string excitation.}
\label{fig_meson_with_string}
\end{center}
\end{figure}

The reason why $\rho(M)$ grows so fast is very nontrivial. Already in 1960's hadronic phenomenology --
the Regge trajectories and Veneziano scattering amplitudes -- could be explained by so called {\em QCD strings}
or flux tubes, connecting the quarks. Important observation is that strings have much more states (configurations) than particles ,
illustrated in Fig.\ref{fig_meson_with_string}.

(Attempts to create {\em effective theory of the QCD strings}  of course later lead to the appearance of the string theory, which with time switched
to much larger space-time dimensions and claimed  a high title of ``theory of everything"
including gravity. Nobody has a clue if this is or is not true.)

%
%
%
%

\section{ Particle-monopoles, including
the real time (Minkowskian) applications } 

How magnetic charges may coexists with quantum mechanics\footnote{Which was just 4 years old then!} 
was explained by \cite{Dirac:1931kp}, who found that
it may only happen when the electric and magnetic charges satisfy a particular relation,
which makes singular lines between monopoles -- the Dirac strings -- $invisible$.

G. t'Hooft and Polyakov discovered monopole solution in Non-Abelian gauge theories with scalars \cite{tHooft:1974kcl,Polyakov:1974ek}.
Existence of monopoles were used in 
famous model of confinement, due to  \cite{Nambu:1974zg,Mandelstam:1974pi,'tHooft:1977hy} who argued that if the monopole density is large enough for their Bose-Einstein condensation,
the resulting ``dual superconductor" will expel electric color field via dual Meissner effect, creating electric flux tubes.
Monopoles were identified on the lattice, starting from 1980's,  their properties, spatial correlations and paths in time $x_m(\tau)$ analysed.
It was in particularly observed that these monopoles do indeed rotate around the flux tubes, producing solenoidal ``magnetic current"
needed to stabilize the flux tubes. Their paths do indeed  indicate their Bose-Einstein condensation at $T<T_c$.
Elimination of monopoles from lattice configurations also kills confinement: thus there are also papers on ``monopole dominance" in the confinement problem.

Note that  since magnetic monopoles are the 3-d topological objects, they are ``particles". Although
lattice simulations can only work with an Euclidean (imaginary) time, nothing prevents one to use 
monopoles in real-time applications. Such applications included studies of the quark and gluon scatterings
on monopoles, significantly contributing to small value of kinetic coefficient (viscosity) of Quark-Gluon plasma.
Recently there were identified contribution of the monopoles to jet quenching.

\section{ Instantons and its constituents, the instanton-dyons  } 
 Finally, non-Abelian gauge theories also have some 4-d solitons with nontrivial topology  \cite{Belavin:1975fg}, known as $instantons$.
They  do not explain confinement in four dimensions, as their fields falls too quickly to generate a Wilson's area law. 
But they induce important effects associated with light quarks. Intantons 
have fermionic zero modes, solutions of the Dirac eqn in such fields with the zero r.h.s.
$$ \gamma_\mu(i\partial_\mu +gA_\mu)\psi=0 $$
Because the contribution for a  gauge field configuration to the partition function is proportional to the
determinant of the Dirac operator, naively zero eigenvalues mean that instantons cannot appear
in the ensemble.  

This is indeed true for a single instanton, however if there is an ensemble of them -- the so called {\em instanton liquid} -- it is possible. The phenomenon can be described as {\em a  collectivization} of these fermionic zero modes
into the so called 
{\em Zero Mode Zone} (ZMZ) of quasi-zero Dirac eigenstates.

\begin{figure}[h!]
\begin{center}
\includegraphics[width=6cm]{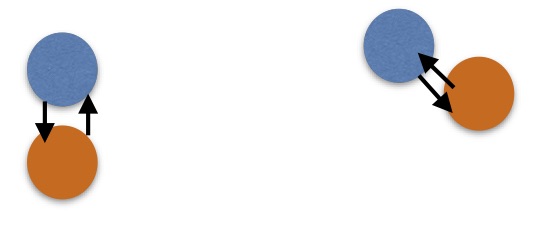}\\
\includegraphics[width=6cm]{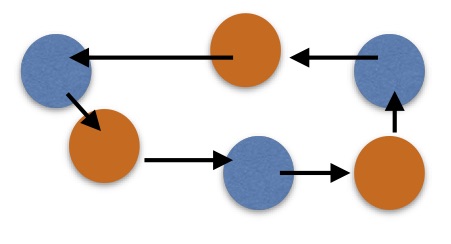}
\caption{short and long loops in the fermionic determinant
}
\label{fig_instanton_loops}
\end{center}
\end{figure}

If the instanton density is sufficiently large, ZMZ has states arbitrary close to zero, which forms the so called
quark condensate. The ZMZ has been observed on the lattice and indeed shown to be made of
linear superposition of zero modes of the individual instantons. Removing ZMZ -- which constitutes only about 
$10^{-4}$ of all fermionic states -- leads to effective restoration of the chiral symmetry, and change in hadronic masses by large amount, typically $20-50$ percents. This is also true for the nucleons, which mass is also that
of all of us and of the whole visible matter -- so it is a very important effect. 

We will also discuss several other instanton-induced effects. One of them is pairing not in the 
$\bar q q$ channel, producing chiral symmetry breaking, but in the diquark $qq$ channel, leading to color superconductivity.  The celebrated  Seiberg-Witten solution of $N$=2 supersymmetric gluodynamics 
is mostly a series of all order instanton effects,
 derived by the explicit calculation of all instanton amplitudes  by N.Nekrasov in 2002.

 Decreasing the temperature below  $2T_c$
one  finds 
a nontrivial average value of the Polyakov line $<P>\neq 1$ , indicating that an expectation value of the gauge potential  
is nonzero $<A_4>=v\neq 0$. This calls for re-defining the boundary condition of $A_4$ at infinity, for all solitons
including instantons. That  lead to
1998 discovery (by Kraan and van Baal, Lee and Li) that nonzero $v$
  splits
instantons  into $N_c$ (number of colors) constituents, the selfdual {\em instanton-dyons}\footnote{
They are  called  ``instanton-monopoles" by Unsal et al, and are similar but not identical
to ``instanton quarks" discussed by Zhitnitsky et al.} .
Since these objects have nonzero electric and magnetic charges and source
Abelian (diagonal) massless gluons, the corresponding ensemble is 
an ``instanton-dyon plasma", with long-range Coulomb-like forces between constituents.  
By tradition the selfdual ones are called $M$ with charges $(e,m)=(+,+)$ and $L$ with charges $(e,m)=(-,-)$, the anti-selfdual antidyons are called  
 $\bar{M}$, $(e,m)=(+,-)$ and  $\bar{L}$, $(e,m)=(-,+)$.

Diakonov and collaborators 
 emphasized that, unlike the (topologically protected) instantons, the dyons interact directly with
 the holonomy field. They suggested that since such dyon (anti-dyon)  become denser
at low temperature, their back reaction  may overcome perturbative holonomy potential and drive it
to its confining value, leading to  vanishing of the mean Polyakov line, or confinement.

In order to study instanton-dyon plasma one needs to know the dyon-antidyon interaction.
This was recently acheieved by Larse and myself, and several works has studied the 
instanton-dyon plasma, both analytically in the mean field approximation, and numerically,
by a direct simulation.

It has indeed been confirmed, that instanton dyons in gauge theory lead to confining phase,
provided their density is large enough. In QCD-like theories with light quarks 
both deconfinement and chiral restoration transition happen at about the same dyon density.
More recent studies focused on QCD deformations by fermion phases, which
were found to modify both phase transitions drastically. 

\section{ Interrelation of various topology manifestations and the generalized phase diagrams}
 The objective of this course can be defined as 
 derivation of effective theories based on topological objects.


Various topological objects present in gauge field configurations are all  inter-related.
For example, intersection of two center vortices where they disappear are the monopoles:
two  fluxes with angle $\pi$ each make one with flux $2\pi$ known as the Dirac string,
ending on a magnetic monopole. Monopole paths may end at the instanton. 
Elimination of  center vortices in lattice gauge configurations also eliminates monopoles, and elimination of the  monopoles
eliminates the instanton-dyons. 

If so, which objects should we study most? My answer is -- from the top down, the instantons first -- is based
on the following arguments. The instantons and their constituents have noticieable action large compared to $\hbar$:
therefore their effective theory based on semiclassical approximation can be self-consistenly constructed.
Furthermore, this effective theory has the form of ``classical statistical mechanics", with integration over certain
collective coordinates (positions of the instanton-dyons). For monopoles their effective theory would include path
integral over their path, corresponding to ``quantum manybody theory". For vortices one would need an analog of 
``quantum string theory", integrating over all of their world sheets: attempts to reformulate gauge theory as such 
were made, but basically  abandoned.

Note however, that even in the absence of consistent theory, one can still 
study effects of these objects qualitatively. 
Probing QFT's in various conditions produce difference response in its different versions,
and comparing to phenomenology (real and lattice experiments) one may identified the best -- if not unique --
explanations. This is what 
we will do in this course a lot.

 Infinitesimal probes
with various quantum numbers excite corresponding elementary excitations.
In QCD those are hadrons -- mesons and baryons. 

Yet we will not be interested that much in hadronic spectroscopy. We will focus on the 
  study of the ``vacuum structure". Generally speaking, those are
   revealed by a multitude of the
 vacuum correlation functions, with varios local or nonlocal operators.
 Yet their detailed discussion will take us too far. 
 
 What we will discuss are vacuum expectation values (VEVs) of various fields
 or field combinations, known as ``condensates". For example,
 we will discuss in detail how and under which conditions a nonzero VEV of the scalar bilinear of the
 quark field 
 $ <\bar{q} q > \neq 0 $
 known as the quark condensate, can be formed, or whether it can coexist with color superconductivity in which
 another bilinear condensate  $ <q q > \neq 0 $. 

Let me at the onset specific strategy we will follow. The vacuum state, with its condensates and strong coupling effects, is very complicated. To understand it gradually, it is convenient
to start with high $T$ setting, in which non-perturbative phenomena are power-suppressed.
Monopoles and instantons appear as rare individual excitations. With decreasing scale $T$
the coupling grows, and non-perturbative phenomena increase their presence, till finally they get dominant and completely reshape the statistical ensemble, creating the world we live in.
Therefore we will focus at deconfinement and chiral transitions.

\section{Which quantum field theories  will we discuss?}
Many books and reviews focus on cataloging QFT's, their properties and interconnections
 in different dimensions.
The  main focus  of this book is  understanding of non-perturbative physical phenomena
occurring {\em in our physical world}.  Therefore we discuss two ingredients of the Standard Model (SM) : \\
(i) the $electroweak$ one based on $SU(2)$ non-Abelian 
gauge theory plus quarks, leptons and (Higgs) scalar;\\
(ii) Quantum Chromodynamics ($QCD$)
 describing $strong$ interactions, based on
 non-Abelian gauge theory with the $SU(3)$ color group and quarks. 
 
 The electroweak theory is in a weak coupling regime, and therefore most
 of its nonperturbative/topological effects are hard to address experimentally.
 We will only study electroweak sphalerons in conjunction with cosmological phase transition
 in sphaleron chapter.
 
 Since in QCD the charge is ``running" (rapidly changes) as a function of the momentum scale considered, from weak to rather strong regime, here nonperturbative phenomena 
 are responsible for many phenomena we will discuss. In particular, 
 the topological effects  appear in it in full glory, being our main focus below.
 
However, in order to understand these phenomena better, it would be sometimes desirable to 
consider not just physical QCD, but wider set of theories with 
and various settings of those, with certain {\em variable parameters},
and investigate how effects considered get modified. 

 The most obvious parameter is the number of colors $N_c$ of the gauge group,
and this is the first parameter of the theory which we will change whenever it is useful to do so. In some cases
we will use the smallest value of it, $N_c=2$, and in some case we would  discuss the opposite limit of large $N_c\rightarrow \infty$.

Much less obvious is the parameter $\theta$, in a CP-odd gauge theories with
$\theta (\vec E \vec B)$ term in the Lagrangian. We will also discuss in some cases the so called $deformed$ QCD, with added (gauge invariant) terms containing powers of the Polyakov loop: this will be needed to affect the confinement phenomenon. 
 
  Furthermore, one can  consider
 the non-Abelian gauge theories possessing  various $fermions$, different from
 the quarks we have in the real world QCD. 
 
 Let us first remind that six known quark flavors are naturally divided in SM
 as follows. The electroweak sector split them in three dublets $(ud),(cs),(tb)$.
 
 In QCD we instead split them into three light $(uds)$ and three heavy $(cbt)$  
 ones. Putting the masses of three light to zero one finds {\em the chiral symmetry} $SU(3)_L \times  SU(3)_R$ for left and right-handed polarizations.
 Expansion in light quark masses is known as {\em chiral perturbation theory}.

Putting the masses of three heavy quarks to infinity $m_q\rightarrow \infty$ one finds the {\em 
heavy quark symmetry} \cite{Shuryak:1981fza,Isgur:1989vq}, and expansion in $1/m_q$ is known as {\em heavy quark effective theory}. 

Of course, one can add more quarks to  QCD, for example
change  the number of light quark flavors to arbitrary integer $N_f,f=u,d,s,...$.

Another deformation or the quark fields to be considered is related with their 
periodicity on the Matsubara circle (in Euclidean time). The quarks, being fermions,
are $antiperiodic$ on this circle. If they instead assumed to be $periodic$, 
quarks effectively become bosons: this setting is used if one would like to
preserve the supersymmetry\footnote{For this to be the case one also would need either
to change the color quark representation from the fundamental to $adjoint$ one, 
in which case those will be called $gluino's$ and the number of the types of those will be denoted by $N_a$; or introduce fundamental scalars or squarks.}.

More generally, we will also use arbitrary
 periodicity phases, which also can be different for each flavor, $\theta_f$.
So these quarks would have intermediate statistics.  As we will see,
the deconfinement and chiral symmetry restoration phase transition
are sensitive to the value of these phases, revealing the 
topological objects which cause them. 

All of these theories can be generically called versions of {\em deformed QCD},
to be used in conjuction with our discussion of this or that phenomena when needed.

One should however carefully single out some of these deformations, the special cases 
in which there are certain {\em new symmetries} , absent in the real world QCD. 
Those may have specific behavior which is $absent$ in QCD. One example is
two-color case, in which antiquarks can be related to quarks by Pauli-Gursey symmetry.

Much more important case are theories with several number of $supersymmetries$  $\cal{N}$. 
They, as well as $supergravity$ theories, occupy large portion of QFT literature, textbooks
and lecture notes in the last three decades. So, let me explain to what extent it
will be discussed in this book. 

As of the time of this writing, we see no evidences that supersymmetry exists in Nature.
We will also $not$ focus on supersymmetry in technical sense, for example we will not use
superspace notations. And 
 yet,  we will turn to
some supersymmetric theories which proved to be useful toy models. In many cases
 they provide valuable lessons, or even analytically derived 
expressions for some quantities,  the analog of which in non-supersymmetric theories are 
only available from numerical simulations if at all. 

To be specific, we will discuss the $\cal{N}$=1 super-gluodynamics in connection with the first
application  in chapter  on the instanton-dyons: in this theory the calculation of the
gluino condensate  resolved some long-standing puzzle.

It would be impossible not to mention many fascinating aspects of the
$\cal{N}$=2 (twice-supersymmetric Yang-Mills theory, two gluinoes, one scalar), also known as  Seiberg-Witten theory, especially in connection
with the issues of electric-magnetic duality in chapter \ref{chap_mono}. Some minimal information about it 
is therefore given in Appendix. 
 Conformal $\cal{ N} $=4 theory (4 gluinos, 6 scalars) leads to many more fascinating features, which we will partly cover in the next chapter and Appendix.

%
%

 
\chapter{Monopoles} \label{chap_mono}
\hspace{2cm} {\em ``One would be surprised if Nature had made no use of it."} {\bf Dirac }
\section{Magnetic monopoles in electrodynamics}
Not discussing the origin of electricity and magnetism in antiquity, let me jump to
pre-Maxwellian XIX century, in which it was clearly stated that electric charge can be divided into positive and negative
charges, while the magnets, if cut, still produce only dipoles, with the so called north and south poles. 
In other words, it was observed that it is not experimentally possible to separate  magnetic charges. And, in 
well known subsequent developments which lead to Maxwellian electrodynamics, magnetism is ascribed to motion
of the electric charges.
Yet by the end of that era some -- notably J.J.Thomson and H.Poincare -- were discussing  possible existence of some hypothetical particles with a magnetic charge.
Since then, 
this idea has a very unusual history: it tends to be dormant for 2-3 decades, then
become very active, and then dormant again. It did so at least 5 times, to my count.

Development of quantum mechanics brought in an issue of quantization. At that stage
the greatest impetus to the whole problem has been provided  in \cite{Dirac:1931kp} and subsequent works.
He argued that the equation $\vec{\nabla} \vec{B}=0$ can be consistent with nonzero magnetic charges if there are
singular lines -- the {\em Dirac strings} -- supplying the magnetic flux from outside. And, under special Dirac quantization condition,  the Dirac strings can be made invisible! Dirac stressed that this condition seems to be the
only known reason explaining why all electric charges are $quantized$. 

\begin{figure}[h]
\begin{center}
\includegraphics[width=8cm]{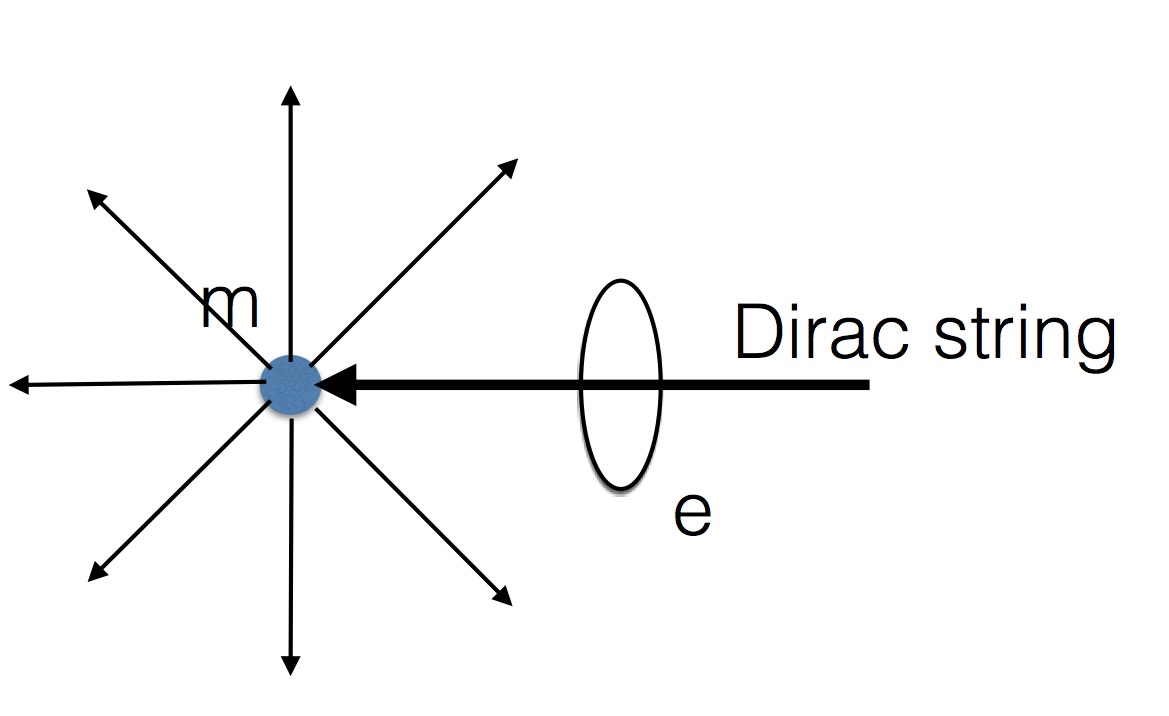}
\caption{The magnetic monopole with the Dirac string (thick line) carried the ingoing magnetic
flux, equal to the total flux carried out by Coulomb-like field (thin lines with arrows). The circle around
the  Dirac string is a path of an electron $e$ moving around it. }
\label{fig_dirac_string}
\end{center}
\end{figure}

An introductory review on electromagnetic magnetic monopoles is e.g. that by Milton \cite{Milton:2006cp}, for an in-depth source 
on monopoles in non-Abelian theories one may consult  the book by Shnir 
\cite{Balian:2005joa}.
Multiple searches for QED magnetic monopoles has produced no convincing candidate events. An argument why this can be the case
follows from the Dirac condition itself. Since the electric fine structure constant\footnote{This section is written
in QED notations in which the Coulomb field is $\vec E=e \vec r /r^3$. In QCD notations, to be used elsewhere,
the field and charge normalization changes, so that $\vec E=e \vec r / 4\pi r^3$
}
 $\alpha=e^2\approx 1/137\ll 1$ is very small, the magnetic one should be very large
$g^2\sim 137 \gg 1$: perhaps at such strong magnetic coupling any separation of the charges is  problematic.

The gauge field configuration with a Dirac string is sketched in Fig.\ref{fig_dirac_string} above.
What one would like to obtain is a Coulomb-like magnetic field
\be \vec B= g {\vec r \over r^3} =\vec \nabla \times \vec A\ee
from some vector potential configuration. Here is one which does so
\be \vec A=g \left(sin(\phi) { 1+cos(\theta) \over r sin(\theta)}, -cos(\phi) { 1+cos(\theta) \over r sin(\theta)}, 0 \right) \ee
where $r,\phi,\theta$ are spherical coordinates. 
Note that for a half-line $\theta=0$ the numerator is 2, and thus the $\vec A$ is singular, 
but for a half-line $\theta=\pi$ it is zero of higher degree and it is in fact zero. 
More general form of the {\em Dirac potential}  is
\be \vec A={g \over r} {[ \vec r \times \vec n] \over r- (\vec r\vec n)} 
\ee
where now $\vec n$ is an unit vector directed in any direction we like. 

The question however remains whether the  Dirac string is or is not visible in any physical experiment.
The answer to it goes back to the so called Aharonov-Bohm effect \cite{Aharonov:1959fk}. An electron making some
closed loop $C$ around the string will pick up a phase $phase=e\int_C A_\mu dx_\mu$. For some thin solenoid\footnote{
The AB effect has even been observed for quantized vortices in a superconductor, in spite of the fact that
quantization mean periodic wave function of the condensate. That is possible because 
the Cooper pairs of the condensate has electric charge $2e$: so if their loop results in an ``invisible phase" $2\pi n$,
 a single electron gets half of the phase  $\pi n$ which is visible for odd $n$. 
}
this phase can take any value, and be observable, even if the electron is never allowed into the region
where $\vec B\neq 0$.  Plugging in the Dirac potential, one finds that the phase is
$$ phase=4\pi e g $$
In order the Dirac string be invisible, the phase needs to be $ 2\pi n$, and so
one thus needs to enforce the following {\em Dirac quantization condition}\footnote{Note that in the sections
below, with different normalization of the field and couplings, the extra factor $1/4\pi$ will appea in the l.h.s. } 
\be e g =  {n \over 2} 
\ee

 \begin{figure}[htbp]
\begin{center}
\includegraphics[width=6cm]{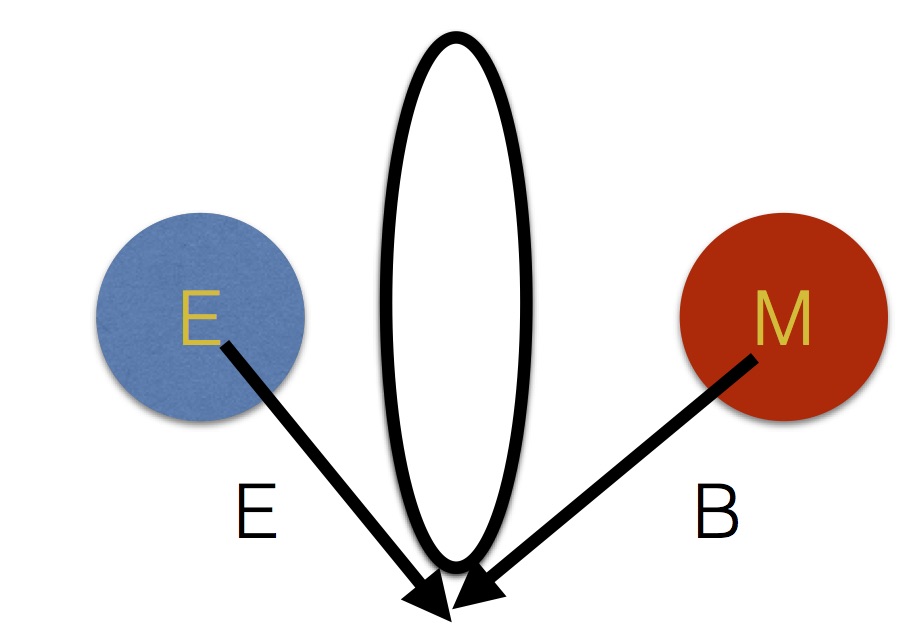}
\caption{J.J.Thomson: static electric and magnetic charges create a rotating field, as indicated by
the rotating Pointing vector $\vec S=[\vec E \times \vec B]$.  }
\label{fig_QandM}
\end{center}
\end{figure}

  There exists also another instructive way toward understanding the Dirac condition. In 1904, J. J. Thomson\footnote{J. J. Thomson was recipient of the 1906 Nobel Prize in Physics,  for measuring  the charge-to-mass ratio of the electron, effectively discovering the first  elementary particle known. His other high distinction was that seven (!) of his students, including his son,
  also became Nobel Prize winners. }
  observed that even $static$  (non-moving) charges, electric $e$ and magnetic $g$ 
  create a rotating electromagnetic field.
   Indeed, two Coulomb fields from such charges meet at a generic point  at some angle
(see Fig.\ref{fig_QandM}) and  thus create a nonzero Pointing vector $$\vec S=[\vec E \times \vec B]\neq 0$$  circling around the line connecting two charges. Thus, the field is rotating, even while the charges themselves
  are not moving! 
 
If one or all charges are allowed to move, the existence of the field angular momentum dramatically changes their trajectories, since only the $total$ angular momentum -- of the particles and the field combined -- is conserved. We will discuss those changes later in the book:
for now it is enough to say that this observation would be the key to understanding of
properties of the Quark-Gluon Plasma.  
  
  In quantum context, the angular momentum carried by the field must be quantized, as usual, to an integer times $\hbar$. The consequences of that are explained in the following exercise:

  {\bf Exercise:} {\it calculate the angular momentum $J$ and show that its quantization
  just mentioned lead to the Dirac quantization condition of the charges.}

\section{The non-Abelian gauge fields and t'Hooft-Polyakov monopole}
The solution, found independently by \cite{tHooft:1974kcl} and  \cite{Polyakov:1974ek} ,
implements these  ideas in the  setting with non-Abelian gauge theory. What is  however required for success of the program, is an ingredient (which QCD-like theories 
do $not$  possess), namely  {\em a scalar field in the adjoint color representation} $\phi$.

 Here is the Lagrangian
of the so called Georgi-Glashow (GG) model
\be L=-{1\over 4} (G^a_{\mu\nu})^2 +{1\over 2} (D_\mu \phi)^2- {\lambda \over 4} (\phi^2 - v^2)^2  \ee
a direct descendant of Ginzburg-Landau
free energy of a superconductor.  
It differs from the electroweak sector of the standard model, the Weinberg-Salam model, exactly by the
fact that $\phi$ has adjoint color representation. Nonzero VEV, shown as $v$, 
 provides different ``Higgsing" of the gauge fields: their mass is proportional to the commutator of their color generator with that of the VEV. In this model the $SU(N_c)$ color group into its $N_c-1$ diagonal subgroups.
%

The simplest case (we will only discuss) is the $N_c=2$ gauge theory. The corresponding algebra has
 3 generators $T^A,A=1,2,3$. Since we will be dealing with $N_c=2$ QCD also, some explanation
of the notations on color representations are in order. 

In the  $N_c=2$ QCD we will be dealing with quarks, which are in $fundamental$ (or spinor)
representation of the $SU(2)$ group. This means that quark color indices run over $a=1,2$  
and color generators are  $T^A=\tau^A/2$, where
$\tau^A$ are three Pauli matrices familiar from quantum mechanics description of spin\footnote{
This is not surprising, since there is a relation between the $SU(2)$ and rotational $O(3)$
groups.}.

 In the GG model the scalar is in $adjoint$ (or vector) representation. This means that scalar color indices run over $a=1,2,3$. The color matrices are then given directly by group
 structure tensor, which in this group is simply $(T^A)_{bc}=\epsilon_{Abc}$.

 Let us take the nonzero VEV to be along the diagonal, so $<\phi^3>=v$.
Then two gauge bosons ($W^{+,-}$) get nonzero
masses, while the boson number 3 (neutral ``photon") remains massless\footnote{The  Georgi-Glashow model
was designed to avoid existence of the $Z$ boson, then unknown. Experiments eventually had shown that it is the Weinberg-Salam version, with fundamental representation of the scalar, which describes the weak interactions of quarks and leptons: it is now known as
{\em The Standard Model}. Glashow was still, quite correctly, awarded the Nobel Prize.}:
this field is called a ``photon".

Since there is such a drastic difference between those components, one may like
to introduce a special notations for the Abelian-projected fields (without color indices) .
\be A_\mu= A^a_\mu \hat \phi^a,\,\,\,\,   \hat \phi^a={\phi^a \over | \phi^a |}\ee
where we introduced  a unit vector indicated by a hat.
In order to define also the Abelian field strength, 
the field definition should not be just the usual Abelian expression based on $A_\mu$ because it should 
 be supplemented by a term  canceling possible
derivatives of the Higgs color direction. The definition is
\be  F_{\mu\nu} =\partial_\mu A_\nu -\partial_\nu A_\mu -{1\over e} \epsilon_{abc}\hat \phi^a \partial_\mu \hat \phi^b \partial_\nu \hat \phi^c.    \label{eqn_abelian_F} \ee
This last term is of course zero for constant Higgs field. Furthermore,  in fact it vanishes for all topologically trivial 
configurations of $\phi^a(x)$: 
 it will be nonzero for topologically nontrivial ones, and the possibility to have a gauge in which there is no Dirac string
 is based on this observation.

The magnetic current can now be defined
  from the definitions given above
\be 
k_\mu=\partial_\nu \tilde{F_{\mu\nu} }=\epsilon^{\mu\nu\rho\sigma} \epsilon_{abc}\partial_\nu \phi^a \partial_\rho  \phi^b\partial_\sigma \phi^c({1 \over 2v^3 e})
\ee  
Here and in many occasions below, we will use tilde for 4-d dual field, given by application of the 4-index epsilon tensor\footnote{Note that magnetic field is dual to electric one, and ``selfdual"
fields we will be discussing later have equal electric and magnetic fields.}.

Unlike the usual Nether currents, this magnetic current $k^\mu$ is conserved by definition, without any underlying symmetry
. The integral of its density is known in mathematics as
the {\em Brouwer degree}. As any other topological quantity, it gives in appropriate normalization an integer,
which defined topologically distinct Higgs field.

How it may happen is clear from  an example, of   a ``hedgehog"-like field\footnote{The relation between
such field configuration and this cute animal -- unfortunately absent in America --
has appeared for the first time  (to my knowledge) in the Polyakov's paper.}
\be \phi^a(r\rightarrow \infty) \rightarrow v {r^a \over r}\ee
in which the ``needles" go radially: the magnetic charge for it is 
\be g=\int d^3x k_0 = {4\pi \over e} \ee

Let us now look
 for a solution consistent with that asymptotical trend, in terms of two
spherically symmetric functions
\be \phi^a= {r^a \over e r^2} H(ver); \,\,\,\,
A^a_n=\epsilon_{amn}{r^a \over e r^2} [1-K(ver)]; \,\,\,\,A^a_0=0
\ee
When those are plugged back into the expression for the Hamiltonian one finds the following expression for the
monopole mass
\be
E={4\pi v \over e} \int_0^\infty {d\xi \over \xi^2}[ \xi^2 \dot{K}^2+ (1/2)(\xi \dot{H}-H)^2\nonumber \\
+(1/2) (K^2-1)^2+K^2 H^2+{\lambda \over 4e^2}(H^2-\xi^2)^2]
\ee
Here we rescaled the radial coordinate $\xi=e v r$ and the derivative over $\xi$ is denoted by a dot.
This expression can be minimized by variational methods. The equations obtained
can be viewed as corresponding to some classical motion in the $K,H$ plane 
of a particle, with $\xi$ being the time. The
standard way to get equations of motion is as described in Classical Mechanics courses.

 The boundary conditions at small $r$ correspond to $H\rightarrow 0$ and $K\rightarrow 1$  
 as then the Higgs field
 is smooth in spite of the ``hedgehog" direction.   At large distances $H\rightarrow \xi$ and Higgs becomes of
 the magnitude $v$, while $K\rightarrow 0$. 
  
  \begin{figure}[t]
\begin{center}
\includegraphics{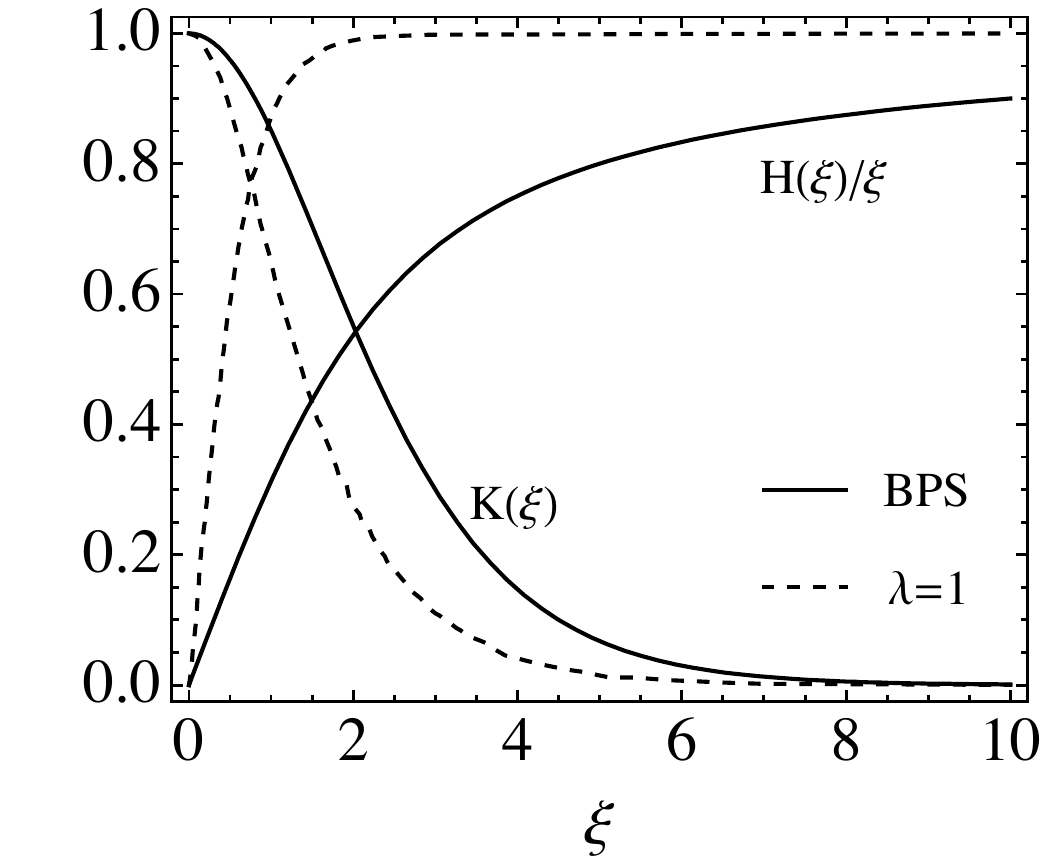}
\caption{
Classical solutions $K(\xi)$ and $H(\xi)$ for the 't Hooft Polyakov monopole at $\lambda=0$ 
(the BPS limit) and $\lambda=1$
}
\label{fig_mono_solutions}
\end{center}
\end{figure}
  
At large distance only Abelan part of the field survives, as it is massless. The obtained soliton happen to be a Dirac
monopole, with the magnetic charge
\be g ={4\pi \over e} \ee 
Note that it is indeed consistent with the Dirac Quantization condition. In fact
we prefer to write it later in a more symmetric form, using electric and magnetic ``fine structure constants", namely
\be \big({g^2 \over 4\pi}\big)\big(  {e^2 \over 4\pi}\big)  =\alpha_e \alpha_m=1
\label{eqn_Dirac}
\ee
which are then inverse to each other.

The solution for the EOM can generally be obtained numerically. Two of them
are shown in Fig.\ref{fig_mono_solutions} taken from \cite{Balian:2005joa}. On general grounds, the monopole mass
can be written as 
\be M={4\pi v\over e} f({\lambda \over e^2}) \ee
with smooth function $f$ depending on the ration of two dimensionless couplings of the GG model. Some of its values are $f(0)=1$ and $f(1)= 1.787 $.

The special case $\lambda=0$, in which scalar has no potential at all
\footnote{At this point one invariably asks: if there is no potential and thus no minimum of it,
how one can select the scalar VEV $v$? The answer is: all values of $v$ is in this case
possible. This means in the BPS limit all $nonequivalent$ vacua with different $v$
values may occur: there is no way to say which one is better than the other.
}
, is the
so called Bogomolny-Prasad-Zommerfeld (BPS) limit.  In this limit
 both profile functions are simplified and 
known analytically
\be K={\xi \over sinh(\xi)}, \,\,\,\,\, H={\xi \over tanh(\xi)} -1 \ee

{\bf Exercise:} {\it Using this definition of the BPS functions, calculate the Abelian (\ref{eqn_abelian_F}) magnetic field for the BPS monopole. Observe the difference in $r$ dependence between the color-diagonal and non-diagonal components. }

Classical monopole solution has 4 symmetries: Three of them is 
a shift in the monopole position, and one is a gauge rotation $U=exp(i\alpha \tau^3)$ which leaves scalar's VEV
unchanged. When quantum corrections are added to classical fields $A\rightarrow A_{cl}+a$,
one finds that quadratic  Lagrangian part $O(a^2)$ has 4 zero modes,
corresponding to 4 directions in Hilbert space of all possible deformations $a$ 
in which the action remains unchanged. 
 Those zero modes, generated by symmetries, create significant problems
 in semiclassical theory: they should be taken out
before calculation of quantum corrections. We will study that using the example of instantons
later.

Solutions with the integer monopole number $M>1$ obviously have $4M$ zero modes. The
corresponding collective coordinate keep the same simple meanings when one speaks about 
well separated monopoles, but the situation changes when they overlap. The shapes of the
solutions with $M>1$ and the metric of the collective coordinates turned out to be a
very nontrivial problem, involving high-power mathematics. For example, the $M=2$ solution deforms from 
two monopoles into a doughnut, and for larger $M$ there appear even more exotic shapes. 
The 4-dimensional ``moduli space" of relative collective coordinates called $M_2$, known as {\em Atiyah-Hitchin
manifold}, has explicitly written metrics and very nontrivial geometry. Manton has further introduced a notion of ``slow motion"
in moduli space {\em along its geodesics}, in nice analogy to motion in general relativity.
However, we will not have time to discuss 
those beautiful results. 
For a pedagogical introduction of them the interested reader can consult \cite{Balian:2005joa}. Some key
elements of it we will discuss later, in connection to selfdual gauge fields in connection to instantons. Two-monopole configurations, bound by fermions, will be discussed
in chapter \ref{sec_unusual_confinement} in connection to the so called ``unusual confinement" issue.

\section{ Polyakov's confinement in three dimensions} \label{sec_Polyakov_conf}
This section belongs 
to this chapter only partly, since in it the monopole solution is not used as a particle, having paths in 4d, but as an instanton
(or pseudo-particle, as Polyakov calls it)
of the 3-dimensional setting. Still, it is a very famous application, from
 classic Polyakov's paper \cite{Polyakov:1976fu}\footnote{ In fact this 1977 paper starts with instantons in the double well quantum-mechanical problem, which we discussed at the beginning of the book,
and ends with acknowledgement to Gribov who suggested tunneling interpretation of them. Needless to say,
as it appeared, I read and re-read it dozens of times since then.}, touching on the confinement problem we will discuss a lot later.

The part of this paper we  discuss here is a chapter on Georgi-Glashow (GG) model, which, after Higgsing, is
renamed as ``compact QED". The monopole is the solution to YM equations of motion
we discussed in the monopole chapter. Now we however will not discuss particle-monopoles in (1+3)dimensional space-time,
as before, but consider this solution to be {\em the instanton} in Euclidean 3-d space.

Recall that generic GG model has two couplings, the gauge one $e$ and the Higgs selfcoupling called $\lambda$.
The classical action for a set of monopoles is
\be S={M_W \over e^2} \epsilon({\lambda \over e^2}) \sum_a q_a^2 + {\pi \over 2 e^2} \sum_{a\neq b} {q_a q_b \over |x_a-x_b|}  
\ee
where $M_W$ is the mass of the non-diagonal gluons, called $W$ by analogy to weak interaction bosons,
and $\epsilon({\lambda \over e^2})$ is the monopole mass. (Recall that for generic couplings it can only be calculated numerically.) 

The main term is the magnetic Coulomb interaction term, so the manybody problem one needs to solve is that
of the 3d Coulomb gas, with the partition function of the type
\be Z= \sum {\xi^N \over N!} \int (\Pi_i d^3x_i )   e^{- {\pi \over 2 e^2} \sum_{a\neq b} {q_a q_b \over |x_a-x_b|}  }
\ee
where fugacity $\xi \sim exp(-{M_W \over e^2} \epsilon({\lambda \over e^2})$ is exponentially small
when the monopole action is large, in weak coupling. 

Polyakov uses standard mean field (or Debye) approximation, introducing a scalar field $\chi$ coupled to charges
\be \int d\chi e^{-{\pi e^2 \over 2}\int (\partial \chi)^2} \sum_N \sum_{q_a=\pm1} {\xi^N \over N!} \int (\Pi_i d^3x_i ) e^{i\sum q_a \chi(x_a)} \ee
The physical meaning of $\chi$ is of course a gauge potential coupled to the charges. Note however, 
that since the charges are magnetic ones, the gauge potential is ``dual" to $A_\mu$, its gradients are not the
electric but magnetic field. 

Note that taking Gaussian integral in $\chi$ will return us to the original Coulomb gas. Instead we will keep $\chi$ as is,
sum over two types of charges $$\sum_{q_a=\pm1}e^{i\sum q_a \chi(x_a)} =2 cos(\chi(x_a))$$ and exponentiate
the series in monopoles. \be Z\sim \int D\chi e^{-{\pi e^2 \over 2} \int d^3x [(\partial\chi)^2 - {4\xi \over \pi e^2}  cos(\chi) ] }\ee 


Expanding the cosine one finds that the (exponentially small!) coefficient in front of it is basically
the mass squared of the field $\chi$. 

Now, the magnetic potential (and thus field)  was the last massless field left after HIggsing: now it is 
also gapped. Polyakov showed that the usual criteria of confinement -- like the area law of the Wilson loop --
hold, and so 3d GG theory is confining. Let me add that the mean field criterium
is that there are many particles in the Debye cloud $n M^{-3} \gg 1$: it is satisfied in weak coupling because
$M$ is exponentially small.

Polyakov then goes on to the 4-d instantons, only to find that their interaction law is not Coulombic ($1/r^2$ in 4d) but
$1/r^4$ or short-range. Thus no Debye screening and no confinement by the same mechanism\footnote{I remember how
 disappointed he was. Four years later I came to see him and told that his instantons beautifully solve another
 famous QCD problem, the chiral symmetry breaking. Polyakov's answer was ``but I have not invented them for that". }.

\section{Electric-magnetic duality}
The term ``duality" generally means that the same theory may have very
different effective descriptions, depending on the dynamical regime in question. 
For example, QCD -- a theory of colored quarks and gluons -- has a dual low-energy description
in terms of the so called chiral effective Lagrangian, describing interaction of the 
lightest particles -- the pions and other Goldstone mesons.

Electric-magnetic duality is basically a similar question: depending which  particles
are the lightest one, effective low-energy description should use appropriate degrees of freedom.
Above we discussed magnetic monopoles in a weak coupling regime $e^2/4\pi \ll 1$, 
in which they are heavy solitons with a large mass $M\sim v/e^2$. The previous section defined 't Hooft-Polyakov monopole as a classical solution. The reader perhaps expect that
we will now develop a semiclassical theory of them, with small fluctuations around classical solutions
included in 1,2, 3 etc number of loops, like we did in the previous chapter for quantum mechanical
extreme paths. And indeed, one can do so: we will in particularly return to the issue of fermions coupled to the monopoles
in one-loop approximation below.

Let us however for now focus on the following question: since the coupling ``runs"
as a function of scale, becoming stronger as momentum scale decreases, one may ask
{\em what happens with the monopoles when $e^2/4\pi \sim 1$ or even become large?}

Due to classic work    \cite{Seiberg:1994rs} we know what happens with
monopoles in the {\cal N}=2 supersymmetric theories (pure gauge or with quark-squark multiplets.
They were able to 
 make a ``quantum leap" over all such steps, to {\em exact} results. Not only those include any number of loops in perturbation theory, but any number of instantons as well. We will not discuss how they figured out
the answer -- for this one has to read their original paper -- but simply present the results.

The theory in question is  the simplest supersymmetric theory which $needs$ to include a complex scalar field in the adjoint representation,
like the Georgi-Glashow model discussed above. 
The  $\cal N$=2 gluodynamics or super-Yang-Mills (SYM) 
theory has
gluons (spin 1), two real gluinoes $\lambda,\chi$ (spin 1/2), and a complex scalar (spin 0) which we will
call $a$. Each of them has two degrees of freedom, thus 4 bosonic and 4 fermionic ones.

The $\cal N$=2 QCD is a theory with additional matter supermultiplets of structure $\psi_f,\phi_f$ with spin 1/2 and 0,
respectively. We will call $N_f$ the number of Dirac quarks, as in QCD, or $2N_f$ Majorana ones.
The {\bf coupling renormalization} in these theories is done only via the
 one-loop beta function, with  the one-loop coefficient of the beta function equal to\footnote{Note that unlike the same coefficient for non-sypersymmetric theories, it is strictly integer. The explanation for that is given in
 the instanton chapter where it will be explicitly derived, see the section on NSVZ beta function.}
\be b=2N_c-N_f \ee
while in the two-loop and higher orders all  coefficients of the beta function are  zeroes.
We will only consider the simplest case with two colors $N_c=2$.
Note further that if 
   $N_f=4$, this version of supersymmetric QCD has  $zero$ beta function, and is therefore a scale-invariant (in fact conformal) theory:
we will however not discuss it.

\begin{figure}[htbp]
\begin{center}
\includegraphics[width=10cm]{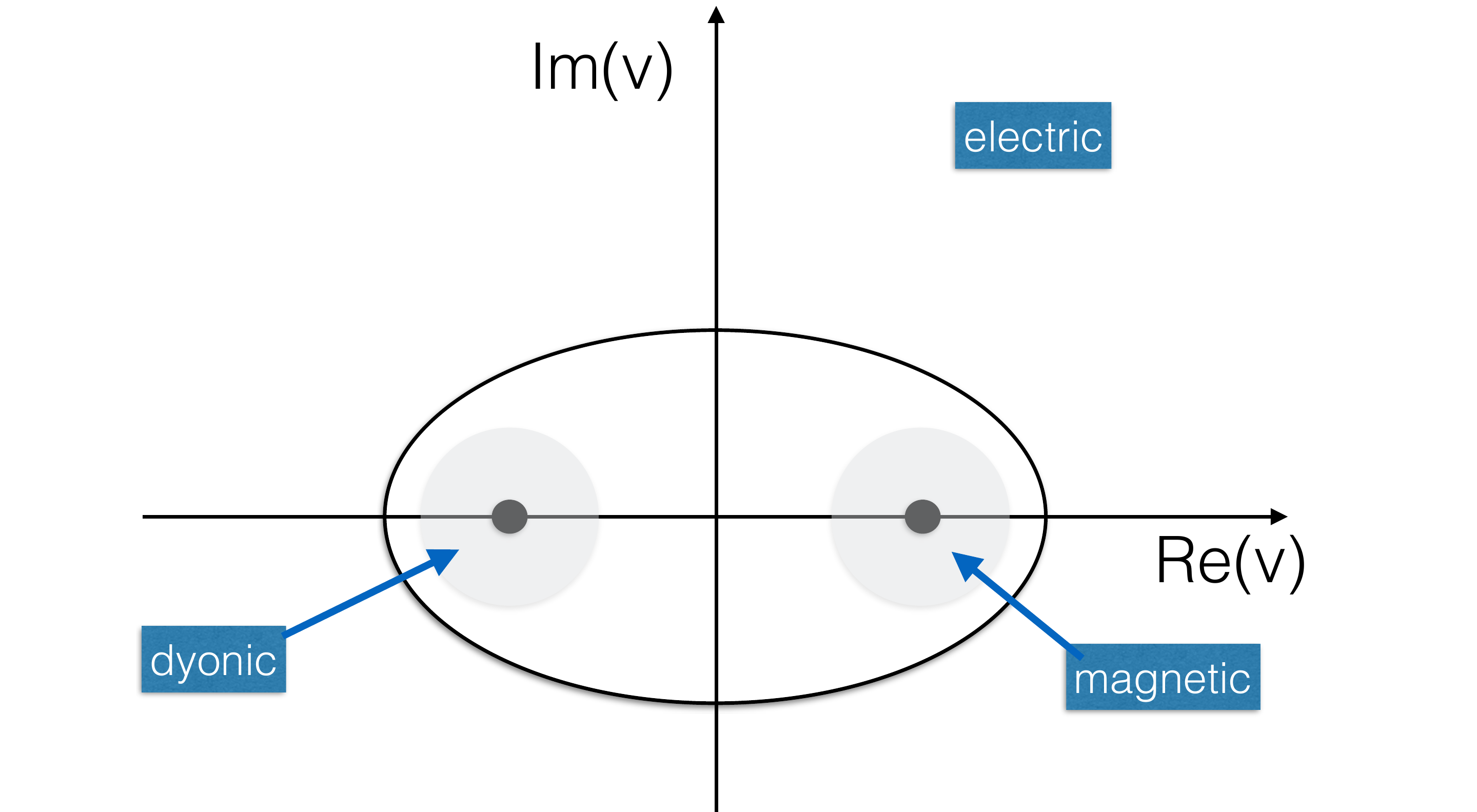}
\caption{The map of the moduli space according to Seiberg-Witten solution.}
\label{fig_SW_moduli}
\end{center}
\end{figure}

 Let us return to the basic supersymmetric gluodynamics, $N_f=0,N_c=2$ or $\cal N$=2 SYM.
The only complex scalar of this theory may have a nonzero VEV, denoted by 
some complex number $v=\langle \phi \rangle$. Supersymmetry require that there is no potential, and thus one has the BPS limit we discussed already.
The set of all possible $v$ fills a complex plane, which is in general called the {\em moduli space} of all possible vacua of the theory.
A schematic plot of this space is given in Fig.\ref{fig_SW_moduli}.

One can also use another variable
\be u=<tr(\phi)^2> \ee
independent of the particular direction of the VEV.

The setting allows for monopoles, as the scalar  is in the adjoint color representaion. 
Supersymmetry does not allow 
 any self-interaction potential for scalars, so (in Georgi-Glashow notations) the coupling $\lambda=0$, so that the monopoles are automatically  in the BPS limit (for which we gave above the explicit classical solution).

Supersymmetry also require that for $any$ value of $v$ the vacuum energy remains zero: thus
there is a whole {\em manyfold of non-equivalent vacua}, known as {\em moduli space}, labeled by a complex number $v$.
All properties of the system are expressed as derivatives of one fundamental holomorphic function 
$F(u)$, in particularly
the effective charge and the theta angle are combined into a variable $\tau$ which is given by its second derivative
\be \tau(u)= {\theta \over 2\pi}+ {4\pi i \over e^2(u)}={\partial^2 F(u) \over \partial u^2} \ee
We will return to the exact form of this function later, in connection with its instanton-based description.

The map of the moduli space is skematically given in Fig.\ref{fig_SW_moduli}
There are three distinct patches on the $v$ plane:\\
(i) at large values of $v\rightarrow \infty$ there is a ``perturbative patch"
, in which the coupling $e^2(v)/4\pi \ll 1$ is weak. It is dominated by electric
particles -- gluons, gluinoes and higgses -- with small masses $O(ev)$, which determine the beta function.
Monopoles have large masses 
$4\pi v/e^2(v)$ there, and can be treated semiclassically.\\
(ii) a ``magnetic patch" around the $v=\Lambda$ point, in which the coupling  is infinitely strong $e^2\rightarrow \infty$, the monopole mass
goes to zero as well as the magnetic charge $g\sim 1/e \rightarrow 0$.  \\
(iii) a ``dyonic patch" around the  $v=-\Lambda$ point, in which a dyon (particle with electric and magnetic charges both being 1) gets massless.

%
%
%

Let us, for definiteness, follow development along the real axes, from right to left. 
At large $|v|$ the coupling is weak $e^2(v)/4\pi\ll 1$ and the effective ``electric" theory resembles electroweak 
sector of the standard model. The lightest particles are gauge bosons (W's) and their superpartners,
with small masses induced by Higgsing $\sim e v$. 

The intermediate region, indicated by an oval on the plot, 
 both the electric and magnetic couplings are  $O(1)$ and thus comparable.  Physics here is very complicated. For example,
non-BPS  bound states appear and disappear there, both electric, magnetic and dyonic.
Positronium-like bound state of a monopole with anti-dyon, with magnetic charge zero and electric charge one,
 can mix with electric states: so the
 classification of even the lowest excitations gets quite complicated. What is however remarkable is that, unlike in the 
case of non-supersymmetric theories, supersymmetry
prevent any phase transitions on the oval.  Exact solution tells us 
 how gradual RG flow of the charge leads to a smooth
 transition.
 
  Reaching the  ``magnetic"  patches on the plot, 
one finds that the lightest degrees of freedom are now monopoles. Obviously here they cannot be treated
by semiclassical approach anymore: but one can easily formulate ``dual" magnetic theory of them, Abelian supersymmetric electrodynamics.
The beta function is that region is indeed  in agreement with such {\em weakly coupled magnetic theory} ,
and it has the opposite sign, as indeed required by the Dirac condition.

We will now terminate our discussion of the {\cal N}=2 supersymmetric theories\footnote{We will return to it
few times later. In this chapter we will 
 discussion how fermions are bound to monopoles.}
and return to the real world where (so far?) supersymmetry is absent.

The topic will be qualitative map of the phase diagram of the QCD-like theories with finite temperature and chemical
potential following the so called ``magnetic scenario" 
\cite{Liao:2006ry} aiming at  the description of quark-gluon plasma (QGP) near the phase boundary. The arguments of this paper 
are based on two very generic ``pillars": (i) the direction of the RG flow and (ii) the Dirac quantization condition.
 Their 
combination require the ``magnetic coupling" (denoted by $g$ in this chapter) to run in the direction opposite to
$e$, thus becoming weak in the IR.

\begin{figure}[h]
\begin{center}
\includegraphics[width=10cm]{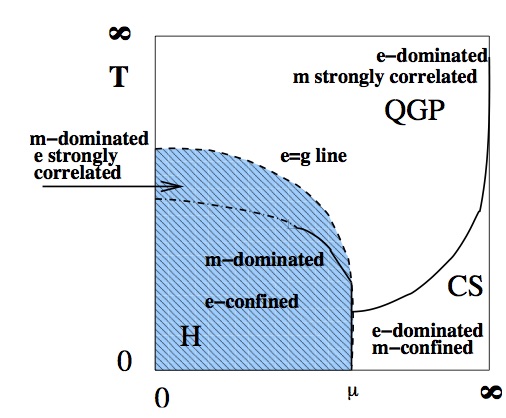}
\caption{A schematic phase diagram on a (``compactified") plane of temperature and baryonic chemical potential $T-\mu$. The (blue) shaded region shows ``magnetically dominated" region $g < e$, which includes the e-confined hadronic phase as well as ``postconfined" part of the QGP domain. Light region includes ``electrically dominated" part of QGP and also color superconductivity (CS) region, which has electrically-charged diquark condensates and therefore obviously m-confined. The dashed line called the ``e=g line" is the line of electric-magnetic equilibrium. The solid lines indicate true phase transitions, while the dash-dotted line is a deconfinement cross-over line.}
\label{fig_magnetic_phase_diagram}
\end{center}
\end{figure}

In Fig.\ref{fig_magnetic_phase_diagram} from that paper one finds the schematic phase diagram on a (``compactified") plane of temperature and baryonic chemical potential $T-\mu$. It resembles upper right part of the plot Fig.\ref{fig_SW_moduli} in that along its periphery the electric coupling is weak and theory is ``$e$-dominated". 
Its physics is weakly coupled QGP, or wQGP for short, describe by the QCD perturbation theory.

The near-circular line marked ``$e=g$ line" is analogous to the oval  in Fig.\ref{fig_SW_moduli}: here 
electric and magnetic quasiparticles become of comparable mass, their interactions are all strong,
and no simple effective description of it is possible. We now know from heavy ion experiments
that in this region QGP is a near-perfect fluid, with small mean free path of all constituents, also known as sQGP.
We will discuss its phenomenology a bit later.

The main difference between the supersymmetric world (and Fig.\ref{fig_SW_moduli}) with the real one (and  Fig.\ref{fig_magnetic_phase_diagram}) is that in the latter case there are phase transitions. The blue shaded
region is the region of confinement. On the plot it is called ``m-dominated" since in this phase
the color-electric field is expelled, into the flux tubes. 

\section{Lattice monopoles in QCD-like theories} \label{sec_mono_latt}
QCD-like theories have gauge fields and fermions, but there are {\em no adjoint scalars}. Therefore, stricktly speaking
there are no 't Hooft-Polyakov magnetic monopoles available in those theories. 

Yet both the pure gauge theory or QCD certainly do possess $some$ quantities which are in adjoint color representation.
One option out is to use $A_4$ (the 4-th component of the adjoint vector field) instead of a scalar. 
At non-zero temperature Lorentz symmetry is broken anyway, and, as already mentioned in the Introduction
, it does have a nonzero VEV (Polyakov loops). This option leads to the so called {\em instanton-dyons}, which we will discuss later in a separate chapter. However, this option would not work outside the Euclidean theory: attempts to return to Minkowski time
would  also transform $A_4\rightarrow i A_4$ which would ruin a would-be monopole solution. 

Another option is to perform lattice simulations with the gauge field, and then look for 
magnetic monopoles in the configurations of the ensemble. A motivation for that, e.g. following from the
example of the  {\cal N}=2 supersymmetric gauge theory, is that in the infrared the effective theory
should be something resembling magnetic Abelian Higgs model.  
't Hooft in particularly argued, that trying to identify physical degrees of freedom from pure gauge ones,
one would inevitably locate monopole-like singularities on the lattice. We already know that
Dirac strings are not observable: but their endpoints are, as 3-cubes in which suddenly magnetic flux
appears\footnote{In one talk I depicted monopoles as some dogs on a leash. While the leash
(the Dirac string)
is unphysical and thus invisible, the corresponding dog's collars (at leach end) can be detected. }. 

Suppose we select some ``composite" (not present in the Lagrangian) adjoint quantity $X$
gauge transforming like $$ X(x) -> U(x) X(x) U^{-1}(X) $$ 
(Examples: a Polyakov line $\hat P$, or some component of the field strength like $F_{12}^{ij}$,
or quark bilinear $\bar\psi^i \psi^j$ but not $A_\mu^{ij}$ as its gauge transform is different.)
One can go to a gauge in which $X$ is diagonal. This separates non-diagonal ``charged" gluons from diagonal
``neutral photons". (We already discussed this for the Polyakov line before.) 

The operator $X$ is local and its eigenvalues depend on the point $x_\mu$. 't Hooft argued that 
the locations at which these  eigenvalues cross lead to singularities of this gauge fixing procedure,
characteristic for the monopoles. The procedure depends on our (rather un-restricted) selection of $X$.

Separating the gauge field into diagonal ``photonic" and non-diagonal ``W" fields (with notation reminding
the Georgi-Glashow model)
\be A_\mu=a_\mu+W_\mu
\ee
one can further define the so called
{\em maximal abelian gauge} (MAG) in which
the local gauge rotations are chosen in such a way as to maximize the following  functional 
\be F_{MAG}=\sum_{\mu,n} Re Tr [U_\mu(n) \tau^3 U^+_\mu(n) \tau^3 ] \ee
%
%
This gauge is widely used by lattice practitioners. Another widely used option is
the so called ``Polyakov gauge" \footnote{ Note however that the Polyakov line is a function of 3-d coordinate,
while the maximal abelian gauge is defined locally for each link in 4d.} in which local value of the Polyakov line
is used to define the Abelian subgroup.

\begin{figure}[h]
\begin{center}
\includegraphics[width=12cm]{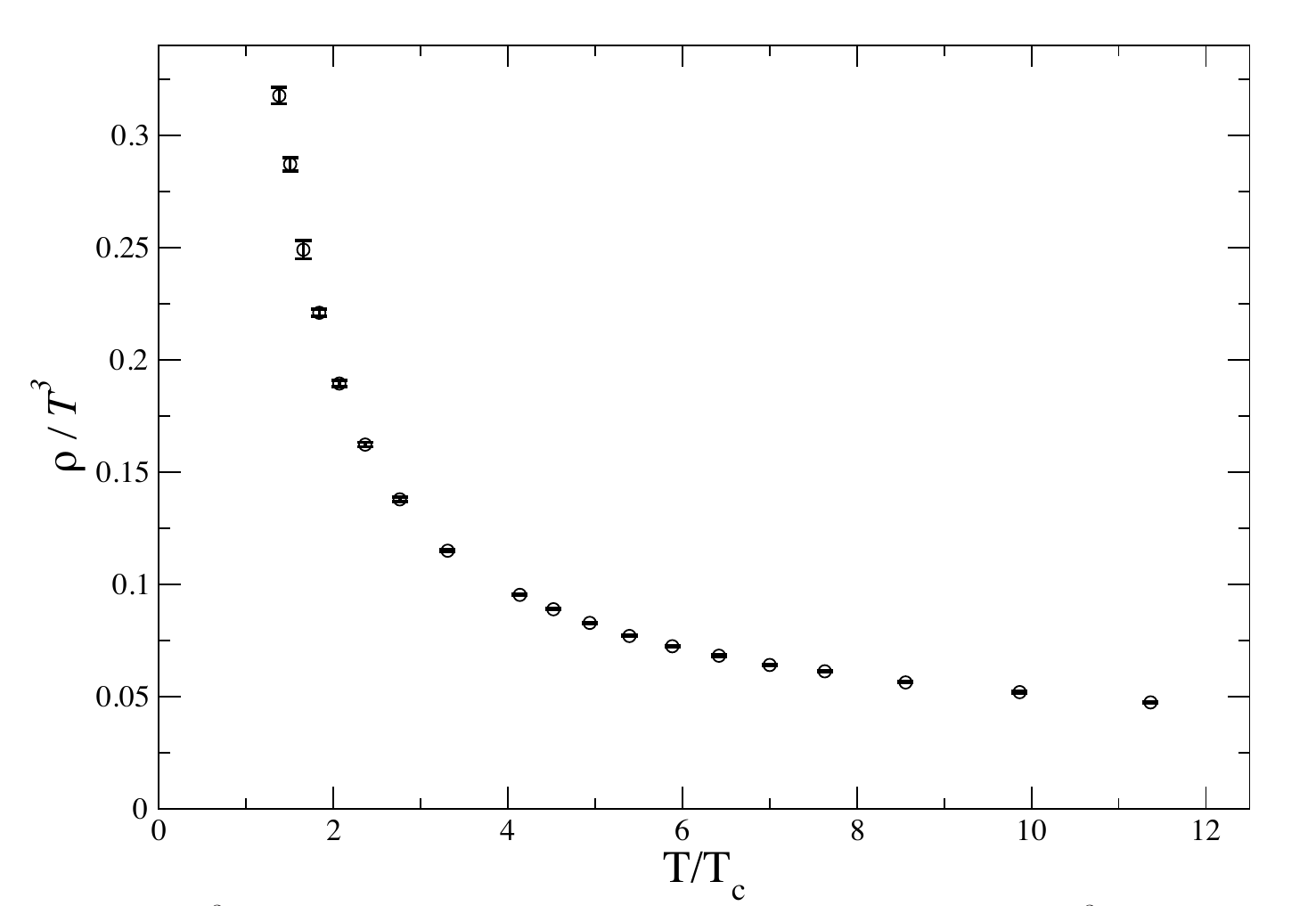}
\caption{The normalized monopole density  $\rho/T^3$ 
for the $SU(2)$ pure gauge theory as a function of the temperature,
in units of the critical temperature $T/T_c$,
above the deconfinement transition.
}
\label{fig_mono_density_T}
\end{center}
\end{figure}

Looking at the ``abelian-projected" part of the field strength (calculated from $a_\mu$ without the commutator) 
one indeed finds what is known as ``lattice monopoles". Some 3-cubes have abelian magnetic fluxes through its surface,
and those cubes make continuous paths in 4-d.
The dimensionless density of monopoles in the $SU(2)$ pure gauge theory at finite temperatures
is shown in Fig.\ref{fig_mono_density_T} from  \cite{D'Alessandro:2007su}.
As expected, the dimensionless density is small at high density but grows rapidly toward
$T_c$. However the best fit to these data are not the inverse power\footnote{This suggests that the effective action of a monopole is not $O(1/g^2)$ 
but rather $O(log(1/g^2))$, in good agreement with the Poisson duality discussion in section \ref{sec_Poisson}.} 
 but the inverse power of the log 
$${\rho(T) \over T^3}\sim {1 \over log\big(T/T_c\big)^{2} }$$

 The question is whether such lattice monopoles {\em do or do not
reflect properties of some real excitations of the system}. 
In fact lattice monopoles -- elementary cubes of size $a^3$ with a magnetic flux through
their surface -- are nothing
but the $endpoints$ of the Dirac strings. Obviously the Dirac string themselves are
gauge artifacts, in different gauges they can go in different directions etc.  
In the continuum limit $a\rightarrow 0$ physical monopoles are expected to have 
finite size, and the question is whether lattice singular monopoles do or do not
correlate with the physical ones. 

It was shown by \cite{Laursen:1987eb} that the lattice monopoles do correlate strongly with such gauge-invariant quantities as squared magnetic field and action. The lattice monopoles
do rotate around electric flux tubes, creating the magnetic ``coil" stabilizing them
\cite{Koma:2003gq}. Eliminating abelian-projected monopoles from lattice configurations does kill confinement \cite{Suzuki:2009xy}, while keeping $only$ monopoles produce nearly
all string tension.   

We will not follow these arguments here, but switch to
another argument, by \cite{Liao:2008jg}, that lattice monopoles are real physical objects:
they do display correlations expected for Coulombic magnetic plasmas.
And, last but not least, we will see that they do respect the famous Dirac condition!

\begin{figure}[h!]
\begin{center}
\includegraphics[width=6.cm]{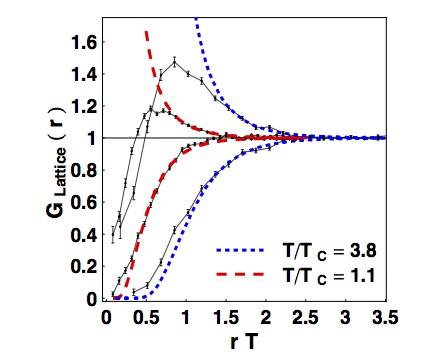}
\includegraphics[width=4.8cm]{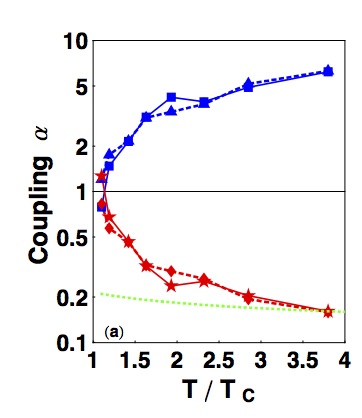}
\caption{
(Left) Monopole-antimonopole (the upper two curves) 
and monopole-monopole (the lower two curves) correlators at $T=1.1 T_c$ (red long dashed) and $3.8 T_c$ (blue dot dashed): points with error bars are lattice data, the dashed lines are used for fits of the coupling strength. 
(Right) The magnetic coupling $g$ (on Log10 scale) versus $T/T_c$ fitted from the monopole-antimonopole (boxes with solid blue curve) and monopole-monopole (triangles with dashed blue curve) correlators. Their inverse, the corresponding to $\alpha_e=e^2/4\pi$ from the Dirac condition, are shown as stars with solid red curve and diamonds with dashed red curve respectively, together with an asymptotic freedom (green dotted) curve
}
\label{fig_magnetic_correlator}
\end{center}
\end{figure}

In Fig.\ref{fig_magnetic_correlator}(left)  from \cite{Liao:2008jg} one see 
two examples of the monopole-monopole and monopole-antimonopole
correlators, as a function of distance between them, calculated from paths of the lattice monopoles by 
D'Alessandro and D'Elia \cite{D'Alessandro:2007su}. Positive correlation for monopole-antimonopole
correspond to attraction, and negative ones for  monopole-monopole pair to repulsion. The shape
of the correlator is exactly what one expects in a Coulomb plasma of charges. The dashed lines
are fits to the part of the correlators where the effect is small and can be treated by a linearized Debye
theory: such fits produce values of the {\em effective magnetic coupling} $g^2/4\pi=\alpha_m$.

In Fig.\ref{fig_magnetic_correlator}(right)  from \cite{Liao:2008jg} the fitted couplings are plotted versus the
temperature. As one can see, they indeed run $opposite$ to the asymptotic freedom, becoming stronger at high $T$.
Furthermore, its reflection  (the bottom of the plot) is in qualitative agreement with the perturbative
asymptotic freedom formula. 

As one can also realize from these plots, by $T=T_c$ magnetic coupling decreases only to become  $\alpha_m\approx 1$, not yet small. This means that the {\em magnetic component of sQGP is also a liquid} -- the title of \cite{Liao:2008jg}.
If it would be otherwise, monopoles would have large mean free paths, in contradiction to heavy ion data!


\chapter{Monopole ensembles}

So far we discussed a single monopole solution: now is the time to consider interacting monopoles and statistical
ensembles of them. This is a complicated subject, and we will approach it in steps, considering:\\
$\bullet$ classical systems made of increasing number of charges and monopoles;\\
$\bullet$  classical plasmas made of many charges and monopoles;\\
$\bullet$ jet quenching and charge-monopole scattering; \\
$\bullet$ quantum-mechanical charge-monopole scattering ;\\
$\bullet$ gluon-monopole scattering ;\\
$\bullet$ kinetic properties of the dual plasmas using the obtained cross sections;\\
$\bullet$ quantum Coulomb Bose gas, with Bose-Einstein condensation (BEC) ;\\

At this point few comments on the term {\em ``plasma"} and other terminology to be used.
Plasma, by definition, is a matter made of particles with a long-range interactions, $1/r$ in three dimensions.
It is mostly used for electromagnetic plasmas made of ionized atoms, or 
quark-gluon plasma in which long-range forces are due to gluon exchange. 
Since monopoles have Coulomb-type Abelian tails, and since they come with
all signs (mono and anti-mono) available, their ensemble fits this definition as well.

Furthermore, we will mostly discuss what we will call a {\em ``dual plasma"},
which includes both charges and monopoles. As we will see,
their mutual interaction is not Coulomb-like, but appears due to Lorentz forces.
While they do not change energy of the particle, they change direction of motion.
Therefore they are not important for thermodynamics of the dual plasmas, but
are central for understanding of their kinetics.

Classical plasmas have one key parameter
\be \Gamma={\langle V \rangle \over T} \ee
characterizing the ratio of the potential and kinetic energies. 

If it is small, $ \Gamma\ll 1$,
the potential is a perturbation, those are called {\em weakly coupled} plasmas, and
their properties are adequately described by the linearized Debye theory.
QGP at high temperatures is an example of this kind.

If $ \Gamma$ is in the range between $\sim 1$ to about $10^3$, the matter is a {\em strongly coupled liquid}. If $ \Gamma$
is even larger, it makes the so called {\em ``ionic solids"}. (The common table salt $Na Cl$ is an example of the latter type.)
 QGP near $T_c$ is a dual plasma with two gamma parameters -- electric and magnetic --
 both in the range 1..10. It is thus an example of strongly coupled dual plasma in a liquid phase.

\section{Classical charge-monopole dynamics}
 Let us start with a very old (19th century) problem in classical electrodynamics:
 an electrically charged
  particle with charge $e$ is moving  in the magnetic field of 
 a static monopole with magnetic coupling $g$.
Classically, one does not need
the vector potential, thus
many subtleties are absent. The interaction
is simply given by the Lorentz force
\be m\ddot{\vec r} = -eg { {[\dot{\vec r}\times \hat{\vec{r}}}] \over r^2}  \ee
where $\times$ indicates the
 vector product and $ \hat{\vec{r}}\equiv\vec r/r$ is the unit vector
along the line connecting both charges.
It is worth noticing that only the product of the couplings ($eg$)  appears.

  \begin{figure}[t]
\begin{center}
\includegraphics[width=8cm]{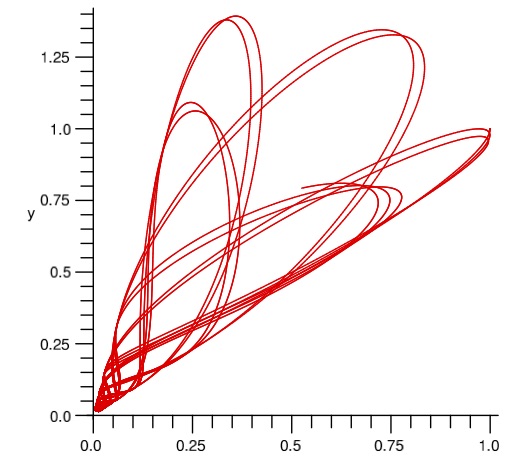}
\caption{Trajectory of a monopole in static electric Coulomb field}
\label{fig_cone1}
\end{center}
\end{figure}

Furthermore, the cross product of the magnetic field of the monopole and
electric field of the charge leads to a nonzero Poynting vector,
which rotates
around $\vec r$: thus, the field 
itself has a nonzero angular momentum. The total conserved
angular momentum  for this problem has two parts
\be 
\vec J= m [\vec r  \times\dot{\vec{r}}]+eg\hat{\vec{r}}.
\ee
The traditional potential scattering only has the first part: therefore,
in that case, 
the motion entirely takes place in the so-called
 ``reaction plane'' normal to $ \vec J$. In the 
charge-monopole problem, however,
the second term restricts the motion to the so-called
 Poincare cone: its
half-opening angle $\pi/2-\xi$ being
\be 
\sin(\xi)={eg\over J} \label{eqn_cone}. 
\ee
Only at large $J$ (large impact parameter scattering) the angle $\xi$
is small and thus the cone opens up, approaching the scattering plane.

Following Ref.~\cite{Boulware:1976tv}, one can project
the motion on the cone to a planar motion, by introducing
\be 
\vec R= {1\over \cos(\xi)} [\vec r - \hat{J} (\vec r\cdot \hat{J})]
\ee 
where the first scale factor is introduced to keep 
the same length for both vectors $\vec R^2=\vec r^{~2}$. 
Now,  
two integrals of motion are
\ba 
\vec J&=& m \vec R \times \dot{\vec{R}}\\
E&=&{m\dot{\vec R}^2 \over 2}- {(eg)^2\over 2m R^2}
\ea
and the problem seems to be reduced to the motion of a particle of mass $m$
in an inverse-square potential.
The scattering angle $\Delta\psi$ for this {\em planar} problem 
can be readily found: it is the variation of $\psi$
as $R$ goes from $\infty$ to its minimum $b$ and back to $\infty$
\be 
\Delta\psi =\pi\left({1\over \cos{\xi}}-1\right) =\pi\left(\sqrt{1+\left({eg\over mvb}\right)^2}-1\right).
\ee 
Note that at large $b$ (small $\xi$) we have $\Delta\psi\sim 1/b^2$,
 as expected for
the inverse-square potential.
Yet  this is $not$ the scattering angle of the original problem,
 because
one has to project the motion back to the Poincar\'e cone.  The
 true scattering angle
-- namely the angle between the initial and final velocities
-- is $\cos{\theta}=-(\hat{v}_i\cdot
\hat{v}_f)$. By relating velocities on the plane and on the cone one can find
it to be
\be 
\left(\cos{\frac{\theta}{2}}\right)^2=(\cos{\xi})^2 \left(\sin{{\pi \over 2 \cos{\xi}}}\right)^2.
\ee
Thus for distant scattering -- small $\xi$ -- one gets 
$\theta\approx 2\xi=2eg/(mvb)$, which is much larger than $\Delta \psi\sim 1/b^2$.
The important lesson that we learn from these formulae is that
the small scattering angle is given by the opening angle
of the cone, rather than the scattering angle in the planar,
 inverse-square effective potential.
Calculating the cross section by $d\sigma=2\pi b db$ one finds that,
at small angles, it is 
\be 
{d\sigma \over d\Omega}=\left({2eg \over mv}\right)^2{1\over  \theta^4}, 
\ee
similar to the Rutherford scattering of two charges.
The difference (apart from different charges) is also the additional
second power of velocity, originating from the Lorentz force.

\begin{figure}[h!]
\begin{center}
\includegraphics[width=14cm]{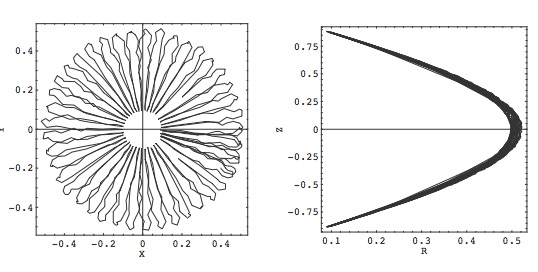}
\caption{Trajectory of monopole motion in a static electric dipole field (with charges at $\pm 1\hat z$) as (left panel)projected on x-y plane and (right panel)projected on R-z plane ($R = \sqrt{x^2 + y^2}$.}
\label{fig_cone2}
\end{center}
\end{figure}

\section{Monopole motion in the field of several charges}

    Liao (my grad student at a time) and myself  \cite{Liao:2006ry} started by investigating curios few-body motion.
Suppose we take two static electric charges, $e$ and $-e$, and put a
monopole into this field. We found that it  can be trapped between them, bouncing from one to the other on a surface which consists of two smoothly connected
Poincare cones with ends on two charges\footnote{So to say, charges can play ping-pong with monopole, without even moving!}. Needless to say, the same happens in a dual settings,
with two static monopoles and a charge bouncing between them\footnote{Note that this last setting
is very similar to famous invention by one of my teachers G.I.Budker, the {\em magnetic bottle}, in which
the magnetic monopole is substituted by a coil with a current. This device traps an electron
and is used a lot in plasma research.}.

\begin{figure}[h]
\begin{center}
\includegraphics[width=5cm]{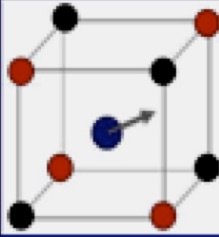}
\includegraphics[width=6cm]{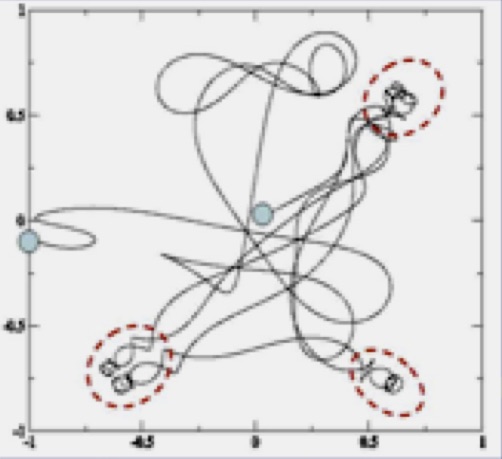}
\caption{A monopole moving in a set of 8 electric charges, with alternating signs.
Left picture shows the setting and right is an example of the monopole path, obtained by
numerically solving classical equation of motion.}
\label{fig_grain_of_solt}
\end{center}
\end{figure}

Our next configuration,  shown in Fig.\ref{fig_grain_of_solt}(a) is called ``a grain of salt",
it consists of 8 charges located at the corners of 3-d cube, with alternating
plus and minus charges. It was found that a monopole can get out of such
``cage", but  with difficulty, suffering multiple collision with
the charges at cube corners, see a typical path in Fig.\ref{fig_grain_of_solt}(b). It happens again,
due to focusing mechanism due to Lorentz force
just described\footnote{I would say the monopole behaves as a proverbial drunkard who cannot go home
because there are several lumpposts on the street, with which he
mysteriously collides all the time. 
}.  


\section{Strongly coupled QGP as a ``dual" plasma}

Discovery  that QGP is a ``perfect liquid", with extremely
 small $viscosity$ had lead to
significant interest in strongly coupled systems in general. 
(It is even now referred to as strong QGP or sQGP for short in literature.)
%
We will now show that electric-magnetic duality, emphasizing the RG flow and transition from weakly coupled electric theory in UV to mixed strongly coupled electric-magnetic sectors around $T_c$, can reproduce the observed unusually high collision rate (small mean free path 
or small viscosity). 

The first one now can be characterized as a mainstream, but it does not
belong to these lectures as it requires a lot of backgrownd knowledge
which the advance undegrad/beginning graduate students do not generally have.
The second one, on the contrary, is rather accessible and pedagogically
fruitful, and so I decided to include it here. 

  In very general terms, we are going to dicuss the behavior of
  quantum systems containing both electrically and magnetically charged 
  particles.  Before we do so, we will start with classical behavior first, and start
  with few particles.

    
    
Classical {\em molecular dynamics} studies of the {\em ``dual plasmas"} 
have been performed by \cite{Liao:2006ry}. What this means is that
 we took few hundred charges, both electric and magnetic, both with positive and negative charges to keep matter neutral, and  solve numerically EOM for all of them. Details are in the original paper: the output are calculation of the diffusion constant
and viscosity, from the corresponding ``Kubo formulae". 
Qualitative conclusion is that each electric charge is trapped by
a cell of magnetic ones around, and each magnetic charge by
electric ones, as described above. Lorentz nature of forces 
between them does decrease particle (diffusion) and momentum 
(viscosity) transport by a lot!

\section{Jet quenching due to jet-monopole scattering} 
 
 The story started by observation \cite{Shuryak:2001me}
 that while theories of jet quenching theory of radiative jet quenching
 \cite{Baier:1996vi}
 did explained the overall magnitude
 of jet quenching, it failed to describe the ellipticity  parameter $v_2(p)$
defined by 
\be E_p{dN \over d^3p}=f(p)\big[ 1+2 v_2(p) cos(2\phi)\big] \ee  
 by a large factor, well beyond the
  experimental accuracy. This issue remained a puzzle till relatively recently:
  to explain it one needed the monopoles! 
 
\begin{figure}[htbp]
\includegraphics[width=5cm]{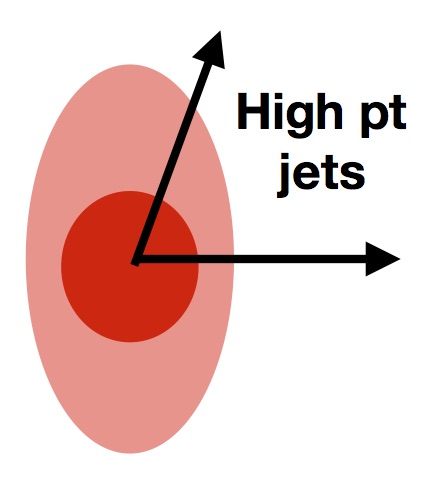}
\includegraphics[width=3cm]{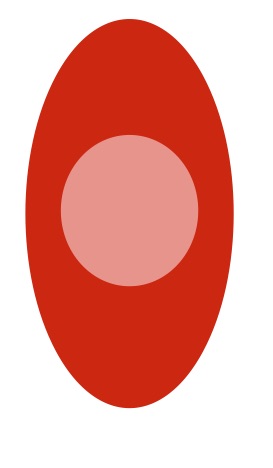}
\caption{ The almond-shape excited system created in non-central
heavy ion collisions: depending on jet direction it finds different amount of matter. Left plot schematically show density distribution,
the right one corresponds to $\hat q$ distribution.}
\label{fig_almond_jet}
\end{figure}

 It was suggested by \cite{Liao:2008dk} that the puzzle can be explained if the
jet quenching be strongly enhanced in the near-$T_c$ matter. 
Fig.\ref{fig_almond_jet} qualitatively explain the idea. The left plot shows
the naive standard picture stemming from the assumption that
the so called jet quenching parameter
\be \hat q\equiv {\langle Q^2\rangle \over length }, \ee
the mean squared transverse momentum kick accumulated by a jet along certain length
of propagation in matter, is simply proportional to matter density. If so, the central darker
region, indicated higher density, of the almond-shape excited system created in non-central
heavy ion collisions, would dominate the jet quenching. However, this darker region
is nearly azimuthally symmetric and contribute little to $v_2(p)$. 

But if the assumption that quenching is proportional to the density $ \hat q\sim n$
is wrong, one may find the way out of a puzzle. For example, if ``by the eyes of the jet"
quenching strength is modified, e.g. as indicated at  the right pictre of Fig.\ref{fig_almond_jet},
then one finds large azimuthal dependence and $v_2(p)$ consistent with experimentally observed values.

But why relatively dilute periphery of the fireball can produce more kicks on the jet
than its central denser part? 
Studies of its possible origin led  
 to {\em jet-monopole scatterings}.  Recall Fig.\ref{fig_mono_density_T}, which shows that the monopole density is large
 only in the vicinity of $T_c$, not in very hot plasma! 
 
\begin{figure}[h!]
\begin{center}
\includegraphics[width=6cm]{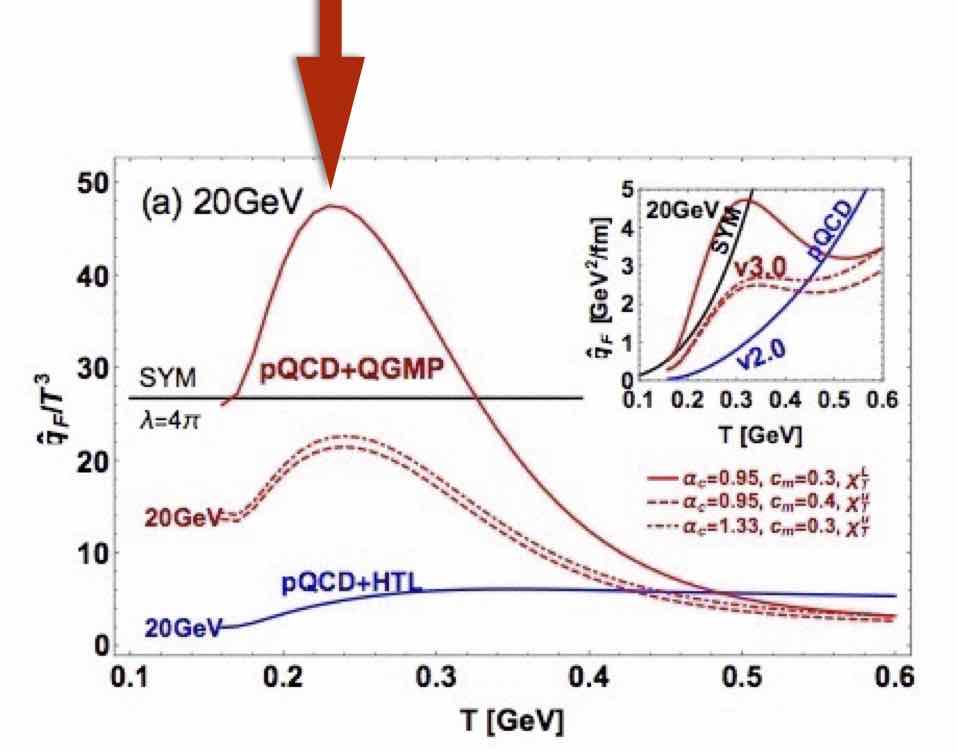}
\includegraphics[width=6cm]{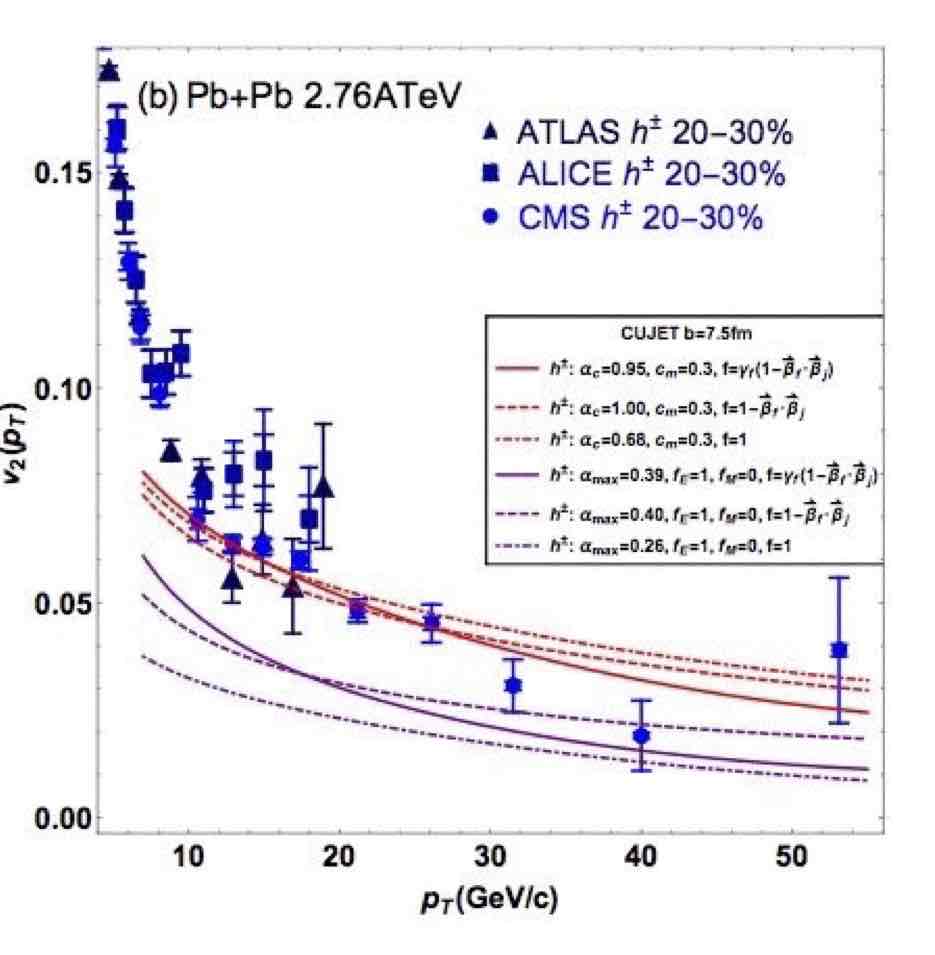}
\caption{Left plot: jet quenching parameter $\hat q/T^3$ versus the temperature $T$.
Red arrow indicate the position of  a peak in the dependence.
Right plot: azimuthal asymmetry parameter $v_2=<cos(2\phi)>$ of jet quenching, as a function of particle transverse momentum $p_T$. 
}
\label{fig_qhat}
\end{center}
\end{figure}

 For detailed discussion of jet quenching see \cite{Xu:2015bbz,Ramamurti:2017zjn}, and we only illustrate
 the points by two plots from this paper Fig.\ref{fig_qhat}. The upper plot shows the so called 
 jet quenching parameter $\hat q$ normalized to $T^3$. The parameter has dimension of the density of scatterers,
 to which it is supposed to be proportional. Its normalization to $T^3$ means that we are looking at the effective number of degrees of freedom. 
 
 The blue line at the bottom of it corresponds to the perturbative QGP: the 
effective number of degrees of freedom is constant at high $T$ and it decreases near $T_c$, being basically proportional to the VEV of the Polyakov line. 
Red lines with a peak at certain temperature near $T_c$ are empirical models which
include scattering on the monopoles. 
Basically those are expected to be proportional to the normalized monopole density shown in FIg.\ref{fig_tension_F_U}(right).

As shown in the lower plot of  Fig.\ref{fig_qhat}, those models
are in much better agreement with the experimental data. 
(Red lines there correspond to those models with a peak, while several purple lines below the data
correspond to the perturbative model without the monopoles.) So, in a way, we have
now some experimental confirmation that a jet-monopole scattering is not only happening, but
even in fact dominates certain observables.

{\bf Summary}: we started this section considering high-$T$ limit, in which
QGP is a weakly coupled plasma, amenable to perturbation theory
(except the suppressed magnetic sector). 
As $T$ decreases and the coupling grows. As a result of it 
the density of magnetic and other topological objects  grows as a power of $T$. By
  $T\approx T_c$, relevent for current experiments, one can view 
  matter as a plasma made of both electric and magnetic particles,
  both with the coupling O(1) and complicated.  Classical and quantum studies of such medium has been successfully done, explaining
  reduction of the mean free path and all forms of kinetic transport.

\section{Quantum-mechanical charge-monopole scattering problem}

The history of {\em quantum monopole-charge scattering problem} is also very interesting, and not widely known, although  in my opinion
it clearly belongs to any good QM textbook.
 There are indications that 
Dirac, after his paper on monopoles in in 1931, discussed it with Tamm and they both tried to solve it, but did not succeed.  It took a long time till it did happened, a year after non-Abelian
monopoles were discovered, by two distinguished teams \cite{Schwinger:1976fr,Boulware:1976tv}.

The problem can be set in two ways. First, one can think of electromagnetic 
monopole and charge, both of size zero, so that only Coulomb fields exist.
Second, we will take 't Hooft-Polyakov monopole solution in full glory (with a charged core and
Coulomb+Higgs tail) and scatter small perturbation of a scalar on it. 

Scattering problem
results are expressed in scattering amplitudes, or scattering phase shifts
as a function of energy. Perhaps, to refresh you memory of the quantum mechanics course,
a couple of general formulae are in order. The wave function 
$\psi=e^{i k z}+ f(\theta) e^{i k r}/r$ contains scattering amplitude $f(\theta)$ which
defines the differential cross section of scattering
$$ d\sigma=| f(\theta) |^2 2\pi sin(\theta) d\theta $$ 
Here are expressions for it in partial waves
$$ f(\theta)={1 \over 2ik} \sum_{l=0}^{\infty} (2l+1)\big(e^{2i\delta_l(k)}-1\big) P_l(cos(\theta)) $$
expanded in Legendre polynomials. General expressions for the total cross section and the transport cross section are
$$ \sigma= \int d\sigma={4\pi \over k^2} \sum_{l=0}^{\infty} (2l+1) sin^2(\delta_l(k)) $$
$$ \sigma_t=\int d\sigma (1-cos(\theta)) $$

Let us think about parameters of the problem.  If particle is point-like, the Coulombic field of a charge has only one 
 dimensionless parameter, the $e^2/\hbar c$ coupling.  It is
scale invariant and cannot provide a scale \footnote{The usual electric Coulomb scattering
does not have finite phase shifts, it leads to logarithmic divergent expression at large $r$
  because of long-range nature of the
Coulomb potential. However, in the charge-monopole problem the force
is the Lorentz force, which is $not$ long range, so the phase shifts are calculable.
}. So, the scattering phase shift cannot depend on collision energy! 

This observation has serious implications. Corrections to total free energy of matter and other 
thermal quantities contain expressions with $\int dk (d\delta_l(k)/dk)... $ which
are in the case considered all zero. As we will see, electric magnetic scattering will not
affect the equation of state of quark-gluon plasma, but be dominant in its kinetic parameters.

Scattering depends on the couplings, the product of electric and magnetic charges. Yet
 according to Dirac condition this product must be an integer 
 \be {eg \over 4\pi}= n=integer. \ee
This integer is {\em the only input} of the problem we try to solve, provided 
  monopole is treated as a point charge.

Let me just give the answer for the scattering phases.
Indeed, they depend on the total angular momentum $j$ and the  integer $n$
from the Dirac condition.  Note that for point charges they
do not depend on energy. 
The expression is where $t'$ is to be found from the r.h.s. of the eqn
\be \delta_t= {\pi \over 2} t', \,\,\,\, t'(t'+1)=t(t+1) - n^2 \ee
The first term in the r.h.s. $t(t+1)$ is the total angular momentum squared, the last term is
subtracted angular momentum of the field $n^2$. The $t'$ thus has the meaning of {\em angular momentum of the particle}. Here comes a shocking fact:
 while $j$ and $n$ in the r.h.s. are both integers\footnote{The second is due to Dirac condition.}, their combination $t'$
entering the scattering phase,
is $not$ an integer. One can easily see it by solving the quadratic equation for it:
it is a square root of an integer.  That is why a quite nontrivial angular distribution of charge-monopole scattering appears!

 Recall that the usual angular basis used $Y_{l m}(\theta,\phi)$ at large quantum numbers $l,m\ll 1$ is in fact planar. Yet in classical charge-monopole problem  the motion happens on the 
{\em Poincare cone} rather than on a plane!
 So, one has to rethink the setting  and select more complicated angular functions, conical in the classical limit. 
 
 The  wave function 
 in spherical coordinates is as usual
 a sum of products of certain $r$-
dependent radial functions, times the angular functions. The former
basically  follow from the
 inverse-square law potential and thus are easily solved
in Bessel functions: they
correspond to  the auxiliary planar projection of the
classical problem of the previous subsection. The nontrivial
part happens to be
 in the unusual angular functions. Before introducing those,
let us hint why the usual set of angular harmonics
$Y_{lm}(\theta,\varphi)$ should $not$ to be used.
 The reason is that their classical limit -- for large
values of the indices  $m\approx l \gg 1$ -- corresponds to flat
planar motion near the $z=0$ plane. Nevertheless, we
have already learned  from the classical limit
 to expect the motion 
to be concentrated around the Poincare
cone instead! 

The functions that we need (for example in the scalar sector)
must satisfy the following set of conditions
  \be
 \left\{ 
 \begin{array}{l}  
 \vec{T}^2 \\  T_3 \\  \vec{I}^2 \\  (\hat{r}\cdot \vec{I} ) \end{array}
 \right\}  
 \phi^{mn}_{ti}(\theta,\varphi)=
  \left\{  \begin{array}{l}  
 t(t+1) \\  m \\  i(i+1) \\  n \end{array}
 \right\}  
 \phi^{mn}_{ti}(\theta,\varphi).  
 \label{eigenvalues}
  \ee
  where $\vec{T}$ is the total angular momentum $\vec{T}=\vec{L}+\vec{I}$, and $\vec{I}$ is the isospin.
  The unusual condition in the above set is the last one, since the vector $\vec{I}$ must
  be projected
  to the (space-dependent) radial unit vector. The functions satisfying
  this requirement \cite{Boulware:1976tv,Schwinger:1976fr} will be introduced below.   Here 
  it is enough to explore  
the large $l,n$ limit of the $D$-functions involved. The result
\be D^l_{nl}\sim e^{i(l-n)\varphi} \exp[-l (\theta-\theta^*)^2/2]\ee
where $\cos(\theta^*)=n/l$, shows that they indeed correspond to the Poincar\'e cone.

Now, we move to second setting, discussing small scalar perturbations scattered on the 't Hooft-Polyakov monopole soliton, in
 the Georgi-Glashow framework
 .
We introduce a total angular momentum operator $\vec{T}$, which is the sum of the orbital angular momentum
$\vec{L}$ and the isotopic spin $\vec{I}$:
\be
\vec{T}=-i\vec{r}\times\vec{\nabla}+\vec{I}
\ee
with $(I^a)_{bc}=-i\epsilon_{abc}$.
In terms of these operators, the wave equation for the scalar fluctuations
can be written in the form 
\ba
&&
\!\!\!\!\!\!\!\!\!\!\!\!\!\!\!\!\!\!\!
\left[\frac{\partial^2}{\partial r^2}-\frac{2}{r}\frac{\partial}{\partial r}-\frac{\left(\vec{T}^2
-\left(\hat{r}\cdot\vec{I}\right)^2\right)}{r^2}-
\partial_0^2\right]\vec{\chi}+\frac{2K(\xi)\left(\vec{I}\cdot\vec{T}-\left(\hat{r}\cdot\vec{I}\right)^2
\right)}{r^2}\vec{\chi}
\label{eea}
\\
&-&\frac{K(\xi)^2\left[\vec{I}^2-\left(\hat{r}\cdot\vec{I}\right)^2\right]}{r^2}\vec{\chi}
-\lambda\left[2\frac{\vec{r}\cdot\vec{\chi}}{e^2r^4}H(\xi)^2\vec{r}+\left(
\frac{H(\xi)^2}{e^2r^2}-v^2\right)\vec{\chi}\right]=0.
\nonumber
\ea
The term which is proportional to $K[\xi]$ induces charge-exchange reactions.

We can define a simultaneous eigenfunction $\phi_{ti}^{mn}(\hat{r})_a$ of the commuting operators 
$\vec{T}^2$, $T_3$, $\vec{I}$ and $\hat{r}\cdot\vec{I}$ (see eq. (\ref{eigenvalues})).
This function depends only on the angular variables specified by $\hat{r}$. A solution to the
equation (a specific partial wave) 
can be written as the product of the angular function $\phi_{ti}^{mn}(\hat{r})$
and a radial function $S_t^n(r)$
\be
\chi(\vec{r})_a=\phi_{ti}^{mn}(\hat{r})_aS_{t}^{n}(r).
\ee
The angular function $\phi_{ti}^{mn}(\hat{r})_a$ is peculiar because the operator $\vec{I}$ is projected
along $\hat{r}$.
Therefore, the angular function must be rotated, from the standard cartesian frame, to a ``radial" frame.
This construction can be achieved by making use of a {\it spatially dependent} unitary matrix
which rotates $\hat{r}\cdot \vec{I}$ into $I_3$:
\be
U(-\varphi,-\theta,\varphi)=e^{-i\varphi I_3}e^{-i\theta I_2}e^{i\varphi I_3}.
\ee
We therefore have
\ba
\hat{r}\cdot\vec{I}\,\,U(-\varphi,-\theta,\varphi)&=&U(-\varphi,-\theta,\varphi)I_3
\nonumber\\
\vec{T}\,\,U(-\varphi,-\theta,\varphi)&=&U(-\varphi,-\theta,\varphi)\vec{\mathcal{T}}
\ea
where
\be
\vec{\mathcal{T}}=-\vec{r}\times\left(i\vec{\nabla}+e\vec{\mathcal{A}}I_3\right)+\hat{r}I_3
\label{jj}
\ee
and
\be
e\vec{\mathcal{A}}=\frac{\hat{r}\times\hat{z}}{r+z}.
\ee
Eqs. (\ref{eigenvalues}) are satisfied by the following function
\be
\phi_{ti}^{mn}(\hat{r})_a=(U(-\varphi,-\theta,\varphi)\chi_{i}^{n})_a\mathcal{D}(\hat{r})
\ee
where $(\chi_{i}^{n})_a$ is an eigenvector of $I_3$ in the cartesian basis
\be
I_3 (\chi_{i}^{n})_a=n(\chi_{i}^{n})_a
\ee
and the function $\mathcal{D}(\hat{r})$ obeys
  \be
 \left\{ 
 \begin{array}{l}  
 \vec{\mathcal{T}}^2 \\ \mathcal{ T}_3  \end{array}
 \right\}  
\mathcal{D}(\hat{r})=
  \left\{  \begin{array}{l}  
 t(t+1) \\  m \end{array}
 \right\}  
 \mathcal{D}(\hat{r})  
 \label{eigenvalues2}
  \ee
where $I_3$ in eq. (\ref{jj}) is now replaced by its eigenvalue, $n$. We have
\be
\mathcal{D}(\hat{r})=\mathcal{D}^{(t)}_{nm}(-\varphi,\theta,\varphi)=\langle t,n|e^{-i\varphi T_3}
e^{i\theta T_2} e^{i\varphi T_3}|t,m\rangle.
\ee
We can write the function $\phi_{ti}^{mn}(\hat{r})_a$ by making use of the following expansion
\ba
\phi_{ti}^{mn}(\hat{r})_a&=&\sum_{n'}(\chi_{i}^{n'})_a(-1)^{n-n'}\times
\label{solution}\\
&\times&\sum_{l=|t-i|}^{t+i}\langle i,-n,t,n|l,0\rangle
\mathcal{D}_{0,m-n'}^{(l)}(-\varphi,\theta,\varphi)\langle l,m-n'|i,-n',t,m\rangle
\nonumber
\ea
where
\be
\mathcal{D}_{0,m}^{(l)}(\alpha,\beta,\gamma)=\sqrt{\frac{4\pi}{2l+1}}Y_{l}^{m}(\beta,\gamma).
\ee
These functions are normalized in the following way
\be
\int_{0}^{2\pi}d\varphi\int_{-1}^{1}d\cos\theta\phi_{t_1i}^{m_1n_1}(\theta,\varphi)^\dagger\phi_{t_2i}^{m_2n_2}(\theta,\varphi)=\frac{4\pi}{2t_1+1}\delta_{t_1,t_2}\delta_{m_1,m_2}\delta_{n_1,n_2}.
\ee
Since the above function is a polynomial in $x/r,~y/r,~z/r$, the function in eq. (\ref{solution})
is analytic everywhere. This clearly shows that there are compensating singularities in 
$U(-\varphi,-\theta,\varphi)$ and $\mathcal{D}(\hat{r})$.
Since the equation (\ref{eea}) in its most general form admits mixing between particles of different charge,
in order to get the equation for the radial functions $S_t^n$ we have to write the order 1 fluctuations of the field around the classical solutions as a superposition of the functions describing 
a particle with definite charge $n$:
\be
\chi(\vec{r},x_0)_a=\sum_{n=-1,0,1}\mathrm{e}^{i\omega x_0}\frac{S_{t}^{n}(r)}{r}\phi_{ti}^{mn}(\theta,\varphi)_a
\ee
where the three functions $S_{t}^{n}(r)$ correspond to the three physical fluctuations with
charge $n=0,~\pm1$. We plug the above expansion for $\vec{\chi}(\vec{r})$
into Eq. (\ref{eea}); through this procedure we obtain the following system of equations for the radial functions
\begin{subequations}
\begin{align}
\nonumber\\
&S_{t}^{0''}(\xi)-\left(\frac{t(t+1)}{\xi^2}+2\frac{K(\xi)^2}{\xi^2}-\omega^2\right)S_{t}^{0}(\xi)-\lambda
\left(3\frac{H\left(\xi\right)^2}{\xi^2}-1\right)S_{t}^{0}(\xi)
\vspace{.3cm}
\nonumber\\
&+\frac{\sqrt{2t(t+1)}}{\xi^2}K(\xi)\left(S_{t}^{1}(\xi)
+S_{t}^{-1}(\xi)\right)=0
\\
\nonumber\\
&S_{t}^{1''}(\xi)-\left(\frac{t(t+1)-1}{\xi^2}+\frac{K(\xi)^2}{\xi^2}-\omega^2\right)S_{t}^{1}(\xi)
-\lambda\left(\frac{H\left(\xi\right)^2}{\xi^2}-1\right)S_{t}^{1}(\xi)
\vspace{.3cm}
\nonumber\\
&+\frac{\sqrt{2t(t+1)}}{\xi^2}K(\xi)S_{t}^{0}(\xi)=0
\\
\nonumber\\
&S_{t}^{-1''}(\xi)-\left(\frac{t(t+1)-1}{\xi^2}+\frac{K(\xi)^2}{\xi^2}-\omega^2\right)S_{t}^{-1}(\xi)
-\lambda\left(\frac{H\left(\xi\right)^2}{\xi^2}-1\right)S_{t}^{-1}(\xi)
\vspace{.3cm}
\nonumber\\
&+\frac{\sqrt{2t(t+1)}}{\xi^2}K(\xi)S_{t}^{0}(\xi)=0.
\\
\nonumber
\label{scalarsystem}
\end{align}
\end{subequations}
where we have introduced the dimensionless variable $\xi=evr$. At the end of this section we will
discuss how to fix the scale and go to physical units.
As it is evident from the above system, a mixing occurs in the monopole core between different charges:
the term $\propto K(\xi)$ involves a mixing between charges that differ by one unit.
The above system of equations has been obtained in the most general case, for generic angular momentum $t$. Nevertheless, we have to keep in mind that there are some restrictions due to
the following requirement:
\be
\hat{r}\cdot\vec{T}=\hat{r}\cdot\vec{L}+\hat{r}\cdot\vec{I}=\hat{r}\cdot\vec{I}=n.
\ee
For this reason, in the case $t=0$ only the $n=0$ scalar fluctuation is allowed. 
The equation for this special case and its solution will be discussed in the following.

For $t>0$, the system of equations (\ref{scalarsystem}) is difficult to solve, due to the mixing
between the different radial functions. This mixing is due to the charge-exchange
reactions that can occur inside the monopole core. If the monopole
core is small (we have seen in Section 2.1 that lattice-based estimates for the monopole size
give $r_m\simeq0.15$ fm)
we can neglect the charge-exchange reactions. 
This corresponds to considering the above system of equations (\ref{scalarsystem}) in the limit
\be
K(\xi)\rightarrow 0~~~~~~~~~~~~~~~H(\xi)\rightarrow\xi.
\ee
In this 
approximation, it reduces to\\
\begin{subequations}
\begin{align}
&S_{t}^{0''}(\xi)-\left(\frac{t(t+1)}{\xi^2}-\omega^2+2\lambda\right)S_{t}^{0}(\xi)=0
\\
\nonumber\\
&S_{t}^{1''}(\xi)-\left(\frac{t(t+1)-1}{\xi^2}-\omega^2\right)S_{t}^{1}(\xi)=0
\\
\nonumber\\
&S_{t}^{-1''}(\xi)-\left(\frac{t(t+1)-1}{\xi^2}-\omega^2\right)S_{t}^{-1}(\xi)=0.
\end{align}
\label{scalarsystem2}
\end{subequations}\\
From the above system it is clear that we can identify the radial functions with 
spherical Bessel functions having index $t'$, which
is {\em the positive root of quadratic equation}
\be
t'(t'+1)=t(t+1)-n^2.
\ee
The r.h.s. is total angular momentum squared minus that of the field: both are integers.
But $t'$, representing angular momentum of the monopole is $not$ an integer number!

This makes scattering rather unusual. 
Namely, in the limit of small monopole core we have $S_{t}^{n}(r)\rightarrow j_{t'}(kr)$.
The corresponding scattering phase will be
$\delta_{t'}=t'\pi/2$, independent of the energy of the incoming particle.

\begin{figure}[t]
\begin{center}
\includegraphics{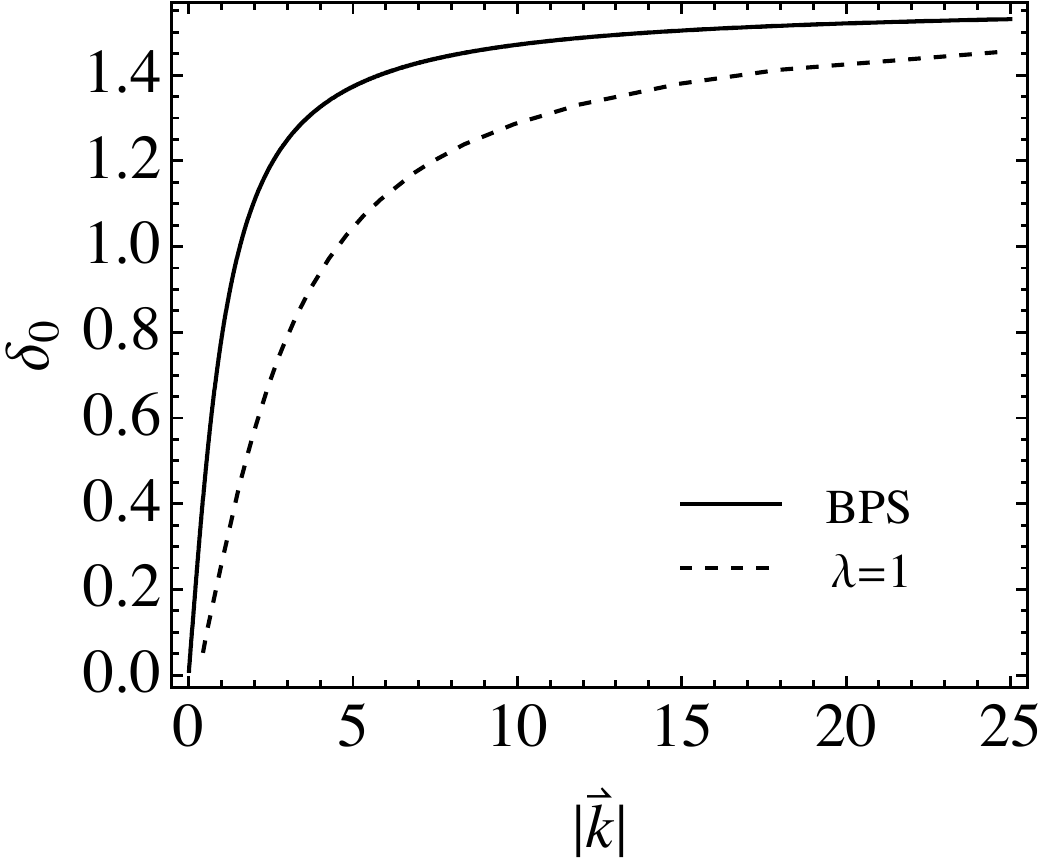}
\caption{Scattering phase $\delta_0$ as a function of $|\vec{k}|=\sqrt{\omega^2-2 \lambda}$. $\delta_0$
is obtained by solving Eq. (\ref{eq0}) with the boundary conditions (\ref{boundary}). The continuous
line corresponds to the BPS limit ($\lambda=0$), while the dashed line corresponds to $\lambda=1$.}
\label{classical}
\end{center}
\end{figure}

For $t=0$ we have only one fluctuation allowed, namely the one having zero-charge: $S_0^0(\xi)$.
It obeys the following
equation
\be
S_{0}^{0''}(\xi)-\left(2\frac{K(\xi)^2}{\xi^2}\right)S_{0}^{0}(\xi)-\lambda
\left(3\frac{H\left(\xi\right)^2}{\xi^2}-1\right)S_{0}^{0}(\xi)=
-\omega^2S_{0}^{0}\left(\xi\right).
\label{eq0}
\ee
In this case, there is no Coulomb potential of the form $1/\xi^2$, which is obvious since
a charge-neutral particle does not feel the Lorentz force. The scattering in this case is entirely
due to the monopole core.
We can solve the above equation numerically, thus obtaining the scattering phase as a function of the
energy of the incoming particle, by imposing the following boundary conditions
\ba
S_{0}^{0}(\xi)&=&\sin\left[\xi\sqrt{\omega^2-2 \lambda}-\delta_0\right]~~~~~~~~~~~~~~~~~~~~~~~~
\xi\rightarrow\infty
\nonumber\\
S_{0}^{0'}(\xi)&=&\sqrt{\omega^2-2 \lambda}\cos\left[\xi\sqrt{\omega^2-2 \lambda}-\delta_0\right]~~~~~~~~~~~~\xi\rightarrow\infty.
\label{boundary}
\ea
 Fig. \ref{classical} shows
the classical solutions
$H(\xi)$ and $K(\xi)$, both in the BPS limit and for $\lambda=1$.

At this point we need to fix the scale in our problem, and to estimate the
scattering length in physical units. In order to do it, we have to connect 
 the physical size of the core of the monopole, $r_m$
 to the dimensionless units that we used in the Georgi-Glashow model.
We recall that we obtained $r_m$ through the width at half height of the ``non-Abelianicity",
which is defined as the square of the non-abelian components of the gauge potential. Therefore,
it is natural to 
fix the scale by imposing that, in the Georgi-Glashow model, $r_m$ coincides with $\xi_m$,
the width at half height of the function $K(\xi)^2$. By looking at 
Fig. \ref{classical}, we can see that, in the BPS limit, we have $\xi_m=1.49$, while for $\lambda=1$ 
we have $\xi_m=0.87$. Thus, $ev=9.94$ fm$^{-1}$ in the BPS limit, while $ev=5.8$ fm$^{-1}$
for $\lambda=1$.

%
%

\section{Quark and gluon scattering on monopoles and viscosity of QGP}

In QGP with a monopole one need to find scattering amplitudes of quarks and gluons. The 
former problem has been solved by another distinguished team \cite{Kazama:1976fm},
but the
latter problem turned out to be especially tedious, because
orbital angular momentum, spin and color-spin of the gluon are all mixed in.
 It has been solved by  \cite{Ratti:2008jz}.  
 
 In the case of vector particles, the generalized angular momentum $\vec{J}$ is made up of three
components: the orbital angular momentum $\vec{L}$, the isotopic spin $\vec{I}$ and the spin 
$\vec{S}$:
\be
\vec{J}=\vec{L}+\vec{I}+\vec{S}=\vec{T}+\vec{S}.
\ee
There are generally two different ways of composing three vectors, depending on 
the two vectors to be added first.
The monopole vector spherical harmonics are eigenfunctions of $\vec{J}^2$ and $J_3$. Due
to the following relation
\be
\hat{r}\cdot\vec{J}=\hat{r}\cdot\vec{L}+\hat{r}\cdot\vec{I}+\hat{r}\cdot\vec{S}=\hat{r}\cdot\vec{I}
+\hat{r}\cdot\vec{S}=n+\sigma,
\label{constraint}
\ee
the allowed values of the total angular momentum quantum number $j$ are $|n|-1,~|n|,...$
except in the case of $n=0$, where $|n|-1$ is absent. 
There are different ways of building the monopole vector spherical harmonics; for example,
they can be constructed by making use of the standard Clebsch-Gordan technique
of addition of momenta.

Another possibility, which we will adopt here, is to build the vector harmonics with $j\geq n$
by applying vector operators to the scalar harmonics, as we will see
in the following \cite{Weinberg:1993sg}. By definition, these harmonics can be introduced as eigenfunctions of the
operator of the radial component of the spin, $\vec{S}\cdot\hat{r}$. We will show that this is
a very useful choice for the ``hedgehog" configuration we are working in:
in fact, there is a natural separation between radial and transverse vectors, making this choice
particularly useful for studying spherically symmetric problems.
The vector harmonics with the minimum allowed angular momentum $j=|n|-1$
cannot be constructed in this way and must be treated specially.
As already mentioned, there is more than one way to obtain a given value of $j$, and thus several
multiplets of harmonics with the same total angular momentum. In the following, we will
classify the multiplets by the eigenvalue of $\hat{r}\cdot\vec{S}=\sigma$. In general,
$\sigma=0,\pm1$, but it is further restricted by the requirement (\ref{constraint}), which
implies that $n+\sigma$ lies in the range $-j$ to $j$. This gives
\begin{itemize}
\item{for $j=0$:
\begin{itemize}
\item{$n=0$ and $\sigma=0$}
\item{$n=1$ and $\sigma=-1$}
\item{$n=-1$ and $\sigma=1$}
\end{itemize}}
\item{for $j=1$ all combinations are allowed, except:
\begin{itemize}
\item{$n=-1$ and $\sigma=-1$}
\item{$n=1$ and $\sigma=1$}
\end{itemize}
}
\end{itemize}
We denote the vector harmonics by $\Phi_{j,n}^{m,\sigma}(\theta,\varphi)_{ai}$. They obey the following eigenvalue equations
\be
 \left\{ 
 \begin{array}{l}  
 \vec{J}^2 \\  J_3 \\  (\hat{r}\cdot\vec{I}) \\  (\hat{r}\cdot \vec{S} ) \end{array}
 \right\}  
 \Phi^{m,\sigma}_{j,n}(\theta,\varphi)_{ai}=
  \left\{  \begin{array}{l}  
 j(j+1) \\  m \\  n \\  \sigma \end{array}
 \right\}  
 \Phi^{m,\sigma}_{j,n}(\theta,\varphi)_{ai}.
 \label{eigenvalues3}
  \ee
In the case {$j\geq 2$}
  all the possible combinations of $n$ and $\sigma$ are allowed.

\section{Transport coefficients from binary quantum scattering}

 In the gas approximation, all transport coefficients
are inversely proportional to the so-called 
transport cross section, normally defined as
\be \sigma_t=\int  (1-\cos\theta) d\sigma\ee
where $\theta$ is the scattering angle.
While the factor in brackets vanishes at
small angles, the Rutherford singularity in the cross section,
for any charged particle,
leads to its logarithmic divergence. 
Since we will be comparing the gluon scattering on monopoles with that
on gluons, let us first introduce those benchmarks, namely
 the well-known (lowest 
order) QCD processes, the $gg$ and $\bar q q$ scatterings:
\be {d\sigma_{\bar q q} \over dt}={e^4  \over  36 \pi} 
\left( { s^4+t^4+u^4 \over s^2 t^2 u^2 } -{8 \over 3 t u  }
\right)  \ee
\be   
{d\sigma_{gg} \over dt}={9 e^4  \over  128 \pi} 
{(s^4+t^4+u^4)(s^2+t^2+u^2) \over s^4 t^2 u^2} 
\ee
where (we remind) the electric coupling is related to
$\alpha_s$ as usual: $e^2/4\pi=\alpha_s$.

While for non-identical particles the transport
cross section is simply given by the cross section weighted 
by momentum transport $t\sim (1-z)$, for identical ones such as $gg$
one needs to introduce the additional factor $(1+z)/2$ 
in order to suppress
backward scattering as well. 
 The
integrated transport cross sections themselves are given by
\be \sigma^t_{gg}   = {3 e^4\over 320\pi s}\left(105 \log(3)-16
+30 \log{\left(4 \over \theta_{min}^2\right)} \right)  
\label{sigmatgg}
\ee
\be \sigma^t_{\bar q q} = {e^4 \over  54\pi s}
\left( 4+7 \log(3)+ 3 \log{\left(4 \over \theta_{min}^2 \right)} \right) \ee
where the smallest scattering angle can be related to the
(electric) screening mass by $\theta_{min}^2=2*M_D^2/s$.
Note that the forward scattering log in the $gg$ case 
has a coefficient which is roughly four times larger, 
as a consequence of the gluon color
being roughly twice that of a fundamental quark.
Note also that the $gg$ scattering is significantly larger at large
angles, as compared to the $\bar q q$ scattering.

\begin{figure}[t]
\begin{center}
\includegraphics[width=4cm]{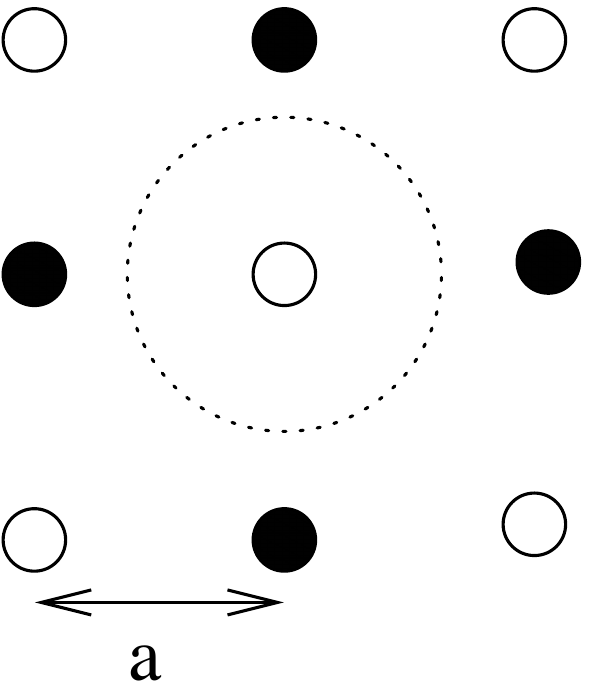}
\caption{A charge scattering on a 2-dimensional array of 
correlated monopoles (open points) and antimonopoles (closed points).
The dotted circle indicates a region of impact parameters for
which scattering on a single monopole is a reasonable approximation.}
\label{fig_monos_2d}
\end{center}
\end{figure}

We explained (already  in the introduction) that the
charge-monopole scattering is Rutherford-like 
 at small angles: this comes 
 from harmonics with
large  angular momenta (large impact parameters).
However, in matter there is a finite density of
 monopoles, so the issue of the scattering should be
reconsidered.
 A sketch of the setting,
assuming strong correlation of monopoles into a crystal-like
structure, is shown
 in Fig. \ref{fig_monos_2d}.  A
 ``sphere of influence of one monopole''(the dotted circle)
gives the maximal impact parameter
to be used. 

As a result,
 the impact parameter is limited from above by some $b_{max}$, which implies that
 only a finite number of partial waves should be included.
The
range of partial waves to be included in the scattering amplitude
can be estimated as follows
\be j_{max}=\langle p_x\rangle n_{mono}^{-1/3}/2\sim aT\sim  1/e^2(T) \sim \log(T)\ee
Since at asymptotically high $T$ the monopole
 density $n_{mono}\sim (e^2 T)^3$ is small compared to
the density of quarks and gluons $\sim T^3$,
 $j_{max}$ asymptotically grows  logarithmically with $T$. So,
only in the academic limit
$T\rightarrow \infty$ one gets $j_{max}\rightarrow \infty$ 
and the usual free-space
scattering amplitudes calculated in \cite{Boulware:1976tv} 
with all partial
waves are recovered.
However, in reality
 we have to recalculate the scattering,
retaining only several lowest partial waves from the sum.
As we will see, this dramatically changes the angular distribution,
by strongly  depleting scattering at small angles and enhancing
scattering backwards. 

\begin{figure}
\includegraphics[width=6cm]{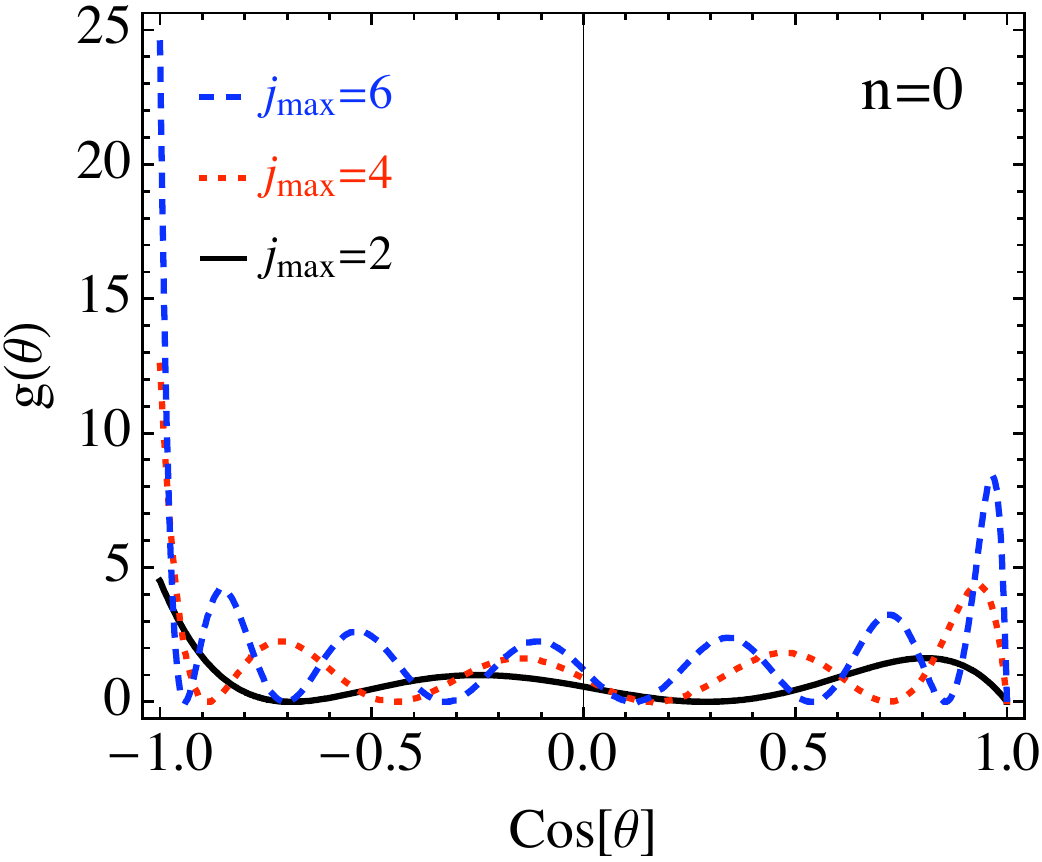}
\includegraphics[width=6cm]{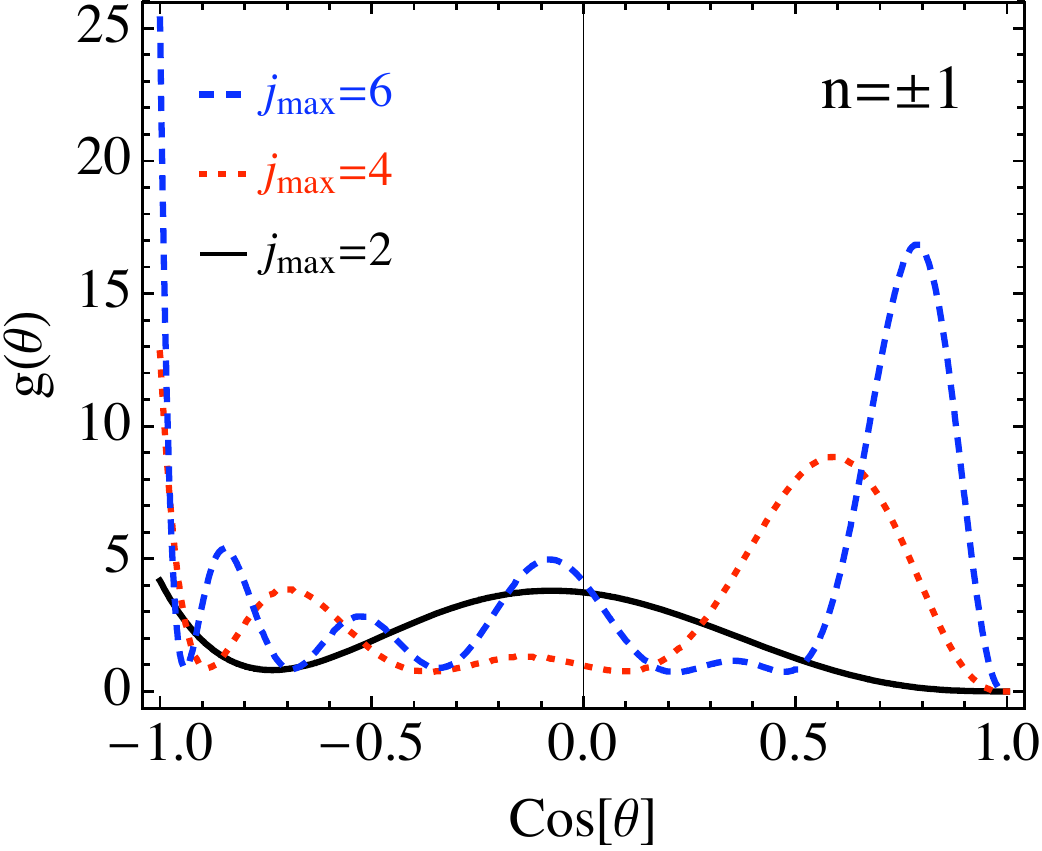}
\caption{Integrand of the trasport cross section 
$g(\theta)=(1-\cos{\theta})|f(\theta)|^2$
with only 2, 4 and 6 lowest partial waves
included, for a scalar particle with $n=0$ (left) and
$n=\pm1$ (right). The curves can be easily recognized by higher $j_{max}$
having more oscillations. }
\label{fig_246}
\end{figure}

\begin{figure}
\begin{center}
\includegraphics[width=7cm]{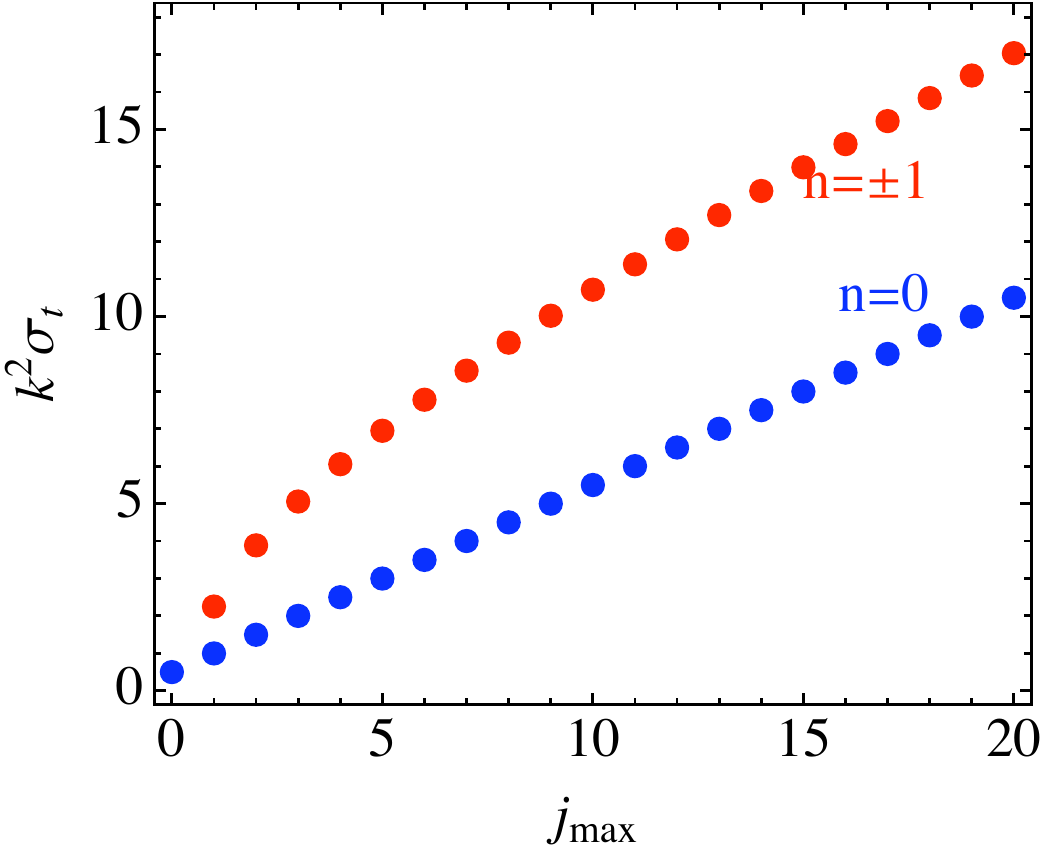}
\caption{Normalized transport cross section as a function of cutoff
in maximal harmonics retained, for $n=0,1$.}
\label{fig_sigma_trans}
\end{center}
\end{figure}

The
integrands of the transport cross section
$(1-\cos{\theta})|f(\theta)|^2$ are shown in Fig. \ref{fig_246} for
$n=0,~j_{max}=2,4,6$ (left panel),
$n=\pm1,~j_{max}=2,4,6$ (right panel).
One can see how much their angular distribution 
is distorted. Strong oscillations of this function occur because
we use a sharp cutoff for the higher harmonics, which represents
diffraction of a sharp edge. This edge in reality does not exist
and can be removed by any smooth edge prescription (for example a gaussian weight).
However, we further found that the transport cross section itself
is rather insensitive to these oscillations, and thus there is no
need in smoothening the scattering amplitude. The transport cross section
as a function of $j_{max}$ is shown in Fig. \ref{fig_sigma_trans}:
it is large and smoothly rising with the cutoff.  

 where one can also
 find the resulting angular distribution and its contribution to transport
 cross section. 
 The upper plot shows angular distribution of the scattering amplitude, with
 a cutoff at certain total angular momentum indicated on the plot. 
 Look for a example at the blue dashed line with the largest $j_{max}=6$: one can see
 a forward peak near zero angle, but also a backward peak in the opposite direction. This backward scattering
 is dominating the transport cross section. 

\begin{figure}
\includegraphics[width=6cm]{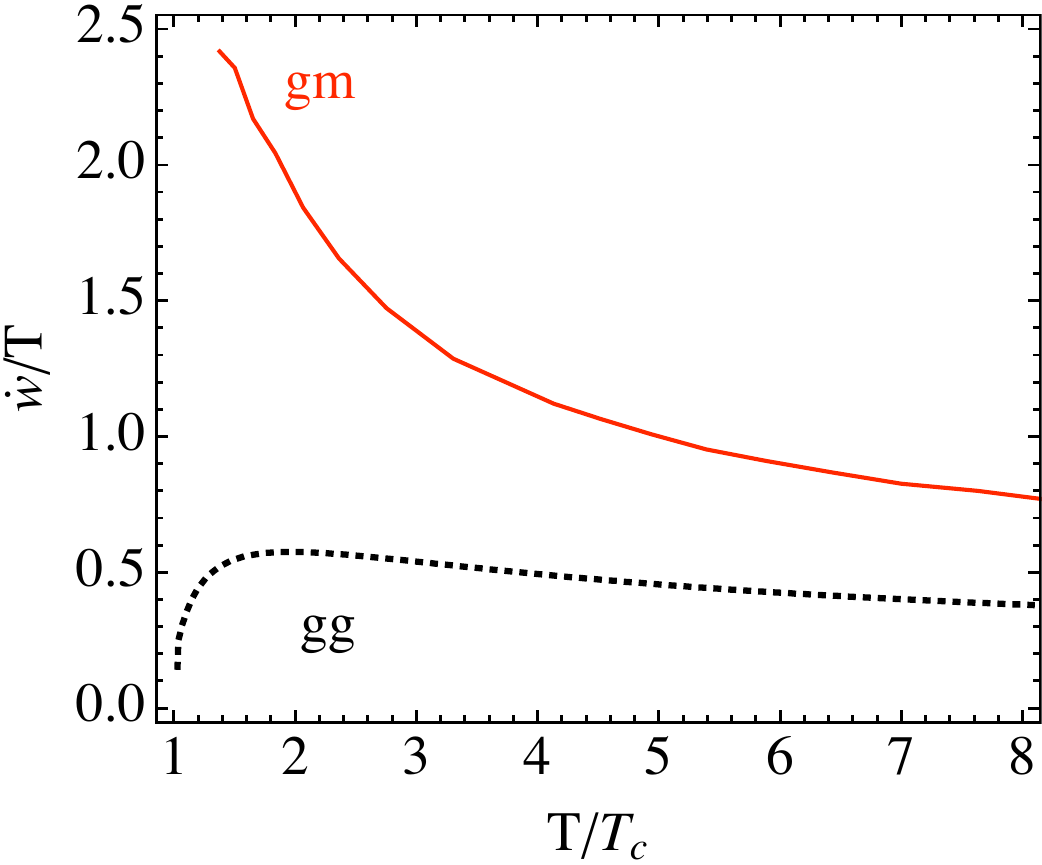}
\includegraphics[width=6cm]{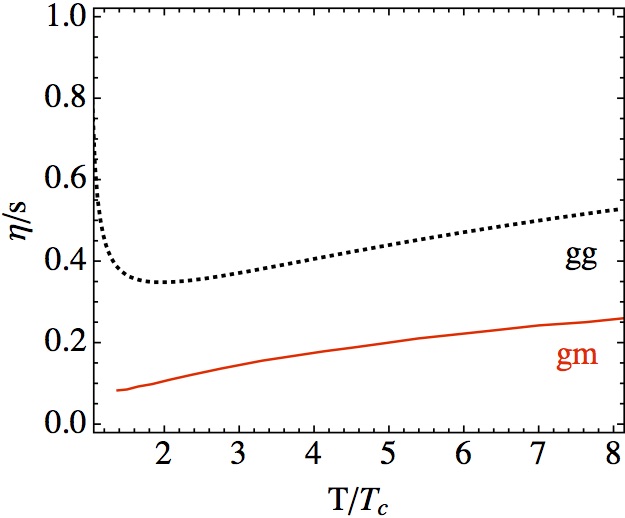}
\caption{Left panel: gluon-monopole and gluon-gluon scattering rate. Right panel: gluon-monopole
and gluon-gluon viscosity over entropy ratio, $\eta/s$.}
\label{fig_viscosity}
\end{figure}

Convoluting the cross sections found with the monopole density 
and gluon momentum distribution,
we plot  the scattering rates $n\sigma_t$ per 
gluon vs $T$ in Fig.\ref{fig_viscosity}.

It follows from this comparison of the gluon-monopole
curve with the gluon-gluon one that the former remains the leading
effect
till very high $T$, although asymptotically it is expected to get subleading.
This maximal $T$  expected  at LHC does not exceed 4$T_c$, where
$\eta/s\sim .2$. This value is well in the region which would ensure
hydrodynamical radial and elliptic flows, although deviations from
ideal hydro would be larger than at RHIC (and measurable!).

The approximate relation of these rates to viscosity/entropy
ratio is 
\be 
{\eta \over s}\approx {T \over 5 \dot w};
\ee
We plot $\eta/s$ in the right panel of Fig. \ref{fig_viscosity}.

\begin{figure}[htbp]
\begin{center}
\includegraphics[width=8.cm]{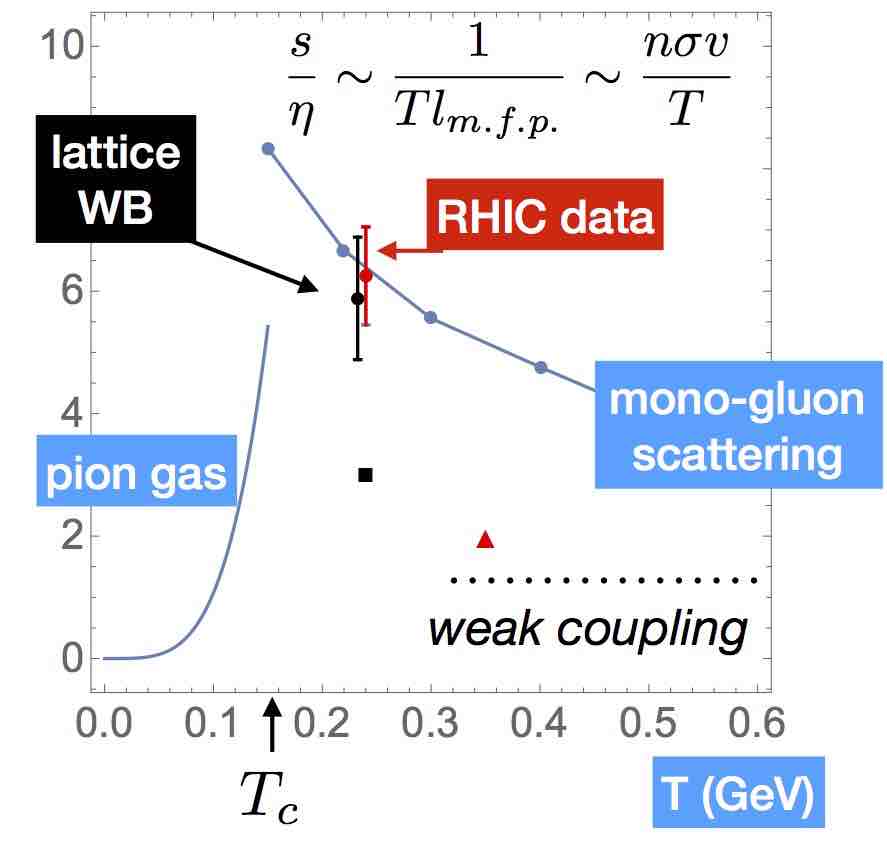}
\caption{Entropy density divided to shear viscosity $s/\eta$ as a function of the temperature
$T$. Two points with error bars at the center are for  lattice and experimental data from RHIC
higher harmonics flow. The points connected by line are for monopole-gluon scattering
described above. }
\label{fig_entropy_to_viscosity}
\end{center}
\end{figure}

The results of this calculation are compared to lattice and experimental data in Fig.\ref{fig_entropy_to_viscosity}. The meaning of this ratio is basically the interparticle distance
$\sim 1/T$ divided by the mean free path. The fact that it is about 6 means that particle
in average collides with something at distance 6 times $smaller$ than the distance to the next particle! This seems impossible: but one should recall that it is not just geometry, the Lorentz force enhances
scattering.

The lower plot compares 
 the scattering rates for gluon-gluon and monopole-gluon scatterings: the latter is clearly dominant in the 
 near-$T_c$ region. 
  The results do indeed indicate that gluon-monopole scattering
 in sQGP  dominates its kinetic properties and explains a small QGP viscosity observed. 

\section{Monopoles and the flux tubes}
Let us start with two popular statements:

{\em (i) Existence of flux tubes between two fundamental charges in QCD-like gauge theories is among the most direct
manifestations of the confinement
phenomenon.

(ii) Confinement  is well described by the ``dual superconductor" model \cite{Mandelstam:1974pi,tHooft:1977nqb} , relating it to
known properties of superconductors via electric-magnetic duality.}

In this section we discuss both of them, arguing that
they are only $partially$ correct. In fact we are going to argue that flux tubes
do exist even above the deconfinement transition temperature, and that
there are much better analogies to confinement than
superconductors. Most facts and considerations needed to understand the phenomena discussed has in  fact been
in literature for some time. The purpose of these comments are simply to remind them, ``connecting the dots" once again, since these questions continue to be asked at the meetings. We will also point out  certain aspects of the phenomena 
which still need to be clarified.

\subsection{Flux tubes on the lattice, at zero $T$ and near $T_c$}

Lattice gauge theory simulations have addressed the confinement issue from their
beginning, and by now there are many works which studied the
electric flux tubes between static charges. Most of those are done at zero/low $T$.
The documented well the profile of the electric field and the  magnetic current
``coiling" around it. 

\begin{figure}[h!]
\begin{center}
\includegraphics[width=7cm]{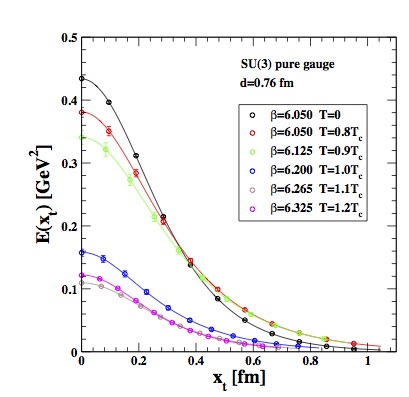}
\caption{ The longitudinal electric field as a function of $transverse$ coordinate is measured,
for a number of temperatures, for pure gauge $SU(3)$ theory, from  \protect\cite{Cea:2017bsa}.}
\label{fig_profile_T}
\end{center}
\end{figure}

The ``dual superconductor" analogy leads to a comparison with Ginzburg-Landau theory
 (also called  ``the dual Higgs model") and
 good agreement with it has been found. These results 
 are well known for two decades, see e.g.  the review \cite{Bali:1998de}.  
 
However, recent lattice studies \cite{Cea:2017bsa} exploring a near-deconfinement range of temperatures
have found that a tube-like profile of the electric field persists even above the critical temperature $T_c$, to
at least  $1.5 T_c$. One of the plots from this work is reproduced in Fig.\ref{fig_profile_T}.
It corresponds to pure gauge $SU(3)$ theory, which has
the first order transition, seen as a jump in the field strength. Note however that at $T>T_c$
the shape of the electric field transverse profile remains about the same, while the   width is $decreasing$ with $T$, making it even more tube-like, rather than expected near-spherical Coulomb behavior. (Note also that the length of the flux tube remains constant, $0.76\, fm$ for this plot.)

Such behavior clearly contradicts the ``dual superconductor" model:
in superconductors, the flux tubes are only observed in the 
superconducting phase.
Why do we observe flux tubes in the ``normal" phase, and do we really have any contradictions with theory here?

\subsection{Does the $T_c$ indeed represent the monopole condensation temperature?} 

Let us start by critically examining the very notion of the deconfinement transition itself,
focusing on whether it is indeed is the transition between the super and the normal phases. 

The $T_c$ itself is defined from thermodynamical quantities, and for pure gauge theories (we will only consider in this note)
its definition has no ambiguities.  

At this point it is worth reminding that in fact the electric-magnetic duality relates QCD not to the BCS superconductors, but rather
to  Bose-Einstein condensation (BEC) of bosons, the magnetic monopoles.
 Multiple lattice studies did confirmed that  $T_c$  does coincide with BEC of monopoles.

One such study I was involved in   \cite{D'Alessandro:2010xg} has calculated the probability
of the so called Bose (or rather Feynman's) clusters, a set of $k$ monopoles  
 interchanging their locations over the Matsubara time period. Its dependence on $k$
 leads to the definition of the  effective chemical potential, which is shown to vanish
 exactly at $T=T_c$. This means that monopoles do behave as any other bosons, and they indeed undergo
Bose-Einstein condensation at exactly $T=T_c$.

Earlier studies  by  Di Giacomo
and collaborators in Pisa group over the years, see e.g.  \cite{Bonati:2011jv}, were based
on the idea to construct (highly non-local) order parameter for monopole BEC. 
 It calculates the temperature dependence of the expectation value of the operator, effectively inserting/annihilating a monopole, and indeed finds a jump exactly at $T_c$.

In summary, it has been shown beyond a reasonable doubt that $T_c$ does indeed separate the ``super" and ``normal" phases.

\subsection{Constructing the flux tubes in the ``normal" phase}
Since electric-magnetic duality relates QCD not to the BCS superconductors, but rather
to BEC, let us at this point emphasize a
 significant difference between them:  the  $un-condenced$ bosons are also
present in the system, both above and even below $T_c$, while the BCS Cooper pairs of  superconductor exist at $T<T_c$ only.

The first construction of the flux tube in the normal phase has been made in
 \cite{Liao:2007mj}. The first point to explain is  that flux tubes as such   do not require supercurrents!
Indeed,  flux tubes are found in various plasmas: e.g. one can even see them in solar corona in a telescope.
What is needed for QCD electric flux tube formation is in fact  the presence of the {\em dual plasma}, a medium with movable magnetic charges. 

\begin{figure}[h!]
\begin{center}
\includegraphics[width=6cm]{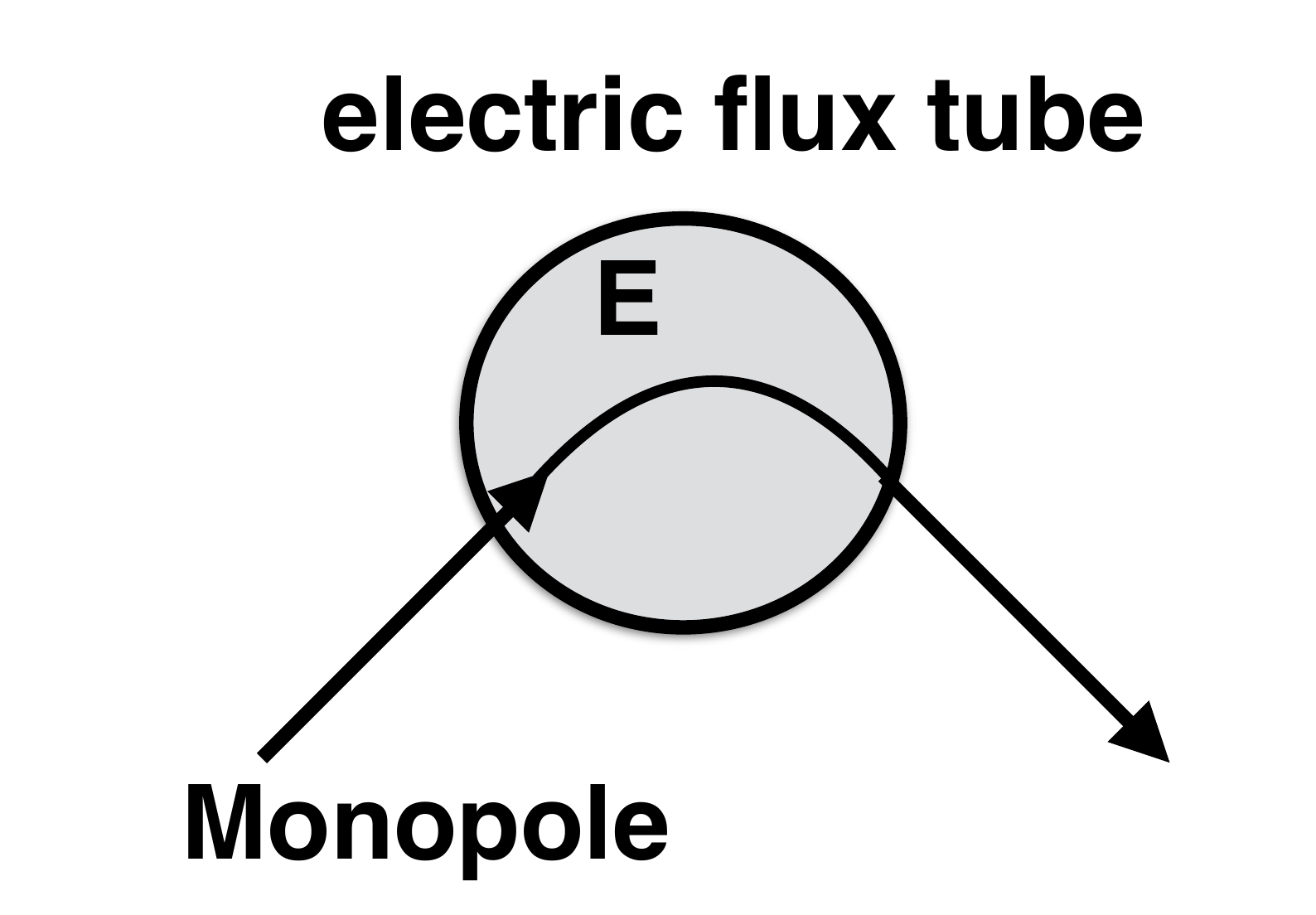}
\caption{A sketch of a monopole traversing the electric flux tube}
\label{fig_flux_tube_mono}
\end{center}
\end{figure}

Their scattering 
on the electric flux tube schematically shown in Fig.\ref{fig_flux_tube_mono} does not change
the monopole energy but changes direction of its momentum, thus
creating a force on the flux tube. If it is strong enough
 able to confine the electric field, a flux tube solution can be constructed. 
 
 For further 
 details see the original paper   \cite{Liao:2007mj}. Let us only comment that (i)
 the ``uncondenced' monopoles exert a $larger$ force than those in the condensate, as their momenta are larger; and (ii)  
   it has in fact been $predicted$
 there that the highest $T$ at which such solution may exist is about $1.5T_c$.   

 \begin{figure}[h]
\begin{center}
\includegraphics[width=5.5cm]{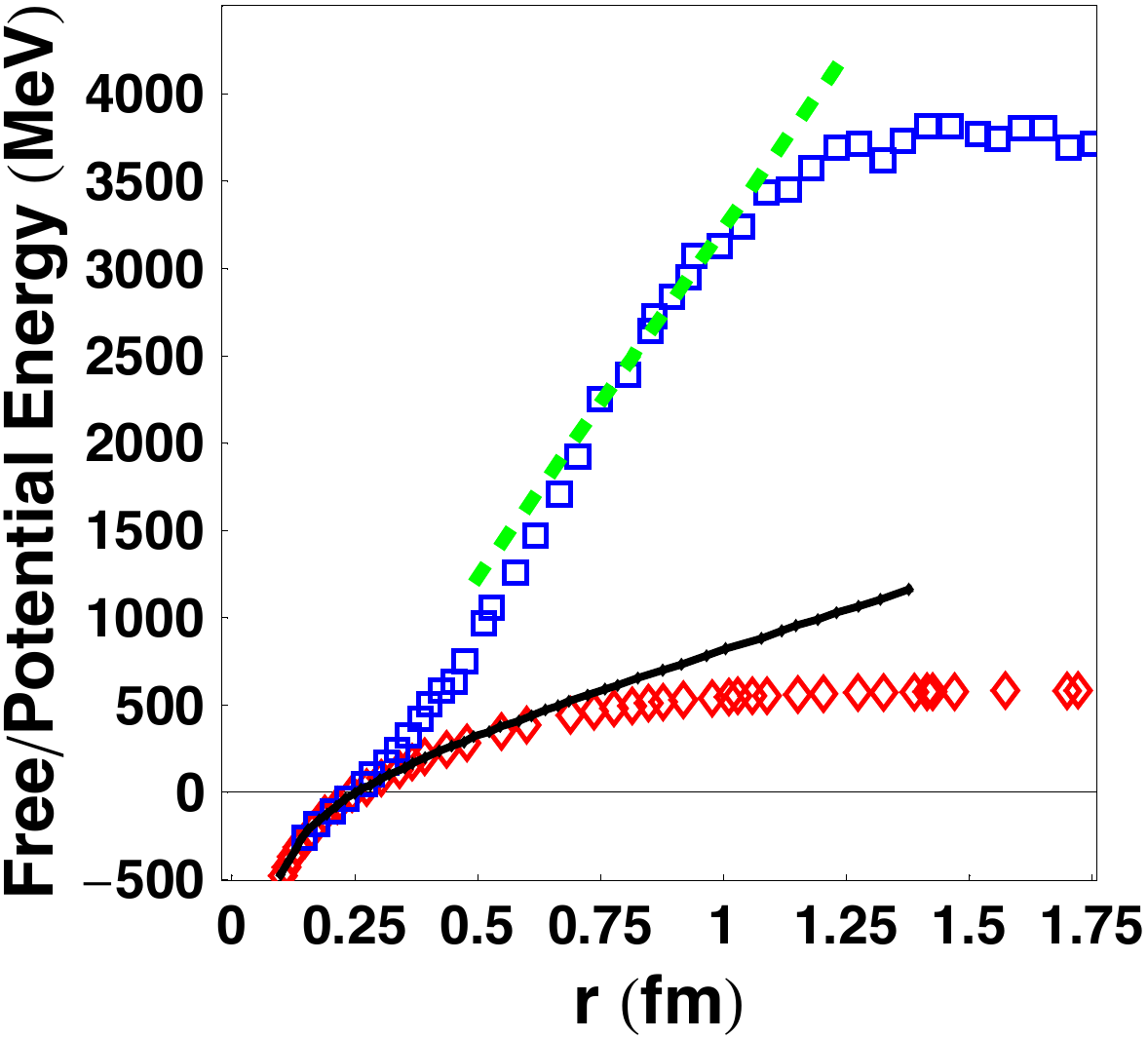}
\includegraphics[width=6cm]{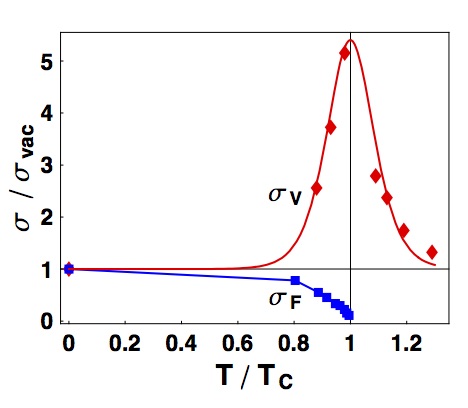}
\caption{Left: Free (red rhombs) energy $F(r)$ and potential (blue squares) energy $V(r)$,
at $T_c$, compared to the zero temperature potential (black line). 
Right: Effective string tension for the free 
 and the internal energy, from \protect\cite{Kaczmarek:2005gi}.} 
\label{fig_tension_F_U}
\end{center}
\end{figure}
 
\subsection{Two static potentials and the flux tube entropy}

So far there were no subtleties involved. But let us  asks now the following question: Provided  these flux tubes at $T>T_c$
 carry some tension (energy-per-length), does it still imply existence of a linear potential between quarks, up to  $T=1.5T_c$? And if they do,  would  it imply that, in a sense, confinement remains  enforced  there?

In order to have proper perspective on  what is going on, let us look back at lattice studies of the static quark (fundamental) potentials, e.g. 
 \cite{Kaczmarek:2005gi}. The key point is that 
 there are $two$ kinds of the potentials. At finite temperatures the natural quantity to calculate, for the observed flux tubes
between static charges, is the {\em free energy}. It can be written as
 \be F(r) =V(r)- T S(r), \,\,\, S(r)={\partial F \over \partial r}   \ee 
where $S(r)$ is the $entropy$ associated with a pair of static quarks. 
Since it can be calculated from the free energy itself, as indicated in the r.h.s., one can 
subtract it and plot also the {\em potential energy} $V(r)$. The derivatives over $r$ are known as  the {\em string tensions}.

These lattice calculations have shown that in certain range of $r$ the tension is
constant (the tension is approximately $r$-independent).  Two resulting tensions, 
shown in Fig.\ref{fig_tension_F_U}(left) have very different temperature dependence.
The tension of the free energy shows the expected behavior: $\sigma_F(T)$ vanishes as
$T\rightarrow T_c$.  But the  tension of the potential energy $\sigma_V(T)$
shows drastically different behavior, with large $maximum$ at $T_c$, and non-zero value above it. This unexpected behavior was hidden in $\sigma_F(T)$, studied in many previous works, because in it a large energy and a large entropy cancel each other. 
  
So, everything would be consistent, provided these novel flux tubes at $T>T_c$ do 
indeed carry the $potential$ energy only, but no $free$ energy tension $\sigma_V\neq 0, \sigma_F=0$.  In other words, we suggest that the potential energy detected (via electric field squared) in \cite{Cea:2017bsa} must be canceled by the  entropy associated with
it, and no actual force between charges would be present in equilibrium.  This conjecture can and should be checked. 

Similar comment applies also to the theoretical calculation of the potential: 
 in \cite{Liao:2007mj} only the mechanical stability of the tube solution was derived.
 The entropy associated with the flux tube still remains to be calculated. 

As a parting comment, while this conjecture sounds like the well known idea of Hagedron string
transition, it cannot be exactly that. Indeed, this idea is known to suggest that at $T>T_c$ 
string gets to be infinitely long. If so, the tube completely delocalizes, and there would be a
Coulomb field rather than what was observed by  \cite{Cea:2017bsa}. The entropy
in question is perhaps related to monopoles  bound to the tube rather than its multiple shapes. 
Also a Hagedorn transition seems to be at odds with the tension increase and size decrease
as a function of $T$ observed.

(Finally, let us for  clarity mention that we only discuss in this note static potentials in thermal equilibrium. We do not discuss potentials in quarkonia, in which quarks are not standing but moving.  This problem is associated with certain time scales, inducing deviation from equilibrium and possible dissipation. It would therefore require a completely separate 
discussion.)

\section{Lattice studies of the Bose-Einstein condensation of monopoles at the deconfinement transition} 

It is convenient to split lattice studies to two kinds: (i) those at $T<T_c$, addressing the bose-condensed state,  and (ii) at   $T>T_c$ exploring the onset of BEC by looking at finite-size
clusters.

At $T<T_c$ (or even $T=0$) the lattice studies focused on direct detection of
the monopole condensate. The idea is to do a {\em insertion} of a monopole
into the vacuum state. This method has been developed by Di Giacomo
and collaborators in Pisa group over the years: see e.g. the most self-contained
articles of that series \cite{Bonati:2011jv}. 

It is based on the idea that  the condensate is a particular quantum state
analogous to the ``coherent state" of a harmonic oscillator, with large
average number of quanta $<n>\gg 1$. Such classical-like states 
are superpositions  of many states $|n>$ with different number of quanta,
created e.g. by exponentiation of the field (coordinate) operator. Therefore
they do not have fixed number of quanta. 
 
Since the quantum mechanical momentum operator acts as a derivative over
the coordinate,  $p=i{d \over dx}$, acting on the wave function.
Its exponential $ exp (ipa)$ can be expanded and considered to be a Taylor series, corresponding
to a shift of the wave function argument by $a$
\be exp (ipa)|\psi(x) > = |\psi(x + a) > \ee
and thus is known as a ``shift operator". Its field theory version generalizes the 
coordinate shift by $a$ into a field shift \be \phi(x)\rightarrow \phi(x) +a(x) \ee  
In a gauge theory (and the $A_0=0$ gauge) the shift of the gauge field we will call $\vec{a}(x,t)$ and the
conjugated canonical momentum is the electric field, so the appropriate operator is
a shift by the monopole field
\be      \mu(x)=exp\left[ i \int d^3y \big(\vec{E}(y,t) \cdot \vec{A}_{mono}(y-x,t)\big) \right] \ee
(In the non-Abelian theory the color index is implied). The object we would like to
add to the vacuum, $\vec{a}(x,t)$, can be of any shape, for a example a magnetic monopole. If the state $|\psi >$ is ``normal" and has a definite (zero) number of monopoles, the average of it would be zero $<\psi | \mu | \psi >=0$, but if the
state has a monopole condensate the average would be nonzero! So the
$<\mu>(T) $ is the order parameter for monopole BEC. 

\begin{figure}[t]
\begin{center}
\includegraphics{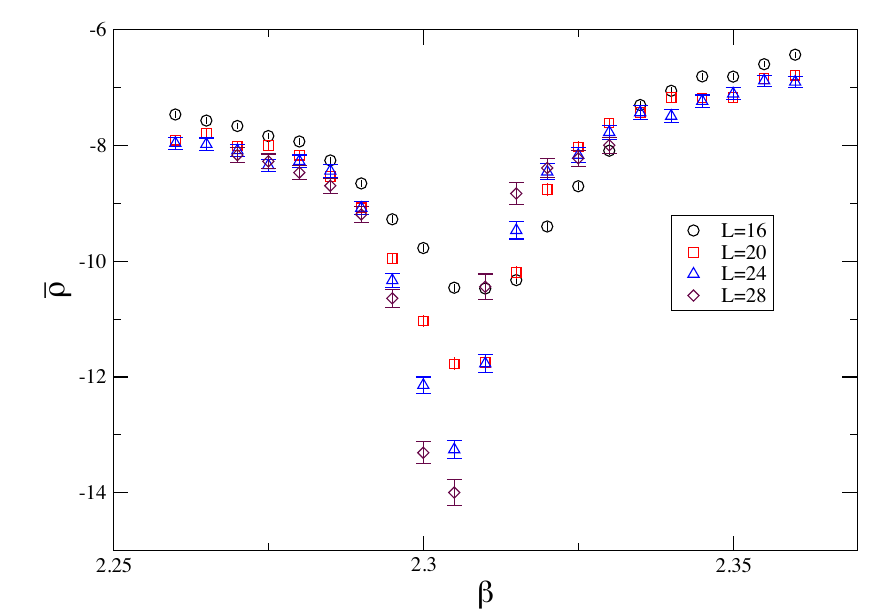}
\caption{The dependence of $\bar\rho$ on $\beta$ for SU(2) gauge theory. The lattice term oral size $N_t=4$ and the spatial sizes are shown as $L$ in the insert for corresponding symbols. A singularity at the critical
deconfinement coupling is revealed. }
\label{fig_conf_order_param}
\end{center}
\end{figure}

The details of the definition/normalization can be found in \cite{Bonati:2011jv}:
we only comment that it is more convenient to measure not the average but its derivative 
$\bar\rho=\partial ln <\mu>/\partial \beta$ over $\beta=2N_c/g^2$, the coefficient
in front of the lattice action.  (The bar has no particular meaning here, just it was
called like this in the original work.) The dependence on the coupling $\beta$ is shown in Fig.\ref{fig_conf_order_param}
( from Ref.~\cite{Bonati:2011jv}): its singularity near $\beta_c=2.2986$ 
known independently  proves that this is indeed singular at the
deconfinement critical coupling.
(The particular monopole inserted is here a Yang-Wu monopole (two half-spaces glued together, no Dirac string) of the charge 4: but similar results are obtained also for other
monopoles.) The expected critical scaling with the known 3d Ising index
for the order parameter  $\bar{\rho}-\bar{\rho}_c\sim L^{1/\nu}, \nu=0.6301$ has also been demonstrated in the same paper.

We already qualitatively discussed  the ``dual superconductor" paradigm
\cite{'tHooft:1977hy,Mandelstam:1974pi}
according to which in the confinement phase at $T<T_c$ monopoles Bose-condense into a -- magnetically charged --
condensate. In this section we will discuss lattice evidences showing that this is indeed what happens.

\begin{figure}[h]
\begin{center}
\includegraphics[width=11cm]{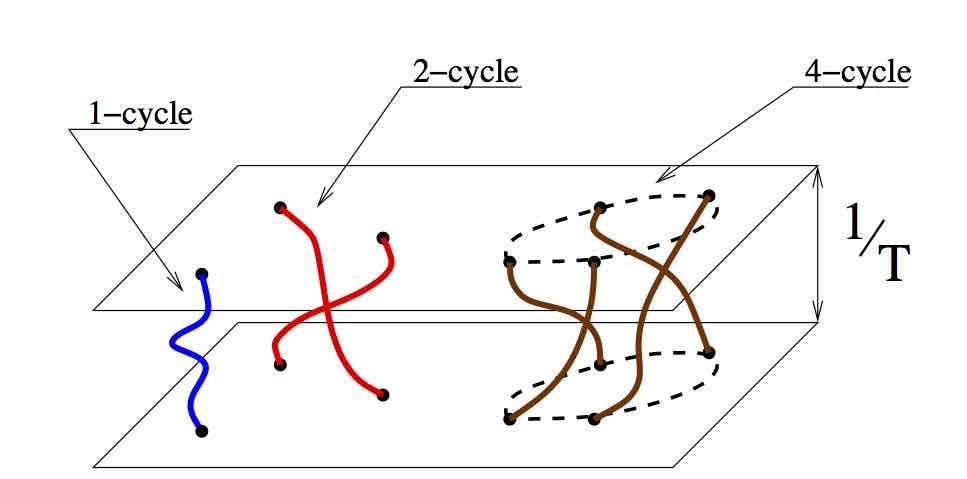}
\includegraphics[width=10cm]{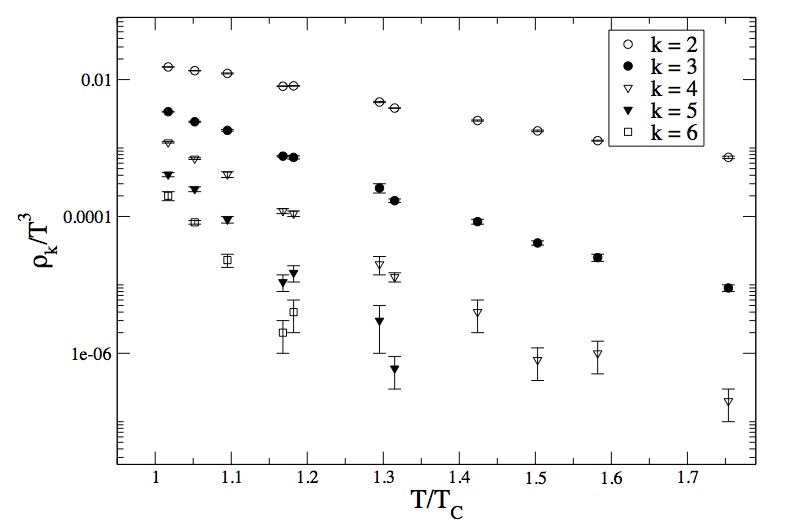}
\caption{(upper) Example of paths of 7 identical particles which undergo a permutation made up of a 1-cycle, a 2-cycle and a 4-cycle. (lower) Normalized densities $\rho_k /T^3$ as a function of $T/T_c$ .
}
\label{fig_mono_permutations}
\end{center}
\end{figure}

The first one is based on study of the so called Feynman clusters. 
As $T\rightarrow T_c$ from above, the ``dual superconductor" paradigm
require that the behavior of monopoles should change, revealing quantum motion and a 
``preparation" to form a Bose-Einstein Condensate (BEC). The idea of identical clusters
is explained in Fig.\ref{fig_mono_permutations}: identical bosons may have ``periodic paths" in which some number $k$ of them exchange places. Such clusters are widely known to community doing manybody path integral simulations
for bosons, e.g. liquid $He^4$. Feynman argued that in order for statistical sum to get singular
at $T_c$, a sum over $k$ must diverge. In other words, one may see how 
the probability to observe $k$-clusters $P_k$ grows as$T\rightarrow   T_c$ from above.

\begin{figure}[t]
\begin{center}
\includegraphics[width=10cm]{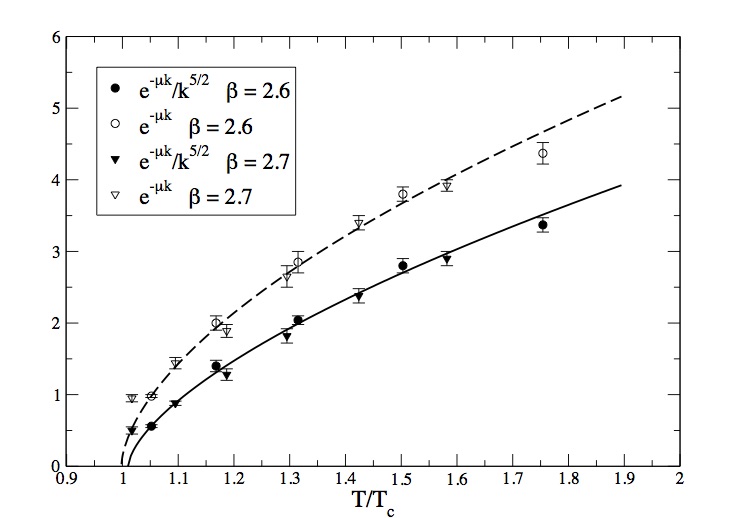}
\caption{The effective chemical potential extracted by two different fits: it should vanish at the 
Bose-Einstein condensation point.}
\label{fig_mono_mu}
\end{center}
\end{figure}

In Fig.\ref{fig_mono_permutations}(lower) from \cite{D'Alessandro:2010xg}
one see the corresponding data for the cluster density. Their dependence on $k$ were fitted by 
the expression
 \be \rho_k\sim  { exp\left(-k\mu_{eff}(T)\right) \over k^{5/2} } \ee
and the resulting effective chemical potential $\mu_{eff}(T)$ is plotted versus temperature at Fig.\ref{fig_mono_mu}.
it vanishes exactly at $T=T_c$. This means that monopoles indeed undergo
Bose-Einstein condensation at exactly $T=T_c$.

\section{Quantum Coulomb gases studied by Path Integral Monte-Carlo (PIMC)} 

Quantum  studies of interacting particles displaying BEC is a very well developed
area of manybody physics. Its traditional applications are  weakly coupled Bose gases and
liquid $^4He$. The former problem was intensely studied in 1950's by C.N.Yang and collaborators, and then, by Feynman diagrams, by S.T.Belyaev. Its theory reactivated at the end of 1990's, after experimental observation of BEC in ultracold atomic gases. Liquid helium
was a frontier of experimental low-$T$ research from the beginning of 20-th century
till 1950's. Its theoretical treatment from the first principles become possible
in 1970's, with the development of numerical PIMC simulations. Here is not a place
to review these works, so let me only comment that while atoms of $^4He$ interact with
each other quite  weakly, by atomic standards, in comparison to the temperature considered
$T\sim 2K$ the interatomic potentials are very large. So, one may call liquid $^4He$
a ``strongly coupled" system. 

In literature one can find studies of BEC for other interactions, e.g. for a Bose gas of 
solid spheres, but not for the Coulomb forces. Therefore, simulations 
(for one and two-component) Coulomb Bose gases
have been made by ourselves \cite{Ramamurti:2017fdn}
for the first time.  using numerical simulation of the manybody path integral method, used previously for such classic systems as liquid $^4He$.
Unlike the previous works, it focused on temperature dependence of the density $\rho_k$ of the Bose-clusters.
Their $T$ dependence for  liquid $^4He$ calculated in this paper looks nearly the same
as the lattice data on monopoles shown in Fig.\ref{fig_mono_permutations}, and the procedure used to locate the BEC critical temperature accurately reproduces the
 value of $T_c$  for  liquid $^4He$,known both from experiments and multiple previous  
 numerical simulations.  
 
The correlations of monopoles has been studied as well, and the coupling strength $\alpha$
defined by 
\be V_{ij}=\alpha{q_i q_j \over r_{ij}}
\ee
has been tuned to reproduce lattice data.
 
\begin{figure}[htbp]
\begin{center}
\includegraphics[width=8cm]{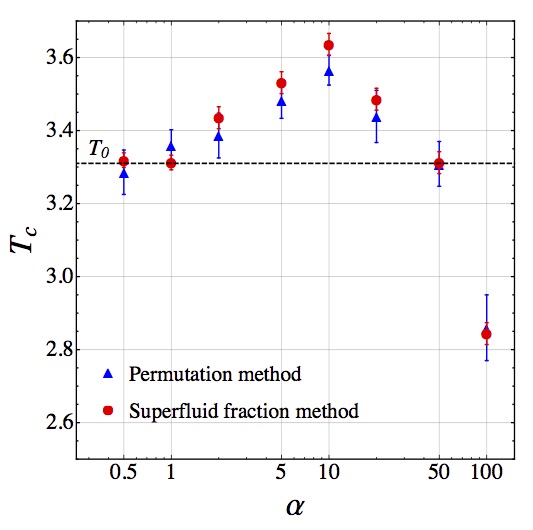}
\caption{The critical temperature for the BEC phase transiion as a function of the coupling, $\alpha$. The red circles are the results of the finite-size scaling superfluid fraction calculation for systems of 8, 16, and 32 particles;  the blue triangles are the results of the permutation-cycle calculation for a system with 32 particles. The black dashed line denotes the Einstein ideal Bose gas critical temperature.
}
\label{fig_Tc_Coulomb_Bose}
\end{center}
\end{figure}

 We will not discuss that, and only show one non-trivial result of this paper, namely the
 dependence of the critical BEC temperature $T_c$ on $\alpha$, see Fig.\ref{fig_Tc_Coulomb_Bose}. Note  that we found the same behavior at small values of the coupling as in the case of low-density hard spheres: the critical temperature for the BEC phase transition grows with the coupling. Yet if the coupling becomes large enough, Tc rapidly drops below the critical temperature for an ideal Bose gas. Eventually, as the particles are ``too repulsive," the BEC phenomenon becomes impossible since it becomes essentially too costly (in terms of the action) to permute them and BEC goes away completely.

\begin{figure}[htbp]
\begin{center}
\includegraphics[width=14cm]{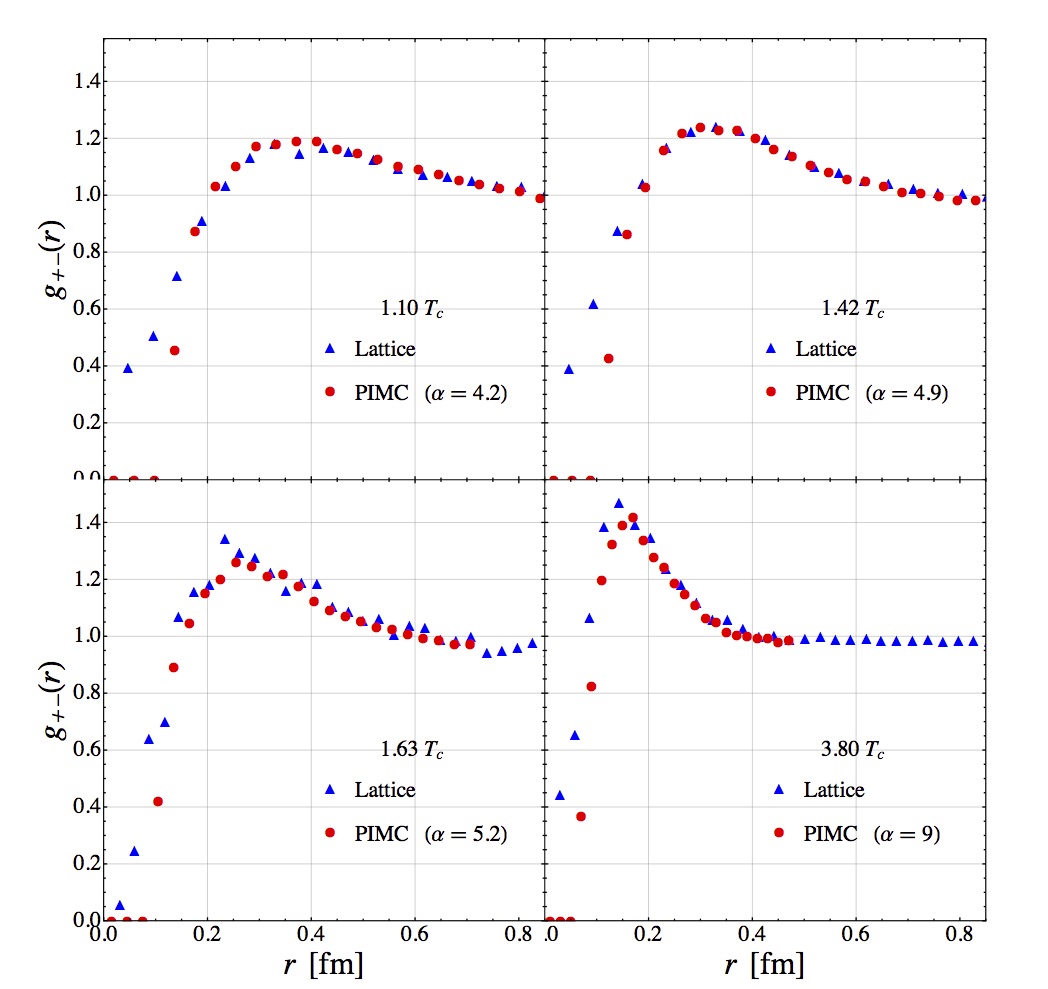}
\caption{Spatial correlations of particles in quantum Coulomb Bose gas, from PIMC
simulations (red circles) compared to lattice data for monopoles.
}
\label{fig_CoulombBoseGas}
\end{center}
\end{figure}

We have discussed at the end of the previous chapter the lattice data on the
spatial monopole-monopole and monopole-antimonopole correlations. In Fig.\ref{fig_CoulombBoseGas} these data are compared with PIMC simulation
for a Coulomb Bose Gas \cite{Ramamurti:2017fdn}. The comparison shows very good agreement, increasing the confidence that a quantum ensemble of monopoles
is described by this model well. It also
allowed  us to fix the effective magnetic coupling rather accurate,
without any reliance on the Debye fits. 

\chapter{Fermions bound to monopoles}

\section{Fermionic zero modes} \label{sec_mono_zeromodes}

Let me start this section with a small historical introduction. In 1974, when the monopole solution
by 't Hooft and Polyakov has been discovered, it was not yet clear whether either Georgi-Glashow or
Weinberg-Salam model is the correct description of the electroweak interactions\footnote{While
some evidences for neutral weak current were known,  e.g. atomic parity violation, direct observation of the $Z$
boson only happened in  1983.}. So some discussion of electroweak monopoles happened.
Afterwards  it shifted to the context of Grand Unification models, most of which do possess such
monopoles. As noticed by Zeldovich in 1978,  their cosmological production versus non-observation became an issue:
A.Guth in 1980 famously  invented cosmological inflation, in order to get rid of the undesired Grand Unification monopoles.
In all of that, scattering of monopoles on ordinary matter was studied: for Grand Unification monopoles
violation of the baryon number was noticed and much discussed.

Not going into discussion of electroweak or Grand Unification models as such, let us return to the
original Georgi-Glashow model, to which we will add some fermions. In fact, we already discussed
the $\cal{N}$=2  supersymmetric models, which have the ``gluinoes", fermionic partners of the gluons,
or quarks, with the fundamental color representation. It is in this context that people wandered what
is the spectrum of the Dirac equation, in the background field of the monopole solution.

The central observation is that there exist the so called zero fermionic modes.
There is rather extensive literature on the   meaning of these states,
starting from \cite{Jackiw:1975fn}.  
Like for other topological solitons, there are topological index theorems, relating
the number of zero modes to   the monopole quantum number $M$. 
They
require $one$ zero mode
for a fundamental fermion (quark), and $N_c$ of them for the adjoint (gluino) fermion.

%
As argued by  \cite{Jackiw:1975fn} in their classic paper, the operator algebra involved
corresponds to a pair of creation/annihilation operators, with the algebra
$\{ a a^+\}=1$,  requiring representation in the form of two states,
 the ``empty''
and ``occupied'' ones. Symmetries of the problem, such as CP conjugation, plus a
requirement that these two states differ by one unit of the fermion number,
let them to (at the time revolutionary) conclusion, that the baryon number of such states
are $semi-integer$, namely $\pm 1/2$.

Let us look at these two states from the viewpoint of the ``Dirac sea" picture of the fermionic vacuum state. All levels
with positive energy are supposed to be empty, and all negative ones occupied. But what to do with extra
two {\em zero energy states} sitting on the monopole? If occupied we still call it a $particle$, if not, a $hole$. 

What about the $spin$ of these zero mode states? Starting in the simplest case of the $N_c=2$ theory, we will use
the term ``isospin''  instead of the color (or ``weak isospin" of electroweak sector).
 Thus the  quarks in fundamental  representation we will call  isospin-1/2 fermion, and gluinoes in the
 adjoint  representation will be called  isospin-1 fermions. The 
 monopole field is a ``hedgehog", relating direction in coordinate space and in the isospin space:
 therefore isospin and spin are not separately conserved. Yet after some observation of the Dirac equation 
 one may prove that
 so called $grand-spin$ $$\vec K=\vec I+\vec S$$ 
 is in this case conserved. Following standard rules of angular momentum representations one finds that
 $K$ can have values $\vec 1/2+\vec 1/2=0,1$ in the case of quark, 
 and $\vec 1+\vec 1/2=1/2,3/2$ in the case of gluino.  Explicit solution of the Dirac equation shows that
in both case zero modes correspond to the lowest values: thus the number of states is $2K+1$. It is 1 for the quark
$K=0$ state and $2$ for
 $K=1/2$ gluino state. Conclusion: a quark bound to a monopole is a single state, thus it is a $scalar$ spin-0 object.
 A gluino bound to a monopole makes 2 states, thus it makes  a $spinor$ state. Standard spin-statistics theorem
 then require that in the former case one produces a boson, and in the latter case a fermion.

Further discussion naturally extends to the problem of counting the number of states and their statistics
when there are several species of the fermions. 
Explicit construction of such states have obtained another motivation 
in mid-1990's, with the discussion of various supersymmetric theories
(for review see \cite{Harvey:1996ur} ). Here is the main idea: all fermions of the theory coupled to the monopole
are expected to produce only ``magnetically charged" multiplets which are {\em consistent with the underlying
symmetries of the theory}. 

Example: consider the $\cal N$=2 SYM, which we already mentioned 
in relation to Seiberg-Witten famous papers. It has two adjoint $real$ gluinoes. 
Each can be bound to a monopole, leading to a fermion state. 

Now, in order to have creation and annihilation operator algebra, with \{ $\hat a^+ \hat a$\} =1 etc, the fileds need to be complex.
So, like in the harmonic oscillator, in which they are built from two hermitian (=real eigenvalued) operators $\hat p,\hat x$,
one need to combine two gluinoes into one complex (Dirac-like) fermion. Thus there is a $single$ set
of  creation and annihilation operators. The maximal spin state is 1/2 -- two partners of a scalar monopole. 

When both are bound at the same time,
it is spin $\vec 1/2  + \vec 1/2=0,1$ states. Spin zero can be combined with the unoccupied monopole, producing
2 scalar states. We in fact do $not$ have a spin-1 state here, as its wave function is not anti-symmetric as
fermions should. As a result, total magnetic supermultiplet gives the ``short"
 representation of the    $\cal N$=2 supersymmetry, with 2 fermions and their scalar partners. 
So, the effective magnetic theory is the $\cal N$=2 electrodynamics.

Further study
reveals, among other things, beautiful examples of theories with complete electric-magnetic
$selfduality$. It has been explicitly shown that   two 
theories,  the $\cal N$=4 SYM and  $\cal N$=2 SQCD with $N_f=4$ quark flavors,
that a magnetic monopole ``dressed'' by all available  fermions of these theories.
  the ``long" supermultiplet, starting from spin-1 states, to  $\cal N$ spin 1/2 states and
 correct number of scalars.  The $\cal N$=4 SYM has 4 real gluinoes, so there are $two$ sets of 
  creation and annihilation operators. Thus the maximal achieavable spin is 1 -- one now has vector
  magntic particle. There are 16 states,  the ``long" supermultiplet, starting from spin-1 states, to  $\cal N$ spin 1/2 states and
 correct number of scalars (6 of them).
 
 What this means is that the effective magnetic theory and the effective electric theory happen to be in these two case
 the same, up to the fact that one has electric and another magnetic -- inverse to electric -- coupling.
 At one hand Dirac condition require that
  their beta functions should have the opposite sign, on the other hand they must be the same,
 as both theories the same Lagrangian. The only solution possible
 here is the following one: both those theories are $conformal$, with their
RG beta function equal to $zero$!  

Let us now return to the non-supersymmetric world and ask what fermion binding to monopoles can 
imply  for various QCD-like theories. Without adjoint fermions, only spin-zero new states 
can be generated, no matter how many quarks are or are not bound to the monopole. 
So the effective magnetic theory can only be a scalar electrodynamics.

The number of magnetic states can be rather large. Both states exist for each flavor, so
in QCD with $N_f$  quark flavors the number of magnetic states we start to consider is thus
 $2^{N_f}$, half with integer and half with semi-integer fermion number.
 Of course, not all of these states can exist: since
 each bound state is a boson, the allowed wave functions must all be symmetric under permutations.
 For example, if all $N_f$ zero modes are occupied -- the states with maximal possible
 number of flavor indices -- it should be symmetric tensor.
 Its baryon number is $N_f/2$, say 5 for $N_f=10$.
 If all quarks have the same mass -- e.g. zero -- there is unbroken flavor symmetry,
 and thus such magnetic states should fall into its proper flavor multiplets. 

\begin{figure}[h!]
 \includegraphics[width=8cm]{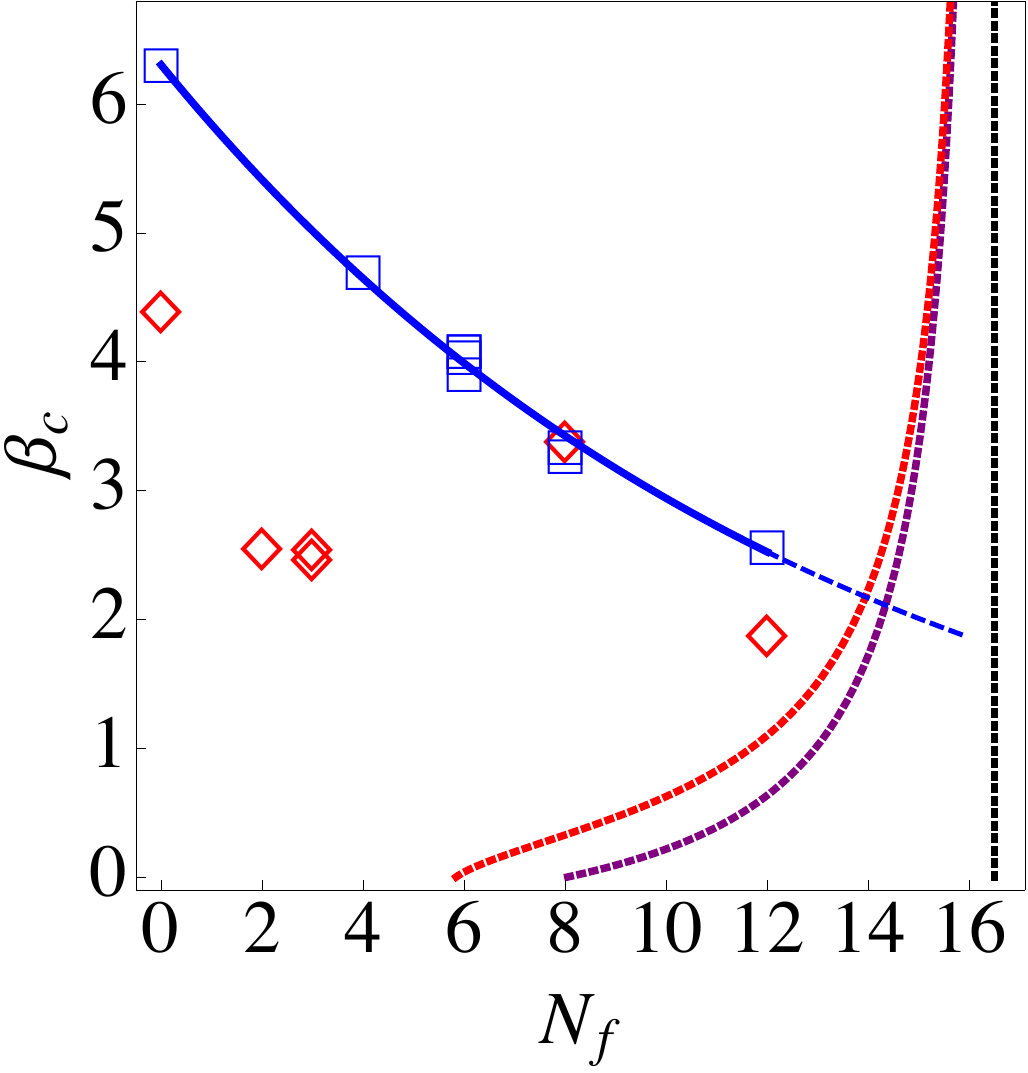}
 \caption{ Dependence of the critical lattice coupling $\beta_c$ at scale $T_c$ versus the number of fundamental quark flavors $N_f$ in QCD-like theories.  The thick blue line is the fitting curve, extended as dashed blue line beyond $N_f=12$. The black/purple/red curves on the right are lines for vanishing beta function at 1,2,3-loop levels: they indicate the boundaries of the so called conformal window at large $N_f$.}
 \label{fig_fundamentals}
 \end{figure}

 Now, how would the deconfinement transition be affected by adding more and more light fermions? 
 As argued by  \cite{Liao:2012tw}, it will dramatically shift downward the transition temperature
 (or, more precisely, shift the transition to stronger coupling). The reason for it is simple:
 since some number of  fermions can be attached to the monopoles via zero modes,
 the resulting states are no longer identical particles. And only the identical ones -- most likely the
 monopoles $without$ fermions -- will undergo the Bose-Einstein condensation, as soon as their density become critical. 
 Shifting to IR direction, with the lower $T_c$ or stronger coupling, decreases the monopole mass and increase the overall  
 monopole density, compensating for ``occupied" monopoles. 

For details see the original paper: let me just show one plot, Fig.\ref{fig_fundamentals}, which shows that
the expected trend is indeed observed, by lattice studies 
of the theories with many quark favors, nowadays up to  $N_f=12 $. Blue boxes are from \cite{Miura:2011mc}: near-coincident boxes being lattice data for the same $N_f$ with different number of lattice cites $N_\tau$ which demonstrate lattice spacing consistency. Red diamonds are from various other lattice studies.

\section{Chiral symmetry breaking by monopoles}
In chapter on instantons we will discuss in detail how a collectivization of the 
4-d fermionic instanton zero modes, resulting in breaking of the chiral symmetry by
a non-zero quark condensate $\langle \bar q q \rangle\neq 0$. 
We will then show in chapter  on instanton-dyons that  this
approach can be directly generalized to finite temperatures, since (some) of the instanton-dyons also
possess the 4-d   zero modes.

Above in this chapter \ref{chap_mono}, we have argued that Euclidean semiclassical theory based on instanton-dyons is {\em Poisson dual} to the monopole approach. If so, one should be able to derive chiral symmetry breaking using monopoles as well. This was indeed accomplished by \cite{Ramamurti:2018hdh}, which we follow in this section.

One obvious difficulty of the problem is the fact that a detailed understanding of the  ``lattice monopoles" is lacking; they are treated as effective objects whose parameters and behavior we can observe on the lattice and parameterize, but their microscopic structure has yet to be understood. In particular, the 't Hooft-Polyakov monopole solution includes a  chiral-symmetry-breaking scalar field, while we know that, in massless QCD-like theories, chiral symmetry is locally unbroken. We assume that the zero modes in question are chiral themselves, like they are in the instanton-dyon theory, and that chiral symmetry breaking can only be achieved by a spontaneous breaking of the symmetry.   

The other difficulty of the problem is the important distinction between fermionic zero modes of (i) monopoles and (ii) instanton-dyons.  As follows from Banks-Casher relation \cite{Banks:1979yr}, the quark condensate is proportional to density of Dirac eigenstates at zero eigenvalue. The monopoles also have fermionic zero modes \cite{Jackiw:1975fn}, which are 3-dimensional. They are, therefore, simply a bound state of a fermion and a monopole. In theories with extended supersymmetries, such objects do exist, fulfilling an important general requirement that monopoles need to come in particular super-multiplets, with fermionic spin 1/2 for ${\cal N}=2$, or spins 1/2 and 1 for ${\cal N}=4$. The anti-periodic boundary conditions for fermions in Matsubara time implies certain time dependence of the quark fields, and (as we will discuss in detail below) the lowest 4-dimensional Dirac eigenvalues produced by quarks bound to monopoles are the values $\lambda =\pm \pi T$, not at zero. 

This, however, is only true for a single monopole. In a monopole {\em ensemble} with non-zero density, the monopole-quark bound states are collectivized and Dirac eigenvalue spectrum is modified. The question is whether this effect can lead to a nonzero $\rho(\lambda=0)\propto \langle \bar q q \rangle $, and if so, whether it happens at the temperature at which chiral symmetry breaking is observed. As we will show below, we find affirmative answers to both these questions. The phenomenological monopole model parameters are such that a non-zero quark condensate is generated by monopoles at $T\approx T_c$. 

Recognizing fermionic binding to monopoles, we now proceed to description of their dynamics in the presence of ensembles of monopoles. The basis of the description is assumed to be the set of zero modes described in the previous section. The Dirac operator is written as a  matrix in this basis, so that $i-j$ element is related to ``hopping" between them. The matrix elements of the ``hopping matrix''
\be
\bf{T} = \begin{pmatrix}
          0 & iT_{ij} \\
	iT_{ji} &0
         \end{pmatrix}
\ee
 where the $T_{ij}$s are defined as the matrix element,
$$
T_{ij} \equiv \bra{i} -i\Dslash \ket{j},
$$
between the zero modes located on monopoles $i$ and antimonopoles $j$. In the SU(2) case we are considering, this is equivalent to
\bea
 T_{ij} = &\braket{\psi_i}{x}\bra{x}-i\Dslash\ket{y}\braket{y}{\psi_j} \nn \\
 = &\int  d^3 x \psi_{kn}^\dagger(x-x_i) (-i \Dslash) \psi_{lm}(x-x_j) \nn\\
 =  &\int d^3 x \psi_{kn}^\dagger(x-x_i) \Big[-i(\vec\alpha\cdot\vec\partial +\vec \alpha\cdot\vec\partial  -\vec \alpha\cdot\vec\partial ) \nn\delta_{nm}\\ 
\nn  &+\frac{1}{2}(A(x-x_i)+A(x-x_j))\sigma^a_{nm}(\vec\alpha\times\vec{\hat{r}})_a 
\\ &+ \frac{G (\phi(x-x_i)+\phi(x-x_j))}{2} \sigma^a_{nm}\hat r_a\beta\Big]  \psi_{lm}(x-x_j) \nn \\
=&\int \sum_m d^3 x \psi_{km}^\dagger(x-x_i) [-i\vec \alpha\cdot\vec\partial ]^{kl}\psi_{lm}(x-x_j)
\eea
where $\psi$s are zero modes with origin at $x_{i,j}$, the locations of the two monopoles, $n,m$ are the isospin/color indices, and we have used the fact that applying the Dirac operator to these wavefunctions gives zero.

Omitting further details, let us explain the quantization procedure adopted in that work. 
 the {\em evolution matrix} $U$, defined as time-ordered integral of the hopping matrix in the previous section over the Matsubara periodic time $\tau\in [0,\beta]$. This matrix will then be diagonalized to find the eigenvalues for the fermion states. Because each eigenstate is still fermionic, each is required to fulfill the fermionic boundary conditions, namely that the state must return to minus itself after one rotation around the Matsubara circle. This defines quantization of the Dirac eigenvalues  by,
$$
\lambda_i+\omega_{i,n} = \left(n+\frac{1}{2}\right) \frac{2 \pi}{\beta} $$
where $\lambda_i$s are the eigenvalues of the hopping matrix $\bf{T}$.
For monopoles that move in Euclidean time, we must integrate over the Matsubara circle to find the fermion frequencies,
$$
U=\oint_\beta d \tau e^{i H \tau} = -{1} \,.
$$
One needs to diagonalize the resulting matrix on the right-hand side and solve to find the quantity $\lambda + \omega$. One can then compute the eigenvalues of the Dirac operator with 
$${\omega_{i,n} = \left(n+\frac{1}{2}\right) \frac{2 \pi}{\beta} -  \lambda_i \,.}$$

\begin{figure*}[h!]
{\includegraphics[width=.44\linewidth]{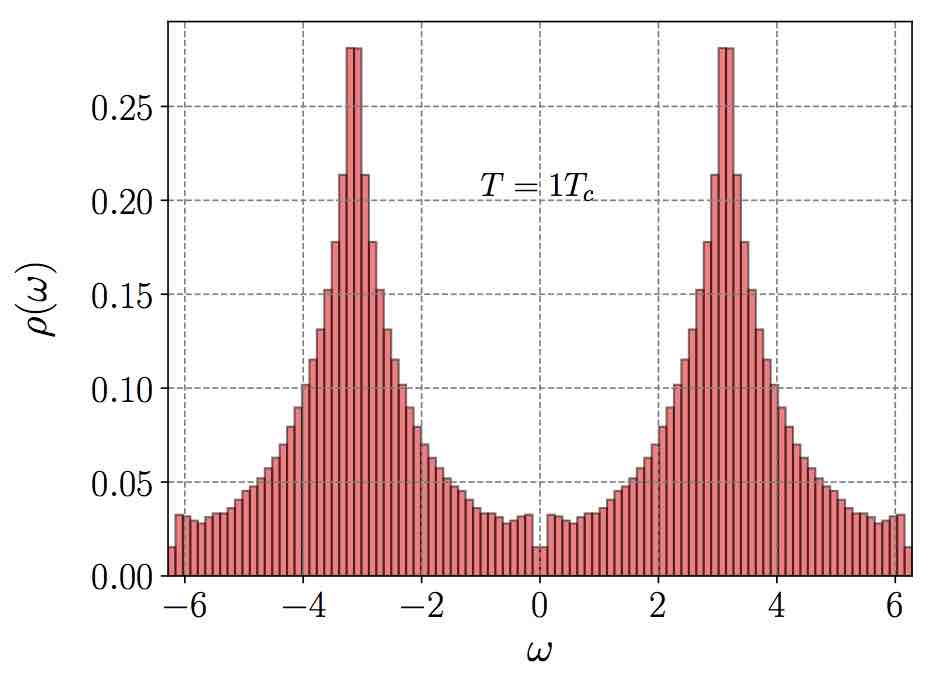} }
{\includegraphics[width=.44\linewidth]{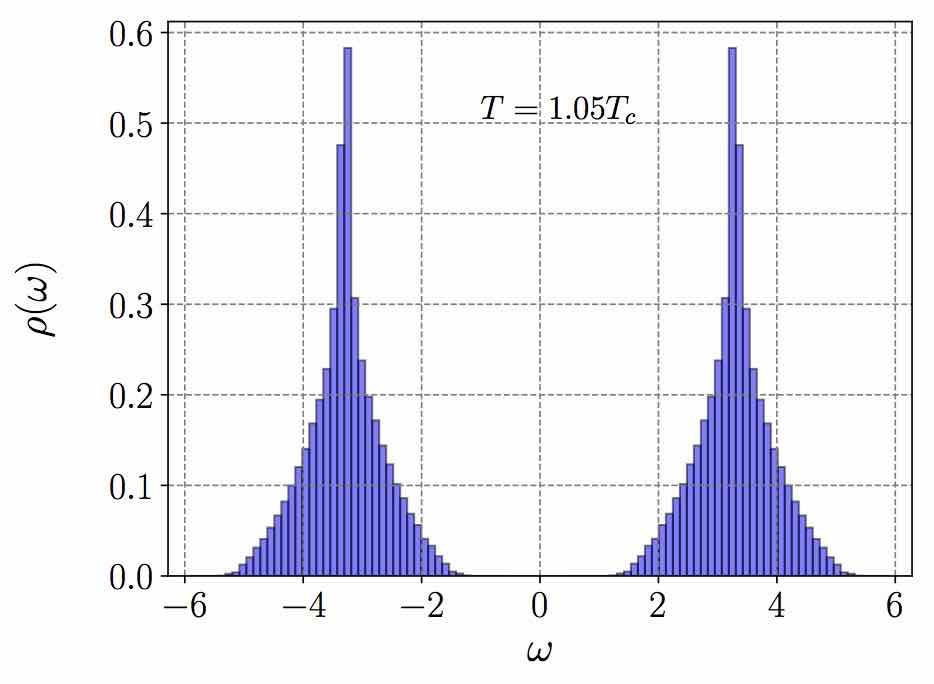}}
\caption{Distributions of Dirac eigenvalues for $T/T_c = $ (a) 1 
 and (d) 1.2, respectively. }
\label{fig_omegas}
\end{figure*}

Considering only the $n=\pm1$ case, so that $\omega_i = \pm \pi T - \lambda_i$, we get the distributions shown in Fig. \ref{fig_omegas} (a), (b), (c), and (d) for $T/T_c = 1,\, 1.05,\, 1.1, { and } 1.2$, respectively. 

The first, and most important, thing to notice is that when $T=T_c$, the eigenvalue distribution has a  finite density at $\omega=0$ (see Fig. \ref{fig_omegas}), which indicates the nonzero value of the chiral condensate; there is no gap in the spectrum present at $T=T_c$. (A small dip seen around zero is a consequence of finite size of the system,  well known and studied on the lattice and in topological models. It should be essentially ignored in extrapolation to zero.)

Let us summarize this section as follows. the mechanism of chiral symmetry breaking based on monopoles is as follows. A single monopole (or anti-monopole) generates additional quark and antiquark bound states. At high temperatures, the monopoles have large mass and the probability of hopping is therefore low. The 4d Dirac operator eigenvalues are well localized near the fermionic Matsubara frequencies $2\pi T(n+1/2)$. Using the condensed matter analogy, one may say that a matter is an insulator.
However, as $T$ decreases toward $T_c$, the amplitudes of quark ``hopping" from one monopole to an antimonopole (and vice versa) grow. Eventually, at some critical density, quarks become ``collectivized" and are able to travel very far from their original locations. The physics of the mechanism is similar to insulator-metal transition in condensed matter under pressure.

\section{More on fermions bound to monopoles, in the SUSY world and perhaps beyond$^*$} \label{sec_unusual_confinement}
  
  This is an advanced topic section on a very specific but fascinating topic, namely the possible role of 
  $multi-monopole$ states and BEC condensation. 
  
  My interest to the subject was inspired by 
  the second Seiberg-Witten work, on $\cal{N}$=2 QCD  \cite{Seiberg:1994aj}, on
%
the supersymmetric QCD with fundamental quarks/scalars.
As the first, it deals with the simplest   $N_c=2$ color group $SU(2)$, but now with $N_f$ flavors of quark-squark multiplets.
 Addition of
massless quarks are restricted to only 4 cases: $N_f=4$ already has zero beta function and is thus
  a conformal theory, as already was mentioned before. 

The case of  $N_f=3$ quark/squark flavors is ``normal", with  the asymptotic freedom. 
Seiberg and Witten predicted two district singularities
on the moduli space, which correspond to the following particles becoming massless: \\
(i)  a quartet  of states with magnetic charge $n_m=1$ and electric charge $n_e=0$;  \\
 (ii) a singlet with $n_m=2,n_e=1$.  
 
  Various SUSY breaking terms would transform those singularities into two non-equivalent vacua,
with two different confinement
phases.
 The second singularity thus induces a ``dual superconductor" in which not just a monopole,
 but a {\em monopole pair} Bose condenses! 
 One may wander what should this state with charges $n_m=2,n_e=1$ be,
 whether it can be identified semiclassically in weak coupling, and why does it only 
 appear in the  $N_f=3$ case?
 
%

  Studies of two- (and
 multi-) monopole states lead me to 
 an  influential paper by Sen \cite{Sen:1994yi} just predated Seiberg-Witten
works.

   Recall that  if there are $k$
   monopoles, there should be 4k-dimensional moduli space, which has the form
   $R^3 S^1 M_{4k-4}$ after the global shifts and phase is separated from the
   relative coordinates.   Let $X^\alpha$ be the collective coordinates and $\lambda^\alpha$
   its fermionic counterparts required by supersymmetries. For the $\cal{N}$=4 theory the Lagrangian of the
   so called {\em supersymmetric quantum mechanics} on the moduli space takes the form
   \be
   L=  ({1\over 2}) g_{\alpha\beta} \partial_0 X^\alpha \partial_0 X^\beta
   +({1\over 2}) g_{\alpha\beta} \bar{\lambda}^\alpha iD_0 \lambda^\beta
   +({1\over 12}) R_{\alpha\beta\gamma\delta} \bar{\lambda}^\alpha \lambda^\beta \bar{\lambda}^\gamma \lambda^\delta
   \ee   
where $g_{\alpha\beta} $ is the metric on the moduli spaces, the covariant derivative
$D_0 \lambda^\beta=\partial_0\lambda^\beta +\Gamma^\beta_{\alpha\gamma} \partial_0 X^\alpha \lambda^\beta$ contains the Crystoffel connection  and the
last term contains the full Riemann tensor, calculated from $g_{\alpha\beta} $ in a standard way. Of course, global coordinate and phase $R^3 S^1$ part is flat and do not
have such additions, which only appear for relative motion.

   The metric is  a non-trivial function of the coordinates $X$, explicitely 
   known for the famous 4d Atiyah-Hitchin manifold of the two monopoles.
     The fermionic part can in principle be 
rewritten in terms of creation-annihilation operators and quantized in a usual way.
There are ``fermion-empty" states, then one-fermion, 2-fermion etc all the way to 
maximal number of fermions one can put in. 
   It has been pointed out by Witten in the index paper of 1982, that such 
    $p$-fermion states for  a supersymmetric sigma model on a manyfold
   correspond to the $p$-differential form. The number of harmonic (zero energy)
   forms is known in mathematics as Betty number $B_p$ and is a topological
   property of the manyfold itself. $\sum (-)^p B_p$ gives the Euler characteristics. 
   Thus the number of bound multi-monopole states depends only on topology
   of the moduli spaces!

It turns out that 
 in the case  of 
  the Atiyah-Hitchin two-monopole manyfold 
the only nonzero Betty number is $B_2=1$: thus there is a $single$ zero energy  state,
corresponding to  such molecular state, with $n_m=2$.
This state was explicitly found by
\cite{Sen:1994yi}, who
  argued that populating gluinoes associated with the  ``trivial" coordinates provides exactly 16 states needed for completing the supermultiplet. Then he noticed that while the maximal
$p$ (and thus number of fermions) is the dimension of the manifold (equal to 4),
one can always convolute with the epsilon and come to $4-p$ form and get a Hodge-dual
solution. Since he knew it would be a single solution, it must be a selfdual case,
with $p=4-p$ or $p=2$.  This means that we need to find an antisymmetric function
of 2 variables, the wave function of two gluinoes. The function itself is a bit technical
and I would not give it here, but only notice that the Riemann term is crucial and at
large distances between the monopoles (when it goes to zero) the fermionc 2-gluino wave function exponentially go to zero as well. 

  Going to less supersymmetric theories and introducing fundamental quarks/squarks
  lead to similar but more complex supersymmetric quantum mechanics on the manyfold,
  see Gauntlett \cite{Gauntlett:1993sh} .  For $\cal{N}$=2 QCD  there is also a ``molecular" state made of two monopoles
 bound together, like the Sen's state, but now by (at least) 3 quarks. Its existence has been shown in \cite{Gauntlett:1995fu} 
 using appropriate index theorem and properties of the Atiyah-Hitchin manyfold:
 but as far as I know, its wave function was not obtained, and thus the normalizability (both at large distances and near the ``bolt" or hole in the manifold) may still be problematic.

What these examples of what I call the ``unusual confinements" tell us? Well, in weak coupling domain the monopoles are heavy, and monopole
molecules like the ones just discussed are of course twice heavier than one monopole.
The molecular state is handicapped at the start.
But, when one moves along the $v$ plane toward the stronger coupling, the binding energy of 
this molecular state seem to grow so much, as to make it the champion! Indeed, it gets
massless and undergo Bose-Einstein  condensation $before$ any of the single-monopole states!
 This binding is so large because of (a bit mysterious) coupling to large curvature of the 
 2-monopole space.
 
 Let me end by asking an (so far unanswered) question: {\em are there 
 other unusual confinement cases in non-supersymmetric theories?}
 The natural place to look for such phenomena is the theories with the 
maximal  $N_f$ close to the conformal window, because they have confinement at
stronger coupling, with longer RG flow.

%
%
%
%
%

\chapter{   Semiclassical theory based on Euclidean path integral} \label{chap_semi}
The quantum mechanics courses include semiclassical methods based on certain
representation of the wave function, starting with the celebrated Bohr-Sommerfeld 
quantization condition, applied to the oscillator and hydrogen atom, and
the Wentzel-Kramers-Brillouin
 (WKB) approximation, developed in 1926.  Unfortunately, subsequent study has shown that
 generalization of those to system with more than one degree of freedom, as well as to
 systematic order-by-order account for quantum fluctuations are difficult.

However, in this book we will only use quantum mechanical examples as pedagogical tools, while
our true interest would be in  applications to systems with many degrees of freedom, and eventually to
QFT's, at zero and finite temperatures. 
Fortunately, 
such generalizable methods exist,  based on  {\em the Feynman path integrals}.  Such methods are
the technical basis of the physics we will focus on below.

For pedagogical reasons, we will 
deviate from historical path of the development and
start with a version of semiclassical theory recently developed in \cite{Escobar-Ruiz:2016aqv}
for the {\em density matrix}, and only later will move to (somewhat technically more involved)
semiclassical theory of the tunneling effects, described by quantum-mechanical $instantons$.

\section{Euclidean path integrals and thermal density matrix} \label{sec_semi}
\subsection{Generalities} 
By definition the Feynman path integral gives the {\em density matrix} in coordinate representation,
see e.g. a very pedagogical book \cite{FH_65}
\be \label{denmat}
\rho(x_i,x_f,t_{tot})\ =\ \int_{x(0)=x_i}^{x(t_{tot})=x_f} Dx(t) e^{i\,S[x(t)]/\hbar} \ .
\ee
This object is a function of the initial and final coordinates, as well as the time needed for the transition between them. Here $S$ is the classical action of the system we study, e.g. for a particle of mass $m$ in a static potential $V(x)$ it is
\[  S \ = \ \int_0^{t_{tot}}dt\, \bigg[ \frac{m}{2}{\bigg(\frac{dx}{dt}\bigg)}^2 - V(x)    \bigg] \ ,\]
Feynman had shown that the oscillating exponent of it on the path provides the correct weight of the paths integral (\ref{denmat}).

For reference, the same object can also be written in forms closer to standard quantum mechanics courses. Heisenberg would write it as a matrix element of the time evolution operator, the exponential of the Hamiltonian\footnote{We here assumed that motion happens in time-independent potential: otherwise it would be time-ordered exponential.
}
\be \rho(x_i,x_f,t_{tot})\ =\langle x_f | e^{i\hat H t_{tot}} | x_i \rangle \ee
between states in which particle is localized at two locations considered. 

Schreodinger set of stationary states $\hat H  | n \rangle = E_n | n \rangle$ can also be used
as the state basis. Because Hamiltonian is diagonal in this basis, there is a single (not double)
sum over them
  \be \rho(x_i,x_f,t)\ =\sum_n \psi_n^*(x_f)  \psi_n(x_i ) e^{i E_n t} \label{psipsi}\ee
with $\psi_n(x)= \langle n | x \rangle$. 

Oscillating weights for different states are often hard to calculate, and one may
wander if it is possible to perform analytic continuation in time to its Euclidean version
with $i$ absorbed into it. For reasons which will soon be clear, we will
also define this imaginary time on a circle with circumference $\beta$
 $$\tau=i\,t\in [0,\beta]$$
In this way we will be able to describe
 $quantum+statistical$ mechanics of a particle in a heat bath with temperature $T$
 related to the
circle circumference $$\beta={\hbar \over T}$$
  Such periodic time is known as the Matsubara time. 
  
  Indeed the expression (\ref{psipsi}) will look as 
    \be \rho(x,x,t)\ =\sum_n |\psi_n(x)|^2 e^{- E_n /T} \ee
    combining quantum-mechanical probability to find particle at point $x$ with the thermal weight. Taking integral over all $x$ and using normalization of the weight functions
    one find the expression for thermal partition function 
    \be Z=\sum_n e^{- E_n /T} \ee
    
 While in this chapter we will focus on the zero temperature limit and the ground state
 (vacuum), it is beneficial to view quantum mechanics as the  limit $T\rightarrow 0,\beta\rightarrow \infty$.

   So, returning to the path integral, 
   the expression  we will use below would be Feynman path integral, but (i) taken over all $periodic$ paths, with the same endpoints, and (ii) with Euclidean or rotated time.
   The probability to find particle at certain point is then  
\be
P(x_0,t_{tot}) =\int_{x(0)=x_0}^{x(\beta)=x_0} Dx(\tau) e^{-S_E[x(\tau)]/\hbar} \ .
\label{P}
\ee


Note here the exponent is not oscillating, including with the minus sign and the so called Euclidean action
 \be S_E \ = \ \int_0^{\beta}d\tau\, [ \frac{m}{2}{(\frac{dx}{d\tau})}^2 + V(x)] \ee
 in which the sign of the potential is reversed and the time derivative are understood to be over $\tau$.

There are $two$ basic approaches to thermodynamics, based 
 of these expressions. While the particular formulae for the statistical sum 
 (and other quantities) obtained by them look different,
 with different dependencies on the temperature and other parameters of the problem,
  if the sums are exact one can prove  that they in fact lead to exactly {\em the same} results.  
 In some applications this proof is related to the Poisson summation formula,
and therefore the phenomenon is known as the {\em Poisson duality} of two approaches.   
 
 One approach, which one may call a $Hamiltonian$ one,
use the standard definition of the density matrix in terms of stationary states,
the eigenstates of the Hamiltonian with definite energy $\hat H | n \rangle =E_n | n \rangle $
\be
P(x_0,\beta)=\sum_n |\psi_n(x_0)|^2 e^{-E_n \beta} \ ,
\ee
The sum over stationary states  is obviously better convergent in the case of large $\beta$, or low $T$.
In the limit $\beta \rightarrow \infty$ only  the lowest -- the ground state dominates
\be P(x_0,\beta\rightarrow \infty) \sim |\psi_0(x_0)|^2 \ee

Another approach, which one may call a $Lagrangian$ one, looks for the periodic paths
with the minimal action. The simplest of such paths is obviously  those for which
particles do not move at all, $x(\tau)=const$! Such path would dominate  
in the  case of small Matsubara circle, $\beta\rightarrow 0$ (or high $T$)\footnote{ Note that it is
opposite to the limit discussed above for the Hamiltonian approach.}
 If one  ignores the time dependence and velocity on
 the paths, there is no kinetic term and only the potential one in the action contributes. So, 
\be
P(x_0,\beta) \sim e^{- \frac{V(x_0)}{T}}\ ,
\ee
which corresponds to classical\footnote{Note that if we would keep $\hbar\neq 1$, the one in
$\beta$ and in the exponent $exp(-S/\hbar)$ would cancel out, confirming the classical nature of this limit. 
} thermal distribution for a particle in a potential $V$. 

In general, the periodic paths on the circle falls into the topological classes, 
depending on the number of rotations -- or {\em the winding number} $n_w$ --
the path makes. The time Fourier transform 
of such paths are described by a $discrete$ Fourier series, with  discrete
{\em Matsubara frequencies } $2\pi n_w/\beta$. 
Needless to say, the general expression for the statistical sum including $all$ Matsubara frequencies is still exact, valid  for any $T$. 

\subsection{The harmonic oscillator} 

The density matrix for this example has been calculated by the path integral
by Feynman himself \cite{FH_65}, and it is impossible not to mention this result. The integrals one
encounter, using the definition of   the path integrals, are all Gaussian, and thus the results
can be obtained exactly, without any approximations.

 The harmonic oscillator is a particle with mass $m$ moving in a one-dimensional
 potential \be V={m^2 \Omega^2 \over 2} x^2 \ee
Feynman's result for the transition amplitude from the initial 
point $x$ to the final poin $y$ rotated into the Euclidean time $\tau$ has the form  
\be G_{osc}(x,y,\tau)= \sqrt{m \Omega \over 2 \pi \hbar \sinh \Omega \tau}
exp\left[-({m \Omega  \over 2 \hbar \sinh \Omega \tau})((x^2+y^2)\cosh(\Omega \tau)-2 x y)\right]\ee

Although it is not very transparent yet at this point, let us note that
the expression in the exponent has a  simple physical meaning: it is the classical action
$S[x(\tau)]/\hbar$ for the classical path, connecting the points.  The pre-exponet factor
includes all quantum/thermal fluctuations around this classical path. All semiclassical
expressions for the amplitude we will get below will have such form, although only for
the harmonic oscillator (and a couple of other related problems, like motion in magnetic field)
such expressions are exact.

The diagonal element of the density matrix, or the 
 probability to find a particle at the point $x$ corresponds to periodic paths, as
 we argued above. 
So, setting $y=x$ and $\tau=\beta=\hbar/T$ one finds that the particle distribution of a harmonic oscillator at $any$
temperature has Gaussian  form
\be P(x)=\sqrt{m \Omega \over 2 \pi \hbar sinh(\hbar \Omega \beta)} exp\left(- {x^2 \over 2\langle x^2 \rangle}\right) \ee
%
The (temperature-dependent) width is given by
\be \langle x^2 \rangle={1\over 2 m\Omega} \coth \left({\Omega\over 2T}\right) \label{eqn_x2}\ee
This expression, which we will meet a lot later in the book, has two important limits. At small $T\rightarrow 0$ 
the width corresponds to the quantum mechanical
 ground state wave function $\psi_0(x)$ of the oscillator. In the opposite limit of high $T\rightarrow \infty$
it corresponds to the {\em classical} thermal result 
$\langle x^2 \rangle={T\over m\Omega^2}$. 
Let me rewrite this expression with $\coth$ once again, in order to elucidate its physical nature. 
Since for harmonic oscillator the total energy is just twice the potential energy,
 which is related to mean $\langle x^2 \rangle$, 
 we also have an expression for the mean energy of the oscillator at temperature $T$.  It can be put into the familiar ``physical" form
\be \langle E \rangle= \Omega \left( {1 \over 2} +  {1 \over e^{\Omega/T}-1}\right) \ee
Now one sees the meaning of the two terms in the bracket: they are  the  energies corresponding to ($T$-independent) 
 zero-point  quantum oscillations (familiar from the QM courses) plus
 the energy of the thermal excitation (familiar from the SM courses).
  Note that we automatically get correct Planck (or Bose) distribution from the transition amplitude in Euclidean time.

\section{Euclidean minimal action (classical) paths: fluctons}  \label{sec_flucton}

Before we go into technical detail, let us clarify the goals and the setting in which semiclassical approximation will be used. Imagine, for pedagogical reasons, a particle
in a potential $V(x)$. Without quantum/thermal fluctuations, a classical particle would be located at its minimum $x_{min}$
(for now, let it be the only one). Including those, one however finds certain nonzero probability
$P(x)$ for a particle to be at any point. In Euclidean time path integral formalism, this probability
is given by an integral over periodic paths, which start and end at $x$.  Since the weight is
$exp\big(-S[x(\tau)]\big)$, the path with the smallest action should give the largest contribution.  Furthermore,
we know how to find such a path: it satisfies classical (Euclidean) equation of motion, and this is what we will do below in this section. The semiclassical approximation  -- the $dominance$ 
 of this path -- would be justified, as soon as the corresponding action is large
\be S_{cl} \equiv S[ x_{cl}(\tau)] \gg \hbar \ee

Such classical paths were called {\em fluctons} in \cite{Shuryak:1987tr}. In order to find them note first
that putting imaginary unit into time flips the sign of the kinetic energy. For the equation of motion it is the same as flipping the sign of the  potential, so that its
minimum becomes a maximum.

\begin{figure}[b]
\begin{center}
\includegraphics[width=10cm]{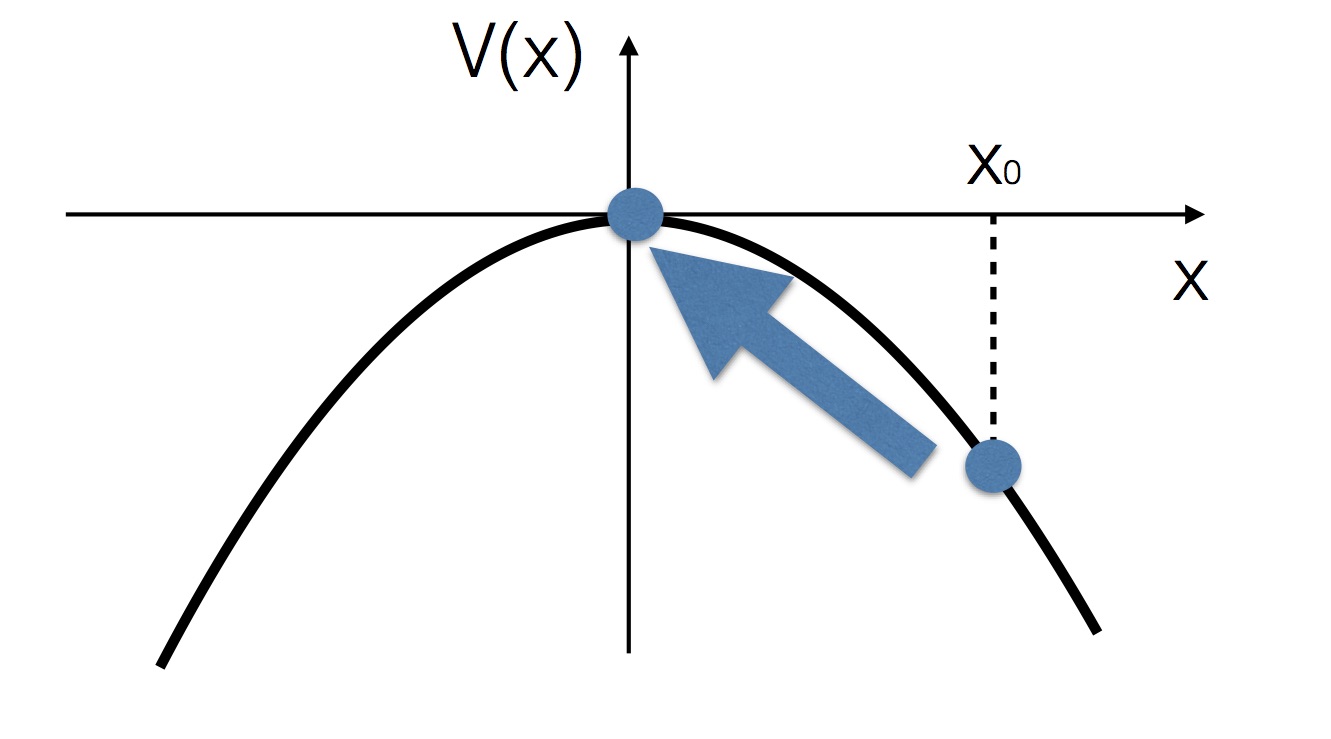}
\caption{Sketch of the flucton path climbing toward the (flipped)  minimum of the potential.}
\label{fig_fluct}
\end{center}
\end{figure}

The paths should have Euclidean time period $\beta=\hbar/T$. For simplicity, 
let us start with ``cold" QM, or vanishingly small $T$,  $\beta\rightarrow \infty$.
Little thinking of how to arrange a classical path with a very long period 
leads to the following solution: the particle should roll to the top of the 
(flipped) potential with exactly such energy as to sit there for very long time, before
it will rall back to the  (arbitrary) point $x_0$ from which the path started. 
The classical paths corresponding to relaxation toward the potential bottom take the form of a path
{\em ``climbing up"}  from arbitrary point $x_0$ to the  maximum, see Fig. \ref{fig_fluct}

Let us start with the harmonic oscillator, as the unavoidable first example.
For simplicity, let us use units 
in which the particle  mass $m=1$ and the
oscillator frequency $\Omega=1$, so that our (Euclidean) Lagrangian
\footnote{Note again the flipped sign of  the potential term: in Minkowski time the potential has sign minus.}
 is written as
\be
  L_E = \frac{\dot{x}(\tau)^2}{2} +  \frac{x(\tau)^2}{2} \ .
\ee
Because of this sign, in Euclidean time $\tau$ the oscillator does not oscillate $e^{it}$ but relaxes $e^{-\tau}$.  For harmonic oscillator, the classical equations
 of motion (EOM) are of course not difficult to solve: but
 it is always easier to get solutions using energy conservation.
 Since we are interested in solution with zero energy $E=0$ they correspond to $\dot{x}^2=2V(x)$. The boundary conditions are $x_0$
 at $\tau=0$ plus periodicity
on a circle with circumference $\beta$. This solution is
\be
   x_{flucton}(\tau)\ =\ x_0\ \frac{\left( e^{\beta - \tau}\ +\ e^\tau \right)}{e^\beta + 1}
    \ .
\ee
defined for $\tau \in [0,\beta]$. The particle moves toward $x=0$ and reach some minimal value, at  $\tau=\beta/2$, and then returns to the initial point $x_0$ again at $\tau=\beta$. 
 Due to periodicity in $\tau$, one may shift its range to  $\tau \in [-\beta/2,\beta/2]$:
The minimal value at $\tau=\beta/2$ $$ x_{min}={x_0 \over cosh(\beta/2)} \rightarrow_{\beta\rightarrow\infty} 0 $$
is exponentially small at low temperature: climbing to the potential top at $x=0$ is nearly accomplished, if the period  is large.

The solution in the zero temperature or   $\beta\rightarrow \infty$ limit
simplifies to  $x_0 e^{-| \tau |}$.
In the opposite limit of small $\beta$ or high $T$, there is no time to move far from $x_0$,
so in this case the particle does not move at all.

The  classical action of the flucton path is
\be
   S_{flucton}\ =\ x_0^2 \ \tanh\bigg(\frac{\beta}{2}\bigg) \ ,
\ee
it tells us that the particle distribution
\be
  P(x_0) \sim \exp\left(- \frac{x_0^2}{\coth (\frac{\beta}{2})}\right) \ ,
\ee
is Gaussian at any temperature. Note furthermore, that the width of the distribution
\be
 <x^2>\ =\ \frac{1}{2} \coth \bigg(\frac{\beta}{2}\bigg)\ =
     \ \frac{1}{2} + \frac{1}{e^\beta -1}\ ,
\ee
can be recognized as the ground state energy plus one due to thermal excitation, which
we already mentioned at the beginning of the chapter. So, we have reproduced 
 well known results for the harmonic oscillator, see e.g. Feynman's Statistical Mechanics \cite{Feynman_SM}.

{\it (II).}\  Our next example is the symmetric power-like potential
\be
\label{V2N}
   V\ =\ \frac{g^2}{2}\  x^{2N}\ ,\quad N=1,2,3,\ldots \ ,
\ee
for which we discuss only the zero temperature $\beta=1/T \rightarrow \infty$ case.
The (Euclidean) classical equation at zero energy $\frac{\dot{x}^2}{2} = V(x)$ has the following solution
\be
   x_{fluct}(\tau) =  \frac{x_0}{\left(1\ +\ g (N - 1) x_0^{N - 1} |\tau| \right)^{N - 1}}
   \ ,\ x_0>0,
\label{eqn_x2N}
\ee
with the action
\be
     S[x_{fluct}]\ =\   \frac{2\,g\,x_0^{N+1}}{N+1} \ ,
\ee
hence
\be
\label{Px_o}
    P(x_0) \sim \exp \left(-\frac{2\,g\,x_0^{N+1}}{N+1} \right) \ ,
\ee
which is in a complete agreement with WKB asymptotics at $x_0 \rar \infty$\

{\it (III).}\ The third example is the anharmonic potential of the kind
\be
\label{AHpot}
     V\ =\ \frac{1}{2}  x^{2}\ (1+g\,x^2)\ , \qquad g>0\ ,
\ee
at zero temperature $\beta=1/T \rightarrow \infty$. The classical flucton solution with the energy $E=0$ is given by
\be
     x_{fluct}(\tau) \ = \ \frac{\sqrt{g}\,x_0}{ \cosh(\tau) +\sqrt{1+g\,x_0^2}\,\sinh(\tau)} \ ,
\ee
which leads to the flucton action
\be
\label{AHaction}
   S[x_0]\ =\  \frac{2}{3}\,\frac{  {(1+g\,x_0^2)}^{\frac{3}{2}} -1}{g}   \ .
\ee
In the limit $x_0\rightarrow \infty$ we obtain
\be
  S[x_{fluct}(\tau)] \ = \   \frac{2 \sqrt{g}}{3}x_0^3+\frac{1}{\sqrt{g}}x_0-\frac{2}{3 g} + O(\frac{1}{x_0}) \ .
\ee
in complete agreement with the asymptotic expansion of the ground state wave function squared .

\begin{figure}[b]
\begin{center}
\includegraphics[width=10cm]{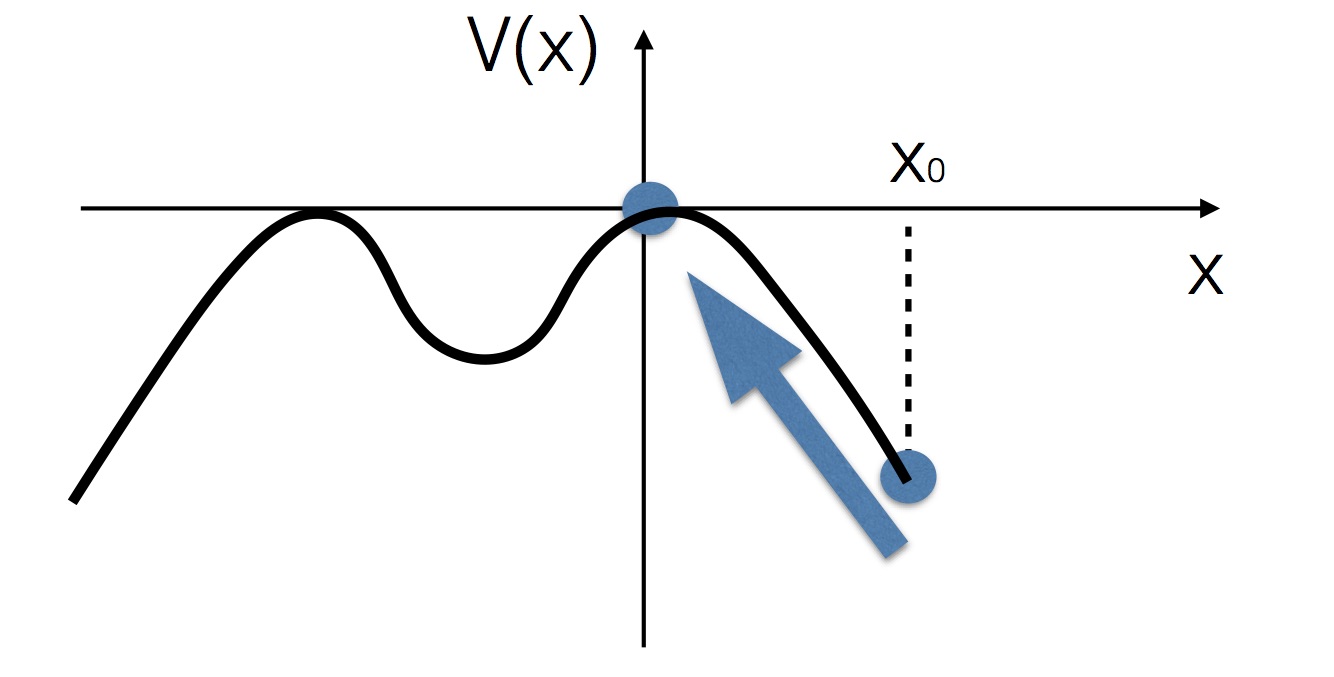}
\caption{Sketch of the flucton path climbing toward the (flipped) double-well potential minimum .}
\label{fig_fluct2}
\end{center}
\end{figure}

 \begin{figure}[htbp]
\begin{center}
\includegraphics[width=10cm]{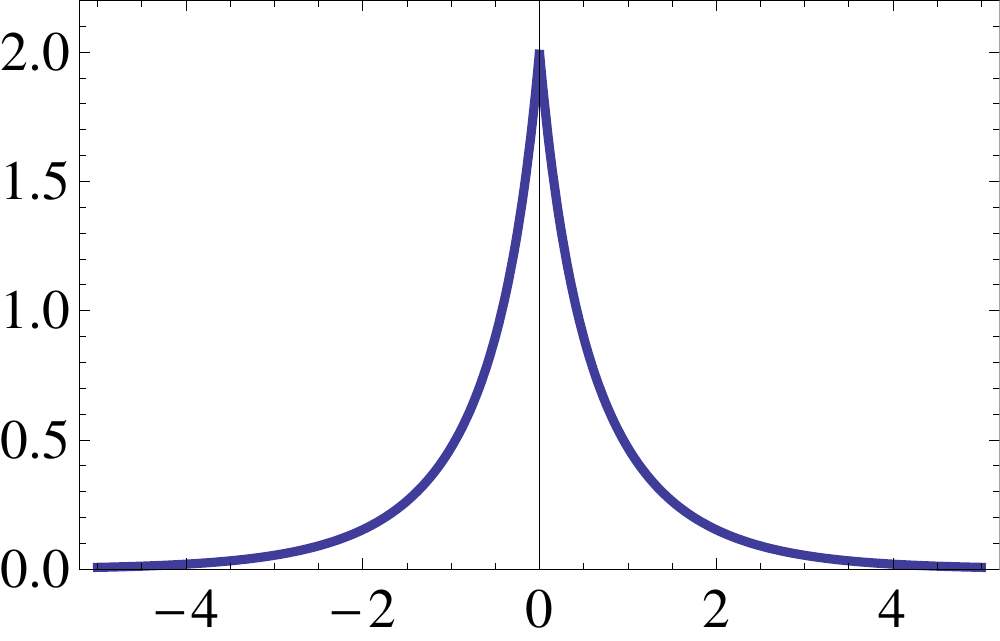}
\caption{Time dependence of the classical flucton solution $y_{fluct}(\tau)$ for $x_0=2,\lambda =0.1$.}
\label{fig_flucton}
\end{center}
\end{figure}

(IV) However, for the most detailed studies we select another example,
 the {\em quartic one-dimensional potential}, also known as the double-well problem
\begin{equation}
   V(x) \ = \  \la\,{(x^2-\eta^2)}^2 \ ,
\end{equation}
with two degenerate minima. Tunneling between them 
will be subject of our subsequent studies later.


Standard steps are selecting units for $\eta$ such that 
motion in a single well are in first approximation like in harmonic 
oscillator with frequency $\om=1$. We will also  shift the coordinate by 
\be
 x(\tau) = y(\tau) + \eta \ ,
\ee
so that the potential (21) takes the form
\be
 V = \frac{y(\tau)^2}{2} \left(1 + \sqrt{2\la} y(\tau) \right)^2 \ ,
\ee
corresponding to harmonic oscillator well at small $y$. The second minimum is
shifted to negative $y$. 

The {\em flucton}  path for the case $x_0$ is outside of a maximum (see Fig.\ref{fig_fluct2})
now takes the form
\be
 y_{fluct}(\tau) = \frac{x_0}
 {e^{|\tau |}(1 + \sqrt{2\la}\ x_0 ) - \sqrt{2\la}\ x_0 } \ ,
\label{eqn_fluct}
\ee
We remind that  in  zero $T$ case, or infinite circle $\beta\rightarrow \infty$, $\tau \in (-\infty,\infty)$,
and solution exponentially decreases to both infinities, see Fig.\ref{fig_fluct}.
Its generalization to finite $T$ is straightforward.

The action of this solution is
\be
  S[y_{fluct}]= x_0^2 (1 + \frac{2 \sqrt{2\la} x_0}{3})\ ,
\ee
and thus in the leading semiclassical approximation the probability to find the particle at $x_0$ takes the form
\be
  P(x_0)\sim \exp\left(-x_0^2  - \frac{2\sqrt{2\la}}{3} x_0^3\right)
\ee
In the weak coupling limit only the first term remains, corresponding to Gaussian ground state wave function of the harmonic oscillator. In the strong coupling limit the second term is dominant, and the distribution then corresponds to well known cubic dependence on the coordinate. These classical-order results are of course the same as one gets from a standard WKB approximation.

%
%

\section{Quantum/thermal fluctuations in one loop}
The paths close to classical ones can be written as
\be y(\tau)=y_{cl}+f (\tau)\ee
Substituting it to the action, one can expand the result in powers of $f$, which is presumed to be small. Since classical paths are extrema of the action, 
the expansion always starts from the second order terms $O(f^2)$.

Taking the path integral over fluctuations around the classical path, in the Gaussian approximation, leads to the following formal expression
\be
   P(x_0) = \frac{ \exp \left( -S[x_{flucton}]\right)}{\sqrt{\bf{Det}\,(O_{flucton})} } \times \left[1+O(two \,\,and \,\, more\,\,  loops) \right]\ ,
\label{Px0}
\ee
with the ``flucton operator" $O_{flucton} $ defined as \be
 O f\ \equiv\ -\ddot{f}(\tau)+ 
  V''(y_{fluct})  f(\tau) \ ,
\label{eqn_operator}
\ee
In the case of the flucton classical solution (\ref{eqn_fluct}) the potential of the fluctuations we put into the form
\[
 V''(y_{fluct})  =1 + W \ ,
\]
where
\be
  W=\frac{6 X (1 + X) e^{|\tau |}}{(e^{| \tau |} -X  + X e^{| \tau |})^2}\ .
\label{fluct_V"}
\ee
The term equal to 1 is taken out, as it correspond to harmonic oscillator. The expression
with the determinant is ``formal" because it is divergent, and get defined $via$ its (re)normalization 
to that of the harmonic oscillator. Note that at $X=0$ we return to the harmonic oscillator case.

 The classical path depends on 3 parameters of the problem, $\la, x_0$ and $\Omega$ (which we already put to 1): but in $W$ the first two appear in one combination only
\be
 X\ \equiv \ x_0 \,\sqrt{2\,\la}\ .
\label{eqn_X}
\ee
This observation will be important later. 

There are several different method to calculate the determinant of the differential operator:
we will use two of them subsequently.

{\bf Method 1} is based on straightforward $diagonalization$ of the operator. Like for a finite matrix,
this includes  tedious calculation of all its eigenvalues and eigenmodes. This was also the first
method we used, and for pedagogical reasons we will start with it.


Note that for $X>0$ we discuss, $W>0$ as well, and it exponentially decreases at large $\tau$. This potential is repulsive, and obviously it has no bound states\footnote{Another classical path, the $instanton$, much more discussed in literature, is different precisely at this point:
it leads to a zero and bound states, which lead to extra complications.}. 
At large $|\tau |$ the nontrivial part of the potential disappears and solutions have a generic form
\be
\psi_p(\tau) \sim \sin(p\,\tau +\delta_p ) \ ,
\ee
with the so called scattering phase $\delta_p$. Thus the eigenvalues of the operator $O$ are, for the double well example (\ref{fluct_V"}),
simply,
\be
   \la_p=1+p^2 \ ,
\ee
and the determinant $\bf{Det}\, O$ is their infinite product. Its logarithm is the sum
\be
    \log \,\bf{Det}\, O\ =\ \sum_n \log(1+p_n^2) \ ,
\ee
where the sum is taken over all states satisfying zero boundary condition
on the boundary of some large box.

The nontrivial part of the problem is not in the eigenvalues themselves, but in the {\em counting of levels}.  Standard 
vanishing boundary conditions at the boundary of some large box, at $\tau=L$, lead to
\be
  p_n L +\delta_{p_n} = \pi\, n \ ,\ n=1,2,\ldots \ .
\ee
At large $L$ and $n$ one can replace summation to an integral, resulting in the generic expression
\be \log \bf{Det}\, O\ =\ \sum_n \log(1+p_n^2)\
 =\ \int_0^\infty \frac{dp}{\pi}  \frac{d\delta_p}{dp}  \log(1+p^2) \ .
\label{eqn_logdet}
\ee

After using few different numerical methods for particular values of the parameter $X$, we discovered that there exist {\it exact} (non-normalized) analytic solution for the eigenfunctions of the form
\be
   \psi_p(\tau)\ \sim\  \sin\left( p\,\tau + \Delta(p,\tau) \right)\ F(p,\tau) \ ,
\label{eqn_sol}
\ee
with the following two functions
\[
 \Delta(p,\tau)= {arctan}\,\bigg[\frac{ -3 p\,(1 + 2 X)}{1 - 2 p^2 + 6 X + 6 X^2}\bigg]
\]
\[
 +\ {arctan} \bigg[\frac{N}{D}\bigg] \ ,
\]
where

\[
  N\ =\ 3 p [1 + 2  X + X^2 - X^2 e^{-2\tau}] \ ,
\]

\[
  D\ =\  (2 p^2-1)(1+X^2) -
  2 e^{-\tau} \big(2 (1 + p^2) -
  e^{-\tau} (2 p^2-1)\big) X + (2 p^2-1)e^{-2\tau} -
     4 e^{-\tau} (1 + p^2) \ ,
\]

$$
  F(p,\tau)\ =\ \frac{1}{(e^{\tau} -X + e^{\tau} X)^2}
 \times
\bigg[e^{4 \tau} (1 + 5 p^2 + 4 p^4) +
     4 e^{3 \tau} (1 + p^2) \bigg(2 - 4 p^2 + e^{\tau} (1 + 4 p^2)\bigg) X + $$ $$
     6 e^{2\tau} \bigg(3 + p^2 + 4 p^4 + 4 e^{\tau} (1 - p^2 - 2 p^4) + 
        e^{2\tau} (1 + 5 p^2 + 4 p^4)\bigg) X^2 +
     4 e^{\tau} \bigg(2 (1 - p^2 - 2 p^4) + $$ $$
        6 e^{2\tau} (1 - p^2 - 2 p^4) + 3 e^{\tau} (3 + p^2 + 4 p^4) + 
        e^{3 \tau} (1 + 5 p^2 + 4 p^4)\bigg) X^3 + \bigg(1 + 5 p^2 + 4 p^4 + $$ $$
        8 e^{\tau} (1 - p^2 - 2 p^4) + 8 e^{3 \tau} (1 - p^2 - 2 p^4) +
        6 e^{2\tau} (3 + p^2 + 4 p^4) +
        e^{4 \tau} (1 + 5 p^2 + 4 p^4)\bigg) X^4\bigg]^{1/2} \ .
$$

\begin{figure}[h]
\begin{center}
\includegraphics[width=3.0in,angle=0]{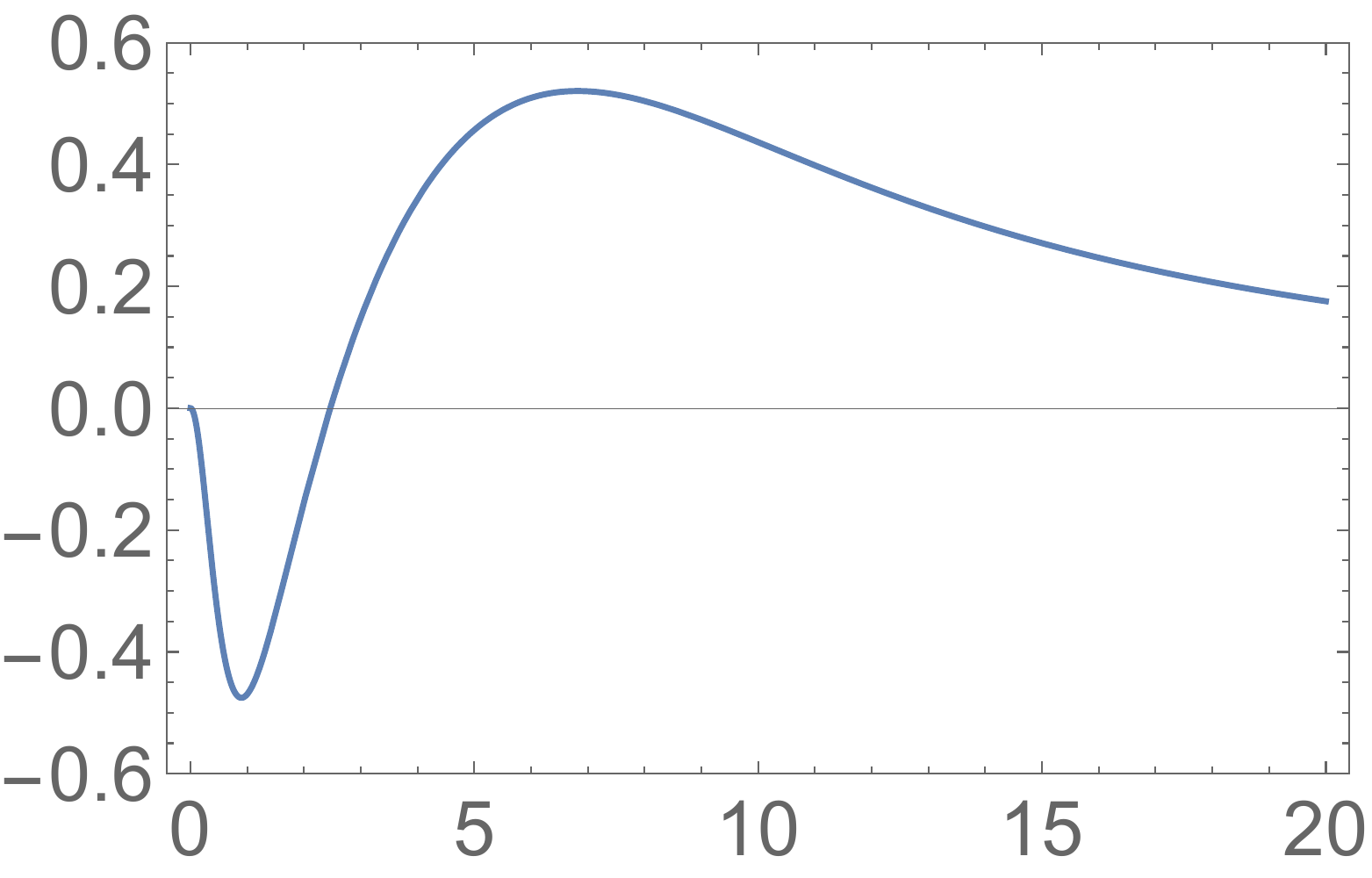}
\includegraphics[width=3.0in,angle=0]{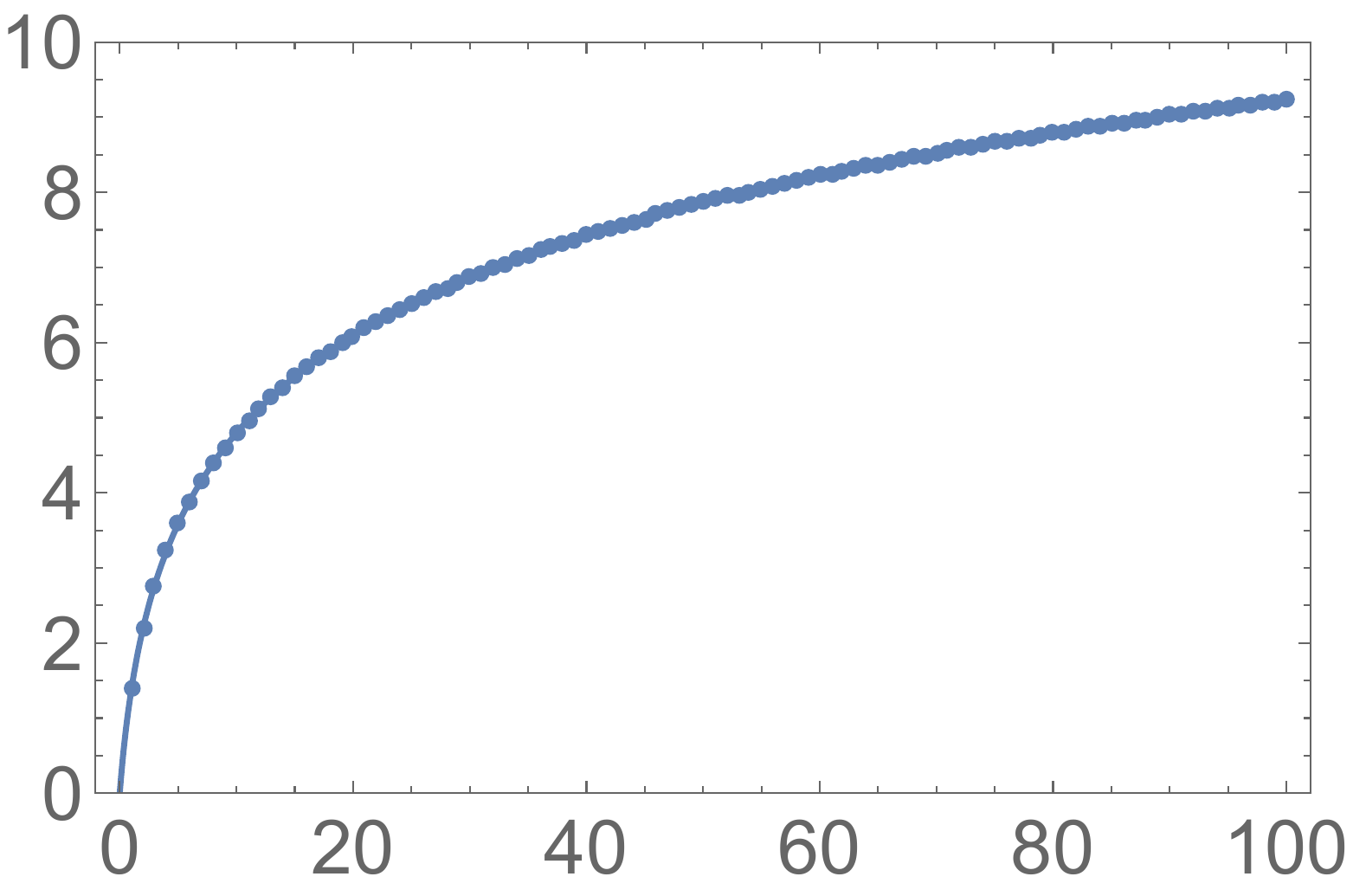}
\caption{(a)The integrand of  (\ref{eqn_logdet}), $\log(1+p^2)d\delta_p/dp$, versus $p$, for $X=4$.
(b) The integral  (\ref{eqn_logdet}) vs parameter $X$: points are numerical evaluation, line is defined in the text.
}
\label{fig_ddeltadp}
\end{center}
\end{figure}

It is important that at $\tau=0$ the solution (\ref{eqn_sol}) goes to zero:  
according to the flucton definition {\em all fluctuations at this time must vanish}. At large time, where all terms with decreasing exponents in $\Delta(p,\tau)$ disappear and
the remaining constant terms define the scattering phase
\be \label{eqn_delta}
  \delta_p = \arctan \bigg[\frac{3 p(1 + 2 X)}{1 - 2 p^2 + 6 X + 6 X^2}\bigg]
     - \arctan\bigg[\frac{3p}{1 - 2p^2}\bigg] \ .
\ee
Comments: \\[5pt]
(i) the scattering phase is $O(p)$ at small $p$; \\
(ii) it is $O(1/p)$ at large $p$ and, thus, there must be a maximum at some $p$; \\
(iii) for $X=0$ two terms in (\ref{eqn_delta})  cancel out. This needs to be the case since in this limit the
nontrivial potential $W$ of the operator also disappears;\\
(iv) at large time the amplitude $F$ (\ref{eqn_sol}) goes to a constant, as it should.\\
The $\arctan$-function provides an angle, defined modulo the period, and thus it experiences jumps by $\pi$.
Fortunately, its derivative $d\delta_p/dp$ entering the determinant (\ref{eqn_logdet}) is single-valued and smooth.
The momentum dependence of the integrand of this expression for $X=4$ is shown in Fig.\ref{fig_ddeltadp}(a).
Analytic evaluation of the integral  (\ref{eqn_logdet}) was not successful, the results of the numerical evaluation are shown by points in Fig.\ref{fig_ddeltadp}(b). However, the  {\it guess} $2 \log(1+X)$, shown by the curve
in Fig.\ref{fig_ddeltadp}(b)  happens to be accurate to numerical accuracy, and thus it must be correct. We will demonstrate that it is exact below.

Since the calculation above includes only a half of the time line, $\tau>0$, and the other half is symmetric, the complete result for the $\log Det\, O$ should be doubled.  Substituting  (\ref{eqn_delta}) to  (\ref{eqn_logdet}) we obtain a (surprisingly simple) exact result
\be
   Det\,(O)\ =\ (1+X)^4 \ .
\label{eqn_Det}
\ee

{\bf Method 2}  to calculate the determinant uses the
Green function of the operator\footnote{For finite matrices,  one also may find it easier to find the inverse matrix
rather than do complete diagonalization.}.
It satisfies the standard equation
\be
  -\ \frac{\pa^2 G(\tau_1,\tau_2)}{\pa \tau_1^2}+  V'' (y_{fluct}(\tau_1)) G (\tau_1,\tau_2)=\ \delta(\tau_1-\tau_2) 
\ee
and one can find two independent solutions of the l.h.s. being zero, and then find the Green function itself.
The result (for $\tau_1,\,\tau_2>0$) is 
\[
    G(\tau_1,\,\tau_2)\ =\ \frac{e^{-|\tau_1-\tau_2|}}{2\ {\big(e^{\tau_1}(1+X)-X\big)}^2\,{\big(e^{\tau_2}(1+X)-X\big)}^2}
    \bigg[ 8\ e^{\frac{1}{2}(\tau_1+\tau_2+3\,|\tau_1-\tau_2|)}\,X^3\,(1+X)
\]
\[
    -\ 8\ e^{\frac{1}{2}(3\tau_1+3\tau_2+|\tau_1-\tau_2|)}\,X\,(1+X)^3
    +\ e^{2\,(\tau_1+\tau_2)}\,{(1+X)}^4      -6\,e^{(\tau_1+\tau_2+|\tau_1-\tau_2|)}\,X^2\,{(1+X)}^2\, |\tau_1-\tau_2|
\]
\[
   +\  e^{(\tau_1+\tau_2+|\tau_1-\tau_2|)}\,\bigg(\,6\,X^4\,(\tau_1+\tau_2)\, +\, 12\,X^3\,(1+\tau_1+\tau_2)
 +\ 6\,X^2\,(3 + \tau_1+\tau_2)  + 4\,X - 1 \bigg)
\]

\be
\label{GDW}
  - e^{2\,|\tau_1-\tau_2|}\,X^4 \ \bigg] \ ,
\ee

\begin{figure}[h]
\begin{center}
\includegraphics[width=3.0in,angle=0]{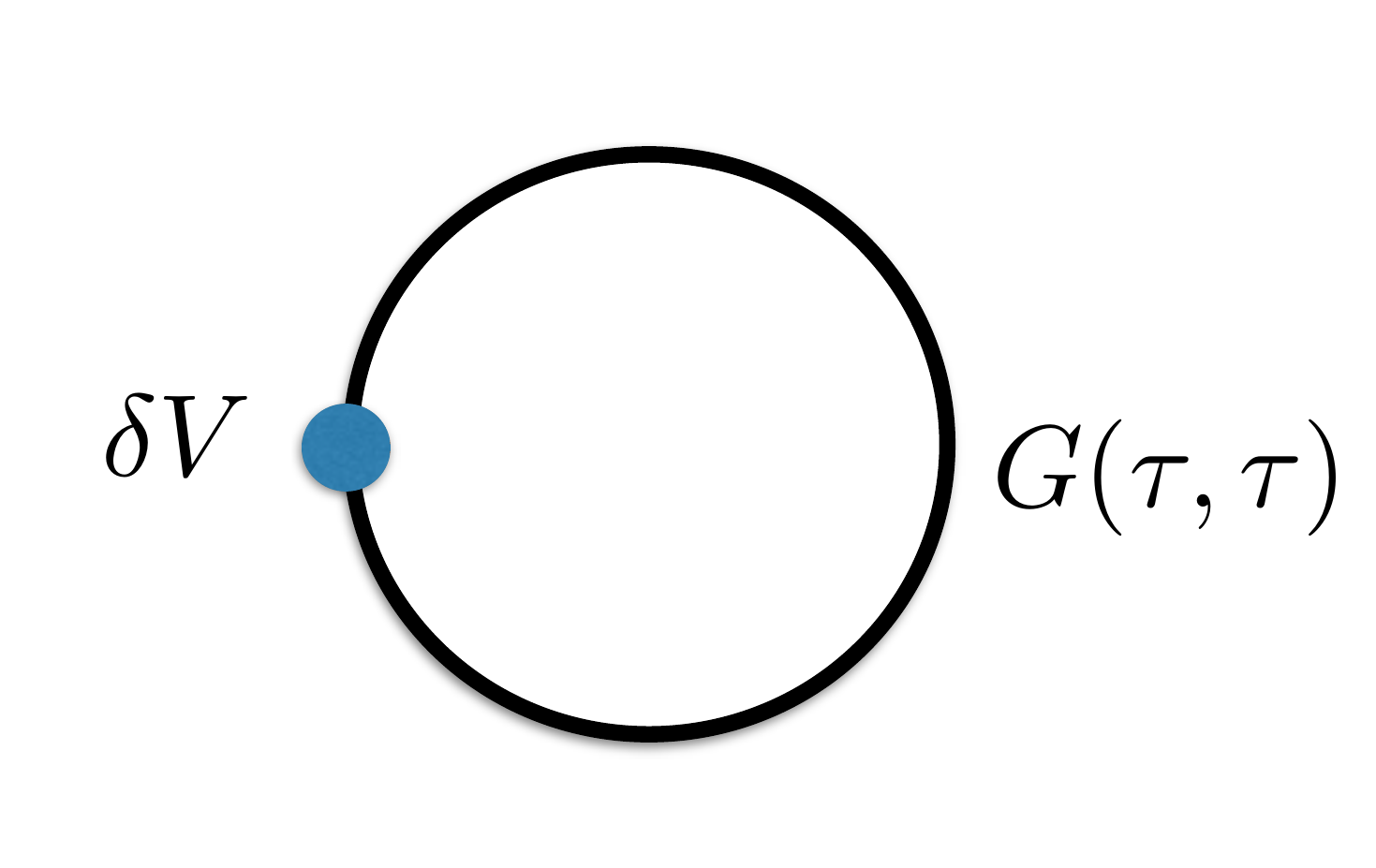}
\caption{Symbolic one-loop diagram, including variation of the fluctuation potential $\delta V$and the simplified ``single-loop"
Green function $G(\tau,\tau)$.
}
\label{fig_oneloop}
\end{center}
\end{figure}

The method  relies on the following observation.
When the fluctuation potential depends on some parameter, it can be varied. In the case at hand (\ref{fluct_V"}), the potential we write as
\[
V_{flucton}=1+W(X,\tau) \ ,
\]
depends on the combination (\ref{eqn_X}). Its variation resulting in extra potential
\be
     \delta  V_{flucton} = \frac{\pa W}{\pa X} \delta X
\ee
which can be treated as perturbation: its effect can be evaluated by the following Feynman diagram
\be
 \frac{\pa \log \bf{Det}\,(O_{flucton})}{\pa X}\ =\ \int d\tau G(\tau,\tau)
 \frac{\pa V_{flucton}(\tau)}{\pa X}\ ,
\label{relation}
\ee
containing derivative of the potential as a vertex and the ``loop",  the  Green function returning to the same time, see Fig.\ref{fig_oneloop}. This idea relates the determinant and the Green function
\footnote{The historical origin of this idea goes back to Brown and Creamer \cite{Brown:1978yj}, see also \cite{Corrigan:1979di}, for gauge theory instanton. Zarembo \cite{Zarembo:1995am} applied it for the monopole  and Diakonov et al, \cite{Diakonov:2004jn} for the calorons at nonzero holonomy. }
: if the r.h.s. of it can be calculated, the derivative over $X$ can be integrated back.

In the quartic double-well problem the ``Green function loop" is
\be
   G(\tau,\,\tau)\ =\ \frac{1}{2 \big(X - e^\tau (1 + X)\big)^4}
\ee
$$ \times \bigg(-X^4 + 8 e^\tau X^3 (1 + X) -
   8 e^{3 \tau} X (1 + X)^3 + e^{4 \tau} (1 + X)^4 +
 e^{2 \tau} (-1 +   4 X +  18 X^2 + 12 X^3 + 12 X^2 (1 + X)^2 \tau)\bigg)\ ,$$
and the ``vertex"
\be
  \frac{\pa V_{flucton}(\tau)}{\pa X}\ =\
 \frac{6 e^\tau \big(X + e^\tau (1 + X)\big)}{\big(-X + e^\tau (1 + X)\big)^3}\ .
\ee

With these expressions  one can evaluate the r.h.s. of the relation (\ref{relation}),
and adding the same expression for negative time, one gets the result
\be
     \frac{\pa \log \bf{Det}\,({O_{flucton} })}{\pa X}\ =\ \frac{4}{1+X}\ ,
\ee
which exactly agrees with the result (\ref{eqn_Det}) from the direct evaluation of the determinant using the phase shift.

\section{Two and more  loops}
Unlike the WKB-like semiclassical theory, the flucton-based version of the semiclassical theory we discuss allows for systematic derivation of higher order quantum corrections,
defined in terms of conventional Feynman diagrams. 
The vertices can be calculated from higher order terms in expansion of the action
in powers of fluctuation $f$, of the order $f^3,f^4$, etc. \footnote{Later in this chapter we will
apply similar approach to the tunneling and ``instanton" paths. In that case the quadratic
operator admits zero modes, and the Green function needs to be defined in a subspace
normal to them. As shown in  \cite{Aleinikov:1987wx,Escobar-Ruiz:2015nsa}, the Jacobian of the orthogonality condition generates additional diagrams, not following from the action. It is for this reason we postpone
discussion of instantons in this text.
}

The other ingredient of the Feynman diagram method -- the Green function inverting the
operator quadratic in $f$ -- should be calculated in a standard way, in the flucton background.  
For some examples discussed above we already have such 
 Green functions, which  has even passed a  nontrivial test  - producing the correct determinant. 
  So, 
using only standard tools from quantum field theory, the Feynman diagrams in the flucton background, one can compute the loop correction to the density matrix (\ref{Px0}) 
of any order. For definiteness, we will show here the results
for the double-well potential. 
%
 The only vertices we will need are the
triple and quartic ones, which follow from the cubic and quartic potential derivatives over $x$
\be
  v_3(\tau)\ =\
   \frac{6 \sqrt{2\,\la} \left(X+e^\tau (1+X)\right)}{-X+e^\tau (1+X)}\ ,
\ee
\be
v_4\ =\ 24\,\la \ .
\ee
The loop corrections in (\ref{Px0}) are written in the form
\[
 \left[1+O(two \,\,and \,\, more\,\,  loops) \right] \ = \  2\,\sum_{n=0}^{\infty}\,B_n\,\lambda^n \ ,
 \quad B_0=\frac{1}{2}\ ,
\]
where $B_n=B_n(X)$. 

In 1+0 dimension of time-space we discuss (1-dimensonal quantum mechanics)
there are no ultraviolet divergences. There are infrared ones, which can be cancelled out 
by  subtraction from each diagram in the flucton background its analog 
for trivial or vacuum path $x(\tau)=0$. This is done by subtracting the same expression with 
the ``vacuum vertices"
\be
  v_{3,0}\ =\  6 \, \sqrt{2\la} \ ,
\ee
\be
  v_{4,0}\ =\ 24\,\la \ ,
\ee
and the ``vacuum propagator"
\be
\label{GDW0}
G_0=G(\tau_1,\,\tau_2)\mid_{{}_{X\rightarrow 0}}\ =\ \frac{e^{-|\tau_1 - \tau_2|}}{2} - \frac{e^{-\tau_1 - \tau_2}}{2} \ .
\ee

\begin{figure}[h!]
\begin{center}
\includegraphics[width=3.0in,angle=0]{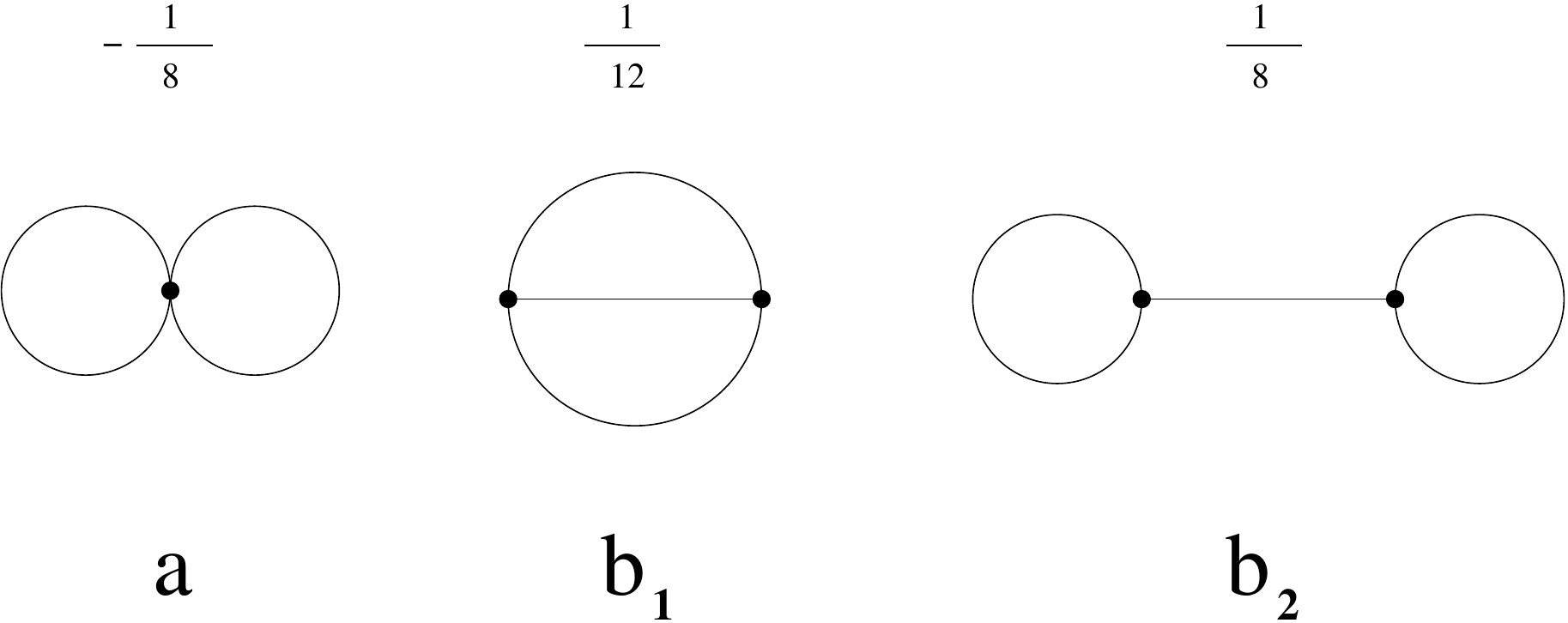}
\caption{Diagrams contributing to the two-loop correction $B_1 = a + b_1 + b_2$. The signs of contributions and symmetry factors are indicated. }
\label{2loopplot}
\end{center}
\end{figure}

The two-loop correction $B_1$ we are interested in can be written as the sum of three diagrams, see Fig.\ref{2loopplot}, diagram $a$ which is a one-dimensional integral and diagrams $b_1$ and $b_2$ corresponding to two-dimensional ones.

Explicitly, we have
\be
 a \ \equiv \ -\frac{1}{8\,\la}\,v_4\,\int_0^{\infty}[G^2(\tau,\,\tau) -
 G_0^2(\tau,\,\tau) ]d\tau  =\ \frac{3}{560 X^2 (1+X)^4}
\ee
$$   \times \bigg(24X-60X^2-520X^3 -1024 X^4 - 832 X^5 - 245 X^6
 +24 (1+X)^2 (1+2 X) (-1+6 X(1+X)) \log(1+X) $$
$$  +288 X^2 (1+X)^4 \bf{PolyLog}\left[2,\frac{X}{1+X}\right]\bigg) \ , $$
here $\bf{PolyLog}[n,z] =\sum _{k=1}^{\infty } z^k/k^n$ is the polylogarithm function and
\be
 b_1 \ \equiv \ \frac{1}{12\,\la}\,\int_0^{\infty}\int_0^{\infty}[v_3(\tau_1)\,v_3\,(\tau_2)G^3(\tau_1,\,\tau_2)
  -\ v_{3,0}v_{3,0}G_0^3(\tau_1,\,\tau_2) ]\,d\tau_1\,d\tau_2
\ee
$$
= \frac{1}{280 X^2 (1+X)^4} \times
  \bigg(-24X+60 X^2+520 X^3+1024X^4+832 X^5+245 X^6
 $$
$$ + 24 (1+X)^2 \left(1-4 X-18 X^2-12 X^3\right) \log(1+X) - 288 X^2 (1+X)^4 \bf{PolyLog}\left[2,\frac{X}{1+X}\right]\bigg) \ ,$$

\be
b_2 \ \equiv \ \frac{1}{8\,\lambda}\,\int_0^{\infty}\int_0^{\infty}\big[v_3(\tau_1)\,v_3\,
     (\tau_2)G(\tau_1,\,\tau_1)G(\tau_1,\,\tau_2)G(\tau_2,\,\tau_2)
   -\ v_{3,0}v_{3,0}G_0(\tau_1,\,\tau_1)G_0(\tau_1,\,\tau_2)G_0(\tau_2,\,\tau_2)\big]
   \,d\tau_1\,d\tau_2
\ee
$$
  = \ -\frac{1}{560 X^2 (1+X)^4} \times
   \bigg(24X-60 X^2+1720 X^3+5136 X^4+4768 X^5+1435 X^6
   $$
$$   +24 (1+X)^2 \left(-1+4 X+18 X^2+12 X^3\right) \log(1+X) +288 X^2 (1+X)^4 \bf{PolyLog}\left[2,\frac{X}{1+X}\right]\bigg) \ . $$

Adding all two-loop corrections one finds  an amazingly simple form,
\be
\label{B1}
   B_1 \, \equiv \ a+b_1+b_2 \ = \ -\frac{X (4+3 X) }{(1+X)^2} \ ,
\ee
in which all $\log$ and $\bf{PolyLog}$ terms $disappear!$\footnote{One way to prove that it must be so was found by \cite{Escobar-Ruiz:2017uhx} using generalized Bloch equation for the wave function. Note however that in explicit QFT calculations of multi-loop diagrams one also 
find  cancellations of irrational contributions, present in individual diagrams but absent 
in their sum.  Such cancellations, to our knowledge, remain mysterious unexplained phenomena, only seen from explicit calculations.} 



The combined results for the probability to find a particle
at point $x_{0}$ in the quartic double-well potential at zero temperature is, with the two-loop
accuracy
\be
\label{PX0}
    P(x_{0})\sim \frac{e^{-\frac{X^2}{2\la} - \frac{X^3}{3\la}}} {(1+X)^2} \left(1-\la \frac{X (4+3 X) }{(1+X)^2} + O(\la^2)\right) \ ,
\ee
where, we remind, $X=\sqrt{2\la}\, x_{0}$. Note that $X=-1$ is indeed a singularity of the potential, located
in the unphysical domain.

The $x_0$ dependence of  (\ref{PX0})  is plotted in Fig.
\ref{fig_compare} by the thick line.  The thin line is asymptotics derived in appendix A:
since $x_{0}$-independent constant remained unknown we  normalized it to our curve at
large distances. Their comparison shows good agreement for $x_{0}>1$.

Although derived semiclassically, and thus formally valid for large flucton action only,
our answer is also obviously correct at small $x_{0}$, where it merges with
the answer for harmonic oscillator.

Let us finally compare the results obtained with those one get from standard asymptotic analysis of the
Schreodinger eqn.
where the double-well potential in shifted coordinates we use is
\be
      V(y) = \frac{y^2}{2} + \sqrt{2\,\la}\, y^3 + \la y^4 \ .
\ee
Note that it smoothly goes to the harmonic oscillator at $\lambda\rightarrow 0$.
Introducing the phase $\phi(y)=-\log \Psi(y)$ we move to the Riccati equation,
\be
     \pa_y^2 \phi - (\pa_y \phi)^2\ =\ 2 E - 2\, V(y)\ ,
\ee
to which one can plug the asymptotic expansion at $|y| \rar \infty$ and obtain all the coefficients (cf. \cite{Turbiner:2009ns})
\[
     \phi\ =\  \frac{1}{3} \sqrt{2} \sqrt{\la} |y| y^2\ +\  \frac{1}{2}y^2\ -\ d \log|y|^2\ +
\]
\be
  \frac{1 +2 E}{2\,\sqrt{2\,\la}}\ \frac{1}{|y|} - \frac{1}{8\, \la \,  y^2}\ +\ \ldots \ ,
\label{phase}
\ee
where $d=1/2$.
The first two terms in the expansion are classical coming from classical Hamilton-Jacobi equation, log-term reflects an intrinsic property of the Laplacian: $y$ is zero mode or kernel, this term comes from determinant, asymptotically the determinant behaves like $|y|^2$, where $d$ is degree with which it enters to the wavefunction. Note that a constant, $O(x_0^0)$ term is absent: it can not be obtained from the Riccati equation containing derivatives only.
Note also that so far the energy remains undefined: to find it one needs to solve the equation to all $x$. The last terms are true quantum corrections, decreasing at large distances.
Intrinsically, this expansion corresponds to the ground state: it implies that the eigenphase $\phi$ has no logarithmic singularities at real $y$.  Quantization for the Riccati equation implies a search for solutions growing at large $y$ with finite number of logarithmic singularities at real finite $y$. For the $n$th excited state the first two growing terms in
(\ref{phase}) remains unchanged while log-term gets integer coefficient, $(n+1) \log |y|$, see \cite{Turbiner:2009ns}.

Multiplying by 2 (path integral is for density matrix, or wave function squared) one finds,
as expected, that the first two terms coincide with the classic action of the flucton.
For the determinant one needs to expand at large $x_0$
\be
\label{eqn_aa}
   \log(1+\sqrt{2\la}x_0)\ =
\ee
\[
\log(x_0) + \log(\sqrt{2\la})+ \frac{1}{\sqrt{2\la} x_0 }\ +\ \dots \ ,
\]
and observe that the leading term agrees with the $\log|y|$ term in the asymptotic expansion (\ref{phase}).

The two-loop correction $B_1\,\la$ found in the text (\ref{B1}) expands in inverse powers of $x_0$ as follows
\be
 -\ \la\, \frac{X (4 + 3 X) }{(1 + X)^2}\ =\ -3 \la + \frac{\sqrt{2\la}}{x_0}\ +\ \dots \ ,
\label{eqn_bb}
\ee
where $X = \,\sqrt{2\,\la} x_0$, see (\ref{eqn_X}).

In order to compare the $1/x_0$ terms in the last two expressions one needs to substitute
the ground state energy to $O(\la)$ accuracy
\be
     E\ =\ \frac{1}{2}\ -\  2\la\ +\ \dots \ ,
\ee
to the $O(\frac{1}{x_0})$ term in (\ref{phase}). After that one finds agreement with both $O(\frac{1}{x_0})$  terms given in (\ref{eqn_aa},\ref{eqn_bb}).

Similar calculations has been made for other quantum-mechanical examples. Let me just mention
Sine-Gordon potential
\be V={1 \over g^2} \left( 1-cos(g x)\right) \ee
for which the expression for the 0-1-2 loop expansion has a simpler form
\be -log( |\psi_0(x)|^2 )={16 \over g^2} sin^2({g x \over 4}) -2 log | cos({g x \over 4})| + {g^2 \over 32} tan^2({g x \over 4})+...\ee 
One can see that even for $g=1$  the second term is few percent and the third few per mill of the classical term,
for all $x$.
So the series over the fluctuations corrections are indeed well convergent. This is in strike contrast with 
the WKB, in which the next-to-classical correction
$1/\sqrt{p(x)}$ (where $p(x)$ is momentum) has an unphysical singularity at the turning point.

\begin{figure}[h!]
\begin{center}
\includegraphics[width=3.0in,angle=0]{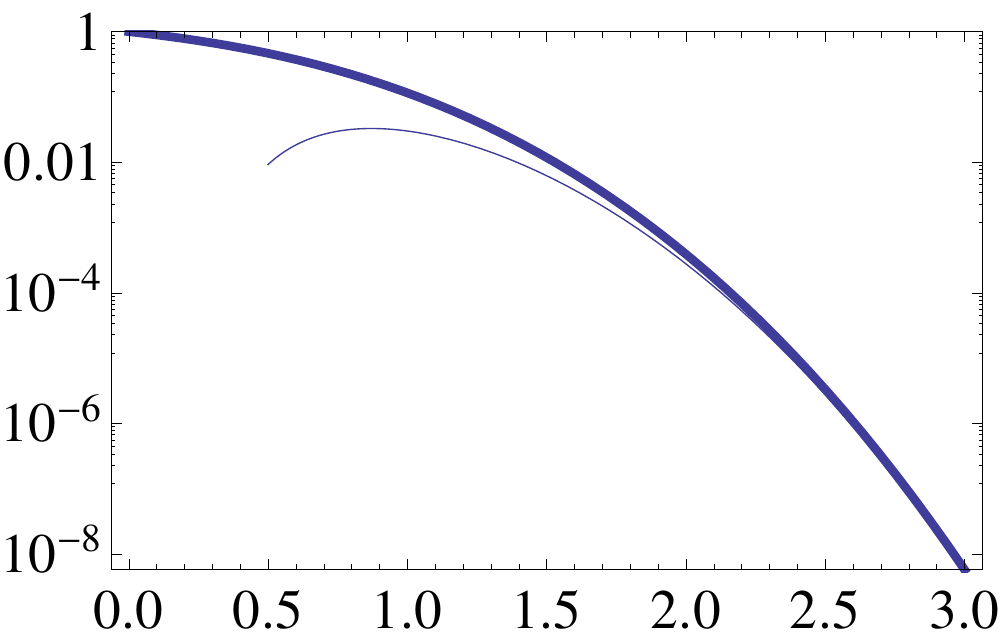}
\caption{The  probability $P(x_{0})$ to find particle at location $x_{0}$ for $\la=0.1$. The thick line
is our result (\ref{PX0}), thin line is its asymptotics. }
\label{fig_compare}
\end{center}
\end{figure}


 \section{Path integrals and the tunneling}
\label{sec_tunn_hist}
   Tunneling through classically impenetrable barrier is perhaps  the
 most
amazing consequence
of quantum mechanics. It was 
discovered already in 1927-1928 when it was just few months old.
First came Friedrich Hund who  in 1927   studied splitting of molecular states, 
and came to the problem of the energy
levels of a  double-well
potential, exactly the problem we will discuss  in detail.  Soon after that came a note \cite{tunneling2} that
(then-new) Schreodinger equation allows particle to go through classically 
forbidden regions. 

But by far the most famous paper\footnote{The paper \cite{CG_28} with a
similar
idea (but without the analytic Gamow factor)
 was submitted the next $day$
after the Gamow's one. My advice to the reader is obvious; never delay
a paper, even for a day.} is that by
 George Gamow \cite{Gamow}, who demystified alpha-decay of 
heavy nuclei. The apparent puzzle was why Rutherford scattering
experiments with alpha particles demonstrated existence of the
repulsive barrier of the height of 10 MeV or more, while
in the decay of the same nuclei alpha particles emerged with
smaller energies, of only few MeV. The
tunneling probability is the square
of the famous 
Gamov factor in the amplitude, which is

\bea \label{eqn_Gamow}
P_{Gamov}\sim exp \left[-{2\pi (2e^2) (Z-2)\over hv}\right] \eea

The $2(Z-2)$ in the numerator is the product of the
electric charges of the alpha-particle and the remaining nucleus.
The main factor in the denominator, $v$, is the $velocity$ of the outgoing alpha particle. This expression thus 
has the characteristic small parameter $v$ in denominator of the exponent,
as many other non-perturbative phenomena we will discuss. It
 explains why it takes up to  billion years
to decay, in spite of the fact that would-be alpha-clusters beat against the
wall 
 about
$10^{22}$ times per second\footnote{Another
advice to the reader; anything  may happen
once somebody is persistent enough, few or even lots of failures 
 never prove that something is impossible.}.
The first semiclassical WKB approximation, which we all learned
in quantum mechanical courses, was developed in 1926. It was sufficient to
explain tunneling in one-dimensional (or rather spherically symmetric radial) case.


Many years later, in 1950's, when one might think the subject
is completely worked out, Feynman  introduced
his formulation of quantum mechanics in terms of path integrals.
He also  realized that  paths integrals are easier to do in an imaginary Euclidean
time $\tau=i t$,  defined (what we no call) a ``Matsubara circle" with the circumference $\beta= {\hbar \over T}$, and proceeded to paths integral formulation
of the finite temperature theory.

Using Euclidean-time paths to evaluate the tunneling rate has been perhaps done
  in 1931, by Sauter \cite{Sauter:1931zz} for electron pair production in constant electric field. This calculation has been confirmed in 1930's by Heisenberg and Euler via their effective ED Lagrangian,
  and finally in 1950's solved exactly by Schwinger. Sauter's path starts as a $e^+e^-$
  pair at the point $x=0$. The energy of an electron
 is $\sqrt{p^2+m^2}-eEx$, and it must be  zero at the initial point: so
 $p=im$, the momentum is Euclidean.  At the final  points of the Euclidean path 
 $x=\pm m/eE$ the momenta can vanish $p=0$: from this point it will be real or Minkowskian. 
 The rest I would leave to the reader as 

{\bf Exercise}: {\em Derive the Euclidean tunneling path and the corresponding action 
 for the  production of a
pair of massive scalar particle
 in
a homogeneous electric field.}

Discussion of the Euclidean particle paths related to tunneling has been revived  
by \cite{Polyakov:1976fu}, following a discovery of the (Euclidean) instanton solution of the gauge theory. For early pedagogical review, including the one-loop (determinant) 
calculation 
by the same tedious scattering-phase method seeABC's of instantons  \cite{Vainshtein:1981wh}
. Two-loop 
corrections were first calculated in \cite{Aleinikov:1987wx},  some technical errors
in it were  corrected by Wohler and myself in
\cite{Wohler:1994pg}.  Three loop corrections have been calculated by 
\cite{Escobar-Ruiz:2015nsa}. 

%
As in the discussion of $fluctons$ above, the main setting is the same. A usage of
Euclidean time leads to successful description of  motion 
in the ``classically forbidden" region under the
barrier. The minimal action -- and thus
classical -- paths are the
most probable ones., since the weight  is exp(-S[x($\tau$)]).
and thus we look for minima of the action.

The Euclidean path method is much more powerful than WKB since it can
be  used in multi-dimensional problems. While our applications will be in QFTs such as pure gauge theory, let us note that they are successfully used in 
chemistry and even biology (e.g. describing the protein folding
paths \cite{Faccioli:2011eg}). 

We will continue to use the usual toy model, a
     double-well potential, with the Euclidean action
\be S= \int d\tau [m{\dot x^2\over 2} + \lambda(x^2-f^2)^2] \ee
in which the dot  means the derivative over $\tau$, not time.
This potential has two  minima at $\pm f$,
 known as  the two "classical vacua".

   The energy levels of this system can be derived
at three levels of sophistication;\\
(i) If one first ignores tunneling, those are given by zero point
oscillations in each well,
with the energy
\bea E_0=\omega/2, \hspace{1cm} \omega=f \sqrt{8\lambda} \eea
In order to have better contact with the gauge theory
later in the chapter, we will eliminate $f$ from all expressions
substituting it by the $\omega$ just defined. 
The maximal hight of the barrier, for example,
 is then $ V_{max}=\omega^4/64\lambda $, etc.\\
(ii) At small value of the only dimensionless parameter 
of the model $\lambda/\omega^3 \ll 1$  (the high barrier limit),
 one can further calculate a whole series of   
{\it perturbative corrections} which go as powers of this parameter,
\bea E_0={\omega\over 2} \left[1+ \Sigma_n C_n (\lambda/\omega^3)^n \right] \eea 
One use Feynman diagrams
to calculate these corrections, corresponding to fluctuations around a ``lazy path" $x(\tau)=0$.\\
(iii) 
Finally   one may take into account the tunneling phenomenon. The left-right 
degeneracy of the levels is then lifted, substituted
by $symmetric$ and $antisymmetric$ wave 
functions under the parity transformation $x\leftrightarrow -x$.  Their energies
\bea \label{eqn_levels_dw}
E_\pm={\omega\over 2}\left[1 \mp \sqrt{{2\omega^3\over\pi\lambda}}
  e^{-{\omega^3\over 12\lambda}} \right]\eea
 are thus separated by exponentially small gap.


Standard textbook description of tunneling goes as follows. 
Since the energy is  conserved, one can read
the Hamiltonian 
 $H=E_{kinetic}+V(x)$ where $$E_{kinetic}=\hat p^2/2m=-{\partial^2_x \psi \over \psi}{1 \over
  2m}$$  In a classically allowed
region, $E_{kinetic}>0$  the wave function resembles a wave 
$\psi \sim exp(ipx)$
 with
{\em real} p. In a classically {\it forbidden} region  $E_{kinetic}<0$, the momentum should be {\it imaginary} and therefore the wave function just decreases
 $\psi \sim exp(-|p|x)$.

The word 'tunneling' hints that one can pass 
the mountain (the barrier due to a repulsive
potential) not by climbing and then descending from it,
but by going through it, {\em as if} there be a path through, the tunnel. 
The point of this section is to show that not only {\it one can imagine}
a path through the mountain,  it is easy and useful to find  them, and even identify
the $best$ ones among them, the $instantons$.

  If $p$ is imaginary, one may
 interpret it as a motion
in {\it  imaginary time}  $\tau=it$. As we already mentioned,
classical equation of motion 
\bea m {d^2x\over d\tau^2}=+{d V \over dx}\eea
correspond to flipping the potential upside 
down!

\begin{figure}[h!]
\begin{center}
\includegraphics[width=10cm]{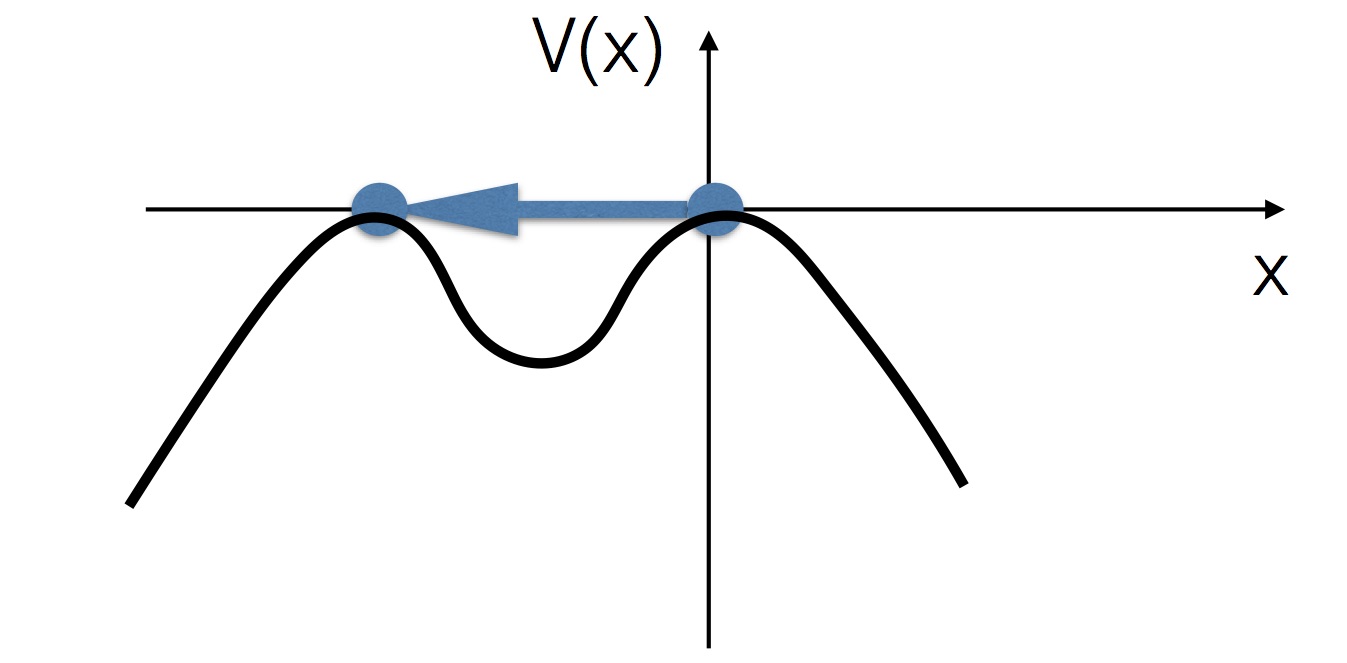}
\caption{Sketch of the instanton path, going from one (flipped)  minimum of the potential to the other.}
\label{fig_instanton}
\end{center}
\end{figure}

The $instantons$ we are looking for
would be classical paths between two classical vacua, in a flipped potential $V\rightarrow -V$, see Fig.\ref{fig_instanton}. As usual, It is  easier to find it using energy conservation. Note that the path
we are looking goes from one minimum to another one (see Fig.\ref{fig_instanton}) must be at the total energy  $E=0$.
 The resulting solution, the {\em instanton },  is thus
\bea x_{cl}(\tau) =f \tanh [{\omega(\tau-\tau_0)\over2}] \eea
\index{instanton in the double well problem}

  The 
   tunneling
probability is given by 
the corresponding action $P \sim exp(-S_{cl})$, which is   twice the average potential energy, so 
\bea S_{cl}=2\int d\tau \lambda [x_{cl}^2-f^2]^2={\omega^3 \over 12\lambda}\eea
Note that it reproduces the exponent in \ref{eqn_levels_dw}.

{\bf Exercise}: Derive the Gamov factor (\ref{eqn_Gamow}) using the classical 
under-the-barrier path
in Euclidean time.

\section{The  zero mode and the dilute instanton gas}
\label{sec_zero_mode_dw}

   The next step, as for the flucton, is to write the general tunneling path as small quantum deviations from the
classical path
\bea x(\tau)=x_{cl}(\tau) + \delta x(\tau) \eea
and expand the action in
powers of quantum corrections  $\delta x(\tau)$.
   In the quadratic order
\bea S=S_{classical}+(1/2)\int d\tau \delta x(\tau)\hat O \delta \delta x(\tau)\eea
 the  differential operator at  the instanton path takes the form
\bea \hat O=-{m\over 2}{d^2\over d\tau^2} +
 {\delta^2 V\over \delta x^2}|_{x=x_{cl}}=-{m\over 2}{d^2\over d\tau^2}+4\lambda  (3 x_{cl}^2-f^2) 
 \label{O_instanton}
 \eea
At time  distant  from the mid-tunneling moment
(to be called also the position of the instanton) $\tau_0$,
 the last term is just constant, $8\lambda f^2$.
But around $\tau_0$  and this last term  is strongly changing
and it is $negative$, unlike that for the flucton. 

This observation leads to
significant modifications of the one-loop theory for instantons. 
  Unlike in the flucton case, now the equation has not only the scattering states, but 
   also two localized (bound states) solutions.  
 The lowest eigenvalue is zero,
 $\epsilon_0=0$, and the next
is positive $\epsilon_1={3\over 4}\omega^2$ but below the scattering state energy.

 The  wave function for the zero mode is 
\bea x_0(\tau)\sim {1 \over \cosh^2(\omega\tau/2)} \eea
and one can find the normalization constant from
the usual normalization condition
$\int d\tau x_n^2=1$ to be $const=\sqrt{3\omega/8} $.

  Every zero should have a simple explanation. And indeed, this one
has a simple origin: it is due to time translational
symmetry
in the problem.  Indeed, 
the action cannot  depend on the instanton displacement in time\footnote{
 Note that in the case of flucton the path is ``pinned" at certain point at $\tau=0$,
 so there was no time shift symmetry and no zero mode.}
 The zero mode
 can be obtained very simply, by differentiation of the
instanton solution over its time $\tau_0$
\bea x_0(\tau)= S_0^{-1/2} {d \over d\tau_0} x_{cl}(\tau-\tau_0)\eea 

{\bf Exercise:} {\em Check this statement by putting this function into the 
fluctuation equation $\hat O \psi_0=0$.}

  Now, if the operator in question has a zero eigenvalue, its determinant is zero too. Since 
  for Gaussian integral over fluctuations the (sqrt) of the determinant  appears in denominator,
the tunneling probability we discuss is in fact infinite!
 Indeed, we encounter here (the simplest case of) the
so called 'valleys' in the functional space:
 one of the integrals is
actually non-Gaussian; zero eigenvalue means that nothing prevents  large
amplitude of fluctuations in the corresponding direction in the Hilbert space.  

  The solution to this problem is 
however quite simple; one may not take this integral at all!
All we have to do is to rewrite the integral over $dC_0$ as the integral over
the collective coordinate $\tau_0$.
  Consider a modification
 of $C_0$ by $dC_0$. The path changes by $dx=x_(\tau)dC_0$
(remember the definition 
$x(\tau)=\Sigma_n c_n x_n(\tau)$). At the same time we 
have another definition  of the zero mode from which it follows
\bea dx={dx_{cl} \over d\tau_0} d\tau_0 = - \sqrt{S_0} x_0(\tau) d\tau_0 \eea
Equalizing two expressions for dx, we have
\bea dC_0= \sqrt{S_0}  d\tau_0 \eea

  Returning to our functional integral over the quantum fluctuation, we now 
have the following form for it
\bea \int D\delta x(\tau) e^{-S}=
e^{-S_{cl}}\prod_{n>0}\sqrt{{2\pi\over \epsilon_n}}
\sqrt{S_0} \int d\tau_0 
 \eea  
The product here is the determinant $without$ the zero mode, 
 often called $det'(\hat O)$. It is finite and its calculation will be done below.
But before we engaged in it, let us
first explain what to do with the divergent
integral over the $\tau_0$. 
Suppose the whole path integrals is taken over some finite time, from
$0$ to $\tau_{max}$; then
the contribution of tunneling -- 
described by the instanton-type path-- grows linearly with  $\tau_{max}$.
One may say that the
finite quantity is the {\it transition probability per unite time},
but it is not a satisfactory solution in the long run.
 If $\tau_{max}$
is very large,  it may overcome the smallness of the exponent.
If so, the amplitude is no longer small and one has to think about ensemble of
{\it many}  instantons. 



   Suppose one has a path with n instantons
(anti-instantons), placed at $\tau_1 < ... < \tau_n < \tau_0$, 
If they are all separated sufficiently far from each other,
(as one says, the instanton gas is {\it dilute}) the action is
the sum of actions. Furthermore, determinants become factorized, and the 
expression for the transition amplitude in this case reads as
\bea G(f,-f,\tau_0) = [\sqrt{\omega/\pi}exp^{-\omega \tau_0/2}]
[\sqrt{6 S_0/\pi}e^{-S_0}]^n \int^{\tau_0}_0 \omega d\tau_n ... 
\int_0^{\tau_2}\omega d\tau_1 \eea
where integrals are done over ordered positions.
The factor which repeatedly appears here 
\be d=\sqrt{{6 S_0\over \pi}}e^{-S_0}\omega \ee
we will call the {\em instanton density} (per unit time).
One can relax the nesting condition on the tunneling moments by
integrate over all interval, but then dividing the result by
the n factorial.
 It looks then like the exponential series, 
but since $n$ is actually odd one gets the following final expression
for the Green function
\bea G(f,-f,\tau_0) = [\sqrt{\omega/\pi}exp^{-\omega \tau_0/2}]
\sinh[ \sqrt{{6 S_0\over \pi}}e^{-S_0} \omega \tau_0] \eea
 
   Now, it  may be used for any  time. If it is very large,
one has another asymptotic of the sinh, the exponential one,
and notice that the total dependence on $\tau_0$ is now again
exponential, with the {\em corrected ground state energy}
\bea E_0={\omega\over 2}-\sqrt{{6 S_0\over \pi}}e^{-S_0} \omega \eea
Note that it is precisely what one also gets 
for level shift from the Schroedinger equation
in the semiclassical approximation.

   Even if the instanton gas is dilute, 
it is still interesting to ask what happens if two instantons are close,
$\tau_k-\tau_{k-1}\sim 1/\omega$. Then they certainly do interact, because
the total action is actually {\it less} than $2S_0$. Numerical studies
indeed show that there is strong positive correlation
between the instanton positions at smaller time intervals.
We will return to interacting instantons later.  
 
   Finally, let me introduce the last issue of this section, related
   with
the so called
correlation functions of the coordinates. Those
 characterize properties of the ground state,
and in particular show the crucial role of the dilute
instanton gas.
The simplest observable we can think of is just the particle 
coordinate, so let us define the correlator of the coordinates as
\bea K(\tau_1-\tau_2)=<x(\tau_1)x(\tau_2)> \eea
where averaging is supposed to be done by appropriately weighted
quantum paths. 
We imagine now the length of all paths $\tau_0$ to be infinite, 
so the correlator depends only on the time {\it difference}.

  Recalling the ``old fashioned'' quantum mechanical expressions, one can
use matrix elements of the coordinate operator, and write this functions as
a sum over  stationary states which can be excited by the operator of
the coordinate
\bea K(\tau)=\Sigma_n |<0|x|n>|^2 exp(-E_n \tau) \eea
This function is not only {\it positive} but it is a
 {\it monotonously 
decreasing} one.
Although the sum runs over all states, only {\it parity odd} ones can be 
excited by the coordinate operator. For example, symmetric ground state
is absent in this sum, because $<0|x|0>=0$.

  Thus, if one knows the 
correlation function at large times, one knows $E_1$, the
lowest parity odd state as well. 
This is analogous to what we will do in chapter 6 for 
calculation of the lowest excitations in QCD, mesons and baryons.

   Now, let us try to understand how the correlation function looks like, if we
do not specifically interested in large times. Clearly, the problem has two
time scales; (i) perturbative scale $\tau_{pert}=1/\omega$ and (ii) tunneling 
scale $\tau_{tunneling}$, the inverse tunneling rate. 

  If one imagines the 
 particle is sitting in the same well all the time,
the correlator is 
$K(\tau)\approx f^2$. 
  However, the tunneling kills the correlation, and eventually
$K(\tau)\rightarrow 0$.
One can write approximately the paths as sequence of steps or kinks
\bea x(\tau) = f \prod_i sign(\tau-\tau_i) \eea
and calculate the correlation 
function
\bea K=f^2 {\Sigma_n  \int \Pi_n d\tau_n d^n sign(\tau-\tau_i)sign(-\tau_i)
\over   \Sigma_n  \int \Pi_n d\tau_n d^n } \eea
where d=$\sqrt{6 S_0/\pi}e^{-S_0} \omega $. Its
 large-time limit is
\bea \label{eqn_coord_cor}
K(\tau) \sim exp(- 2 \tau d) \eea
Comparing it to the expression above, one gets the $gap$,
the energy splitting
between the vacuum and the first excited state, $E_1-E_0=2d$.
Intuitively 2 appears because of instanton plus the anti-instantons.

   Summarizing this discussion; if the classical action
is large $S_0 \gg 1$ (actually, much larger  than the Plank constant), 
the
ensemble of paths is an exponentially dilute gas
of instantons and anti-instantons. 
 All such paths together lead to 'exponentiation'
of tunneling correction, and 
corresponding to the negative shift of the ground state energy.
They also randomize the sign of the coordinate and create the gap
between the vacuum and the excited states.
As we will see later in the chapter, exactly the same thing
happens in gauge theories.

{\bf Exercise} Derive the  expression (\ref{eqn_coord_cor}) for the correlator
of coordinates in dilute instanton gas
 approximation. 

\section{  Quantum fluctuations around the instanton path }
\label{sec_twoloop}


The Feynman rules, as usual, are based on $vertices$, calculated from
the derivatives of the potential over $\delta x$ in various powers (large than 2),
and $propagators$(Green functions), inverting the quadratic form $\hat O$. 
There are two complications:\\
$\bullet$  this inversion should be made
{\em in the ``primed" subspace excluding the zero mode}. \\
$\bullet$  jacobian related with collective coordinate leads to
new type of Feynman diagrams, not following from the action. 

Starting with the one-loop calculation, one needs to 
 calculate  the ``primed" determinant of the operator (\ref{O_instanton})
 over all non-zero modes. Fortunately, since the quadratic form $\hat O$ 
 in this case corresponds to exactly solvable equation, one may proceed with the 
 most direct method
 of its calculation -- the direct diagonalization.
 One can determine the scattering phase 
 $\delta(k)$ and summing log of all eigenvalues with the appropriate level counting.
 We would not reproduce this tedious calculation here, the interested reader can find it 
 in  \cite{Vainshtein:1981wh}. 
 
 The second method -- using a
 loop diagram and the Green function -- does not work in this case, although it provides a very nontrivial test for 
 the Green function:  discussion of this can be found in Appendix in \cite{Escobar-Ruiz:2016aqv} .
 We will skip calculation of the instanton determinant, and proceed directly to 
 the Green function and higher order corrections, following \cite{Wohler:1994pg,Escobar-Ruiz:2015nsa}.
%
%
We will of course not give full details of these calculations here, and only present
the results, with some comments.


  To next order in $1/S_0$, the tunneling amplitude can be 
written as
\bea
\label{eq_twoloopexp}
\langle -f | e^{-H\tau} | f \rangle 
 = \left|\psi_{0}(f)\right|^{2}  
 \left(1 + \frac{2A}{S_{0}} + \ldots\right) \\ \nonumber 
 e^{\left[-\frac{\omega\tau}{2} \left(1 + \frac{B}{S_{0}} + \ldots
 \right)\right]} 2d \left(1 + \frac{C}{S_{0}} +\ldots\right) \tau,
\eea
where  corrections, $A$ and $B$ are those to the wave function at the
 minimum
and to the energy due to anharmonicity of the oscillations.
These  two corrections are 
unrelated to tunneling and we can get rid of them by dividing the 
amplitude by $\langle f|\exp(-H\tau)| f\rangle$, in which they appear
in the same way.
We are interested in the coefficient $C$, the next order correction
to the tunneling amplitude (instanton density $d$)
and eventually to the level splitting.

%
%
%

Step one is the calculation of the propagator (the Green
function) of the differential operator $\hat O$ we defined above. It however has the following
complication: unlike in the case of the flucton path, 
the instanton operator  has a zero mode. Thus, strickly speaking,
it {\em cannot be inverted!}
 
Yet the inversion is
still uniquely defined in a Hilbert subspace
{\em orthogonal to the  zero mode}. Note that delta function
in the r.h.s. of the Green function equation can be written as follows\footnote{We write the sum over zero modes for generality,
assuming  there may be several of them. In the problem we study now there is only one zero mode, corresponding to time-shift symmetry of the instanton path.}
\be \delta(\tau-\tau')=\sum_{\lambda \neq 0} \psi^*_\lambda(\tau) \psi_\lambda(\tau')+
\sum_{\lambda = 0} \psi^*_\lambda(\tau) \psi_\lambda(\tau') \ee
which expresses completeness of the set of all eigenvalues, zero and nonzero.

In a subspace orthogonal to zero modes one can define a new Green function 
$G_\perp(\tau,\tau')$ satisfying a modified equation  with a new r.h.s.
\be \hat{O}G_\perp(\tau,\tau')=\delta(\tau-\tau')- \psi_0(\tau) \psi_0(\tau') \ee
Although we will from now on drop the subscript $\perp$ on the Green function,
this is the one which is to be used below.
Its analytic form is 
\bea
G(x,y) = G_{0}(x,y)\left[ 2-xy+ \frac{1}{4} |x-y| (11-3xy) +
  (x-y)^2\right] \\ \nonumber
  + \frac{3}{8} \left(1-x^2\right)\left(1-y^2\right)
   \left[\log\left(G_{0}(x,y)\right) - \frac{11}{3}\right] \\ \nonumber
\eea
where $x =\tanh(\tau/2),\, y=\tanh(\tau'/2)$ and the perturbative Green function near trivial vacuum in this notations is $$G_{0}(x,y)\;=\; \frac{1-|x-y| - xy}{1 +|x-y|- xy} $$

{\bf Exercise}: {\em Check that it is indeed orthogonal to the zero mode, in respect to both of its variables.} 

\begin{figure}[h]
\begin{center}
\includegraphics[width=3.5in,angle=0]{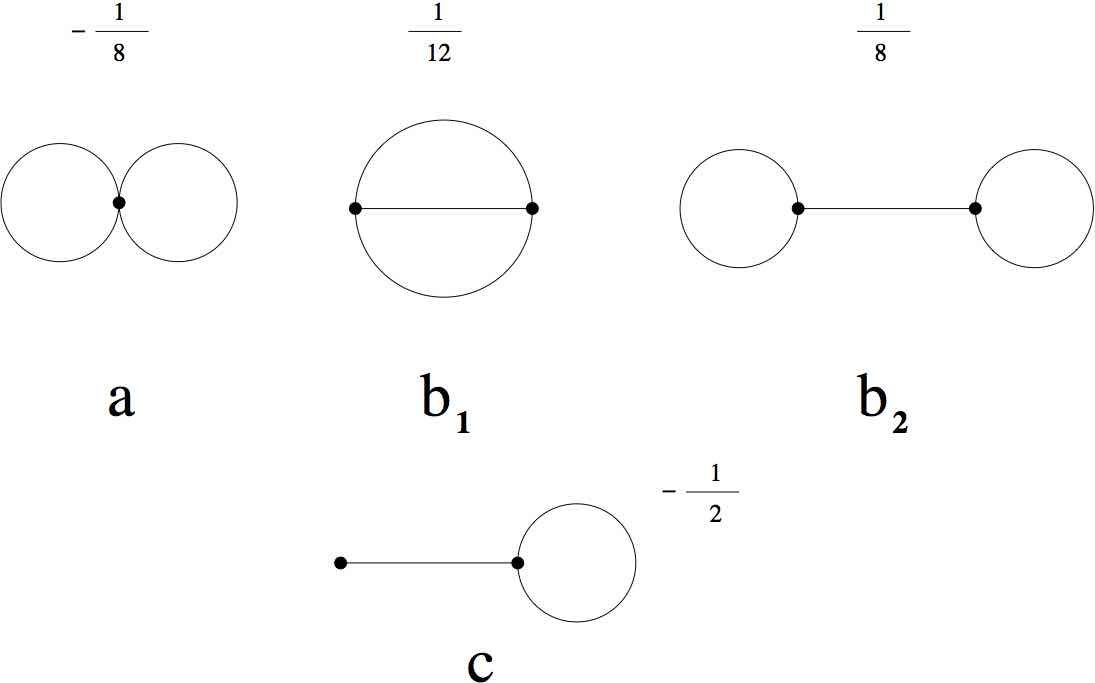}  \vspace{0.8cm}  \\
\caption{Two-loop diagrams contributing to the instanton amplitude. The signs of contributions and
symmetry factors are indicated. The only difference with the flucton case is the  appearance of the new diagram $c$.}
\label{F1}
\end{center}
\end{figure}

There are four diagrams
at the two loop order, see Fig.\ref{F1}. The first three
diagrams are of the same form as for the flucton path. 
 Subtracting the contributions of the same diagrams around the trivial path $x(\tau)=0$ , one gets rid of the 
effects far from the tunneling event: this is so-to-say infrared renormalization of the diagrams.
 Their  contributions are
\be
a = -3  \int^{\infty}_{-\infty}dt\, \left( G^{2}(t,t) - 
  G^{2}_{0}(t,t)\right) \;=\; -\frac{97}{1680} \ee
$$b_1 = 3 \int^{\infty}_{-\infty} \int^{\infty}_{-\infty}
  dt dt'\,\left(\tanh(t/2)\tanh(t'/2)G^{3}(t,t') - G^{3}_{0}(t,t')\right) 
  \;=\;  -\frac{53}{1260} $$
$$ b_2= \frac{9}{2}\int^{\infty}_{-\infty}
 \int^{\infty}_{-\infty} dt dt'\,\big(\tanh(t/2) \tanh(t'/2)
 G(t,t)G(t,t')G(t',t')  \\
- G_{0}(t,t)G_{0}(t,t')G_{0}(t',t') \big) 
 = -\frac{39}{560}.
$$

However, unlike for the flucton path, now there appear a  {\em new series of diagrams} 
introduced in \cite{Aleinikov:1987wx} and
not following from classical action. Indeed, all fluctuations under consideration
should be orthogonal to the zero mode
\be \int d\tau \delta x(\tau) x_0(\tau)=0 \ee
Inserting into the functional integral the delta function
with this condition {\em a la} Faddeev and Popov, one finds  
  the Jacobian which generates 
 tadpole graphs proportional to a new vertex $\ddot x_{cl}$ which needs to appear
 only once. These diagrams 
 have no counterpart in the expansion around the trivial vacuum, and thus they need no subtraction. Its two-loop contribution is only one diagram contributing
\bea
\label{eq_2loopdiagrams}
c= -9 \beta \int^{\infty}_{-\infty} \int^{\infty}_{-\infty}
 dt dt'\,\frac{\tanh(t/2)}{\cosh^{2}(t/2)}\tanh(t'/2)G(t,t')G(t',t') 
 = -\frac{49}{60}.
\eea
The sum of the four diagrams is $$C =a+b_{1}+b_{2}+c=-71/72$$
is in agreement the result obtained from Schreodinger eqn, see e.g. review 
\cite{ZinnJustin:2004cg}.

\begin{table}[th]
\begin{center}
\setlength{\tabcolsep}{20.0pt}
\begin{tabular}{|c  ||c  | c |}
\hline
Feynman                     &    Instanton        &       Vacuum
\\[1pt]
diagram                     &     $B_2$           &      $A_2$
\\[4pt]
\hline
$a_{1}$           \quad     &   $\, -0.06495185$      &   \quad  $\frac{5}{192}$
\\[3pt]
\hline
$b_{12}$          \quad     &   $\quad 0.02568743$    &     $-\frac{1}{64}$
\\[3pt]
\hline
$b_{21}$          \quad     &   $\quad 0.04964284$    &     $-\frac{11}{384}$
\\[3pt]
\hline
$b_{22}$          \quad     &   $\,-0.13232566$       &    \quad $\frac{1}{24}$
\\[3pt]
\hline
$b_{23}$          \quad     &   $\quad 0.28073249$    &     $-\frac{1}{8}$
\\[3pt]
\hline
$b_{24}$          \quad     &   $\,-0.12711935$       &  \quad   $\frac{1}{24}$
\\[3pt]
\hline
$e$               \quad     &   $\quad 0.39502676$    &     $-\frac{9}{64}$
\\[3pt]
\hline
$f$               \quad     &   $\,-0.35244758$       &  \quad   $\frac{3}{32}$
\\[3pt]
\hline
$g$               \quad     &   $\,-0.39640691$       &  \quad   $\frac{3}{32}$
\\[3pt]
\hline
$h$               \quad     &   $\quad 0.31424977 $   &      $-\frac{3}{32}$
\\[3pt]
\hline
$c_1$             \quad     &   $\,-0.3268200$        &     $-$
\\[3pt]
\hline
$c_2$             \quad     &   $\quad 0.63329511$    &     $-$
\\[3pt]
\hline
$c_3$             \quad     &   $\quad 0.12657122$    &     $-$
\\[3pt]
\hline
$c_4$             \quad     &   $\quad 0.29747446$    &     $-$
\\[3pt]
\hline
$c_5$             \quad     &   $\,-0.77100484$       &     $-$
\\[3pt]
\hline
$c_6$             \quad     &   $\,-0.80821157$       &     $-$
\\[3pt]
\hline
$I_{2D}$           \quad    &   $\quad  0.0963$   &    -$\frac{7}{384}$
\\[3pt]
\hline
$I_{3D}$           \quad    &   $\, -0.0158$      &   \,  $\frac{19}{64}$
\\[3pt]
\hline
$I_{4D}$           \quad    &   $\, -0.8408$      &   -$\frac{155}{384}$
\\[3pt]
\hline
\end{tabular}
\caption{
Contribution of diagrams in Fig. (\ref{F2}) for the three-loop corrections $B_2$ (left)
and $A_2$ (right). We write $B_{2} = (B_{2loop} + I_{2D} +I_{3D} +I_{4D}) $ where $I_{2D},I_{3D},I_{4D}$
denote the sum of two-dimensional, three-dimensional and four-dimensional integrals, respectively.
Similarly, $A_{2}=I_{2D}+I_{3D}+I_{4D}$. The term $B_{2loop}=39589/259200\approx 0.152735$ (see text).}
\end{center}
\end{table}

\begin{figure}[htp]
\begin{center}
\includegraphics[width=3.in,angle=0]{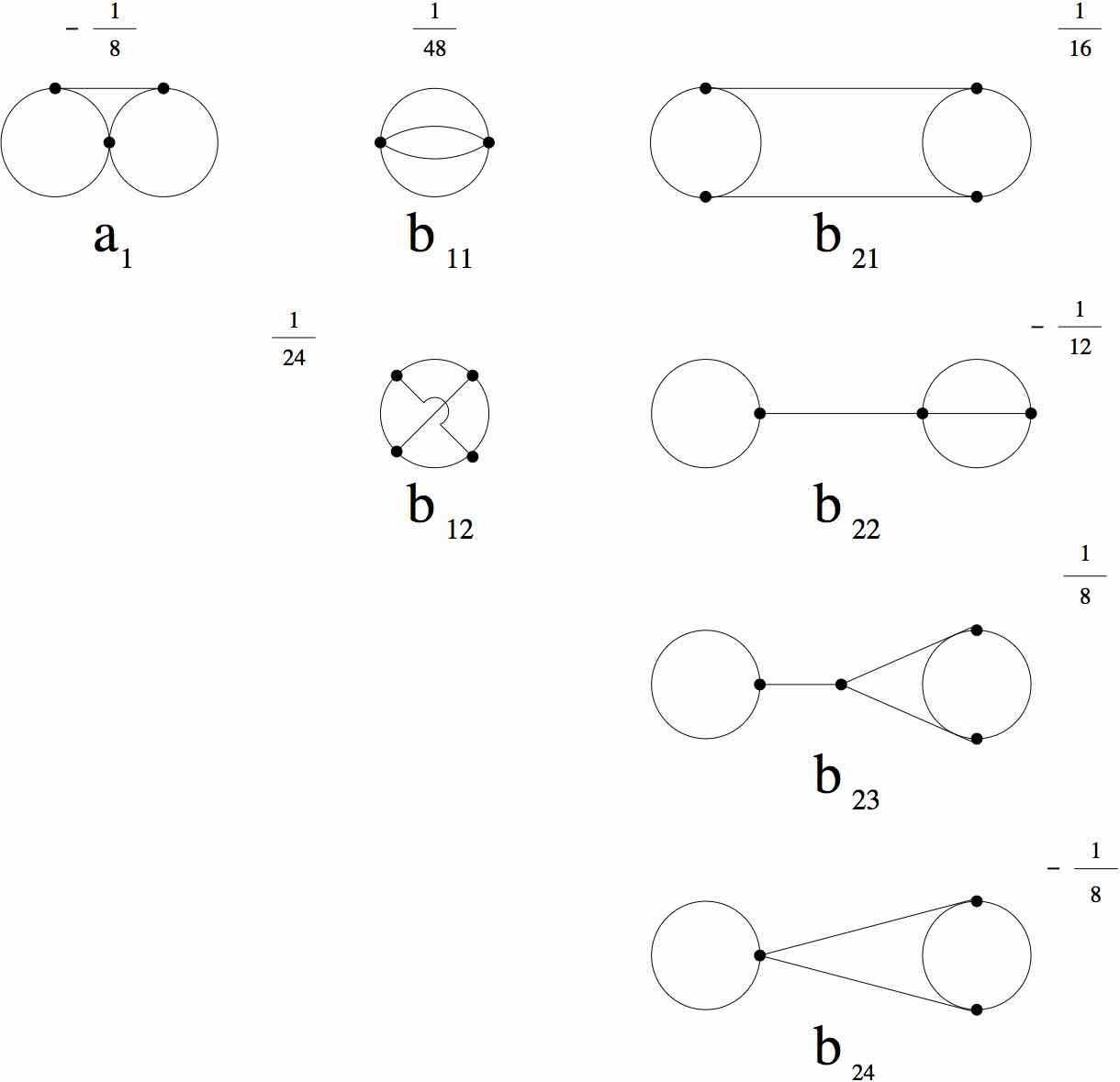}  \vspace{0.8cm}  \\
\includegraphics[width=4.in,angle=0]{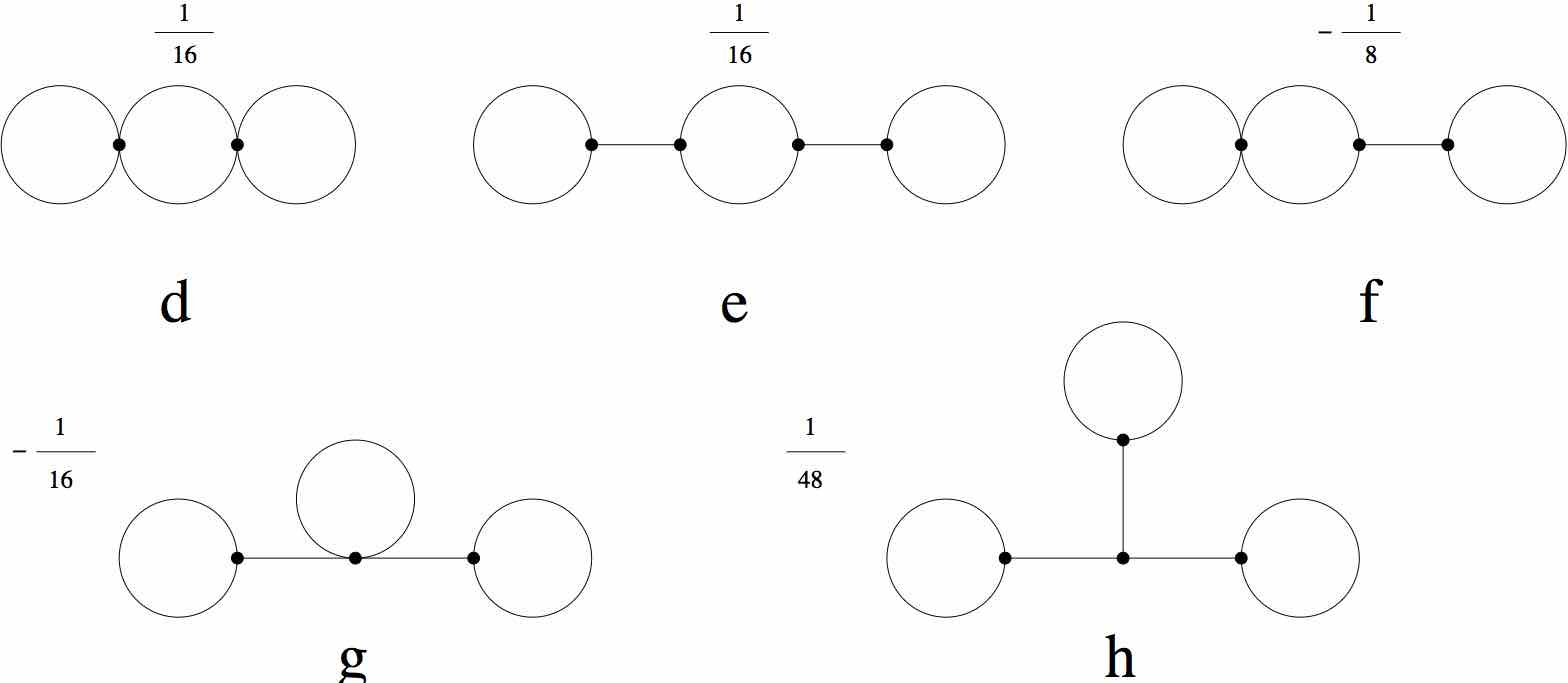} \vspace{0.8cm}  \\
\includegraphics[width=4.in,angle=0]{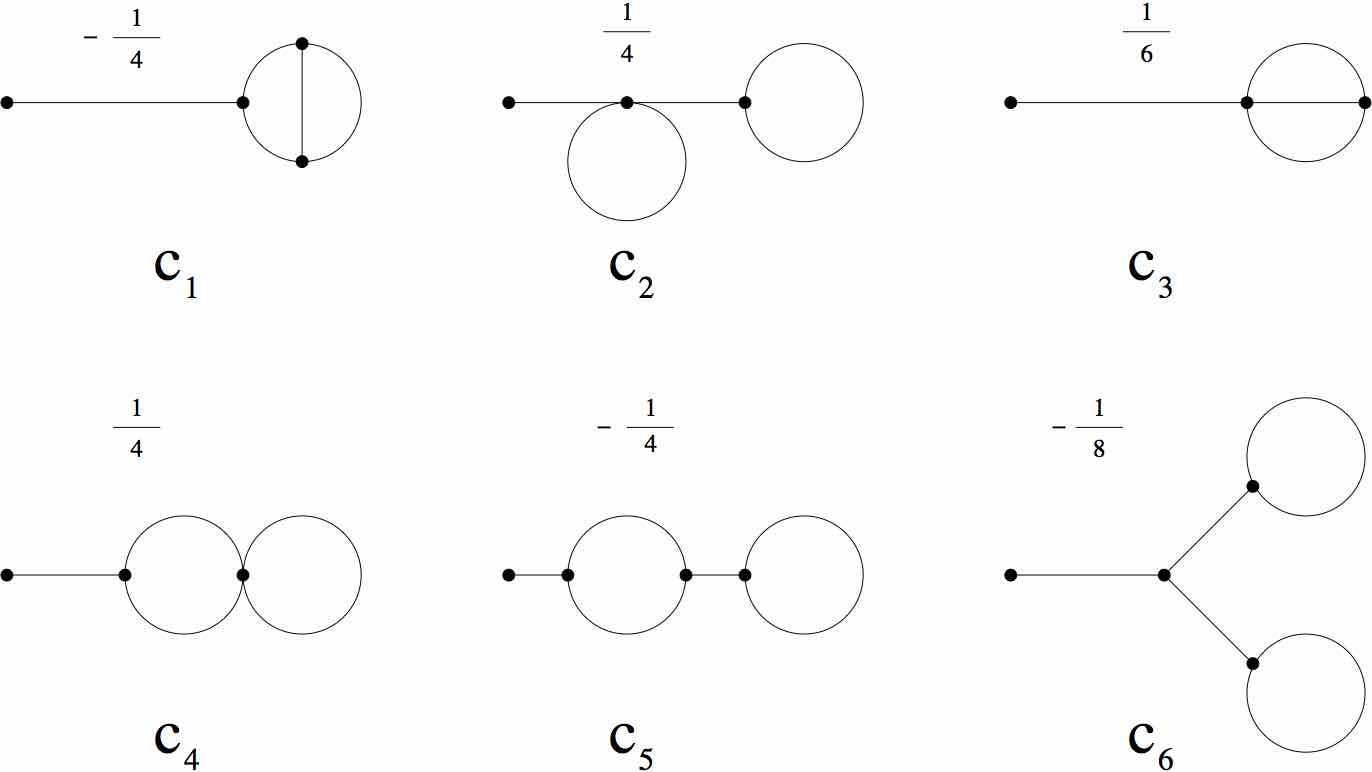}
\caption{Diagrams contributing to the three loop corrections to the instanton amplitude. The signs of contributions and
symmetry factors are indicated.}
\label{F2}
\end{center}
\end{figure}

 Three loop corrections  \cite{Escobar-Ruiz:2015nsa} come from 
  the diagrams shown in Fig.\ref{F2}, their  numerical contributions are given in the following table.
As one can see, unfortunately not all diagrams we were able to calculate analytically.  
Those which we could contain irrational parts, related with Riemann zeta function. 
Yet the final result for three loop coefficient agrees well with the answer from Schreodinger equation
we mentioned before in (\ref{ZJ_inst}). This answer is
$rational$, so all  terms with special functions must somehow cancel.

 The reason it does happen 
must be the existence of 
 certain
 {\em resurgent relation} between perturbative and instanton series we discussed above.
The first of such relation has been proposed in \cite{ZinnJustin:2004cg}. 

\section{ Transseries and resurgence}
In the previous sections we have shown how one can calculate corrections
due to quantum/thermal fluctuations, order by order, to the density matrix (fluctons)
and the vacuum energy (instantons). We did not paid much attention to similar
perturbative series, on top of the trivial vacuum path $x_{cl}(\tau)=0$, except to use it
to subtract infrared divergencies. 

Now we shift our focus and in this section discuss more general theoretical questions. 
We start with the unavoidable issue, namely the fact that the coefficients
of all such series are increasing with their order, so that all of them are asymptotic
-- strictly speaking divergent -- series. 

{\it ``Divergent series are the invention of the devil, and it is
  shameful
to base on them any demonstration whatsoever"} wrote N.H.Abel in 1828.
But  modern physicists tend to be ``non-Abelian'', and they use the
perturbation theory widely,  its divergent series notwithstanding.

The main idea why the perturbative series $must$ diverge was 
suggested by Dyson \cite{Dyson:1952tj}. In short, the argument
went as follows. In QED one makes expansion in $e^2$, order by order.
Let us think a bit what would happen if one analytically continue QED into negative $e^2<0$
domain. Then protons and
electrons
would no longer attract each other, so atoms would dissolve. On the
other hand, electrons would attract each other and congregate in large
number, and so would  positrons and protons. A bit of thought  shows
that the binding energy in this case can be made arbitrary large; so
there would be a complete collapse of the theory, as it lacks any stable ground
state. Now, if the positive and negative vicinity of the $e^2=0$
 point are that different, there must be singularity at this point. Therefore
 any power expansion around
 it obviously cannot be nice 
 (convergent).

We already mentioned in the introduction a distinction between perturbative and non-perturbative phenomena. In the toy models we studied those 
can be thought of as effects proportional to $powers$ and {\em inverse exponents}
of the coupling $g^2$. 
The
more definite mathematical formulation of those is given by the so called $transseries$.
French mathematician J.Ecalle in 1980's defined those as triple series (``standing on three whales")
including not only powers of the parameter (=coupling) as in perturbation theory,
but also exponential and logarithmic functions
\be f(g^2)=\sum_{p=0}^\infty\sum_{k=0}^\infty\sum_{l=1}^{k-1} c_{pkl} g^{2p} 
\left(exp(-{c\over g^2} )\right)^k \left(ln(\pm {1\over g^2} )\right)^l
\ee

His argument was based on symmetries: such set of functions is $closed$, under manipulations including
the so called Borel transform plus analytic continuations. 
Examples studied in mathematics usually were limited to some integrals with parameters,
defining special functions. QM path integrals are infinite-dimensional, but simpler than those in QFT's,
and, as we will see, in this case one can indeed  build the   transseries explicitly and demonstrate
that indeed those two functions do appear. Furthermore, there are 
certain relations between the coefficients, known as $resurgence$: in some cases those can be derived,
but its very existence in the QFT context remains a mystery. 

Inter-relations of perturbative and non-perturbative effects are very deep. One reason
of why they $must$ be connected can be demonstrated using Borel transform as follows. Consider the following example:  for $x>0$ the integral of the r.h.s. generates the factorially divergent series in the l.h.s., clear from expanding the r.h.s. in $x$ in geometric series
\be \sum_{n=0}^\infty (-1)^n n! x^n =\int dt e^{-t}{ 1 \over 1+x t} \ee
 Borel
suggested doing the integral over $t$ instead: since for $x>0$ the pole is at $t=-1/x<0$ is outside the integration
region, the result is a  good ``Borel improved" result
\footnote{Note that Sum operation in Mathematica
now have an option to try Borel-based and other regularizations, try to do this sum with $ Regularization \rightarrow "Borel" $ prescription added.}
But for $x<0$ the series do not have the alternated sign
\be \sum_{n=0}^\infty  n! x^n =\int dt e^{-t}{ 1 \over 1-x t} \ee
the Borel pole is at $t=1/x$, right on the integration line. Shifting 
the integration cone up or down
results in ambiguous imaginary part $$ \pm {i\pi \over g} e^{-1/g} $$
The quantity of interest -- say the energy of our QM system -- cannot have
imaginary part, and yet its perturbative series, via Borel re-summation,  
seem to get it!  

These puzzling questions are potentially resolved by a generalization of the perturbative series to the $transseries$. The idea is that the exponential terms in them are such that {\em all ambiguous/unphysical effects} -- like
unwanted imaginary parts  -- {\em must safely cancel out} to all orders. Furthermore, 
one expects that correct transseries define a $unique$ function of the parameters.
 
 Since we will study in detail tunneling in the double-well potential, let me use this particular example to 
 elucidate the trasseries issue. A standard reference for results obtained using Schreodinger eqn is a summary by Zinn-Justin and Jentschura
\cite{ZinnJustin:2004cg}. Their definition includes particle with mass $m=1$ in  the double well potential 
 \be V(q,g)={q^2\over 2} (1-\sqrt{g}q)^2\ee
The quantity of interest will be the ``vacuum energy" (that of the ground state). The beginning of its
{\em perturbative expansion} in powers of $g$ reads 
\be E_0^{pert}={1 \over 2} -g -{9\over 2} g^2 -{89 \over 2} g^3 -{5013 \over 8} g^4 ... \ee
Note that the coefficients grow rapidly (in fact factorially) and that all signs are the same (minus):
thus we indeed recognize an example of ``bad" Borel  nonsummable series. Several hundreds of
those terms has been generated by some recursive relation: they confirm this conclusion.

It is important to note that the  perturbation theory has no knowledge of the existence of the second well:
thus the two lowest levels $E_0$ and $E_1$ are degenerate.
The {\em nonperturbative effects} are in this case representing by splitting of those levels
due to tunneling effects. The first contribution is given by non-analytic exponent in $g$ with a particular coefficient
\be E_0=E_0^{pert}-{2 \over \sqrt{\pi g}} exp(-{1 \over 6 g})(instanton\, series)\ee 
times another series in $g$ called the ``instanton series". 

Full transseries for $ E_0$ have the form of multi-instantons times new singular function, the log, appearing starting from the two-instanton term
\be E_0=E_0^{pert}+
\sum_{n=1} ({2 \over g})^n \left( - { e^{-{1 \over 6 g}} \over \sqrt{\pi g}} \right)^n \sum_{k=0}^{n-1}
\left( ln(-{2\over g} )\right)^k E_{0,nkl}g^l \ee
We will calculate below several terms in the expansion, so let me now give for reference their values
\be E_{0,100}=1, \,\,\,  E_{0,101}=-{71 \over 12}, \,\,\,  E_{0,102}=-{6299 \over 288} 
\label{ZJ_inst}
\ee

The question of $resurgence$ is whether there are any relations between the perturbative series and those with
different instanton number $n$. In \cite{ZinnJustin:2004cg} the series generating functions $A(E,g),B(E,g)$ were shown
to be related by some exact relation, emanating from the condition that the 
ground state wave function should be symmetric and thus its derivative must be zero at the middle of the potential. 
Dunne and Unsal \cite{Dunne:2013ada} had found even more direct expression of the series, in terms of one function.
We will not however discuss these developments here, following the principle that
only material generalizable to QFT's is inside the scope of these lectures.

The instanton series, by themselves, also literally make no sense:  negative argument of the log
leads to imaginary part, which is physically meaningless for the ground state energy: there can not be any decays as we discuss
the lowest (ground) state of the system. Properly defined  transseries however make all those unphysical imaginary parts
to cancel among themselves, producing the correct real answer. 
If one would be able to show that some well-defined  transseries appear, at least for some QFT's, 
this would be a dramatic shift toward strict mathematical formulation of what these QFT's are.
Yet for now no such examples are known, and all this remains just a theorist's dream.

\section{Complexification and Lefschetz  thimbles }
\subsection{Elementary examples explaining the phenomenon}
Naive approach may suggest simple analytic continuation, from positive to negative $x$, and 
the pioneering papers (Dyson, Bogomolny, Zenn-Justin et al) argued so. But it is not really justified
and does not generally lead to the correct results: see e.g. the following \\
 
{\bf  Excersise}: two functions are defined by the following integrals
$$Z_1=\int_{-\infty}^{\infty} dx e^{-{1 \over 2 \lambda}sinh^2(\sqrt{\lambda}x)}, \,\,\,\, Z_2=\int_{-\infty}^{\infty} dx e^{-{1 \over 2 \lambda}sin^2(\sqrt{\lambda}x)} $$
Expand them in powers of $\lambda$ and show that one leads to Borel summable and another
to non-summable series. Naively they are related by analytic continuation $Z_1(-\lambda)=Z_2(\lambda)$
 but this is not true. Expressing both integrals in terms of Bessel functions, derive
 a correct relation between them which includes an imaginary part:
 $$ Z_1(e^{\pm i \pi} \lambda)= Z_2(\lambda)\mp i e^{-1  \over 2 \lambda} Z_1(\lambda) $$
 
Examples like that show that naive analytic continuations in parameters often 
lead to wrong answers. In general, these are consequences of the so called Stokes phenomenon. The general theory to be discussed elucidate how and why the very geometry
of the integration contours can be abruptly changed, bringing in or out new
saddle points and thus new asymptotic representation of the function.
 
In general the idea to complexify the integration variable of some integrals and change the integration contour
is very old, used in particularly in ``saddle point" method. In quantum mechanics and QFT's applications, with path
integrals, it was tried also in cases with complex action (real time path
integrals, Euclidean theories with finite chemical potential or theta angle, etc) in form of the so called
``complex Langevin quantization". Many people (including myself) experimented with it in 1970's
and found that it worked for some integrals and failed for the others. Recent wave of using 
 complexification are based on more solid ground
 related with Lefschetz -Picard theory.
 
 It  has been introduced
in QM setting by Witten in \cite{Witten:2010zr}, which we now follow. Suppose 
we study a one-dimensional integral
\be  I(a)=\int_{-\infty}^\infty dz e^{S},\, \,\,\,\,\,\,\,\,\S=a*x^2-x^4  \ee
where the function in exponent $S$  is chosen to resemble actions with typically have in QFT's, and 
the ``mass" $a$ is in general some complex parameter.
 
Let us promote $x$ to a complex variable $z=x+i y$ and replace real axes of integration by some
contour $\Gamma$ in the complex plane. Since this particular function has no singularities, any integrals over closed contours $\Gamma$ must 
vanish: and thus only the open contours (with different endpoints) are of interest. 

\begin{figure}[htbp]
\begin{center}
\includegraphics[width=6cm]{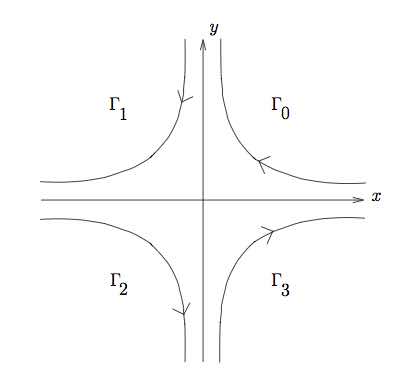}
\caption{Four basic contours on which integral is well defined: since their sum is zero, only 3 are independent.}
\label{fig_4contours}
\end{center}
\end{figure}

What those contour can be?
At large $|z|$ the quartic term is dominant, and the integral is well defined 
only provided the integration contour $C$ ends up at some $z^4>0$ lines, or $z_n\sim e^{i 2\pi {n\over 4}}$ where
$n=0,1,2,3$, see Fig.\ref{fig_4contours}. So $\Gamma$ should approach one of those four lines for the integral to be well defined.

 There are 3 extrema of this action, solutions to
\be 0=S'(z_m)=2*a*z_m -4x_z^3 \ee
which are at the locations $$z_0=0, z_1=\sqrt{a \over 2},z_2=-\sqrt{a\over 2}$$
The simplest to discuss is the first one at the origin: action near it can be approximated by the first
term. For $a>0$, it is clear that the real axis has a minimum at it, while along the imaginary axis it is 
the maximum. For complex $a$ this direction rotates accordingly.

So, rotating the integral to the imaginary axes makes sense. The other two
also have directions on which the integral has a maximum.
All of this is well known, as ``saddle point" method to do the integrals.

 Further improvements start with the question whether there are lines along which
$ReS$ are increasing (or decreasing) $monotonously$.
Those can be followed by ``gradient flow" equation
\be \dot z={\partial \bar{S} \over \partial z} \ee 
where dot stands for the derivative over some ``computer time". The asymptotics of the lines approaches four ``good" directions, along which the integral is convergent.

{\bf Exercise:} Solve this equation and find these 6 lines, originating from all three extrema in both directions, for $a=i$

These lines are called  {\em Lefschetz thimbles}: on them are  the real part of the action grows monotonously, from its 
values at the extreme points to infinity. 
The important statement of the Lefschetz  theory is that the imaginary part of the action, $Im(S)$, {\em remains constant on these lines}. Instead of proving it, let me suggest to check it for the example at hand

{\bf Exercise:} Find the thimble lines for the example at hand using this theorem.

Important consequence of this theorem 
is that since any  integral over some contour $C$ can be 
rewritten as a superposition of  integrals over the thimbles
$$I=\sum_i c_i \int_{\Gamma^i} dz e^{S(z)} $$
 with some coefficients $c_i $, basically $\pm 1$ or 0. Each of them has fixed $Im(S)$ 
 which can be taken out of the remaining integrals, which are therefore all real and well convergent. 

Further consequence is that Lefschetz thimbles can only cross each other if their phases  match,  \be Im(S_i)=Im(S_j), \,\,\,\, (i\neq j) \ee 
Those can be easily found at the extremal points. Rapid changes of the integral -- the ``phase transitions" -- can be caused
by crossing and change of the thimble geometry and thus $c_i$. Thus matching of the phases is a very useful tool.
 
 If the integral has more than one dimension, the geometry of thimbles gets more complicated. And yet there
 are successful practical applications of that, e.g. for models similar to finite density QCD \cite{Alexandru:2015xva}.

Generic contributions of the thimbles have distinct complex phases, and this prevent their crossing.
However at some values of parameters the phases can be the same. Often 
this happens when phases are multiple of $\pi$ with some integer $n$, and the corresponding contributions are real.
Those cases obviously split into two groups: the even $n$ allows for addition of such contributions, but odd $n$
lead to subtractions, sometimes to outright cancellation of them!  
A prototype model (from \cite{Behtash:2015kna})  nicely illustrate   this last point.

\begin{figure}[htbp]
\begin{center}
\includegraphics[width=10cm]{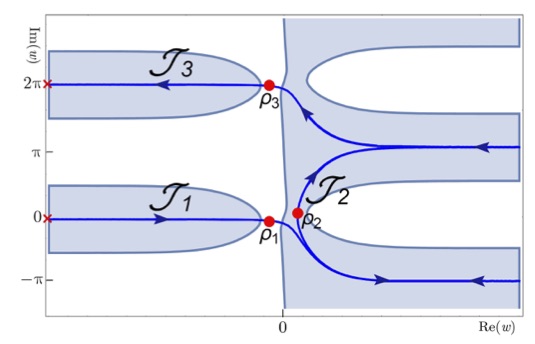}
\caption{The blue areas 
show ``good regions" in which the integrand falls sufficiently rapidly at infinity to guarantee convergence. The red dots give the locations of three saddle points, and the blue contours show  the ``Lefschetz  thimble" paths. 
}
\label{fig_3thimbles}
\end{center}
\end{figure}

 Consider the integral
 \be I(k,\lambda) = \int_\Gamma dw e^{2\lambda sinh(w)+kw } \ee
 with complex parameters $k,\lambda$. The three extrema of it and  thimbles are shown in Fig.\ref{fig_3thimbles},
 and the contribution of the leading 1 and 3 thimbles have the same modulus of the exponential factor and form the following combination
 $$ I\sim  (1+e^{2\pi i(k+1/2)})e^{-S_1} $$
 which, for integer $k$, vanishes because of cancellation between thimbles. It was then shown by these authors
 that in QES and supersymmetric cases when no instanton contribution was present a similar cancellation -- between the
instantons and  some {\em complex saddles} -- take place!

 \subsection{Quasi-exactly solvable models and the necessity of complex saddles}
 Let me start with the definition: quasi-exactly solvable (QES) quantum mechanical problems
 are those which allow {\em some group of states} to be solved explicitly, with their wave functions
 and energies exactly known. (They are different from exactly solvable models in which
 all states can be found.)  References to the original works can be found in  a review \cite{Turbiner:2016aum}, in which also
 the underlying algebraic structure of QES problems is described in detail. We will not go into general discussion, focusing below on just  one
 example .
 
  For general orientation, let me give the simplest example of a familiar general statement: all problems with supersymmetry, which remains unbroken, have the ground state energy {\em exactly equal to zero}.
   That means, that not only
  perturbative series needs to be canceled term by term (which is often possible to see), but also all
   nonperturbative effects should somehow get cancelled. How exactly that happens needs to be understood. 
 
 We will return to QES example a bit later, but we will first address a problem which do not even 
 belong to this class, and yet it provides a very instructive puzzle.  Its proposed resolution is from
  \cite{Kozcaz:2016wvy,Behtash:2015zha} which we here follow. The problem is known as 
  {\em tilted double well potential } (TDW), which is the double well problem with added linear term  $ p\cdot  x$,
  with some parameter $p$.

  Of course, at small $p$ one can proceed perturbatively. The TDW is related to supersymmetric quantum mechanics with the action
  \be S=\int dt \left( {\dot{x}^2 \over 2} + {(W')^2 \over 2}+ \bar \psi \psi + p W" \bar \psi \psi \right)  \ee 
 where $W(x)$ is known as superpotential and $\psi$ is time-dependent Grassmanian field. We will use
 \be W={x^3 \over 3}-x ,\,\,\,  W'= x^2-1, \,\,\,\, W" =2x \ee 
 reproducing the TDW. For  $p=1$ the action is supersymmetric, and perturbatively the ground state energy
 vanishes. However, following a famous Witten's argument, supersymmetry is broken dynamically and
 the actual ground state energy is $nonzero$ (and of course $positive$, since the Hamiltonian is a square). 
 
  For non-tilted case, $p=0$, we developed above a nice theory of instantons and antiinstantons, in full glory and up
  to quantum corrections with one, two and three  loop diagrams. 
  The non-perturbative correction to ground state energy is $negative$,
  But when $p\neq 0$ there are no classical solutions
  going from one maximum to the other, since their hight is now different and energy on classical paths is conserved. 

The authors of  \cite{Kozcaz:2016wvy,Behtash:2015zha} argue that such classical solution is obtained
if one complexify the coordinate, $x\rightarrow z=x+i\cdot y$ and look for solution of the {\em holomorphic
Newton's equation} for inverted complexified potential
\be {d^2 z \over dt^2}= + {\partial V \over \partial z} \ee
Since the solution is going to have finite action, it must start from the highest point of the (inverted) potential
we call $x_{max}$: but where can it go? Some thinking leads to the solution: at $p\neq 0$ the lowest
maximum splits into a pair of two turning points, and our quartic potential can be re-written 
in a form convenient for motion with the maximal energy $E(x_{max})$ as
\be V=E(x_{max})+ (z-x_{max})^2(z-z_1)(z-z_2) \ee
where $z_1^*=z_2$. At the turning point the velocity is zero, and so recoiling back (or going into an
entirely new direction) is possible: so the paths we need to construct should go between $x_{max}$ and $z_1$.

A numerical example of such path is shown in Fig.???, one can also find its analytic form, see    \cite{Behtash:2015zha}.
Its proposed name is ``complexified bion", or $cb$ for short. The
action of this path is complex
\be S_{cb}= {8 \over 3g} + p \cdot log({16 \over pg}) +... \pm i p\pi \ee
 but for $p$ integer $e^{-S}$ is real. Furthemore, for $p=1$ one get the sign opposite from what
 is normally expected from a semiclassical expression (in the Euclidean time).  A number of numerical evidences
 that this is the right solution are given in  \cite{Kozcaz:2016wvy}, but those include using Dunne-Unsal relation
 instead of a direct evaluation of the determinant. 
 
 Now we turn to the second example from \cite{Kozcaz:2016wvy}:  its Hamiltonian
 is 
 \be H={g\over 2} p^2+ {1 \over 2g}( W'(x))^2+{p\over 2} W''(x) \ee
 with $W=-\omega cos(x)$. 
 
 Note that one can put $\omega=1$, recovering all answers by dimensional arguments later.
 The first term in Hamiltonian $\sim sin^2(x)$ has half natural period, while the
  last  $\sim cos(x)$ has a natural period and the coefficient containing an extra parameter
 $p$. So at a generic value of   $p$ every second max/min is different from the first: that is why it is called Double Sine Gordon (DSG) problem.
 
 The  interesting thing about this particular problem is that it is an example of QES problems.
Let us start by following its usual logics, by proposing an Ansatz for the wave functions 
\be \psi(x)=u(x)e^{-{W(x)\over g}} \ee  
 Schreodinger eqn for $u(x)$ is
 \be \left[ -{g \over 2} {d^2 \over dx^2} + sin(x) {d\over dx} - {(p-1) \over 2} cos(x)  \right] u=Eu \ee
 The operator in brackets can be represented in terms of generators of the SU(2) algebra
 \be J_+= e^{ix} (j-i{d \over dx}),\,\,\,\,  J_-= e^{ix} (j+i{d \over dx}),\,\,\,\, J_3=i{d \over dx}
 \ee

 The operator in bracket for our problem can be written in terms of those operators
 \be [ ]= (g/2) J_3^2-(1/2)(J_++J_-) \ee  
 provided $j=(p-1)/2$. The representations of SU(2) require $j$ to be integer or semi-integer:
 thus the trick works only when $p$ is integer. The number of states $2j+1=p$ can be separated from the rest,
 and using $p\times p$ matrix representation of the operators 
 \be J_{\pm}= \sqrt{(j\mp m)( j\mp m+1)} \ee
 one can get exact Hamiltonian.
 Its diagonalization should produce the energies of the $p$ states in question. In the paper 
  \cite{Kozcaz:2016wvy} that is all done for $p=1,2,3,4$.
  
   {\bf  Exercise }: {\em check that all the commutators are indeed those for  SU(2) algebra.
 Also check that
 the Casimir operator is just number, $(1/2)J_+ J_- +(1/2)J_- J_+ J_3^2=j(j+1)$. For $j=1/2,p=2$
 calculate the hamiltonian in the matrix form and find the energy eigenvalues. Expand them in
 powers of $g$ and see the convergent series without $exp(-1/g^2)$ terms.}
 
 Summarizing this discussion: if $p$ is some integer, the problem at hand belongs to the QES set.
Indeed,  there are exactly $p$ levels whose energy and
 wave functions can be exactly calculated. 
 
 Since the energy has no instanton-like terms,
 one needs to explain a puzzle {\em how this may be possible}, since physically the tunneling between different minima
 of the potential cannot possibly disappear. The direct indication of that is that energies of all levels {\em other than the chosen $p$}
 do have the characteristic tunneling corrections!
  
 The resolution of the puzzle proposed in  \cite{Kozcaz:2016wvy} is based on the interference of the contributions of the two
 tunneling paths. One, is real tunneling between $x=0$ and $x=2\pi$ minima: its contribution
 is straightforward to calculate. Also clear is that it shifts the ground energy downward.
 
 New element is the ``complex bion" solution, with the action having imaginary part, as in the case we discussed
 at the beginning of this section. The solution
 \be z_{cb}= 2\pi\pm 4 \left[ arctan (e^{-\omega_{cb} (t-t_0)}) + arctan (e^{-\omega_{cb} (t+t_0)}) \right] \ee 
 can be obtained from the analytic continuation from the real bounce by $p\rightarrow p e^{i\theta}$
 and taking $\theta=\pi$. Here $\omega_{cb}=\sqrt{1+p g/8}$ and complex $t_0 \approx {1\over 2\omega_{cb}} log(-{32\over pg})$ is such that its real part is related to the duration of the bounce.  
 
 The exponent of its action $exp(-S_{CB})=exp(-Re S_{CB}-i \pi p )$ for odd $p$
 gets additional minus sign, meaning the energy is modified upward. It is this contribution which
 has {\em the potential\footnote{So far it was shown that both real parts of the action
 are the same. For two amplitudes to cancel each other $exactly$ one need to 
 be sure that the real and complex paths generate exactly the same perturbative corrections: to see this is beyond 
 current technical means. 
 }  to cancel} that of the real tunneling.  It is argued indirectly, from asymptotic of the perturbative series,
 that such cancellation happens for $p$ levels, but not others. Why it is the case is still not clear. 
 
 Let us end with a wider question. Two examples discussed in this section show a $necessity$
 to include some complex extrema of the path integral,
  beyond the obvious real tunneling paths. Most likely the complex ones 
 proposed so far in the papers mentioned
 do indeed fit the bill, do the right job, explaining these particular puzzles. Yet the general question --
 which of all possible complex classical paths (or solitons for QFTs) one should include, and which
 one should not -- remains unaswered. The Picard-Lefshetz theory provide some light on this issue, so to say ``in principle",
 since the original integration contour can be deformed only to  some 
 particular combination of their thimbles. Yet to use it outside some single (or few-)dimensional integrals  
 is at the moment well beyond our abilities.

%
%
%

\chapter{Gauge field topology and instantons}

\section{Chern-Simons number and topologically nontrivial gauges} 
Topological invariants is a traditional field in mathematics, and we will
need those in a form  discovered by  \cite{Chern:1974ft}.
Generally, they exist in a different form 
in odd and even dimensional spaces, and are  related in a curious way.

 We will start with $d=3$
topology, physically relevant  for a gauge theory defined in 4 dimensions\footnote{ Interesting gauge theories in 3 dimensions  can be defined using $N_{CS}$ 
as the Lagrangian: such construction was introduced by Witten in 1988 and is called the {\em topological field theory} .
While it has applications in physics, e.g. in quantum Hall effect, we will not discuss it. 
}.
The so called 3-d Chern-Simons number density is defined as the 4-th component of the following topological current
\be K_\mu= {1 \over 16\pi^2} \epsilon^{\mu\alpha\beta\gamma} \left( A^a_\alpha \partial_\beta A^a_\gamma +{1\over 3}\epsilon^{abc}A^a_\alpha A^b_\beta A^c_\gamma \right)
\ee
$$ N_{CS}=\int d^3 x K_4 $$

Let us select $t_1=-\infty, t_2=\infty$, and think of 
 the gauge field at such times being ``pure gauge", with zero field strength:
 \be A_i=U^+(\vec x) i\partial_i U (\vec x)
 \ee 
Substituting it to $N_{CS}$ one finds the following expression
\be N_{CS}={1\over 24\pi^2} \int d^3x 
\epsilon^{ijk} (U^+ \partial_i U)  (U^+ \partial_j U)  (U^+ \partial_k U) \ee
Now, $U(\vec x)$ is a map from a 3-d space to a the gauge group. If it is $SU(2)$, with 3 generators
and 3 Euler angles, the group is basically the 3-sphere. The expression above is in fact nothing else
but the topological invariant of such map, known as the {\em winding number}: it is the integer 
 number of times the map covers the group. \\
 
{\bf Exercise:} {\it Consider a ``hedgehog" form for $$U = exp[i{(\vec r  \vec\tau) \over r}  P (r)]$$ with $\tau$ being Pauli matices, $P(0)=0$ and $P(\infty)=\pi n$ with integer $n$.  Note that only with such $P(r)$ the map of the point $r=\infty$
is smooth on the group.
Substitute it into the previous expression
and show that the result is equal to $n$. }\\

What we learned is that pure gauge fields can be split into some topologically distinct classes, and, because of 
the relation (\ref{3_4_relation}), if before and after of certain gauge field configurations the pure gauge
unphysical fields change this class, there must be some 4d topological charge in between.
Usually we do not think much about pure gauge fields, considering them to be unphysical and 
basically completely irrelevant. Now we learned that the so called ``large gauge transformations",
which change the winding number, cannot be  irrelevant because their change is related to 
the 4-d topological charge $Q_T$ which is expressed in terms of gauge fields and is clearly gauge invariant and physical.

\begin{figure}[h!]
\begin{center}
\includegraphics[width=8cm] {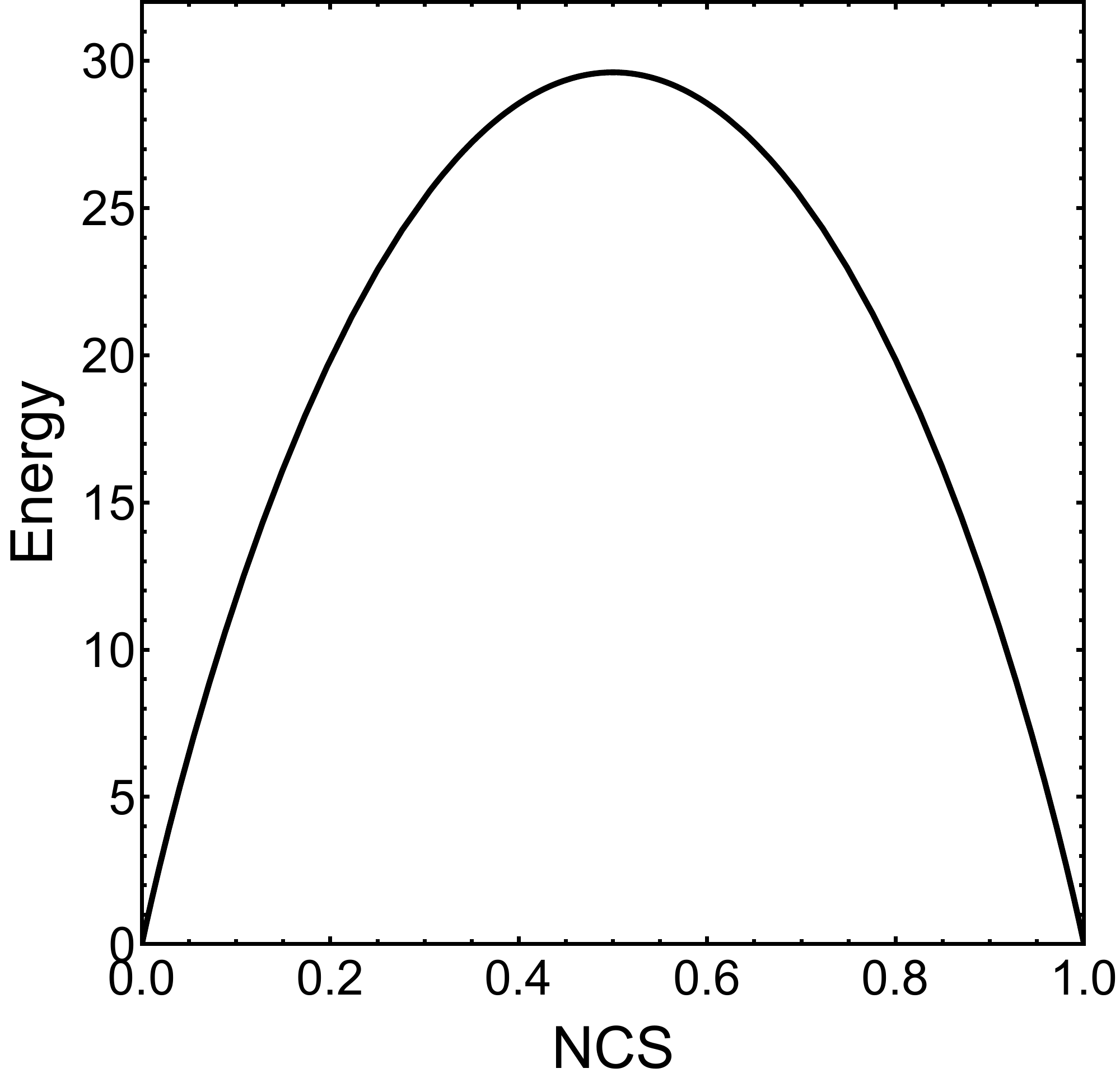}
\caption {\label{fig_sphaleron_path}
The potential energy $E$ (in units of $1/g^2\rho$)
 versus the Chern-Simons number $\tilde{N}_{CS}$, for the 
 ``sphaleron path" solution to be derived in Sphaleron chapter.
}
\end{center}
\end{figure}

Suppose one would like to start with a classical vacuum in a
trivial gauge, with $N_{CS}(-\infty)=0$, and interpolate it somehow with time-dependent  intermediate field to the one
with  $N_{CS}(\infty)=1$. This relation tells us that in doing so one necessarily has to go though
intermediate fields with a nonzero field strength, and thus energy: topologically distinct vacua must be
separated by some kind of a {\em physical barrier}. Since there is no reason transition from 0 to 1 is different from $n$ to $n+1$, we come to the conclusion that the 3-d gauge field configurations have a periodic 
potential as a function of $N_{CS}$. The optimal path, leading from one minimum to another,
is known as the {\em sphaleron path}. In  chapter \ref{chap_sphalerons}  we will derive explicit set of configurations
along which $N_{CS}$ changes from 0 to 1. Their energy $ E_{stat}$ will be calculated as well, explaining the shape and
in particularly the $hight$
of the barrier separating different topological sectors. Let me here give the answer
 \be \label{eqn_dimas_pot}
 E=3\pi^2(1-\kappa^2)^2/(g^2\rho) \\ \nonumber
 {N}_{CS}={\rm sign}(\kappa)(1-|\kappa|)^2(2+|\kappa|)/4 \ee
in parametric form. The corresponding plot $E(N_{CS})$,  Fig.\ref{fig_sphaleron_path}
shows the energy of the ``sphaleron path" configurations between two subsequent
values of Chern-Simons number, 0 and 1. Here $\rho$ is arbitrary r.m.s. size of
the field distribution: it appears because classical Yang-Mills is scale invariant and
the energy is simply inversely proportional to the size. 

Since $N_{CS}\in [integers]$ takes all integer values, from $-\infty $ to $\infty$,
this potential repeats itself infinitely many times. This tells us that,
as a function of this topological coordinate $N_{CS}$, the  gauge theory
 resembles an {\em infinitely long crystal}. Therefore the states in it can be written as
 plain waves
 \be  \langle \theta | N_{CS} \rangle =\sum e^{i\theta N_{CS}} \ee
with quasimomentum $\theta\in [-\pi,\pi]$. We will return to this in the next section.

\section{Tunneling in gauge theories and the BPST instanton}
So, we already know that there is infinite set of pure gauge fields -- therefore with zero field strength and zero energy, ca;;ed {\em classical vacua} -- classified by the integer  $N_{CS}$. We also know that
there are also field configurations with non-integer $N_{CS}$: but those do have nonzero energy, and therefore form a kind of a barrier separating classical vacua.

This barrier separating adjusted classical vacua, e.g. $N_{CS}=0,1$,
turns out to be penetrable for quantum tunneling. Furthermore, as we will see,
the tunneling rate can be found even without knowledge of the detailed
shape of the potential. 

This is possible because of the following  key relation, between the  topological current and the topological charge
in 4d
\be \partial_\mu K_\mu={1 \over 32 \pi^2} G^a_{\mu\nu} \tilde{G}^a_{\mu\nu}   \label{eqn_K_GG} \ee
The integral over the r.h.s. 
\be Q_T=  {1 \over 32 \pi^2} \int d^4 x G^a_{\mu\nu} \tilde{G}^a_{\mu\nu} \ee
 is the so called 4-d topological charge. 

Let us think about the consequence of this relation. Assuming one deals with some gauge field which
decays well at spatial infinity, in a spirit of Gauss theorem, let us consider two time surfaces
and integrate this relation in 4-volume $V_4$ between them
\be Q_T(V_4)= \int_{t_1}^{t_2} dt {\partial N_{CS}\over \partial t} = N_{CS}(t_2)-N_{CS}(t_1) \label{3_4_relation} \ee
It means that the topological charge in a volume between two time surfaces is
equal to the $difference$ of $N_{CS}$ defined at those time moments. 

(This is completely analogous to what one learns studying static electrodynamics: if there is a difference
between electric field fluxes through two planes, you know how much charge is enclosed in between.)

In this section  we are going to look for a tunneling path in gauge theory, which connects
topologically different classical vacua, found in  the famous work
by Belavin, Polyakov, Schwartz and Tyupkin and thus known as the BPST instanton \cite{Belavin:1975fg}. 

 To find classical solution corresponding to tunneling, BPST  used  the following 4-dimensional spherical ansatz  depending on $radial$ trial function $f$
\be g A_\mu^a=\eta_{a\mu\nu} \partial_\nu F(y), \,\,\,\, F(y)=2\int_0^{\xi(y)} d\xi'   f(\xi')     \ee
with $\xi=  ln(x^2/\rho^2)$ and $\eta$ the 't Hooft symbol
defined by 
\be 
\label{eta_def}
\eta_{a\mu\nu}&=&\left\{ \begin{array}{rcl}
 \epsilon_{a\mu\nu} &\hspace{0.5cm}& \mu,\,\nu=1,\,2,\,3, \\
 \delta_{a\mu}      &              & \nu=4,  \\
-\delta_{a\nu}      &              & \mu=4.
\end{array}\right.
\ee
We also define $\overline\eta_{a\mu\nu}$ by changing the sign of
the last two equations. Further properties of $\eta_{a\mu\nu}$ are 
summarized in appendix \ref{app_eta}. 
Upon substitution of the gauge fields in  the gauge Lagrangian $(G_{\mu\nu})^2$ 
one finds that the effective Lagrangian has the form
\be L_{eff}=    \int d\xi \left[{\dot{f}^2\over 2}+2f^2(1-f)^2 \right]
\ee   
corresponding to the motion of a particle in a double-well potential. For the exploding sphaleron we
used it in real (Minkowski) time, by changing $\xi \rightarrow i\xi_M$ and $non$-flipping the potential
to its Minkowski version, with an infinite energy at $x\rightarrow \infty$. 
Now we need much simpler Euclidean solution, the same as for the quantum mechanical instanton,
connecting the $maxima$ of the flipped potential.  The corresponding field is 
\be  
\label{BPST_inst} 
A^a_\mu(x)= {2\over g}{\eta_{a\mu\nu}x_\nu \over x^2+\rho^2}
\ee 
Here $\rho$ is an arbitrary parameter characterizing the size of
the instanton. Like in the potential we discussed in the preceding
section, its appearance is dictated by the scale invariance of classical
Yang-Mills
equations.

The ansatz itself perhaps needs some explanation. The 't Hooft symbol projects to sefl-dual fields.
The reason we selected it is related with the following identity 
\be 
\label{Bog_ineq}
S &=& \frac{1}{4g^2} \int d^4x\, G^a_{\mu\nu} G^a_{\mu\nu} \;=\;
 \frac{1}{4g^2}\int d^4x\, \left[\pm G^a_{\mu\nu} \tilde G^a_{\mu\nu}
 + \frac{1}{2} \left( G^a_{\mu\nu}\mp \tilde G^a_{\mu\nu}\right)^2
  \right],
\ee
where $\tilde G_{\mu\nu}=1/2\epsilon_{\mu\nu\rho\sigma}G_{\rho\sigma}$
is the dual field strength tensor (the field tensor in which the roles
of electric and magnetic fields are interchanged). Since the first term
is a topological invariant (see below) and the last term is always 
positive, it is clear that the action is minimal if the field 
is (anti){\em  self-dual}\footnote{This condition written in Euclidean
notations; in Minkowski space extra $i$ appears in the electric field.}
\be 
\label{self_dual}
G^a_{\mu \nu}&=&\pm\tilde G^a_{\mu \nu},
\ee
In a simpler language, it means that Euclidean electric and magnetic
fields are the same\footnote{In the BPST paper the selfduality condition (1-st order differential equation)
was solved, rather than (2-nd order) EOM for the quartic oscillator. }. The action density is given by
\be 
\label{g2_inst}
(G^a_{\mu\nu})^2 &=&\frac{192\rho^4}{(x^2+\rho^2)^4} .
\ee 
and one can see that it is indeed spherically symmetric and very very well localized, 
at large distances it is $\sim x^{-8}$.
The action depends on scale only via the running coupling
\be  S={8\pi^2 \over g^2(\rho) }
\ee
which we will discuss more in the next section.

Note that while the gauge potential is long-range, 
$A_\mu\sim 1/x$, in the field strength the gradient and 
the commutator terms canceling each other. So physical
effects are not long-range: it suggests
that the tail of the potential is gauge artifact. 
 
Using invariance of the Yang-Mills equations under 
coordinate inversion  implies that the singularity of 
the potential can be shifted from infinity to the origin by means 
of a (singular) gauge transformation $U=i\hat x_\mu\tau^+$. The 
gauge potential in singular gauge is given by
\be 
\label{inst_sing} 
A^a_\mu(x)= {2\over g} 
 \frac{x_\nu}{x^2}\frac{\overline\eta_{a\mu\nu}\rho^2}{x^2+\rho^2}. 
\ee 
This singularity at the origin is unphysical, pure gauge, like the one for regular gauge at infinity.
While there is only one infinity, each instanton has its own zero, and so the singular gauge is 
better suited to make a superposition of many instantons. 

What about the multi-instanton solutions? Its a long story, and the solution known as 
ADHM construction  (for Atiah,Drinfeld,Hitchin,  and Manin \cite{Atiyah:1978ri}) in principle solved it. 
What ``in principle" means is that if one can solve all equations, the number of the parameters
in the solution is equal to the number of zero modes, that $4N_c$. 
Let me just mention that when the solution is found, one also gets Green functions, Dirac zero modes
etc for free, automatically. 
It all started from one
brilliant idea. If the field is pure gauge, $A_\mu=\Omega^+ \partial \Omega$ the unitary gauge matrix
can be eliminated from long derivatives in a standard way.  The unitary gauge matrix can be written as
$\Omega=exp(i \vec n \vec \tau)$with real $\vec n$. If one take $complex$ $\vec n$ instead,
$\Omega$ is not unitary and the field is therefore not a pure gauge. And yet, in many
equations one can still get rid of this  $\Omega$
as if it would be a gauge matrix. For example, the quark Green function 
is still $S(x,y)=\Omega(x) S_0(x,y) \Omega(x)^+$.

\subsection{The theta-vacua}

The fact that the action for the instanton is finite, means that the barrier separating valleys in
the topological landscape, with different $N_{CS}$, is penetrable. Since the potential as a function of 
$N_{CS}$ is periodic, the complete set of  states  $\psi_\theta$, characterized by a phase $\theta\in [0,2\pi]$  is the so called ``theta-vacua", with the theta
parameter -- ``quasimomentum" -- defined by the 
periodicity condition
\be \psi_\theta(x+n)
=e^{i\theta n}\psi_\theta(x) \ee
   Let us see how this band arises from tunneling events. If instantons
are sufficiently dilute, then the amplitude to go from one topological
vacuum $|i\rangle$ to another $|j\rangle$ is given by
\be 
\label{sum_SIA}
 \langle j | \exp(-H \tau) | i\rangle  &=& \sum_{N_+} \sum_{N_-} 
 \frac{\delta_{N_+-N_- -j+i}}{N_+! N_-!} 
 \left(K \tau e^{-S}\right)^{N_+ +N_-},
\ee 
where $K$ is the pre-exponential factor in the tunneling amplitude
and $N_\pm$ are the numbers of instantons and ant-instantons. Using 
the identity
\be  
\delta_{ab} &=& \frac{1}{2\pi }\int_0^{2\pi} d\theta\, e^{i \theta (a-b)} 
\ee 
the sum over instantons and anti-instantons can rewritten as 
\be 
 \langle j | \exp(-H \tau) | i\rangle  
 &=& \frac{1}{2\pi} \int_0^{2\pi} d\theta\, 
 e^{i \theta (i-j)} \exp\left[2K \tau \cos(\theta) \exp(-S)\right] .
\ee
This result shows that the quantum energy eigenstates are the theta vacua 
$|\theta\rangle =\sum_n e^{in\theta}|n\rangle$. Their energy is 
\be 
\label{E_theta}
E(\theta) &=&  - 2 K \cos(\theta) \exp(-S) 
\ee
As usual, the width of the zone is on the order of the tunneling rate. The lowest 
state corresponds to $\theta=0$ and has $negative$ energy. This is as
it should be, since the tunneling lowers the ground state energy.

At nonzero $\theta\neq 0$ the vacuum is not $T$ or $CP$ invariant: indeed it has ``an arrow
of time". The instanton amplitude has complex phase $e^{i\theta}$, and anti-instanton gets
the conjugate phase $e^{-i\theta}$. 
In a world with nonzero $\theta$ there exists the so called Witten effect:
electric and magnetic fields get admixed. For example, a magnetic monopole
obtains some electric charge as well. Neutrons, together with their
usual magnetic moment, obtain also an electric dipole, etc.

Experiments\footnote{Specifically, the hunt for a nonzero  electric dipole moment of the neutron.}  show that $CP$ symmetry is satisfied in strong interactions, so
$ |\theta | < 10^{-10}$. So we do live in the bottom (the lowest state) of the $\theta$ zone\footnote{Why?
The value of  $\theta$ cannot be changed within the QCD. Hypothetical new particles, called $axions$ were suggested,  to  relax any theta-vacuum to the bottom of the zone.
Multiple searches for axions were made, so far without success.}.

\begin{figure}[h!]
\begin{center}
\includegraphics[width=7cm]{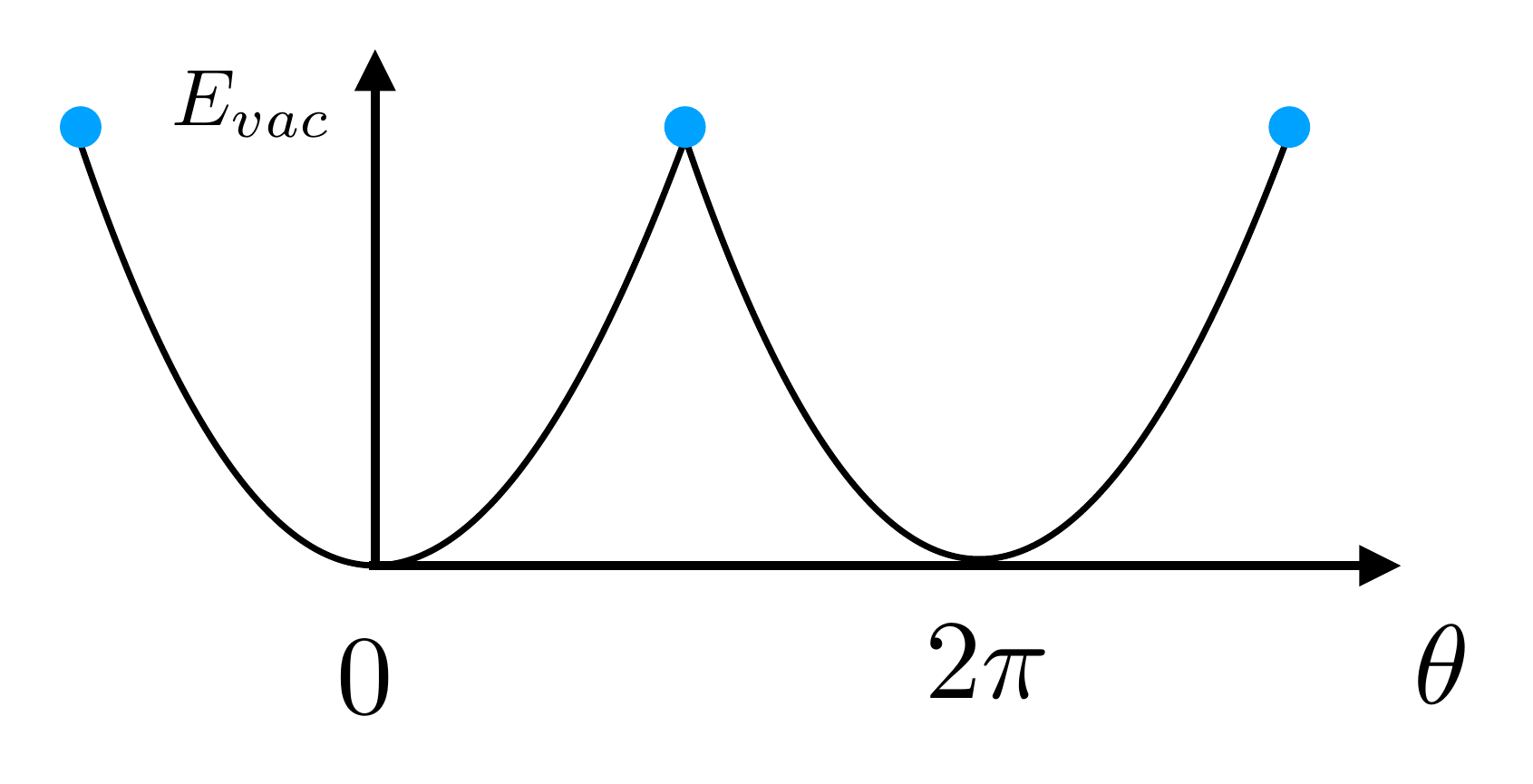}
\caption{Vacuum energy versus theta, schematically.}
\label{fig_theta_vacua}
\end{center}
\end{figure}

It is obvious that all effects -- e.g. the vacuum energy --are periodic in  $\theta$ with period $2\pi$. 
An interesting fact is that two brunches of the vacuum meet at $\theta=\pi$, crossing as
shown schematically in Fig.\ref{fig_theta_vacua}. The world with $\theta=\pi$ is 
$T$ and $CP$ even, as instantons and anti-instantons get the same phase and factor $-1$.
The two branches lead however to the double-degenerate vacua, and selecting one of them
breaks $T$ and $CP$ symmetry $spontaneously$. If one arranges domains of both types of vacuum, they are separated with the 1d topological object, the {\em domain wall}.   
The excitations living of this wall have been studied, and provided interesting window into
the  QCD-like theories compactified to 3 dimensions. 

Presence of light quarks affect theta-vacua dramatically. It is enough to say that
if $any$ quark flavor be truly massless, the fermionic determinant
of all gauge fields configurations with global $Q\neq 0$ would vanish.
This would mean that the whole theta-zone would collapse into a single vacuum,
as theta-depepdency would be erased.  In the real world, however, this is not the case.

\subsection{The one-loop correction to the instanton: the bosonic determinant}
  The next natural step is the one-loop calculation of the pre-exponent 
in the tunneling amplitude. In gauge theory, this is a rather tedious 
calculation which was done in the classic paper by 't Hooft\cite{tHooft:1976snw}.
Basically, the procedure is completely analogous to what we did in the 
context of quantum mechanics. The field is expanded around the 
classical solution, $A_\mu=A_\mu^{cl}+\delta A_\mu$. In QCD, we have
to make a gauge choice. In this case, it is most convenient to work
in a background field gauge $D_\mu (A_\nu^{cl})\delta A_\mu=0$. 
 
  We have to calculate the one-loop determinants for gauge fields,
ghosts and possible matter fields (which we will deal with later). The
determinants are divergent both in the ultraviolet, like any other 
one-loop graph, and in the infrared, due to the presence of zero modes. 
As we will see below, the two are actually related. In fact, the QCD 
beta function is only $partly$ determined by the zero modes, while in certain 
supersymmetric theories, the beta function is completely determined by 
zero modes, as we will discuss later.

  First one has to deal with the $4N_c$ zero modes of the system.
 (For two groups one has to deal with in practice, electroweak $SU(2)$ and QCD $SU(3)$,
 let us enumerate them explicitly. In both cases there are 4 coordinates plus one size $\rho$.
 For   $SU(2)$ there are 3 Euler angle for rotations, either in space or in color space -- does not matter
 as those are directly related: thus 5+3=8.  For   $SU(3)$ group one simply do imbedding of the
 2-color solution into some subgroup of  $SU(3)$. Out of $N_c^2-1=8$  rotation angles,
 one can see that only $one$ does not affect the  $SU(2)$ instanton, so there are 7 angles, and 5+7=12.)
  
The integral over the zero mode is traded for an integral over 
the corresponding collective variable. For each zero mode, we get one
factor of the Jacobian $\sqrt{S_0}$. The group integration is compact,
so it just gives a factor, but the integral over size and position we 
have to keep. As a result, we get a differential tunneling rate 
\be
 dn_I \sim  \left(\frac{8\pi^2}{g^2}\right)^{2N_c}
 \exp\left(-\frac{8\pi^2}{g^2}\right) \rho^{-5}d\rho d^4z,
\ee
where the power of $\rho$ can be determined from dimensional considerations\footnote{ 
Note that we for the first time meet here 5-d Anti-de-Sitter space
with the 5-th coordinate being a scale; it is the same as in famous
Maldacena duality.} 

  The ultraviolet divergence is regulated using the Pauli-Vilars
scheme, the only known method to perform instanton calculations (the
final result can be converted into any other scheme using a perturbative 
calculation). This means that the determinant $\det O$ of the differential 
operator $O$ is divided by $\det(O+M^2)$, where $M$ is the regulator mass. 
Since we have to extract $4N_c$ zero modes from $\det O$, this gives a 
factor $M^{4N_c}$ in the numerator of the tunneling probability. 

   In addition to that, there will be a logarithmic dependence on $M$
coming from the ultraviolet divergence. To one loop order, it is just
the logarithmic part of the polarization operator. For any classical
field $A^{cl}_{\mu}$ the result can be written as a contribution to 
the effective action \cite{Brown:1978yj}
\be 
\delta S_{NZM}= {2\over 3} {g^2 \over 8\pi^2} \log(M \rho)S(A^{cl})
\ee
In the background field of an instanton the classical action cancels 
the prefactor ${g^2 \over 8\pi^2}$, and $\exp(-\delta S_{NZM})\sim 
(M\rho)^{-2/3}$. Now, we can collect all terms in the exponent of
the tunneling rate
\be
 dn_I \sim  \exp\left(-\frac{8\pi^2}{g^2}+4N_c\log(M\rho)
  - \frac{N_c}{3}\log(M\rho) \right) \rho^{-5}d\rho dz_\mu \\ \nonumber
  \;\equiv\; \exp\left(-\frac{8\pi^2}{g^2(\rho)}\right) 
\rho^{-5}d\rho dz_\mu,
\ee
where we have recovered the running coupling constant $(8\pi^2)/
g^2(\rho) = (8\pi^2)/g^2 - (11N_c/3)\log(M\rho)$. Thus, the infrared 
and ultraviolet divergent terms combine to give the coefficient 
of the one-loop beta function, $b=11N_c/3$, and the bare charge 
and the regulator mass $M$ can be combined into to a running coupling 
constant. At two loop order, the renormalization group requires the
miracle to happen once again, and the non-zero mode determinant 
can be combined with the bare charge to give the two-loop beta function
in the exponent, and the one-loop running coupling in the pre-exponent. 

  The remaining constant from the determinant of the non-zero modes
was calculated in \cite{tHooft:1976snw}. The 
result is
\be 
\label{eq_d(rho)} 
dn_I &=& \frac{0.466\exp(-1.679 N_c)}{(N_c-1)!(N_c-2)!}
 \left(\frac{8 \pi^2}{g^2}\right)^{2 N_c} 
 \exp\left(-\frac{8\pi^2}{g^2(\rho)}\right) 
 \frac{d^4zd\rho}{\rho^5}. 
\ee
The tunneling rate $dn_A$ for anti-instantons is of course identical. 
Using the one-loop beta function the result can also be written as
\be 
\frac{dn_I}{d^4z} &\sim&  \frac{d\rho}{\rho^5} (\rho \Lambda)^b 
\ee
and because of the large coefficient $b=(11 N_c/3)=11$, the 
exponent overcomes the Jacobian and small size instantons are
strongly suppressed. On the other hand, there appears to be 
a divergence at large $\rho$, although
the perturbative beta function is not applicable in this regime.


\subsection{Propagators in the instanton background}

 In the chapter on semiclassics we discussed a systematic semiclassical loop expansion,
 allowing to calculate the tunneling probability and vacuum energy order by order, using
 Feynman diagrams. The same method may of course be used for QFTs and the gauge theory in
 particular. Unfortunately, the gauge theory instanton amplitudes has not yet been calculated even 
 to the two-loop order\footnote{For a comparison, some QCD quantities like vector current polarization operator
 is now calculated to 5 loops.}. In this section we will discuss technical difficulties related with
 such calculation, as well as some ideas how to go around those.
 
The central objects we would need for  Feynman diagrams are of course the propagators of the relevant fields,
the quarks and gluons and perhaps ghosts, in the instanton background. As always, the 
fields are written as classical plus quantum ones, $$A^\mu=A_{cl}^\mu+a^\mu$$ and the action expanding in powers of the latter. The
propagators (or Green functions) are
the inverse of the corresponding differential operator defined  by the quadratic form $O(a^2)$.
The technical problems mentioned are related to the fact that inversion can only be
performed in a part of the functional space ``normal" to zero modes.

But let us follow the problem methodically. The fist step is the calculation of a propagator for $scalar$ particles,
both in fundamental and adjoint color representations. They satisfy the equation
\be -D_\mu^2 G(x,y)=\delta(x-y) \ee  
with the covariant derivative 
containing the background field of the instanton
\be  i D_\mu= i \partial_\mu + T_a A_{\mu}^a  \ee 
and so, symbolically the scalar propagator is 
$$ \Delta(x,y)=\langle x | {1 \over - D^2 } | y \rangle $$

There are no zero modes and the explicit solution 
was found by 
\cite{Brown:1977eb}. It is instructive to explain why they were able to do so.
The  instanton potential in the singular gauge can be written as 
\be A_{\mu}^a(x)=-\bar \eta^a_{\mu\nu} \partial_\nu ln\big[1+{\rho^2 \over (x-z)^2 }\big] \ee
If the quantity in the square bracket is some general function $log[\Pi(x)]$ and the field is supposed to be self-dual, the condition on $\Pi$ turns out to be  the Laplacian
\be \partial_\mu^2 \Pi=0 \ee
in which the usual derivative, {\em not the covariant one}, appears. Thus, one can use it to generate
a multi-instanton solution of the form\footnote{Counting parameters one would find that
this cannot be the most general multi-instanton solution.}
\be \Pi=1+\sum_i{\rho^2_i \over (x-z_i)^2 } \ee
The crucial observation is that the instanton potential  has the form
$\Omega(x-z)^{-1} i\partial_\mu  \Omega(x-z)$  which looks like pure gauge,
except that $\Omega$ is $not$ an unitary matrix (and therefore the field strength is {\em not zero}). 
This features allows to factor the $D_\mu^2$ operator and therefore also factor out its inverse.
Using ansatz
\be \Delta(x,y) = \Pi(x)^{-1/2}  {F(x,y) \over 4\pi^2 (x-y)^2 }  \Pi(y)^{-1/2} \ee
with $\Pi(x)$ defined above,
one can finally find the form for the function\footnote{We use here and elsewhere
the 4-d Pauli matrix extension $\tau_\mu=(\vec \tau,i)$.} 
\be F(x,y)=1+\sum \rho_i^2 {\big(\tau_\mu (x-z_i)_\mu \big) \over (x-z_i)^2}{\big(\tau^+_\mu (y-z_i)_\mu \big) \over (y-z_i)^2} \ee
Note, that it defines the scalar propagator not just for one instanton, but for multi-instanton solution of the form considered. Note also, that at large distances $x,y \gg \rho$ all the correction factors are 
small compared to one in them, and the propagator reduces to 
free (no background) scalar propagator $$\Delta \rightarrow  {1 \over 4\pi^2 (x-y)^2 } $$

{\bf Exercise:} {\em check that it satisfies the equation} \\ 

The next step is getting the isospinor spinor (quark) propagator, or more precisely its part
normal to the zero mode. 
This was achieved by using the fact that zero mode resign for one chirality only.
This leads to the form of the propagator
\be S_{nz}=(\gamma_\mu D_\mu) {1 \over -D^2} ({1+\gamma_5 \over 2})+ 
{1 \over -D^2} (\gamma_\mu D_\mu) ({1-\gamma_5 \over 2}) \ee
and since the inverse of $D^2$ we know already, the scalar propagator, 
the task is accomplished just by differentiation of it.

At this point the reader perhaps expects the gauge propagator to be
obtained by similar tricks. This is correct: the symbolic expression for the
vectror field propagator can indeed be written as 
\be D^{nz}_{\mu\nu} \sim D_\mu ({1\over D^2 }) ({1\over D^2 })  D_\nu \ee
The long derivatives at the left and right of the expression act on the middle of it.
The product of two inversions of $ D^2$ should be understood as 
 the $convolution$ of two scalar propagators, which includes the integration over some
intermediate point\footnote{Not to be confused with the instanton center we also called $z$ above. Currently we have a single instanton with the center at the origin.} 
$z_\mu$, namely
$$ \sum_z\langle x | ({1\over D^2 })  | z \rangle \langle z |({1\over D^2 }) | y \rangle = $$
\be \int {d^4z \over 4\pi^2} 
{ (1/2)tr\big[ \tau_a (\tau^+_x \tau_z+\rho^2)  (\tau^+_z \tau_y+\rho^2) \tau_b (\tau^+_y \tau_z+\rho^2)  (\tau^+_z \tau_x+\rho^2) \big] \over (x^2+ \rho^2)(y^2+ \rho^2) (z^2+ \rho^2)^2  (x-z)^2(y-z)^2      }  \ee
where we used shorthand notations, $\tau_z=(\tau_\mu z_\mu)$, etc. The trace is a certain
polynomial in components of $x,y,z$ vectors.
So, if the integral over $z$ can somehow be calculated, the propagator is obtained by differentiation. The problem remains, that
 only some analytic limits of the convolution integral is
analytically known, not the complete integral.

The equation for the non-zero mode vector Green function has the  explicit form
\be -\big( D^2 \delta_{\mu\lambda} +2 G_{\mu\lambda}\big) D^{nz}_{\lambda\nu}(x,y)=P_\perp(x,y)=\delta_{\mu\nu}\delta(x-y)-\sum_{i} \phi_\mu^i(x)\phi^i_\nu(y) \ee
in which the r.h.s. is the projector to all $non-zero$ modes.
It is a complete delta funciton minus projector to $all$ zero modes.

The SU(2) instanton has 8 of them: 4 translations, 1 scale transformation and 3 Euler
angles of rotation. All modes can be obtained by differentiation over corresponding collective coordinates.

Already from the equation itself one can see a coming problem. Let us look at it at
large distances from the instanton. The operator in the l.h.s. becomes the ordinary Laplacian
$\partial^2$ and the r.h.s. has the large-distance tails of the modes.
Let us take one of them, the scale one, as example. $$\phi_{scale} \sim {\partial A_\mu \over \partial \rho^2}\sim {x \over (x^2+\rho^2)^2} \sim {1 \over x^3}$$
Since $\partial^2 D^{nz}(x,y)\sim x^{-3}$, the tail of the vector Green function
must be $D^{nz}\sim 1/x$. If so, the Feynman diagrams involving 4-d integrals of
it can have bad infrared divergences.  

It has then been pointed out by
 \cite{Levine:1978ge} that  this difficulty can be eliminated by a redefinition of
 ``orthogonality": in Hilbert space it can be defined with some weight function,
 which can be chosen to decay with distance appropriately. Their ``improved"
 propagator was shown to have no $D^{nz}\sim 1/x$ tail. And yet, neither these authors
 no anybody else used their improved propagator for four decades.

 one of the methods 
for evaluation of the determinant, namely its relation to the Green function.
The reader may be reminded that the method was based on differentiation
of the classical solution over some parameter, and relating the results to a one-loop
Feynman diagram including the propagator, which can be calculated
if the latter is known.

A comment about ADHM construction above sheds some light on
how exact propagators in the field of instanton (or many instantons)
were calculated -- see the actual derivations in the original paper \cite{Brown:1977eb}.
Now,In the case of the instanton there is indeed such parameter -- the size $\rho$ -- 
and the method can be applied. In fact that method was used by Brown and  Creamer
\cite{Brown:1978yj} for this purpose for the first time. 
One may expect that following this route one can cut off many difficulties 
of the problem, resolved by 't Hooft by brute force diagnalization.

Brown and  Creamer were able to show that all UV divergencies occur as expected, leading
to the correct renormalized charge. But, attempting to calculate the finite part, Brown and  Creamer unexpectedly
found infrared divergencies stemming from projector normal to zero modes. 

Although the setting of next order calculations is in principle quite 
analogous to those we have discussed in quantum mechanical
context in chapter Semiclassical, and  even the diagrams are
the same,  in all the years passed since 1976
 the two-loop correction to 't Hooft formula has not yet
been
calculated. 

\subsection{The exact NSVZ beta function for supersymmetric theories}
\label{sec_NSVZ}

  At first glance, instanton amplitudes seem to violate supersymmetry:
the number of zero modes for gauge fields and fermions does not match, 
while scalars have no zero modes at all. 
We will however not discuss translation of the problem to a superspace, and other related issues here,  sticking to the standard notation. The
remarkable fact is that the determination of the tunneling 
amplitude in SUSY gauge theory is actually $simpler$ than in QCD.
Furthermore,  one can determine the 
complete perturbative beta function from a generic calculation of the tunneling amplitude!

  The tunneling amplitude in question is given by
\be
\label{susy_amp_1l}
 n(\rho) \sim  \exp\left(-\frac{2\pi}{\alpha(M)}\right)
 M_0^{n_g-n_f/2} \left( \frac{2\pi}{\alpha(M)} \right)^{n_g/2}
 d^4x \frac{d\rho}{\rho^5} \rho^k \prod_f (fermions)_f ,
\ee 
where all factors can be understood from the 't Hooft calculation 
discussed above. There are $n_g=4N_c$ bosonic zero 
modes that have to be removed from the determinant and give one power of 
the regulator mass $M$ each. Similarly, each of the $n_f$ fermionic zero 
modes gives a factor $M^{1/2}$. Introducing collective coordinates 
for the bosonic zero modes gives a Jacobian $\sqrt{S_0}$ for every
zero mode. Finally, the last factor in the integral is related to fermionic 
collective coordinates and zero modes (to be discussed a bit later) 
 and $\rho^k$ is the power of $\rho$ needed 
to give the correct dimension. 

Here is the key observation: {\em supersymmetry ensures that spectra of eigenmodes
for bosonic and fermionic fluctuations around the instanton are related}. 
As a result, one can show that the
non-zero mode contributions in bosonic and fermionic determinants exactly cancel:
therefore there is no need to calculate them!

 More precisely, the 
subset of SUSY transformations which does not rotate the instanton 
field itself, mixes fermionic and bosonic modes non-zero modes 
but annihilates zero modes. This is why all non-zero modes cancel
but zero modes can be unmatched. Note another consequence of the 
cancellation: {\em the power of $M$ in the tunneling amplitude is 
an integer}.

   Renormalizability demands that the tunneling amplitude is 
independent of the regulator mass. This means that the explicit 
$M$-dependence of the tunneling amplitude and the $M$ dependence
of the bare coupling have to cancel. As in QCD, this allows us to 
determine the one-loop coefficient of the beta function $b=(4-N)
N_c-N_f$. Again note that $b$ is an integer, a result that would 
appear very mysterious if we did not know about instanton zero modes.

   In supersymmetric theories one can even go one step further 
and determine the beta function to all loops \cite{Novikov:1985rd}. 
For that purpose let us write down the renormalized instanton 
measure
$$
\label{susy_amp}
 n(\rho) \sim  \exp\left(-\frac{2\pi}{\alpha(M)}\right)
 M_0^{n_g-n_f/2} \left( \frac{2\pi}{\alpha(M)} \right)^{n_g/2}
 Z_g^{n_g} \left( \prod_f Z_f^{-1/2} \right)
 d^4x \frac{d\rho}{\rho^5} \rho^k \prod_f (fermions)_f ,
$$
where we have introduced the field renormalization factors $Z_{g,f}$
for the bosonic/fermionic fields. Again, non-renormalization theorems
ensure that the tunneling amplitude is not renormalized at higher
orders (the cancellation between the non-zero mode determinants 
persists beyond one loop). For gluons the field renormalization
(by definition) is the same as the charge renormalization $Z_g=\alpha_R/
\alpha_0$. Furthermore, supersymmetry implies that the field 
renormalization is the same for gluinos and gluons. This means
that the only new quantity in (\ref{susy_amp}) is the anomalous
dimension of the quark fields, $\gamma_\psi=d \log Z_f/d \log M$. 

   The renormalizability demands that all physical quantities -- such as the amplitude 
   under consideration -- are independent
of $M$. All powers of $M$ we found should thus be compensated by the $M-$dependence of the
charge. Indeed the cutoff of the integrals at $M$ implies that
the original charge was in fact the ``bare one", $g(M)$.
This condition gives the charge dependence on the scale, which can be re-formulated as the
so called {\em NSVZ beta function} \cite{Novikov:1985rd}
which, in the case $N=1$, reads
\be
\label{NSVZ_beta} 
\beta(g)&=&-\frac{g^3}{16\pi^2} 
  \frac{3N_c-N_f+N_f\gamma_\psi(g)}{1-N_c g^2/8\pi^2}.
\ee
The anomalous dimension of the quarks has to be calculated 
perturbatively. To leading order, it is given by
\be
\gamma_\psi(g) &=&-\frac{g^2}{8\pi^2} \frac{N_c^2-1}{N_c} +O(g^4).
\ee 
As far as I know,
the result (\ref{NSVZ_beta}) was checked by explicit calculations
up to three loops\footnote{ Note that the beta function is 
scheme dependent beyond two loops, so in order to make a comparison
with high order perturbative calculations, one has to translate 
from the Pauli-Vilars scheme to a more standard perturbative 
scheme, e.g. $\overline{MS}$.}.

   In theories without quarks, the NSVZ result determines the 
beta function completely. For $N$-extended supersymmetric 
gluodynamics, we have
\be
\label{NSVZ_beta_N}
\beta(g) &=& -\frac{g^3}{16\pi^2}\frac{N_c(N-4)}{1+(N-2)N_c g^2/(8\pi^2)}.
\ee
Let us recognize several  interesting special cases:\\
(i)  For $N=4$,
the beta function vanishes and the theory is conformal. 
The reason for that we already discussed in the chapter on monopoles,
where it was shown that this theory is electric-magnetic $self-dual$.\\
(ii) The case 
$N=2$ shows another curious phenomenon: the non-trivial part of the
denominator vanishes, so that {\em the one loop result for the 
beta function becomes exact}. This theory is the one partially solved by Seiberg and Witten:
we will follow the charge running in it in the next section.\\
(iii)
  The next interesting case is the $N=1$ SUSY QCD, where we add $N_f$ 
matter fields (quarks $\psi$ and squarks $\phi$) in the fundamental 
representation. Let us first look at the NSVZ beta function. 
 The beta function vanishes 
at $g^2_*/(8\pi^2)= [N_c(3N_c-N_f)]/[N_f(N_c^2-1)]$, where we have 
used the one-loop anomalous dimension. This is reliable if $g_*$ is 
small, which we can ensure by choosing $N_c\to\infty$ and $N_f$ in 
the conformal window $3N_c/2<N_f<3N_c$.  Seiberg showed 
that the conformal point exists for all $N_f$ in the conformal 
window (even if $N_c$ is not large) and clarified the structure of 
the theory at the conformal point. This is a phenomenon
which also exists in QCD-like theories with many fermions.

 Another observation is that
for $N=1$
the beta function blows up at $g^2_*=8\pi^2/N_c$, so the renormalization 
group trajectory cannot be extended beyond this point. The meaning of it remains mysterious (to me).

\subsection{Instanton-induced  contribution to the renormalized charge}

Specific running of the coupling in non-Abelian gauge theories -- the {\em asymptotic freedom} --
gave birth to QCD, and so it is not surprising that higher order corrections to the 
original celebrated one-loop result had attracted a lot of attention. Current knowledge
of perturbative corrections can be found in the Appendix ??.

But one should  not necessarily think only about the UV divergences in perturbative diagrams while considering the renormalized
coupling constant. Let me give two examples: one very simple and one very complex.
My first example are Feynman diagrams in quantum-mechanical 1+0 dimensional path integrals, say with 
$g^2 x^4$ interaction term. There are many loop diagrams renormalizing $g^2x^4$ operators, but
there are no divergences and UV logarithms. 

The second example --
conceptually simple but technically challenging -- is the case of 4-d {\em superconformal field theories}, SCFT's.
We mentioned two of them before:  $\cal N$=4
SYM and also $\cal N$=2,$N_c=2$,$N_f=4$ supersymmetric QCD. SCFT's have zero beta functions and thus -- unlike QCD --
no $\Lambda$ parameter.
There is a bare coupling $g_0$ 
in the Lagrangian, which is however different from the ``true" coupling $g$ in certain exactly known observables. 
The relation between them can be expanded in the instanton series: for details see \cite{Alday:2009aq,Marshakov:2009kj}. 

Let us however return to QCD-like theories with the running coupling.
A charge is defined, by general OPE rules ,
as a coefficient $1/g^2(\mu)$ in front of the operator $G_{\mu\nu}^2$ in the action, 
where general rules define the field $G$ as the ``soft" one, containing only Fourier harmonics with $p<\mu$.
All ``hard" phenomena, with $p>\mu$, are supposed to be delegated to the coefficient.
All non-perturbative phenomena -- and in particular instantons with sufficiently small size $\rho \mu<1$ -- 
need to be included in the running charge. 
The difference between the perturbative and non-perturbative series is that the former are
series in $1/log(\mu)$ while the latter in (inverse) powers of $\mu$.

 \begin{figure}[htbp]
\begin{center}
\includegraphics[width=10.cm]{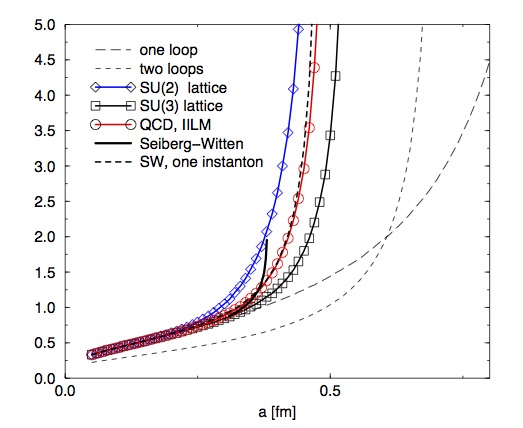}
\caption{The effective charge $b g_{eff}(\mu)/8\pi^2$, where $b$ is the one-loop  
coefficient of the  beta function) versus normalization scale ? (in units of its value at which the one-loop charge blows up). The thick solid line correspond to exact solution [20] for the N=2 SYM, the thick dashed line shows the one-instanton correction. Lines with symbols (as indicated on figure) stand for N=0 QCD-like theories, SU(2) and SU(3) pure gauge ones and QCD itself. Thin long-dashed and short-dashed lines are one and tho-loop results.}
\label{fig_RRS_charge}
\end{center}
\end{figure}

 Let us discuss the running coupling in a number of theories, following the paper by Randall, Ratazzi
 and myself  \cite{Randall:1998ra}.
 The best known case is the  $\cal N$=2 SYM theory, in which one knows both the exact 
 analytic expression for the charge
dependence on the scalar VEV $a$ from Seiberg-Witten elliptic curve (see Appendix ??) , and
also its perturbative-nonperturbative series which start as follows
\be 
{8\pi^2 \over g^2(a)}= 2 log({2a^2 \over \Lambda^2})- 6\left( {\Lambda \over a}\right)^4+...
\ee 
where the dots do not include high loop logs -- they vanish in this theory, as shown in the previous section --
but higher powers of the instanton terms $\left( {\Lambda \over a}\right)^{4k}$. The number 4 in power,  the same as
the coefficient of $log(a)$, is nothing but the one-loop coefficient of the beta function $b$: see NSVZ result discussed above. All terms have been explicitly calculated by Nekrasov \cite{Nekrasov:2002qd}, and they confirm
the expansion of the Seiberg-Witten elliptic curve.

This expression of course should only be used when the second term is much smaller than the first,
but one can still make  a tempting observation: since they are of the opposite sign, perhaps at some scale
the r.h.s. vanishes, which means that the coupling gets infinite! According to Seiberg-Witten it is indeed
the case, but the singularity is at a place slightly misplaced compared to what one would get from those two terms. 

In Fig.\ref{fig_RRS_charge} from  \cite{Randall:1998ra}   the exact answer (solid think line) is compared with the
one-instanton expression (the thin dashed line). The QCD curves  correspond to the OPE definition
\be
{8\pi^2 \over g^2(a)}=b log({a \over \Lambda}) - {4\pi^2 \over N_c^2-1} \int_0^{1/a} dn(\rho) \rho^4 ({ 8\pi^2 \over g^2(\rho)})^2
\ee
including the instanton density with the size $\rho$ extracted either from lattice simulations or
models -- both to be discussed below. The bottom line is that in all cases one finds 
a very similar behavior: at certain scale the instanton-induced power term is rapidly switched in,
and increases the coupling. This explains why the transition from weak to strong coupling 
happens rather abruptly. 

%

\section{Single instanton effects }
\subsection{Quarkonium potential and scattering amplitudes} 


The simplest effects we will describe simply utilize relatively strong
gluonic fields of the instantons. Perhaps the simplest effect is instanton contribution
to the effective gluon mass\footnote{The effective quark mass is related to
chiral symmetry breaking and is thus multi-instanton effect, to  be discussed 
in next chapters.}, obtained by averaging of the gluon propagator. As shown by
\cite{Musakhanov:2017erp}, with the original parameters of the instanton liquid model, it is $M_g^{inst}\approx 360\, MeV$.

\begin{figure}[htbp]
\begin{center}
\includegraphics[width=6.cm]{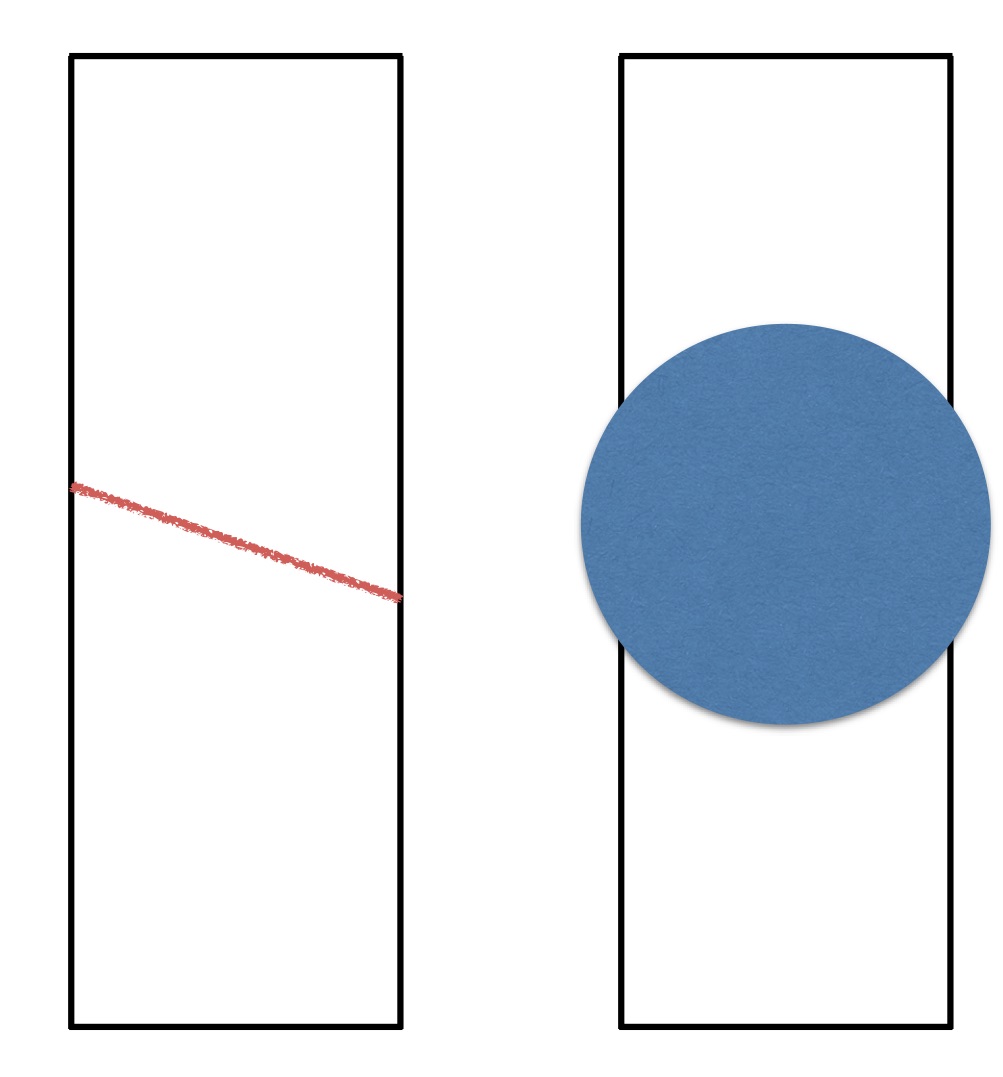}
\caption{The one-gluon exchange and instanton-induced settings for the quarkonium potential}
\label{fig_pot_pert_inst}
\end{center}
\end{figure}

The
 first example would be instanton contribution to the quarkonium potentials:
 it is one of the oldest ideas, suggested by Callan et al \cite{Callan:1978ye}.
Quarkonium is substituted by a color dipole, a pair of Wilson lines.
In Fig.\ref{fig_pot_pert_inst} we show a perturbative and instanton-induced settings.

Spin-independent (or central) potential is represented as 
\be V=\int d\rho {1 \over \rho^2}{dn(\rho) \over d\rho} W(x/\rho) 
\ee
where density $dn(\rho)$ includes both instantons and antiinstantons, and the last factor 
 is the convolution of two Wilson lines, each done exactly
$$ W(x/\rho) ={1 \over 3\rho^3} \int d^3r \, tr \left[ 1- W(\vec x-\vec r)W^+(-\vec r)\right] $$
$$ W(\vec r)=cos({\pi r \over \sqrt{r^2+\rho^2}})+ {\vec r \vec \tau \over r } sin({\pi r \over \sqrt{r^2+\rho^2}}) $$
Since these authors were also interested  in  magnetic effects as well -- spin-spin and spin-orbit ones -- they make the lines 
a bit tilted relative to the time direction, so they looked as follows
\be Pexp(i \int A_\mu dx_\mu)=exp(i \int d\tau { 
 -(\vec \tau  \vec x)+(\vec \tau  \vec v \times \vec x)
  \over 
\tau^2 + \rho^2+(\vec x +\vec v \tau)^2  })
\ee
and then expanded in velocity to the needed order. More details can be found in the paper.

\begin{figure}[htbp]
\begin{center}
\includegraphics[width=14cm]{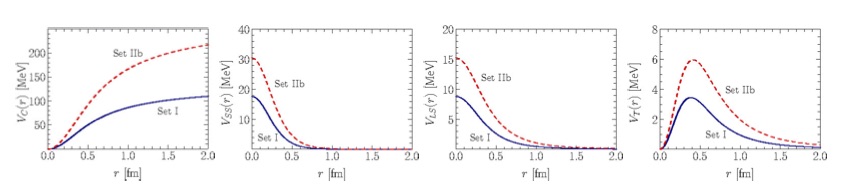}
\caption{The instanton-induced heavy quark potentials. Solid (dashed) curves are for $\rho=1/3\, fm, R=1 \, fm$  ($\rho=0.36\, fm, R=0.89\, fm$) instanton parameters.}
\label{fig_charm_potentials}
\end{center}
\end{figure}

One can also calculate other relativistic corrections, namely the spin-spin, spin-orbit and tensor potentials, defined by
\be V=V_C(r)+V_{SS}(r)\big( \vec S_Q  \vec S_{\bar{Q}}\big) +V_{LS}(r)\big( \vec S \vec L\big)
+V_T(r)\big[ 3 (\vec S_Q \vec n )(\vec S_{\bar{Q}}\vec n)-(\vec S_Q  \vec S_{\bar{Q}})\big] \ee
As shown by \cite{Yakhshiev:2018juj,Musakhanov:2018sdu}, the instanton model \cite{Shuryak:1981ff} with the original parameters,
$\rho=1/3\, fm, n=1 \, fm^{-3}$ describes well spectrum of known charmonium states,
including $L=0,1,2$ states. The corresponding potentials are shown in Fig.\ref{fig_charm_potentials} for two sets of the instanton parameters.

Similarly, Zahed and myself \cite{Shuryak:2000df}  have  generalized this calculation to 
instanton-induced static dipole-dipole potential, or the interaction $between$ two
quarkonia. Before describing it, let me remind the situation in QED:
according to famous Casmir-Polder paper the interaction of two distant dipole is
\be V(r)=-{ \alpha_1 \alpha_2 \over r^7} \ee
where $\alpha_i$ are the so called polarizabilities. Note that it is not a square of the dipole field $\sim 1/r^6$:
the difference comes from the  time delay.

Let me briefly reproduce this result in Euclidean setting we use.
For simplicity let us only consider the case when dipoles are small $d\ll R,\rho$. In this case 
te dipole approximation is justified, both gluons (photons)
are emitted at close time and the correlator to consider is
\be  <\vec E^2(\tau_1)\vec E^2(\tau_2)> \sim {1 \over \left[ R^2+(\tau_1-\tau_2)^2\right]^4 }\ee
Its integral over relative time leads to Casmir-Polder answer just mentioned.

\begin{figure}[t]
\begin{center}
\includegraphics[width=6.cm]{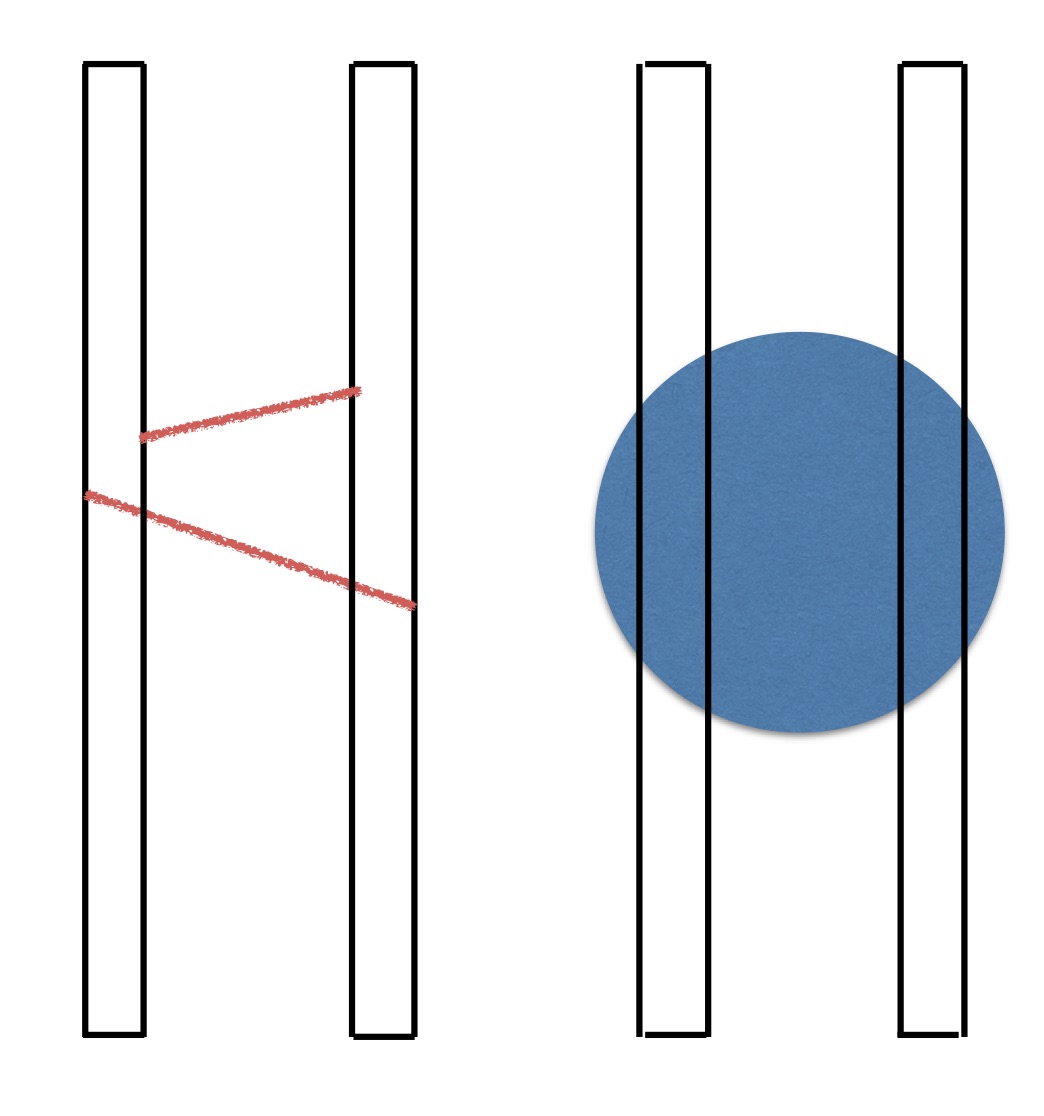}
\caption{The setting of the dipole-dipole potential calculation: two-gluon exchange and instanton-induced }
\label{fig_dipole_dipole}
\end{center}
\end{figure}

The correlator of two fields squared in the instanton background  is calculated using its expression (\ref{g2_inst}).
The averaging over the instanton location can be carried out analytically: we then get the so called correlation function
(to be discussed systematically in the next chapter)

\begin{figure}[h]
\begin{center}
\includegraphics[width=10.cm]{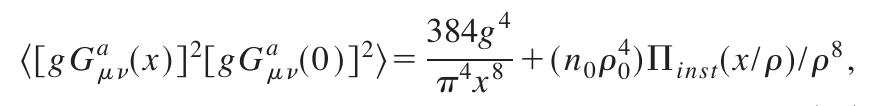}\\
\includegraphics[width=10.cm]{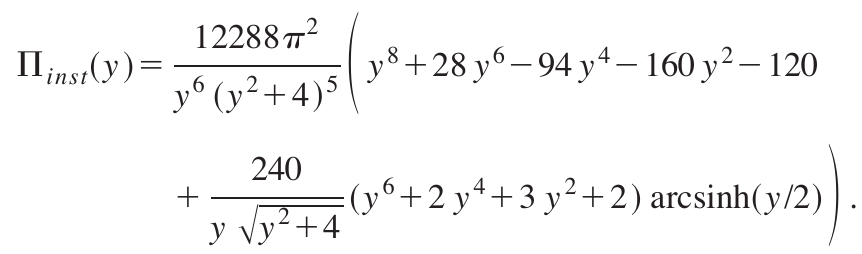}\\
\includegraphics[width=10.cm]{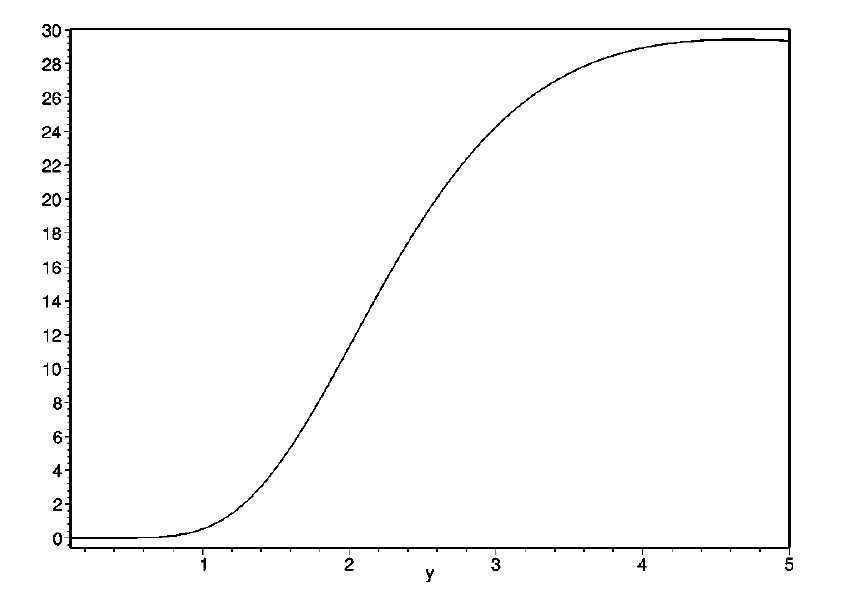}
\caption{The correlator of two gauge fields squared, perturbative and instanton-induced. The
plot is the ration of the former to the latter, as a function of distance (in units of the instanton seize $\rho$}
\label{fig_dipole_dipole_2}
\end{center}
\end{figure}

 It is important to
point out that while in the quarkonium potential the instanton-induced effect is relatively
 small compared to
the perturbative one-gluon exchange, it is in fact dominant in the
dipole-dipole case for $R>\rho$, see Fig.\ref{fig_dipole_dipole_2}
. This is a   manifestation of a general trend: the higher is the order of the effect
in perturbation theory, the more important instanton-induced effects become. Indeed, because
the instanton field is $O(1/g)$, the coupling to a quark is cancelled out and extra exchanges go ``for free",
without any additional penalty. The only small factor in the problem is then the instanton amplitude itself.

Introduction of the non-zero angle
$\theta$ between the Wilson lines promotes the calculation into that of the
scattering
amplitude. A continuity between static potential and the low-energy
scattering amplitude $\theta\rightarrow 0$ for two very heavy dipoles is 
then apparent. 
 The untraced
and tilted Wilson line in the one-instanton background reads
\begin{equation}
{\bf W} (\theta, b) = {\rm cos}\,\alpha -i\tau\cdot\hat{n}\,{\rm sin}\,\alpha
\label{eq_QQI1}
\end{equation}
where
\begin{equation}
n^a ={\bf R}^{ab}\,\eta^b_{\mu\nu}\,\dot{x}_\mu (z-b)_\nu ={\bf R}^{ab}\,{\bf n}^b
\label{eq_QQI2}
\end{equation}
and $\alpha=\pi\gamma/\sqrt{\gamma^2+\rho^2}$ with
\begin{eqnarray}
\gamma^2=n\cdot n={\bf n}\cdot{\bf n}\nonumber=
(z_4{\rm sin}\theta - z_3 {\rm cos}\theta )^2 + (b-z_\perp)^2\,\,.
\label{eq_QQI3}
\end{eqnarray}

 \begin{figure}[htbp]
\begin{center}
\includegraphics[width=4.cm]{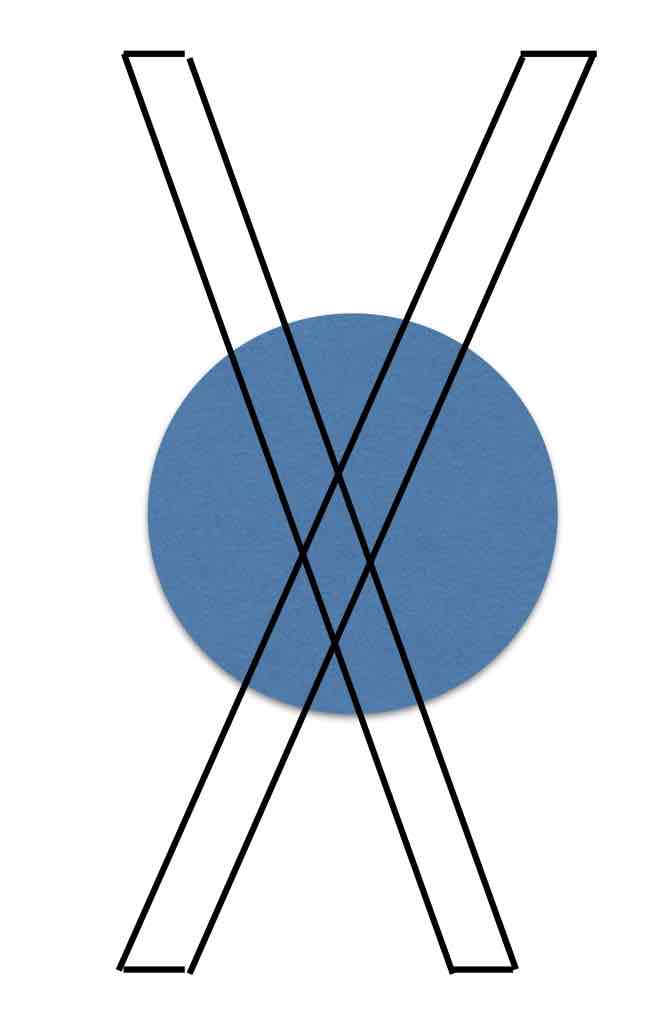}
\caption{default}
\label{fig_dd_inst}
\end{center}
\end{figure}

In case of high energy collisions eikonalized expressions 
for the scattering amplitude in terms of
a correlator of two Wilson-lines (quarks) 
or Wilson-loops (dipoles) are well known, developed 
first  systematic  step toward a semi-classical but
non-perturbative formulation of high-energy scattering
in QCD was suggested by \cite{Nachtmann:1996kt},
relating the scattering amplitude  to the
correlator of pairs of Wilson lines.

 One can use these
expressions,
in Euclidean space-time with  the angle $\theta$ between the Wilson lines, and  
later  analytically continue the result into the Minkowski world by the substitution
 \be \theta \rightarrow iy \ee 
where $y$ is the Minkowski rapidity difference between the colliding objects. It has been checked in \cite{Meggiolaro:1997dy,Shuryak:2000df}   and elsewhere that
in 
that it works correctly for perturbative amplitudes. We  used it for quark-quark and dipole-dipole scattering
amplitudes as well
\cite{Nowak:2000de,Shuryak:2003rb}
: the setting for the latter case is shown in Fig.\ref{fig_dd_inst}. 

Strictly speaking,a mutial scattering of two small dipoles  correspond to {\em double deep-inelastic scattering}.
For example, future lepton collider can be used as a collider of two virtual photons $\gamma^*\gamma^*$.
 The quark-antiquark pair produced by the photons have 
small sizes $d\sim 1/Q$ provided each photon is highly virtual  $Q\gg \Lambda_{QCD}$.
But in practice the parton enesemble is often represented as a set of dipoles, and even the
proton itself sometimes is treated as a color dipole made of quark and diquark. For the details about instanton-induced
effect in scattering amplitudes see the papers mentioned.

\section{ Fermionic transitions during changes of gauge topology}
\subsection{The fermionic zero mode of the instanton}
\label{sec_zero_modes}

The so called {\em index theorems} connect topology of the manifold with
the {\em number of zero modes} of the Dirac operator defined on them. 
Without going into details, let me just say that
if the gauge field has the topological charge $Q$, then \\
(i) the  fundamental (isoscalar) fermions (e.g. 
quarks)  should have  $Q$ zero modes; \\
(ii) the adjoint (isovector) fermions (e.g. gluinoes of supersymmetric gauge theories)
should have $Q N_c$ zero modes.

The meaning of these facts are very profound, with important consequences for instanton-induced effects we will be discussing.  

Let us first look at the explicit form of the zero mode
 originally discovered by 't Hooft \cite{tHooft:1976snw}.  It is a solution of
 the Dirac equation $iD\!\!\!\!/\,\psi_0(x)=0$
in the instanton field. For an instanton in the singular gauge, the
zero mode wave function take the form
\be  
\label{eq_zm}
\psi_0(x)={\rho \over \pi} \frac{1}{(x^2+\rho^2)^{3/2}} 
 \frac{\gamma\cdot x}{\sqrt{x^2}}\frac{1+\gamma_5}{2} \phi 
\ee
where $\phi^{\alpha m}=\epsilon^{\alpha m}/\sqrt{2}$ is a constant 
spinor\footnote{It can therefore be called ``spinor hedgehog". 
The norm is such that this
mode is normalized to $\int d^4x \bar\psi_0\psi_0=1$.} in which the $SU(2)$ color index $\alpha$ is coupled to the 
spin index $m=1,2$. Let us briefly digress in order to show that
(\ref{eq_zm}) is indeed a solution of the Dirac equation. First
observe that\footnote{We use Euclidean Dirac matrices that 
satisfy $\{\gamma_\mu,\gamma_\nu\}=2\delta_{\mu\nu}$. We also
will use the following combinations
of gamma matrices $\sigma_{\mu\nu}=i/2[\gamma_\mu,\gamma_\nu]$ and 
$\gamma_5=\gamma_1\gamma_2\gamma_3\gamma_4$.}
\be
\label{dslash2}
 (iD\!\!\!\!/\,)^2 &=& 
      \left( -D^2+\frac{1}{2}\sigma_{\mu\nu}G_{\mu\nu} \right) .
\ee
We can now use the fact that $\sigma_{\mu\nu}G_{\mu\nu}^{(\pm)}=\mp
\gamma_5 \sigma_{\mu\nu}G_{\mu\nu}^{(\pm)}$ for (anti) self-dual fields 
$G_{\mu\nu}^{(\pm)}$. In the case of a self-dual gauge potential the 
Dirac equation $iD\!\!\!\!/\,\psi=0$ then implies ($\psi=\chi_L+\chi_R$)
\be
\label{dslash_LR}
 \left( -D^2 +\frac{1}{2}\sigma_{\mu\nu}G_{\mu\nu}^{(+)}\right)\chi_L\;=\;0,
 \hspace{1cm}  -D^2\chi_R \;=\; 0,
\ee
and vice versa ($+\leftrightarrow -,\,L\leftrightarrow R$) for 
anti-self-dual fields. Since $-D^2$ is a positive operator, $\chi_R$
has to vanish and the zero mode in the background field of an instanton 
has to be left-handed, while it is right handed in the case of an 
anti-instanton.
This result is not an accident. Indeed, 
there is a mathematical theorem (the Aliyah-Singer index theorem), 
that requires that $Q=n_L-n_R$ for every species of chiral fermions. 
A general analysis of the solutions of (\ref{dslash_LR}) was given in
\cite{tHooft:1976snw}.  For (multi) instanton gauge 
potentials of the form $A_\mu^a=\bar\eta_{\mu\nu}^a\partial_\nu
\log\Pi(x)$  the solution is of the
form 
\be
\label{zm_ansatz}
 \chi_\alpha^m &=& \sqrt{\Pi(x)}\partial_\mu 
   \left(\frac{\Phi(x)}{\Pi(x)}\right) (\tau_\mu^{(+)})^{\alpha m}.
\ee
The Dirac equation requires $\Phi(x)$ to be a harmonic function,
$\Box\Phi(x)=0$. Using this result, it is straightforward to verify 
(\ref{eq_zm}). Again, we can obtain an $SU(3)$ solution by embedding 
the $SU(2)$ result. 

 It is also important that
 there is no chirality partner for zero modes:
 the "pairing" theorem for $\lambda$ and $-\lambda$ modes 
 holds  for non-zero modes only. So what was wrong with 
the proof? Of course the assumption that $\gamma_5 \psi_\lambda$
can be used as $another$ eigenvector: it would not work for purely chiral
solutions. 
So, what the existence of the zero mode mean for the tunneling rate?
  At the first glance, zero mode is a problem since it vanishes the
fermionic determinant in the partition function. Indeed,
the determinant is of the operator $i\Dslash+im$, and since the former
term gives zero on a zero mod 
 one has  to conclude that for massless fermions the tunneling
probability vanishes.
Not necessarily, argued 't Hooft, since the mass term can be
supplemented by external scalar current. What it all means, there is
no tunneling {\em unless} 
 a $\bar q_R q_L$ pair for each massless flavor is produced. 

  Still the whole process looks very mysterious.
 The final "demystification" of the anomaly 
%
 is as follows: one can follow the tunneling configurations adiabatically,
and for each value of time we are looking for static energy 
levels of the Dirac
particle and ignoring all time derivatives. 
One then finds that the levels move
in such a way, that all left-handed states make one step down,
to the next level, and all 
right-handed ones make one step up.
A hint that this is the case can be explained as follows:
in the adiabatic approximation (slow change in time) the time-dependent
solution is 
\be \psi(t,x)=\psi_{static}(t,x) exp[-\int^t_0 dt' 
\epsilon_{static}(t')] \ee
If energy is positive for large t and negative for t $\rightarrow  \infty$,
the corresponding time-dependent wave function is 4-dimensionally normalizable.
The explicit 't Hooft zero mode is such a normalizable
 solution. Thus, if  only one such 
solution exists, it means that only $one$ state has passed 
the zero energy mark. 

 So, when tunneling is finished, the spectrum is
of course the same, but it is the {\em level  occupation}
which is different! 

\subsection{Electroweak instantons violate baryon and lepton numbers!}
Briefly about the electroweak instantons : Very little changes in terms of formulae are need, while the numbers
involved are drastically different. The Higgs VEV sets a scale and therefore
instanton size also get fixed \cite{tHooft:1976snw}. The charge at
  electroweak scale is small, and therefore
 the probability of tunneling
is now extremely small
\bea P \sim exp(-{16\pi^2\over g_w^2}) \sim 
10^{-169} \ee
so it seems out of question that one can observe any manifestations
of it. 

However, it is still important to discuss  what should happen if one observes
the consequences of electroweak instanton. Unlike QCD, weak gauge fields 
are only coupled to left-handed fermions. So there is no cancellation of the
anomaly (by right-handed fermion loop) for vector current: it is $not$ therefore conserved,
as well as the axial current.
 
 Let us think what it implies: unlike chirality in QCD, we now have nonconservation
 of the $baryon$ $B$ and $lepton$ $L$ numbers! More specifically, an instanton must generate
production of 9 quarks (3 colors times 3 generations) and 3 leptons, or \be \Delta B=\Delta L=3 \ee 
The difference $B-L$ is thus conserved. 

While electroweak instantons (tunneling) are unobservable, the corresponding
  {\em sphaleron transitions} at finite temperature electroweak plasma
  are much more probable. They have quark and lepton zero modes, and generate similar process, to be discussed 
in the sphaleron chapter.

\subsection{Instanton-induced  ('t Hooft) Effective Lagrangian}
\label{sec_tHooft_L}

\begin{figure}[htbp]
\begin{center}
\includegraphics[width=10.cm]{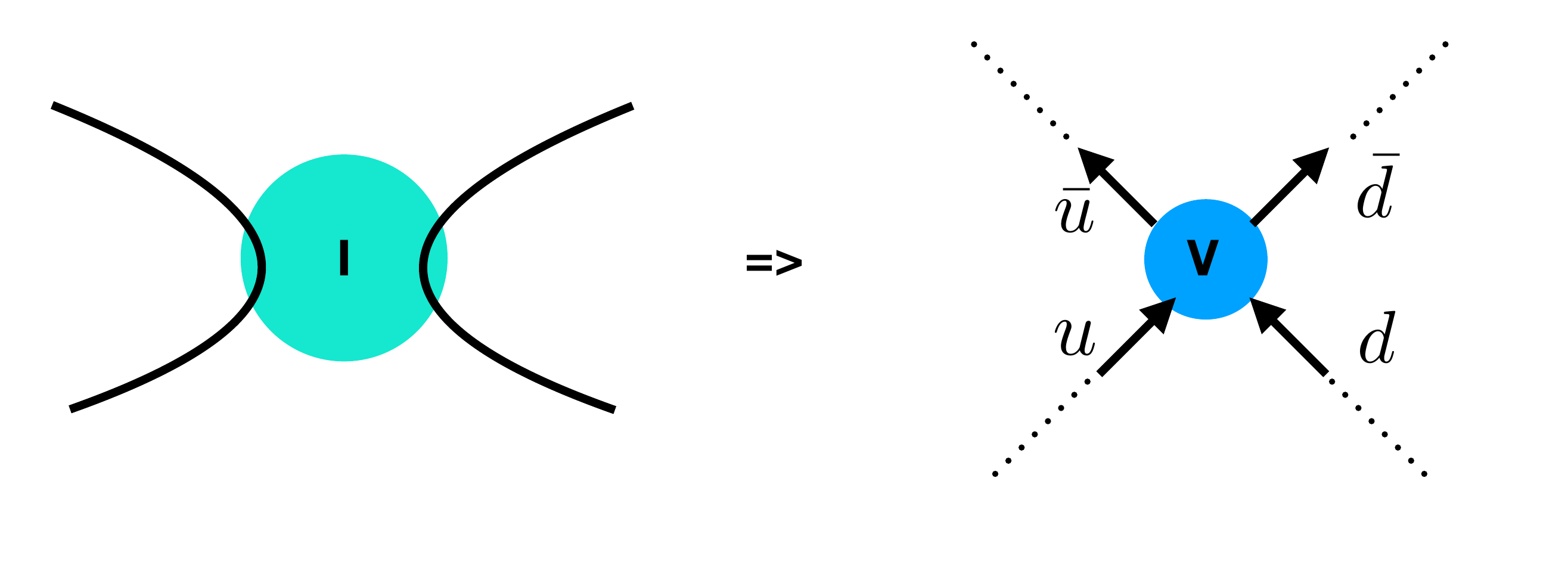}
\caption{The $2N_f$-leg Green function (left, for $N_f=2$) at large distances
can be seen as a local $2N_f$ operator $V$ with free propagators (shown by the dotted lines). The free propagators can be amputated.}
\label{fig_tHooft_vertex}
\end{center}
\end{figure}

Let us introduce quark sources $j_f(x)$ via auxillary
terms in Lagrangian $\int d^4x (j^+_f (x)\psi_f(x)+cc)$ and 
calculate a $2N_f$-quark 
Green's function $$G(x_1...x_f y_1 ...y_f)=\langle \prod_f \bar j_f(x_f) j_f(y_f)
\rangle$$ containing a quark and an anti-quark of each flavor once. Contracting 
all the quark fields, the Green's function is given by the tunneling 
amplitude multiplied by $N_f$ fermion propagators. It is shown schematically in Fig.\ref{fig_tHooft_vertex}

It would be important to introduce also quark mass $m$ (same for all) as a IR regulator.
Every propagator 
has a zero mode contribution with one power of the fermion mass in 
the denominator: 
\be
\label{S_inst}
S(x,y)&=&\frac{\psi_0(x)\psi^+_0(y)}{im}
 +\sum_{\lambda\neq 0}\frac{\psi_\lambda(x)\psi^+_\lambda(y)}{\lambda+im} 
\ee
where I have written the zero mode contribution separately. 
Note that if both points $x,y$ are far from the instanton center
(relative to $\rho$), one can use asymptotic expression for $\psi_0$
which at large arguments behave as constant spinor times
 $1/x^3$. Since  this behavior is nothing else but just free
 propagator
for a massless fermion, one sees that in this limit the first term
can be interpreted as two free propagators, from $x$ to $z$ and from 
  $y$ to $z$, times some constant vertex. The procedure we have
 described
is in fact standard ``amputation of external legs'' of the Green functions,
used when one would like to derive the effective vertex or Lagrangian.

Let us now look at the dependence on the light quark masses. Suppose 
there are $N_f$ light quark flavors, and all masses are the same.
The instanton amplitude 
is proportional to $m^{N_f}$ (or, more generally, to $\prod_f m_f$)
due to the fermionic determinant in the weight. But contributions of the zero modes
in the propagators give us  $1/m^{N_f}$ !
 As a result, the zero mode contribution to the 
Green's function is finite in the chiral limit\footnote{Note that 
Green's functions involving more than $2N_f$ legs are not singular
as $m\to 0$. The Pauli principle always ensures that no more than 
$2N_f$ quarks can propagate in zero mode states.} $m\rightarrow 0$.

\begin{figure}[ht]
\centering
\includegraphics[width=9.cm]{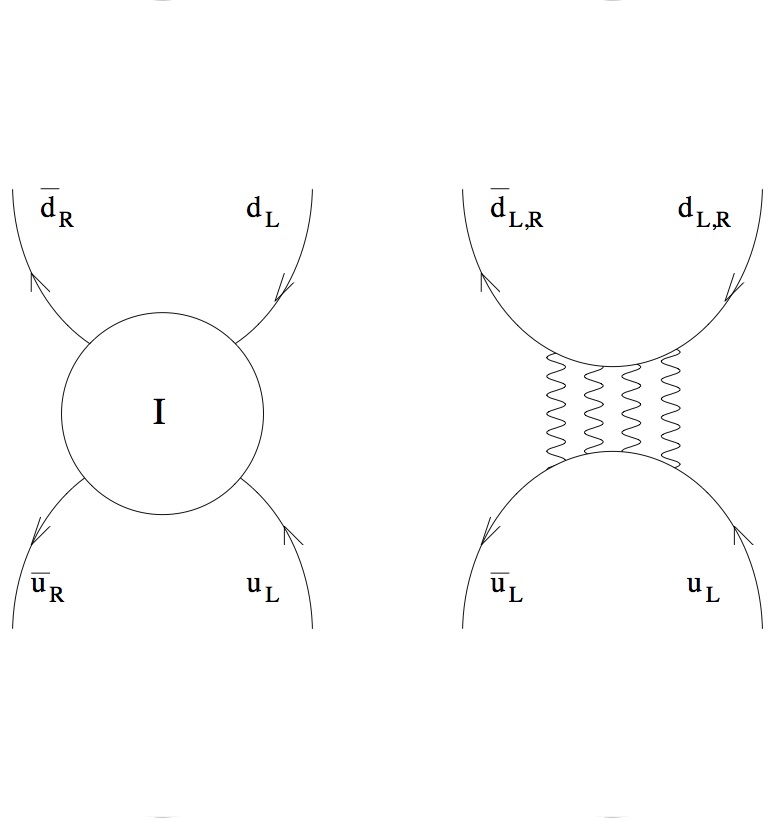}
\caption{\label{fig_tHooft_1storder}
The instanton-induced 't Hooft vertex (a) for 2 flavor QCD
 versus the ordinary
gluon exchange diagrams (b). Note a very different chiral structure
of the two: the latter does not violate any chiral symmetry
because chirality is conserved along each line.
}
\end{figure}
    The result can be written in terms of a new effective Lagrangian 
\cite{tHooft:1976snw}.  It is a local $2N_f$-fermion vertex, where  
the quarks are emitted or absorbed in the (x-independent) spinor states 
obtained after the ``amputation of the free motion" from
the zero mode wave functions  $$\chi=lim_{x\rightarrow \infty} {\psi_0(x) \over S_0(x_c,x)}$$ 
as shown in the right of Fig.\ref{fig_tHooft_vertex}.

One may wish to simplify this 
general fairly complicated vertex.  First, if instantons are uncorrelated, one can average over their 
orientation in color space. For $SU(3)$ color and $N_f=1$ the result 
is 
\be 
\label{Leff_nf1}
{\cal L}_{N_f=1}= \int d\rho\, n_0(\rho) \left( m\rho- \frac{4}{3}
 \pi^2\rho^3 \bar q_R q_L \right), 
\ee
where $n_0(\rho)$ is the tunneling rate without fermions. Note that 
the zero mode contribution acts like a mass term. This is quite natural, 
because for $N_f=1$, there is only one chiral $U(1)_A$ symmetry, which is 
anomalous. Unlike the case $N_f>0$, the anomaly can therefore generate 
a fermion mass term.

For $N_f=2$, the result is 
\be
\label{Leff_nf2}
{\cal L}_{N_f=2}= \int d\rho\, n_0(\rho) \left[  \prod_f
 \left( m\rho- \frac{4}{3}\pi^2\rho^3 \bar q_{f,R} q_{f,L} \right) \right.
 \nonumber \\ \left.
 +  \frac{3}{32}\left(\frac{4}{3}\pi^2\rho^3\right)^2
 \left( \bar u_R\lambda^a u_L \bar d_R\lambda^a d_L
  - {3\over 4}\bar u_R\sigma_{\mu\nu}\lambda^a u_L 
    \bar d_R\sigma_{\mu\nu}\lambda^a d_L \right) \right] + (L \leftrightarrow R)
\ee
where the $\lambda^a$ are color Gell-Mann matrices. One can easily 
check that the interaction is $SU(2)\times SU(2)$ invariant, but 
$U(1)_A$ is broken. This means that the 't Hooft Lagrangian provides 
another derivation of the $U(1)_A$ anomaly. Furthermore, 
 we will argue below that the importance of this interaction goes much 
beyond the anomaly, and that it explains the physics of chiral symmetry 
breaking and the spectrum of light hadrons. 

Any multi-fermion operator can be identically rewritten using the so called 
Fierz identities, grouping fermions into various possible pairs. For 4-fermion
operators there are 3 pairings possible (fermion 1 paired with any of 3 remaining). 
Therefore,  $N_f=2$ 't Hooft Lagrangian can be written in 3 different (but identical)
forms. One or the other may be more or less convenient for particular problem, but various forms in literature may create some confusion. Following  
\cite{Rapp:1999qa} we give all three of them here. Here $t^a$ are color generators (Gell-Mann matrices for $N_c=3$), $\tau^- = (\vec \tau,i)$ and $\vec \tau$ is an isospin matrix, and momentum-dependent formfactor $F(k)$ (which is also a matrix in the Dirac
indices) is related to zero mode solution.

$$ L_1={G \over 4 (N_c^2-1)} \big[   {2N_c -1 \over 2 N_c}  (\bar q F^+ \tau_\alpha^- F q)^2
+ {2N_c -1 \over 2 N_c} (\bar q F^+ \gamma_5 \tau_\alpha^- F q)^2 
$$
\be+{1 \over 4N_c} (\bar q F^+ \sigma_{\mu\nu} \tau_\alpha^- F q)^2  \big] \ee
\be L_2={G \over 8 N_c^2} [(\bar q F^+ \tau_\alpha^- F q)^2+(\bar q F^+ \gamma_5 \tau_\alpha^- F q)^2  \ee
$$+{N_c-2 \over 2(N_c^2-1)} \big( (\bar q F^+ \tau_\alpha^- t^a F q)^2+(\bar q F^+ \gamma_5 \tau_\alpha^- t^a F q)^2  \big)   - {N_c\over 4 (N_c^2-1)} (\bar q F^+ \sigma_{\mu\nu} \tau_\alpha^- t^a F q)^2  \big] $$

$$  L_3={G \over 8 N_c^2}  \big[
-{1 \over N_c-1}(q^T F^T C \tau_2 t^a_A F q) (\bar q F^+ \tau_2 t^a_A C F^* \bar q^T) 
- {1 \over N_c-1}(q^T F^T C \tau_2 \gamma_5 t^a_A F q) (\bar q F^+ \tau_2 \gamma_5 t^a_A C F^* \bar q^T) $$
\be +{1\over 2(N_c+1)}(q^T F^T C \tau_2 \sigma_{\mu\nu}  t^a_A F q) (\bar q F^+ \tau_2 \sigma_{\mu\nu}  t^a_A C F^* \bar q^T) 
\big]
\ee
The beginning of $L_1,L_2$ is often called the ``mesonic form", since they are squares of 
colorless  $(\bar q ...q)$ ``meson currents". These mesons include four different
species, scalars and pseudoscalars 
with isospin zero  $\sigma, \eta'$ and one $\pi,\delta$. In the first order in $L_i$ the masses
of these four mesons get shifted, up or down according to the signs of the terms: $\pi\sigma$ down, $ \delta,\eta'$ up.

The Lagrangian $L_3$ is called the ``diquark form", because the brackets include $(qq)$ and $(\bar q \bar q)$ pairs. Here $T$ means transposed, $C$ is charge conjugation
which in reality is the antisymmetric isospin matrix $\tau_2$. Color generators $t^a_A$ in it
are only those which are antisymmetric in indices, this is what subscript $A$ indicate. 
Like mesons, the first two terms have a sign opposite to the last one: it implies that
to first order the scalar and vector diquarks get mass corrections of the opposite sign.
(The spin 0 is going down, spin 1 up.).

   The effective Lagrangian is a kind of non-local vertex, in which
   either quarks scatter on each other, or 
a pair of quarks  is produced from initial quark. 
Note that this interaction explicitly violates $U_a(1)$ chiral symmetry.
Instanton-induced production of the axial charge is of course
just a specific case of a ``sphaleron process": whether the motion in
the topological landscape is by real or virtual fields does not matter,
general relation between $N_{CS}$ and $Q_5$ holds in any process.

\begin{figure}[h!]
\begin{center}
\includegraphics[width=8cm]{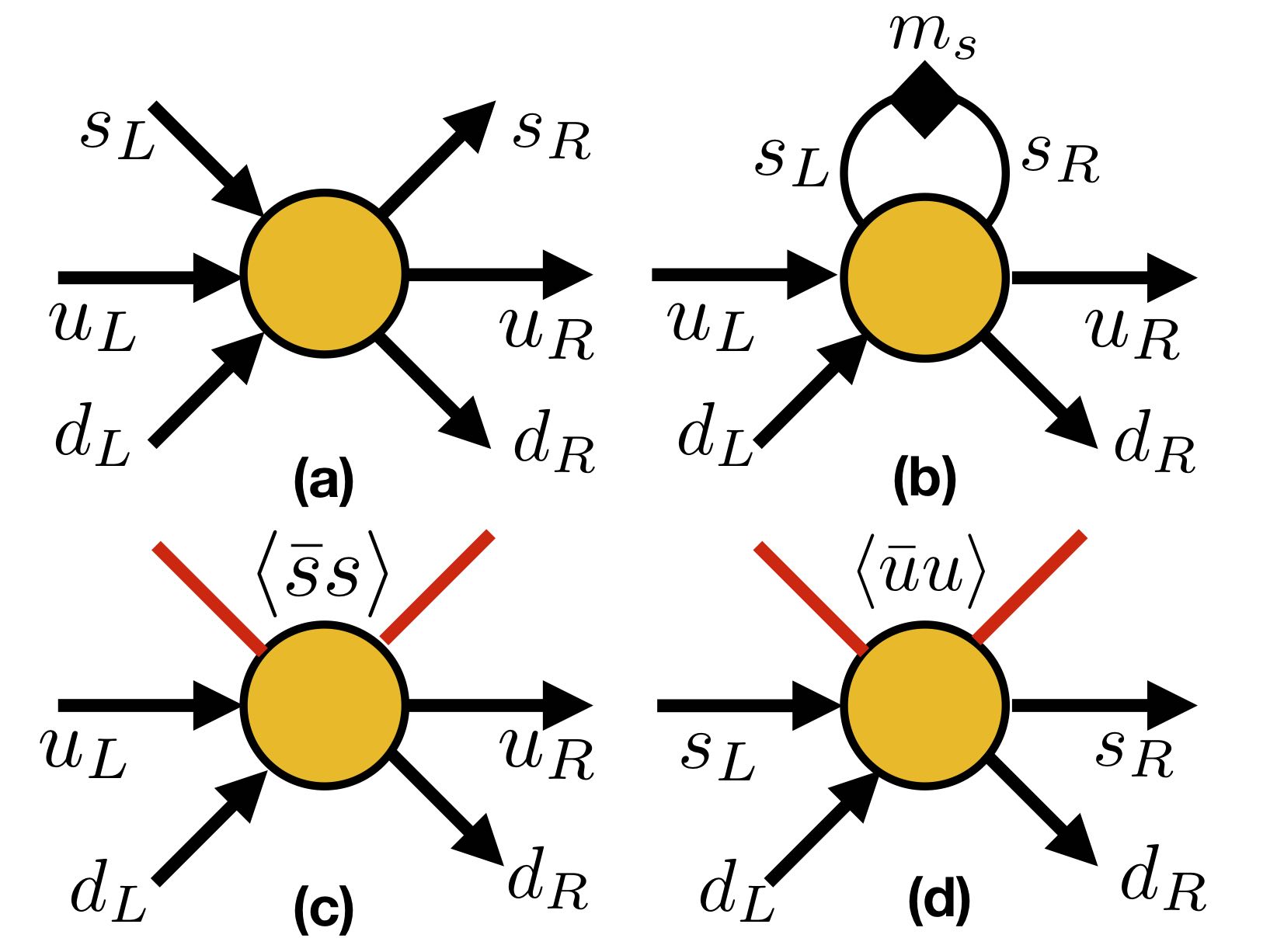}
\caption{Schematic form of the 6-quark 't Hooft effective Lagrangian is shown in fig (a). If quarks are
massive, one can make a loop shown in (b), reducing it to 4-fermion operator. Note a black rhomb indicating  the mass insertion into a propagator. We only show it for $s$ quark, hinting that for $u,d$ their masses are too small to make such
diagram really relevant. In (c,d) we show other types of effective 4-fermion vertices, appearing because 
some quark pairs can be absorbed by a
nonzero quark condensates (red lines). }
\label{fig_multiquark}
\end{center}
\end{figure}

In the real-world QCD one has not two but $three$ light quark flavors, $u,d,s$.
Therefore 't Hooft Lagrangian is in fact 6-quark operator. However, at $T<T_c$ chiral symmetry is spontaneously broken, and therefore one (or two) pairs of quarks can be substituted by
their condensate: this leads to 4- (or 2-) quark effective operators. Both are extremely 
important for understanding of hadronic physics. The 2-quark operator leads to nonzero ``conatituent
quark mass". The resulting 4-fermion quark operators, shown in Fig.\ref{fig_multiquark} leads to many effects we will discuss below.

A comment about violation of flavor $SU(3)$ symmetry by strange quark mass, which is much larger than those of light $u,d$ quarks
$$ m_s\sim 120 \, MeV \gg m_{u,d} \sim few\, MeV $$
If one ignores the latter, one finds that operator of the type $(\bar u u)(\bar d d)$ has two contributions
(diagrams (b) and (c) in the last figure), while operators of the type $(\bar u u)(\bar s s)$ and  
$(\bar d d)(\bar s s)$ has only contribution of the diagram (d). All of them are comparable, so
it turns out that the $(\bar u u)(\bar d d)$ interaction is about twice larger than others. So, 
in effects caused by these operators the flavor $SU(3)$ symmetry is violated at level $O(1)$.

  Finally, in order to complete the effects of light quarks
on the tunneling,
we need to include the effects of non-zero modes.
 One effect is that the coefficient in the 
beta function is changed to $b=11N_c/3 -2N_f/3$. In addition to 
that there is an overall constant that was calculated in 
\cite{tHooft:1976snw,Carlitz:1978yj}
\be 
 n(\rho) &\sim& (1.34 m\rho)^{N_f} \left( 1 + N_f (m\rho)^2 \log
 (m\rho) + \ldots \right),
\ee 
where we have also specified the next order correction in the quark 
mass. Note that at two loop order, one not only gets the {\em two-loop 
beta function} in the running coupling, but also 
the {\em one-loop anomalous
dimensions} of the quark masses.

\subsection{Instanton-induced quark anomalous chromomagnetic moment}
This derivation follows \cite{Kochelev:1996pv} and rely on the color-spin term
in the 't Hooft effective Lagrangian including external gluon field, written for a quark of type $q$  as
\be  \Delta L=\int d\rho { n(\rho) \over m^*_q } {\pi^4 \rho^4 \over g}( i\bar q \sigma_{\mu\nu} {\lambda^a\over 2} q) G^a_{\mu\nu} 
\ee 
where, in the mean field approximation, the instanton density is written as 
\be n(\rho)=n_{gluodynamics}(\rho) \prod_q (m_q^* \rho) ,\,\,\, m_q^*=m_q-2\pi^2\rho^2 \langle \bar q q \rangle \ee
and contains the product of  effective quark masses of all light quark flavors $q=u,d,s$.
Note that in the previous formula the mass $m_q^*$ of the  quark under consideration
is in denominator and thus drops out: only masses of $other$ flavors appear. 

Comparing this with the definition of quark anomalous chromomagnetic moment
\be  \Delta L=-i \mu_q {g \over 2 m_q^*} (\bar q \sigma_{\mu\nu} {\lambda^a\over 2} q) G^a_{\mu\nu} 
\ee
one obtains its value 
\be \mu_q=-{\pi^3 n_{inst} \rho^4 \over 2 \alpha_s(\rho) } \ee
The numerator contains the instanton diluteness combination $n_{inst} \rho^4\sim 10^{-2}$ 
but $ \alpha_s$ is in denominator as it should be, due to non-perturbative nature of the instanton field. The absolute magnitude for light quarks is  $\mu_{u,d} \approx -0.2$,
and it is used in effective quark models of hadronic spectra.

\begin{figure}[h]
\begin{center}
\includegraphics[width=14cm]{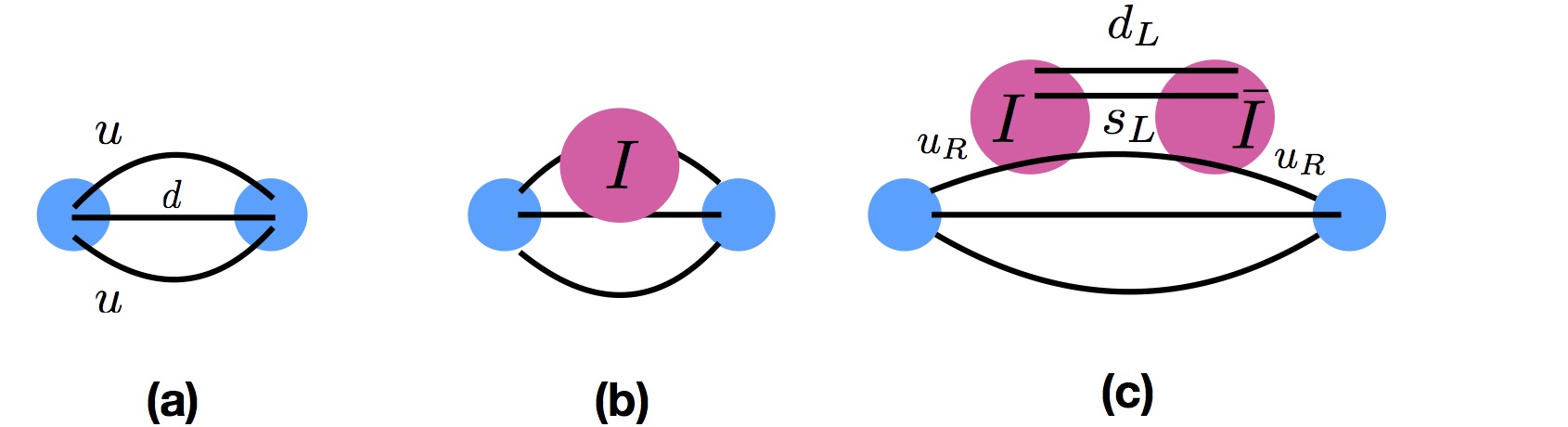}
\caption{Quark configurations of the nucleon, including the lowest order instanton-induced ones.}
\label{fig_inst_in_nucl}
\end{center}
\end{figure}

\subsection{Instanton-induced diquark- quark configurations in the nucleon} 

This subsection just introduce this important subject: more detailed discussion of it will
be continued in chapter \ref{sec_light_front} devoted to hadronic wave functions on the light front. 

In Fig.\ref{fig_inst_in_nucl} we show the simplest valence-quark nucleon configuration
(a), together with the lowest-order instanton-induced effects. The diagram (b)
illustrate the $ud$ diquark correlation, appearing in the first order in 't Hooft Lagrangian.
Since the diquark has spin zero, the $d$ quark in it does not contribute to the total spin of the nucleon. This conclusion is supported by lattice studies.

The attention to the last diagram (c) comes from the paper \cite{Dorokhov:1993fc},
where it was noted that ``sea quarks"  produced by instantons, and resulting in the 5-quark configuration, are highly polarized both in spin and isospin. Indeed, the valence $u$ quark
can only produce $d,s$ ones (flavor polarization). Furthermore, if this quark happens to be right-handed, the sea quark pair would be left-handed (and vice versa). In that paper
this configuration was proposed as an explanation of observed deviations from Ellis-Jaffe and Gottfried sum rules, related to the famous ``spin crisis" of the nucleon.

Note that the instanton-induced production of sea quarks is very different from the usual one-gluon vertex creating $\bar q q$ pairs, which is obviously flavor and chirality-blind. Thus the 
usual pQCD evolution of structure functions, while dominant at very small $x$,  $cannot$ start from simple valence quark 
distributions and needs asymmetric phenomenological input. 

\subsection{Instanton-induced decays of $\eta_c$ and scalar/pseudoscalar glueballs}

\begin{figure}[htbp]
\begin{center}
\includegraphics[width=10cm]{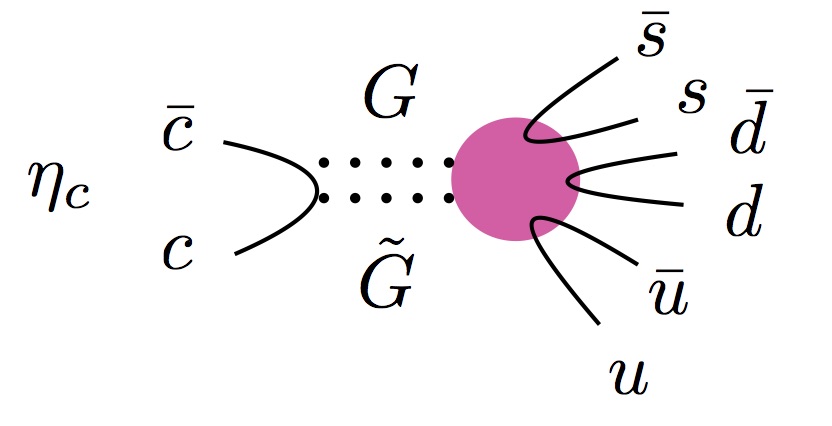}
\caption{The instanton-induced decay of the pseudoscalar $\eta_c$.}
\label{fig_etac_decay}
\end{center}
\end{figure}

Let me here provide one example (which we would not discuss
in detail). It was noticed by Bjorken \cite{Bjorken:2000ni}
that decays of $\eta_c$ has 3 large 3-body modes, about 5\% each
of the total width: $$\eta_c \rightarrow KK\pi;\,\,\, \pi\pi\eta;\,\,\, \pi\pi\eta'$$ Note that there is
no $\pi\pi\pi$ decay mode, or many other decay modes one may think of: that is because
't Hooft vertex $must$ have all light quark flavors including the $\bar{s}s$, see Fig.\ref{fig_etac_decay}.
More generally, 
in fact the average multiplicity of $J/\psi,\eta_c$ decays is significantly larger than 3, so
large probability of these 3-body decays is a phenomenon by itself. 
Bjorken pointed out
that the vertex seems to be $\bar u u\bar d d \bar s s$ and suggested
that these decays proceed via 't Hooft vertex.

\begin{figure}[h]
\begin{center}
\includegraphics[width=12cm]{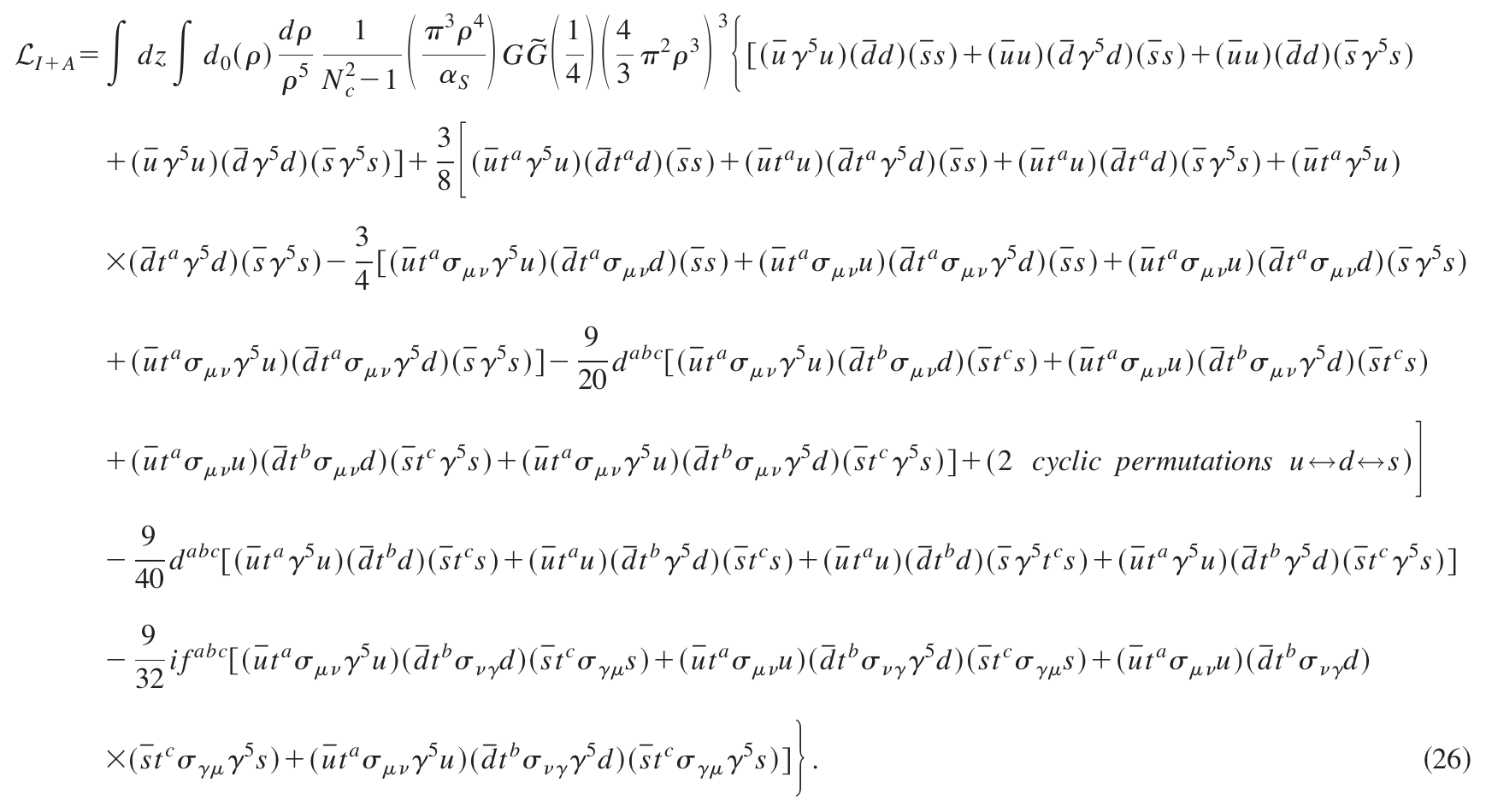}
\caption{The form of the $N_f=3$ effective 't Hooft Lagrangian}
\label{fig_3flavor_Lagrangian}
\end{center}
\end{figure}

The actual calculations were done by \cite{Zetocha:2002as}, it included the following
  two and three mesons decays channels of the lowest charmonium state
 \be \eta_c\rightarrow \pi\pi, KK,\eta\eta, KK\pi, \eta\pi\pi, \eta'\pi\pi \ee 
using the 3-flavor Lagrangian shown in Fig.\ref{fig_3flavor_Lagrangian}.

 Their results contain rather high power of the instanton radius and therefore strongly depend
 on its value. So the authors used the inverted logic, evaluating from each data point the
 corresponding value of the mean instanton size $\bar\rho$.
  The results reasonably well
reproduced the ratios between the channels and even the absolute width.
Furthermore, these calculations provide about the most accurate
evaluation of the average instanton size available,  in the range of $\bar\rho=0.29-0.30\, fm$.

\subsection{Instanton-induced spin polarization in heavy ion collisions}

Heavy ion collisions create new form of matter -- Quark-Gluon Plasma (QGP) -- which
epands and cools according to relativistic hydrodynamics. This field is rich and will
be discussed in a separate lecture volume. However one particular phenomenon
fits here so well that I decided to include it. The point is $noncentral$ collisions lead to
large orbital momentum and consequently to fluid undergoing rotational motion, with a nonzero $vorticity$
\be [{\partial \over \partial \vec x} \times \vec v] =\vec \omega \neq 0 \ee
If one goes to rotational frame, vorticity act as magnetic field, and in thermal equilibrium
the spin states get polarized 
\be P\sim exp\big[(\vec \omega \cdot \vec S)/T\big] \ee
The observation of $\Lambda$ hyperon polarization has been observed by STAR collaboration at RHIC 
\cite{STAR:2017ckg}.

This story may look boring, unless one asks the question 
of $how$ exactly a spin can get equilibrated by rotating medium. Investigation of this
by \cite{Kapusta:2019ktm} started with  spin-flip reactions induced by thermal flow fluctuations.
With disappointment it was found that 
this mechanism can only equilibrate spins at times longer than $>10^3\, fm/c$,
obviously not available in practice in the collisions. Then it ws shown that perturbative amplitude of strange quark spin flip --
proportional to strange quark mass $m_s$ -- also leads to unreasonably long equilibration time.

Therefore \cite{Kapusta:2019ktm}, in desperation, eventually turn their attention to nonperturbative QCD.
  They used 4-quark operators obtained from 6-quark 't Hooft operator and ``instanton liquid model" parameters, which does indeed produce resonable time for spin equilibration.
  
  Note however, that reduction of  6--to-4 quark operator involves 
  nonzero quark condensates, and therefore is only possible not in the QGP phase,
  but only at $T<T_c$, in hadronic phase. This implies that spin polarization only occurs late
  in the collision. 
  
  In principle, spin polarization can be observed not only for  $\Lambda$ hyperon but for many other hadron species, like vector mesons or $S=3/2$ baryons like $\Delta^{++}$. In a comment
  \cite{Shuryak:2016hor} I suggested the latter to be a good way to tell apart the effect of vorticity
  from that of magnetic field. Eventually, one should be able to compare polariization of light-to-strange quarks and compare it to predictions from 't Hooft operator. At the time of this writing
  there appeared measurements of spin polarization for strange vector mesons $K^*,\phi$:
  but there is no understandingof how are they related with local vorticity and/or
  magnetic field  yet .

\chapter{Topology on the lattice}

\section{Global topology: the topological susceptibility and the interaction measure}

 ``Global topology" we will discuss in this first section is the total topological charge
 of the whole lattice: in the subsequent sections we will review more local observables
 aimed at revealing the topological substructure of the gauge theory vacuums.

The overall fluctuations of the topological charge can be 
deduced from lattice measurements of the so called topological susceptibility, 
defined by
\be \chi_{top}={<Q^2 > \over V_4} \ee
where $V_4$ is the 4-volume in which the topological charge $Q$ is measured. 
We will not go into technical details about the topological charge definitions on the
lattice, and just notice that with proper definition $Q$ is always integer-valued.
Lattice configurations provide histograms of the probabilities $P(Q)$ used
in the averaging implied in the definition.
If it would be Gaussian, the only parameter would be $\chi_{top}$, related to its width.

The non-Gaussian distributions possess higher moments. The
so called {\em interaction parameter} 
is defined by the ratio of the following moments of the $Q$ distribution, via
\be b_2=-{<Q^4>-3 <Q^2>^2 \over 12 <Q^2>^2 }\ee

If one thinks that topological charge is contained in an  ensemble of instantons,
$\chi_{top}$ provides the combined density of instantons and antiinstantons.
Below we will focus on its  dependence 
  on the temperature $T$, but one should also study its dependence on other parameters
  such as the quark masses $m_f$ and the number of colors $N_c$.  

Dilute instanton gas approximation (DIGA)  provides some predictions. 
We will discuss the total density a bit later, and now notice 
that since DIGA
predicts the theta dependence of the vacuum energy to be
\be F(\theta,T)- F(0,T)=\chi_{top}(T) \left(1- cos(\theta) \right) \ee
Expanding this to powers of theta, one finds prediction for higher order coefficients, in particular 
the value of $b_2$ is defined by
$$ b_2^{DIGA}=-{1\over 12} $$
The interaction between instantons will influence this parameter, and so
there were devoted measurements on the lattice. 

Of course, it would be nice to directly simulate the ensemble at nonzero $\theta$:
but it is not possible because of the complex weight. One method used
to go around this is to do simulations for imaginary $\theta=i \tilde\theta$ (the chemical potential
conjugated to the topological charge) and then extrapolate back from imaginary to real one.

Significant efforts have been invested in studies of the dependence of these parameters on the number of colors $N$.
In particular, 
$\chi_{top}/\sigma^2, b_2$ for three, four and six colors \cite{Bonati:2016tvi}. 
The results provide a robust evidence of the large-N behavior predicted by standard large-N scaling arguments, i.e. 
$ b_{2n} = O(N^{-2n})$. In particular,  \be b_2 = { \bar{b}_2 \over N^2} + O(1/N^4), \,\,\,\,   \bar{b}_2  =- 0.23(3).\ee

In QCD with light quarks one can study the dependence of these parameters on the quark masses.
Recall that gauge fields with nonzero $Q$ have fermionic zero modes: so if $m\rightarrow 0$ the determinant is zero
and thus those configurations are impossible. So, in the massless limit $\chi\rightarrow 0$ as well. We have noticed before that to vanish $\chi_{top}$
in fact it is enough that a single quark flavor be massless!
For $m_u=m_d$ and ignoring strangeness, the chiral perturbation theory gives 
\be \chi=m_\pi^2 f_\pi^2 /4\approx (75. \, MeV)^4\ee
which does indeed vanish in the massless limit $m_\pi^2\sim m_q f_\pi$. All of that is
reproduced by lattice measurements\footnote{Note however, that DIGA
would predict much stronger vanishing, since the instanton amplitude is proportional
to the $product$ of all quark masses $\prod_f q_f$. It is not the case in the CD vacuum
because of chiral symmetry breaking, but becomes true at high $T$.}.

\begin{figure}[h]
\begin{center}
\includegraphics[width=6.5cm]{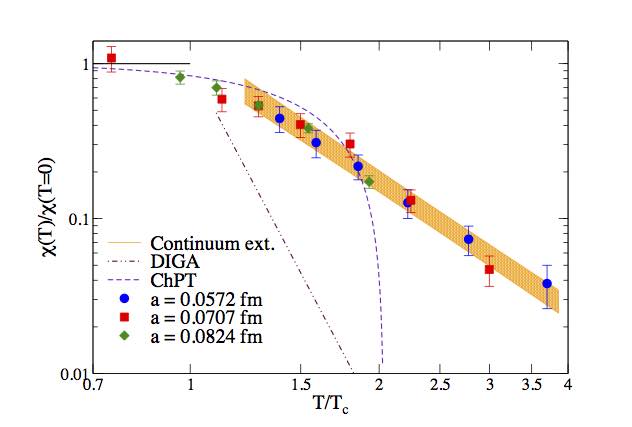}
\includegraphics[width=6.5cm]{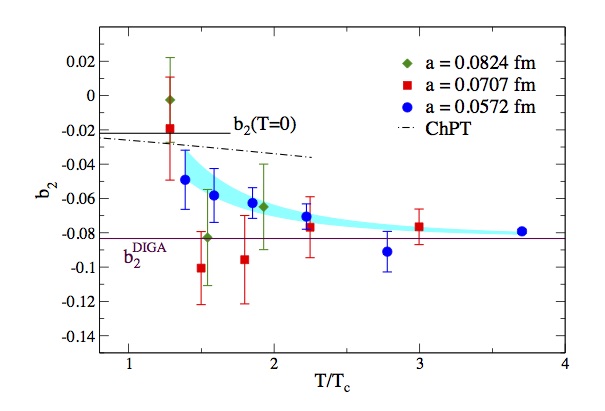}
\caption{(A) Ratio of the topological susceptibilities $\chi(T )/\chi(T = 0)$, evaluated at fixed lattice spacing. The horizontal solid line describes zero $T$-dependence, while the dashed line is the prediction from finite temperature ChPT and the dashed-dotted line shows the slope predicted by DIGA computation. The band corresponds to the continuum extrapolation using the function $\chi(a,T)/\chi(a,T = 0) = D_0(1 + D_1a^2)(T/Tc)^{D_2}$, only data corresponding to $T>1.2T_c$ have been used in the fit.
(B) $b_2$ evaluated at three lattice spacing. The horizontal solid lines correspond to the value
of $b_2$ at T = 0 predicted by ChPT (which is about -0.022) and to the instanton-gas expected value
$b_{DIGA} =-1/12$. The dotted-dashed line is the prediction of finite temperature ChPT, while the 2
light blue band is the result of a fit to the smallest lattice spacing data. 
}
\label{fig_top_of_T}
\end{center}
\end{figure}

Topological susceptibility in QCD at finite $T$ has been studied in
\cite{Bonati:2015vqz}, from which we took the following Fig. \ref{fig_top_of_T}.
Let us start with the second plot,
displaying the interaction parameter $b_2$.
Note that at low $T$ it is very different from the DIGA prediction  $-1/12$, but 
at high $T$  it does go to an agreement  with to the dilute gas prediction.

The first plot shows that $\chi(T)$ in the QGP phase does have a power dependence on $T$,
but the fitted power $D_2\approx 3$. On the other hand, DIGA predicts
\be \chi^{DIGA}(T)\sim T^4 \left({\Lambda \over T}\right)^b  \left({m \over T }\right)^{N_f} \ee
with $N_f=3,b=11N/3-2N_f/3\approx 9$ the power of the temperature thus should be $D_2=8$
(shown on the plot by dash-dotted curve):
 thus there is a very serious disagreement. We will return to its discussion
 in the next chapters. 
 
 At the other hand, another lattice group \cite{Borsanyi:2016ksw} had measured $\chi(T)$ up to much higher $T\approx 2300\, MeV$.
 In this range  $\chi(T)$ drops by about 10 orders of magnitude! 
 This group finally does see at high $T$ the dependence corresponding to DIGA:
  their fitted power of $T$ is 8.16, close to 
 the power from the beta function $(11N_c/3)-(2N_f/3)=8.666$ (for 4 quark flavors, $u,d,s,c$.
 
  Since their method  basically
 included only  configurations with total charge $Q=\pm 1$, which means they kept a single instanton in a box, 
no doubt the result agrees with a single-instanton 
amplitude. 

What would be important to test is the coefficient, given by the semiclassical
calculation. To my knowledge the coefficient does not agree with expectations, being about
an order of magnitude $larger$ than the prediction. In view of high power of 
$\Lambda_{QCD}$ involved, and the fact that it is only one-loop prediction available,
it is perhaps not yet a problem, but the issue clearly needs more work. 

\begin{figure}[h]
\begin{center}
\includegraphics[width=9cm]{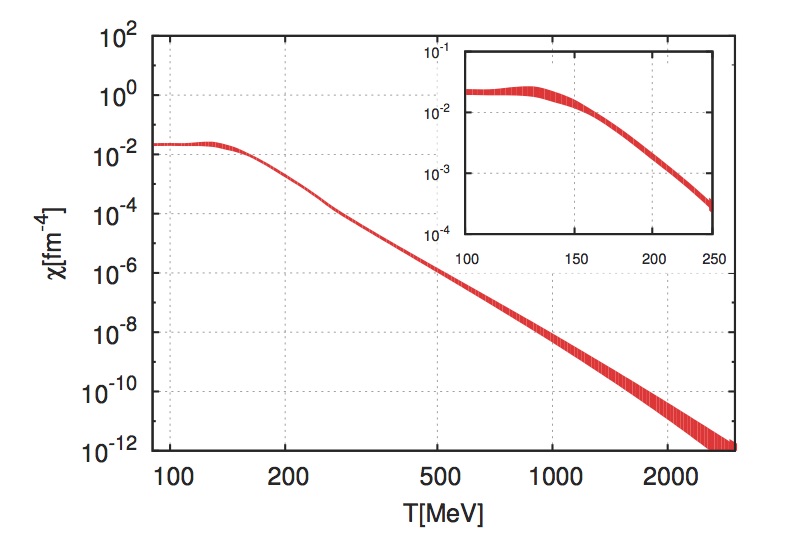}
\caption{The topological susceptibility versus temperature, for QCD with physical quark masses. The insert shows the dependence
around the deconfinement phase transition. 
}
\label{fig_chitop_highT}
\end{center}
\end{figure}

\section{Cooling and instantons}

Lattice field configurations are not of course classical: they do include 
the full extent of quantum fluctuations of all fields. But the tools used to reveal the underlying topology are
basically the same, ``cooling" and ``gradient flow".

The simplest method to look for topological object on the lattice is ``cooling",
an iterative procedure reducing local values of the gauge action. Indeed, the
topological solitons represent local extrema (minima) of the action, and thus
those should be approached under cooling. The first works 
\cite{Teper:1985ek,Ilgenfritz:1985dz}
 applying cooling
had indeed found topological lumps, with their action being equal (up to a sign)
to the topological charge, $S=|Q|$. 

Further works, have however found certain technical issues, which made hard to make really quantitative results from cooling. One is that the results obtained depend on the number of cooling steps: the reason is that close instanton-antiinstanton pairs move toward each other
and eventually annihilate. Trying to locate and include such annihilation events made
the cooling studies more an art than a regular method. 

Of similar nature is the question
of ``topological defects", instanton-like objects with a size comparable to lattice spacing $a$.
In order to define the topological charge, lattice discretized fields should be made continuous
by some extrapolation procedures, which, unfortunately, by definition are not unique.  

\begin{figure}[h!]
\begin{center}
\includegraphics[width=10cm]{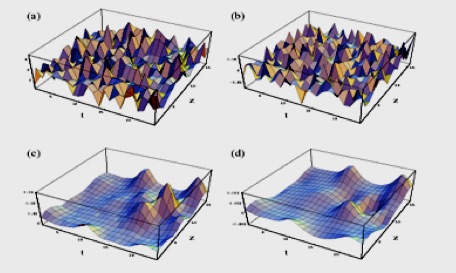}
\includegraphics[width=7cm]{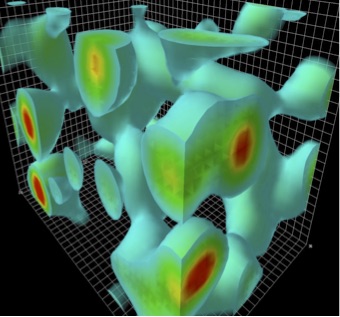}
\caption{The topological structure revealed by ``cooling" of lattice gauge configurations. Four upper plots are from 
the MIT group by Negele et al, they show the distributions of the action $\sim GG$ and topological charge $\sim G \tilde G$ (left and right). The upper plots are before and the lower ones after cooling. The lower 3d picture of
the topological charge is from the Adelaide group (Leinweber et al) (lower).
 }
\label{fig_cooled_inst}
\end{center}
\end{figure}

The other issue is the fact that only in the continuum limit the
instanton action becomes scale-independent. On the lattice there is no scale invariance,
and, depending on particular choice of the lattice action, one can force instantons either
to grow or shrink, as the cooling proceeds. This issue has been addressed  using
the $impoved$ lattice actions \cite{deForcrand:1997esx}. Without going into detail,
let me just say that with tuning it one can tune the $O(a^2)$ term in the instanton
action to prevent changes of the instanton size during cooling. This allowed to measure
the instanton density with the size distribution\footnote{
The measurements started from
some threshold value of $\rho_{min}=2.3a$, but semiclassical theory tells us that only very negligible number
of instantons can be below this threshold. 

} $dn(\rho)/d^4x d\rho$, The average size 
was found to be $\langle \rho \rangle\approx .43 \, fm$, somewhat larger than in the instanton liquid model  . The distribution have also clear peak at $\rho_{max}\approx 0.4\, fm$. 
Distribution of separation between instantons have a clear peak at $0.9\, fm$, close to what 
the instanton liquid model value. 
In Fig.\ref{fig_cooled_inst} we show two more examples of the vacuum topological structure 

Later works, included extrapolation back to ``no cooling" time, allowed careful
determination of the density and size distributions, in pure gauge $SU(2),SU(3)$ and
even physical QCD with quarks \cite{Hasenfratz:1999ng}. The average size 
was found to be $\langle \rho \rangle\approx .30 \pm 0.01\, fm$, a bit smaller than in the ILM.
The mean distance was found instead to be $0.61\pm 0.02 \, fm$.

\begin{figure}[t]
\begin{center}
\includegraphics[width=6cm]{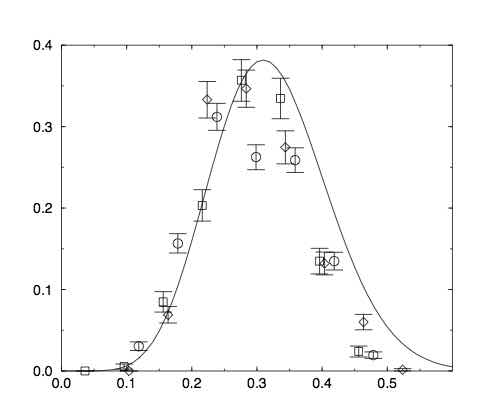}
\includegraphics[width=6cm]{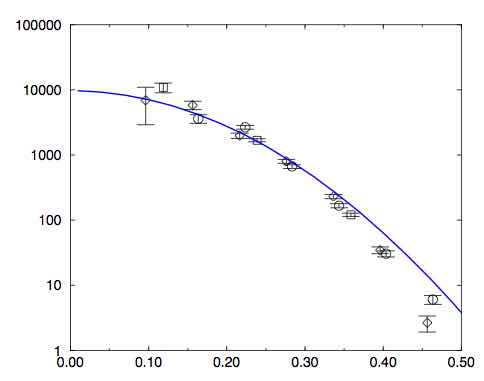}
\caption{(left) The instanton size $\rho$ [fm]
distribution  $dn/d\rho d^4z$. (right) The combination $\rho^{-6}dn/d\rho d^4z$, in which the main one-loop behavior drops
 out for $N_c = 3,N_f = 0$. 
 The points are from the lattice work 
 for this theory, with $\beta$=5.85 (diamonds), 6.0 (squares) and 6.1 (circles). Their comparison should demonstrate that results are lattice-independent. The line corresponds to the proposed expression  , see text.
 }
\label{fig_inst_sizes}
\end{center}
\end{figure}

  Lattice data on the instanton size distribution 
  are shown in Fig.\ref{fig_inst_sizes}
(the figure is taken from \cite{Shuryak:1999fe}, the lattice data from
A.Hasenfratz et al). The left plot shows the 
size distribution itself. Recall that the semiclassical theory predict it to be
$dn/d\rho\sim \rho^{b-5}$ at small sizes, with $b=11Nc/3=11$ for pure gauge $N_c=3$  theory. 
The right plot -- in which this power is taken out --is constant at small $rho$, which agrees with
the semiclassical prediction.

The other feature is a peak at $\rho\approx 0.3 \, fm$ -- the value first proposed phenomenologically
in \cite{Shuryak:1981ff}, long before these lattice data. 
The reason for the peak is a suppression at large sizes.

 Trying to understand its origin, one may factor out all known effects. The right plot shows that 
 after this is done, a
rather simple suppression factor $\sim exp(-const*\rho^2)$ describes it well, for about 3 decades.
What can the physical origin of this suppression be?

There is no clear answer to that question. One option is that it is due to mutual repulsion between instantons and antiinstantons: we will return to it in section ?? where we will discuss theory of the instanton liquid.

Another one, proposed in ref.\cite{Shuryak:1999fe} already mentioned, is that the coefficient is proportional to
the dual magnetic condensate, that of Bose-condensed monopoles. It has been further argued there that
it can be related to the string tension $\sigma$, so that the suppression factor should be 
\be {dn \over d\rho}= {dn \over d\rho}|_{semiclassical} \cdot e^{-2\pi \sigma \rho^2} \ee
If this idea is correct, this suppression factor should be missing at $T>T_c$, in which the dual magnetic condensate
is absent. But, on the other hand, here quantum/thermal fluctuations generate at high $T$ a similar factor \cite{Pisarski:1980md} 
\be  {dn \over d\rho}= {dn \over d\rho}|_{T=0} \cdot e^{- ({2N_c+N_f \over 3}) (\pi \rho T)^2}\ee
 related to scattering of quarks and gluons of QGP on the instanton \cite{Shuryak:1994ay}. 
Empirically, the suppression factor at all temperature looks Gaussian in $\rho$, interpolating between those limiting 
cases.

Another example of lattice study focusing on instanton contribution to
 certain Green functions, is Ref.\cite{Athenodorou:2018jwu},
 in full quantum vacuum and with cooling.
The original motivation has been extraction of the gluon coupling $\alpha_s(k)$,
so the observable on which this work was focused id the
following ratio of 3-point to 2-point Green function (in configurations transformed to Landau gauge)
\be \alpha_{MOM}(k)={k^6 \over 4\pi} {\langle  G^{(3)}(k^2)  \rangle^2 \over    \langle  G^{(2)}(k^2)    \rangle^3 }  \label{ratioG3toG2}\ee

\begin{figure}[h]
\begin{center}
\includegraphics[width=10cm]{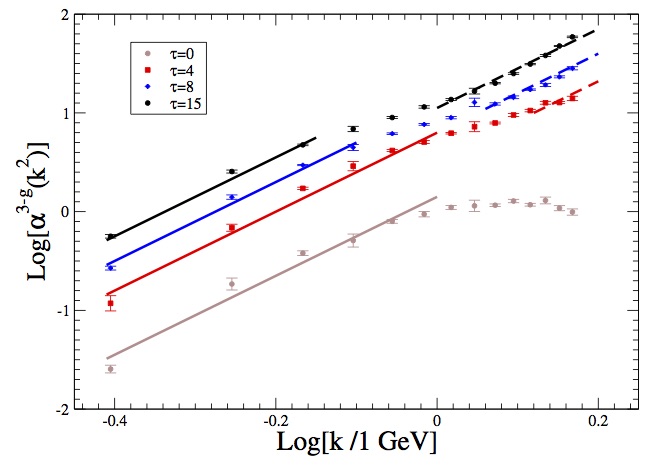}
\caption{The ratio of Green functions $\alpha_{MOM}(k)$ versus momentum scale $k$.
Different color of points correspond to different cooling time. The lines correspond to
the instanton ensembles with fitted densities. }
\label{fig_alpha_of_k}
\end{center}
\end{figure}

In Fig.\ref{fig_alpha_of_k} the results are plotted versus the momentum scale $k$.
At the lower curve (corresponding to uncooled quantum vacuum with gluons)
at large  $k>1\, GeV$ the effective coupling  starts running downward, as 
asymptotic freedom requires. However in infrared, at
 low $k\rightarrow 0$, one finds certain positive power. Its slope was found to be
 exactly what one would obtain from an instanton ensemble. Furthermore, after cooling
 for different time $\tau$ (shown by three other curves) it was seen that
 the same power also persists at high $k> 1\, GeV$ as well (see dashed curves
 at the right).  This corresponds well to the expectation that cooling eliminates perturbative 
gluons (the plain waves) but, for some time, preserve instantons.

The authors notice that this power corresponds well to that calculated from instantons.
The ratio defined above (\ref{ratioG3toG2}) calculated for en ensemble of instantons
reads 
\be \alpha_{MOM}(k)={k^4 \over 18\pi n} {\langle  \rho^9 \big(I(k\rho)\big)^3  \rangle^2 \over    \langle  \rho^6 \big(I(k\rho)\big)^2    \rangle^3 } \ee
where the function containing the instanton size $\rho$ is 
$$ I(s)={8\pi^2 \over s}\int_0^\infty dz z J_2(sz) \phi(z) $$
a certain Bessel-transform of the shape function $\phi$. ( To compare, the BPST instanton has $\phi=1/(1+z^2)$.)

\begin{figure}[h]
\begin{center}
\includegraphics[width=6cm]{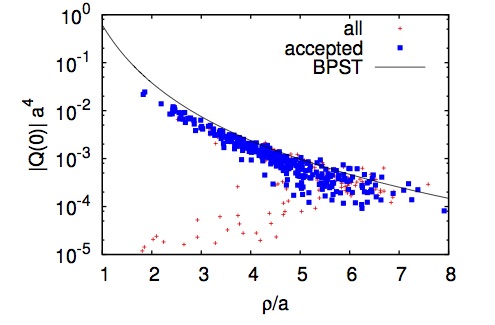}
\includegraphics[width=6cm]{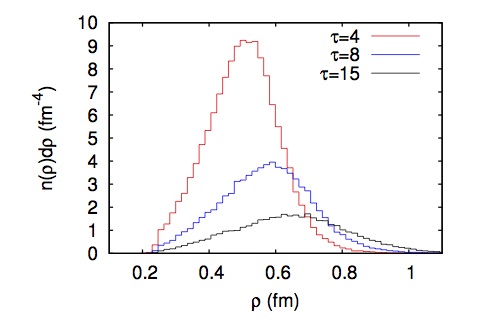}
\caption{(left) The topological charge at the origin vs the fitted radius, compared to
the BPST profile shown by the line.
(right): evolution of the instanton size distribution with the cooling time $\tau$}
\label{fig_qof0_vs_rho}
\end{center}
\end{figure}

The gradient flow cooling allowed to identify instantons quite well. As shown in Fig.\ref{fig_qof0_vs_rho},
the topological charge density at the center depends on the  fitted size in a way quite close
to that of BPST instantons. 

 While with the cooling time the instanton sizes grow, one should remember to
extrapolate to {\em zero cooling time} $\tau\rightarrow 0$ to recover how they were
in the original quantum vacuum. Therefore, this density is not what one can calculate
from the right plot of  Fig.\ref{fig_qof0_vs_rho}, but significantly larger:
their extrapolated total density is $\sim 10 \, fm^{-4}$, an order of magnitude larger than
in the original ILM. 

Extrapolating  these results back to {\em zero cooling}, the authors  used these data for an estimate of 
the vacuum instanton density. Their conclusion is that it is much higher than
the value suggested by the instanton liquid model, by about an order of magnitude.
If so, it erases the diluteness parameter of that ensemble, making
the ``instanton liquid" really dense. 

\section{ "Constrained cooling", preserving the Polyakov line}

From the Introduction chapter we emphasized that the VEV of the Polyakov line
does play a very important role in non-perturbative phenomena. It is used as the
most practical confinement measure, and, as we discussed in the Introduction,
it  influences the instantons, effectively splitting them into the instanton-dyons (we will discuss later).

\begin{figure}[htbp]
\begin{center}
\includegraphics[width=6cm]{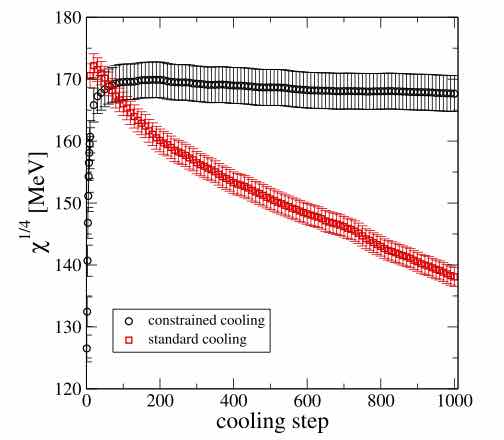}
\includegraphics[width=6cm]{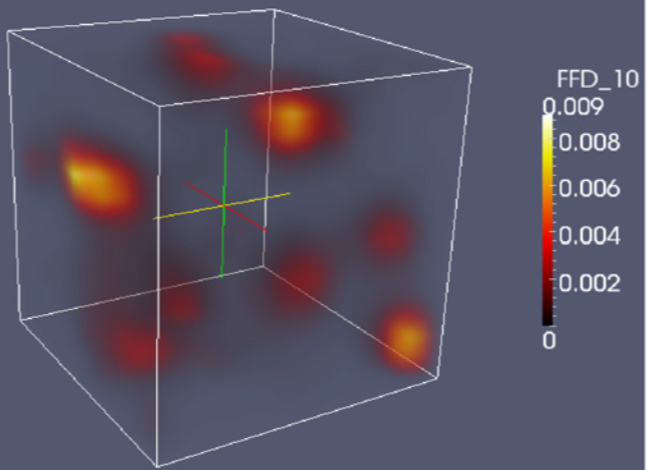}
\caption{The left plot shows the topological susceptibility $\chi_{top}$ as a function of the number of cooling sweep $N_{cool}$, for standard (red points  decreasing with $N_{cool}$)
and constrained (black points). The right plot shows 
the distribution of the topological charge in configurations obtained by the constrained cooling
. }
\label{fig_constr_cooling}
\end{center}
\end{figure}

So it is rather natural, following
\cite{Langfeld:2010nm}, to work out the so called the so called {\em constrained cooling},
with the local values of the Polyakov line  preserved. For technical definition how it is
achieved one need to see the original paper: let me show only two plots in Fig.\ref{fig_constr_cooling} . The left figure compares the behavior of the topological 
susceptibility $\chi_{top}$ on the amount of the cooling steps. While the standard
method shows a global topology disappearing, the  constrained cooling
does preserve it.

Let me briefly  summarize the main findings of \cite{Langfeld:2010nm} as follows:
while the global topological charge $Q$ is quantized to integer values, the actions and 
the topological charges of  local 
topological clusters do not: So, they cannot  really be instantons! 

On the other hand,
they satisfy another major requirements: the fields in these clusters
is  {\em locally selfdual}. The picture in the right side of Fig.\ref{fig_constr_cooling} 
shows also that they are well localized and thus their fields are quite strong.
Those features still suggest that the objects are topological, and that their semiclassical treatment looks promising.

%
%
%
%
%

\chapter{Instanton ensembles}

\section{Qualitative introduction to the instanton ensembles}

In the previous chapter we have discussed the semiclassical theory of 
a single instanton, together with some applications of it. However, since
the translational zero modes are substituted by an integral over collective coordinates -- the instanton
location in 4-d -- the tunneling amplitude for an instanton  is proportional to the 4-volume $V_4$ considered.
This means that if $V_4$ is large enough to overcome the exponential tunneling suppression,
naive probability of tunneling exceeds 1, and re-summing (unitarization) of the probability is necessary.
As was shown in the previous chapter, simple exponentiation of the amplitude leads to 
``ideal instanton gas". The instanton density is directly related to the 
{\em non-perturbative downward shift in vacuum energy density}.

 The so called ``diluteness parameter"\footnote{Here the discussion is qualitative, which
 means we ignore
 numerical factors, as well as  the difference between $<\rho>^4$ and  $<\rho^4>$.}
\be \kappa=n_{I+\bar{I}} \rho^4 \ee   
describes how far instantons are from each other in units of their size. When it is large enough, this ideal gas approximation does not hold, and one naturally needs to
take into account instanton interactions. 

This section is about different regimes
in which the instanton ensemble may exist.
Before we describe them in detail, let us just enumerate them (in historical order):\\
 (i) a gas of individual instantons\\
(ii) a gas of pairs, the {\em instanton-antiinstanton molecules} \\
(iii)  the {\em ``instanton liquid"}   \\
(iv)  the   {\em ``instanton polymers"}, producing quasibound Cooper quark pairs, of color superconding phases \\

The first two are ``dilute" ensembles, as their density can be arbitrarily small, so that the interaction
between constituents can be safely neglected: we will discuss those in the next subsection.

The last two are not only ``dense", in the sense that the interaction needs to be accounted for, 
but is in fact ``dense enough" for certain qualitative changes in the system to take place. The   
``instanton liquid"  is a phase in which the chiral symmetry is spontaneously broken, so that there is a
nonzero quark condensate $<\bar q q> \neq 0$. The color supercondor phase has 
nonzero condensate of diquarks $<q q> \neq 0$ (it may or may not also include  $<\bar q q> \neq 0$,
depending on details of the setting). 

The value of the condensate influences back the
instanton density:  these parameters are all found from
minimization of the free energy.   In a mean field approximation
the free energy is approximated by certain analytic expressions: its derivatives over parameters,
set to zero, are known as the ``gap equations": their solution may or may not be done analytically.
We will discuss those in the rest of this chapter.

\begin{figure}[h!]
\begin{center}
\includegraphics[width=10.cm]{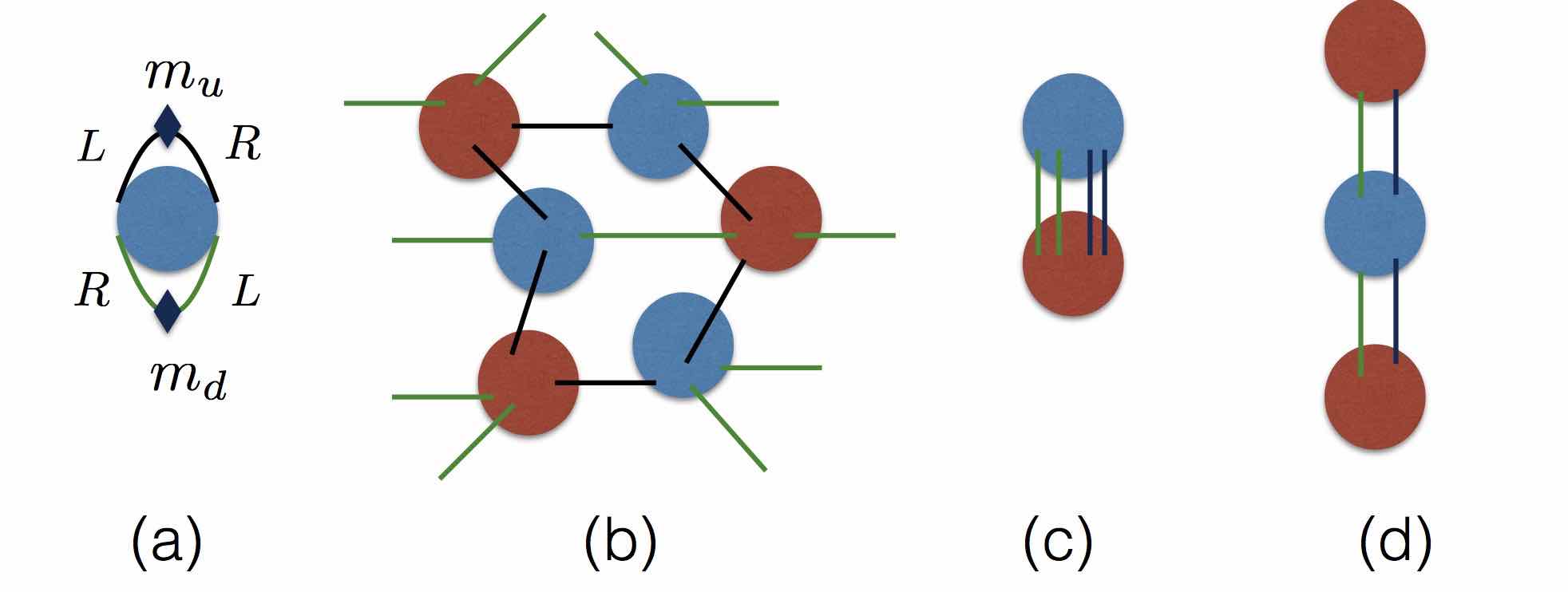}f
\caption{ Schematic representation of four instanton ensembles for $N_f=2$, see text.
Red and blue circles correspond to instantons and anti-instantons, black and green lines to $u,d$ quarks.
Black diamonds are mass insertions, flipping chirality from left (L) to right (R).}
\label{fig_ensembles}
\end{center}
\end{figure}

\section{The dilute gas of individual instantons}

At high $T$ the larger size instantons are screened out, so only those with sizes
$$ \rho T < 1 $$
are present\footnote{A question at this point usually comes, if the size
should be  instead limited by the electric Debye mass, $M_D\sim g T$.
However, there is no $g$ in the condition above, because classical instanton field
is large $O(1/g)$.}.  That is why the dimensionless diluteness of the ensemble
should depend on $T$ as large inverse power
\be n \rho^4 \sim \big({\Lambda_{QCD} \over T}\big)^b \ee

Furthermore,
in QCD with light quarks the dilute instanton gas is made far more diluted by the
existence of fermionic zero modes. In the chiral limit $m_q\rightarrow 0$, any object with nonzero topological
charge $Q\neq 0$, taken by itself, should have $Q$ fermionic zero modes. Recall that in the QCD partition function
the fermionic determinant is in the numerator:   so zero eigenvalue make it to vanish. Thus, in the chiral limits,
there simlply cannot exist any  dilute instanton gas.

If one take into account finiteness of quark masses, the dilute instanton gas is possible, but with
rather small density.
We will call it ``the 't Hooft regime": as we already
discussed in the previous chapter, in this case the instanton amplitude is additionally -- on top of the tunneling exponent --
suppressed by product of all light quark masses.  This factor is numerically about
\be \prod_{f=u,d,s} (m_f\rho) \approx \big({ 2.5\, MeV \over 600\, MeV} \big)  \big({ 5.\, MeV \over 600\, MeV} \big) \big({ 100\, MeV \over 600\, MeV} \big)\approx 6\cdot 10^{-6}  \ee
where we put the mean instanton size $\rho$ for dimension. (Its numerical value will be extensively discussed below in this chapter.) If all instanton effect in QCD would be that small, perhaps there would not be a need to discuss
them. Fortunately, this regime only occurs at high temperatures $T>T_c$: first indications for that has been recently
demonstrated on the lattice.

Recall that these mass insertions appear in the single instanton amplitude because all quarks emitted via zero modes {\em need to be reabsorbed back}, Fig.\ref{fig_ensembles}a. Zero modes for a quark and an antiquark have opposite chiralities,
and the price for flipping it is the mass insertion. But such {\em chirality flips are not needed if a quark emitted by an instanton
is absorbed by an anti-instanton!} 

In  other ensembles,  a quark emitted 
by an instanton can be  absorbed by  antiinstantons. Therefore these other phases can exist even in the chiral limit $m_f\rightarrow 0$.

If the density is high and a quark emitted 
by an instanton can,
with comparable probability, be absorbed by multiple antiinstantons (see Fig.\ref{fig_ensembles}b):
this ensemble is called {\em the instanton liquid"} \cite{Shuryak:1981ff}.
The fermionic determinant is a sum of closed loops, so, if one follows one particular (say $u$) quark, it will always come back. In the instanton liquid ensemble a typical length of the loop is very long, scaling with the volume $V_4$.
What this means is that it gets infinite in the $V_4\rightarrow \infty$ limit, and thus $SU(N_f)$ chiral symmetry
gets {\em spontaneously broken}. We will have a chapter devoted to it soon.

If the density is low, the simplest loop is a travel to the nearest antiinstanton and back, see Fig.\ref{fig_ensembles}c.
This ensemble is called a ``molecular phase" made of instanton-antiinstanton pairs \cite{Ilgenfritz:1994nt}.
Transition from the option (ii), the instanton liquid, to (iii), $\bar I I $ pairs,  as a function
of $T$ is the instanton-based explanation for the chiral symmetry restoration transition\footnote{
Its physics is rather similar to the so called Beresinsky-Kosterlitz-Thouless transition with
the 2-dimensional vortices, for which the 2016 Nobel prize was awarded.
}

The last ensemble (iv) produces condensates of {\em  quark Cooper pairs} in cold but dense quark matter:
 thus it is called (the instanton-induced) color superconductor   in which there is a nonzero VEV $<qq> \neq 0$. Depending on the parameters, it may exist in several interesting phases
 unfortunately we will not have time for its discussion in these lectures.

\section{The ``instanton liquid model" (ILM)  }

   Complementing the theory built from first principles,
one may also look at the phenomena under consideration from empirical
point of view, searching for hints and combining those into a simple approximate model. 

It is precisely what happened in the early 80's.
   By that time, the QCD sum rules \cite{Shifman:1978bx} has been widely used,
and they
provided some  understanding of the behavior
of the QCD correlation functions. Combining partonic description at
small distances with hadronic description at large ones, one learned
what happens in between. Furthermore, using Wilsonian 
Operator Product Expansion (OPE), one was able to
qualitatively understand the correlator phenomenology in terms of
VEV's of few ``vacuum condensates", the gluonic  $<G_{\mu\nu}^2>$
and quark $<\bar q q>$ ones. 
(Some of that we will discuss in chapter on CD correlation funcitons.)
Furthermore,  as it was pointed out by the authors of the method themselves, not ``all
hadrons are alike'': for spin-zero channels
the OPE-based theory apparently failed to predict the magnitude of the non-perturbative 
effects, even qualitatively.

 
  In order to explain available ``phenomenology of the vacuum'',
   a  qualitative model
 was proposed in my work \cite{Shuryak:1981fza},  
the so called ``instanton liquid'' model (ILM) of the CD vacuum.
%
%

   There were two parameters  --
the (mean) instanton
size $\rho$ and the instanton density $N$, to be determined
phenomenologically. 
The 
instanton size distribution was assumed to be just 
\bea {dn\over d\rho}=n_0 \delta(\rho-\rho_0) \eea

An idea what instanton 
density in the QCD vacuum can be was taken from 
 the empirical value of the gluon 
condensate:
\bea n_0 < n_c = <(gG^a_{\mu \nu})^2>/32\pi^2 \sim (1 fm)^{-4} \eea
(where the density means both pseudosparticles together, $I$ and $\bar I$).


To evaluate the typical size one more observable was used, the  quark condensate
$<\bar q q>$, in a version of mean field estimate. The result
was 
\bea \rho_0 \sim {1\over 3} fm = {1\over 600 \, MeV} \eea
It was soon found that this model can reproduce well also other properties
of chiral symmetry breaking, such as the pion decay constant $f_\pi$. I then
 proceeded forward calculating many different
 correlation functions
 \cite{Shuryak:1982qx},  including those where the OPE sum rules failed,  and found that 
the model reproduces phenomenology in those channels as well.


While this model  historically originated from studies of the hadronic phenomenology and related correlation functions,
most of its elements have been studied directly from gauge field configurations generated in
lattice numerical simulations. Therefore, we will not follow the historical path, but discuss first
certain lattice results. Yet, to have certain picture in mind, let me 
still
 summarize here some important qualitative features of  this model:

\noindent 1.The {\bf diluteness parameter} is small
\index{diluteness parameter}
\bea n_0\rho^4_0=(\rho/R)^4 \sim (1/3)^4 \eea 
where R is the typical distance between the
pseudoparticles. 
 So only few per cent of the space-time is occupied by
strong field. The factorization hypothesis is thus violated, by
the inverse diluteness
\\
2.{\bf semiclassical formulae are applicable}. The action of the typical instanton
is large enough
 \bea S_0 = 8 \pi^2/g(\rho)^2
\sim 10 \gg 1\eea
Quantum corrections go as $1/S_0$ and are presumably small enough.\\
3.{\bf The interaction does not destroy instantons}. Estimated by the dipole formula,
interaction was found to be typically
\bea |\delta S_{int}| \sim (2-3) \ll S_0 \eea \\
4.{\bf It is a liquid, not a gas.}
 The interaction is not small in the
statistical mechanics of instantons, on the contrary
correlations are strong
\bea exp |\delta S_{int}| \sim 20 \gg 1 \eea

\section{Statistical mechanics of the instanton ensembles}


   Early attempts to
relate instantons with  practical applications to QCD
were summarized by important paper by Callan, Dashen and Gross
\cite{Callan:1978bm}.  Incorporating the dipole-like forces between 
instantons and antiinstantons 
they have tried to create a self-consistent theory of interacting
instantons\footnote{  But soon this groups switched to so called ``merons'' -- half-instantons --
using which they have attempted to explain
confinement. This idea did not work but in some way predated the instanton-dyons to be discussed below
.}. After \cite{Witten:1978bc} pointed out difficulties in
approaching the large $N_c$ limit using instantons\footnote{It took many years till 
the issue was clarified by instanton ensemble simulations, see discussion to follow later.}
, this
Princeton group abandoned this direction and switched to this limit of QCD.

   In  1982   a simple model \cite{Shuryak:1981fza} with
the fixed instanton
size and density was propose.
Since the instantons in it were considered uncorrelated, it was called $Random$ Instanton Liquid Model (RILM for short).
While it was phenomenologically quite successful, 
 some consistent many-body theory of the
instanton ensemble was of course needed.

 The first step was a simplified  hard-core model by Ilgenfritz and
 Mueller-Preussker \cite{Ilgenfritz:1980bm}. The second
was the variational approach  by Diakonov and Petrov
\cite{Diakonov:1983hh}. For a  ``sum
ansatz'' configurations--
the gauge potential equal to a sum of those for
individual instantons and anti-instantons, in a singular gauge-- classical interaction was
calculated 
 and the mean-field approximation (MFA) 
of the statistical mechanics. The interaction was quite repulsive, leading to
 quite dilute       equilibrium ensemble.
Inclusion of light quarks has followed \cite{Diakonov:1985eg}, 
which lead to the first calculation of the quark condensate, also in the mean field approximation.
The accuracy of the mean field  approach
 remained however unclear, as the quark-induced interactions
 between instantons are very strong and expected to produce strong correlations
 between them.

More direct and accurate
methods have to be developed to 
  treat the statistical ensemble, which was done  in a
series of  papers started with
\cite{Shuryak:1988rf}: this approach was called the $Interacting$ Instanton Liquid Model (IILM).
 It uses the combined fermionic determinant for all instantons, which is 
is equivalent to including the diagrams containing
 't Hooft effective Lagrangian to {\it all oders}\footnote{Well, strictly speaking, for all orders which included all  instantons+antiinstantons in a box, typically $N\sim$ few hundreds.
 Note that the number of diagrams is of the order $N!$ and is huge. }.
  
First, the experimentally known correlation functions 
were reproduced by the model at {\em small distances} at a quantitative level. 
Then, for many mesonic channels \cite{Shuryak:1992jz,Shuryak:1992ke} significant
numerical efforts were made, allowing to calculate the 
relevant correlation functions  till larger distances (about 1.5 fm),
where they decay by
few decades. As a results, the predictive power of the model has been
explored in substantial depth. Many
the coupling constants and even hadronic masses
were calculated, with good
 agreement with experiment and lattice. 
 
  Subsequent calculations of baryonic correlators \cite{Schafer:1993ra}
has revealed further surprising facts. 
In the instanton vacuum the nucleon was shown to be made 
of a ``constituent quark" plus a deeply bound $diquark$, with a mass nearly the same
as that of constituent quarks. On the other hand, decuplet baryons (like
$\Delta^{++}$) had shown no such diquarks, remaining  weakly bound
set of three  constituent quarks. To my knowledge, it was the first dynamical
explanation of deeply bound scalar diquarks. While being a direct consequence
of 't Hooft Lagranigian, this phenomenon has been missed for a long time.
It also lead to realization that diquarks can become Cooper pairs in dense quark matter,
see \cite{Schafer:2000et} for a review on ``color superconductivity".

Further elaboration  of such analysis for vector and axial 
has been made \cite{Schafer:2000rv}:
RILM happens to reproduce
 rather accurately the ALEPH data on $\tau$ decay
for both vector and axial correlators.
A comparison between the correlators
  calculated in RILM and on the lattice
\cite{Chu:1994vi} have also found good agreement,
including the baryonic channels unreachable by
usual phenomenology . 
Studies of hadronic ``wave
functions''  
and even glueball  correlation funcitonshas followed, again with 
results very close to what lattice measurements had produced: for a review see \cite{Schafer:1996wv}.

\subsection{The mean field approximation (MFA)}
The main assumption of the mean field approach is that a particle interact with
its many neighbors via their common ``mean" field, rather than developing specific correlations
with few of them. More technically, the multibody distributions are then approximated by
Ansatz, made of a product of independent 1-body distributions\footnote{ In this respect,
this assumption is similar to {\em Boltzmann hypothesis} behind his kinetic equation.
While true for gases, it does not hold for stronger correlated systems such as liquids. 
As we will see below, it is not a good approximation for instanton liquid as well.
}.
Whether this assumption is or is not applicable needs of course be studied on a case
by case basis.

In this section we follow \cite{Diakonov:1983hh} . 
Let us start with 
 the partition function 
for a system of instantons in pure gauge theory
\be
\label{Z_glue}
 Z &=& \frac{1}{N_+!N_-!}\prod_i^{N_++N_-}\int [d\Omega_i\, n(\rho_i)]
 \, \exp(-S_{int}).
\ee
Here, $N_{\pm}$ are the numbers of instantons and anti-instantons, 
$\Omega_i=(z_i,\rho_i,U_i)$ are the collective coordinates of the 
instanton $i$, $n(\rho)$ is the semi-classical instanton distribution
function (\ref{eq_d(rho)}) and $S_{int}$ is the bosonic instanton 
interaction. 

%

So, MFA assumes that the partition function is evaluated
 via a
  $product$ of single instanton 
distributions $\mu(\rho)$   
\be
\label{var_ans}
 Z_1 &=& \frac{1}{N_+!N_-!}\prod_i^{N_++N_-}\int d\Omega_i\, 
 \mu(\rho_i)= \frac{1}{N_+!N_-!}(V\mu_0)^{N_++N_-}
\ee
where $ \mu_0 = \int d\rho\,\mu(\rho)$. The distribution $\mu(\rho)$
 is then determined from the variational principle,
or maximization of the $Z$ $\delta\log Z_1/\delta\mu
=0$. In quantum mechanics a variational wave functions always provides
an upper bound on the true ground state energy. The analogous 
statement in statistical mechanics is known as Feynman's variational
principle. Using convexity 
\be 
\label{convex}
   Z &=& Z_1 \langle\exp(-(S-S_1))\rangle  
   \;\geq\; Z_1 \exp(-\langle S-S_1\rangle ),
\ee
where $S_1$ is the variational action, one can see that the
variational vacuum energy is always higher than the true one.
 
  In our case, the single instanton action is given by $S_1=\log
(\mu(\rho))$ while $\langle S\rangle $ is the average action calculated 
from the variational distribution (\ref{var_ans}). Since the 
variational ansatz does not include any correlations, only 
the average interaction enters. For the sum ansatz
\be
\label{av_int}
 \langle S_{int}\rangle  &=& {8\pi^2\over g^2}\gamma^2\rho_1^2\rho_2^2, \hspace{1cm}
  \gamma^2 = \frac{27}{4}\frac{N_c}{N_c^2-1}\pi^2
\ee
the same for both $IA$ and $II$ pairs. Note that (\ref{av_int}) is of the 
same form as the hard core mentioned above, with a different 
 dimensionless parameter $\gamma^2$. Applying the variational 
principle, one finds \cite{Diakonov:1983hh}
\be
\label{reg_dis}
\mu(\rho) &=& n(\rho)\exp\left[ -\frac{\beta\gamma^2}{V}N
\overline{\rho^2}\rho^2\right],
\ee
where $\beta=\beta(\overline{\rho})$ is the average instanton
action and $\overline{\rho^2}$ is the average size. We observe
that the single instanton distribution is cut off at large sizes
by the average instanton repulsion. 
Note also that the cutoff has a Gaussian dependence on $\rho$, the same as
seen in lattice studies we mentioned above.

The average size $\overline{
\rho^2}$ is determined by the self consistency condition $\overline{
\rho^2}=\mu_0^{-1}\int d\rho\mu(\rho)\rho^2$. The result is
\be
\label{rho2_av}
\overline{\rho^2} &=& \left(\frac{\nu V}{\beta\gamma^2 N}\right)^{1/2},
\hspace{1cm}\nu = \frac{b-4}{2},
\ee
which determines the dimensionless diluteness of the ensemble,
$\rho^4(N/V)=\nu/(\beta\gamma^2)$. Using the pure gauge 
beta function $b=11$, $\gamma^2\simeq 25$ from above and $\beta
\simeq 15$, we get the diluteness $\rho^4(N/V)=0.01$, even more dilute than
phenomenology requires. The instanton density can be fixed from
the second self-consistency requirement, $(N/V)=2\mu_0$ (the factor
2 comes from instantons and anti-instantons). one gets 
\be
\label{dens_av}
\frac{N}{V} &=& \Lambda_{PV}^4 \left[ C_{N_c}\beta^{2N_c} \Gamma(\nu)
(\beta\nu\gamma^2)^{-\nu/2}\right]^{\frac{2}{2+\nu}},
\ee
\be
\label{MFA_res}
\chi_{top}\simeq\frac{N}{V} = (0.65\Lambda_{PV})^4,\hspace{0.5cm}
(\overline{\rho^2})^{1/2}= 0.47 \Lambda_{PV}^{-1} \simeq{1\over 3} R,
\hspace{0.5cm} \beta=S_0\simeq 15,
\ee
It may be
consistent with the phenomenological values if $\Lambda_{PV}
\simeq 300$ MeV. It is instructive to calculate the free energy
as a function of the instanton density. Using $F=-1/V\cdot\log Z$, 
one has
\be
\label{F_MFA}
F &=& \frac{N}{V} \left\{ \log\left(\frac{N}{2V\mu_0}\right)
 -\left(1+\frac{\nu}{2}\right) \right\}.
\ee
The instanton density is determined by the minimizing the free energy,
$\partial F/(\partial (N/V))=0$. The vacuum energy density is given by 
the value of the free energy at the minimum, $\epsilon=F_0$. We find 
$N/V=2\mu_0$ as above and
\be
\label{eps_MFA}
 \epsilon &=& -\frac{b}{4} \left(\frac{N}{V}\right)
\ee
Estimating the value of the gluon condensate in a dilute instanton gas,
$\langle g^2G^2\rangle =32\pi^2(N/V)$, we see that (\ref{eps_MFA}) is 
consistent with the trace anomaly. 

  The second derivative of the free energy with respect to the instanton 
density, the compressibility of the instanton liquid, is given by
\be
\label{comp_MFA}
 \left.\frac{\partial^2 F}{\partial (N/V)^2}\right|_{n_0} 
 &=& \frac{4}{b}\left(\frac{N}{V}\right) ,
\ee
where $n_0$ is the equilibrium density. This observable is also determined 
by a low energy theorem based on broken scale invariance 
\be
\label{scal_let}
\int d^4x\; \langle g^2G^2(0)g^2G^2(x)\rangle  
 &=& (32\pi^2)\frac{4}{b}\langle g^2G^2\rangle .
\ee
Here, the left hand side is given by an integral over the field strength
correlator, suitably regularized and with the constant disconnected term
$\langle g^2G^2\rangle^2$ subtracted. For a dilute system of instantons, 
the low energy theorem gives 
\be
\label{n_fluc}
 \langle N^2\rangle -\langle N\rangle ^2
 &=&\frac{4}{b}\langle N\rangle .
\ee
Here, $\langle N\rangle$ is the average number of instantons in 
a volume $V$. The result (\ref{n_fluc}) shows that 
density fluctuations in the instanton liquid are not Poissonian. 
Using the general relation between fluctuations and the 
compressibility gives the result (\ref{comp_MFA}). This 
means that the form of the free energy near the minimum is 
determined by the renormalization properties of the theory.
Therefore, the functional form (\ref{F_MFA}) is more general 
than the mean field approximation used to derive it. 

   How reliable are these results?
   
   First of all, it cannot be better than the underlying interaction, obtained from the particular set of gauge field configurations.
   Later studies had shown that the
sum ansatz used indeed is $not$ a good representation of instanton-antiinstanton valley
( which is not surprising since it was
chosen without any justification other than simplicity). The interaction based on streamline configurations found by  \cite{Verbaarschot:1991sq}
can be used instead:  but the results also also not satisfactory, because the ensemble contains too many close 
pairs and too many large instantons. Stabilization of the density
is reached by an exclusion of configurations with small action -- 
presumably already included in the perturbation theory -- with is a  repulsive core weaker than in the 
sum ansatz, but still present.

  Another issue is 
 the accuracy of the MFA itself.
 As the density decreases, the binary correlations are building up, which are ignored in the MFA.
Its accuracy can be checked by
doing statistical simulations of the full partition function.

\subsection{Diquarks and color superconductivity}
The second most attractive channel
is the interaction of two quarks in the scalar S=I=0 channel.
 It also follows from 't Hooft effective
Lagrangian, and suppressed by a factor $1/(N_c-1)$ relative to
the most attractive scalar $\bar q q$ channel.
It was pointed out in two simultaneous papers\footnote{They were
submitted to the archive
 on the same day} \cite{Alford:1997zt,Rapp:1997zu} in 1997:  the same interaction
leads to a very robust Cooper pairing in high density QCD.
In few years this field have boomed and has now a bibliography of
about 500 papers but we will not discuss it here.

The issue of diquarks may require  some further discussion. Since there exists only one zero mode per instanton, only diquarks with $different$ flavors can be formed: $ud,us,ds$\footnote{
Also Fermi statistics requirement needs to be satisfied: with total spin zero the 
spin wave function is anti-symmetric, color part is antisymmetric as well: so flavor should also
be antisymmetric. A product of three minuses is minus, as Fermi statistics requires. }.  
 
 One well known argument for a  nucleon-like hadrons being made of quark-diquark pair
 comes from Regge trajectories. As  we will discuss in chapter on the flux tubes,
 the slope of the nucleon Regge trajectory is the same as for mesons. This means
 there are two color objects connected by a $single$ flux tube.
 
 Another interesting consequence is for a nucleon spin structure. If it is made of $u$+$(ud)$ scalar diquark (and never $d+(uu)$) then there is no way $d$ quark contribute anything to the nucleon spin. Recent lattice works did indeed confirmed this. $d$ contributes half of $u$
 into the total momentum, but (withing decreasing errors) nothing to spin.

\subsection{Instantons for larger number of colors}
Recall that the 
(one-loop)  instanton action is given by $S_0=(8\pi^2)/
g^2=-b\log(\rho\Lambda)$ where $b=(11N_c)/3$ is the first 
coefficient of the beta function in pure gauge QCD. In the 
't Hooft limit $N_c\to\infty$ with $$\lambda_{tHooft}\equiv g^2N_c=const$$ we expect 
$S_0=O(N_c)$ and $\rho=O(1)$.  It lead to argument by \cite{Witten:1978bc}
that instantons do not survive the large-$N_c$ limit. 

 Our discussion of finite-$T$ theory will show rather simple realization of that
 idea: instantons will be indeed split to $N_c$ instanton-dyons. The action per dyon
 is thus finite in the 't Hooft limit. 

However, 
it was also shown that the IILM in fact also has a reasonable large $N_c$ limit,
although it is reached in a non-trivial way.
We will discuss that now following
a detailed study by  \cite{Schafer:2002af}, who 
managed to show that few known
paradoxes of the dilute gas approximation do disappear  in
the interacting instanton liquid. In fact,  a
  self-consistent picture of the ensemble emerges, 
which well agrees
with the pre-existing theoretical expectations, including the Witten's
conjectures about the  topological susceptibility and
the $\eta'$ mass. Another remarkable feature of this regime is that
the difference with what we know about instanton ensemble in QCD
is not really drastically changed, even in the large $N_c$ limit.

 In brief, main features of this regime is as follows.
The density of instantons is predicted to
grow as $N_c$, whereas the typical instanton size remains
finite. The effective diluteness (accounting for the fact that
instantons not overlapping in color do not interact) remains constant.
Interactions between instanton are important 
and suppress fluctuations of the topological charge.
As a result the $U(1)_A$ anomaly is effectively restored
even though the number of instantons increases. Using mean 
field approximation and then numerical IILM simulations 
one finds that this scenario does not require fine tuning but 
 arises naturally if the instanton ensemble is stabilized by 
a classical repulsive core. Although
 the total instanton density is large but the instanton
liquid remains effectively dilute because instantons are not strongly
overlapping in color space. 

 Since the {\em total instanton density} is related 
to the non-perturbative gluon condensate 
\bea 
\label{glue}
 \frac{N}{V} = \frac{1}{32\pi^2}
   \langle g^2 G^a_{\mu\nu} G^a_{\mu\nu} \rangle .
\eea
the $N_c$ counting suggests that $\langle g^2G^2\rangle =
O(N_c)$ and we are lead to the conclusion that $(N/V)=O(N_c)$. 
This is also consistent with the expected scaling of the vacuum 
energy. Using equ.~(\ref{glue}) and the trace anomaly relation 
\bea 
\label{trace_anom}
\langle T_{\mu\mu} \rangle = -\frac{b}{32\pi^2}
  \langle g^2 G^a_{\mu\nu} G^a_{\mu\nu} \rangle ,
\eea
the vacuum energy density is given by
\be
\label{eqn_epsilon_vac}
\epsilon = -\frac{b}{4} \left(\frac{N}{V}\right).
\eea 
Using $(N/V)=O(N_c)$ we find that the vacuum energy scales as 
$\epsilon = O(N_c^2)$ which agrees with our expectations for a 
system with $N_c^2$ gluonic degrees of freedom. 

  Note that $(N/V)=O(N_c)$ implies that the effective 
diluteness of instantons remains constant in the 
large $N_c$ limit. Indeed, in spite of large density
most instantons do not see each other: 
 the number of mutually commuting $SU(2)$ subgroups of 
$SU(N_c)$ scales as $N_c$.

 If instantons are distributed randomly then fluctuations
in the number of instantons and anti-instantons are expected 
to be Poissonian. This leads to the predictions
\bea
\label{n2}
 \langle N^2 \rangle - \langle N\rangle^2 
  = \langle N \rangle , \\
\label{q2}
 \langle Q^2 \rangle  =  \langle N \rangle ,
\eea
where $N=N_I+N_A$ is the total number of instantons and 
$Q=N_I-N_A$ is the topological charge. Equ.~(\ref{q2}) 
implies that 
\be
 \chi_{top} = \frac{\langle Q^2\rangle}{V} = 
   \left(\frac{N}{V}\right) .
\eea
Using $(N/V)=O(N_c)$ we observe that $\chi_{top}=O(N_c)$ which is 
{\em in contradiction to Witten's conjecture} $\chi_{top}=O(1)$. However,
as we shall see, the interactions between instantons suppress
the fluctuations and invalidate  equs.~(\ref{n2},\ref{q2}). 

We now include the fermion-related dynamics, and ask
how the chiral condensate scales with $N_c$, using first analytic MFA
\footnote{
 For definiteness, we will consider
the case $N_f=2$ but the conclusions are of course independent 
of the number of flavors.}. 
 After averaging over the color
orientation of the instanton the effective Lagrangian is 
\be
\label{l_nf2}
{\cal L} = \int n(\rho)d\rho\, 
   \frac{2(2\pi\rho)^4\rho^2}{4(N_c^2-1)}
 \epsilon_{f_1f_2}\epsilon_{g_1g_2} 
 \left( \frac{2N_c-1}{2N_c}
  (\bar\psi_{L,f_1} \psi_{R,g_1})
  (\bar\psi_{L,f_2} \psi_{R,g_2}) \right. \\ \nonumber 
\left. - \frac{1}{8N_c}
  (\bar\psi_{L,f_1} \sigma_{\mu\nu} \psi_{R,g_1})
  (\bar\psi_{L,f_2} \sigma_{\mu\nu} \psi_{R,g_2})
  + (L \leftrightarrow R ) \right) \nonumber .
\ee
We observe from it that the explicit $N_c$ dependence is given by 
$1/N_c^2$. This is again related to the fact that instantons
are $SU(2)$ objects. Quarks can only interact via instanton 
zero modes if they overlap with the color wave function of
the instanton. As a result, the probability that two quarks 
with arbitrary color propagating in the background field of
an instanton interact is $O(1/N_c^2)$. 

 The MFA gap equation for the spontaneously generated constituent 
quark mass is
\bea 
M = GN_c \int \frac{d^4k}{(2\pi)^4}\frac{M}{M^2+k^2},
\eea
where $M$ is the constituent mass and $G$ is the effective 
coupling constant in equ.~(\ref{l_nf2}). The factor $N_c$
comes from doing the trace over the quark propagator. The 
coupling constant $G$ scales as $1/N_c$ because the density of 
instantons is $O(N_c)$ and the effective Lagrangian contains an 
explicit factor $1/N_c^2$. We conclude that the coefficient in 
the gap equation is $O(1)$ and that the dynamically generated
quark mass is $O(1)$ also. This also implies that the quark
condensate, which involves an extra sum over color, is 
$O(N_c)$.

\begin{figure}[t]
\centering
\includegraphics[width=6.cm]{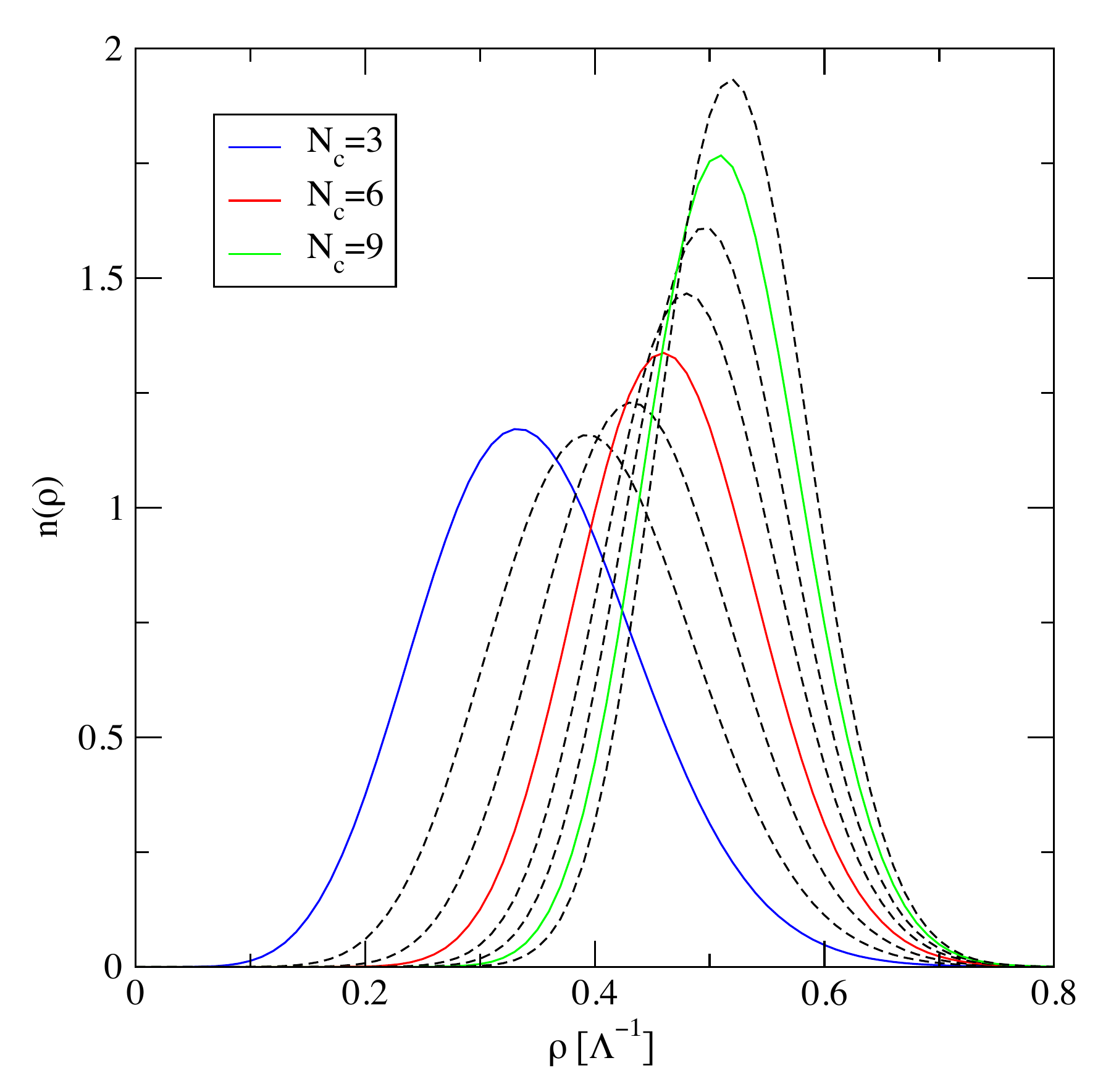}
\includegraphics[width=6.cm]{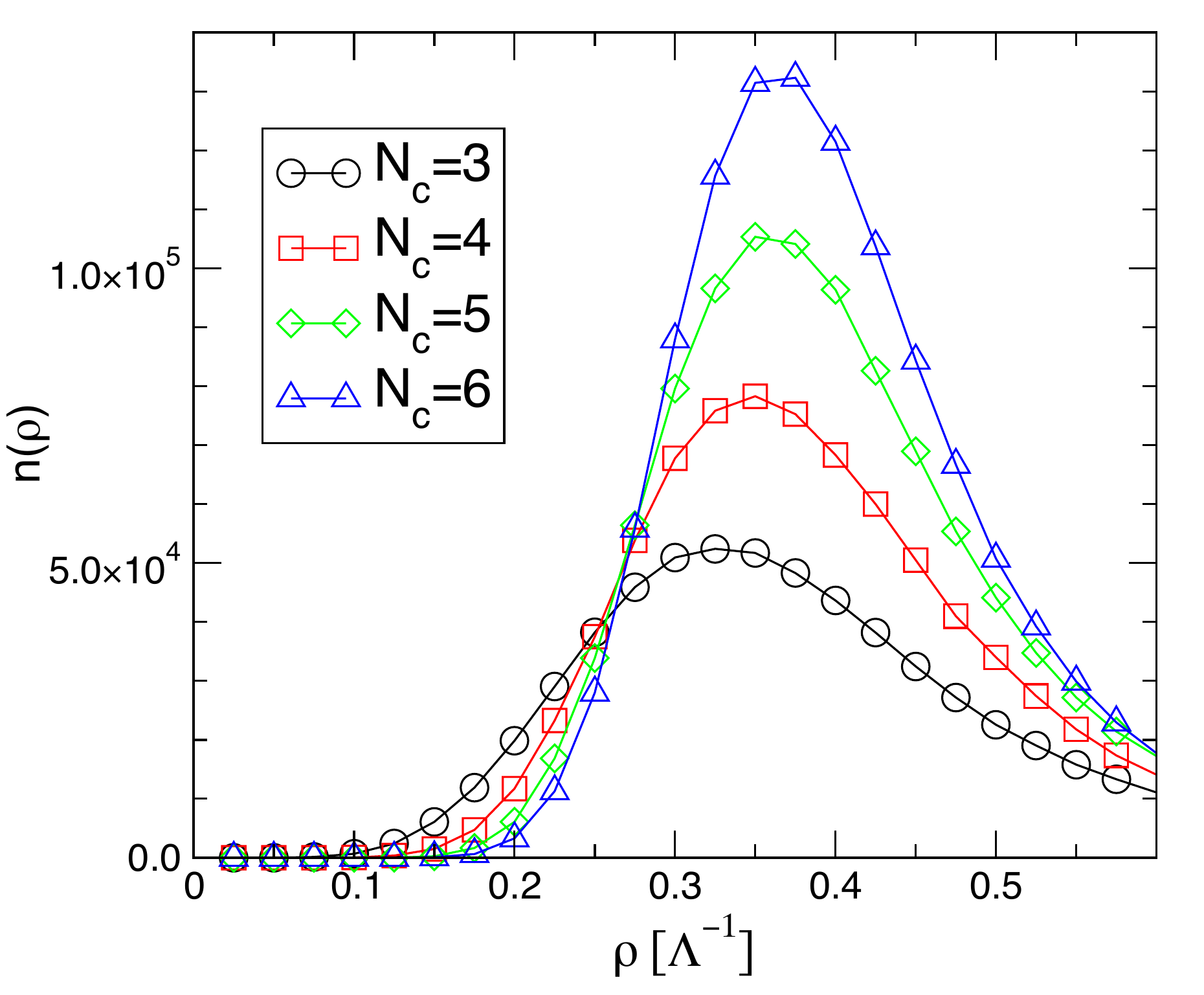} 
\caption{\label{fig_mfa}
 Instanton size distribution $n(\rho)$ for different numbers
of colors: (left) $N_c=3,\ldots,10$ calculated using mean field approximation, (right) 
from numerical simulations with $N=128$
instantons.}
\end{figure}

The results in the mean field approximation are summarized in the
  Fig.~\ref{fig_mfa} which
  shows that for $N_c>4$ the average instanton
size is essentially constant while the instanton density grows
linearly with $N_c$. This can also be verified by  
expanding $\log(N/V)$ in powers of $N_c$
and $\log(N_c)$: one observes that independent of the details
of the interaction the instanton density scales at most as 
a power, not an exponential, in $N_c$. 

  Another way to see  why the instanton
density scales as the number of colors is as follows: the 
size distribution is regularized by the interaction between
instantons. This means that there has to be a balance between    
the average single instanton action and the average interaction
between instantons. If the average instanton action satisfies
$S_0=O(N_c)$ we expect that $\langle S^{tot}_{int}\rangle =O(N_c)$ 
also. Using $\langle S^{tot}_{int}\rangle = (N/V) \langle S_{int} 
\rangle$ and the fact that the average interaction between 
any two instantons satisfies $\langle S_{int} \rangle = O(1)$
we expect that the density grows as $N_c$.

  Fig.~\ref{fig_mfa} (b) shows the instanton size distribution
for different numbers of colors. We observe that the number 
of small instantons is strongly suppressed as $N_c\to\infty$ 
but the average size stabilizes at a finite value $\bar{\rho}
<\Lambda^{-1}$. We also note that there is the critical size
$\rho^*$ for which the number of instantons does not change 
as $N_c\to\infty$. The value of $\rho^*$ is easy to determine
analytically. We write $n(\rho)=\exp(N_cF(\rho))$ with $F(\rho)
=a\log(\rho)+b\rho^2+c$ where the coefficients $a,b,c$ are 
independent of $N_c$ in the large $N_c$ limit. The critical
value of $\rho$ is given by the zero of $F(\rho)$. We find
$\rho^*=0.49\Lambda^{-1}$. The existence of a critical
instanton size for which $n(\rho^*)$ is independent of $N_c$ 
was discussed by \cite{Teper:1979tq,Neuberger:1980as,Shuryak:1995pv} and
indeed observed on the lattice 
\cite{Lucini:2001ej}.
Fluctuations in the net instanton number are related to the 
second derivative of the free energy with respect to $N$ (\ref{n_fluc}). 
This result  is in agreement with a low energy theorem
(\ref{scal_let}) based 
on broken scale invariance,
 based solely on the 
renormalization group equations. The left hand side is 
given by an integral over the field strength correlator, 
suitably regularized and with the constant term $\langle 
G^2\rangle^2$ subtracted. For a dilute system of instantons 
equ.~(\ref{scal_let}) reduces to equ.~(\ref{n_fluc}).
The result (\ref{n_fluc}) shows that fluctuations of 
the instanton ensemble are suppressed by $1/N_c$. This
is agreement with general arguments showing that  
fluctuations are suppressed in the large $N_c$ limit. We 
also note that the result (\ref{n_fluc}) clearly shows that 
even if instantons are semi-classical, interactions between 
instantons are crucial in the large $N_c$ limit.

 Fluctuation in the topological charge can be studied 
by adding a $\theta$-term to the partition function. The mean square
is just the average instanton number
\be
\label{q2_mfa}
 \langle Q^2 \rangle = \langle N \rangle ,
\eea
which is identical to the result in the random instanton 
liquid and not in agreement with Witten's hypothesis
$\chi_{top}=O(1)$. However, Diakonov et al.~noticed
that equ.~(\ref{q2_mfa}) is a consequence of the fact 
that in the sum ansatz the average interaction between
instantons of the same charge is identical to the average
interaction between instantons of opposite charge. 
 In general there is no reason for this
to be the case and more sophisticated instanton interactions do 
not have this feature.
If $r$ denotes the ratio of the average interaction between
instantons of opposite charge and instanton of the same
charge, $r=\langle S_{IA}\rangle/\langle S_{II}\rangle$, 
then 
\be
 \langle Q^2 \rangle = \frac{4}{b-r(b-4)}\langle N\rangle.
\eea
This result shows that for any value of $r\neq 1$ fluctuations
in the topological charge are suppressed as $N_c\to\infty$.
We also note that $\chi_{top}=O(1)$, in agreement with 
Witten's hypothesis.

 In \cite{Schafer:2002af}  the instanton size distribution,
the topological susceptibility and the spectrum of the 
Dirac operator for different numbers of colors 
have been determined in IILM numerically. 
The instanton
size distribution obtained shows that small instantons are strongly suppressed as
the number of colors increases. We observe a clear fixed 
point in the size distribution at $\rho^*\Lambda \simeq 
0.27$.
 The simulations were carried out in the total topological charge
$Q_{top}=0$ sector of the theory. One can nevertheless determine
the topological susceptibility by measuring the average $Q^2_{top}$ 
in a sub-volume $V_3\times l_4$ of the euclidean box $V_3\times L_4$ . 
 The topological susceptibilities 
 for $N_c=3$  agree
well with the expectation based on Poissonian statistics, 
$\chi_{top}\simeq (N/V)$. For $N_c>3$, however, fluctuations
are significantly suppressed and the topological susceptibility 
increases more slowly than the density of instantons,  consistent
with a scenario in which $\chi_{top}$ remains finite as 
$N_c\to\infty$.

\begin{figure}[h]
\centering
\includegraphics[width=8cm]{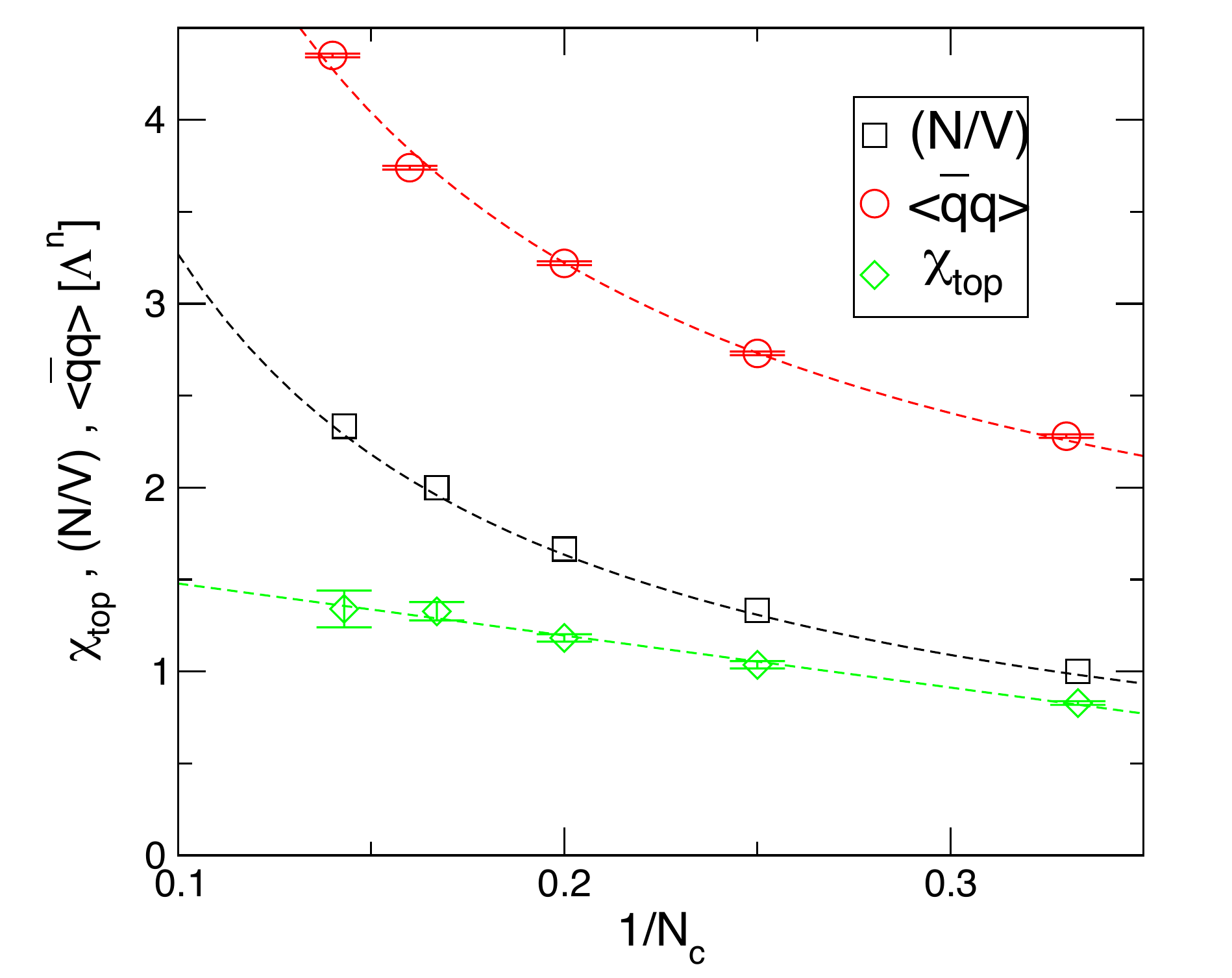}  
\caption{\label{fig_obs}
Dependence of the chiral condensate $\langle\bar{\psi}\psi\rangle$
and the topological susceptibility $\chi_{top}$ on the number
of colors. The instanton density $(N/V)$ was assumed to scale 
as $(N/V)\sim N_c$. The dashed lines show fits of the form 
$a_1N_c+a_2$ (for $\langle\bar{\psi}\psi\rangle$ and $N/V$)
and $a_2+a_3/N_c$ (for $\chi_{top}$).}
\end{figure}

 The chiral
condensate 
for $m_q=0.1\Lambda$ 
and topological susceptibility are shown in Fig.~\ref{fig_obs}.
We clearly see that $\langle\bar{q}q\rangle$ is linear in $N_c$
while $\chi_{top}$ approach a constant.


\chapter{QCD correlation functions and topology}
\section{Generalities}
\label{sec_cor_gen}
\subsection{Definitions and an overall picture}
Before we embark on technical discussion of the correlation function, let us first summarize in the non-technical terms
some pictures of the vacuum and hadronic structure, which result from what we learned about instantons.

In particularly, we know that the QCD-like theories with light quarks have spontaneously broken $SU(N_f)$ chiral
symmetry. In terms of hadronic spectroscopy, this phenomenon manifests itself 
in two ways:\\
 (i) quarks obtain ``constituent quark masses" $M\sim 300-400 \,\, MeV$, \\
  (ii)
there exist multiplets of (near)massless Goldstone modes -- the pions. 

\begin{figure}[h!]
\begin{center}
\includegraphics[width=6.5cm]{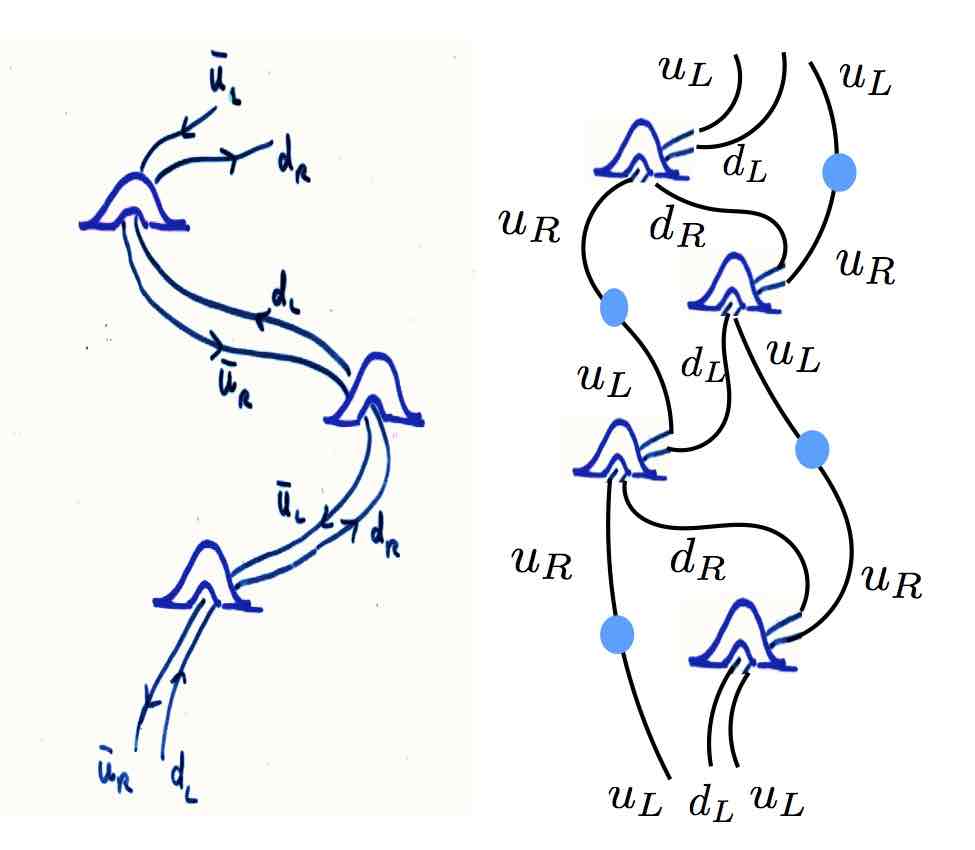}
\caption{
The  pion (left) and the proton (right), depicted as a sequence of tunneling events. The blue circle indicate mass insertions. Note $ud$ diquarks inside the proton.}
\label{fig_pion_as_tunneling}
\end{center}
\end{figure}

We have further learned that chiral
symmetry breaking is induced by collectivization of the quark zero modes, associated with nontrivial topology
of the instantons. In Fig.\ref{fig_pion_as_tunneling}
we sketch a picture of that:  the pions exist because light quark undergo
tunneling\footnote{Once in a talk, T.D.Lee was explaining 't Hooft interaction by saying that 
it is like cars go in parallel lanes in the tunnel, like from Queens to Manhattan. I happen to be there, and
said that it is more like the tunnel between England and France: whoever was left-handed become right-handed, and vice versal.
} events (the instantons) in pairs.  In terms of the path integral,
such correlated quark paths are interpreted as existence of some  attraction between $\bar u$ and $d$,
strong enough to cancel twice the constituent quark mass $2M$ and make it (near)zero.

Another manifestation of the of the correlated tunneling is in the $ud$ di-quark pair. In two-color QCD
its strength is exactly the same as in the pion channel, and thus this diquark (and anti-diquark) are also
(near)massless Goldstone modes, 5 in total for $N_f=2$. In the three-color world we live in,
the $ud$ di-quark is not color neutral and thus is not a hadron: but it exists as a correlated diquark\footnote{
Pauli principle works for zero modes: so the third quark is prohibited from tunneling together with $ud$ pair:
it needs to ``drive around" rather than take a tunnel, and so its constituent mass is not reduced.
}
inside the nucleons! 

Note that, because of topological index theorem, quark zero modes have specific chiralities: therefore  ' t Hooft 
interaction has very specific chiral structure. 
For example, the picture above cannot hold for vector or tensor mesons.
So, the aim of this section is to elucidate
-- using phenomenological or lattice correlation functions -- the role of the topology in the vacuum and hadronic structure.

In fact, one could have done it even simpler, without quarks and their zero modes, in pure gauge theories.
Indeed, the topological solitons themselves -- being selfdual or antiselfdual -- are made of so-to-say ``chirally
polarized gluonic fields. Scalar and pseudoscalar glueballs are strongly affected by tunneling events,
while other ones -- e.g. tensor ones -- are not.

   Correlation functions are 
 the main tools used in studies of 
 structure of the QCD vacuum.  They can be obtained in several
ways. First, they 
 can in  many cases be deduced phenomenologically, using 
vast set of data accumulated in hadronic physics. 
Second, they can be directly 
calculated {\em ab initio}  using quantum field theory methods, such as lattice gauge
 theory,
or semiclassical methods. Significant amount of work has also been
done in order to understand their small-distance behavior,  based
on the Operator Product Expansion (OPE). The large distance limit can also
be understood using effective hadronic approaches
or
  various quark models of hadronic structure.
In this section we focus on 
 available {\it
phenomenological information} about
the correlation functions, emphasizing the most important observations,
which are then compared 
 with {\it predictions of
   various theoretical approaches}; lattice numerical
simulations,
the operator product expansion  and 
interacting instantons approximation.
  As a ``common denominator" for our discussion we have chosen
 {\it the point-to-point
 correlation 
functions in coordinate representations}.  

We will discuss
 two types of the
operators, mesonic and baryonic ones 
\be O_{mes}(x)=\bar\psi_i  \delta_{ij} \psi_j , \,\,\,\,\,O_{bar}(x)=\psi_i \psi_j \psi_k \epsilon^{ijk}
\ee 
( Here the color indices are explicitly shown, 
but they will be suppressed below).
As  all color indices are properly contracted and all quark fields are taken
 at the same point x, these operators are 
manifestly gauge invariant.

The correlation function is  the vacuum expectation value (VEV) of
the product
of two (or more) of them 
 at different points
\be K(x-y)= \langle 0 | O(x) O(y)|0 > \ee
Since
 the vacuum is homogeneous,  it depends on the 
relative distance. 
The distance  is assumed to be {\it space-like}:
 we prefer to deal with
{\it virtual} (instead of real) propagation of quarks or 
hadrons from one point to another, so  
one deals with   decaying (instead of
 oscillating) 
functions\footnote{
At this point I was inevitably asked an
 old question;
how can be any correlation between the fields
 {\it outside of the
light cone}? It was 
essentially answered by Feynman long ago, who had to defend his
 propagator and $i\epsilon$ prescription following from Euclidean definitions.
In the path integral the particles can propagate along
{\it any}    
path going from x to y. The correlations outside the light cone
 {\em do not contradict  causality} because one cannot use
it for the signal transfer. 
}
.
 One can look at a pair of  points separate by Euclidean distance  in
terms of two distinct
limits.
Either they can be  two different 
points in space, taken at the same time moment,
 or be two events at the same spatial point 
  separated by non-zero interval 
 of the {\it imaginary (Euclidean) time}; $ix_0-i y_0=\tau$.
Below we will use both, depending on which is more convenient at the moment.
Due to Lorentz invariance (rotational O(4) in Euclid) the answers are
of course the same.

\subsection{Small distances:  perturbative normalization of the correlators}

At small  distances
 $| x | \Lambda_{QCD} \ll 1$ (remember,  the other argument
 of the correlator we take at the origin $y$=0) the ``asymptotic
freedom" of QCD tells us, that (up to small and calculable radiative 
corrections) 
  quarks and gluons propagate freely. Therefore\footnote{Strictly speaking, small $x$ includes
the zero distance $x=0$, and thus the reader should be aware that some correlators may have
 $local$ terms, proportional to $\delta(x)$ and/or its derivatives. We
 will mostly
 ignore such terms, not showing them in the plots etc, unless the integral over x is done.}
 $K(x)\approx K_{free}(x)$,
where free quark correlator in 
mesonic (baryonic) case  is essentially the square (cube) of the free massless
quark propagator, 
\be S_{free} (x) = <\bar q(x) q(0)>=(\gamma_\mu x_\mu)/2\pi^2
 x^4 \ee

 From the dimension of the free correlators they can only 
be for the mesonic (baryonic) channels
 $K_{free}(x) \sim x^{-6}$ ($\sim x^{-9}) $.
Of course, QCD does have a dimensional parameter $\Lambda_{QCD}$,
 which shows up in physical (non-free) correlators $K(x)$ and cause
deviations from  $K_{free}(x)$.
 However, in 
the perturbation theory it only comes in via the radiative corrections.
Therefore, at small
$x\Lambda<<1$, those produce corrections to our estimates above containing 
powers of $\alpha_s(x)\sim 1/log(1/x\Lambda)$.
If quarks are allowed to propagate to larger 
 distances, they
 start interacting  with the non-perturbative vacuum fields.
If corrections  are not too large, one can take these effects into
account using the operator product expansion (OPE) formalism. 
At intermediate distances
 description of the correlation functions becomes in general very complicated,
 and one may only evaluate them
either using lattice numerical simulations or some
 ``vacuum models" (e.g., the instanton ensembles we discuss now).

   At {\it large distances} the
 behavior of the correlation 
functions is given 
  using  the time evolution of an operator
$O(t)=e^{iHt}O(0)e^{-iHt}$
where H is a Hamiltonian, and then 
insert a complete set of
{\it physical intermediate states}
between the two operators. In Minkowski time this means
\be K(t)=\Sigma_n |\langle 0 |O(0)| n\rangle|^2 e^{-itE_n}. \ee
Now one can analytically continue the correlation function into
 the Euclidean domain $\tau=it$, and get a sum over 
decreasing exponents.

   Physically, application of such relation in QCD means that one
 considers propagation of physical excitations, or
{\it hadrons} between  two Euclidean points, so
$K(x) \sim exp(-mx)$ at large x, where m is
 the  mass of the lightest particle with the corresponding quantum numbers.
Note, that this is essentially
 the idea of Yukawa, who had related the
range of the nuclear forces to the (then hypothetical) meson mass.


   After we have recollected these general facts, let us try to explain
 {\it why the correlation functions are so important in
non-perturbative QCD and hadronic physics.}    
The answer is; it is the most effective way
to study the {\it inter-quark effective interaction}.
    Models of hadronic structure -- various bags, skyrmions etc --
   resemble the state of the
nuclear physics in its early days, when only limited information about the 
nuclear forces 
were known from properties
of the simplest the bound states (e.g. the deuteron). 
Those are sufficient 
to reproduce
qualitative features of the interaction. But  it is the
extensive studies of the NN scattering phases in all relevant channels
 at all relevant energies which
 had eventually revealed the details of nuclear forces, with
their complicated spin-isospin structure.

Quite similarly, applications of quark models are mostly
are averaged over few lowest states, and the precise dependence
of the inter-quark
interaction on distance and momenta remains unknown. A confining
potential
with few additions (like spin forces) fit the spectrum. 
Since,  due to confinement, the  $qq$ or 
the $\bar q q$ scattering is  experimentally impossible,
 a set of  correlation functions $K(x)$ per channel substitute for the
 the phase shifts in nuclear physics.
 Roughly speaking, the correlator tells us about
 {\it virtual $\bar q q$ or $ qq$ scattering}, using
 instead of physical hadrons the {\it wave packets
of a variable size}.

%
%
%

\subsection{Dispersion relations and sum rules}

If one makes a Fourier transform of $K(x)$, the resulting function
$K_{mom}(q^2)$
 depends
on the momentum transfer q flowing from one operator to another. For
clarity we use the following notations, introducing momentum squared with
negative sign
 $Q^2=-q^2$, so  for {\it virtual}
 space-like momenta $q^2<0$ we are interested in
(like in scattering experiments) 
 $Q^2>0$.

 Due to causality, it
 satisfies standard dispersion relations
\be \label{eq_disp_rel}
K_{mom}(q^2)= \int {ds\over \pi} {ImK_{mom}(s) \over (s-q^2)} \ee
where the r.h.s. contains the so called {\it physical spectral density}
$ImK_{mom}(s)$. It  describes 
the squared matrix elements of the operator in question
 between the vacuum and
 all  hadronic states with the invariant mass $ s^{1/2} $, and certainly
is non-zero only for  {\it positive} $s$. 
Note, that because
we will only consider {\it negative} $q^2$, or the semi-plane 
without singularities, we never come across a
 vanishing denominator and therefore ignore  $i\epsilon$
 which is usually put in denominator. This simplifications are
possible because 
 our discussion is restricted to 
{\it virtual} processes (although in the r.h.s. we do use
information coming from the {\it real} experiments of annihilation
type)\footnote{In principle, virtual processes contain all the information, but
of course in practice it is much more
 difficult to go in the opposite direction, and
reproduce the physical spectral density from the correlators considered.}.

   The dispersion relation can be a  basis of the so called {\it
sum rules}. Their general idea is
as follows; suppose one knows 
the l.h.s. $K_{mom}(q^2)$ in some region of the argument; it means that
  some integral in the r.h.s. of 
the physical spectral density is known. It can be used  to relate a set of
physical parameters.
Unfortunately, the so called {\it finite energy sum rules}
using directly momentum space
are not very productive;  most of the dispersion 
integrals are divergent,  leading to usable 
  sum rules only after some subtractions, 
which introduce extra parameters and  undermine their prediction 
power.

For example,
 we have mentioned above that at small x
mesonic
correlators  are just given by $K_{free}\sim 1/x^6$, 
the free quark propagator squared.
Its  Fourier transform
is $K_{mom}(q^2)\sim q^2 \log q^2$
and the imaginary part of the log gives for  the 
spectral density  in the r.h.s.
 $Im K_{mom}(s) \sim s$.
Therefore, putting it into
 dispersion relation given above one finds an
ultraviolet divergent
integrals\footnote{This  signal that in the last  argument
something is missing. In this particular example it is clear
what is it; the constant under the log is lost.}.
A simple way to go around this  is to 
consider the second derivative
  over $Q^2$ of both sides of (\ref{eq_disp_rel}); then one finds
 a  convergent dispersion relation. However, while going back to
the original function  $K_{mom}(s)$ from its derivative,
 one has to fix the
integration constant\footnote{Note that polynomials in $s$ generate
local terms in $K(x)$
we ignore, assuming that $x$ is never zero.}.

%
%
 We will use the coordinate space. By applying Fourier transformation to (\ref{eq_disp_rel})
one obtains the following 
nearly self-explanatory form \cite{Shuryak:1993kg};
\be K(x) =  \int  {ds\over \pi} Im K_{mom}(s) D(s^{1/2},x) \ee 
the  former function describe the amplitude of production
of all intermediate states of mass $\sqrt{s}$, while the latter function
\be D(m,x)=(m/4 \pi^2 x) K_1(mx) \ee 
is nothing else  but the Euclidean propagation amplitude of this states 
 to the distance 
x. 
  The difference between this expression and Borel sum rules is not
really  very
significant; at large x the propagator goes as $exp(-mx)$, therefore one
also has an exponential cutoff, only $exp(-\sqrt{s}x)$
substitutes the $exp(-s/m^2)$. The space-time one can be
calculated numerically on the lattice or in instanton liquid, and
also analytic formulae are simpler to derive.

For completeness, let me also mention one more type of the correlation 
function,
the one
traditionally used in LGT. This is the so called 
{\it plane-to-plane} correlation function,  obtained from K(x) by an integration
over the 3-dimensional plane;
\be   K_{plane-to-plane}(\tau) = < \int d^3x O(x,\tau) O(0,0) > \ee 
In other terms, spatial integration
makes the  momentum of intermediate   states
to be zero, so dispersion relation are done in energy only. This function,
respectively, can be
related to physical spectral density by
\be   K_{plane-to-plane}(\tau) =  \int {dm\over \pi} Im K_{mom}(m) exp(-\tau m) \ee 
%
%
%
%
%
%
%

\begin{figure}[t]
\centering
\includegraphics[width=8cm]{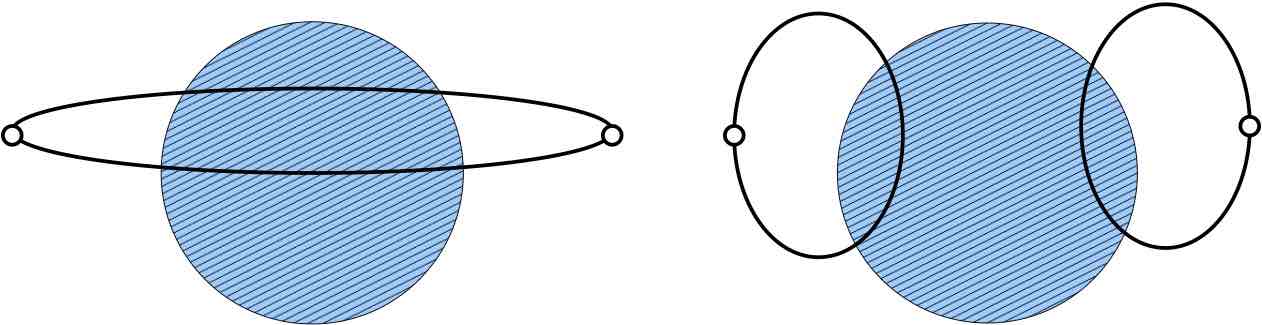}
\caption{\label{fig_one_two_loops}
Two diagrams for mesonic correlators in a some gluonic background
(indicated by
the dashed circle). The flavored currents have only the single loop 
diagram (a) 
 while the unflavored currents have also
the two-loop contributions of the type (b).
}
\end{figure}

\subsection{Flavor and chirality flow: combinations of correlators} 

Let us now follow the quark flavor flow in the correlators.
 The $flavored$ and $unflavored$ currents have different
 types of quark diagrams; as shown in Fig.\ref{fig_one_two_loops}
the former ones have only
the one-loop contributions (a), while the latter have the two-loop
diagrams (b) as well.
For example, we will discuss
the flavored (I=1) channels like $\pi^+,\rho^+$ etc which have 
 operators of the type
$\bar u \Gamma d$ (where $\Gamma$ is the appropriate
 Dirac matrix). Since two quark lines are in this case
of different flavor, $\bar u$ and $d$ respectively,
obviously one cannot have a double-loop. 
On the other hand, considering similar unflavored I=0 channels
like $\eta,\omega...$ one has the $\bar u u,\bar d d,\bar s s$
terms which can be ``looped''. Thus, if one would be interested in
a $difference$ between say $\rho$ and $\omega$ channels, that would be
entirely due to the two-loop diagram (b).
Furthermore, one can be interested in the so called $non-diagonal$ correlators,
for example with a different flavors such as
 $<\bar u(x)\Gamma u(x)\bar d(0)\Gamma d(0) >$. Again, 
one is restricted to the two-loop diagrams only.

  The main focus of lattice work deals with the 
one-loop diagrams and therefore
with the flavored channels; the reason is
technical to which we would not go into.

It is often instructive to use specific combination
of correlation functions, which focus on the
 phenomena we would like to study. In particular, one would like to
understand how breaking of the $SU(N_f)$ and $U(1)_A$ manifest
themselves in the correlation functions. 
%
Let me give two examples of such choices. The first example is 
the following linear combinations
 $\Pi_{V-A}=\Pi_V-\Pi_A$ of the I=1 vector and axial
correlation functions. 
All effects of the interaction in which quark chirality
is preserved throughout the loop cancels out, because
two $\gamma_5$ produced $(\pm)^2=1$. 
  Furthermore, $\Pi_{V-A}$ is
non-zero
only due to the
effects of chiral symmetry breaking. This can be most clearly
expressed
if one writes the two currents in terms
of left and right handed currents, as $\bar q_L\gamma_\mu q_L \pm 
\bar q_R\gamma_\mu q_R$; then this combination
 includes the chirality flip $twice$
\be \Pi^{V-A}_{\mu\nu}=4<\left(\bar q_L\gamma_\mu(x) q_L \right)\left(\bar q_R\gamma_\mu q_R(y)\right)> \ee
Therefore in the chiral limit $m_f\rightarrow 0$ the charged component of it  (one loop diagram) is obviously zero
to any order of the perturbation theory.

The second example is 
a similar difference, but between the I=1 scalar (called $a_0$ or
$\delta$) 
and the I=1 pseudoscalar (the pion $\pi$).
\be
\label{RNS}
R^{NS}(\tau);=\frac{A^{NS}_{flip}(\tau)}{A^{NS}_{non-flip}(\tau)}=
\frac{\Pi_\pi(\tau)-\Pi_\delta(\tau)}
{\Pi_\pi(\tau)+\Pi_\delta(\tau)},
\ee
where $\Pi_\pi(\tau)$ and $\Pi_\delta(\tau)$ are pseudo-scalar and scalar NS
two-point correlators related to the currents $J_\pi(\tau);=\bar{u}(\tau)
\,i\gamma_5\,d(\tau)$ 
and $J_\delta(\tau);=\bar{u}(\tau)\,d(\tau)$.
If the propagation is chosen along the (Euclidean) 
time direction, $A^{NS}_{flip(non-flip)}(\tau)$ represents 
the probability amplitude for a 
$|q\,\bar{q}> $ pair with iso-spin 1 to be found after a 
time interval $\tau$ in a state
in which the chirality of the quark and anti-quark 
is interchanged (not interchanged)
Notice that the ratio $R^{NS}(\tau)$ 
must vanish as $\tau\to 0$ (no chirality flips),
 and must approach 1 as  
$\tau\to \infty$ (infinitely many chirality flips).
Again, this amplitude receives no leading perturbative contributions.
 The difference
with the example 1 is that 
this ratio is proportional to the chirality structure $\bar R L \bar R
L+(R\leftrightarrow L)$, same as for the
't Hooft vertex, while
in the example 1 it was  $\bar L L\bar R R$.

\begin{figure}
\includegraphics[width=5.9cm]{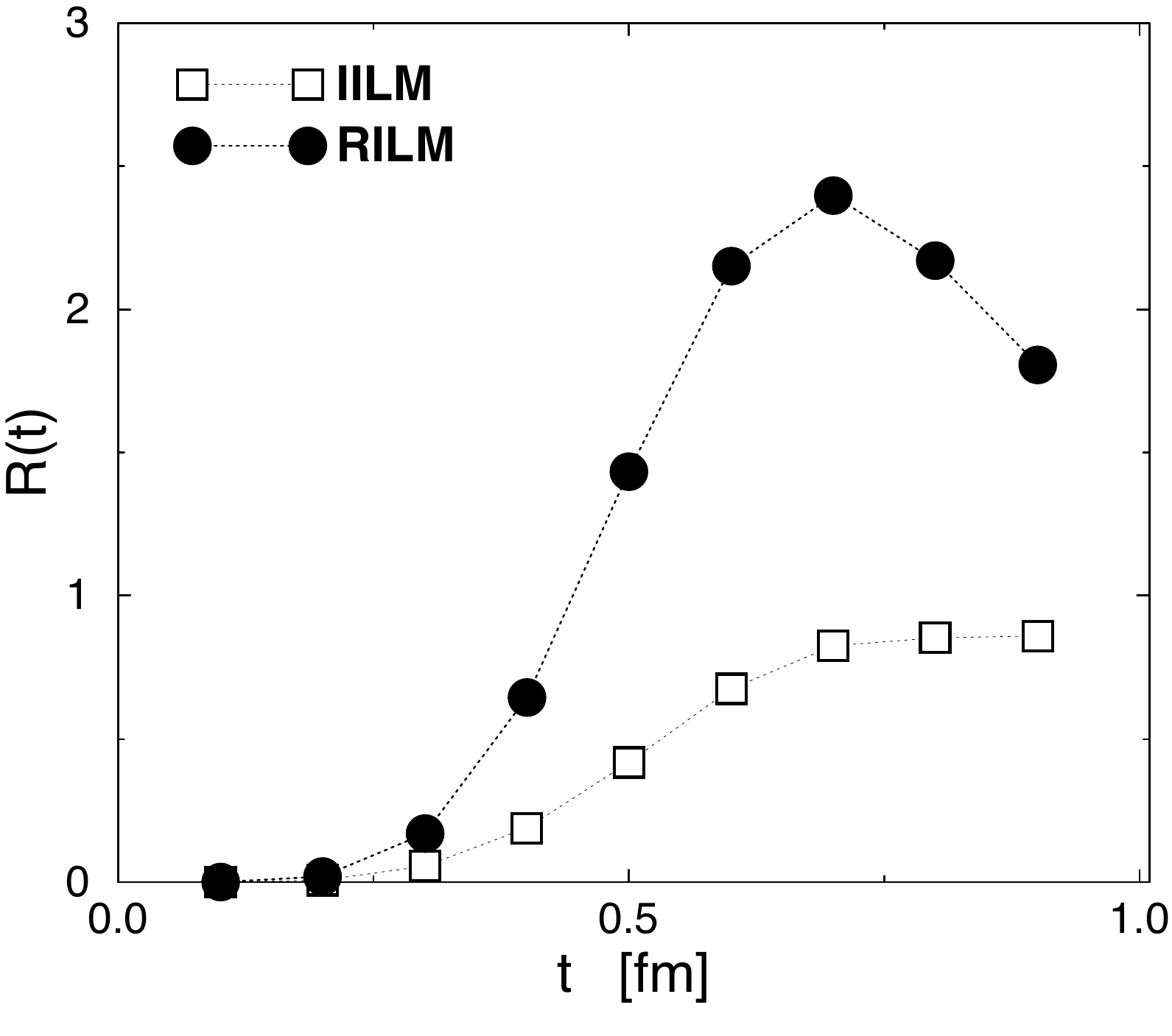}
\includegraphics[width=5.9cm]{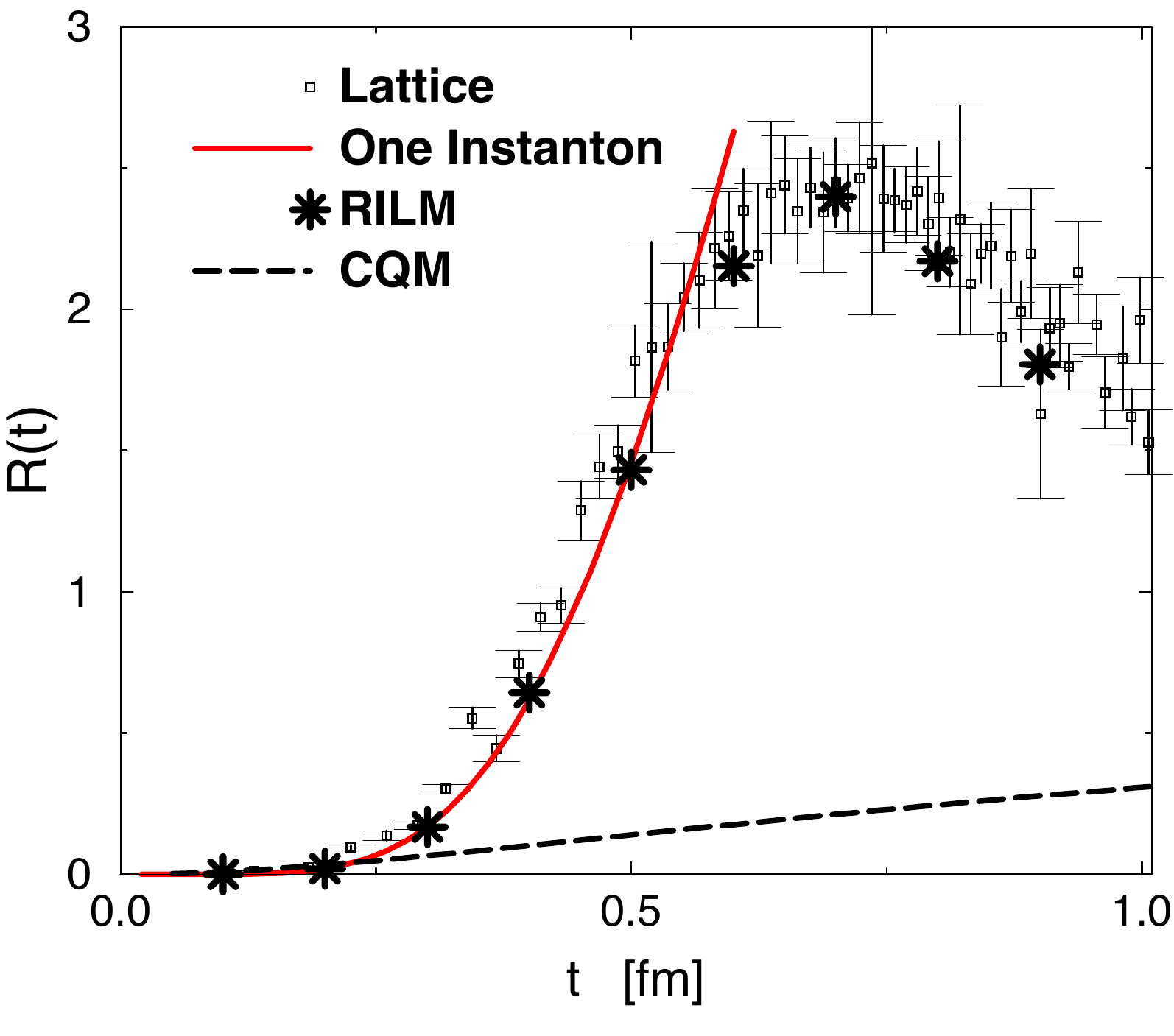}
\caption{\label{fig_pietro_ratio}
The chirality-flip ratio, $R^{NS}(\tau)$, in lattice and
in two phenomenological models.(a)  Circles are RILM (quenched) results, squares are IILM 
(unquenched) results. (b)
Squares are lattice points of previous DeGrand simulation. 
Stars are RILM points
obtained numerically from an ensemble of 100 instantons of 1/3 fm size in a 
$5.3\,\times\,2.65^3\,\textrm{fm}^4$ box.
The solid line is the contribution of a single-instanton, 
calculated analytically.
The dashed curve was obtained from  two free ``constituent''
quarks with a mass of
400 MeV. Such a curve  qualitatively resembles the
prediction of a model in which chiral symmetry is broken through a vector
coupling (like in present DSE approaches).}
\end{figure}

Unlike the former combination, 
the latter one
 is growing surprisingly rapidly already at small
distances,
$\sim .3-.6 \, fm$, see Fig. \ref{fig_pietro_ratio}(a) from  the lattice study
 \cite{Faccioli:2003qz}.
The reason  is that this correlator
 admits
a direct instanton-induced 't Hooft vertex, while the V-A does not.

Two instanton liquid ensembles used are: the random one RILM which has no
fermionic determinant, and the interacting IILM, in which it is included.
Note that in RILM the chirality flip ratio rises rapidly and exceed 1, while
the unquenched IILM follows the unitarity requirement,
$R^{NS}(\tau)\le 1$. The result of quenched
lattice
calculation shown in Fig.(b) follow very accurately the
same ``overshooting'' of 1 as the RILM results.
(The unquenched lattice ones are not yet available.)
At the other hand, the naive constituent quark models, which only include
chirality flips due to constituent quark mass, show very slow rise
shown by the lowest curve in the Fig.(b).

\subsection{General inequalities between the one-quark-loop correlators}

For correlators given by  one-quark-loop diagrams
there exist some important general inequalities between them.
In order to derive those
  following  \cite{Weingarten:1983uj},one  uses the following
relation for the propagator in backward direction    
\be S(x,y)=-\gamma_5 S^+(y,x)\gamma_5 \ee  
Second, one can decompose it into a sum over all 16 Dirac matrices
$ S=\Sigma a_i \Gamma_i$ where 
$\Gamma_i=1, \gamma_5, \gamma_\mu, i\gamma_5\gamma_\mu, i\gamma_\mu\gamma_\nu$
where in the last term is anti-symmetric only and $(\mu 
\not= \nu)$. The third step; write all one-loop correlators
of the type $\Pi=Tr(S(x,y) \Gamma_i S(y,x)\Gamma_i)$, perform the
traces, 
and write them explicitly in terms of the coefficients.\\\\


{\bf Exercise}: 
\label{ex_Weingarten} {\em Prove the Weingarten expression for the inverse
propagator.
Derive the expressions for all the diagonal
 correlators with $\Gamma_i=1, \gamma_5, \gamma_\mu,
 i\gamma_5\gamma_\mu, i\gamma_\mu\gamma_\nu$ in terms of the
propagator decomposition of the same type.  
Use Weingarten expression
 for the inverse propagator.}

  The resulting expression for the I=1 pseudoscalar (pion)
 correlator contains a simple sum of all coefficients squared;
\be \Pi_{PS} \sim |a_1|^2 + |a_5|^2 +|a_\mu|^2+|a_{\mu 5}|^2
+|a_{\mu\nu}|^2) \ee
while others have some negative signs, 
e.g. the scalar one is instead
\be \Pi_{S} \sim -|a_1|^2 - |a_5|^2 +|a_\mu|^2+|a_{\mu 5}|^2
-|a_{\mu\nu}|^2 \ee
As a result, the {\em Weingarten inequality}
\index{Weingarten inequality}
 follows;  
\be \Pi_{PS}(x)>\Pi_{S}(x) \ee
the 
pseudoscalar correlator should  exceed the scalar one at all distances.
  The non-trivial consequence is that the masses of the lowest
states should also then have the inequality, $m_{PS}<m_{S}$. 
This of course is satisfied
in real world, as the physical pion is indeed  lighter than any
scalars. Note however that we did not said a word about chiral
symmetry breaking and Goldstone theorem here; the result is more general.
Note also that
 at large $x$ the scalar correlator must be much much smaller than the
pseudoscalar one, since the different lowest masses are in the exponent. 
 It means that there must be a very delicate cancellation
between different components of the quark propagator in all channels
except the pseudoscalar one (in which all terms appear as squares with
positive coefficients).

More information  is provided by similar
 relations for vector ($\rho$) and axial ($A_1$) channels;
\be \Pi_{V} \sim (2|a_1|^2 - 2|a_5|^2 +|a_\mu|^2-
|a_{\mu 5}|^2) \ee
\be \Pi_{A} \sim (-2|a_1|^2 +2 |a_5|^2 +|a_\mu|^2
-|a_{\mu 5}|^2) \ee
and the Verbaarschot inequalities follow;
\be \Pi_{PS}/\Pi_{PS}^{free} > (1/2)(\Pi_{V}/\Pi_{V}^{free}+
\Pi_{A}/\Pi_{A}^{free}) \ee
  
\be \Pi_{PS}/\Pi_{PS}^{free} > (1/4)(\Pi_{V}/\Pi_{V}^{free}-
\Pi_{A}/\Pi_{A}^{free}) \ee

More information about other general
inequalities can be found in the review  \cite{Nussinov:1999sx}. 
Note that these inequalities are identities, to be  satisfied for {\it any
configuration} of the gauge fields, not just for their ensembles.
%
%

\section{ Vector and axial correlators}
 \label{sec_cor_vect}  
We start the discussion of the correlation 
functions with the vector and axial currents. 
These currents  really exist in nature,
  as  the
 {\em electromagnetic} ones coupled to photons
and  (the parts of) the {\em weak} current coupled to $W,Z$ bosons .
 Therefore, we know a lot about such correlators. In fact,
quite
complete spectral  densities
have been  experimentally measured, in
$ e^+e^-$ annihilation into hadrons and in weak decays of heavy
lepton $\tau$.
The  currents and their correlation functions
 will be denoted by the name of the lightest
meson in the corresponding channel, in particular
 \be j^{\rho0}_\mu = {1\over \sqrt{2}} [ \bar u \gamma_\mu u - \bar d
 \gamma_\mu d]
 \hspace{1cm} 
 j^{\rho-}_\mu =\bar u \gamma_\mu d \ee 
 \be j^{\omega}_\mu = {1\over \sqrt{2}} [ \bar u \gamma_\mu u +
 \bar d \gamma_\mu d] \hspace{1cm} 
  j^{\phi}_\mu =  \bar s \gamma_\mu s  \ee 
(see more definitions in the Table \ref{tab_cur_def}).

\begin{table}[t]
\begin{tabular}{llll}
channel         &    current         
                & matrix element     
                & experimental value    \\  \hline
$\pi$           &  $j^a_\pi=\bar q\gamma_5\tau^a q$           
                &  $\langle 0|j_\pi^a|\pi^b\rangle=\delta^{ab}\lambda_\pi$             
                &  $\lambda_\pi\simeq (480\,{\rm MeV})^3$     \\
                &  $j^a_{\mu\,5}=\bar q\gamma_\mu\gamma_5\frac{\tau^a}{2}q$
                &  $\langle 0|j^a_{\mu\,5}|\pi^b\rangle
                                    =\delta^{ab}q_\mu f_\pi$
                &  $f_\pi=93$ MeV \\
$\delta$        &  $j^a_\delta=\bar q\tau^a q$           
                &  $\langle 0|j_\delta^a|\delta^b\rangle
                                    =\delta^{ab}\lambda_\delta$         
                &       \\
$\sigma$        &  $j_\sigma=\bar q q$           
                &  $\langle 0|j_\sigma|\sigma\rangle=\lambda_\sigma$                   
                &       \\
$\eta_{ns}$     &  $j_{\eta_{ns}}=\bar q\gamma_5  q$           
                &  $\langle 0|j_{\eta_{ns}}|\eta_{ns}\rangle
                                    =\lambda_{\eta_{ns}}$           
                &       \\
$\rho$          &  $j^a_{\mu}=\bar q\gamma_\mu\frac{\tau^a}{2}q$
                &  $\langle 0|j^a_{\mu}|\rho^b\rangle
                      =\delta^{ab}\epsilon_\mu\frac{m_\rho^2}{g_\rho}$ 
                &  $g_\rho=5.3$  \\
$a_1$           &  $j^a_{\mu\,5}=\bar q\gamma_\mu\gamma_5\frac{\tau^a}{2}q$
                &  $\langle 0|j^a_{\mu\,5}|a_1^b\rangle
                      =\delta^{ab}\epsilon_\mu\frac{m_{a_1}^2}{g_{a_1}}$
                &  $g_{a_1}=9.1$  \\
$N$             &  $\eta_1 = \epsilon^{abc}(u^aC\gamma_\mu u^b)\gamma_5
                   \gamma_\mu d^c$
                &  $\langle 0|\eta_1 |N(p,s)\rangle =\lambda_1^N u(p,s)$ 
                &             \\
$N$             &  $\eta_2 = \epsilon^{abc}(u^aC\sigma_{\mu\nu} u^b)
                   \gamma_5\sigma_{\mu\nu} d^c$
                &  $\langle 0|\eta_2 |N(p,s)\rangle =\lambda_2^N u(p,s)$ 
                &             \\
$\Delta$        &  $\eta_\mu = \epsilon^{abc}(u^aC\gamma_\mu u^b) u^c$
                &  $\langle 0|\eta_\mu |N(p,s)\rangle 
                           =\lambda^\Delta u_\mu(p,s)$ 
                &             \\
\end{tabular}
\caption{Definition of various currents and hadronic matrix elements  
referred to in this work.}
\label{tab_cur_def}
\end{table}

Let us remind that the  electromagnetic current is  the following
combination 
 \be \label{eqn_em_current}
j^{em}_\mu= (2/3)\bar u \gamma_\mu u-(1/3)\bar d \gamma_\mu d  +... =
(1/2^{1/2})j^{\rho}_\mu-(1/2^{1/2}3)j^{\omega}_\mu +..   \ee 
The correlation functions are defined as
 \be \Pi_{\mu \nu}(x)= \langle 0 | j_{\mu}(x)j_{\nu}(0) | 0 \rangle \ee 
 and the Fourier transform (in Minkowski space-time) is
 traditionally written as
\be i \int d^4x e^{iqx} \Pi_{\mu \nu}(x)=
\Pi(q^2)(q_\mu q_\nu-q^2 g_{\mu \nu})  \ee 
The r.h.s. is explicitly ``transverse"  (it vanishes if multiplied by
momentum q), because all vector currents are conserved.

 The dispersion relations
for the scalar functions $\Pi(q^2)$ has the usual form
 \be \Pi(Q^2 = -q^2) =  \int{ds\over \pi} {Im \Pi(s) \over (s+Q^2)}  \ee 
where the physical spectral density
$Im \Pi_i(s)$ is  directly related with the cross section
of $e^+e^-$ annihilation into hadrons. 
As this quantity is dimensionless, it is proportional to the 
normalized cross section
\be R_i(s) = {\sigma_{e^+e^- \rightarrow i}(s)\over \sigma_{e^+e^- \rightarrow \mu^+
\mu^-}(s) }  \ee
where the denominator
includes the cross section of 
the muon pair production\footnote{The  muon mass is neglected in this expression.}
 $\sigma_{e^+e^- \rightarrow \mu^+
\mu^-} = (4\pi\alpha^2/3 s)$ and $\alpha$ is the fine structure constant. 
If the current considered
has only one type of the quarks (like e.g. $\phi$ one) one gets
 \be Im \Pi_s(s)= {R_s(s) \over 12\pi e^2_s} = {R_s(s) \over 12\pi (1/9)}\ee 
where
$e_s$ is the s-quark electric charge. Generalization to
$\rho,\omega$ channels is straightforward; instead of the charge there appear
the corresponding
coefficients  in the expression for the
            electromagnetic current (\ref{eqn_em_current});
 \be Im \Pi_\rho(s) = {1 \over 6\pi}R_\rho(s)  \hspace{1.cm} Im \Pi_\omega(s) = {3 \over 2\pi}R_\omega(s)\ee 
   (The reader may wonder how 
 the
   experimental selection of the channels is actually made. It is clear enough
for heavy flavors (c and b); if the final state has a pair of such quarks, 
there are much more chances that they were directly produced in the 
electromagnetic current than that these are produced by strong ``final state
interaction".
Below we use this idea for the strange quark as well, although  some
corrections should, in principle, be applied in this case.
It is also possible to separate
  light quark $\rho,\omega$ channels; they have 
a different isospin
 I=1,0, which is conserved by any strong final state interaction.
As it is well known, C-parity plus isotopic invariance leads to the so called
G-parity conservation, and pions have {\it negative G-parity}.
 Therefore, strong interactions do not mix
 states with even and odd number of pions. The currents $j_\rho,j_\omega$
have fixed G-parity as well, and therefore pionic states created by 
them can have  
only  even or  odd number of pions, respectively.)

  Let us start the simple and well known predictions of QCD; 
all the ratios $R_i(s)$  have  very simple limit at high
energies s. It is conjugate to the small-distance
limit $x\rightarrow 0$ in which $\Pi\rightarrow \Pi_{free}$ because 
 quarks and anti-quarks
propagate there as free particles. 
 For currents containing only one quark flavor q  the only
difference with muon is a different electric  charge and a color factor; 
 \be lim_{s \rightarrow \infty} R_q(s)= e^2_q N_c \ee
which for $\phi$ case give $lim_{s \rightarrow \infty} R_\phi
(s)=1/3$.
For $\rho$ and $\omega$ cases one may use the following decomposition of the
electromagnetic
current;
 \be lim_{s \rightarrow \infty} R_\rho(s)= 3/2; \hspace{1cm}
lim_{s \rightarrow \infty} R_\omega(s)= 1/6 \ee 
As we will show shortly, these relations are well satisfied
experimentally (being historically one of the first and simplest
justification for QCD).

   Coming back to coordinate representation of the dispersion relation one
obtains; 
\be \Pi_{i,\mu\nu}(x)=(\partial^2 g_{\mu\nu}-\partial_\mu\partial_\nu)
{1 \over 12\pi^2} \int^\infty_0 ds R_i(s)D(s^{1/2},x)  \ee 
were, we remind, D(m,x) is just the propagator of a scalar mass-m particle 
 to distance x.
Convoluting indices and using the equation $-\partial^2 D(m,x)=m^2 D(m,x)+
$ contact term (which we disregard), one finally obtains 
the following for   the dispersion relation  
\be \label{eq_vect_cor_disp}
\Pi_{i,\mu\mu}(x)=
{1\over 4\pi^2} \int^\infty_0 ds s R_i(s)D(s^{1/2},x)\ee
Since the r.h.s. is experimentally
available, this equation
would serve as our  ``experimental definition" of
the l.h.s., the vector correlation functions in Euclidean space-time.
 
    As we are interested in quark $interactions$,
  it  is    convenient  to  plot  all correlators 
normalized  to free motion of a massless quarks
$ \Pi_{\mu\mu}(x)/\Pi_{\mu\mu}^{free}  (x)$ 
 where $\Pi_{\mu\mu}^{free}(x)$
   corresponds to perturbative loop diagram without interaction.

\begin{figure}[h]
 \centering 
\includegraphics[width=14.cm]{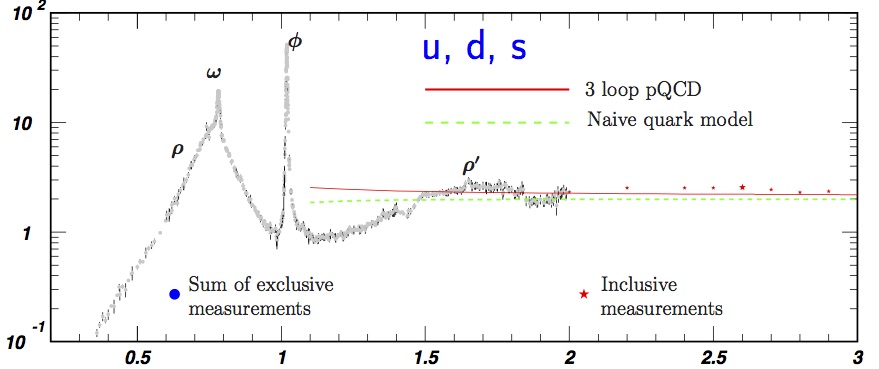}
\caption{ \label{fig_rho_data}
The ratio of $R=\sigma(e^+e^- \rightarrow hadrons)/
\sigma(e^+e^- \rightarrow \mu^+ \mu^-)$ versus the  total invariant mass
of the hadronic system $\sqrt{s}$ in GeV. 
}
\end{figure}


    Fig.\ref{fig_rho_data} from Particle Data Group shows a sample of
experimental data on $e^+e^-$ annihilation into hadrons. One can see that this function consists of two 
 quite different parts ; (i) at $\sqrt{s}<1.1\,  GeV$ the
 prominent $\rho,\omega,\phi$-meson resonances (which all have very distinguishable
 decay channels, $2\pi, 3\pi$, and $\bar K K$, respectively, measured but not shown in the plot);
 and (ii)  ``primed" resonances (of which only
  $\rho' $ is indicated on the plot, decaying
  mainly into   the 4 pion channel,etc.
  
  The distinction between the lowest resonances and the primed one can be seen from a
  striking fact, that the contribution of the latter's  are in good agreement
  with the horizontal curves   (with and without perturbative corrections).
Adding  the next ones ($\rho''$ seen in the 6 pion channel) etc. creates a
 rather smooth
 non-resonance ``continuum", corresponding to perturbative quark propagation.
 As seen form this plot, this happens 
   at energies $\sqrt{s}> 1.5 \, GeV$.  It is not shown on this plot, but still true,
   that this in fact happen not just in the sum, but in each  $\rho,\omega,\phi$ channel individually.

\begin{figure}[h]
 \centering 
\includegraphics[width=9.cm]{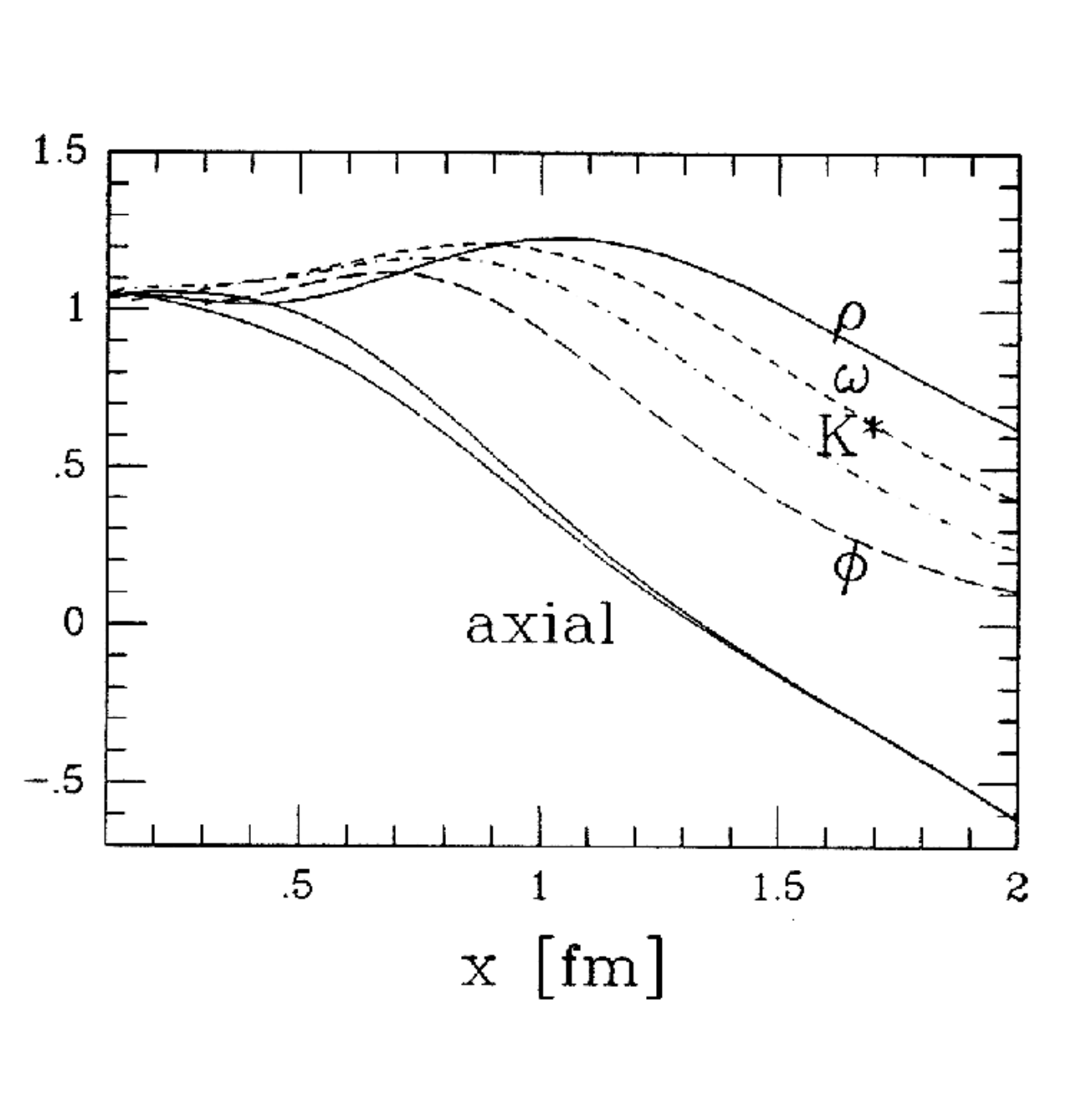}
\caption{\label{fig_vectors_cor}
All vector correlators together, plus the $a_1$ axial
correlator for comparison.
}
\end{figure}

     Using parametrizations of the data in each channel,  and the dispersion relation 
(\ref{eq_vect_cor_disp}), 
 one can calculates the Euclidean correlation function.
The resulting curve is shown in  Fig.\ref{fig_vectors_cor}.
Note that, starting with rather complicated 
spectral densities, containing high peaks and low dips,
 one arrives at a very smooth function of the 
distance. Clearly, the way back, from the Euclidean 
space to the
physical spectral density would be next to impossible task! 
   
  Quite striking observation 
 made in \cite{Shuryak:1993kg} (which is specific to vector currents only, as
  we will see later in this section) is that
 the resonance and continuum contributions
complement each other very accurately.
As a result the ratio $\Pi(x)/\Pi_{free}(x)$
 remains close to one {\it up to the distances as large as 1.5
 fm!}, while each functions falls by orders of magnitude.
This
``fine tuning"  was called in \cite{Shuryak:1993kg} a {\it 
superduality}; let me explain why this is indeed a remarkable
(and so far unexplained) fact.
 
At small distances it is very natural to
expect the so called ``hadron-parton duality" between the sum over hadronic 
states and the pQCD quark-based description; basically
it is a simple consequence of the ``asymptotic
freedom" up to about
$x < 1/\, GeV=.2 fm$. In the interval $x= .2 - 1.5\, fm$ 
 the correlator itself drops by more 
than 4 orders of magnitude, while its ratio to the free loop
(free quark propagation) remains close to 1 within  10-15 percents!
What this remarkable phenomenon means is that in this channel 
all kind of interactions -- perturbative, instanton and
confinement-related
ones -- cancel out in wide range of distances. 

Omitting the details of how other correlators has been extracted from phenomenology, let us 
show in Fig.\ref{fig_vectors_cor}
how all 4 vector correlators and also the axial one look like, as a function of Euclidean distance
and normalized at small distances as explained above.
  Comparing them  one can see, that
in spite of completely different widths of the resonances and different decay states
(even or odd pion numbers),
all four vector correlators look {\it remarkably similar}.  The main difference between them
is the strange quark mass, which systematically suppress the correlators at larger distance.
This demonstrates deep consistency between 4 independent sets of data
which is rather impressive.

To complete our discussion of the vector channels, let us comment on the 
the difference between the  $\rho^0$ and $\omega$ correlators: what induces it?
Smallness is in particular due to the  following facts: 
(i)The rho-omega mass difference is only 12 \, MeV; 
(ii) The omega-phi mixing angle is only 1-3 degrees.
Those were the basis of the so called ``Zweig rule",
forbidding the
flavor-changing transitions.

Let us look at the transition correlator itself.
The former current  $\rho\sim \bar u u - \bar d d$  while  the latter is
$\omega \sim \bar u u + \bar d d$. Thus the difference between them 
is the  {\it flavor-changing} correlator
\be K^{\omega-\rho}(x) 
=2(\Pi_{\omega,\mu\mu}-\Pi_{\rho,\mu\mu}) = < \bar u \gamma_\mu u (x)\bar d \gamma_\mu d (0)> \ee 
which can only come from the two-loop diagram. 
%
%
The reason for such strong suppression of 
 is 
 that in  vector  channels 
there are no direct instanton contribution in the first order in 't Hooft
interaction, and the effects of the second order tend to cancel.

\section{The pseudoscalar correlators }
\label{sec_ps_cor_phen}
  Now we move to the pseudoscalar  $SU(3)$ octet $\pi,K,\eta$ channels, 
  and the   $SU(3)$ singlet $\eta'$   where we see a completely different picture.
  Their definitions are 
 \be j_\pi = {i\over \sqrt{2}}(\bar u \gamma^5 u - \bar d \gamma^5 d)\hspace{1cm}
 j_K=i\bar u \gamma^5 s \hspace{1cm} \\ \nonumber
 j_\eta=  {i\over \sqrt{6}}(\bar u \gamma^5 u + \bar d \gamma^5 d
-2\bar s \gamma^5 s) \,\,\,\,\,\
 j_{\eta'}=  {i\over \sqrt{3}}(\bar u \gamma^5 u + \bar d \gamma^5 d
+\bar s \gamma^5 s)
\ee

%

\begin{figure}[h]
 \centering 
\includegraphics[width=8cm]{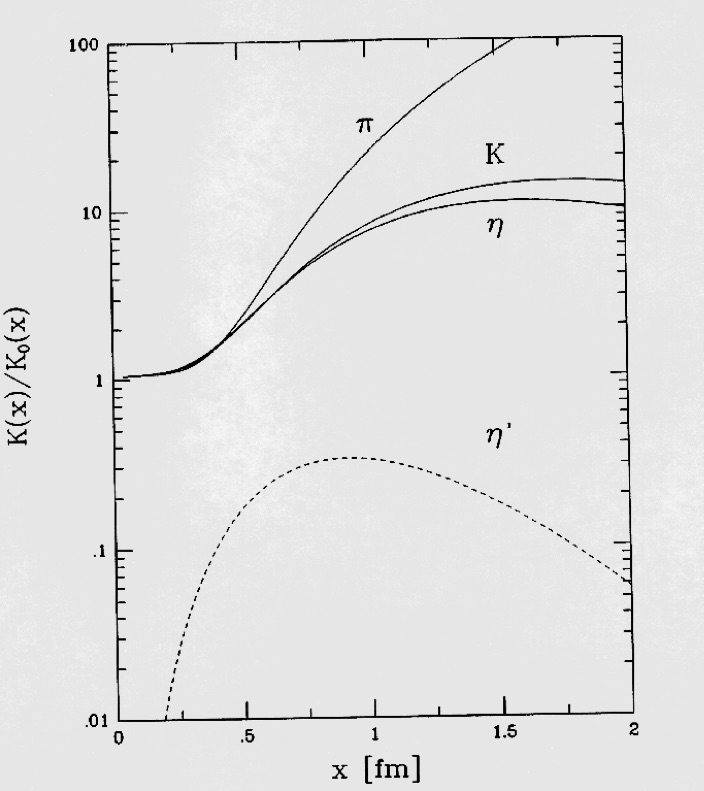}
\caption{\label{fig_ps_cor}
(a) The correlation functions for the pseudoscalar nonet. Note
that in contrast to the preceding figures it is now shown with the
logarithmic scale.
}
\end{figure}

   Again, omitting the details of the phenomenological inputs, the resulting $\pi,K,\eta,\eta'$
pseudoscalar correlators are shown in Fig.\ref{fig_ps_cor}. 
 Note huge difference
compared to the vector correlators considered above; instead of changes within
10-20 $\%$
in the interval of distances x $\sim$ 1 fm considered,
 for pseudoscalars the ratio $K/K_{free}$
has changed by up to two orders of magnitude!

  One reason for that behavior is
 of course small masses  of the pseudoscalar mesons.
In terms of the $\bar q q$ interaction, this
 implies a very strong attraction.
Note also, that up to distances of the order of .5 fm there is {\it no
marked difference between  the three curves}, which means that
all effects proportional
to the strange quark mass are  irrelevant in this
region. 

What is really
  surprising, is that  the {\it asymptotic freedom is violated at very small 
distances}, about 1/5 fm, and that this happens
  due to the contributions of the
lowest mesons themselves. This fact, noticed first by Novikov et al 
\cite{Novikov:1981xi},
shows  that the pseudoscalar channels differ substantially
from vector and axial ones, already at very
 small distances.


    The SU(3) singlet  meson  $\eta$' 
can be associated with tho important operators. One is the singlet axial
 current mentioned above
and another is the gluonic pseudoscalar $G \tilde{ G}$.

The relation between matrix elements of these 3 operators
is given by sandwiching (between vacuum and the $\eta'$ state) of
the famous  Adler-Bell-Jackiw anomaly relation we discussed in chapter 1;
\be \partial_\mu j^\mu_{\eta'} = \sqrt{3}(2im_s \bar s \gamma_5 s +
{3g^2 \over 16 \pi^2}G \tilde{ G}) \ee
where the contributions proportional to the light quark masses are  ignored.

Omitting details, we comment that the latter matrix elements can be extracted from charmonium decay,
with the result\footnote{One may wander why they grow, rather decrease, with the mass: the reason is
the pseudoscalar glueball state is expected to be in the range $M_{0^- glueball}=2-3\, GeV$,
and the closer $\eta's$ are to it, the larger is its admixture. }
\be \langle 0 | G \tilde{ G}|\eta>
\approx .9 \, GeV^3,\, \langle 0 | G \tilde{ G}|\eta' \rangle \approx 2.2 \, GeV^3,\\ \nonumber
\langle 0 | G \tilde{ G}|\eta(1440)> \approx 2.9 \, GeV^3 \ee

  To complete discussion of this section let us now do some
    preliminary 
estimates of the contribution of three $\eta$ states to 
  pseudoscalar  gluonic correlator. As in all other channels, at
small distances the correlator is dominated by
perturbative ``asymptotically free" gluonic contribution,
which is equal to
\be K(x)=\langle 0 | G \tilde{ G}(x) G\tilde{ G}(0) | 0 \rangle, \hspace{1cm} 
 K_{free}(x)= {48 (N_c^2-1) \over \pi^4 x^8} \ee
It is thus instructive to ask a simple question, at what $x$
the contribution of  $\eta,\eta',\eta(1440)$ together to
$K(x)$ become equal to $K_{free}$. The answer is, at $x\approx .2 \, fm$: a very short distance
 compared to  $\sim 1\, fm$ for all vector resonances.
And,  keep in mind, that we still have not seen
the contribution of
 ``the {\it true pseudoscalar glueball}" yet, expected to
dominate the $\eta$'s. 

The main conclusion one can draw from this discussion  is that  the boundary between pQCD and
non-perturbative physics are very much channel dependent. They go to larger momenta (smaller distances)
in pseudoscalar channels, and are even higher (smaller distances)
in the spin-zero gluonic channels. All of this point to the topological solitons and their zero modes,
the subject of the next subsection.

\section{The first order in the 't Hooft effective vertex}
\label{sec_tHooft_1storder}
We have seen in section \ref{sec_zero_modes} that 
 the zero modes of the Dirac operator in the instanton field
play a special role in the chiral limit $m_q\rightarrow 0$.
For flavored currents
 the  single-loop diagram in which
only the  zero mode parts of both propagators is used leads to
 expression
of the type
\be 
K(x-y)={n\over m_u m_d} \int d^4z \bar \psi_0 (x-z) \Gamma \psi_0 (x-z) \bar \psi_0 (y-z) \Gamma \psi_0 (y-z)
\ee
where for the time being $N_f=2$ and we ignore the strange quark entirely,
n is the instanton density and the matrix $\Gamma$ is the gamma matrix in
the vertex.

Now recall that for the instanton background field
the quark zero mode is right-handed, while the antiquark one is
left-handed
only. (It is flipped $L \leftrightarrow R$ for the anti-instanton.)
This leads to the following general conclusions;

$\bullet$ Such contribution in the vector or the axial channels, when
$\Gamma=\gamma_\mu,\gamma_\mu \gamma_5$, is $zero$
 since these currents are nonzero only
if
both spinors have the same chirality.

$\bullet$ For scalar and pseudoscalar\footnote{Note that we put $i$ 
into the current, in order to keep the zeroth order
free quark loop the same in both cases.} channels $\Gamma=1,i \gamma_5$,it is non-zero
and have the {\em opposite sign}. 

$\bullet$ For flavored current one can see that the pseudoscalar
 $\pi$ gets positive
relative correction while the scalar $\delta$ or $a_0$ gets a negative one.    

$\bullet$ Analogous calculation for flavor-singlet scalar $\sigma$ (or
$f_0$) and pseudoscalar we would still call $\eta'$ gets split as well,
with the former getting positive and the latter negative contribution.

$\bullet$ The absolute magnitude of these corrections for all 4 cases
considered is the same. 

So, all the signs are in perfect agreement with phenomenology,
which (as discussed in the previous chapter) does indeed suggest
light $\pi,\sigma$ and heavy $a_0,\eta'$. 

Now, as we are satisfied that signs are correct,
what about the absolute magnitude of these corrections?
The product of small quark masses appear in the denominator, 
which however can be 
tamed in a single-instanton background; its density $n$
also has the product of these masses due to the fermion determinant.

However, a consistent evaluation of any effect
cannot proceed without account for
broken chiral symmetry  in the QCD vacuum and condensates.
In the single instanton approximation (SIA) discussed in chapter 4  in
place of the bare quark masses one should substitute properly
defined effective
masses. 
In the first paper on the subject \cite{Shuryak:1982qx} it was done more
crudely,
in the MFA,
for $\pi,K,\eta,\eta'$ correlators.


 Let us return to a single instanton background and 
proceed to  the   vector  channel. 
We have shown above that the zero mode term does not contribute
in this case, so the correlator is actually finite in the
chiral limit without $m_u*m_d$ in the fermionic determinant.
The calculation itself was done
  by Andrei and Gross \cite{Andrei:1978xg}, and this paper created
 at the time significant controversy.
  
  Non-vanishing contributions come from 
the non-zero mode propagator, and from the interference between 
the zero mode part and the  mass correction.
The latter term survives even in the chiral limit, because the factor 
$m$ in the mass correction is canceled by the $1/m$ from the zero
mode.
\be
\Pi^{AG}_\rho(x,y) &=& {\rm Tr}\left[ \gamma_\mu S^{nz}(x,y)
\gamma_\mu S^{nz}(y,x) \right] + \nonumber \\
 2{\rm Tr}\left[ \gamma_\mu
\psi_0(x)\psi_0^\dagger(y)\gamma_\mu \Delta(y,x)\right]
\ee
 After averaging over the 
instanton coordinates,
the result is\footnote{There is a mistake by an
overall factor 3/2 in the original work, originated from color traces.
In other words, the result is correct in SU(2).} 
\be
\label{vec_SIA}
\Pi^{SIA}_\rho(x) &=&  \Pi^0_\rho + \int d\rho\, n(\rho)
 \frac{12}{\pi^2}\frac{\rho^4}{x^2}\frac{\partial}{\partial (x^2)}
 \left\{ \frac{\xi}{x^2} \log\frac{1+\xi}{1-\xi}\right\}
\ee
where $\xi^2=x^2/(x^2+4\rho^2)$.

The reason we discuss this result  is its relations to the
OPE.
    Expanding (\ref{vec_SIA}), 
we get
\be
\label{vec_SIA_exp}
\Pi^{SIA}_\rho(x) &=& \Pi^0_\rho(x) \left( 1 + \frac{\pi^2x^4}{6}
 \int d\rho n(\rho) \right). 
\ee
This agrees exactly with the OPE expression, provided we use the 
average values of the operators in the dilute gas approximation
\be
 \langle g^2G^2\rangle  \;=\; 32\pi^2 \int d\rho\, n(\rho)\, ,\hspace{1cm} 
 m\langle\bar qq\rangle \;=\;  - \int d\rho\, n(\rho) \, .
\ee 
Note, that the value of $m\langle\bar qq\rangle$ is ``anomalously" 
large in the dilute gas limit. This means that the contribution from
dimension 4 operators is attractive, in contradiction with the OPE
prediction based on the canonical values of the condensates.

%
%

\section{Correlators in the instanton ensemble}
\label{sec_prop_ens}

  In this section we generalize the results of the last section to 
the more general case of an ensemble consisting of many pseudo-particles.
The quark propagator in an arbitrary gauge field can always be expanded as 
\be
\label{prop_exp} 
 S &=& S_0 + S_0 A\!\!\!/ \, S_0 + S_0 A\!\!\!/\,
  S_0 A\!\!\!/ S_0 + \ldots ,
\ee
where the individual terms have an obvious interpretation
as arising from multiple gluon exchanges with the background
field. If the gauge field is a sum of instanton contributions,
$A_\mu = \sum_I A_{I\,\mu}$, then (\ref{prop_exp}) becomes
\be
\label{prop_sum} 
 S &=& S_0 + \sum_I S_0 {A\!\!\!/}_I S_0
 + \sum_{I,J} S_0 {A\!\!\!/}_I
  S_0 {A\!\!\!/}_J S_0 + \ldots \\
\label{prop_hop}
 &=& S_0 + \sum_I (S_I - S_0) + \sum_{I\neq J} 
 (S_I - S_0) S_0^{-1} (S_J - S_0) \\
 & & + \sum_{I\neq J,\,J\neq K}  (S_I - S_0) S_0^{-1} 
 (S_J - S_0)S_0^{-1}(S_K-S_0) + \ldots \; .\nonumber
\ee
Here, $I,J,K,\ldots$ refers to both instantons and anti-instantons. 
In the second line, we have re-summed the contributions corresponding 
to an individual instanton. $S_I$ refers to the sum of zero and 
non-zero mode components. At large distance from the center of the 
instanton, $S_I$ approaches the free propagator $S_0$. Thus Eq.
(\ref{prop_hop}) has a nice physical interpretation; Quarks propagate 
by jumping from one instanton to the other. If $|x-z_I|\ll \rho_I,\,
|y-z_I| \ll \rho_I$ for all $I$, the free propagator dominates. At 
large distance, terms involving more and more instantons become 
important.

  In the QCD ground state, chiral symmetry is broken. The 
presence of a condensate implies that quarks can propagate 
over large distances. Therefore, we cannot expect that 
truncating the series (\ref{prop_hop}) will provide a useful 
approximation to the propagator at low momenta. Furthermore, 
we know that spontaneous symmetry breaking is related to small 
eigenvalues of the Dirac operator. A good approximation to the 
propagator is obtained by assuming that $(S_I-S_0)$ is dominated 
by fermion zero modes
\be
\label{zm_dom}
 \left(S_I-S_0\right)(x,y) \simeq 
   \frac{\psi_I(x)\psi_I^\dagger(y)}{im}.
\ee
In this case, the expansion (\ref{prop_hop}) becomes
\be
 S(x,y) \simeq S_0(x,y) + \sum_I\frac{\psi_I(x)
  \psi_I^\dagger(y)}{im} \nonumber \\
   + \sum_{I\neq J} \frac{\psi_I(x)}{im}
      \left( \int d^4r\,\psi_I^\dagger(r)
      (-i\partial\!\!\!/ -im)\psi_J(r) \right)
     \frac{\psi_J^\dagger(y)}{im} + \ldots ,
\ee
which contains the overlap integrals $T_{IJ}$ . 
This expansion can easily be summed to give
\be
\label{prop_zmz}
 S(x,y) &\simeq& S_0(x,y) + \sum_{I,J}\psi_I(x)
    \frac{1}{T_{IJ}+im D_{IJ} -im \delta_{IJ}}
     \psi_J^\dagger(y).
\ee
Here, $D_{IJ}= \int d^4r\,\psi_I^\dagger(r)\psi_J(r)-\delta_{IJ}$ 
arises from the restriction $I\neq J$ in the expansion (\ref{prop_hop}). 
The quantity $mD_{IJ}$ is small in both the chiral expansion and in 
the packing fraction of the instanton liquid and will be neglected in 
what follows. Comparing the re-summed propagator (\ref{prop_zmz}) with 
the single instanton propagator (\ref{zm_dom}) shows the importance of 
chiral symmetry breaking. While (\ref{zm_dom}) is proportional to $1/m$, 
the diagonal part of the full propagator is proportional to $(T^{-1})_{II}
=1/m^*$.

    The result (\ref{prop_zmz}) can also be derived by 
inverting the Dirac operator in the basis spanned by the 
zero modes of the individual instantons
\be
\label{prop_zmz2}
 S(x,y) &\simeq& S_0(x,y) + \sum_{I,J}|I\rangle \langle I|
    \frac{1}{iD\!\!\!\!/ + im}|J\rangle \langle J| .
\ee
The equivalence of (\ref{prop_zmz}) and (\ref{prop_zmz2})
is easily seen using the fact that in the sum ansatz, the 
derivative in the overlap matrix element $T_{IJ}$ can be
replaced by a covariant derivative. 

  The propagator (\ref{prop_zmz}) can be calculated either
numerically or using the mean field approximation . We will discuss the mean field propagator 
in the following section. For our numerical calculations, we have
improved the zero mode propagator by adding the contributions
from non-zero modes to first order in the expansion (\ref{prop_hop}). 
The result is 
\be
\label{s_num}
 S(x,y) &=& S_0(x,y) + S^{ZMZ}(x,y) + 
  \sum_I (S^{NZM}_I(x,y)-S_0(x,y) ) .
\ee
How accurate is this propagator? We have seen that the propagator 
agrees with the general OPE result at short distance. We also
know that it accounts for chiral symmetry breaking and spontaneous 
mass generation at large distances. In addition to that, we have
performed a number of checks on the correlation functions that
are sensitive to the degree to which (\ref{s_num}) satisfies the
equations of motion, for example by testing whether the vector 
correlator is transverse (the vector current is conserved).

\subsection{Mesonic Correlators }
\label{sec_IL_mes}

\begin{figure}[t]
\begin{center}
\includegraphics[width=10cm]{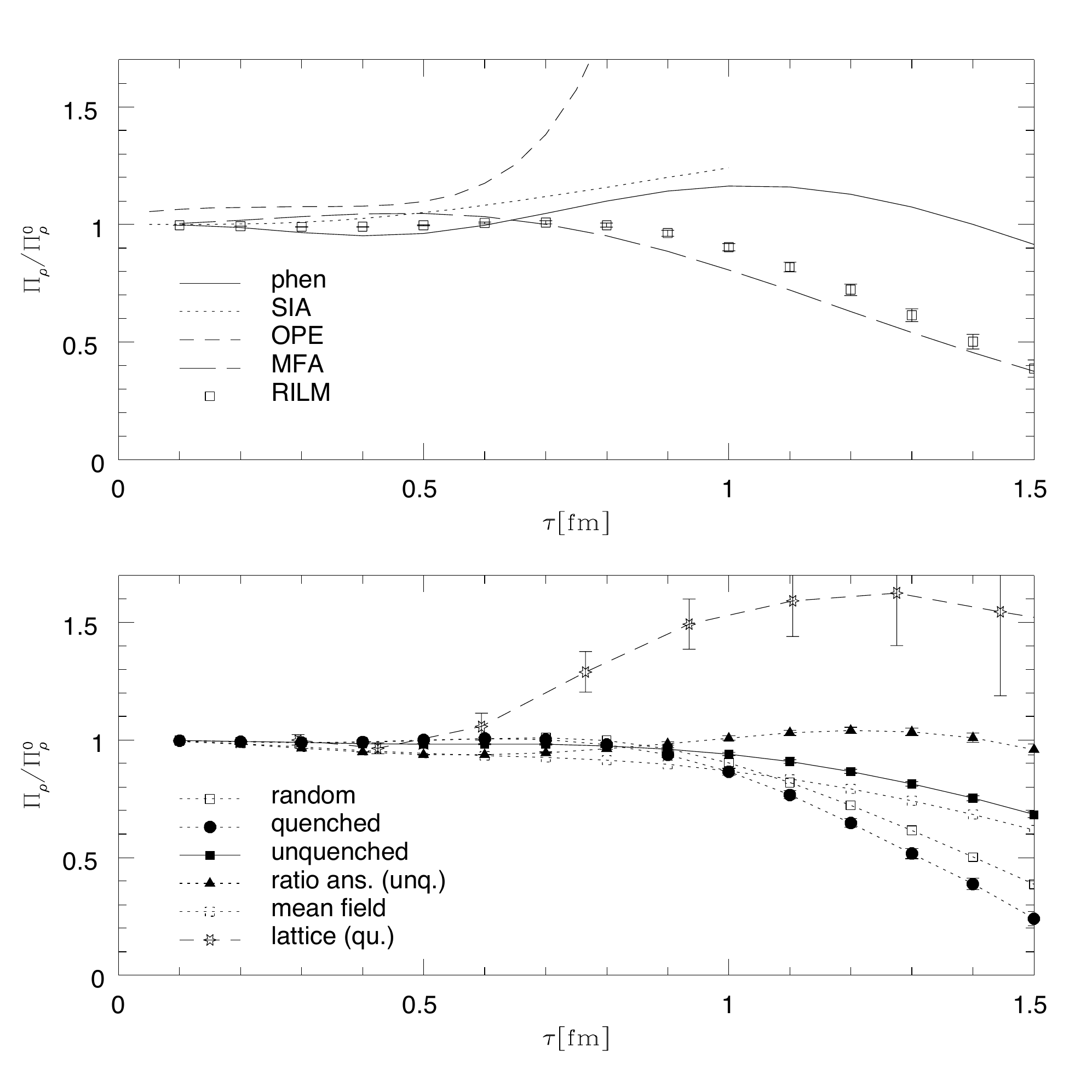}
\end{center}
\caption{\label{fig_rho_cor}
Rho meson correlation functions. The dashed squares show
the non-interacting part of the rho meson correlator in the interacting
ensemble.} 
\end{figure}

  In the following we will therefore discuss results from 
numerical calculations of hadronic correlators in the instanton
liquid. These calculations go beyond the RPA in two ways; (i) 
the propagator includes genuine many instanton effects and 
non-zero mode contributions; (ii)  the ensemble is determined 
using the full (fermionic and bosonic) weight function, so it 
includes correlations among instantons. In addition to that,
we will also consider baryonic correlators and three point
functions that are difficult to handle in the RPA.

   We will discuss correlation function in three different
ensembles, the random ensemble (RILM), the quenched (QILM)
and fully interacting (IILM) instanton ensembles. In the random
model, the underlying ensemble is the same as in the mean field
approximation, only the propagator is more sophisticated. In 
the quenched approximation, the ensemble includes correlations
due to the bosonic action, while the fully interacting ensemble 
also includes correlations induced by the fermion determinant.
In order to check the dependence of the results on the instanton
interaction, we study correlation functions in two different
unquenched ensembles, one based on the streamline interaction
(with a short-range core) and one based on the ratio ansatz  
interaction. The bulk parameters of these ensembles are 
compared in Tab. \ref{tab_liquid_par}. 

\begin{table}
\caption{\label{tab_liquid_par}
Bulk parameters of  different instanton ensembles.}\vskip .3cm
\begin{tabular}{crrrr}
               & Streamline         & quenched           & Ratio ansatz       
               & RILM              \\  \hline
$n$            & 0.174$\Lambda^4$   & 0.303$\Lambda^4$   & 0.659$\Lambda^4$   
               & 1.0  ${\rm fm}^4$ \\
$\bar\rho$     & 0.64$\Lambda^{-1}$ & 0.58$\Lambda^{-1}$ & 0.66$\Lambda^{-1}$ 
               & 0.33 ${\rm fm}$   \\
               & (0.42 fm)          & (0.43 fm)          & (0.59 fm)          
               &                   \\
$\bar\rho^4 n$ & 0.029              & 0.034              & 0.125              
               & 0.012             \\
$ \langle \bar q q \rangle$   & 0.359$\Lambda^3$   & 0.825$\Lambda^3$   & 0.882$\Lambda^3$   
               & $(264\,{\rm \, MeV})^3$ \\
               &$(219\,{\rm \, MeV})^3$&$(253\,{\rm \, MeV})^3$&$(213\,{\rm \, MeV})^3$
               &             \\
$\Lambda$      & 306 \, MeV            &    270 \, MeV         &   222 \, MeV          
               &  -          \\
\end{tabular}
\end{table}

   Correlation functions in the different instanton ensembles 
were calculated in \cite{Shuryak:1992ke,Schafer:1993ra,Schafer:1995pz} to which we refer 
the reader for more details. The results are shown in 
Fig. \ref{fig_rho_cor} and summarized 
in Tab. \ref{tab_mes_res}. The pion correlation functions 
in the different ensembles are qualitatively very similar. The
differences are mostly due to different values of the quark
condensate (and the physical quark mass) in the different
ensembles. Using the Gell-Mann, Oaks, Renner relation, one can 
extrapolate the pion mass to the physical value of the quark 
masses, see Tab. \ref{tab_mes_res}. The results are consistent with 
the experimental value in the streamline ensemble (both quenched 
and unquenched), but clearly too small in the ratio ansatz ensemble. 
This is a reflection of the fact that the ratio ansatz ensemble is 
not sufficiently dilute.

   In Fig. \ref{fig_rho_cor} we also show the results in the $\rho$ 
channel. The $\rho$ meson correlator is not affected by instanton zero 
modes to first order in the instanton density. The results in the 
different ensembles are fairly similar to each other and all fall 
somewhat short of the phenomenological result at intermediate distances 
$x\simeq 1$ fm. We have determined the $\rho$ meson mass and coupling 
constant from a fit, the results are given 
in Tab. \ref{tab_mes_res}. The $\rho$ meson mass is somewhat too heavy 
in the random and quenched ensembles, but in good agreement with 
the experimental value $m_\rho=770$ \, MeV in the unquenched ensemble.

    Since there are no interactions in the $\rho$ meson channel 
to first order in the instanton density, it is important to study 
whether the instanton liquid provides any significant binding. In 
the instanton model, there is no confinement, and $m_\rho$ is close 
to the two (constituent) quark threshold. In QCD, the $\rho$ meson 
is also not a true bound state, but a resonance in the 2$\pi$ continuum. 
In order to determine whether the continuum contribution in the 
instanton liquid is predominantly from 2-$\pi$ or 2-quark states 
would require the determination of the corresponding three point 
functions, which has not been done yet. Instead, we have compared 
the full correlation function with the non-interacting (mean field)
correlator , where we use the average (constituent 
quark) propagator determined in the same ensemble, see Fig. 
\ref{fig_rho_cor}). This comparison provides a measure of the 
strength of interaction. We observe that there is an attractive 
interaction generated in the interacting liquid
due to correlated instanton-anti-instanton pairs.
This is consistent with the fact that the interaction is considerably 
smaller in the random ensemble. In the random model, the strength of 
the interaction grows as the ensemble becomes more dense. However, 
the interaction in the full ensemble is significantly larger than 
in the random model at the same diluteness. Therefore, most of the 
interaction is due to dynamically generated pairs.

\begin{figure}[tbh]
\begin{center}
\leavevmode
\vspace*{-1.7cm}
\includegraphics[width=10cm]{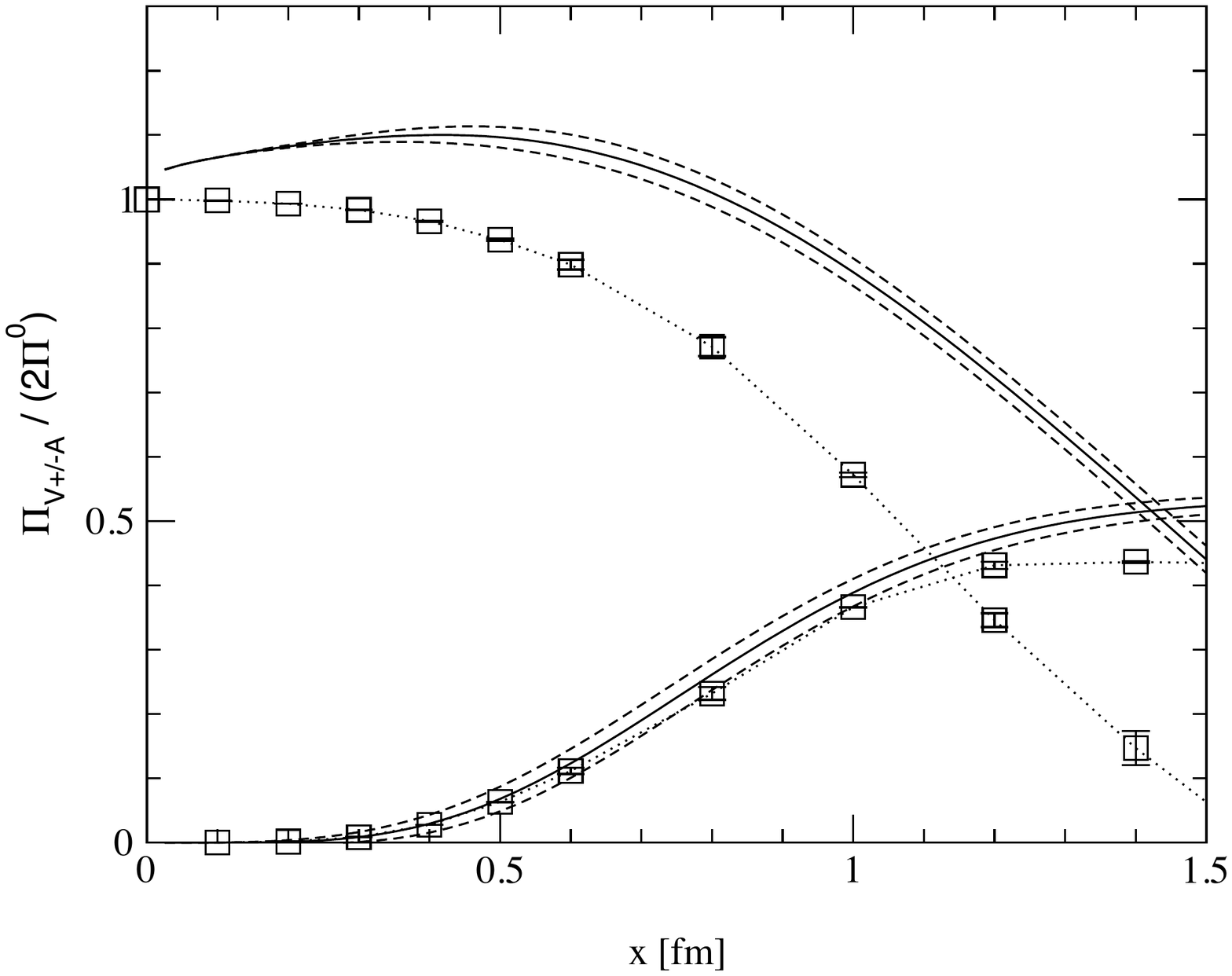}
\end{center}
\vspace*{-.5cm}
\caption{\label{fig_cor1}
Euclidean coordinate space correlation functions $\Pi_V(x)
\pm \Pi_A(x)$ (upper and lower points and curves, respectively)  normalized to free quark correlator. The solid lines
show the correlation functions reconstructed from the ALEPH spectral
functions and the dotted lines show the corresponding error band.
The squares show the result of a random instanton liquid model
and the diamonds the OPE fit described in the text.}
\end{figure}

We have already discussed ALEPH $\tau$-decay data  and have shown  the data compared to OPE and the calculation
in the random instanton liquid model (RILM)  by Schafer and myself
\cite{Schafer:2000rv}. Another figure from this work, Fig.\ref{fig_cor1},
shown here shows larger-x part of the correlator studied.
As one can see, RILM works for the whole $V-A$ curve, and, with 10\%
radiative correction $\alpha_s/\pi$, it works very well for 
 $V+A$ as well. There is no fit of any parameter here, and in fact
the calculation preceded the experiment by few years.

   The situation is drastically different in the $\eta'$ channel. Among
the $\sim 40$ correlation functions calculated in the random ensemble, 
only the $\eta'$ (and the isovector-scalar $\delta$ discussed in the 
next section) are completely unacceptable; The correlation function 
decreases very rapidly and becomes negative at $x\sim 0.4$ fm. This 
behavior is incompatible with the positivity of the spectral function. 
The interaction in the random ensemble is too repulsive, and the 
model ``over-explains" the $U(1)_A$ anomaly. 

    The results in the unquenched ensembles (closed and open points) 
significantly improve the situation. This is related to dynamical
correlations between instantons and anti-instantons (topological charge
screening). The single instanton contribution is repulsive, but 
the contribution from pairs is attractive. Only if 
correlations among instantons and anti-instantons are sufficiently strong, 
the correlators are prevented from becoming negative. Quantitatively,
the $\delta$ and $\eta_{ns}$ masses in the streamline ensemble are still 
too heavy as compared to their experimental values. In the ratio ansatz,
on the other hand, the correlation functions even shows an enhancement 
at distances on the order of 1 fm, and the fitted masses is too light. 
This shows that the $\eta'$ channel is very sensitive to the strength 
of correlations among instantons.

    In summary, pion properties are mostly sensitive to global
properties of the instanton ensemble, in particular its diluteness.
Good phenomenology demands $\bar\rho^4 n\simeq 0.03$, as originally
suggested in \cite{Shuryak:1981ff}. The properties of the $\rho$ meson are 
essentially independent of the diluteness, but show  sensitivity 
to $IA$ correlations. These correlations become crucial
in the $\eta'$ channel.

 Let us add about dependence of the correlators  on the number
of colors $N_c$, studied in \cite{Schafer:2002af}.
 The results were
obtained from simulations with $N=128$ instantons 
in a euclidean volume $V\Lambda^4=V_3\times 5.76$.
$V_3$ was adjusted such that $(N/V)=(N_c/3)\Lambda^4$.
In order to avoid finite volume artifacts the current
quark mass was taken to be rather large, $m_q=0.2
\Lambda$. We observe that the rho meson correlation 
function exhibits almost perfect scaling
with $N_c$ and as a result the rho meson mass is 
practically independent of $N_c$. The scaling is 
not as good in the case of the pion. As a consequence
there is some variation in the pion mass. However,
this effect is consistent with $1/N_c$ corrections
that amount to about 40\% of the pion mass for
$N_c=3$. Finally, we study the behavior of the $\eta'$
correlation function. There is a clear tendency 
toward $U(1)_A$ restoration, but the correlation
function is still very repulsive for $N_c=6$. 
It was also found  that the $\eta'$ correlation
function only approaches the pion correlation for fairly
large values of $N_c$. For example, the $\eta'$ correlation
function does not show intermediate range attraction unless
$N_c>15$. 

\begin{table}[t]
\caption{\label{tab_mes_res}
Meson parameters in the different instanton ensembles. All
quantities are given in units of \, GeV. The current quark mass is $m_u
=m_d=0.1\Lambda$. Except for the pion mass, no attempt has been made to
extrapolate the parameters to physical values of the quark mass.}\vskip .3cm
\begin{tabular}{crrrr}
               & unquenched         & quenched           & RILM       
               & ratio ansatz (unqu.)  \\  \hline
$m_\pi$        &  0.265             &   0.268            &  0.284            
               &  0.128      \\
$m_\pi$ (extr.)&  0.117             &   0.126            &  0.155
               &  0.067      \\
$\lambda_\pi$  &  0.214             &   0.268            &  0.369             
               &  0.156      \\
$f_\pi$        &  0.071             &   0.091            &  0.091            
               &  0.183 \\
$m_\rho$       &  0.795             &   0.951            &  1.000             
               &  0.654      \\
$g_\rho$       &  6.491             &   6.006            &  6.130            
               &  5.827      \\
$m_{a_1}$      &  1.265             &   1.479            &  1.353             
               &  1.624      \\
$g_{a_1}$      &  7.582             &   6.908            &  7.816             
               &  6.668      \\
$m_{\sigma}$   &  0.579             &   0.631            &  0.865             
               &  0.450      \\
$m_{\delta}$   &  2.049             &   3.353            &  4.032            
               &  1.110      \\
$m_{\eta_{ns}}$&  1.570             &   3.195            &  3.683             
               &  0.520 \\
\end{tabular}
\end{table}
   After discussing the $\pi,\rho,\eta'$ in some detail we only 
briefly comment on other correlation functions. The remaining 
scalar states are the isoscalar $\sigma$ and the isovector
$\delta$ (the $f_0$ and $a_0$ according to the notation of the 
particle data group). The sigma correlator has a disconnected 
contribution, which is proportional to $\langle\bar qq\rangle^2$
at large distance. In order to determine the lowest resonance 
in this channel, the constant contribution has to
be subtracted, which makes it difficult to obtain reliable 
results. Nevertheless, we find that the instanton liquid favors
a (presumably broad) resonance around 500-600 \, MeV. The isovector
channel is in many ways similar to the $\eta'$. In the random
ensemble, the interaction is too repulsive and the correlator 
becomes unphysical. This problem is solved in the interacting 
ensemble, but the $\delta$ is still very heavy, $m_\delta>1$
\, GeV. 

   The remaining non-strange vectors are the $a_1,\omega$ and $f_1$. 
The $a_1$ mixes with the pion, which allows a determination of
the pion decay constant $f_\pi$ (as does a direct measurement of
the $\pi-a_1$ mixing correlator). In the instanton liquid, disconnected
contributions in the vector channels are small. This is consistent
with the fact that the $\rho$ and the $\omega$, as well as the $a_1$
and the $f_1$ are almost degenerate.   
 
   Finally, we can also include strange quarks. $SU(3)$ flavor
breaking in the 't Hooft interaction nicely accounts for the 
masses of the $K$ and the $\eta$. More difficult is a correct 
description of $\eta-\eta'$ mixing, which can only be achieved
in the full ensemble. The random ensemble also has a problem
with the mass splittings among the vectors $\rho,K^*$ and $\phi$
\cite{Shuryak:1992ke}. This is related to the fact that flavor symmetry 
breaking in the random ensemble is so strong that the strange 
and non-strange constituent quark masses are almost degenerate. 
This problem is improved (but not fully solved) in the interacting 
ensemble. 

\subsection{Baryonic correlation functions}
\label{sec_bar_cor}

 As emphasized few times above,
 the existence of a strongly attractive interaction
in the pseudo-scalar quark-anti-quark (pion) channel also implies an
attractive interaction in the scalar quark-quark (diquark) channel. 
This interaction is phenomenologically very desirable, because it
not only explains why the spin 1/2 nucleon is lighter than the 
spin 3/2 Delta, but also why Lambda is lighter
than  Sigma.

The vector components of the diagonal correlators receive perturbative 
quark-loop contributions, which are dominant at short distance. The 
scalar components of the diagonal correlators, as well as the 
off-diagonal correlation functions, are sensitive to chiral symmetry 
breaking, and the OPE starts at order $\langle\bar qq\rangle$ or higher.
Instantons
introduce additional, regular, contributions in the scalar channel
and violate the factorization assumption for the 4-quark condensates. 
Similar to the pion case, both of these effects increase the amount of 
attraction already seen in the OPE. 

\begin{figure}[t]
\begin{center}
\includegraphics[width=10cm]{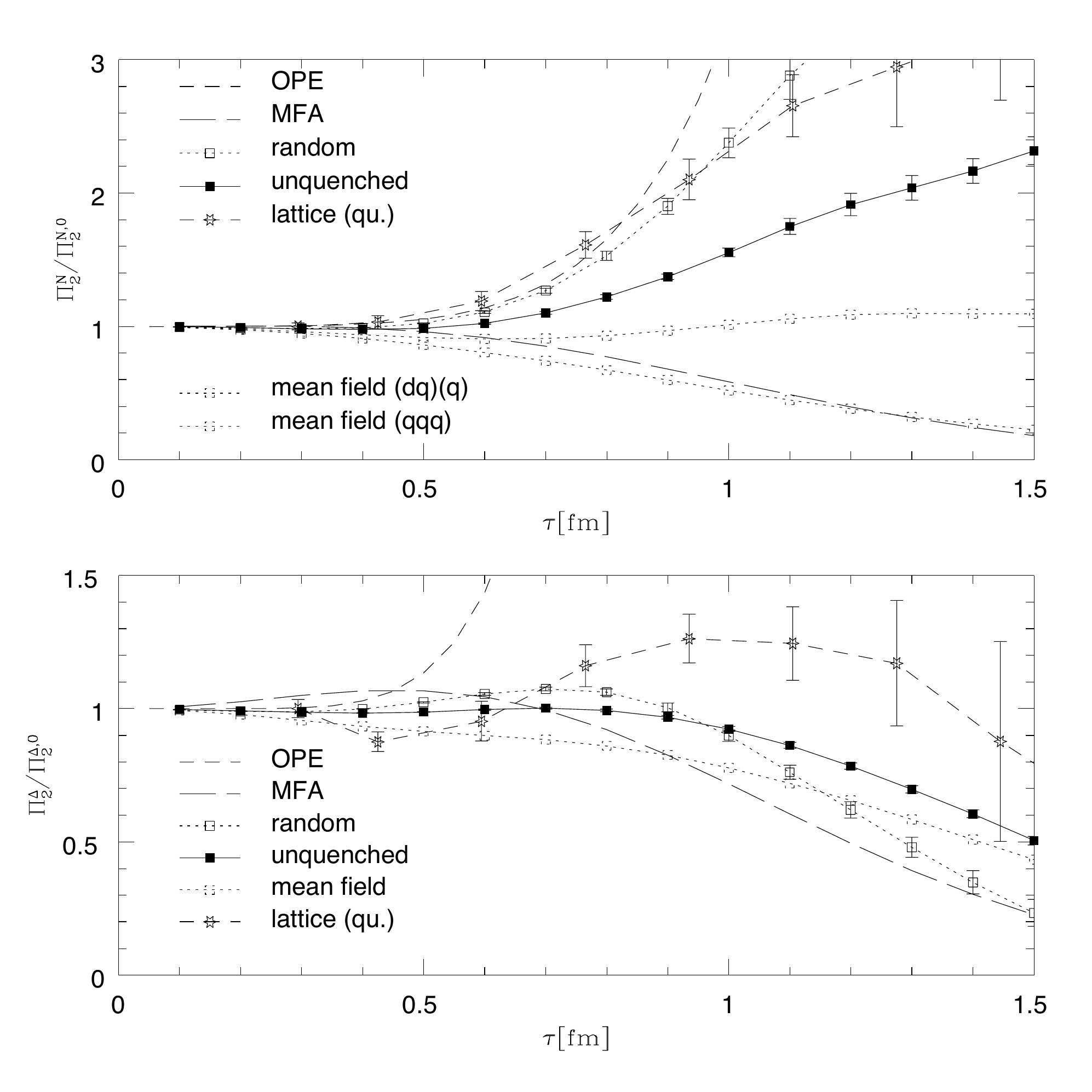}
\end{center}
\caption{\label{fig_bar_cor}
Nucleon and Delta correlation functions $\Pi_2^N$ and $\Pi_2^\Delta$.
}
\end{figure}

  The correlation function $\Pi_2^N$ in the interacting ensemble
is shown in Fig. \ref{fig_bar_cor}. There is a significant enhancement
over the perturbative contribution which corresponds to a tightly 
bound nucleon state with a large coupling constant. Numerically, 
we find\footnote{Note that this value corresponds to a relatively 
large current quark mass $m=30$ \, MeV.} $m_N=1.019$ \, GeV (see Tab. 
\ref{tab_bar_res}). In the random ensemble, we have measured the 
nucleon mass at smaller quark masses and found $m_N=0.960\pm 0.30$ 
\, GeV. The nucleon mass is fairly insensitive to the instanton ensemble. 
However, the strength of the correlation function depends on the 
instanton ensemble. This is reflected by the value of the nucleon 
coupling constant, which is smaller in the interacting model.

   Fig. \ref{fig_bar_cor} also shows the nucleon correlation 
function measured in a quenched lattice simulation \cite{Chu:1994vi}.
The agreement with the instanton liquid results is quite impressive,
especially given the fact that before the lattice calculations 
were performed, there was no phenomenological information on the 
value of the nucleon coupling constant and the behavior of the 
correlation function at intermediate and large distances. 

   The fitted position of the threshold is $E_0\simeq 1.8$ \, GeV,
larger than the mass of the first nucleon resonance, the Roper 
$N^*(1440)$, and above the $\pi\Delta$ threshold $E_0=1.37$ \, GeV.
This might indicate that the coupling of the nucleon current to 
the Roper resonance is small. In the case of the $\pi\Delta$
continuum, this can be checked directly using the phenomenologically
known coupling constants. The large value of the threshold energy
also implies that there is little strength in the (unphysical)
three-quark continuum. The fact that the nucleon is deeply bound can 
also be demonstrated by comparing the full nucleon correlation function 
with that of three non-interacting quarks, see Fig. \ref{fig_bar_cor}). 
The full correlator is significantly larger than the non-interacting 
(mean field) result, indicating the presence of a strong, attractive 
interaction. 

    Some of this attraction is due to the scalar diquark content
of the nucleon current. This raises the question whether the nucleon 
(in our model) is a strongly bound diquark very loosely coupled to a 
third quark. In order to check this, we have decomposed the nucleon
correlation function into quark and diquark components. Using the
mean field approximation, that means treating the nucleon as a 
non-interacting quark-diquark system, we get the correlation 
function labeled (diq) in Fig. \ref{fig_bar_cor}. We observe that 
the quark-diquark model explains some of the attraction seen in 
$\Pi_2^N$, but falls short of the numerical results. This means 
that while diquarks may play some role in making the nucleon bound, 
there are substantial interactions in the quark-diquark system. 
Another hint for the qualitative role of diquarks is provided 
by the values of the nucleon coupling constants $\lambda^{1,2}_N$.
One can translate these results into the 
coupling constants $\lambda^{s,p}_N$ of nucleon currents built 
from scalar or pseudo-scalar diquarks. We find that the coupling
to the scalar diquark current $\eta_s= \epsilon_{abc}(u^a C\gamma_5 
d^b)u^c$ is an order of magnitude bigger than the coupling to the
pseudo-scalar current $\eta_p=\epsilon_{abc}(u^a Cd^b)\gamma_5u^c$.
This is in agreement with the idea that the scalar diquark channel
is very attractive and that these configurations play an important
role in the nucleon wave function. 

\begin{table}[t]
\caption{\label{tab_bar_res}
Nucleon and delta parameters in the different instanton ensembles. All
quantities are given in units of \, GeV. The current quark mass is $m_u
=m_d=0.1\Lambda$.}\vskip .3cm 
\begin{tabular}{crrrr}
                & unquenched         & quenched           & RILM       
                & ratio ansatz (unqu.)  \\  \hline
$m_N$           &    1.019           &    1.013           & 1.040          
                &    0.983    \\
$\lambda_N^1$   &    0.026           &    0.029           & 0.037              
                &    0.021    \\
$\lambda_N^2$   &    0.061           &    0.074           & 0.093             
                &    0.048    \\
$m_\Delta$      &    1.428           &    1.628           & 1.584             
                &    1.372    \\
$\lambda_\Delta$&    0.027           &    0.040           & 0.036             
                &    0.026    \\
\end{tabular}
\end{table}


  The Delta correlation function in the instanton liquid is shown
in Fig. \ref{fig_bar_cor}. The result is qualitatively different 
from the nucleon channel, the correlator at intermediate distance
$x\simeq 1$ fm is significantly smaller and close to perturbation 
theory. This is in agreement with the results of the lattice 
calculation \cite{Chu:1994vi}. Note that, again, this is a quenched
result which should be compared to the predictions of the random 
instanton model. 
  
  The mass of the delta resonance is too large in the random model,
but closer to experiment in the unquenched ensemble. Note that similar
to the nucleon, part of this discrepancy is due to the value of the 
current mass. Nevertheless, the Delta-nucleon mass splitting in the
unquenched ensemble is $m_\Delta-m_N=409$ \, MeV, still too large as 
compared to the experimental value 297 \, MeV. Similar to the $\rho$ 
meson, there is no interaction in the Delta channel to first order 
in the instanton density. However, if we compare the correlation 
function with the mean field approximation based on the full
propagator, see Fig. \ref{fig_bar_cor}, we find evidence for
substantial attraction between the quarks. Again, more detailed
checks, for example concerning the coupling to the $\pi N$ 
continuum, are necessary.

\section{Comparison to correlators on the lattice}

 The study of hadronic (point-to-point) correlation functions on the
lattice was pioneered by the MIT group \cite{Chu:1994vi} which 
measured correlation functions of the $\pi,\delta,\rho,a_1,N$ and 
$\Delta$ in quenched QCD. The correlation functions were calculated 
on a $16^3\times 24$ lattice at $6/g^2=5.7$, corresponding to a lattice 
spacing of $a\simeq 0.17$ fm. 
We have already shown some of the results of the MIT group in Figs. 
\ref{fig_rho_cor}-\ref{fig_bar_cor}. The correlators were measured 
for distances up to $\sim 1.5$ fm. Using the parametrization introduced 
above, they extracted ground state masses and coupling constants and
found good agreement with phenomenological results. What is even more
important, they found the {\em full correlation functions} to agree
with the predictions of the instanton liquid, even in channels (like
the nucleon and delta) where no phenomenological information is 
available. 

  In order to check this result in more detail, they also studied 
the behavior of the correlation functions under cooling \cite{Chu:1994vi}. 
The cooling procedure was monitored by studying a number of gluonic 
observables, like the total action, the topological charge and and 
the Wilson loop. From these observables, the authors conclude that 
the configurations are dominated by interacting instantons after 
$\sim 25$ cooling sweeps. Instanton-anti-instanton pairs are continually 
lost during cooling, and after $\sim 50$ sweeps, the topological charge 
fluctuations are consistent with a dilute gas. The characteristics of 
the instanton liquid were already discussed above.
After 50 sweeps the action is reduced by a factor $\sim$300 while
the string tension (measured from $7\times 4$ Wilson loops) has 
dropped by a factor 6.

\begin{figure}[h] 
\begin{minipage}[c]{5.7cm}
 \centering 
\includegraphics[width=5.7cm]{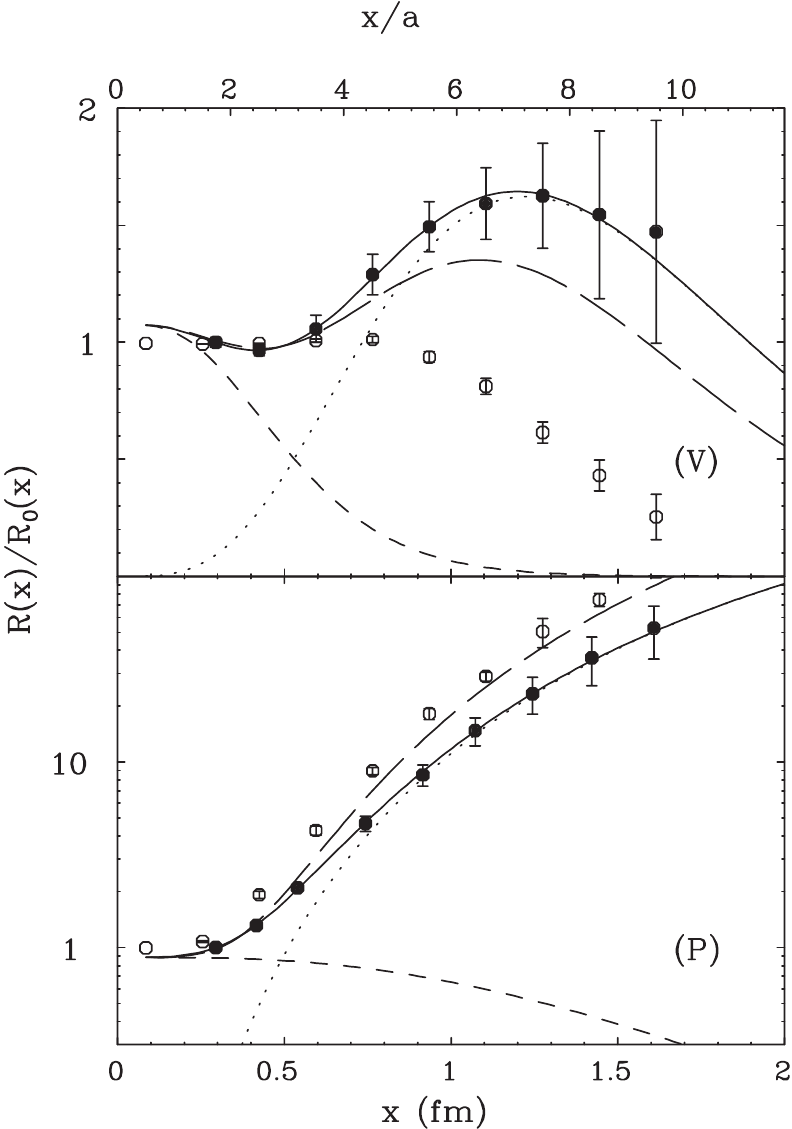}
 \end{minipage}
\begin{minipage}[c]{6.3cm}
 \centering
\vskip .4cm 
\includegraphics[width=6.9cm]{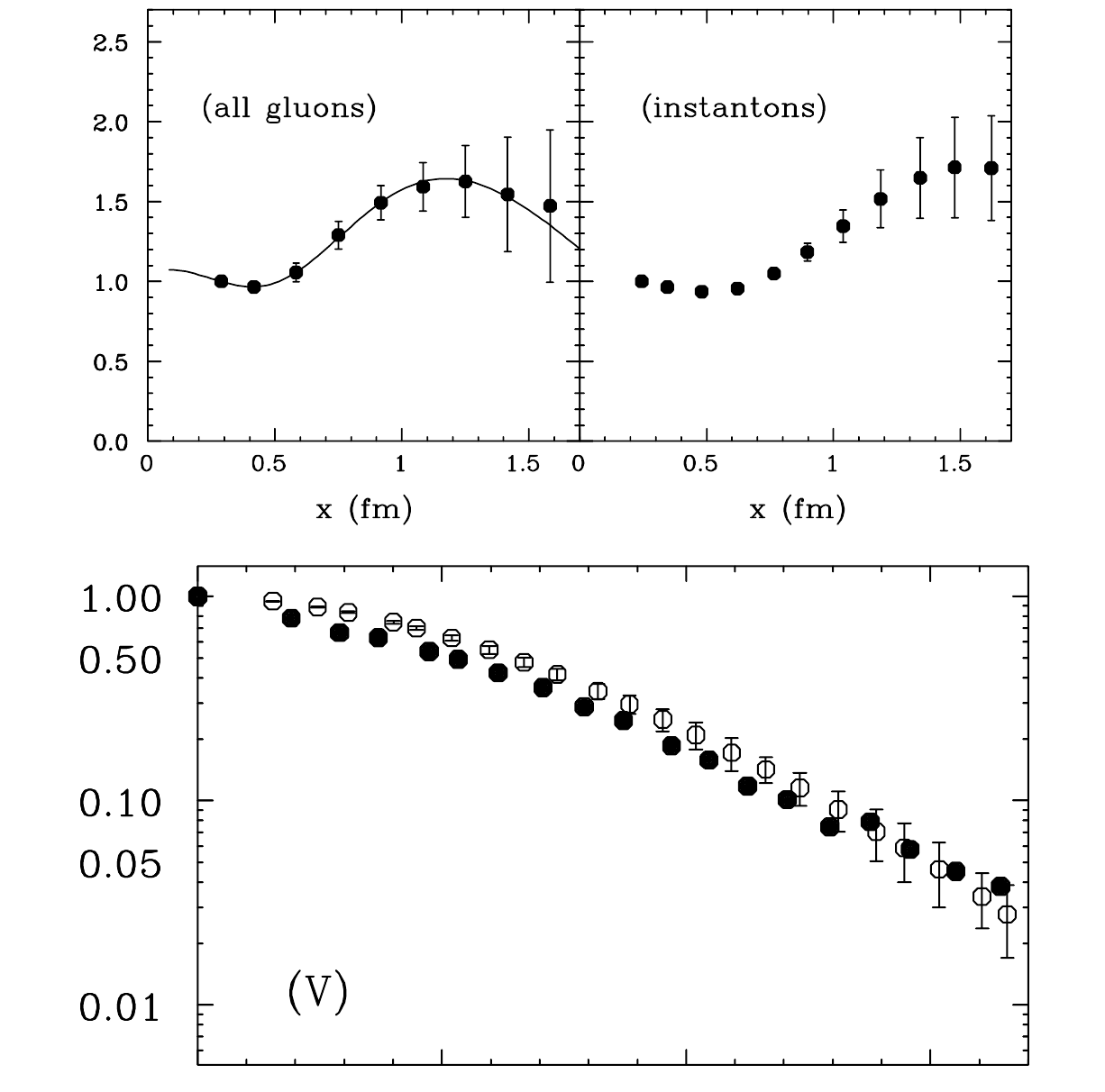}
 \end{minipage}
\caption{The left panel shows the
correlation functions for the vector (marked (V)) $\rho$ channel
and the pseudoscalar  (marked (P)) $\pi$ channel. The 
long-dashed lines are 
  phenomenological ones, open and closed circles stand for RILM
 \protect\cite{Shuryak:1992jz} and lattice calculation \protect\cite{Chu:1994vi}, respectively.
The upper right panel compares vector correlators before and after
``cooling''. The lower part
shows the same comparison for the $\rho$ wave function; the closed and
open
points here correspond to ``quantum'' and ``classical'' vacua, respectively.
\label{fig_negele_data1}}
\end{figure}

The first
comparison made
between the instanton liquid results and those obtained
on the lattice \cite{Chu:1994vi} are shown in
Fig.\ref{fig_negele_data1}(a),
for $\rho$ vector (V) and $\pi$ pseudoscalar (P) channels.

\begin{figure}[t]
\begin{center}
\includegraphics[width=10cm]{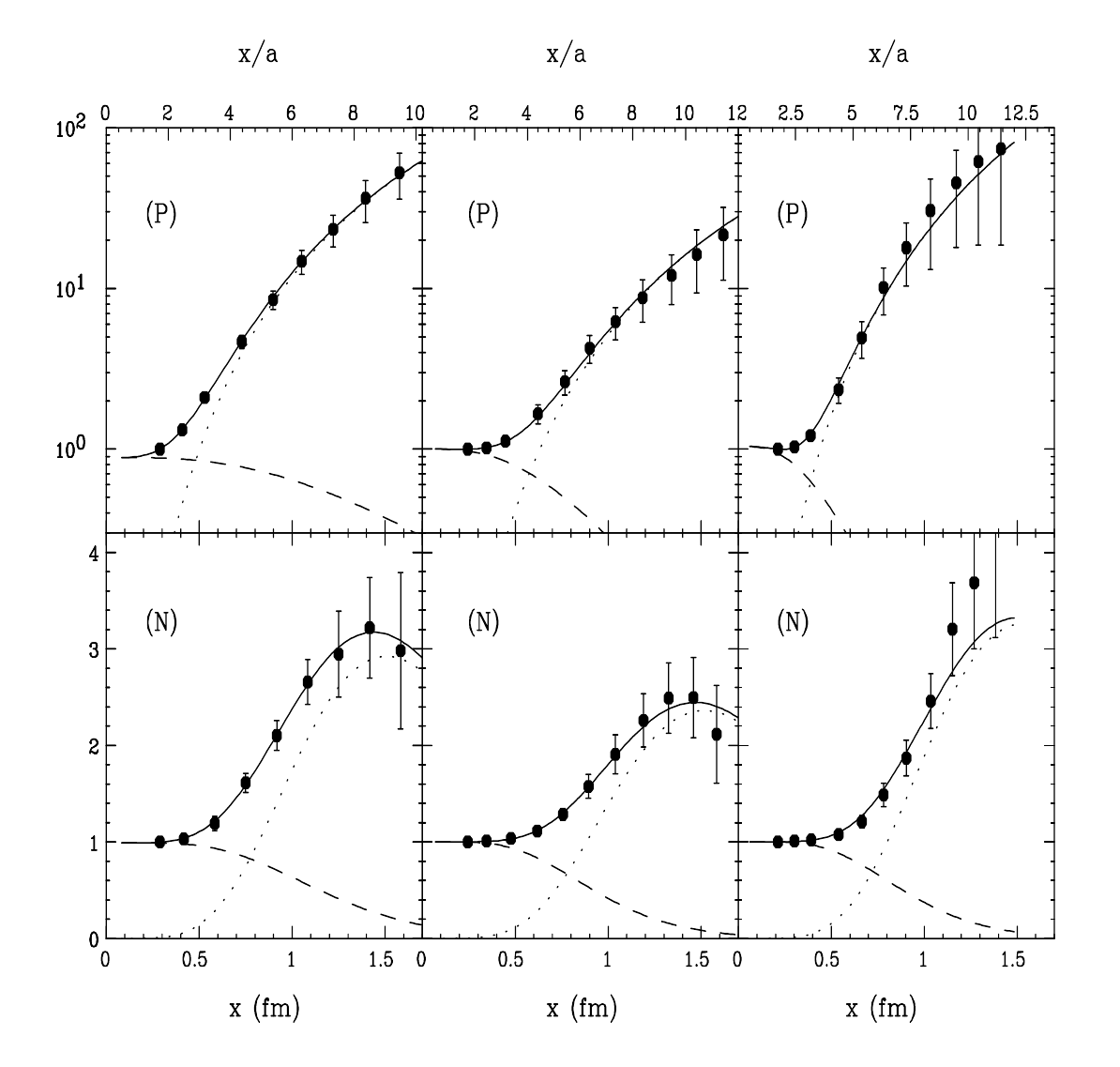}
\end{center}
\caption{\label{fig_cool_cor}
Behavior of pion and proton correlation functions under cooling, from
\protect\cite{Chu:1994vi}. The left, center, and right panels show the 
results in the original ensemble, and after 25 and 50 cooling sweeps. 
The solid lines show fits to the data based on a pole plus continuum
model for the spectral function. The dotted and dashed lines show
the individual contributions from the pole and the continuum part.}
\end{figure}

Even more direct comparison was between the correlator calculated
on the ``quantum'' configurations, as compared to ``cooled'' or
``semiclassical'' lattice configurations  \cite{Chu:1994vi}.
  The behavior of the pion and nucleon correlation functions under 
cooling is shown in Fig. \ref{fig_cool_cor}. The behavior of the 
$\rho$ and $\Delta$ correlators (not shown)
was  quite similar. During the cooling 
process the scale was readjusted by keeping the nucleon mass fixed\footnote{ 
This introduces only a small uncertainty, the change in scale is 
$\sim$16\%. We observe that the correlation functions are {\em stable 
under cooling}, they agree almost within error bars. This is also 
seen from the extracted masses and coupling constants. While $m_N$ 
and $m_\pi$ are stable by definition, $m_\rho$ and $g_\rho$ change
by less than 2\%, $\lambda_\pi$ by 7\% and $\lambda_N$ by 1\%. 
Only the delta mass is too small after cooling, it changes 
by 27\%. }.

\section{Gluonic correlation functions}
\label{sec_glue_cor}

   One of the most interesting problems in hadronic spectroscopy
is whether one can identify glueballs, bound states of pure glue, 
among the spectrum of observed hadrons. This question has two aspects. 
In pure glue theory, stable glueball states exist and they have 
been studied for a number of years in lattice simulations. In full 
QCD, glueballs mix with quark states, making it difficult to 
unambiguously identify glueball candidates. 
 
   Even in pure gauge theory, lattice simulations still require
large numerical efforts. Nevertheless, a few results appear to be
firmly established 
(i)   The lightest glueball is the scalar $0^{++}$, with a mass in the
      1.5-1.8 \, GeV range.
(ii)  The tensor glueball is significantly heavier $m_{2^{++}}
      /m_{0^{++}}\simeq 1.4$, and the pseudo-scalar is heavier still,
      $m_{0^{-+}}/m_{0^{++}}=1.5$-$1.8$.
(iii) The scalar glueball is much smaller than other glueballs. The size 
      of the scalar is $r_{0^{++}}\simeq 0.2$ fm, while $r_{2^{++}}\simeq 
      0.8$ fm \cite{deForcrand:1992ww}. For comparison, a similar measurement for the 
      $\pi$ and $\rho$ mesons gives 0.32 fm and 0.45 fm, indicating that 
      spin-dependent forces between gluons are stronger than between quarks.

\begin{figure}[t]
\begin{center}
\includegraphics[width=8cm]{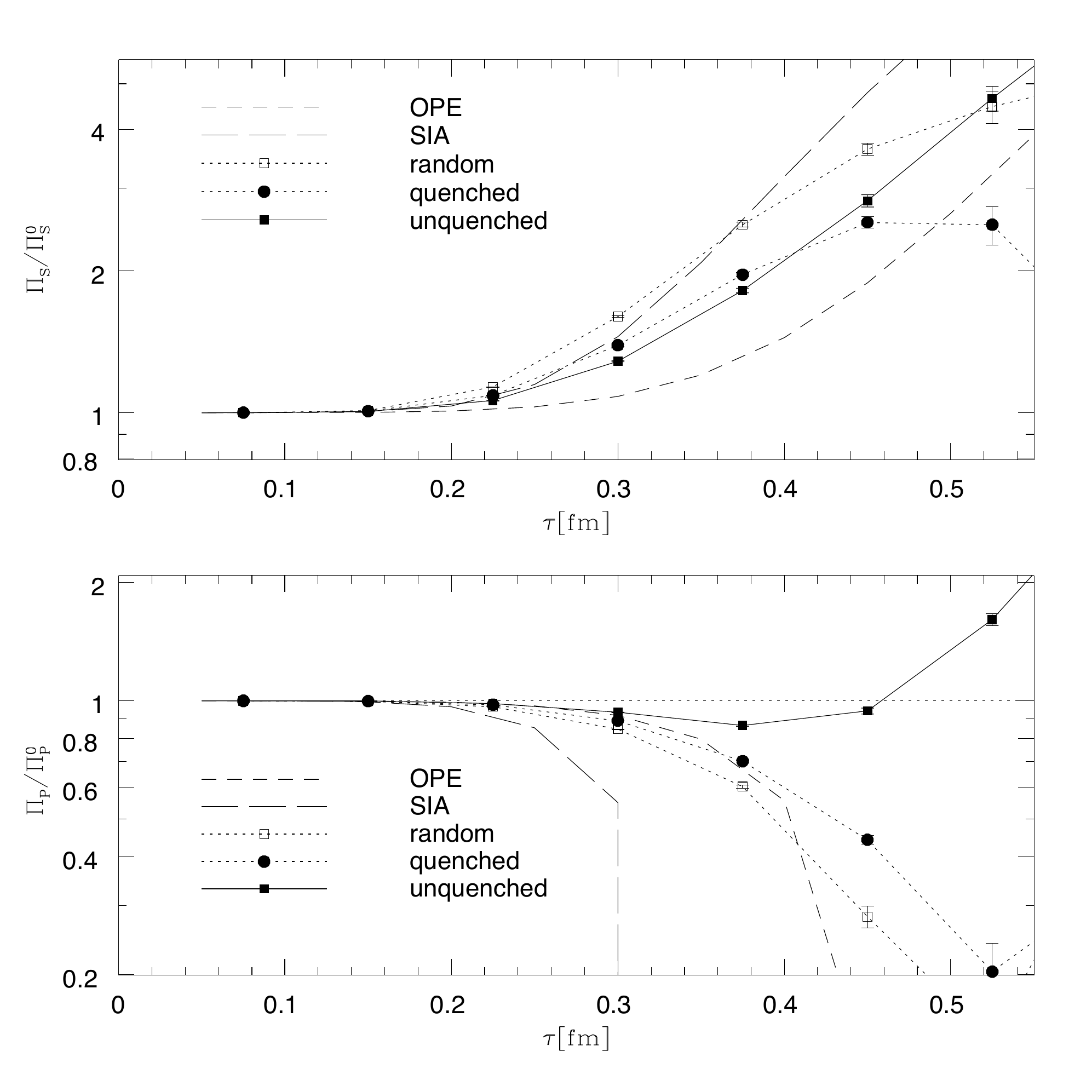}
\end{center}
\caption{\label{fig_glue}
Scalar and pseudo-scalar glueball correlation functions. Curves labeled
as in Fig. \ref{fig_rho_cor}.}
\end{figure}

   Gluonic currents with the quantum numbers of the lowest glueball 
states are the field strength squared ($S=0^{++}$), the topological 
charge density ($P=0^{-+}$), and the energy momentum tensors ($T=2^{++}$);
\be
\label{glue_cur}
j_S=  (G^a_{\mu\nu})^2, \hspace{0.5cm}
j_P= \frac 12 \epsilon_{\mu\nu\rho\sigma} G^a_{\mu\nu}G^a_{\rho\sigma},
     \hspace{0.5cm}
j_T= \frac 14 (G^a_{\mu\nu})^2- G^a_{0\alpha}G^a_{0\alpha}\, .
\ee
The short distance behavior of the corresponding correlation
functions is determined by the OPE 
\be
\label{sglue_ope}
\Pi_{S,P}(x) &=& \Pi_{S,P}^0 \left( 1 \pm \frac{\pi^2}{192g^2}
  \langle f^{abc}G^a_{\mu\nu}G^b_{\nu\beta}G^c_{\beta\mu}\rangle x^6
 + \ldots \right) \\
\label{tglue_ope}
\Pi_{T}(x) &=& \Pi_{T}^0 \left( 1 + \frac{25\pi^2}{9216g^2}
 \langle 2{\cal O}_1 - {\cal O}_2\rangle \log (x^2) x^8 +\ldots \right)
\ee
where we have defined the operators ${\cal O}_1 = (f^{abc}
G_{\mu\alpha}^bG_{\nu\alpha}^c)^2,\;{\cal O}_2 = (f^{abc}
G_{\mu\nu}^bG_{\alpha\beta}^c)^2$ and the free correlation 
functions are given by
\be
\label{gb_cor_free}
  \Pi_{S,P}(x)\;=\; (\pm)\frac{384g^4}{\pi^4 x^8}, \hspace{1cm}
  \Pi_T(x)\; = \;\frac{24g^4}{\pi^4 x^8}.
\ee
Power corrections in the glueball channels are remarkably small. 
The leading-order power correction $O(\langle G_{\mu\nu}^2\rangle/x^4)$ 
vanishes\footnote{There is a $\langle G_{\mu\nu}^2 \rangle
\delta^4(x)$ contact term in the scalar glueball correlators
which, depending on the choice of sum rule, may enter momentum
space correlation functions.}, while radiative corrections of the 
form $\alpha_s \log(x^2)\langle G_{\mu\nu}^2\rangle/x^4$ (not 
included in (\ref{sglue_ope})), or higher order power corrections 
like  $\langle f^{abc}G_{\mu\nu}^a G_{\nu\rho}^b G_{\rho\mu}^c
\rangle/x^2$ are very small.

   On the other hand, there is an important low energy theorem 
that controls the large distance behavior of the scalar correlation
function \cite{Novikov:1981xi}
\be
\label{sglue_let}
\int d^4x\,\Pi_S(x)&=&\frac{128\pi^2}{b}\langle G^2\rangle,
\ee
where $b$ denotes the first coefficient of the beta function. In 
order to make the integral well defined and we have to subtract 
the constant term $\sim \langle G^2\rangle^2$ as well as singular 
(perturbative) contributions to the correlation function. Analogously, 
the integral over the pseudo-scalar correlation functions is given by 
the topological susceptibility $\int d^4x\,\Pi_P(x)=\chi_{top}$. In  
pure gauge theory $\chi_{top} \simeq (32\pi^2)\langle G^2\rangle$, while 
in unquenched QCD $\chi_{top}=O(m)$. These 
low energy theorems indicate the presence of rather large non-perturbative 
corrections in the scalar glueball channels. This can be seen as follows; 
We can incorporate the low energy theorem into the sum rules by using a 
subtracted dispersion relation
\be
\label{sglue_sr}
  \frac{\Pi(Q^2)-\Pi(0)}{Q^2} &=& \frac{1}{\pi}
  \int ds\, \frac{{\rm Im}\Pi(s)}{s(s+Q^2)} .
\ee
In this case, the subtraction constant acts like a power correction. 
In practice, however, the subtraction constant totally dominates over
ordinary power corrections. For example, using pole dominance, the 
scalar glueball coupling $\lambda_S = \langle 0|j_S|0^{++}\rangle$ 
is completely determined by the subtraction, $\lambda_S^2/m_S^2 
\simeq (128\pi^2/b)\langle G^2\rangle$.
 
   For this reason, we expect instantons to give a large contribution
to scalar glueball correlation functions. Expanding the gluon operators
around the classical fields, we have
\be
\label{gluecor_exp} 
  \Pi_S(x,y)=\langle 0| G^{2\, cl}(x) G^{2\, cl}(y) |0\rangle  +
   \langle 0| G^{a\, ,cl}_{\mu\nu}(x) \left[ D_\mu^x D_\alpha^y
   D_{\nu\beta}(x,y)\right]^{ab} G^{b\, ,cl}_{\alpha\beta}(y) |0\rangle 
   +\ldots ,
\ee
where $D^{ab}_{\mu\nu}(x,y)$ is the gluon propagator in the classical
background field. If we insert the classical field of an instanton, 
we find 
\be
\label{gb_SIA}
\Pi_{S,P}^{SIA}(x)=\int \rho^4 dn(\rho) {12288\pi^2\rho^{-8}\over 
y^6(y^2+4)^5}
   \left[ y^8+28y^6-94y^4-160y^2-120  \right. \nonumber \\ \left.
    + \frac{240}{y\sqrt{y^2+4}}
       (y^6+2y^4+3y^2+2){\rm asinh} (\frac{ y}{2}) \right]
\ee
with $y=x/\rho$.

 There is no classical 
contribution in the tensor channel, since the stress tensor in the 
self-dual field of an instanton is zero. Note that the perturbative 
contribution in the scalar and pseudo-scalar channels have opposite sign, 
while the classical contribution has the same sign. To first order in the 
instanton density, we therefore find the three scenarios discussed 
in Sec. \ref{sec_IL_mes}; {\em attraction} in the scalar channel, 
{\em repulsion} in the pseudo-scalar and {\em no} effect in the tensor 
channel. The single-instanton prediction is compared with the OPE 
in Fig. \ref{fig_glue}. We clearly see that classical fields 
are much more important than power corrections.

\begin{table}
\caption{Scalar glueball parameters in different instanton ensembles.
\label{tab_glue_res}}\vskip .3cm
\begin{tabular}{crrrr}
                  &                 &  random   & quenched   & unquenched
                \\  \hline
$m_{0^{++}}$      & $[{\rm \, GeV}]$   &  1.4      &   1.75     &  1.25  \\
$\lambda_{0^{++}}$& $[{\rm \, GeV}^3]$ &  17.2     &   16.5     &  15.6  \\
\end{tabular}
\end{table}

  Quantum corrections to this result can be calculated from the  
second term in (\ref{gluecor_exp}) using the gluon propagator in
the instanton field 
The singular contributions
correspond to the OPE in the instanton field. There is an analog
of the Dubovikov-Smilga result for glueball correlators; In a
general self-dual background field, there are no power corrections
to the tensor correlator 
This is consistent with 
the result (\ref{tglue_ope}), since the combination $\langle 
2{\cal O}_1-{\cal O}_2\rangle$ vanishes in a self-dual field.
Also, the sum of the scalar and pseudo-scalar glueball correlators
does not receive any power corrections (while the difference
does, starting at $O(G^3)$).

   Numerical calculations of glueball correlators in different 
instanton ensembles were performed in \cite{Schafer:1994fd}. At short distances, 
the results are consistent with the single instanton approximation. At 
larger distances, the scalar correlator is modified due to the presence
of the gluon condensate. This means that (like the $\sigma$ meson), the
correlator has to be subtracted and the determination of the mass is 
difficult. In the pure gauge theory we find $m_{0^{++}}\simeq 1.5$ \, GeV and
$\lambda_{0^{++}} = 16\pm 2\,{\rm \, GeV}^3$. While the mass is consistent
with QCD sum rule predictions, the coupling is much larger than expected
from naive calculations that do not enforce the low energy theorem. 

   In the pseudo-scalar channel the correlator is very repulsive and there
is no clear indication of a glueball state. In the full theory (with 
quarks) the correlator is modified due to topological charge screening.
The non-perturbative correction changes sign and a light (on the glueball
mass scale) state, the $\eta'$ appears. Non-perturbative corrections
in the tensor channel are very small. Isolated instantons and anti-instantons 
have a vanishing energy momentum tensor, so the result is entirely due 
to interactions. 

   In \cite{Schafer:1994fd} we also evaluated the glueball wave functions. The most
important result is that the scalar glueball is indeed small, $r_{0^{++}}
= 0.2$ fm, while the tensor is much bigger, $r_{2^{++}}=0.6$ fm. The
size of the scalar is determined by the size of an instanton, whereas
in the case of the tensor the scale is set by the average distance 
between instantons. This number is comparable to the confinement
scale, so the tensor wave function is probably not very reliable. On 
the other hand, the scalar is much smaller than the confinement scale, 
so the wave function of the $0^{++}$ glueball may provide an important
indication for the importance of instantons in pure gauge theory.

\section{Wave functions}
   One  quantity which is close to
{\it hadronic wave function} is the so called 
Bethe Salpeter amplitude (which is different from the so called
light-cone
one).Such Bethe-Salpeter amplitudes
have been measured in a number of lattice gauge simulations.
%
 In the pion case 
 this quantity is defined by 
\be
 \label{pion_bs}
 \psi_\pi(y)&=&\int d^4x\, \langle 0 |\bar d(x)Pe^{i\int_x^{x+y}A(x')dx'}
 \gamma_5 u(x+y)|\pi>\, .
\ee
%
   In practice, it is 
extracted from the three point correlator
\be
\label{pion_cor}
 \Pi_\pi(x,y)=\langle 0 |T(\bar d(x)Pe^{i\int_x^{x+y}A(x')dx'}
 \gamma_5 u(x+y)\bar d(0)
 \gamma_5 u(0)) | 0 \rangle \nonumber  \\ \sim \psi(y) e^{-m_\pi x}
\ee 
where $x$ has to be a large space-like separation in order to 
ensure that the correlation function is dominated by the 
ground state and $y$ is the separation of the two quarks in 
the transverse direction ($(x\dot y)=0$). In practice it is convenient
to divide the 3-point by 2-point function, canceling the x-dependent
part and the coupling constants.

\begin{figure}[t]
\begin{center}
\includegraphics[width=6cm]{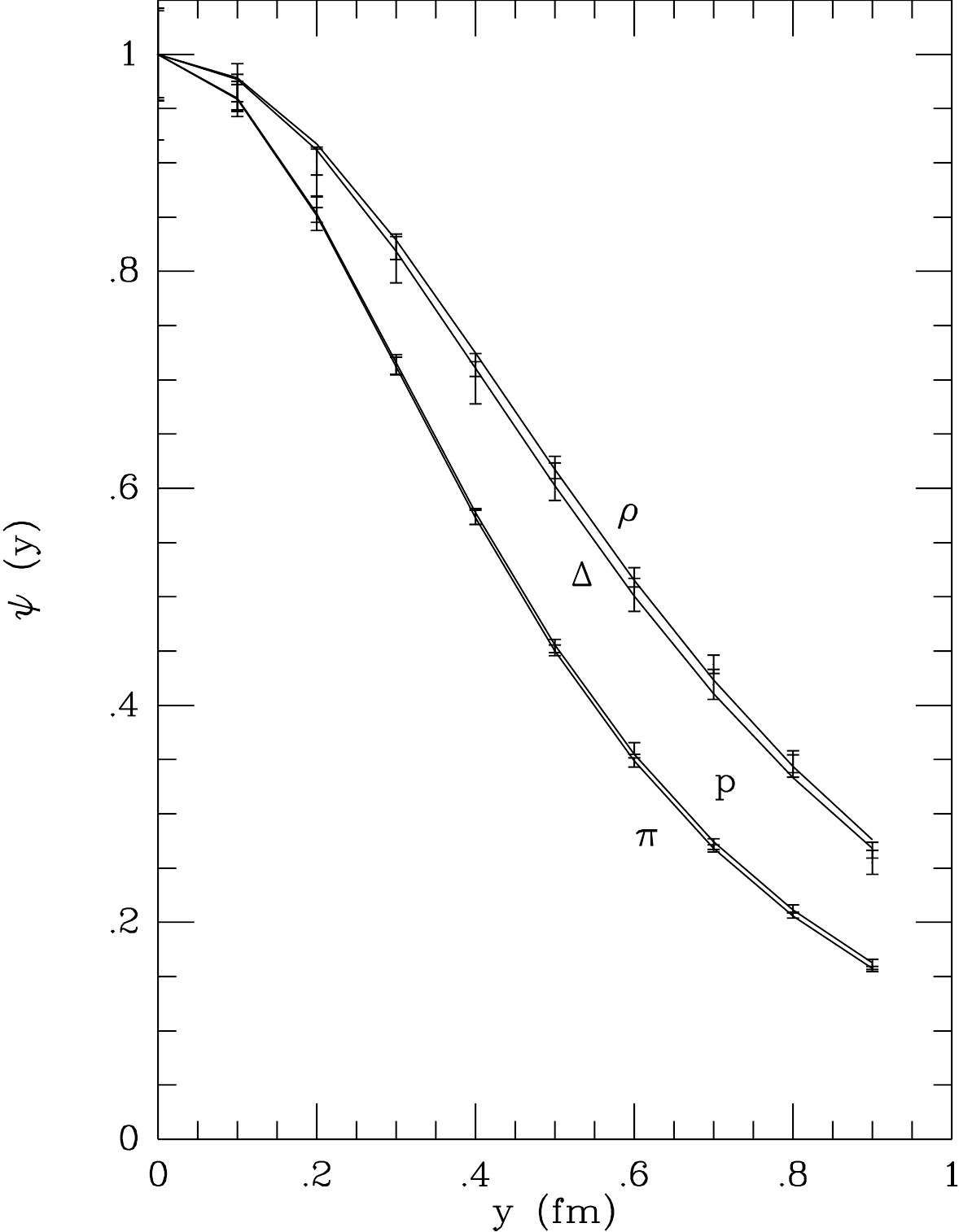}
\end{center}
\caption{\label{fig_wfct}
Hadronic wave functions of the pion, rho meson, proton and delta
resonance in the random instanton ensemble.}
\end{figure}

  Like the two point correlation functions, the Bethe-Salpeter 
amplitudes are calculated from the light quark propagator\footnote{
 In general, 
the inclusion of the Schwinger Pexp 
factor is expected to give an important contribution to the measured wave 
functions, since it corresponds to an additional string type
potential, but not in  
the instanton model.}

   A qualitative 
understanding of the wave functions can be obtained 
 using the single-instanton 
approximation. For for small transverse separations $y$ and
$x\to\infty$ we get a very simple result
\be
  \psi_\pi(y) = 1-y^2/(2\rho)^2 + \ldots
\ee
indicated that a pion radius (as determined by Bethe-Salpeter
amplitude)
is directly related to the instanton radius.


The wave functions in the random ensemble was calculated by
 \cite{Schafer:1994fd}.
 Those for
$\pi,\rho,N$ and $\Delta$ are shown in fig.\ref{fig_wfct}.
We observe that the pion and the proton as well as the rho meson and 
the delta resonance have very similar wave functions,
but the sizes for pion and 
the proton are  $smaller$ than the rho meson and 
the delta resonance. We have already argued that the scalar diquark 
in the nucleon is linked with the instanton-induced
 attraction.

\begin{figure}[htbp]
\begin{center}
\includegraphics[width=8.5cm]{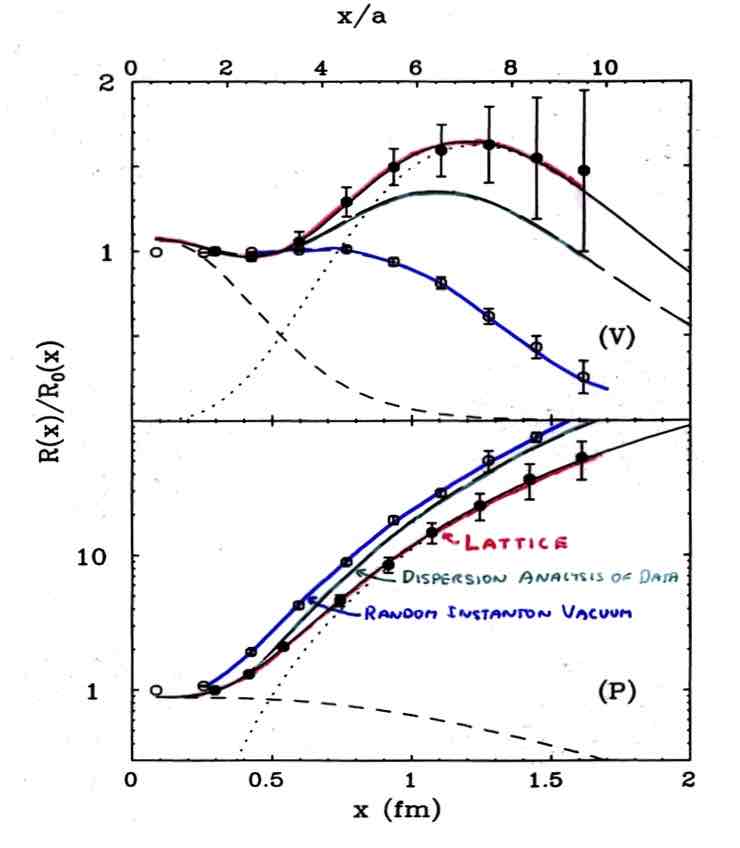}
\includegraphics[width=8.5cm]{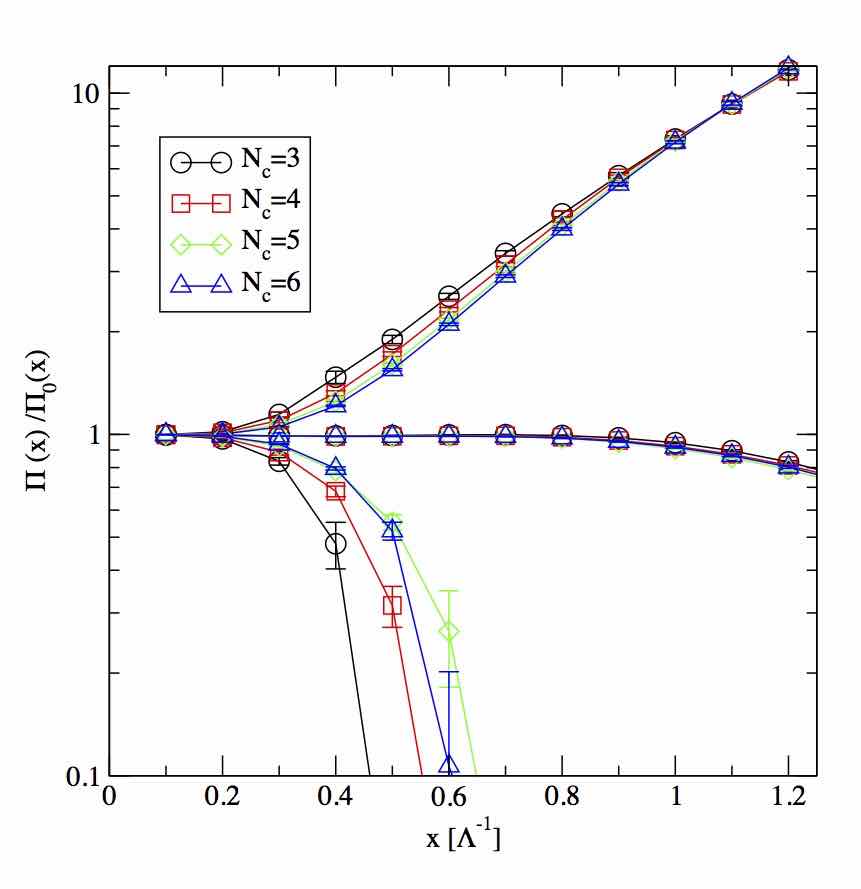}
\caption{(A) 
(B) }
\label{fig_corrs}
\end{center}
\end{figure}

\chapter{Light front wave functions of hadrons and instanton-induced effects} \label{sec_light_front}
\section{Quark models of hadrons}
Historically, understanding of quark substructure of hadrons started in 1960's with
two theoretical breakthroughs, both related with the corresponding symmetries. 

One was related to properties of strange partners of better known particles, $K$ partners to pions,
$K^*$ to $\rho$ mesons, $\Lambda, \Sigma$ to nucleons, etc. It lead Gell-Mann and Neeman
to the so called ``eightfold way",  their classification via the $octet$ representation of the flavor $SU(3)$  group. 
Three flavors were, of course, $u,d,s$ quarks, and spectra were well described via 
$SU(3)$-symmetric expression, perturbed by symmetry-breaking mass terms. 
This approach culminated by prediction  of the baryon decuplet -- another representation of 
$SU(3)$ group -- with subsequent experimental discoveries of all its members, completed by the $sss$ $\Omega^-$
baryon.  Needless to say, with the discovery of $c$ and then $b$ quarks, nonrelativistic
description of corresponding hadrons followed.

The description of masses were complemented by similar description of other properties, such as baryon magnetic moments. Note that Dirac particle has magnetic moment proportional to ``magneton" unit $\sim e/m$, inversely proportional to mass. Baryon magnetic moments 
suggested that quarks have mass about 1/3 of the nucleon mass. 
Quark models of hadrons indeed use such ``constituent quark masses".

Another insight into hadronic and quark properties was proposed even a bit earlier, in 1961, in
 \cite{Nambu:1961tp}. This model, called NJL, pointed out  the notion of chiral symmetry and its
spontaneously breaking. The authors postulated existence of 4-fermion interaction, with some coupling $G$, strong enough to make a superconductor-like gap even in  vacuum, at the surface of the Dirac sea. The second (and the last)
parameter of the NJL model was the cutoff $\Lambda \sim 1\, GeV$, below which their hypothetical
attractive  4-fermion interaction operates. 

Although there are many ``quark models"\footnote{In this chapter we will not discuss
$glueballs$: we however will discuss them in later chapters, in connection to flux tubes and Pomerons.
}, of different degree of sophistication, their
common elements are:\\
(i) the constituent quark masses;\\
(ii) interquark potential, including confining linear $\sim r$ and Coulomb $\sim 1/r$ potentials\\
(iii) and various ``residual interactions", mostly of 4-fermion type\\
The method of choice, like in atomic and nuclear physics, is selection of
certain large basis of the wave functions, representing the Hamiltonian as a matrix
in it, with its subsequent diagonalization. 

While the constituent quark masses constitute most of the hadronic masses, the potentials (ii) 
contribute typically $O(1/10)$ of them, and the residual interactions are very important only
in some ``exceptional" hadrons, such as $\pi,\eta'$ mesons.

The field of hadronic spectroscopy was, for a long time, reduced to mesons ($\bar q q$)
and baryons ($qqq$) states, while larger systems -- such as pentaquarks $q^4 \bar q $
and dibaryons $q^6$, predicted by such models, were never seen. 
Only with inclusion of heavy quarks, pentaquarks were  relatively recently discovered. 

Completing this introductory section, let me also mention the ``QCD cousins" with the number of colors $N_c\neq 3$. Obviously, there are two directions in which $N_c$ can be changed, up and down, and both are rather interesting.

If  $N_c\rightarrow \infty $, the baryons containing $N_c$ quarks need to become very heavy.
Skyrme's view on baryons, as classical solitons made of mesonic fields, become justified in this limit. This direction we will further discuss in chapter on holographic models \ref{sec_holo}.

One can also reduce $N_c$ from 3 to 2, considering ``two-color QCD". Due to specific
properties of the   $SU(2)$ group (two-index $\epsilon_{\alpha\beta}$) quarks and antiquarks
are not really different, and it has additional Pauli-Gursay symmetry relating them.
This symmetry has multiplets in which mesons and baryons enter together. In particular,
with $u,d$ quarks the chiral symmetry breaking leads to 5 massless Goldstone modes:
3 pions, diquark $ud$ and its antiparticle. 

One may wander how parameters of quark models depend on $N_c$, and whether
extrapolation from $N_c=2$ to the real world with
$N_c=3$ would be possible. It was pointed out \cite{Rapp:1997zu}
in connection to color superconductivity,  that ``residual forces", the Coulomb and instanton-induced ones, scale as 
\be {V_{qq} \over V_{\bar q q}} \sim  {1 \over N_c-1 } \ee
so they are: (i) negligible in the $N_c\rightarrow \infty $ limit; (ii) equal to -1 in QED when $N_c=0$, (iii)
equal when  $N_c=2$; and, finally,  (iv)  for $N_c=3$ 
they are different by a factor 1/2. So, the real world is equidistant, from small and large number of color limits!

\section{Light-front observables}
Unlike atomic or nuclear structure, the structure of hadrons is mostly experimentally probed in relativistic kinematics. Take as an example the pion formfactor, at large momentum transfer. The most symmetric frame is in which initial pion has momentum $Q/2$ and outgoing one
$-Q/2$. If $Q\sim few\, GeV\gg m_\pi$ both momenta are large compared to the pion mass,
so it moves relativistically. 

The iconic Deep Inelastic Scattering (DIS) of highly virtual photon
on a nucleon gives us the so called Parton Distribution Functions (PDFs) 
$f_i(x,Q^2)$, where $x$ is fraction of the total momentum carried by the quark absorbing the photon, and $Q^2$ is the photon virtuality,
and $i$=gluons, quarks and antiquarks of different flavors.  Traditionally,
DIS is compared to a microscope, the better is its spatial resolution $\sim 1/Q$ the more
partons it can resolve. 

Technical tool to describe the $Q$-dependence of PDFs
was developed in pQCD and is known as Dokshitzer-Gribov-Lipatov-Altarelli-Parisi
 (DGLAP) evolution equation.  It includes certain ``splitting functions", describing probablity of a parton splitting into two. Also important to note, as any other single-body  Boltzmann equation, DGLAP is based on implicit assumption that higher correlations between
 bodies are small and can be neglected. 
 
 Let me only comment here that PDFs are, by their nature, single-parton observables,
 in a many-parton system.  DGLAP
 has the form of a kinetic equation, and, surprisingly recently \cite{Kharzeev:2017qzs}
 noticed that as for any  kinetic equation one can introduce the notion of $entropy$, its inclrease with time, with eventual
   approach of its maximum, the equilibrium state. 

Entropy is property of the {\em single-parton density matrix}, but the
hadrons themselves are of course pure quantum states, with  wave functions.
They too can be defined using light-front coordinates ( LFWF) as $\psi(x_i,\vec p_\perp^i)$ with $i=1..N$, $N$ total number of partons.  As we will discuss below, the wave function includes sectors with different $N$: e.g. baryons have the simplest sector with $N=3$, and then $ N=5,7...$ as extra quark pairs can be added\footnote{Note that this feature is not new or exclusive
to hadrons: e.g. atomic and nuclear states are also described as ``closed shells", plus some
number of ``quasiparticles", plus any number of particle-hole pairs excited by the 
``residual" part of the Hamiltonian.}.

Completing this introductory section, let me mention several physics phenomena we will 
be discuss below in the chapter. Perhaps the most famous of them is the so called ``spin crisis":
related to the question how the total spin of the proton, of course equal to 1/2, is shared between
different kinds of partons. 

I would prefer to reformulate this puzzling issue into several: (i) What is the contribution of  the valence $d$ quark to the spin? (ii) What is the contributions of the sea, of the different flavors of quarks and antiquarks?  (iii) Why is the sea of the polarized proton polarized at all? (Iv) Why is
the sea polarized in the isospin? 

\section{Quark models on the light front: mesons in the $\bar q q$ sector}
Let me start with the general kinematics and normalization condition of the light front wave functions.
They are defined in momentum representation, with $x_i$ being momentum fractions along the 
beam direction $\vec p_{||}=x P$. The integration measure, which defines the orthogonality
condition in the sector with $i=1..N$ partons is defined as
\be 
\int \big(\prod_i dx_i {\vec{dp_\perp^i} \over (2\pi)^2 }\big)
 \big( \Pi_i x_i  \big) \delta\big(\sum_i x_i-1 \big) \delta(\sum_i \vec p_\perp^i) \psi_n^*(x_i, \vec p_\perp^i ) \psi_m(x_i,\vec p_\perp^i )=\delta_{mn}  \label{eqn_orthogonality}
\ee 
Note that apart of two delta functions expressing momentum conservation, there is also a 
product of all momentum fractions, introduced for convenience. The so called ``asimptotic" wave funciton
corresponds, in this normalization, is independent of $x_i$.
Note also that states
in sectors with different $N$ are assumed to be orthogonal to each other by definition. 

The simplest case is the 2-body sector of the mesons.  In this case the restriction on momentum fraction becomes just
$ x_1+x_2=1 $,  and therefore the wave function is a function of a single
longitudinal  variable. It is convenient to select the {\em asymmetry} variable $s$ defined via
\be s\equiv  x_1-x_2 \ee
The inverse relations, expressing momentum fractions via it are
$$ x_1={1+s \over 2},\,\,\,  x_2={1-s \over 2} $$

The basic setting we will follow is due to \cite{Jia:2018ary}.  The Hamiltonian 
has
 four  terms including (i) the effective quark masses coming from chiral symmetry breaking; (ii) the longitudinal confinement;
 (iii) the transverse motion and confinement;   and, last but not least, (iv) the NJL 4-quark effective interaction
 \be H=H_M+H_{conf}^{||}+ H_{conf}^\perp+H_{NJL} \ee
 $$ H_M={M^2 \over x_1}+{\bar M^2 \over x_2} $$
 $$ H_{conf}^{||}={\kappa^4 \over (M+\bar M)^2} {1 \over J(x) }\partial_x J(x) \partial_x $$
 $$ H_{conf}^\perp=\vec k_\perp^2 \big({1 \over x_1}+{1 \over x_2}\big) +\kappa^4 x_1x_2 \vec r_\perp^2 $$
where $M,\bar M$ are masses of quark and antiquark, $\kappa$ is the confining parameter, and the integration measure is $J(x)=x_1x_2=(1-s)^2/4$,
 $\vec k_\perp,\vec r_\perp$ are transverse momenta and coordinates. 
 If the masses are the same ( no strange quarks) however,
 one can simplify it $${1\over x_1}+{1 \over x_2}={1 \over x_1x_2}={4 \over (1-s^2)}  $$
 and therefore the matrix element of $H_M$ simply lack the factor $(1-s^2)$ normally present in the integration measure. The value of their parameters are
 \be M_q = .337 \, GeV , \,\,\,\,\,
\kappa = .227 \, GeV
 \ee

In Ref.\cite{Jia:2018ary}  the basis is designed so that  the part of Hamiltonian, other than $H_{NJL}$, is diagonal. 
In subsequent paper \cite{Shuryak:2019zhv}  this requirement is dropped and the basis is simply 
given by Jacobi
polynomials $P^{1,1}_n(s)$. Other simplifications include dropping transverse momentum dependence and using only
topology-induced 't Hooft 4-fermion operator. Therefore, it is absent in the $\rho$ meson, and enter in effective Hamiltonian for
$\pi$ and $\eta'$ with the  opposite signs (we already seen it in the chapter on correlation functions).

\begin{figure}[h!]
\begin{center}
\includegraphics[width=6cm]{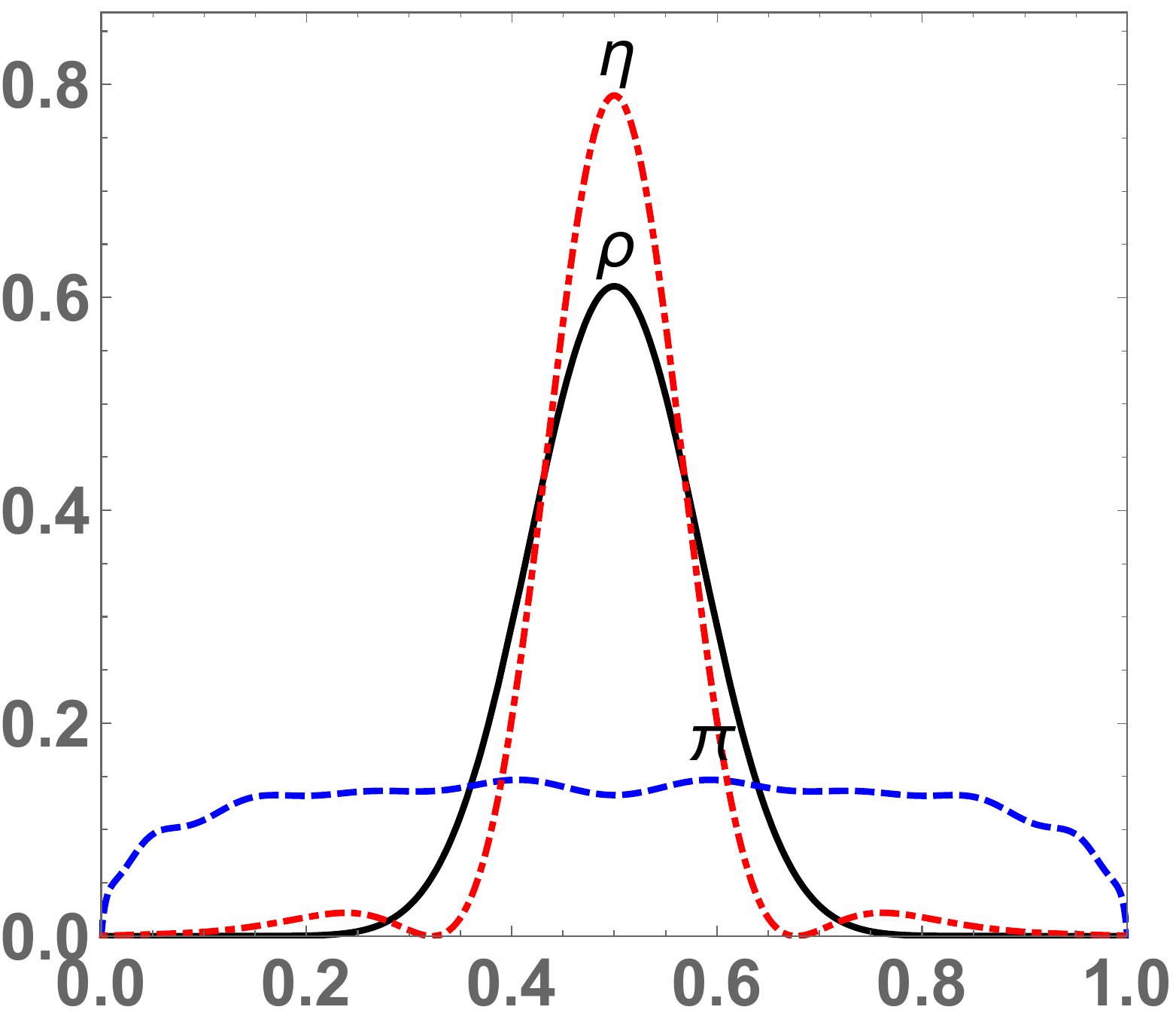}
\includegraphics[width=7.1cm]{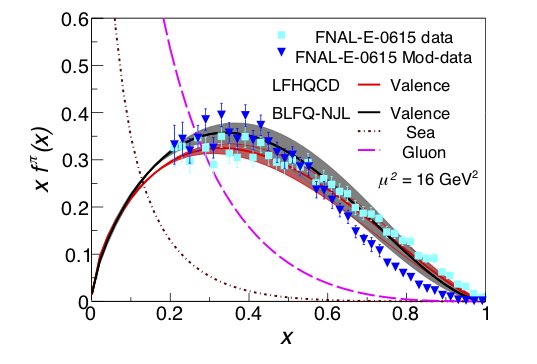}
\caption{Upper: momentum distribution for pion, rho and eta-prime mesons, calculated in the model. 
Lower (from \protect\cite{Geesaman:2018ixo}) comparison between the measured pion PDF (points)  and the JV model (lines).}
\label{fig_pi_rho_eta}
\end{center}
\end{figure}

The results are shown in 
Fig.\ref{fig_pi_rho_eta}. Note that the predicted PDF for mesons is just the wave function squared
(there are no extra variables to integrate). The quark momentum distribution for
$\rho$ meson (in which the 4-fermion interaction is presumed to be absent) is peaked near the symmetric point $x=1/2,s=0$. The pion one, in contrast, has a completely different flat shape. The $\eta'$ PDF is 
deviating to the opposite direction, getting even more concentrated in $x$\footnote{Note that it is probability in momentum representation. The coordinate one does the opposite, making pion very compact and $\eta'$, with a local repulsive potential at its core, of larger size.}.

As one compares the lower plot to our result, one should keep in mind the fact that
 the PDF include also contribution from sectors with the quark number larger than 2,
 while ours (so far) do not.

\section{Quark models on the light front: baryons as $qqq$ states}
The quark wave functions we will be discussing are defined as, for Delta 
baryon 
\be | \Delta^{++} \rangle \sim \psi_\Delta(x_i)  |u^\uparrow(x_1) u^\uparrow(x_2) u^\uparrow(x_3)  \rangle \ee
and for the proton
\be | p \uparrow \rangle \sim \psi_p(x_i) \big( |u^\uparrow(x_1) u^\downarrow(x_2) d^\uparrow(x_3)  \rangle \label{def_p_wf} \ee
$$ - | u^\uparrow(x_1) d^\downarrow(x_2) u^\uparrow(x_3)  \rangle
\big) $$
and in what follows we will focus on the former component, in which the $d$ quark has the last momentum fraction $x_3$. For a review see \cite{Braun:2006hn}.

Here we follow \cite{Shuryak:2019zhv} paper. Three $x_i,i=1,2,3$ sums to one, so one needs
 two kinematical variables. The ones used 
    in this case are defined as
   \be  s= {x_1-x_2 \over x_1+x_2}, \,\,\,\,\, t=x_1 + x_2 - x_3 \ee
   in terms of which 
   \be x_1={ (1+s) \over 2} {(1+t )\over 2}, \,\,\, x_2={ (1-s) \over 2} {(1+t) \over 2},    \ee
 and the integration measure   
     \be \int (\prod_i dx_i ) \delta(\sum_i x_i-1) (\prod_i x_i )...= \ee $${1\over 2^5} \int_{-1}^1 ds (1-s) (1+s)  \int_{-1}^1 dt (1-t) (1+t)^3 ... $$
     is  factorized. Therefore one can split it into two and select appropriate  functional basis of two Jacobi functions $$\psi_{n,m}(s,t)\sim P^{1,1}_n(s) P_m^{1,3}(t)$$
The Hamiltonian is simple generalization of that by Jia-Vary
\be H_{conf}=-{\kappa^4\over J(s,t) M_q^2} \big[ {\partial \over  \partial s} J(s,t) {\partial \over  \partial s}+{\partial \over  \partial t} J(s,t) {\partial \over  \partial t}\big]
\ee
with the measure function $J(s,t)$  appearing in the $s,t,$ integration. Note that coefficient 4 in denominator is missing: this is cancelled by factor 4
coming from a difference between derivatives in $x$ and $s,t$ variables. 

The third (and the last) effect we incorporate in this work
is the topology-induced 4-quark interactions.  Note that topological 't Hooft Lagrangian
 is flavor antisymmetric. This means that it does not operate e.g. in 
 baryons made of the same flavor quarks, like the $\Delta^{++} =uuu$.
 Another reason why the 't Hooft vertex should be absent is in any states in which all chiralities of quarks are the same, like $LLL $. For both these reasons, $\Delta^{++}$ is not affected by topology effects,
  therefore serving as a  benchmark (like the $\rho$ meson did in the
 previous sections). 
 
\begin{figure}[h!]
\begin{center}
\includegraphics[width=6cm]{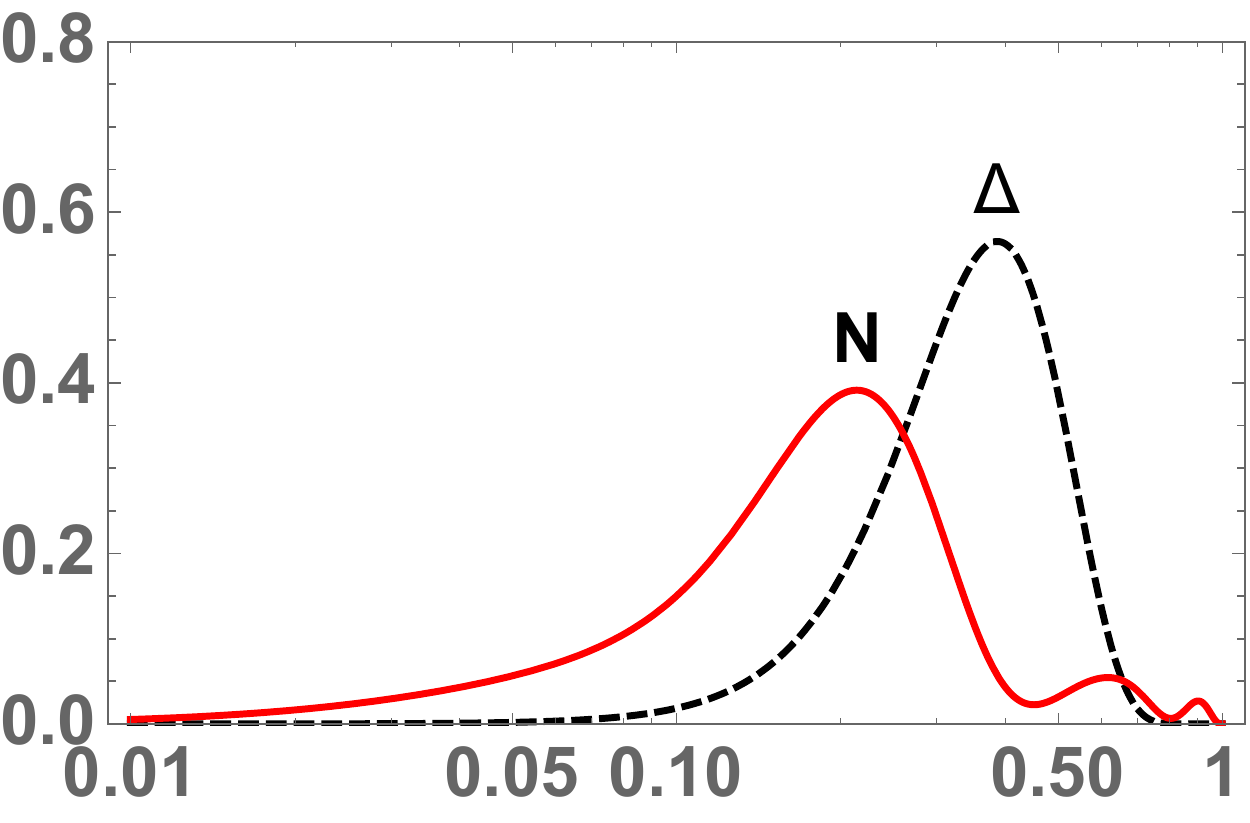}
\includegraphics[width=6.3cm]{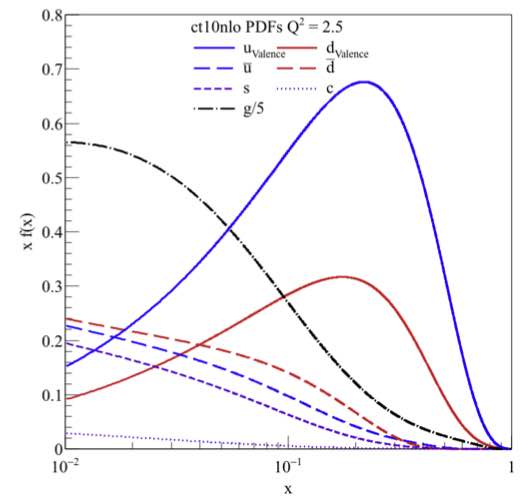}
\caption{
Upper: our calculation of the $d$ quark distribution in the Nucleon times $x$, $x d(x)$ (red, solid)
and Delta (black, dashed) states. For comparison, the lower plot shows empirical structure functions
(copied from \protect\cite{Geesaman:2018ixo} ), where the valence $xd(x)$ is also shown in red. }
\label{fig_d_structure}
\end{center}
\end{figure}

 For discussion of the masses etc see the original paper. The main results are the PDF and the wave functions. The distribution in momentum of the $d$ quark is obtained by integration over $s$
\be d(x_3=1-2t)=\int_{-1}^1 ds J(s,t) \Psi_N^2(s,t)
\ee  
 and it is
  shown in Fig.\ref{fig_d_structure}(upper), for Delta and Nucleon
wave functions we obtained. Two comments: (i) the peak in the Nucleon distribution moves to lower $x_d$, as compared to $x=1/3$ expected in $\Delta$ and
non-interacting three quarks; (ii)  there appears larger tail toward small $x_d$ in the nucleon,
but also some peaks at large $x_d$
Both are unmistakably the result of local $ud$ pairing (strong rescattering) in a diquark cluster.
The Fig.\ref{fig_d_structure} (lower), shown for comparison, includes the empirical valence $xd(x)$  distribution, also shown by red solid line. The location of the peak (i) roughly corresponds to data, and (ii)
the presence of small-$x_d$ tail is also well seen (recall that what is plotted is
distribution times $x$). The experimental distribution is of course
much smoother than ours.
 It is expected feature: our wave function 
 is expected to be ``below the pQCD effects", at resolution say $Q^2 \sim 1\, GeV^2$, while
 the lower plot is at $Q^2=2.5\, GeV^2$, ant it includes certain amount of pQCD gluon radiation.
 contains higher order correlations between quarks. 
 
In order to reveal the structure, 
one can of course compare the wave functions without any integrations, as they depend
on  two variables $s,t$ only.  Such plots, of $J(s,t) \Psi^2(s,t) $, we show  In Fig.\ref{fig_Delta_N_CZ}  for 
the Delta, the nucleon, and some model discussed  in \cite{Chernyak:1987nv} (see Appendix). While the Delta
shows a peak near the symmetry point $x_1=x_2=x_3$ as expected, without any other structures, 
our Nucleon WF indicate more complicated dynamics. Indeed, there appear several bumps, most prominent near $s\approx 1,t\approx 1$ which is $x_1\approx 1$. Such strong peaking corresponds to large momentum transfer inside the $ud$ diquark clusters. Yet there is also the peak in the middle, roughly
corresponding to that in Delta. So, the nucleon wave function is a certain coherent mixture of a three-quark 
and quark-diquark components.

\begin{figure}[h!]
\begin{center}
\includegraphics[width=5cm]{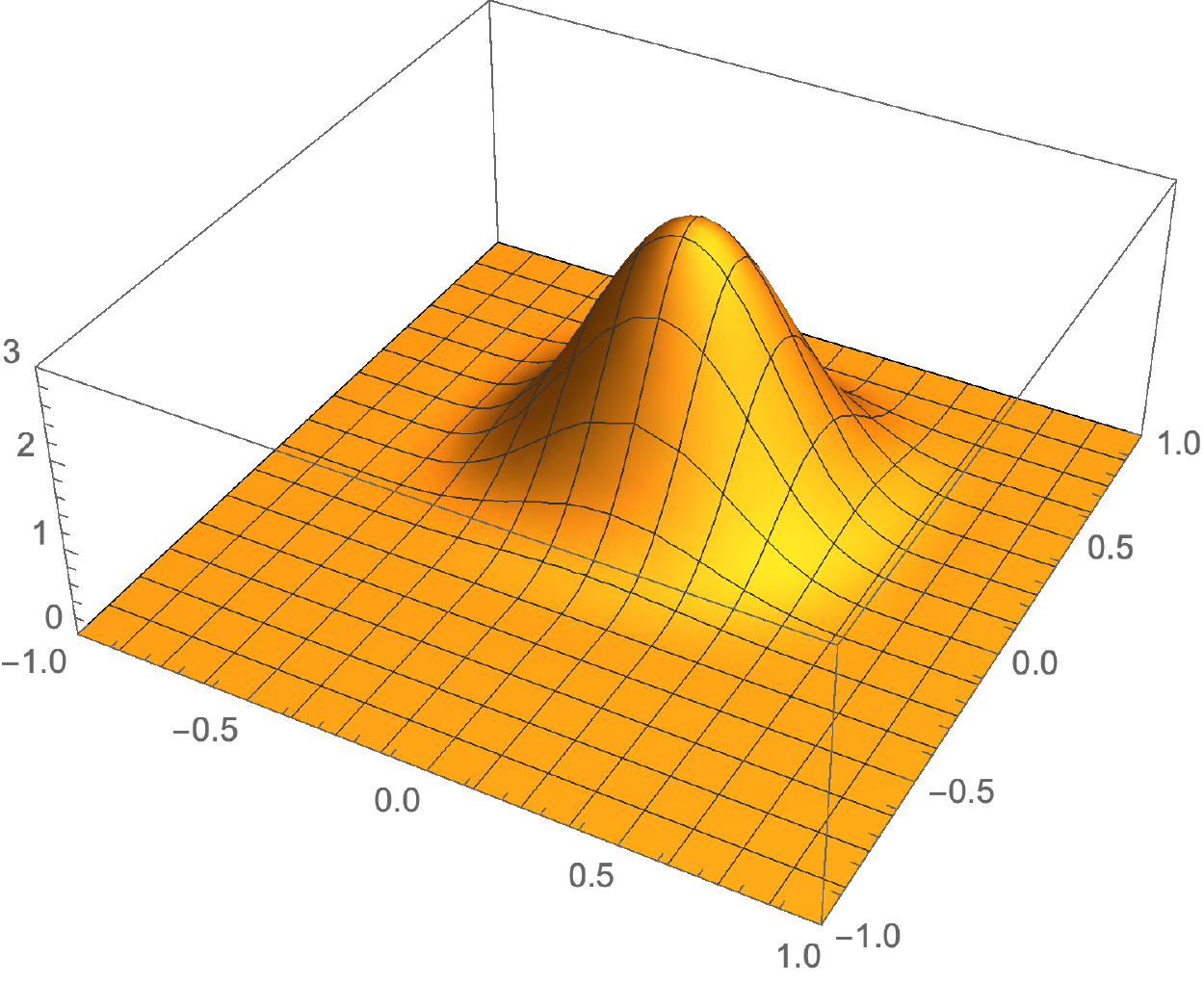}
\includegraphics[width=5cm]{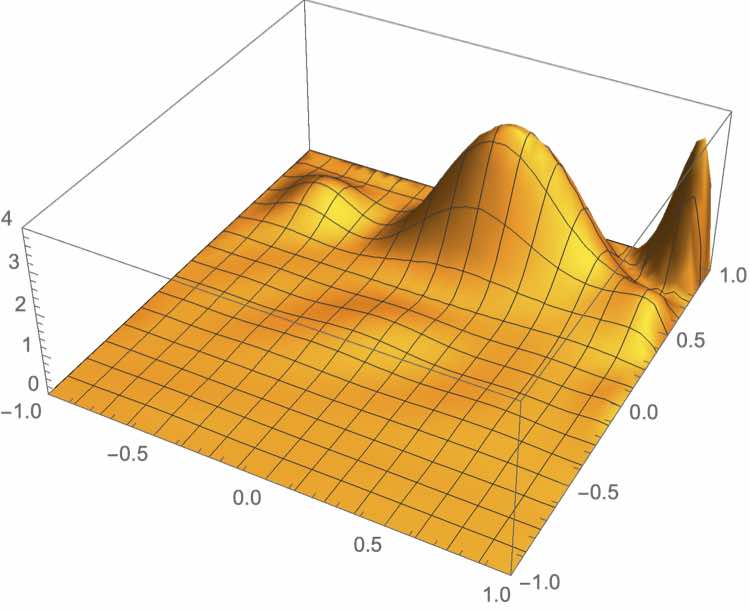}\\
\caption{The probability distribution $J(s,t)\Psi^2(s,t)$ in $s,t$ variables, for the Delta
and Nucleon lowest states, as calculated from the model. 
 }
\label{fig_Delta_N_CZ}
\end{center}
\end{figure}
\begin{figure}[h!]
\begin{center}
\includegraphics[width=7cm]{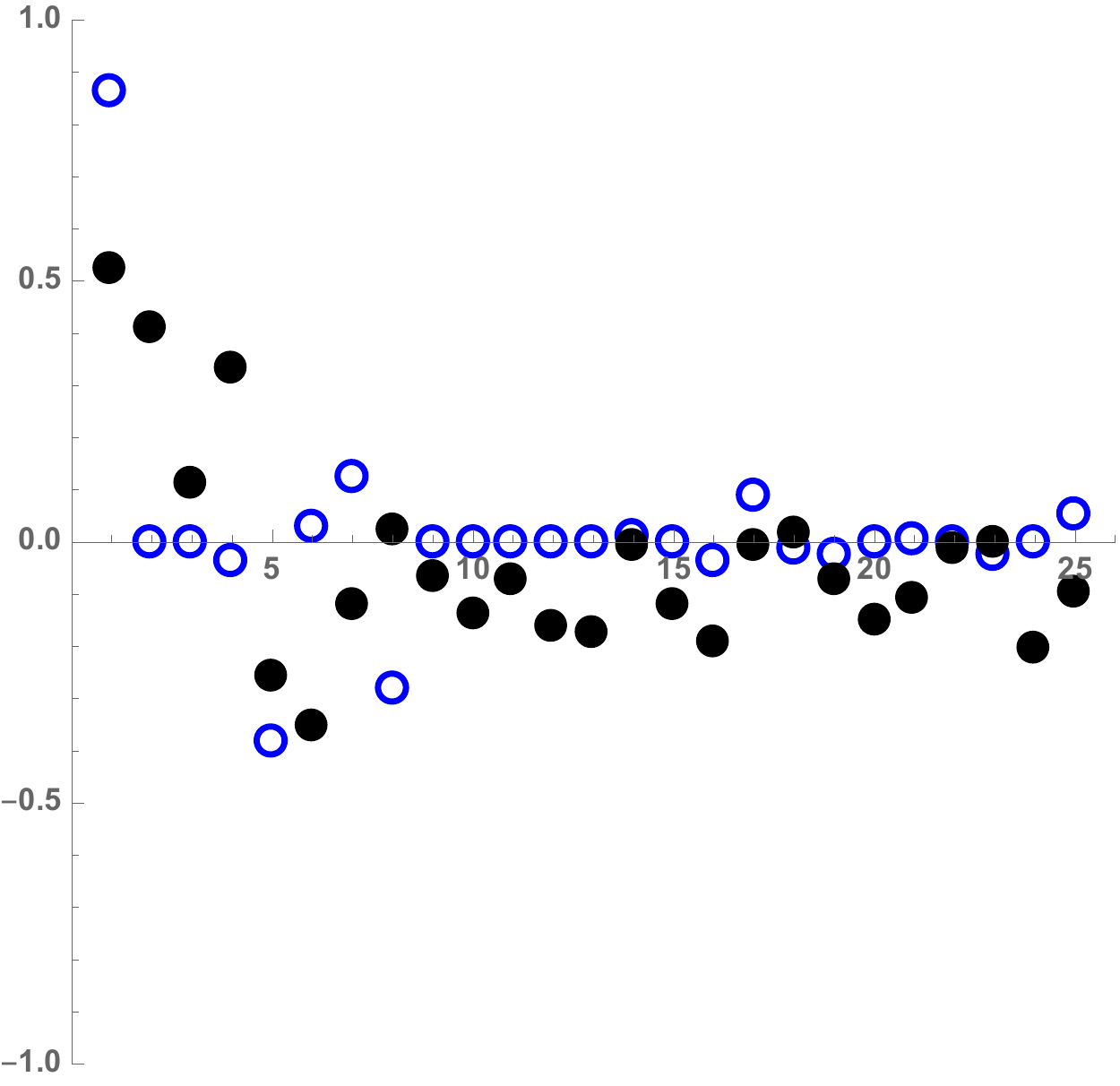}\\
\includegraphics[width=5cm]{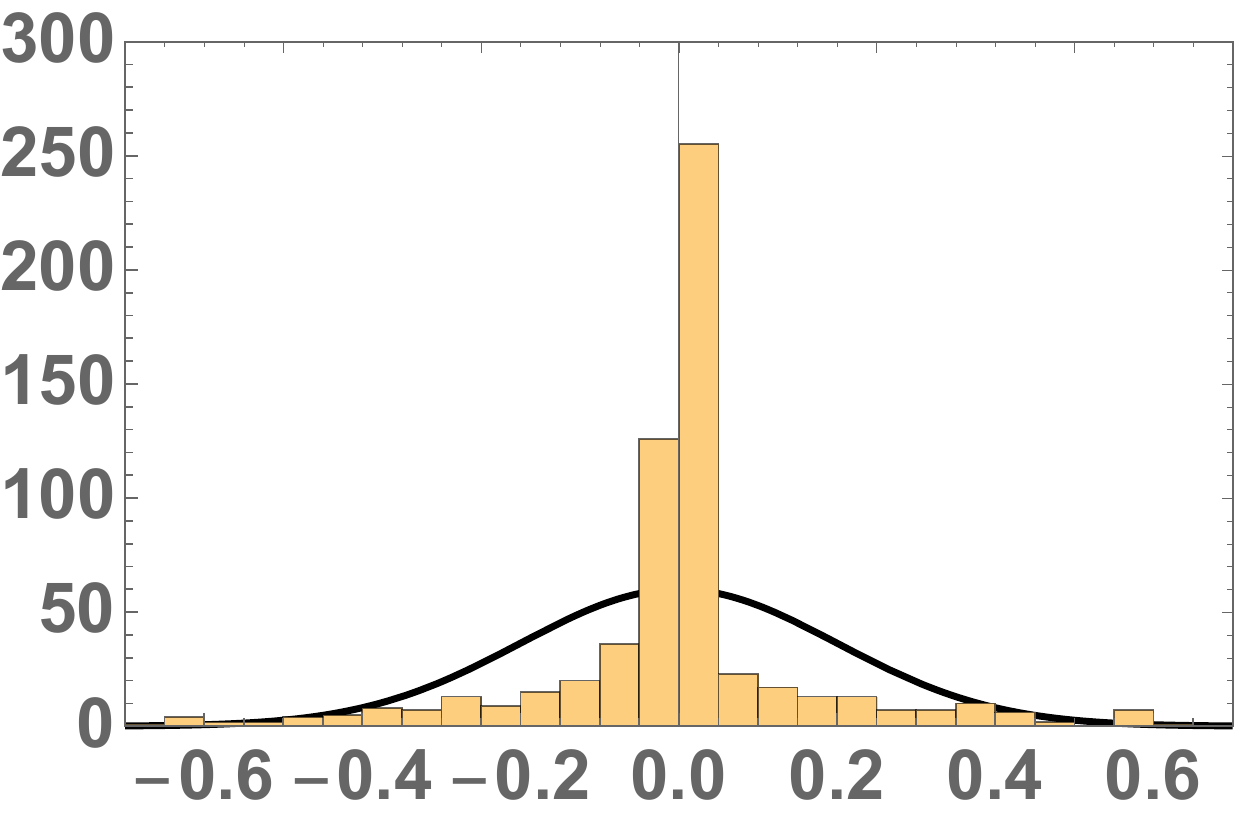}
\includegraphics[width=5cm]{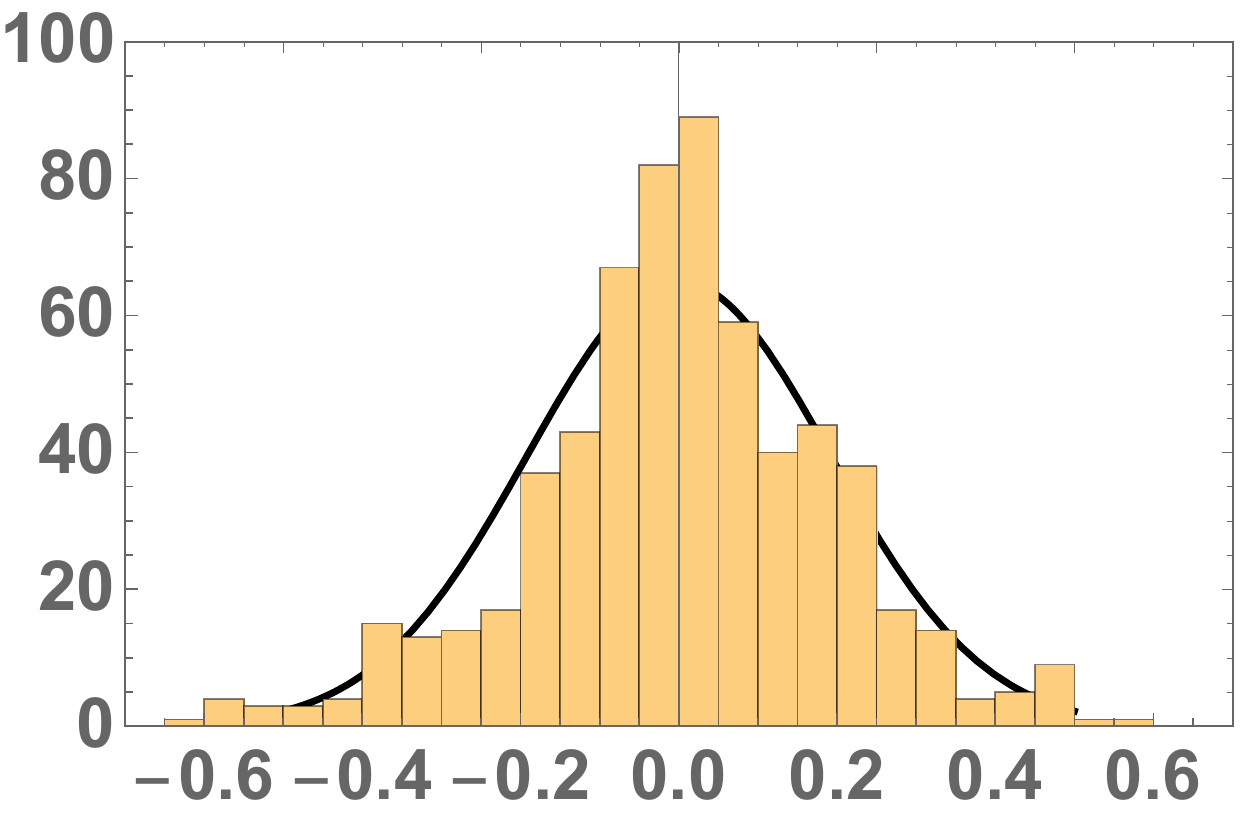}
\caption{Upper: The coefficients of the wave function $C^\alpha_n $ (the order of basis functions is specified in Appendix), for the ground state Delta (open points) and the Nucleon (closed points) states. Two lower plots contain histograms of these coefficients for 25 lowest
 Delta and Nucleon states, respectively. The black lines in the background are the Gaussians with
 the appropriate width. }
\label{fig_harmonics}
\end{center}
\end{figure}

Additional information about the nucleon wave function may be obtained from
the amplitudes corresponding to particular basis states.  
The wave function coefficients $C^\alpha_n $, 
defining expansion in the basis functions
$$ | \alpha >=\sum_n C^\alpha_n | n >$$ 
  are shown in Fig. \ref{fig_harmonics}.
  The upper plot compares those for the ground state
    Delta and Nucleon channels.
Note first that the largest coefficients are the first (corresponding to trivial $\psi_{0,0}(s,t)\sim const(s,t)$). Furthermore, for the Delta it is close to one, while it is only $\sim 1/2$ for the nucleon. The fraction of ``significant coefficients" is much larger  for $N$.   The nucleon wave function has a nontrivial tail toward  
 higher $n,l$ harmonics
which does not show any  decreasing trend.
One may in fact  conclude that  convergence of the harmonic expansion is not there. This 
 can be traced to apparent peak  near $x=1$, perhaps the pointlike residual interaction leads
 to true singularity there. 

Two lower plots of  Fig. \ref{fig_harmonics} address the distribution of the wave function coefficients $C^\alpha_n $ in these two channels, without and with the residual 4-quark interaction. It includes not the ground state but the lowest 25 states in each channel.
As it is clear from these plot, in the former (Delta) case the distribution is 
very non-Gaussian, with majority of coefficients being small. The latter (Nucleon) 
case, on the other hand, is in agreement with Gaussian. In other quantum systems,
e.g. atoms and nuclei, Gaussian distribution of the wave function coefficients $C^\alpha_n $  is usually taken as 
a  manifestation of ``quantum chaos". In this language, we conclude that
our model calculation shows that  the residual 4-quark interaction
 leads to chaotic motion of quarks, at least
inside the  Nucleon resonances. (If this conclusion surprises the reader,  we remind that
the same interaction was shown to  produce chaotic quark condensate in vacuum. In particular,
numerical studies of Interacting Instanton Liquid in vacuum has lead to 
Chiral Random Matrix theory of the vacuum Dirac eigenstates near zero,
accurately confirmed by lattice studies.)  

\section{Quark models on the light front: pentaquarks and the 5-quark sector of baryons }
Here we continue to follow \cite{Shuryak:2019zhv} paper.
The kinematics in the
{\bf 5-particle sector} has four variables for longitudinal momenta $s,t,u,w$, collectively called $z_i,i=1,2,3,4$ are defined by 
\be   s = { x_1 - x_2 \over x_1 + x_2}, \,\,\,\, 
t = {x_1 + x_2 - x_3 \over x_1 + x_2 + x_3} , \ee $$
 u = {x_1 + x_2 + x_3 - x_4 \over x_1 + x_2 + x_3 + x_4 } , \,\,\, w=x1 + x2 + x3 + x4 - x5 $$
 The principle idea can also be seen from the inverse relations

$$ x_1  = {1 \over 2^4} (1 + s) (1 + t) (1 + u) (1 + w) $$
$$ x_2  =  {1 \over 2^4} (1 - s) (1 + t) (1 + u) (1 + w) $$
$$ x_3  =  {1 \over 2^3}(1 - t) (1 + u) (1 + w), $$
\be x_4  =  {1 \over 2^2} (1 - u) (1 + w) \ee
$$ x_5  = 1 - x_1 - x_2 - x_3 - x_4= {1 \over 2} (1-w) $$

 The integration measure follows the previous trend, has the form of product of factors depending on single variables of the set, namely
\be  \int_{-1}^1 {ds dt du dw \over 16777216} (1 - s) (1 + s) (1 - t) (1 + t)^3  (1 - u) (1 + u)^5 (1 - 
   w) (1 + w)^7... \ee
   The orthonormal polynomial basis to be used is then provided by 
   the product of appropriate Jacobi polynomials
   \be  \psi_{lmnk}(s,t,u,w)\sim P_l^{1,1}(s)P_m^{1,3}(t)P_n^{1,5}(u)P_k^{1,7}(w) \label{eqn_basis} \ee
   with normalization constants defined to satisfy (\ref{eqn_orthogonality}). 
   
  Without inclusion of 4-quark residual interaction we get multiple penrtaquark states, with the lowest mass  
\be  M_{min.penta}=2.13 \, GeV \ee
To get this number in perspective, let us briefly remind the history of the light pentaquark search.
In 2003 LEPS group reported observation of the first pentaquark called $\Theta^+=u^2d^2\bar s$  with surprisingly  light mass, of only $1.54 \,GeV$,
$0.6 \, GeV$ lower than our calculation (and many others) yield. Several other experiments were also quick to report observation of this state, till
other experiments (with better detectors and much high statistics) show this pentaquark candidate does not really exist. 
 Similar sad experimental status persists for all  6-light-quark dibaryons
, including the flavor symmetric $u^2d^2s^2$ spin-0 state much discussed in some theory papers. 
Yet neither instanton liquid nor lattice studies never found any indication for light pentaquarks or dibaryons:
in fact $ud$ diquarks strongly repel each other. 

The 4-quark interaction relates the 3-q of the baryons with the 5-q pentaquark states. 
\begin{figure}[htbp]
\begin{center}
\includegraphics[width=5cm]{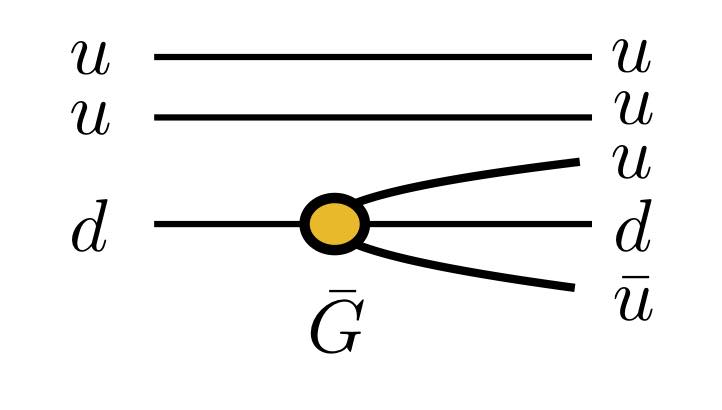}
\caption{The only diagram
in which 4-quark interaction connects the 3 and 5 quark sectors, generating
 the $\bar u$ sea.}
\label{fig_3_to_5}
\end{center}
\end{figure}

\begin{figure}[h!]
\begin{center}
\includegraphics[width=6cm]{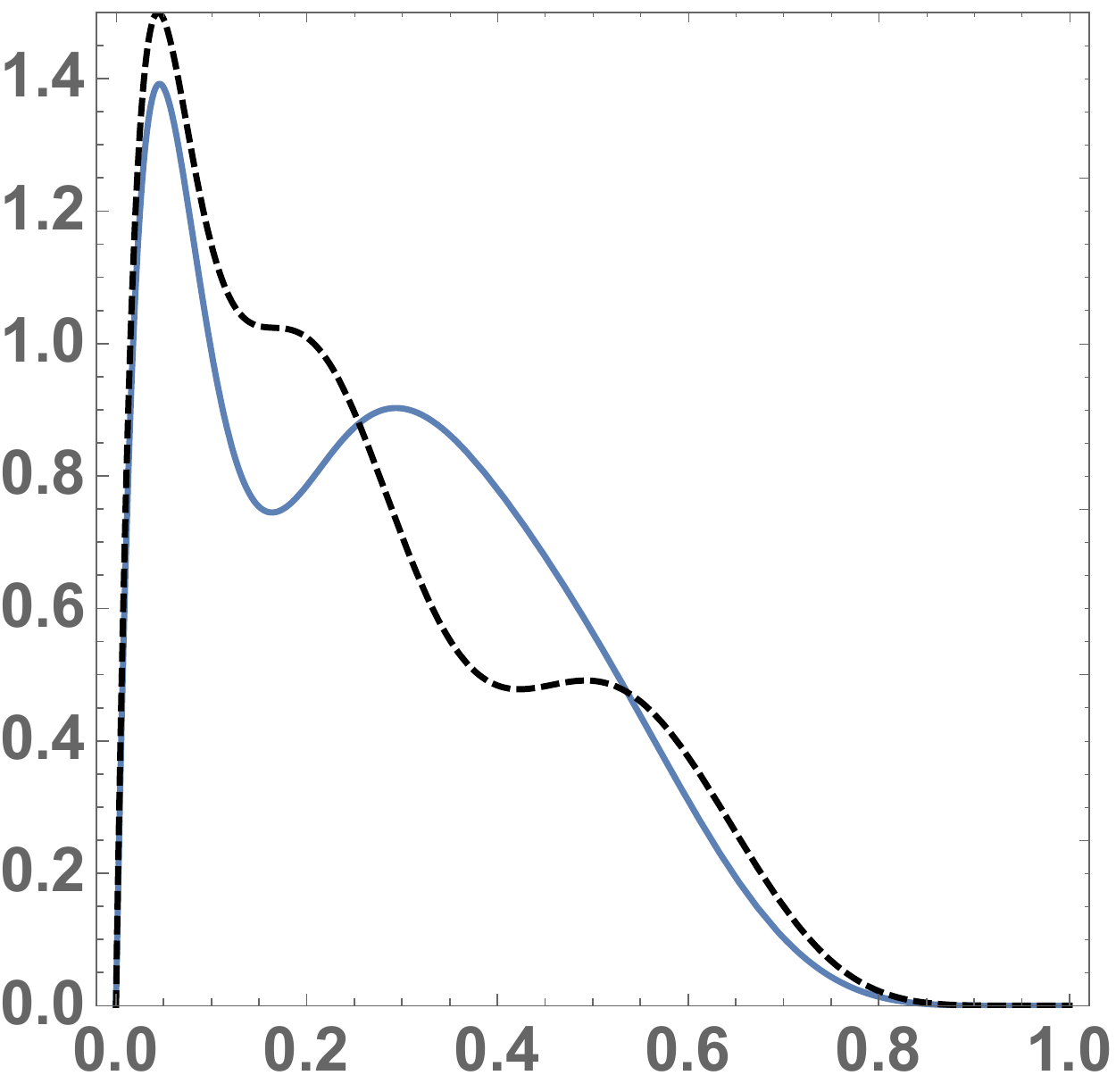}
\caption{The distribution over that $\bar u$ in its momentum fraction, for the
Nucleon and Delta 5-quark ``tails" (solid and dashed, respectively).}
\label{fig_antiu}
\end{center}
\end{figure}

The Hamitonian matrix element corresponding to the diagram shown in Fig\ref{fig_3_to_5}
we calculated between the nucleon and each of the pentaquark wave functions, defined above, 
by the following 2+4 dimensional integral
over variables in 3q and 5q sectors, related by certain delta functions
 \be \langle N | H | 5q,i  \rangle=\bar G \int ds dt J(s,t) ds' dt' du' dw' J(s',t',u',w') \ee
$$ \psi_N(s,t) \delta(x_1-x_1') \delta(x_2-x_2') 
 \psi_i (s',t',u',w') $$
The meaning of the delta functions is clear from the diagram, they are of course expressed via proper integration variables and  numerically
approximated by narrow Gaussians. After these matrix elements are calculated, the
5-quark ``tail" wave function is calculated via standard perturbation theory expression 
\be \psi_{tail}(s',t',u',w')= -\sum_i { \langle N | H | 5q,i \rangle \over M_i^2-M_N^2} \psi_i (s',t',u',w')  \label{eqn_tail} \ee
The typical value of the overlap integral itself for different pentaquark state is $\sim 10^{-3}$,
and using for effective coupling $\bar G$ the same value as we defined for $G$ 
from the nucleon, namely $\sim 17 GeV^2$, one finds that admixture of several pentaquarks
to the nucleon is at the level of a percent.  The normalized distribution
of the 5-th body, namely $\bar u(x)$, over its momentum fraction is shown in Fig.\ref{fig_antiu}.
One can see a peak at $x_{\bar u}\sim 0.05$, which looks a generic phenomenon. The oscillations at large $x_{\bar u}$ reflect strong correlations in the wave function
between quarks, as well as perhaps indicate the insufficiently large functional basis used.
This part of the distribution is perhaps numerically unreliable.

As originally emphasized by Dorokhov and Kochelev \cite{Dorokhov:1993fc},
The 't Hooft topology-induced 4-quark interaction leads to processes
$$ u \rightarrow u (\bar d d), \,\,\,\,  d \rightarrow d (\bar u u) $$ but not
$$u \rightarrow u (\bar u u),\,\,\,\,  d \rightarrow d (\bar d d) $$
which are forbidden by Pauli principle applied to zero modes. Since there are
two $u$ quarks and only one $d$ in the proton, one expects 
this mechanism to produce twice more $\bar d$ than $\bar u$.

\begin{figure}[htbp]
\begin{center}
\includegraphics[width=6cm]{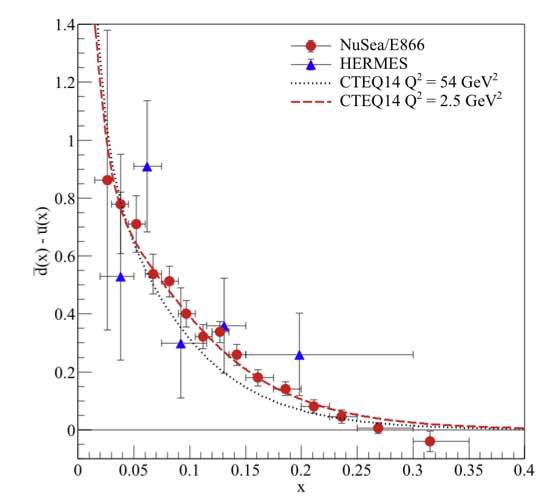}
\caption{The difference of sea antiquarks momentum distributions, $\bar d-\bar u$, from }
\label{fig_antid-antiu}
\end{center}
\end{figure}

The available experimental data, for the $difference$ of the sea antiquarks distributions  $\bar d-\bar u$ (from  \cite{Geesaman:2018ixo}) is shown in Fig.\ref{fig_antid-antiu}. In this difference
the symmetric gluon production should be cancelled out, and therefore it is
sensitive only to a non-perturbative contributions. 

Few comments: (i) First of all, the sign of the difference is indeed  as
predicted by the topological interaction, there are more anti-d than anti-u quarks;\\
(ii) Second, since 2-1=1, this representation of the data directly give us the nonperturbative antiquark production per valence quark, e.g. that of $\bar u$. This means it can be directly compared to the distribution we calculated from
the 5-quark tail of the nucleon and Delta baryons, Fig.\ref{fig_antiu}. \\
(iii)  The overall shape is qualitative similar, although our
calculation has a peak at $x_{\bar u}\sim 0.05$ while the experimental PDFs 
do not indicate it. Of course, there exist higher-quark-number sectors
with 7 and more quarks in baryons, which our calculation does not yet include: those should populate
the small $x$ end of the PDFs. \\
(iv) The data indicate much stronger decrease toward large   $x_{\bar u}$ than the calculation.\\
(v) There are other theoretical models which also reproduce the flavor asymmetry of the sea,
e.g. those with the pion cloud. In principle, one should be able to separate those and 
topology-induced mechanism (we focused here) by further combining flavor and
spin asymmetry of the sea. In particular, as also noticed in \cite{Dorokhov:1993fc}, if $d$ 
quark producing $u\bar u$ pair has positive helicity, the sea quark and antiquark from 't Hooft 4-quark operators must
have the $opposite$ (that is negative) helicity. The spin-zero pion mechanism,
on the other hand, cannot transfer spin and would produce flavor but not spin sea asymmetry.

\section{Puzzles of the antiquark sea can be understood by  topological forces}

The results of our calculation cannot be directly compared to the   sea quark and antiquark PDFs, already plotted in Fig.\ref{fig_d_structure}, as those  include large perturbative contributions,
from gluon-induced quark pair production $g\rightarrow \bar u u, \bar d d $, dominant at very small $x$. However, these processes are basically flavor and chirality-independent,
while the observed flavor and spin asymmetries of the sea indicate that there must also
exist some nonperturbative mechanism of its formation. For a general recent review see
\cite{Geesaman:2018ixo}.

As originally emphasized by \cite{Dorokhov:1993fc},
The 't Hooft topology-induced 4-quark interaction leads to processes
$$ u \rightarrow u (\bar d d), \,\,\,\,  d \rightarrow d (\bar u u) $$ but not
$$u \rightarrow u (\bar u u),\,\,\,\,  d \rightarrow d (\bar d d) $$
which are forbidden by Pauli principle applied to zero modes. Since there are
two $u$ quarks and only one $d$ in the proton, one expects 
this mechanism to produce twice more $\bar d$ than $\bar u$.

The available experimental data, for the $difference$ of the sea antiquarks distributions  $\bar d-\bar u$ (from  \cite{Geesaman:2018ixo}) is shown in Fig.\ref{fig_antid-antiu}. In this difference
the symmetric gluon production should be cancelled out, and therefore it is
sensitive only to a non-perturbative contributions. 

Few comments: (i) First of all, the sign of the difference is indeed  as
predicted by the topological interaction, there are more anti-d than anti-u quarks;\\
(ii) Second, since 2-1=1, this representation of the data directly give us the nonperturbative antiquark production per valence quark, e.g. that of $\bar u$. This means it can be directly compared to the distribution we calculated from
the 5-quark tail of the nucleon and Delta baryons, Fig.\ref{fig_antiu}. \\
(iii)  The overall shape is qualitative similar, although our
calculation has a peak at $x_{\bar u}\sim 0.05$ while the experimental PDFs 
do not indicate it. Of course, there exist higher-quark-number sectors
with 7 and more quarks in baryons, which our calculation does not yet include: those should populate
the small $x$ end of the PDFs. \\
(iv) The data indicate much stronger decrease toward large   $x_{\bar u}$ than the calculation.\\
(v) There are other theoretical models which also reproduce the flavor asymmetry of the sea,
e.g. those with the pion cloud. In principle, one should be able to separate those and 
topology-induced mechanism (we focused to in this paper) by further combining flavor and
spin asymmetry of the sea. In particular, as also noticed in \cite{Dorokhov:1993fc}, if $d$ 
quark producing $u\bar u$ pair has positive helicity, the sea quark and antiquark from 't Hooft 4-quark operators must
have the $opposite$ (that is negative) helicity. The spin-zero pion mechanism,
on the other hand, cannot transfer spin and would produce flavor but not spin sea asymmetry. 

Summarizing this chapter, let me first comment that application of some model Hamiltonians to light front wave functions of hadrons
should have been done long time ago. Yet it is just starting, with
the model presented in this paper still being (deliberately) rather schematic, but at least including
constituent quark masses, confinement and some residual interactions due to gauge topology.
We have shown that local 4-fermion residual interaction does indeed modify 
meson and baryon wave in a very substantial way. In \cite{Jia:2018ary}
  it was shown that the $\pi$ and $\rho$ meson have very different wave functions.
We now show that it is also true for
baryons: the proton and the $\Delta$ have qualitatively different wave functions.
 The imprint left by strong diquark ($ud$) correlations on the light-front wave functions
 is now established. We furthermore found evidences that Nucleon resonances 
 show features of ``quantum chaos" in quark motion. 
 We also seen the first steps toward the ``un-quenching" the  light-front wave functions, estimating multiple
 matrix elements of the mixings between 3-q and 5-q (baryon-pentaquark) components of the nucleon wave function. 
This puts mysterious spin and flavor asymmetries of the nucleon sea inside the domain
of consistent Hamiltonian calculations.


\chapter{The topological landscape and the sphaleron path } \label{chap_sphalerons}
\section{The sphalerons}

\begin{figure}[h]
\begin{center}
\includegraphics[width=8cm]{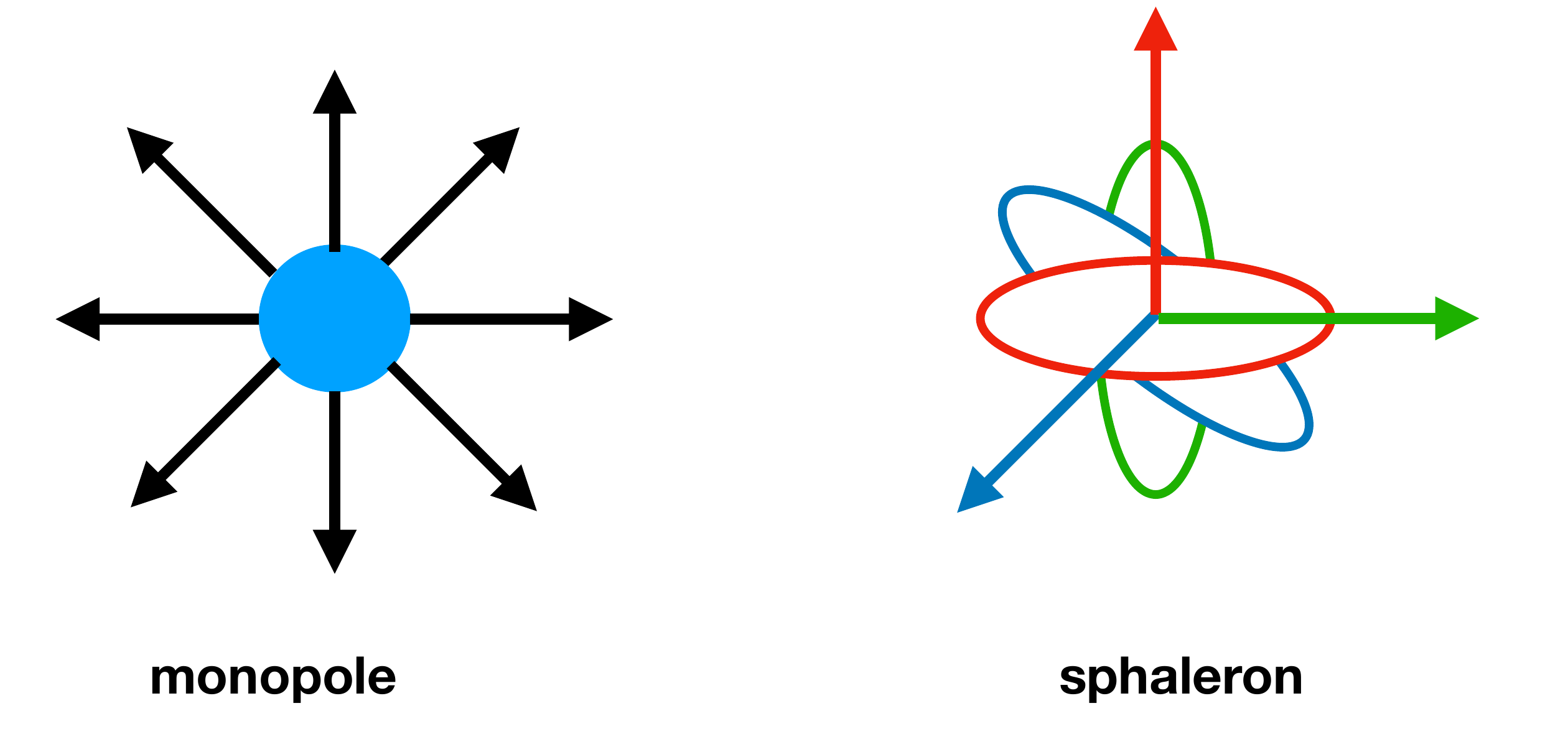}
\caption{Qualitative distinction between the magnetic field in monopole and spalerons}
\label{fig_mono_sph}
\end{center}
\end{figure}

Let me begin with a qualitative picture. Like monopoles, sphalerons are 3-d magnetic objects
made of $SU(2)$ non-Abelian gauge fields. They are ``solitons", solutions of classical Yang-Mills eqn. 
Unlike monopoles, their magnetic field lines do not go radially but in circle, see Fig.\ref{fig_mono_sph}. Three colors (red,blue and green) on the picture
 depict not the number of colors (which is two) but three adjoint fields (due to three generators of $SU(2)$). As the picture suggests, each field type rotates around a different axis.

Historically existence of such solution  have been suggested by \cite{Dashen:1974ck}
as an extended soliton for the Young-Mills-Higgs equations of motion in the
electroweak sector of the Standard Model (Weinberg-Salam theory), as  a 3d
soliton alternative
to the t'Hooft-Polyakov monopole in Georgi-Glashow model.
A decade later 
 \cite{Klinkhamer:1984di} (KM) have looked for an explicit solution, variationally
and numerically (in note-added-in-proofs). As all other non-perturbative solitons, the field is $O(1/g)$ and the mass  $O(1/g^2)$ times the relevant scale. The electroweak coupling is small and the mass of the KM sphaleron solution 
turned out to be as large as\footnote{Since it is smaller than current $\sqrt{s}$ of LHC
one may ask whether one can expect its production there. 
Unfortunately, even if one would build collider with much higher energy available, there are no chances
to produce this object experimentally, due to vanishingly small overlap with two colliding proton state.
} 
\be M_{KM}\approx 8 \, TeV   \label{eqn_KM} \ee
The fact that only numerical solution was found is  related to the fact that it is not represented by any simple spherically symmetric shape, 
as it gets deformed  by a specific direction 
taken by the Higgs field in broken electroweak vacuum with nonzero Higgs VEV. Indeed one of the components, $Z$ 
boson, does not have the same mass as the other two, $W^\pm$, which makes it ``elliptic".
For early review 
consult e.g.  \cite{McLerran:1988ja}.

It has been subsequently found that, while this configuration solves the equation of motion, it is unstable. 
Unlike other solitons discussed so far -- fluctons and instantons -- which are
 $minima$ of the action, the sphalerons are 
  {\em saddle points} in a space of all configurations. While for fluctons the  quadratic operator 
  of the fluctuations had only positive modes, and for instantons positive and zero modes,
  for sphalerons 
  one (or more)  $negative$ mode appears in the spectrum.
Thus their name, $sphaleron$, which (according to its discoverers) means  in Greek ``ready to fall".

In this chapter we will not follow these original papers. First of all, our main focus will be on QCD 
applications, where complications related with Higgs and its VEV are absent. Even in 
section  devoted to electroweak sphalerons and cosmological baryogenesis, it would be 
 sufficient to much simpler analytic solution found later for pure gauge sector 
\cite{Ostrovsky:2002cg} and known as the ``COS sphaleron".  
The way toward it was historically related to the instanton-antiinstanton
configurations, which we need to consider first.


\section{Instanton-antiinstanton interaction and the ``streamline" set of configurations} 
In the chapter devoted to interacting instanton ensembles we skipped the
discussion of the instanton-antiinstanton interaction, which we will discuss now. 
\footnote{Interaction between instantons does not exist at the classical level, as is clear from the
relation between the action and the topological charge.  }
interaction. 

Returning to the double-well potential, let us start with a simple ``sum ansatz" 
\bea
\label{qm_sum}
 q_{sum}(\tau) &=& \frac{1}{g} \left(
 \frac{1}{\tanh(\tau-\tau_I)}-\frac{1}{\tanh(\tau-\tau_A)}-1
 \right).
\eea
This path has the action $S_{IA}(T)=1/g^2(1/3-2e^{-T}+O(e^{-2T}))$,
where $T=|\tau_I-\tau_A|$. It is qualitatively clear that if the two 
instantons are separated by a large time interval $T \gg 1$, the action 
$S_{IA}(T)$ is close to $2S_0$. In the opposite limit $T\rightarrow 0$, 
the instanton and the antiinstanton annihilate and the action $S_{IA}(T)$ 
should tend to zero. In that limit, however, the IA pair is at best 
an approximate solution of the classical equations of motion and it
is not clear how the path should be chosen. 

  The best way to deal with this problem is the ``streamline" 
method \footnote{Although suggested independently, it is
actually  belong to a class of configurations known
in mathematics as  ``Lefschetz  thimble" which we briefly discussed in chapter on semiclassics.
}\cite{Balitsky:1986qn}. To define them, one starts with an extremum of the action
(
in this case, infinitely separated IA pair) 
and lets the system evolve with the ``gradient flow" to smaller action. 
The ``force" is defined as $f(\tau)=\partial S[x]/\partial x(\tau)$ and the ``streamline equation"
is sliding along the direction of the force
\be {dx(\tau) \over dt}=f(\tau) \ee
Here $t$ is extra ``sliding time", not to be confused with the Euclidean time $\tau$ on which all paths depend. At zero $t$ there are well-separated instanton and antiinstanton, at $t\rightarrow \infty $ they annihilate each other to a trivial path $x(\tau)=0$.
A set
of such paths  for the double-well instantons were obtained numerically
 in my paper
\cite{Shuryak:1987tr}.

 For the gauge theory instantons  one can start with the simplest {\em sum ansatz}
 \be {g \over 2} A^a_\mu={\bar 
 \eta_{a\mu\nu} y_I^\nu \rho^2 \over  y_I^2 ( y_I^2 +\rho^2)} +
   { \eta_{a\mu\nu} y_{\bar {I}}^\nu \rho^2 \over  y_{\bar{I}}^2 ( y_{\bar{I}}^2 +\rho^2) }  
\ee
 where, for simplicity, we selected the same radii and orientation of both solitons. Note however that
 while for a single instanton the pure gauge singularity at the origin cancels in the expression
 for the fields, it does not do so for the sum ansatz. This can be cured in the following {\em ratio ansatz}
  \be {g \over 2} A^a_\mu=
  {\bar 
 \eta_{a\mu\nu} y_I^\nu \rho^2 /  y_I^2 + \eta_{a\mu\nu} y_{\bar {I}}^\nu \rho^2 / y_{\bar{I}}^2 
 \over
 1+ \rho^2/y_{\bar{I}}^2 +\rho^2/y_{{I}}^2 }  
\ee
In search for better approximation,
\cite{Verbaarschot:1991sq}, using conformal invariance of classical Yang-Mills equation, 
mapped this problem into a $co-central$ instanton and antiinstanton with different radii $\rho_1,\rho_2$. 
This introduced a notion that the action dependence on the 4-d distance between the centers  and 
these two sizes may only come in a conformal invariant dimensionless combination\footnote{Which has the meaning of the 
geodesic  distance between points in the $AdS_5$ space, with the sizes $\rho_i$ identified with the 5-th coordinates.}
\be X^2_{conf}={ (z^I_\mu-z^{\bar I}_\mu)^2+(\rho_1-\rho_2)^2 \over \rho_1\rho_2} \ee

The gradient flow (streamline) equation has the same meaning as in QM: the
force is substitute by the current $j_\mu^a =\partial S /\partial A_\mu^a$ and
then
\be {dA_\mu^a(x)\over dt}=j_\mu^a \ee
The ``streamline" set of configurations, going along the gradient, is known in mathematics as {\em Lefschetz  thimbles},
special lines connecting extrema in the complex plain.  

In co-centrical setting it has been reduced to a single-variable and solved numerically.
Verbaarschot also discovered that the so called {\em Yung ansatz} 
\begin{eqnarray} 
ig{\cal A}_{\mu}^{Yung}(x) &=& 
ig{\cal A}_{a\mu}^{Yung}(x)\frac{\tau^a}{2} \nonumber \\
&=& {\bar {\tilde {y_2}} \over \sqrt{\tilde y_2} }
 {R \over \sqrt{R^2} } 
 {(\bar \sigma_\mu y_1-y_1^\mu)\rho_1^2 \over y_1^2 ( y_1^2+\rho_1^2)}
 {\bar R \over\sqrt{R^2}}{\bar {\tilde {y_2}} \over \sqrt{\tilde y_2}}  
 \nonumber \\
&& + {(\bar \sigma_\mu y_2-y_2^\mu)\rho_2^2 \over  y_2^2+\rho_2^2}+ 
 {\rho_1\rho_2 \over z y_1^2( y_2^2+\rho_2^2)}\nonumber \\
&& \Bigg[ (\bar \sigma_\mu y_1-y_1^\mu)-
{\bar {\tilde y_2} \over \sqrt{\tilde y_2}}\nonumber \\
&& {R \over \sqrt{R^2}} (\bar \sigma_\mu y_1-y_1^\mu) 
{\bar R \over\sqrt{R^2}}{\bar {\tilde {y_2}} \over \sqrt{\tilde y_2}}
\Bigg] , \label{eqn_yung_ansatz}
\end{eqnarray} 
is numerically close to his streamline solution.

A word of explanation of notations here: all vectors without an indicative index are SU(2) matrices
obtained by their contraction with the vector $\sigma_\mu=(1,-i\vec\tau)$, 
for example $R=x_1-x_2=R_\mu \sigma_\mu$.  
An overbar similarly denotes contraction with $\bar\sigma= (1,i\vec\tau)$. 
Note that barred and unbarred matrices always alternate, in all terms; 
this is because one index of each matrix is dotted and the other not, 
in spinor notation. The additional coordinate with tilde is
\begin{equation} 
\tilde y_2=x_2- \frac{R \rho_2}{z\rho_1-\rho_2} \,.
\end{equation} 
The  $u$ here stands for relative orientation SU(2) color matrix parameterized
by $u_\mu \sigma_\mu$, and $u \cdot \hat R$ is its projection to unit relative distance
vector. For same orientation of the instanton and antiinstanton $u=(0,0,0,1)$.  

This expression
represents a very good approximation to the streamline equation not only at large distances $ (z^I_\mu-z^{\bar I}_\mu)^2\gg \rho^2$, as claimed by Yung in the original paper, but also at $all$ distances as well. In fact at distance zero
the formula produces a very complicated field $A_\mu$, which  however after inspection was found to be a pure gauge, with zero field strength!

The interaction for this ansatz  \cite{Verbaarschot:1991sq} is
\be
\label{S_IA,Yung}
 S_{IA} &=& \frac{8\pi^2}{g^2}
   \frac{1}{(\lambda^2-1)^3}  \bigg\{
    -4\left( 1-\lambda^4 + 4\lambda^2\log(\lambda) \right)
         \left[ |u|^2-4|u\cdot\hat R|^2 \right] \\
   & & \hspace{1cm}\mbox{} +
     2\left( 1-\lambda^2 + (1+\lambda^2)\log(\lambda) \right)
         \left[ (|u|^2-4|u\cdot\hat R|^2)^2
                + |u|^4 + 2(u)^2(u^*)^2 \right]
       \bigg\}\nonumber ,
\ee
where \be
\label{conf_par}
\lambda &=& \frac{R^2+\rho_1^2+\rho_2^2}{2\rho_1\rho_2}
+ \left( \frac{(R^2+\rho_1^2+\rho_2^2)^2}{4\rho_1^2\rho_2^2}-1
\right)^{1/2}.
\ee
is related to the conformal distance parameter defined above. Large distance $R\gg \rho$
is large $\lambda$, zero distance at $\rho_1=rho_2$ is $\lambda=1$.

\begin{figure}[h!]
\begin{center}
\includegraphics[width=6cm]{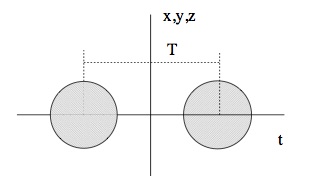}
\caption{The instanton and antiinstanton are placed at $\tau=-T/2$ and $\tau=T/2$
along the Euclidean time axes. The sphaleron path configurations are located at
the middle plane $\tau=0$.}
\label{fig_I_Ibar_plane}
\end{center}
\end{figure}
 
\section{From the instanton-antiinstanton configurations to the sphaleron path}
In the previous section we defined 
 some set of configurations describing an instanton and an antiinstanton, placed at a certain (Euclidean time) positions $\tau=-T/2$ and $\tau=T/2$
 (see Fig.\ref{fig_I_Ibar_plane}). 

Using those, we will deduce some special set of 3d configurations possessing
only $magnetic$ fields.
Indeed,
the instanton fields are self-dual, $\vec E=\vec B$, while the antiinstanton ones are anti-selfdual, $\vec E=-\vec B$. 
By symmetry, at the 3-d plane $\tau=0$ in between,
 the electric field must vanish.  Therefore, at this 3-d plane the field
  is  purely magnetic. 
  
\begin{figure}[h]
\begin{center}
\includegraphics[width=10cm]{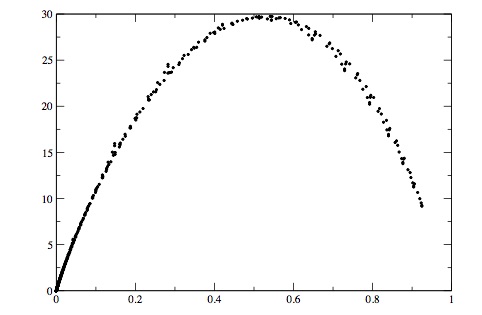}
\caption{The energy times the r.m.s, radius $E\cdot R$ 
 versus Chern-Simons number $\tilde{N}_{CS}$, for the 3-d configurations obtained from the instanton-antiinstanton Yung ansatz as described in the text.
 The curve should be compared to  the sphaleron path we soon obtain from constrained minimization.}
\label{fig_sphaleron_Yung}
\end{center}
\end{figure}

  Furthermore, these configurations make a one-dimensional set, parameterized by the time distance $T$ between the instanton and the anti-instanton. For each of those, one can define
  their energy normalized by size
  \begin{equation} 
ER=\frac{1}{2}
\left[\int d^3r r^2 {\cal B}^2 \times \int d^3r {\cal B}^2\right]^{1/2}.
\end{equation} 
   and the   Chern-Simons number $N_{CS}(T)$.
It has been calculated by \cite{Ostrovsky:2002cg} for
 the ``streamline" Yung ansatz, and shown   in Fig.\ref{fig_sphaleron_Yung}.
Note that at $N_{CS}=0$ (corresponding to large distance $T$)
the energy is zero, and it is also tend to zero at $N_{CS}=1$.
The maximum is in the middle, at $N_{CS}=1/2$, when the instanton and anti-instanton ``half-overlap".  
These configurations constitute the so called
{\em sphaleron path}, leading from one classical vacuum to the next.


Let us now look at this problem from a different point of view. 
In the $A_0=0$ gauge the electric field is given by the time derivative of  $A_m$.
Therefore the electric field has a meaning of momenta, conjugated to coordinates $A_m$,
and $E^2$ term in energy interpreted as field kinetic energy.
Pure magnetic field configurations with zero electric field,  have thus 
only the potential  energy. 

One can view those as some 
 {\em turning points},
at which momentum is zero.  Indeed, such points connect
the tunneling part of the path at $E<V$ with the real time evolution at $E>V$. 
%
In fact the action on the plane can be viewed as given the
$probability$ (rather than the amplitude) of such forced tunneling, with a vertical line
in the  Fig.\ref{fig_I_Ibar_plane} literally interpreted as the {\em unitarity
cut} describing  the field  state ready to propagate into the Minkowski time.

\section{The sphaleron path from a constrained
minimization}

By ``sphaleron path" we mean a one-parameter set of configurations, interpolating between the
nearest values of Chern-Simons number, e.g. from $N_{CS}=0$ to $N_{CS}=1$. 
Because of symmetry of the barrier we are trying to calculate, it is expected that the maximum -- the sphaleron --
corresponds to $N_{CS}=1/2$.

Naively, what needs to be done is to 
fix
$N_{CS}$ to some value, and then find among such configurations the one with  {\em
minimal possible
energy}. This is indeed what was done
 in electroweak theory.  The expected mass  scale 
 is $\sim v/\alpha_{EW}$ is
 defined by the Higgs VEV $v\approx 1/4\, TeV$ and the electroweak coupling constant.

 We however mentioned above, that in all application we will discuss the Higgs VEV, or even existence of the
 Higgs fields itself, will not be important. The problems we want to address are dominated by
 the gauge fields: thus it is natural to ask whether one can identify sphaleron path in {\em pure gauge theory}.
 
  Unlike electroweak theory, classical pure gauge theory is scale invariant, it
has no dimensional parameters. Naively
this would imply that all minima are at zero value, because one can
always reduce the energy by rescaling the size of the configuration
upward. To
 break this unwanted  scale symmetry, one need to  set an additional requirement, basically
fixing the size of the soliton in question. It can
e.g. be  defined as r.m.s. radius 
\be
<r^2>=\frac{\int d^3x r^2 {\cal B}^2}{\int d^3x {\cal B}^2} \label{eqn_rr} \ee 
If it is fixed, there will be a 
particular  static solution with the minimal energy we are looking for. The product of
the energy times r.m.s. size is dimensionless, and that will be what we will evaluate below.
%
We will follow in this section the work by Ostrovsky, Carter and myself
\cite{Ostrovsky:2002cg}, and will find the shape of the barrier.

Since we are interested in static 3d configurations, those would be purely made up of magnetic fields,
since the electric fields have negative $T$-parity and thus not allowed.
As a starting simplifying assumption, we will consider a  spherically
symmetric 3-d configuration of the gauge field. 
Indeed, often  
 the field configurations 
with the
minimal energy  have the  maximal  possible symmetry. Of course, 
we expect the energy density $(\vec B^a)^2$ -- and not the gauge fields themselves -- be spherically symmetric.

For the SU(2) color subgroup in which we are interested, 
configurations of the gauge field ${\cal A}_\mu^a$ can be expressed 
through the following four space-time (0, $j=1..3$) 
and color ($a=1..3$) structures 
\begin{eqnarray}
\label{sph_ansatz}
{\cal A}^a_j &=& A(r,t)\Theta^a_j + B(r,t) \Pi^a_j + C(r,t)\Sigma^a_j
\nonumber\\
{\cal A}^a_0 &=& D(r,t) \frac{x^a}{r}
\end{eqnarray}
with three mutually orthogonal projectors
\begin{equation}
\Theta^a_j=\frac{\epsilon_{jam}x^m}{r}\,, \quad \Pi^a_j =
\delta_{aj}-\frac{x_ax_j}{r^2}\,, \quad \Sigma^a_j = \frac{x_ax_j}{r^2}\,.
\label{projectionops}
\end{equation}
While for the sphaleron path problem the four functions should be static (independent on time $t$),
the attentive reader would notice that we included the time. The reason for it is that later on
we will also discuss a dynamical problem of the {\em sphaleron explosion}.

One may rewrite a problem with $r$ and $t$- dependent functions
as some 1+1 dimensional Lagrangian. In fact this is true for the problem at hand, and our four
 functions of Eq.~(\ref{sph_ansatz}) can be rewritten as 
four
fields of the Abelian gauge-Higgs model
($A_{\mu=0,1},\,\phi,\, \alpha$)
on a hyperboloid :
\begin{equation}
A=\frac{1+\phi\sin\alpha}{r}, \quad B=\frac{\phi\cos\alpha}{r},\quad
C=A_1,\quad D=A_0.
\end{equation}

One can express the field strengths in these terms as
\begin{eqnarray}
{\cal E}^a_j = {\cal G}^a_{0j} 
&=&\frac{1}{r} [\partial_0\phi\sin\alpha +
\phi\cos\alpha(\partial_0\alpha-A_0)]\Theta^a_j \nonumber\\
&&+ \frac{1}{r} [\partial_0\phi\cos\alpha -
\phi\sin\alpha(\partial_0\alpha-A_0)]\Pi^a_j
\nonumber\\&&
+ (\partial_0A_1-\partial_1A_0)\Sigma^a_j
\label{electric}\end{eqnarray}
and
\begin{eqnarray}
{\cal B}^a_j = \frac{1}{2}\epsilon_{jkl}{\cal G}^a_{kl} 
&=&\frac{1}{r} [-\partial_1\phi\cos\alpha +
\phi\sin\alpha(\partial_1\alpha-A_1)]\Theta^a_j \nonumber\\
&&+ \frac{1}{r} [\partial_1\phi\sin\alpha +
\phi\cos\alpha(\partial_1\alpha-A_1)]\Pi^a_j 
\nonumber\\ &&
+ \frac{1-\phi^2}{r^2}\Sigma^a_j, 
\label{magnetic}\end{eqnarray}
where $\partial_0\equiv \partial_t \mbox{ and } \partial_1\equiv\partial_r$.
Putting those expression into 
the usual   3+1 dimensional Minkowski action and integrating over angles one find reduced 1+1d action
\begin{eqnarray}\label{action}
S &=& \frac{1}{4g^2}\int d^3x dt\left[\left({\cal B}^a_j\right)^2 
- \left({\cal E}^a_j\right)^2\right]
\nonumber\\
&=& 4\pi\int dr dt \Bigg[\left(\partial_\mu\phi\right)^2+
\phi^2\left(\partial_\mu\alpha-A_\mu\right)^2\nonumber\\
&&+\frac{(1-\phi^2)^2}{2r^2}-
\frac{r^2}{2}\left(\partial_0A_1-\partial_1A_0\right)^2\Bigg] \,,
\end{eqnarray}
with the summation now over the 1+1 dimensional indices. The $t,r$ space is with the $(-,+)$ metric.
Note that $\phi$ and $\alpha$ now have a meaning of the modulus and phase of some charged scalar.
The charge is Abelian, as seen from the last term containing the ``field strength" squared.

What exactly does this elegant re-writing of the Lagrangian in new fields give us? Well,
it helps to understand better the remaining symmetries of the model.
The spherical ansatz is preserved by a set of gauge transformations
generated by unitary matrices of the type
\begin{equation}
\label{gauge_transform}
U(r,t)=\exp\left(i\frac{\beta(r,t)}{2r}\tau^ax^a\right).
\end{equation}
These transformations naturally coincide with the gauge symmetry of 
the corresponding abelian Higgs model:
\begin{equation}
\phi'=\phi, \quad \alpha'=\alpha+\beta, \quad A'_\mu=A_\mu+\partial_\mu
\beta\,.
\end{equation}
This freedom can be used to gauge out, for example, one component of $A_\mu$: we will use the gauge 
${\cal A}_0=0$ from now on.

Before proceeding any further, let us express  the
topological current
\begin{equation}
K_\mu=-{1 \over 32 \pi^2} \epsilon^{\mu\nu\rho\sigma}
\left({\cal G}^a_{\nu\rho} {\cal A}^a_\sigma -
{g \over 3}\epsilon^{abc} {\cal A}^a_\nu  {\cal A}^b_\rho  {\cal A}^c_\sigma
\right)\,.
\end{equation}
in the reduces form,  in the ${\cal A}_0=0$ gauge: 
\begin{eqnarray}
K^0 &=& \frac{1}{8\pi^2r^2}
\left[(1-\phi^2)(\partial_1\alpha-A_1)-\partial_1(\alpha-\phi\cos\alpha)\right]
\nonumber\\
K^i &=& \frac{x^i}{8\pi^2r^3}
\left[(1-\phi^2)\partial_0\alpha-\partial_0(\alpha-\phi\cos\alpha)\right] \,,
\end{eqnarray}
while the topological charge becomes
\begin{eqnarray}
\partial_\mu
K^\mu &=& \frac{1}{8\pi^2r^2}\left\{-\partial_0\left[(1-\phi^2)
(\partial_1\alpha-A_1)\right]\right.
\nonumber\\
&&+\left.\partial_1\left[(1-\phi^2)(\partial_0\alpha-A_0)\right]\right\}\,.
\end{eqnarray}
Note that only gauge-invariant combinations of field derivatives 
appear here.

As a ``topological coordinate'' marking the tunneling paths and
the turning states one can use the Chern-Simons number
\begin{eqnarray}
N_{CS} = \int d^3x K_0
&=& -\frac{1}{2\pi}\int dr
(1-\phi^2)(\partial_1\alpha-A_1)
\nonumber\\&&
+\frac{1}{2\pi} \left.(\alpha-\cos\alpha)\right|_{r=0}^{r=\infty}
\label{csdef}
\end{eqnarray}
The first, gauge-invariant term is sometimes called
the {\em corrected} or {\em true} Chern-Simons number 
$\tilde{N}_{CS}$, while the second (gauge-dependent) term is referred to as
the {\em winding number}. It is the change in $\tilde{N}_{CS}$ which
is equivalent to the integral over the local topological charge.

  Now we are done with the digression of re-writing spherically symmetric 3+1 problem into a 1+1 form,
  and return to the static sphaleron path. 
To keep both {\em the
Chern-Symons number} and {\em the mean radius} constant, we introduce two
Lagrange multipliers $1/\rho^2,\eta$
 and search for the minimum of the following functional 
\be
\tilde{E}=\frac{4\pi}{g^2}
          \int dr \left(1+\frac{r^2}{\rho^2}\right)
                  \left[
                          (\partial_r \phi)^2
                        + \phi^2 (\partial_r\alpha)^2
                        + \frac{(1-\phi^2)^2}{2r^2}
                  \right]\nonumber\\
+\frac{\eta}{2\pi}\int dr (1-\phi^2)\partial_r\alpha \ee

It is convenient to introduce new variable $\xi=2arctan(r/\rho)-\pi/2$. Then
\be
\tilde{E}=\frac{8\pi}{g^2}\left\{
          \int\limits_{-\pi/2}^{\pi/2} d\xi
                  \left[
                          (\partial_\xi \phi)^2
                        + \phi^2 (\partial_\xi\alpha)^2
                        + \frac{(1-\phi^2)^2}{2\cos^2\xi}
+\kappa(1-\phi^2)\partial_\xi\alpha\right]\right\}
\ee where $\kappa=\eta\rho g^2/(32\pi^2)$
The
Euler-Lagrange equations  are \be\label{EL1}
\partial^2_\xi\phi-\phi(\partial_\xi\alpha)^2+
\frac{(1-\phi)^2\phi}{cos^2\xi}+2\kappa\phi\partial_\xi\alpha=0 \\ \nonumber
\partial_\xi(\phi^2\partial_\xi\alpha)+\kappa\partial_\xi(1-\phi^2)=0
\ee Finiteness of the energy demands the following boundary conditions
$\phi^2(\xi=-\pi/2)=\phi^2(\pi/2)=1$

The second Eq. \ref{EL1} gives
\be
\partial_\xi\alpha=-\kappa\frac{1-\phi^2}{\phi^2}
\ee with integration constant equals 0 as it follows from the form of
energy. After substitution $\partial_\xi\alpha$ to the eq. \ref{EL1} one has
\be
\partial^2_\xi\phi+\frac{(1-\phi^2)\phi}{cos^2\xi}=
\kappa^2\frac{1-\phi^4}{\phi^3} \ee 

The solution to this equation exists for
$-1<\kappa<1$, it is $ \phi^2=1-(1-\kappa^2)cos^2\xi. $.
Assuming $\phi$ to be positive one finds finally
\be
\phi(r)=\left(1-(1-\kappa^2)\frac{4\rho^2 r^2}{(r^2+\rho^2)^2}\right)^{1/2}
\ee
\be
\partial_r\alpha(r)=-2\kappa\frac{1-\phi^2}{\phi^2}\frac{\rho}{r^2+\rho^2}.
\ee

For any $\kappa$ mean radius of the solution is
the same $<r^2>=\rho^2$, and the energy
density, the total energy, and  
the (corrected) Chern-Symons number
are respectively
 \be 
B^2/2=24(1-\kappa^2)^2\rho^4/(r^2+\rho^2)^4 \\ \nonumber 
 E_{stat}=3\pi^2(1-\kappa^2)^2/(g^2\rho) \\ \nonumber
 \tilde{N}_{CS}={\rm sign}(\kappa)(1-|\kappa|)^2(2+|\kappa|)/4 \ee


\begin{figure}[h!]
\begin{center}
\includegraphics[width=8cm] {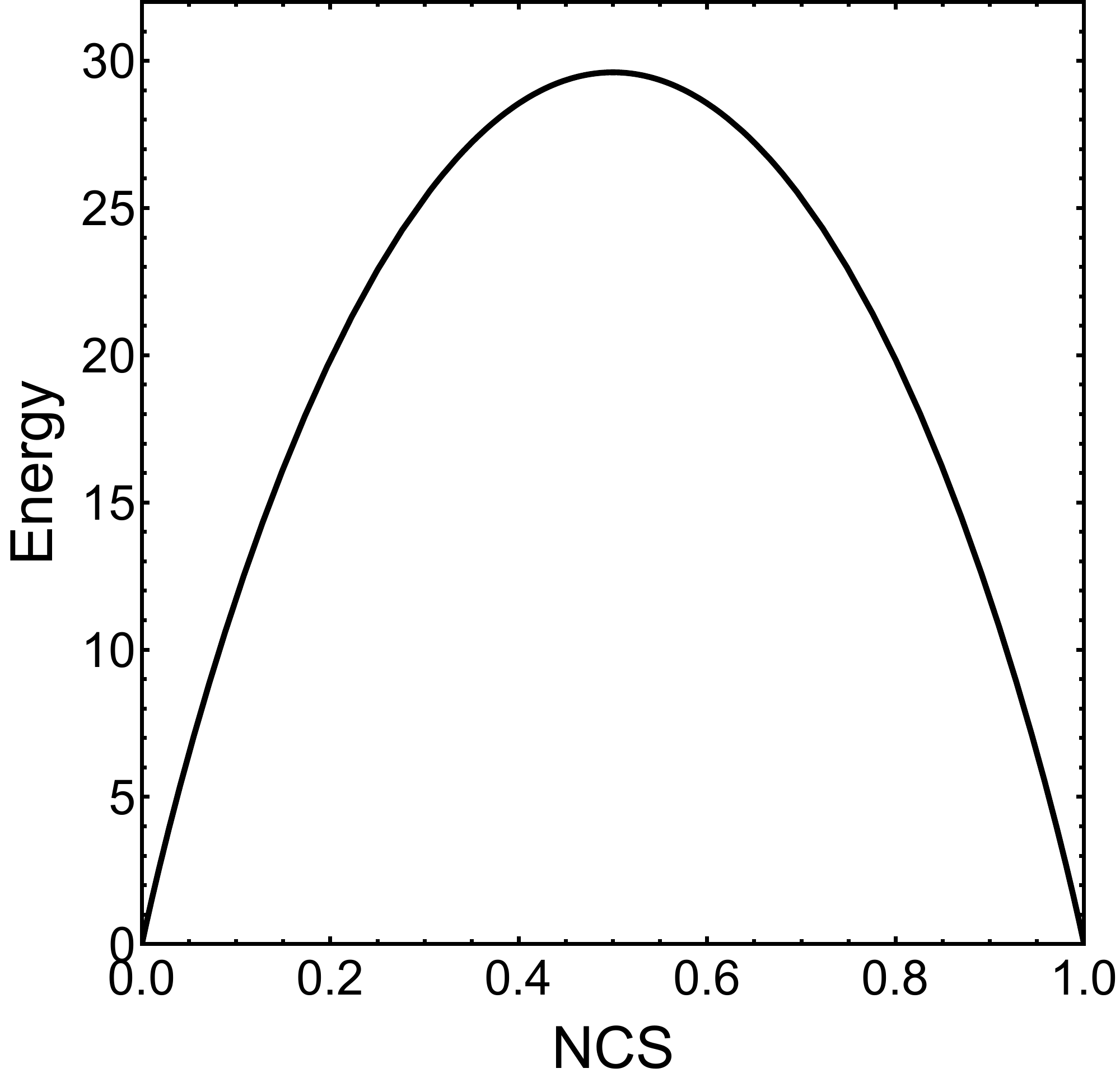}
\caption {
The potential energy $E$ (in units of $1/g^2\rho$)
 versus the Chern-Simons number $\tilde{N}_{CS}$, for the 
 ``sphaleron path" solution to be derived in Sphaleron chapter.
} \label{fig_sphaleron_path2}
\end{center}
\end{figure}

Two last equations define the parametric form of the potential,
see Fig.\ref{fig_sphaleron_path2}. 
 The same 
profile obviously continues from 1/2 to 1, and so on, as a  periodic potential with zeros at all integer
values of $N_{CS}$, as a chain of mountains separated by valleys.
In fact there are mountains of any hight, but tall ones are narrow.
If energy is expressed in units of
$1/g\rho$, it becomes unique. 
Note that the maximum is about parabolic but near zero energy
the behavior is linear; so valleys are actually more like a deep canyons.
The maximum is the 
{\em  sphaleron solution} corresponding to $\kappa=0$ and $N_{CS}=1/2$
\be
\phi=\frac{|r^2-\rho^2|}{r^2+\rho^2}, \hspace{1cm} \alpha=\pi\theta(r-\rho). \ee

Let me finish this technical part with qualitative description of the solution. Remember that
states of the sphaleron path are static balls of purely magnetic field. Unlike the monopole, in which case the field is radial $\vec B\sim \vec r$, now the magnetic field rotates around in
circles. Three color components of the field (recall, we are in SU(2) and there are three generators) rotate around axes $x^1,x^2,x^3$, repesctively. The total sum squared
is spherically symmetric. Summarizing, they are neat magnetic bombs, ready to explode!  


\section{Sphaleron explosion}
As we already mentioned above, the sphalerons are saddle point
solution which are unstable.  Add a small perturbation
to a ball placed on the top of the mountain pass,
and will start rolling down into one or the other valley. 


The same paper \cite{Ostrovsky:2002cg} came up with an  analytic solution
for this problem as well. The static sphaleron field configuration,
found in previous section, is used as the initial condition for
real-time, Minkowski evolution of the gauge field. 
Let us first consider the equations of motion in the 1+1 dimensional
dynamical system.
Variation of the action, Eq.~(\ref{action}), gives
\begin{eqnarray}
\label{ELt1}
\partial_\mu\partial^\mu\phi+\phi(\partial_\mu\alpha-A_\mu)^2
+\frac{(1-\phi^2)\phi}{r^2}&=&0
\\
\label{ELt2}
\partial^\mu\left[\phi^2\left(\partial_\mu\alpha-A_\mu\right)\right] &=& 0
\\
\phi^2(\partial_1\alpha-A_1)-
\partial_0\left[\frac{r^2}{2}(\partial_0A_1-\partial_1A_0)\right]&=&0
\nonumber\\
\label{ELt3}
\phi^2(\partial_0\alpha-A_0)-
\partial_1\left[\frac{r^2}{2}(\partial_0A_1-\partial_1A_0)\right]&=&0.
\end{eqnarray}
The solution of Eq.~(\ref{ELt2}) has the form
\begin{eqnarray}
\phi^2(\partial_0\alpha-A_0) &=& -\partial_1\psi \nonumber\\
\phi^2(\partial_1\alpha-A_1) &=& -\partial_0\psi \,,
\end{eqnarray}
where $\psi(r,t)$ is an arbitrary smooth function.
Eqs.~(\ref{ELt3}) are consistent with this solution if
\begin{equation} 
\partial_0A_1-\partial_1A_0=-\frac{2\psi}{r^2}
\end{equation} 

Now, combining Eq.~(\ref{ELt2}) and Eqs.~(\ref{ELt3}) one has
\begin{equation} 
\label{psi_rt}
\partial^\mu\left(\frac{\partial_\mu\psi}{\phi^2}\right)=
\partial_0A_1-\partial_1A_0=\frac{2\psi}{r^2},
\end{equation} 
which can be viewed as a necessary and sufficient condition
for $\psi$ to be a solution for Eq.~(\ref{ELt2}) and Eqs.~(\ref{ELt3})
simultaneously.
Eq.~(\ref{ELt1}) is now
\begin{equation} 
\label{phi_rt}
\partial_\mu\partial^\mu\phi-\frac{(\partial_\mu\psi)^2}{\phi^3}
+\frac{(1-\phi^2)\phi}{r^2}=0\,.
\end{equation} 

The initial conditions for Eqs.~(\ref{psi_rt}) and (\ref{phi_rt}) are
\begin{eqnarray}
\phi(r,0)&=&\phi(r)\,,
\nonumber\\
\partial_0\phi(r,t)|_{t=0}&=&0\,,
\nonumber\\
\partial_1\psi(r,0)&=&-\phi(r)^2\partial_0\alpha(r)=0 
\Rightarrow \psi(r,0)=0,
\nonumber\\
\partial_0\psi(r,t)|_{t=0}&=&-\phi(r)^2\partial_1\alpha(r),\nonumber
\end{eqnarray}
where the $t$-independent fields on the right sides of the equations
are the static solutions of $\phi$ and $\alpha$ from the previous section.

As with static solutions, it is more convenient to discuss the time-evolution
equations in hyperbolic coordinates. 
Let us choose $\omega$ and $\tau$ such that
\begin{equation}
\label{tau-omega}
r=\frac{\rho\cos\omega}{\cos\tau-\sin\omega} \,, 
\quad t=\frac{\rho\sin\tau}{\cos\tau-\sin\omega} \,.
\end{equation}
The physical domain of $0<r<\infty$ and $-\infty<t<\infty$ 
is covered by $-\pi/2<\omega<\pi/2$ and $-\pi/2+\omega<\tau<\pi/2-\omega$.
For $t>0$, the corresponding domain is
$-\pi/2<\omega<\pi/2$ and $0<\tau<\pi/2-\omega$. 
This change of variables (\ref{tau-omega}) is a conformal one.

In the new variables Eqs.~(\ref{psi_rt}) and (\ref{phi_rt}) become
\begin{eqnarray}
\label{eqs_ot}
-\pt^2\phi+\po^2\phi-\frac{(\pt\psi)^2-(\po\psi)^2}{\phi^3}
+\frac{(1-\phi^2)\phi}{\cos^2\omega} &=& 0 
\nonumber\\
-\pt\frac{\pt\psi}{\phi^2}+\po\frac{\po\psi}{\phi^2}-
\frac{2\psi}{\cos^2\omega} &=& 0 \,.
\end{eqnarray}

Before solving these equations let us note that it is possible to predict
the large-$t$ behavior of gauge field from the form of 
the conformal transformation (\ref{tau-omega}). 
Indeed, the $t\rightarrow\infty$ limit corresponds to the line
$\tau=\pi/2-\omega$ on the $(\omega,\tau)$ plane. 
If one now takes the limit $|r-t|\rightarrow\infty$ 
(regardless of the limit for $|r-t|/t$),
the position on $(\omega,\tau)$ plane is either 
$\omega\rightarrow -\pi/2\,,\,\tau\rightarrow 0$ or $\omega\rightarrow \pi/2
\,,\, \tau\rightarrow\pi$. 
This means that the entire line $\tau=\pi/2-\omega$
corresponds to space-time points with finite differences between
$r$ and $t$ and, therefore, if $\phi$ and $\psi$ are smooth functions
of $\omega$ and $\tau$, then for asymptotic times the field is concentrated
near the $r=t$ line.
This corresponds to the fields expanding as a thin shell in space.

We must now supply Eqs.~(\ref{eqs_ot}) with initial conditions, which are
\begin{eqnarray}
\label{init}
\phi(\omega,\tau=0)^2 &=& 1-(1-\kappa^2)\cos^2\omega
\nonumber\\
\pt\phi(\omega,\tau)|_{\tau=0} &=& 0
\nonumber\\
\psi(\omega,\tau=0) &=& 0
\nonumber\\
\pt\psi(\omega,\tau)|_{\tau=0} &=& \frac{\rho}{1-\sin\omega}
\partial_t\psi(\omega,\tau)|_{t=0} 
\nonumber\\
&=&\kappa(1-\kappa^2)\cos^2\omega \,.
\end{eqnarray}
One of the solutions of Eqs.~(\ref{eqs_ot}), first found in 1977 by
L\"uscher  and Schechter , is
\begin{eqnarray}
\phi(\omega,\tau)^2 &=& 1-(1-q^2(\tau))\cos^2\omega \nonumber\\
\psi(\omega,\tau) &=& \frac{\dot{q}(\tau)}{2}\cos^2\omega \,,
\end{eqnarray}
with a function $q(\tau)$ that satisfies
\begin{equation}
\label{ddotq}
\ddot{q}-2q(1-q^2)=0 \,.
\end{equation}
This is the equation for a one-dimensional particle
moving in double-well potential of the form $U(q)=(1-q^2)^2/2$.

We now have to check that the solution
satisfies the initial conditions, (\ref{init}).
This is indeed the case if one identifies $q(0)=\kappa$ 
and takes $\dot{q}(0)=0$.
For the initial condition of this type
({\em i.e.} for energy $\varepsilon=\dot{q}^2/2+U(q)<1/2$),
the solution of Eq.~(\ref{ddotq}) is
\begin{equation}
\label{solution_q}
q(\tau)=\tilde{q} dn\left(\tilde{q}(\tau-\tau_0), k\right),
\end{equation}
where $dn$ is Jacobi's function and
$\tilde{q}=\sqrt{2-\kappa^2}$ is the second stopping point
for a particle in the potential $U(q)$.
We have also defined
\[
k^2=2\frac{1-\kappa^2}{2-\kappa^2} \quad {\rm and}\quad
\tau_0\tilde{q}=\frac{T}{2} \,,
\]
where $T$, the period of oscillations in the potential $U(q)$, is 
$T=2K(k)$, with $K(k)$ being the complete elliptic integral of the first kind.
The idea is, of course, that ``oscillations'' in $\tau$ begin
from the rest point, close to $\tau = 0$.

Let us now look at several properties of the solution for large times.
The solution (\ref{solution_q}) is apparently regular in the $(\omega, \tau)$
plane, and therefore for large times the field is concentrated near $r=t$. 
At asymptotic times the energy density, $e(r,t)$, is given by
\begin{equation}
4\pi e(r,t) = \frac{8\pi}{g^2\rho^2}(1-\kappa^2)^2
\left(\frac{\rho^2}{\rho^2+(r-t)^2}\right)^3 \,.
\end{equation}
The change in topological charge is
\begin{eqnarray}
\Delta Q &=& \int\limits_0^{\infty} d^3x dt \,\partial_\mu K^\mu\nonumber\\
&=& \frac{1}{2\pi}\int dr dt
\left[-\partial^2_t\psi+\partial^2_r\psi-\frac{2\psi}{r^2}\right]\nonumber\\
&=& \frac{\pi}{2}\kappa(3-\kappa^2)
-{\rm sign}(\kappa)\arccos\left(
\frac{\mbox{\rm cn}(\tilde{q}\pi, k)}{dn(\tilde{q}\pi, k)}\right) .
\end{eqnarray}
The evolution of $\tilde{N}_{CS}$ begins from time $t=0$, where
\begin{equation}
\tilde{N}_{CS}(0)=\frac{1}{4}{\rm sign}(\kappa)(1-|\kappa|)^2(2+|\kappa|) \,,
\end{equation}
and as $t\rightarrow\infty$ its limit is
$\tilde{N}_{CS}(\infty)=\tilde{N}_{CS}(0)+\Delta Q$.

We now estimate number of gluons produced by the described evolution. 
In $\phi,\psi$ language the chromoelectric and chromomagnetic
fields are
\begin{eqnarray}
{ E}^a_j  &=&
\frac{1}{r} \left(\partial_t\phi\sin\alpha -
\frac{\partial_r\psi\cos\alpha}{\phi}\right)\Theta^a_j \nonumber\\
&&+ \frac{1}{r} \left(\partial_t\phi\cos\alpha +
\frac{\partial_r\psi\sin\alpha}{\phi}\right)\Pi^a_j +
\frac{2\psi}{r^2}\Sigma^a_j \,,
\label{electric2}
\end{eqnarray}
\begin{eqnarray}
{ B}^a_j &=&
-\frac{1}{r} \left(\partial_r\phi\cos\alpha +
\frac{\partial_t\psi\sin\alpha}{\phi}\right)\Theta^a_j \nonumber \\
&& + \frac{1}{r}\left(\partial_r\phi\sin\alpha -
\frac{\partial_t\psi\cos\alpha}{\phi}\right)\Pi^a_j +
\frac{1-\phi^2}{r^2}\Sigma^a_j \,.
\label{magnetic2}
\end{eqnarray}

Terms proportional to $\Sigma^a_j$ are longitudinal and die out as 
$t\rightarrow\infty$. 
The remainder is a purely transverse field.
The main result becomes apparent when we choose a gauge where
\[\phi\partial_r\phi\cos\alpha + \partial_r\psi\sin\alpha=0\,, \]
in which 
\begin{eqnarray}
{ E}^a_j &\rightarrow&
\frac{1}{r}
\sqrt{\frac{(\partial_r\psi)^2}{\phi^2}+(\partial_r\phi)^2}\Theta^a_j 
\nonumber\\ 
&\rightarrow&
\frac{1-\kappa^2}{r\rho}\left(\frac{\rho^2}{\rho^2+(r-t)^2}\right)^{3/2}
\Theta^a_j \,,
\label{electric4}
\end{eqnarray}
\begin{equation}
{ B}^a_j \rightarrow
\frac{1-\kappa^2}{r\rho}\left(\frac{\rho^2}{\rho^2+(r-t)^2}\right)^{3/2}
\Pi^a_j \,.
\label{magnetic4}
\end{equation}

We now perform a Fourier transform, finding
\begin{eqnarray}
{ E}^a_j(\vec{k})
&=& 4\pi\rho(1-\kappa^2)K_1(\omega\rho)\Theta^a_j\nonumber\\
{ B}^a_j(\vec{k})
&=& 4\pi\rho(1-\kappa^2)K_1(\omega\rho)\Pi^a_j\,,
\end{eqnarray}
where $\Theta^a_j$ and $\Pi^a_j$ are the color/space 
projectors in momentum space analogous to those in coordinate space
(\ref{projectionops}), the frequency $\omega = |\vec{k}|$,
and $K_1$ is a Bessel function.
One can easily verify that ${ B}^a_j=\epsilon_{jlm}k_l{ E}^a_m/k$, as 
is required for a radiation field. 

Completing this section, let us mention that alternative derivation of the solution for the sphaleron explosion 
has been found by
Zahed and myself \cite{Shuryak:2002qz}. It starts with Euclidean 4-d symmetric ansatz
\be g A_\mu^a=\eta_{a\mu\nu} \partial_\nu F(y), \,\,\,\, F(y)=2\int_0^{\xi(y)} d\xi'   f(\xi')     \ee
with $\xi=  ln(x^2/\rho^2)$ and $\eta$ the 't Hooft symbol. 
Upon substitution of the gauge fields in  the gauge Lagrangian $G_{\mu\nu}^2$ 
one finds that the effective Lagrangian has the form
\be L=    \int d\xi \left[{\dot{f}^2\over 2}+2f^2(1-f)^2 \right]
\ee   
corresponding to the motion of a particle in a double-well potential.

The off-center conformal transformation in question
has the form
\be (x+a)_\mu={2 \rho^2 \over (y+a)^2} (y+a)_\mu
\ee
with $a_\mu=(0,0, 0, \rho) $.  While the original solution depends only on the radial coordinate in 4 dimensions
$y^2$, in terms of $x_\mu$ this symmetry is broken and there is separate dependence on
$x_4$ and the 3-dimensional radius $r=\sqrt{x_1^2+x_2^2+x_3^2}$. 

The last step in getting the final solution is the analytic continuation to Minkowski time $t$, via $x_4\rightarrow i t$. 
It has the explicit form
\be gA_4^a=-f(\xi) { 8 t\rho x_a \over [(t-i\rho)^2-r^2]  [(t+i\rho)^2-r^2]  } \label{eqn_field} \ee
$$ gA^a_i=4\rho f(\xi) { \delta_{ai}(t^2-r^2+\rho^2)+2\rho \epsilon_{aij} x_j +2 x_i x_a \over [(t-i\rho)^2-r^2]  [(t+i\rho)^2-r^2]  } $$
which includes $i$ but still is manifestly real.  Expressions for the gauge fields can be easily generated from it.

The only comment worth making is about the original parameter $\xi$ in terms of 
these   Minkowskian coordinates, which we still call  $x_\mu$, has the form
\be \xi ={1\over 2}  log{y^2\over \rho^2}={1\over 2} log\left( {(t+i\rho) ^2-r^2 \over (t-i\rho) ^2-r^2 } \right) \ee
which is pure imaginary.To avoid carrying the extra $i$, we use the real 
\be \xi_E \rightarrow -i \xi_M = arctan\left( { 2 \rho t \over t^2-r^2-\rho^2 } \right)   \label{arctan}  \ee 
and in what follows we will drop the suffix $E$.
Switching from imaginary to real $\xi$ corresponds to switching from the Euclidean to
 Minkowski spacetime solution. It changes the sign of the acceleration
in equation of motion , or, equivalently, the sign of the effective potential $V_M=-V_E$,
to that of the normal double-well problem. The static sphaleron solution corresponds
to the particle standing on the potential maximum at $f=1/2$, and the needed sphaleron decay
to ``tumbling" paths.  Since the start from exactly
 the maximum takes a divergent time, we will start  nearby the turning point.

\begin{figure}[t]
\begin{center}
\includegraphics[width=8cm]{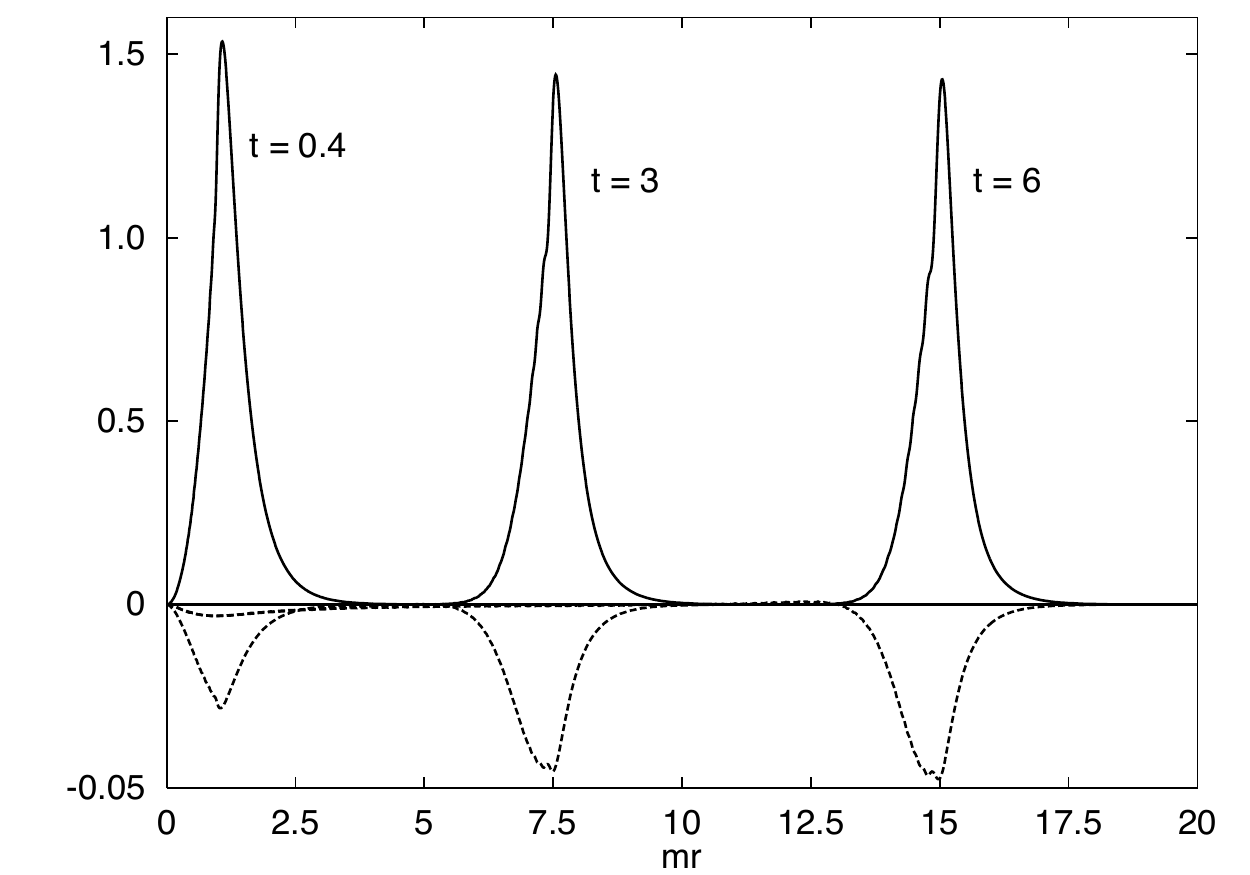}\\
\includegraphics[width=8cm]{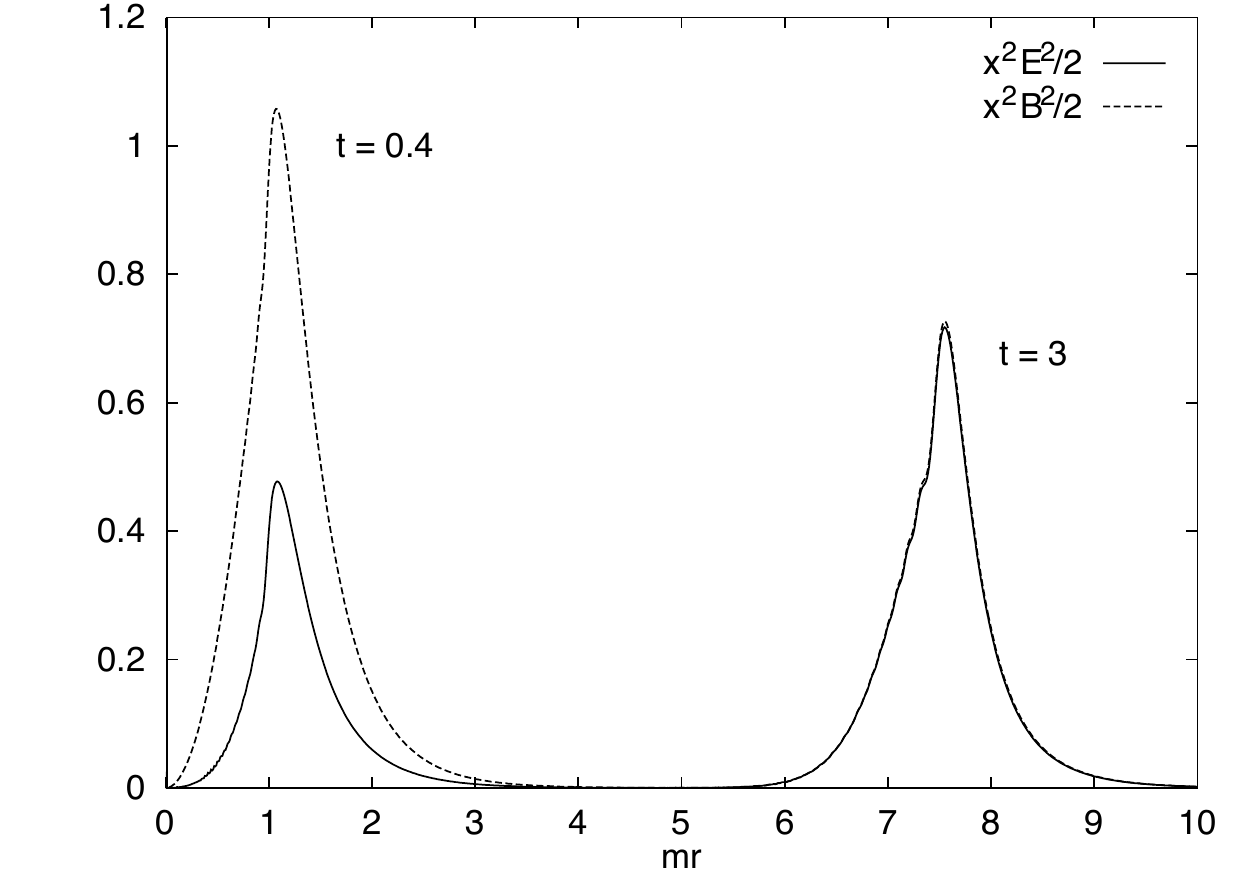}
\end{center}
\caption{
(a) The energy (solid lines) and Chern-Simons number densities (dashed lines) 
for three times during the explosion, $t =$0.4, 3, and 6 fm.
(b) the electric and magnetic fields squared.
}\label{ecsfig}
\end{figure}

\section{Chiral anomaly  and sphaleron decay} \label{sec_anomaly}
We had already discuss the relation between  the 3d and 4d topological charges: the
divergence of the current containing Chern-Simons number is equal to
\be \partial_\mu K_\mu={1 \over 32 \pi^2} G^a_{\mu\nu} \tilde{G}^a_{\mu\nu}   \label{eqn_CS} \ee
the 4d topological charge 
\be Q_T=  {1 \over 32 \pi^2} \int d^4 x G^a_{\mu\nu} \tilde{G}^a_{\mu\nu} \ee
 We used Gauss theorem connecting
 the volume integral of the r.h.s., the 4d topological charge, to the change
 of the 3d topological charge, the Chern-Simons number: 
 \be N_{CS}(\tau\rightarrow \infty)-  N_{CS}(\tau\rightarrow -\infty)=Q \ee

The r.h.s. of this relation
also appears in another important relation, known in QCD-like theories as Adler-Bell-Jackiw (ABJ)  or {\em axial anomaly}:
\be  \partial_\mu j_\mu^5= {1\over 32\pi^2} \epsilon^{\alpha\beta\gamma\delta}G_{\alpha\beta}  G_{\gamma\delta}
\ee 
including 
divergence of  the axial quark current
 \be  j_\mu^5=\bar q\gamma_\mu\gamma_5 q\ee 
It was historically obtained from triangular diagrams with one axial and two vector currents:
its derivation we will not discuss, yet the physics related to it we will discuss in next chapters. 

Since the r.h.s. of both equations is the same topological charge, their change (in appropriate units) is the same. Say, one instanton $Q=1$ produces change in axial density $\Delta n_a=2 N_f$. The classical minima with different Chern-Simons number are associated with
different axial charge. The difference between them is gauge invariant and given by
$Q$. Another combination
of two currents $is$ in fact conserved, but not gauge invariant. 

Integrating the zeroth component of the currents over space one finds
that, in QCD with $N_f$ light quarks
\be \Delta Q_5=   (2N_f) \Delta N_{CS}\,\,\,\,\,(QCD)  \ee
Therefore the Chern-Simons number of the gauge field configuration is {\em rigidly locked to the axial charge},
the number of left minus right-polarized fermions! For example, if $N_f=3$, a transition over the sphaleron barrier
from $N_{CS}=0$ to  $N_{CS}=1$ must be accompanied by 6 units of the axial charge.

\begin{figure}[htbp]
\begin{center}
\includegraphics[width=4cm]{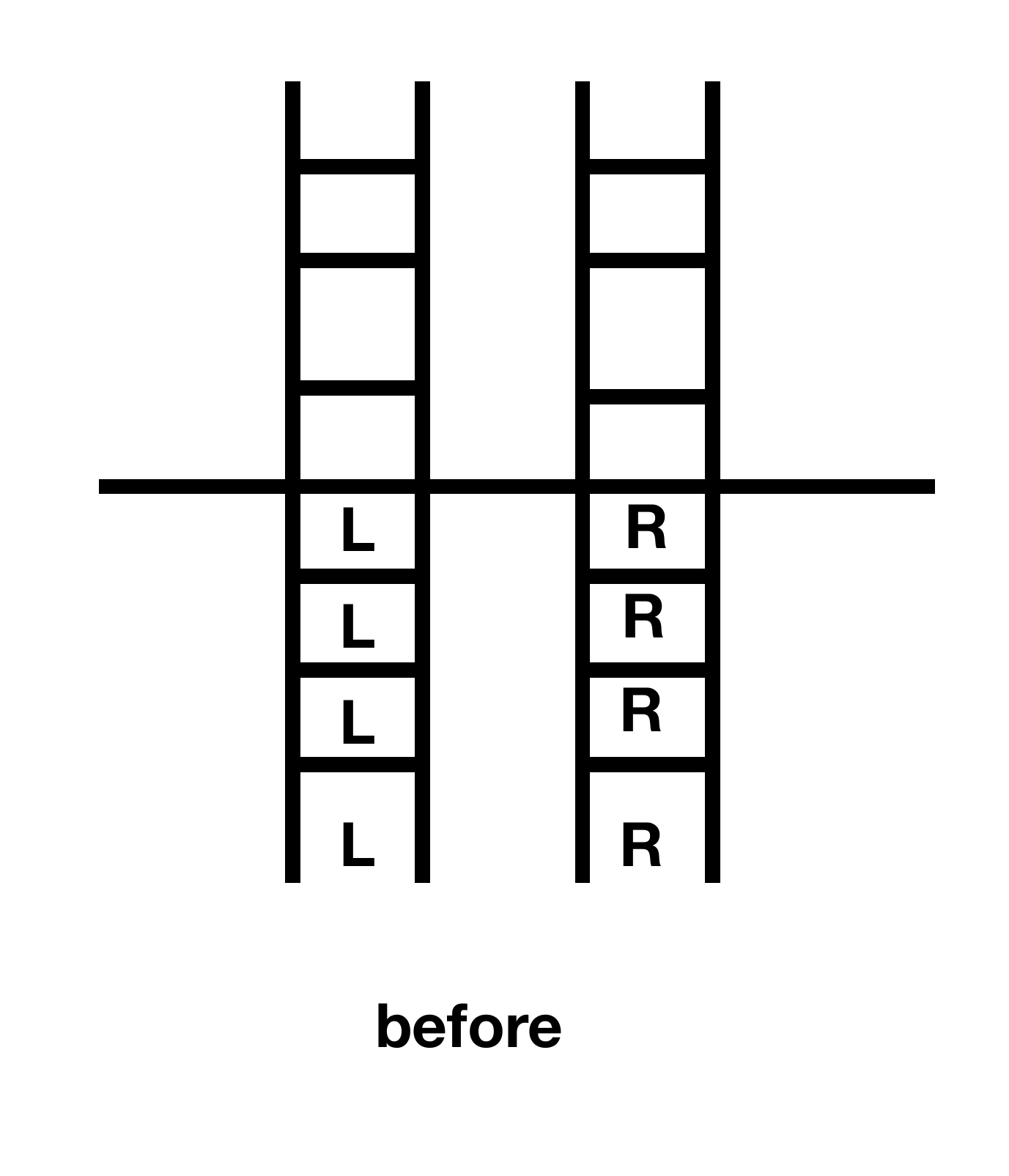}
\includegraphics[width=4cm]{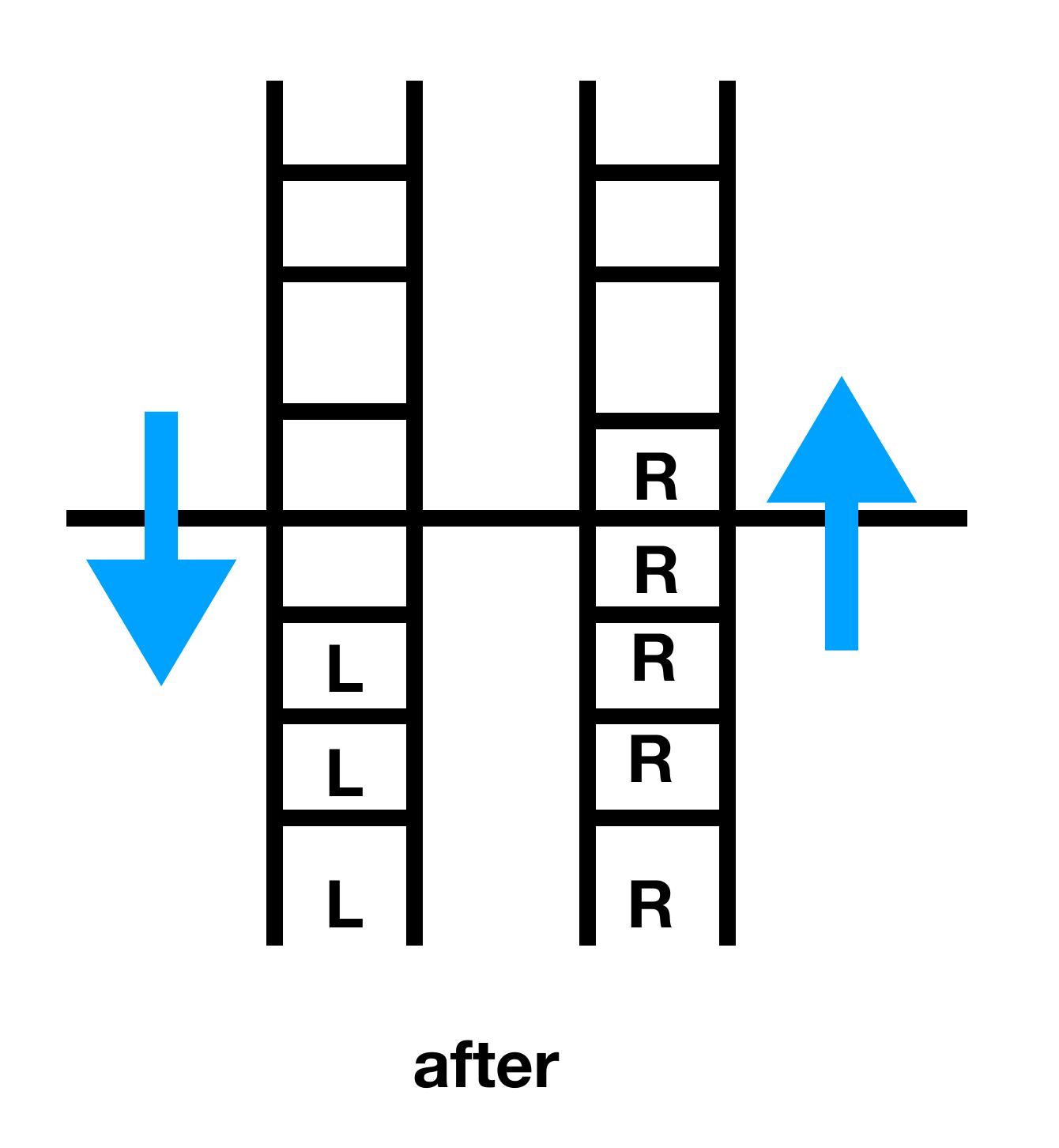}
\caption{Occupancy of the ``Dirac sea hotel", before and after the earthquake}
\label{fig_infinite_hotel}
\end{center}
\end{figure}

Let me out this into a small story, so you better remember it. There was a sea-side hotel called the Dirac Sea Hotel. It was so big that we think of it extending indefinitely, both
above the ground and below.
It has a bit strange policy to put occupants at as low level as possible, and to mark which of the occupants are right and left-handed, which were put in two
separate towers, see the left figure. A strange earthquake happened one day, leading
to the shift as indicated in the right figure: all lefties went down by one floor, and all righties one up, see Fig.\ref{fig_infinite_hotel}.

The same anomaly relation in electroweak theory leads to even more drastic consequences. Since only the
left-polarized quarks and leptons interact with the electroweak gauge field, there are no right-hand fields in the
equation. Furthermore, there is no distinction between vector and axial current, as only left-handed part matters.
Therefore, the $number$ (rather than just chiralities) of quarks and leptons are changed in electroweak theory.
A travel from $N_{CS}=0$ to  $N_{CS}=1$ must lead to simultaneous production of 9 quarks and 3 leptons! 
If the polarization of the sphaleron field is ``up", these 12 fermions are $t_rt_bt_g c_rc_bc_g  u_ru_bu_g \tau\mu,e$
where $r=red,b=blue,g=green$ are three colors of the quarks. 
The baryon and lepton numbers increase by 3 units,  and their difference is zero
\be \Delta B=\Delta L =3 \Delta N_{CS} \,\,\,\, (electroweak \, part \, of  \, the \,  SM) \ee

Can those drastic statement be put to experimental tests? 

As far as the electroweak theory is concern, the chances to do so are not there. In the Higgs-broken vacuum we live today one has to go over the barrier with the hight $M_{KM}\approx 8\, TeV $. And the energy is not the main problem:
one needs to produce $O(100)$ $W$ bosons, fit all of them in a small volume of an electroweak scale, keeping all of them
in a form of  coherent configurations of the sphaleron path.
Electroweak sphaleron transitions may however happen above and near the electroweak phase transition,
because in this case Higgs VEV is absent or small. 
 
In QCD sphalerons have energy of only several GeV, so one might think those are well studied experimentally.
Unfortunately, this is not the
case: it still remains to be done. One proposal  \cite{Shuryak:2003xz}  to do that 
in hadron-hadron collisions is via double-diffractive production of hadronic clusters with
strong left-right quark asymmetry, 
say 6 units of the axial charge. These clusters should resemble hadronic channels which we discussed in connection to $\eta_c$ decays.

In heavy ion collisions one expect multiple sphaleron processes at the initial stage of the collision, 
leading to fireballs with a disbalanced chirality. Observation of this can be done with the help of Chiral Magnetic Effect
(CME), to be discussed at the end of this chapter


While we have shown that general Adler-Bell-Jackiw (ABJ)
anomaly relation require locking of the Chern-Simons number to the (axial) charge of the
fermions, it has not yet been explained  how exactly it happens.
In fact one can  follow this phenomenon using the ``exploding sphaleron" solution
derived previously.

Let us start at time $t=0$, from the static (but unstable) sphaleron. 
Dirac equation in the field of the sphaleron has a fermionic
zero mode. Like we previously discussed it for the monopole,  it should be i interpreted as a zero 
energy bound state of a massless fermion, which can be occupied or empty.


Sphaleron explosion solution includes
some radial electric fields (at intermediate time
only, at the late time it is transverse). This field
can accelerate a fermion, from the initial zero energy state to some excited state with positive
energy.  Thus a qualitative picture of fermion production does not mean their produciton
``from nowhere", but
includes level motion, out of occupied levels in
the ``Dirac sea" to physical states with positive energy.


What about antifermions? In order to have {\em the same} 
zero mode (or other) solution as fermions, one needs the same form of the Dirac
equation, with the same colormagnetic moment. Since the charge of antifermion is opposite to that
of the fermion, one has to flip the spin to compensate.

In QCD there is no problem with that, just chirality of the produced positive energy antiquark  would be
opposite to that of the quark. As a result, 
one has  production
of a particle-hole pair. The baryon charge is not changed, but the axial one -- the difference
between left and right -handed fermions -- is changed by two units.

In electroweak theory there is a problem: only left-handed fermions interact, while
right-handed do not. As a result, there is no antifermion solution mirroring the fermion one,
and the sole fermion is produced. Both axial and baryon (or lepton) charge is changed.

This reasoning is simple and consistent with  what one expects from the
anomaly, so there is no doubts in its validity. However attempts to follow
dynamically the outlined scenario had encounter the following problem.
Physical fermions are either produced or not, so change in the axial (or total) charge 
 can only be an integer. The change
 of the Chern-Simons number during the sphaleron explosion is expected to be 1/2,
 and complemented by a similar  process of sphaleron formation it is expacted to give 1,
 also an integer.

However, the first numerical solutions of the KM electroweak sphaleron ???
explosion revealed a trouble: the Chern-Simons
number apparently refused to settle at the expected change of 1/2.
It looked like a ball rlling down from the saddle point
of the potential does not rall all the way to the valley's bottom!
It was first taken as numerical problem, but
 the same feature was confirmed in  subsequent numerical solutions. 
 Moreover, it is there in the COS analytic  solution presented above:
the Chern-Simons number 
 stabilizes at late times is $ N_{CS} \simeq 0.12$.
 The not-quite completed transition in terms of $N_{CS}$ is consequence of the 
classical approximation, in which the gauge boson mass is neglected.



Fortunately, 
  the analytic solution to the
Dirac equation in the background of the exploding sphaleron was found in  \cite{Shuryak:2002qz}
%
%
%
%

\begin{figure}[h!]
\begin{center}
\includegraphics[width=4.5cm]{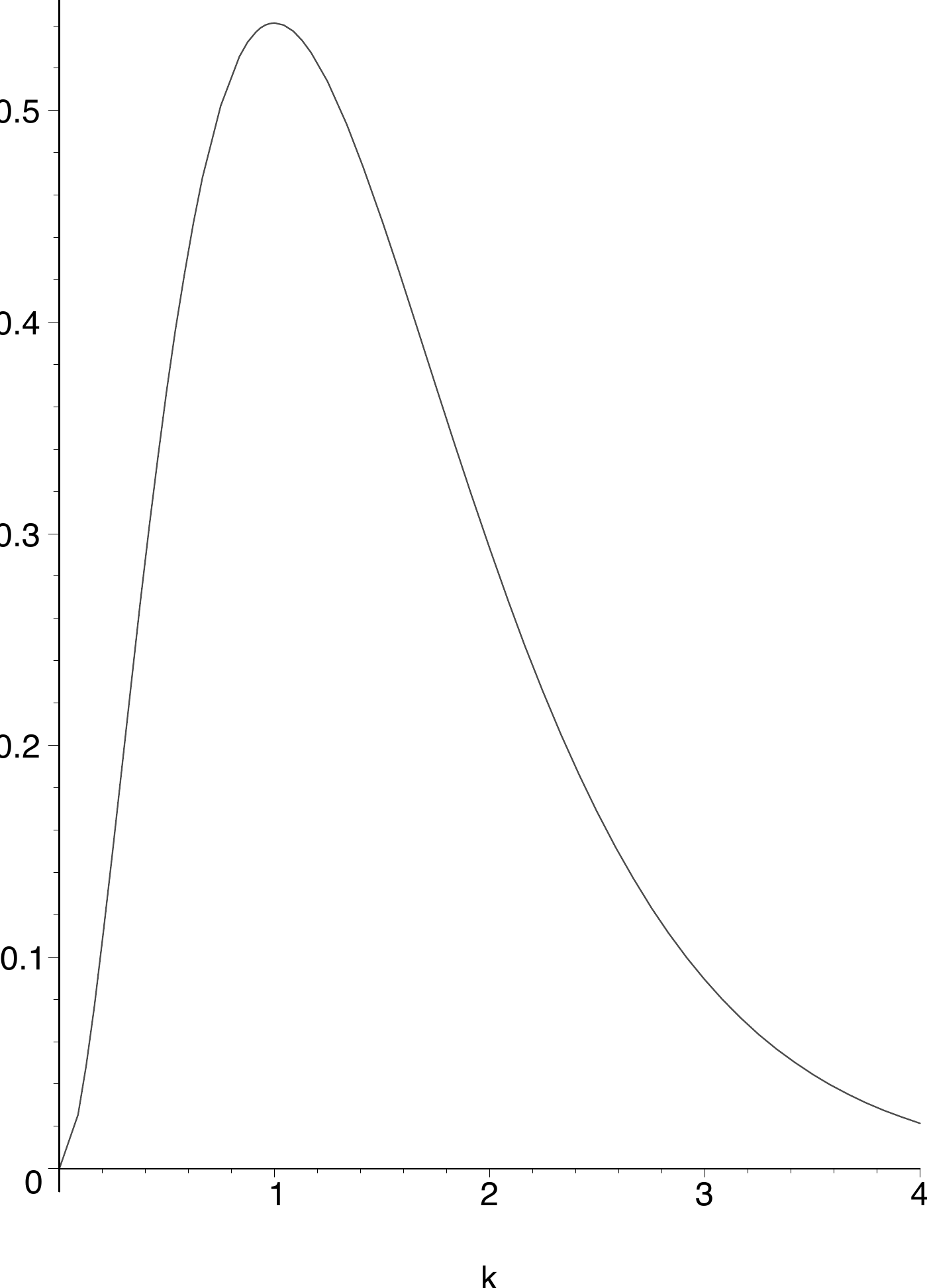}
\caption{\label{fig_exploding_quarks}
The spectrum  of {\bf R} quarks released at the Sphaleron point
versus its momentum k in units of $1/\rho$.}
\end{center}
\end{figure}

 The solution to 
the massless Dirac equation  in the Minkowski background field of
exploding sphalerons can also be obtained by
the same conformal mapping, from the O(4) Euclidean zero
modes. We explicitly construct these  states and show that at the initial
Minkowski time t=0 those are zero energy states, while at
 asymptotically large time they reduce to a free quark
or free antiquark of specific chirality. We also calculate the
spectrum of the produced fermions.

The solution itself is obtained by the following inversion formula
\be
{\bf Q}_+ (x) = \gamma_4\,\frac{\gamma_\mu \,(y +a)_\mu}{1/(y+a)^2}
\,{\bf \Psi}_+ (y)
\label{x1}
\ee
and it solves the ($\gamma_4\times$)
Dirac equation in the (Euclidean) 4-d spherically symmetric gauge configuration.


Omitting the technical details, let us proceed to the results.
The quark spectrum is close to Planckian with an effective
temperature $T=2/\rho$.
The distribution integrates exactly to $one$ produced fermion of each kind in electroweak theory
(with also antiquarks in QCD) as the anomaly relation requires.

\chapter{Sphaleron transitions in Big and Little Bangs}
\section{Electroweak sphalerons and primordial baryogenesis}

 \subsection{Introduction to Cosmological Baryogenesis}

 The observed Baryonic Asymmetry of the Universe (BAU) is usually expressed as the ratio of the baryon
 density to that of the photons. This key parameter enters calculations of primordial nucleosynthesis of such nuclei as $d,t,He^4,Li^7$: all of them agree that its value is
  \be n_B/n_\gamma\sim 6\times 10^{-10}   \ee

The question   how it was produced is 
  among the most difficult open questions of physics and cosmology. Sakharov 
 had formulated three famous {\em necessary  
conditions} for its generation: \\ (i) the baryon number violation\\
(ii)  the CP violation \\
(iii)  deviations from thermal equilibrium.

Although all of them are formally satisfied by
 the Standard Model (SM) and standard Big Bang cosmology, we do not yet know
a specific mechanisms creating the baryon asymmetry.
Smooth crossover electroweak transition does not generate large deviations from
thermal equilibrium, the baryon number violation is suppressed by huge sphaleron barrier,
and $CP$ violation via $CKM$ quark mixing matrix is quite small. 
It is therefore widely accepted that BAU is due to some so far hypothetical mechanisms,
e.g. heavy neutrino decays or large CP violation in Higgs sector, for review of the field see e.g. \cite{Dine:2003ax}.
This common view
is from time to time challenged by some suggested scenarios: one of them we will discuss
later in the chapter.


In the electroweak theory   semiclassical description of the tunneling through the barrier, separating
topologically distinct gauge fields, is given by
  the (by now much discussed) $instanton$ solution. However, unlike in QCD, the coupling is weak, the action large, and the tunneling
probability in the broken phase we live in is  extremely low
\be \Gamma_{tunneling}/T^4 \sim \exp(-4\pi/\alpha_w) \sim 10^{-170} \ee

%
What about thermal excitation of electroweak sphalerons?
In electroweak theory the coupling is  small, thus the sphaleron energy large
is $E\sim v/\alpha_w\sim  O(10\, TeV)$. Naive comparison with the highest
temperature at which the broken phase exists, namely the electroweak critical temperature $T_{EW}\approx 0.16 \, TeV$ give Boltzmann factor is not as small as the probability of tunneling just mentioned, but it is still
very small, $O(exp(-100))$. However, close to $T_{EW}$  the Higgs VEV is 
 small and the sphaleron rate should not be suppressed that much. We will return to this issue shortly.


However, this estimate is too naive, as the parameters of the electroweak theory are strongly renormalized near $T_c$.
The main  reason of the increase of the rate, from the 
KM sphaleron to revised one is of course the reduction of the Higgs  
 VEV from its vacuum value  $v=246\, GeV$ to near-zero. 

More accurate calculation of the  equilibrium
 sphalerons rates
 at the electroweak cross over region give much larger rates.   For an update see e.g.\cite{Burnier:2005hp} who estimated those in the range 
 \be \Gamma/T^4\approx  10^{-20}\ee
 including rather large preexponent calculated  semiclassically in
Refs. \cite{Arnold:1987mh,Carson:1990jm,Moore:1995jv}. 
Such rates are however still too  small for the solution of the  
baryogenesis puzzle.

   On the other hand, in the symmetric phase  at $T>T_c$, in which there is no Higgs VEV,
the sphaleron size $\rho$ can be much larger than the electroweak scale $T_c\sim 100\, GeV $, with respectively much higher
sphaleron rates. There is a limit to $\rho$ set by
 the inverse  magnetic screening mass \be \rho< {1\over m_{mag}}\sim {1 \over g^2 T}\ee
 and the dimensional arguments thus put the sphaleron transition rate to
 $ \Gamma \sim \alpha_w^4 T^4$. More complicated
 analysis of the problem \cite{Arnold:1996dy} shows that it is suppressed by one more power, so
 \be {\partial (\Delta N_{CS})^2 \over \partial t \partial V}=\Gamma \sim \alpha_w^5 T^4 \ee
Although there appears a rather high power of weak coupling,  the exponential suppression is gone, and so 
in the symmetric phase the sphaleron rates obtained are $larger$ than the expansion rate
of the Universe at the corresponding era. This means appearance of another
problem: all asymmetries which may be primordially generated
would then be wiped out above the electroweak critical temperature! 
These consideration force us to search for resolution of the  baryon asymmetry puzzle
at narrow temperature interval at or right below the electroweak phase transition.

But before we return to recent works on electroweak sphalerons, let us briefly describe the  
 history of cosmological electroweak transition.
 
\subsection{Electroweak phase transition} 
Originally, this transition was assumed to be 1-st order, in which the broken phase
(in which Higgs field obtains a nonzero VEV) first appear in form of bubbles.
Their coalescence and excistence of bubble walls, moving in the process, have
created hopes in 1980's-1990's for large local deviations from thermal equilibrium.  

However further studies of the transition have shown that the  1-st order transition can only happen if the Higgs mass is relatively small, less than 70\, GeV or so. After CERN discovery of Higgs particle with a mass of 125\, GeV, it became obvious that SM ony predicts relatively smooth crossover transition.
After this has fact has been acknowledged, people either looked at various phenomena $beyond$ the
SM, such as its supersymmetric extensions, which may still allow the
1st order transitions. We will also briefly discuss the so called ``cold" or ``hybrid" scenario,in which electroweak transition starts right after inflation, with strong deviation from equilibrium. 

Let us however return to standard cosmology and SM framework. The 
  best value of electroweak transition today is calculated on the lattice, e.g. 
\cite{DOnofrio:2014rug} produce the following value
\be 
T_{EW}=(159 \pm 1) \, {\rm GeV }
\ee
The temperature of the Universe today is $T_{\rm now}=2.73 K$. The ensuing redshift 
z-factor is

\be 
z_{EW}={T_{EW}\over T_{\rm now}}\approx 6.8\cdot10^{14} 
\ee
During the radiation dominated era, the  relation of time to temperature  is given by Friedmann
relation
\be 
t=\bigg({90 \over 32\pi^3 N_{\rm DOF}(t)}\bigg)^{\frac 12}{M_P \over T^2} 
\ee
Inserting the Planck Mass $M_P=1.2\cdot 10^{19}\, {\rm GeV}$, the transition temperature and the effective number
of degrees of freedom $N_{\rm DOF}=106.75$, we find  the time after the Big Bang to be

\be t_{EW}\sim 0.9\cdot 10^{-11} s, \,\,\,\,
c t_{EW}\approx 2.7\,{\rm  mm} \ee
Thus Universe was really tiny then!

 The rate of 
thermal sphaleron transitions, both semiclassically  and on the lattice, especially in the
symmetric phase
  see \cite{Klinkhamer:1984di,Arnold:1987mh,Carson:1990jm,Moore:1995jv}.
  The main fact is that the sphaleron rate before the transition, at $T>T_{EW}$,
  is suppressed by $\alpha_{EW}^5$ or about 7 orders of magnitude. And yet, it 
  is still many orders of magnitude $higher$ than the expansion rate of the Universe 
  (the Hubble ``constant" at the time)! So, ``large" sphaleron rate creates a problem of 
  potentially {\em washing out}
  any BAU created before the electroweak transition. Looking at it in more positive way,
  there appears a possibility to transfer some lepton number asymmetry into BAU,
  since sphalerons produce quarks and leptons in fixed combination. 
  
  Recent lattice studies gave us better data on the sphaleron rate. We will discuss it 
  a bit later, and now just note that there is a specific moment, known as {\em sphaleron freezeout}, when this rate $is$ equal to the Universe expansion rate. From 
 \cite{DOnofrio:2014rug} we know that it corresponds to temperature
 $T_{FO}\approx 130\, {\rm GeV}$. The corresponding cosmological time is then 
\be t_{FO}\sim 1.36\cdot 10^{-11} s, \,\,\,\,
c t_{FO}\approx 4\cdot \,{\rm  mm} \ee
Quite amusingly, these numbers nearly exactly 1000 time larger than critical and freezeout
temperatures in heavy ion collisions: $T_c\approx 155\, MeV$ and hadronic freezeout
$T_{FO}=117-140 \, MeV$, depending on the collision energy.

\subsection{Sphaleron size distribution}
At this point, let us interrupt the discussion of CP violation, and discuss whether 
large size (and small momentum scale fields) can indeed be found, in cosmological setting 
of electroweak phase transition.  

\begin{figure}[h]
\begin{center}
\includegraphics[width=14.cm]{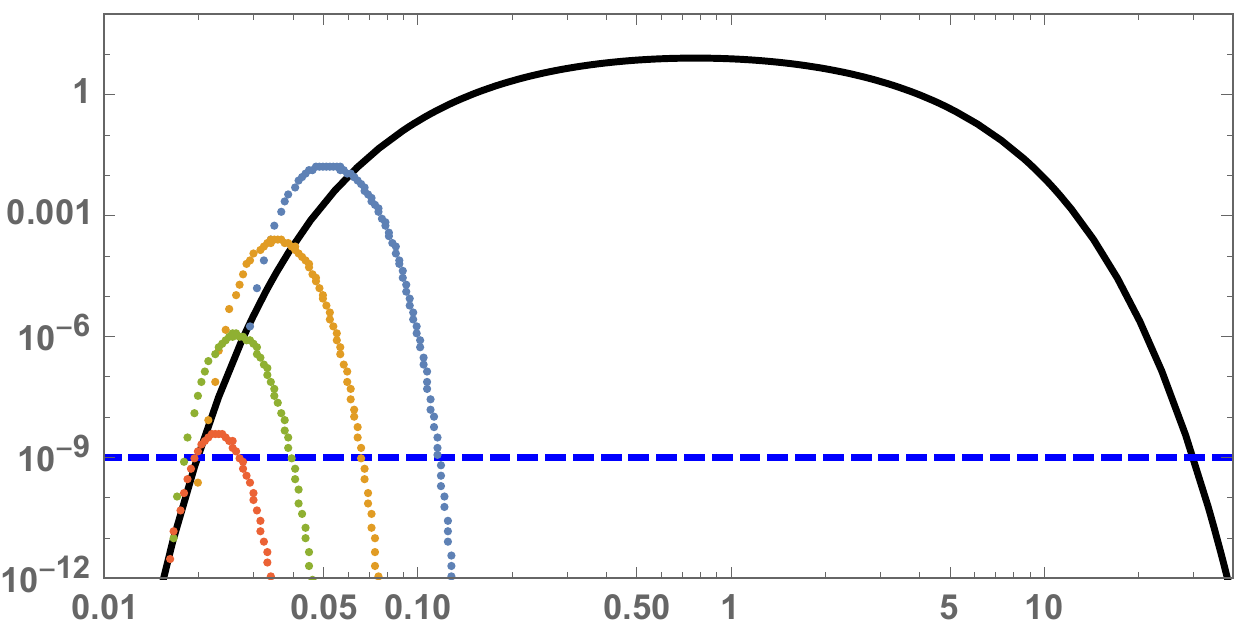}
\caption{The sphaleron suppression rates as a function of  the sphaleron size $\rho$ in ${\rm GeV}^{-1}$. 
The solid curve corresponds to the unbroken phase $v=0$ at $T=T_{EW}$. 
Four sets of points, top to bottom, are for well broken phase, at
$T=155,150,140,130\, {\rm GeV}$. They are calculated via Ansatz B described in Appendix C, and normalized to lattice-based rates. 
 The horizontal  dashed line indicates the Hubble expansion rate relative to these rates.}
\label{fig_P_of_rho}
\end{center}
\end{figure}

 The details can be found in our original paper \cite{Kharzeev:2019rsy}, let me just show the result
 in Fig. \ref{fig_P_of_rho}. In short, at $T>T_{EW}$ (as shown by solid line) the sphaleron
 tend to be rather large size, on a scale of electroweak magnetic screening length. But,
 as the temperature shifts below $T_{EW}$ and nonzero Higgs VEV appears, their sizes
 rapidly shrink (as shown by the colored points). The mass of the sphaleron grows, as with
 it the Boltzmann factor $exp(-M/T)$ rapidly falls, till at $ T < T_{fo}
 \approx 130\, {\rm GeV}$  the rate goes smaller than the Universe expansion rate
 (the horizontal  dashed line) and sphelarons become unimportant.  

 \subsection{The hybrid (cold) cosmological model and sphalerons} \label{sec_cold_baryo}

 A scenario which we will discuss would then be
the so called hybrid (cold) cosmological scenario   in which
the end of inflation coincides with the electroweak transition, so that equilibration happens at $T<T_c$.
This ensures {\em large deviations from equilibrium}.  While based on some fine tuning of the  unknown physics of the inflation, it avoids many  pitfalls of the standard  cosmology, such as ``erasure" of asymmetries generated before the electroweak scale.

The {\em baryon number violation} in it is due to the sphaleron transitions occur inside the bubbles of certain
size $\rho$ : and those can be studied numerically and semiclassically in detail. The {\em CP violation}
is not yet calculated accurately, but its magnitude near the sphalerons will be estimated.

 In such a scenario there are coherent oscillations of the gauge/scalar fields 
  studied in detail in  real-time lattice simulations \cite{GarciaBellido:2003wd,Tranberg:2003gi}. 
The simulated models include two scalars - the inflaton and
the Higgs boson -- and the electroweak gauge fields of the
SM, in the approximation that the Weinberg angle is zero
($Z$ is degenerate with $W$). All fermions of the SM are
ignored:  the effect of the top quark in particular
is the subject of the present paper.
After inflation ends, all bosonic fields are engaged
in damped oscillations
 for relatively short time, at the end of which
the Higgs VEV and gauge fields stabilize to their equilibrium
values, with the bulk temperature  $T_{bulk}\sim 50 \, GeV$, 
well below the critical (crossover) temperature. 

\begin{figure}[t!]
\includegraphics[width=6cm]{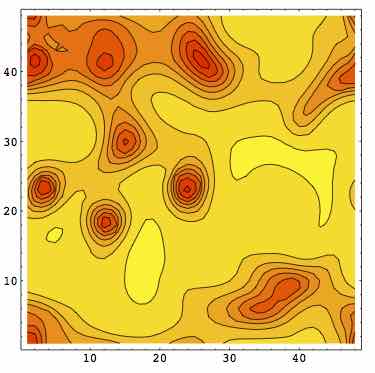}
\includegraphics[width=6cm]{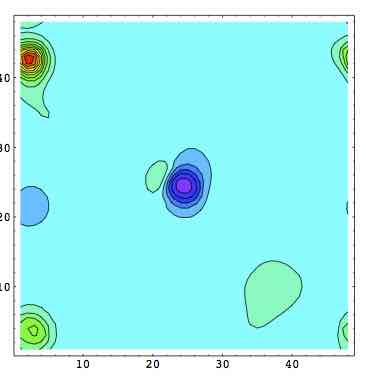}
\caption{(Color online). The
contour plots from \protect\cite{GarciaBellido:2003wd}
of the modulus of the Higgs field $|\phi(x)|^2$ (upper plot) and the topological charge density  $Q(\vec x,t)$   at 
time  $mt= 19$, for the 
model A1, $N_s=48$. Red (dark) areas in the upper plot correspond to small VEV, while yellow (light) bulk
corresponds to the broken phase. On the lower plot
 lumps of the topological charge
 density appear as red regions (dark in black and white display). While
 most of the no-Higgs spots do not have the topological transitions, all transitions seem to be inside the spots.
}
\label{contour1}
\end{figure}

(i) One important finding of the simulations is that the initial coherent oscillations of scalars soon give way
to the usual broken phase. The most important  feature is persistence of 
 ``no-Higgs spots'' 
in which Higgs VEV is very far from the equilibrium value $v$ and
is instead close to zero. The gauge fields in them have however rather high magnitude.
 Fig.\ref{contour1} (from \cite{GarciaBellido:2003wd})
show an example of a snapshot 
  of the Higgs field modulus. Typically
the volume fraction occupied by such ``no-Higgs spots"  is of the order of several percents and is decreasing with time. 

(ii) The second
important findings is that of
 topologically nontrivial fluctuations
of the gauge fields. As shown in the lower Fig.\ref{contour1}, those are well localized only $inside$ 
 the ``no-Higgs spots" mentioned above. Indeed, this becomes apparent from  the distribution
of the topological charge shown in the lower part of Fig.\ref{contour1},
 for the 
same time  configuration of the Higgs as shown in the upper part of the  Fig.\ref{contour1} .
Of course, it is only one snapshot, but the authors found from the simulations that it is true for the whole sample.

The fraction of ``hot spots" (no-Higgs-VEV)  which induced   topological transitions
is also in the range of few  percent. More precise measure is the so called
 sphaleron rate is defined by the mean square deviation from zero of the Chern-Simons number 
\be \Gamma(t) = {1 \over m^4 V}{d \Delta N_{CS}^2\over dt}\ee
(here and below all quantities are defined via one characteristic mass parameter $m$: for the
simulations its value is about $264\,\, GeV$, close to $v$. .)

Since the process only exist for finite period of time -- too early there are no gauge fields and too late there are no no-Higgs spots --
 its time integral is well converging. Its value
 \be I(m t)=\int^t_{t_i} d(mt) \Gamma(t) \ee
 is in the range
\be I\sim 10^{-4}\ee
for more details about different parameters sets see Table II of \cite{GarciaBellido:2003wd}. 
This quantity, as well as of course snapshots like those shown in Fig.\ref{contour1},
directly give the spatial distance between the
 topological fluctuations 
 $R_{sph}m=  20..30$.

\begin{figure}[htb]
\includegraphics[width=6.cm]{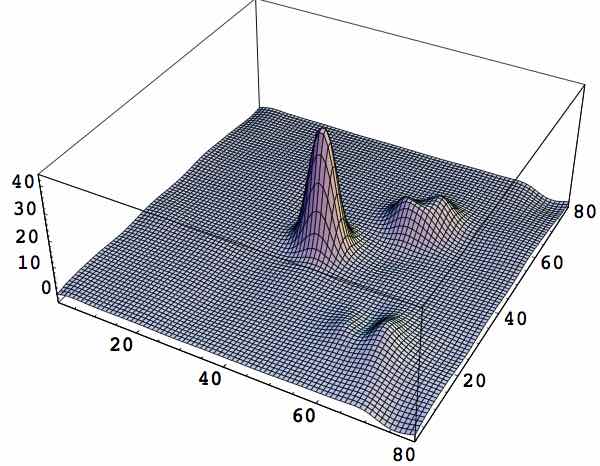}
 \includegraphics[width=6.cm]{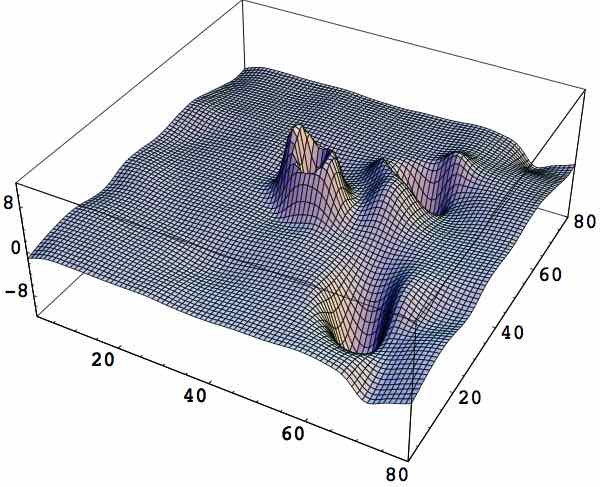} 
\caption{(Color online). From \protect\cite{GarciaBellido:2003wd}.
Two snapshots of the topological charge, at times $mt=18$ and $19$.}
\label{fig_top_explo}
\end{figure}

(iii) Space-time evolution of the  topological charge Q is shown 
in Fig.\ref{fig_top_explo} as two snapshots.
 One can see that the fluctuation at some time moment 
is very much concentrated in a small spherical cluster (the upper one in Fig.\ref{fig_top_explo})
is followed by an expanding spherical shell (the lower one in Fig.\ref{fig_top_explo})
which gets near-empty inside. 

The decomposition into electric and magnetic components of the field
 shows  that the fluctuations starts as nearly (90\%) 
magnetic object at $mt\sim 17$,  with the electric field
and thus the topological charge $\sim \vec{E}\vec{B}$
peaking some time later. Then there appears an expanding
shell, followed by the magnetic field  rebounding  to its secondary
maximum of smaller amplitude. We will return to discussion of all those features in the next chapter.

%

Before discussing these numerical results, let us point out an important issue which
are discussed in \cite{Flambaum:2010fp} but are outside our current interest:
 the  ``no-Higgs spots" seen in numerical simulations can be identified with the $non-topological$
solitons of the electroweak theory known as the  "WZ-top bags" \cite{Crichigno:2010ky}.

What we will consider now is
  the role of the quarks, especially the top quarks, in baryogenesis: due to high cost
 of inclusion of the fermions those up to now were not included in lattice numerical simulations.
 
%
%

Now we turn to the next question:
how much the {\em sphaleron rate} can be affected by their presence, as compared to purely bosonic ones in
the simulations \cite{GarciaBellido:2003wd}?

  The  Adler-Bell-Jackiw anomaly  require
 12 fermions to be produced. Particular
fermions depend on orientation of the gauge fields in the electroweak SU(2): since we are interested in
utilization of top quarks, we will assume  it to be ``up". In such case the 
produced set contains $t_rt_bt_g c_rc_bc_g  u_ru_bu_g \tau\mu,e$ , where $r,b,g$ are quark colors.
 We refer to it below as the  $0\rightarrow 12$ reaction.
Of course, in matter with a nonzero fermion
 density  many more reactions
of the type $n\rightarrow (12-n)$  are allowed, with $n$ (anti)fermions captured from
the initial state.

Evaluating the back reaction of the fermions on gauge field the
(analytic) solution to the
Dirac eqn  of the ``expansion stage'' 
 \cite{Shuryak:2002qz} we discussed above is very useful. A new element we are adding now
is  that its
time-reflection  can
 also describe the compression stage,
from free fermions captured by a  convergent
 spherical wave of gauge field at  $t\rightarrow-\infty$
 and ending at the sphaleron zero mode at $t=0$. The details will be omitted, but
 the main result is that by ``eating the top quarks" already present in he bags,
 one effectively lower the barrier and thus reach further increase of the sphaleron rate.
 
\subsection{Effective Lagrangian for CP violation }

The only known\footnote{Perhaps similar phase of the neutrino mass mixing matrix will be
discovered soon.} source of CP violation in the Standard Model is that induced by the phase of the 
of the  Cabbibo-Kobayashi-Maskawa (CKM)  \index{CKM matrix} quark mixing matrix. 
CP violation, originally discovered in decays of neutral $K$ mesons, is now studied mostly by LHCb CERN experiment at LHC in $B$ meson decays. It is a very large and involved field,
into which we have no time to go. 

Still, we need some brief introduction here. Let me start with QED and remind that 
4-th order ``electron box" diagram generates effective Lagrangian $L=O( F_{\mu\nu}^4)$,
originally calculated by Heisenberg in Euler in 1930's. Similarly, one can imagine
some ``background electroweak fields" $W_\mu,Z_\mu$ and effective Lagrangians generated by quark loop. 
On general ground, the CP-odd effects  require at least 4  CKM matrices, so the effect may in principle appear
starting from the  4-th order $O(W^4)$ ``quark box" diagram. So, one may think that
the calculation of CP-violating effective Lagrangian is just an exercise, most likely
done half century ago when electroweak theory was found. 

I am sorry to tell, that it is rather involved calculation and
the answer {\em remains unknown} (to me) even now.
For some reason, plenty of field theorists who do diagrams 
 of all possible kind in all possible theories failed to solve this important issue.
 
According to one explicit calculations  of the effective CP-odd Lagrangian
\cite{Hernandez:2008db} in the leading $W^4$ order there is no CP violation
(and we sill soon check that it is true in our approach ).  
These authors then observe that the effect appear in next-to-leading order diagram leads to 
the following dimension-6 operator
\be
L_{CP}=C_{CP}\epsilon^{\mu\nu\lambda\sigma}
\left[ Z_\mu W_{\nu\lambda}^+W_\alpha^-\left(W_\sigma^+W_\alpha^-+W_\alpha^+W_\sigma^-\right)+{\rm c.c.}\right]
 \label{eqn_LCP}  \ee
containing four charged gauge boson fields $W$ fields and one neutral $Z$.
Yet subsequent investigations in \cite{GarciaRecio:2009zp}  have not confirmed a non-zero coefficient
for this operator, but came up instead with a set operators  
of dimension 6 possessing with a completly different structure. 
Another group \cite{Brauner:2012gu} confirmed their finding. Remarkably, all the
13 operators $O_i$ found  are C-odd and P-even while the
above-given (\ref{eqn_LCP}) is P-odd and C-even. 

Zahed and myself  \cite{Shuryak:2016ipj} proposed to simplify the problem, by splitting it 
into two separate parts. Quark in the loop moves in background field, so the natural
basis in which one should approach the problem, is that of Dirac eigenstates
\be \Dslash \psi_\lambda(x)=\lambda \psi_\lambda(x) \ee
for the Dirac operator in this field. The propagators are in this representation diagonal,
as they are inverse to the Dirac operator, and contain quark masses in a standard manner.
The vertices are matrix elements of fields projected on currents made of $\psi_\lambda(x) $.
One may start with diagonal vertices.  

The heart of the problem is in performing flavor summation, with 4 CKM matrices and needed
number of propagators.  The lowest order box is given by
\begin{eqnarray}
\label{box}
 \int \prod_{i}^4 d^4 x_i  \, {\rm Tr}\bigg(W\!\!\!\!\!/(x_1) \hat V  \hat S_u(x_1,x_2) 
 W\!\!\!\!\!/(x_2)  \hat V^\dagger  \hat S_d (x_2,x_3) W\!\!\!\!\!/(x_3)  \hat V  \hat S_u  (x_3,x_4)
 W\!\!\!\!\!/ (x_4)   \hat V^\dagger   \hat S_d(x_4,x_1)\bigg)
\nonumber
\end{eqnarray}
Here hats indicate that the CKM matrix $V$  and the propagators
  are $3\times 3$ matrices in flavor space. The propagator also have
labels $u,d$, indicating up or down type quarks.   

The trace is over Dirac  and color indices as well: slash 
with $W$ field as usual mean convolution of vector potential with gamma matrices. 
The point is: one should look first at the flavor trace, multiplying 8 matrices explicitly
(in Mathematica) .
And indeed, the leading box diagram produces no CP violation!
We also found that if one add one $Z$ field, we still get zero. Adding two $Z$, one on $u$
and one on $d$ quark line, does produce the expression
\be
F(\lambda) =
\lambda^4 {\rm Tr}\left(  \hat V S_d  \hat V^\dagger  S_u  \hat V S_d  \hat V^\dagger  S_u \right)
\ee

%
Let us use representation in which the main operator is diagonalized, so that 
\be i\Dslash  \psi_\lambda(x)=\lambda  \psi_\lambda(x) \ee
where the notations with the slash here and below mean the convolution with the Dirac matrices $\Dslash=D_\mu \gamma_\mu$.
The corresponding  (Eucidean time) propagator -- describing a quark of flavor $f$ propagating  in the background -- can thus be written as
the sum over modes 

\be S(x,y) = \sum_\lambda {\psi^*_\lambda(y) \psi_\lambda(x) \over \lambda +M \partslash^{-1} M^+} \ee
The generic fourth-order diagram in the weak interaction, containing necessary four CKM matrices, 
takes in the coordinate representation the form
$$ \int d^4 x_1  d^4 x_2  d^4 x_3  d^4 x_4 Tr[ \Wslash(x_1) \hat V  \hat S_u(x_1,x_2)  \Wslash(x_2)   $$
$$ \hat V^+  \hat S_d (x_2,x_3)\Wslash(x_3)  \hat V  \hat S_u  (x_3,x_4)\Wslash(x_4)   \hat V^+  \hat S_d(x_4,x_1)] $$
where $V$ is the CKM matrix, the hats on them and the propagators 
with $u,d$ subscripts indicate that they are the $3\times 3$ matrices in flavor subspace,
and the trace is implied to be over flavor indices.
If one consider next order diagrams, with $Z,\phi$ field vertices, the expressions are generalized straightforwardly. 

Spin-Lorentz structure of the resulting effective action is very complicated. To understand the
scale dependence we will
now we make  strong simplifying assumptions. First,  we will focus on the diagonal matrix elements of the operator
$ \Wslash$ and assume it to be approximately proportional to $\lambda$ (with some coefficient $\xi$)
\be  <\lambda | \Wslash | \lambda' > \approx \xi \lambda \delta_{\lambda \lambda'} \ee
Second, we assume that right-handed operator $i\partslash$ can similarly be represented by
diagonal matrix element we will call $\pslash$.
If so, one can use the orthogonality condition of different modes and perform the integration over coordinates,
producing much simplified expression, with a single sum over eigenvalues
$ \sum_\lambda F(\lambda)$ 
where
\be
F(\lambda) =
\lambda^4 Tr\left(  \hat V S_d  \hat V^+ S_u  \hat V S_d  \hat V^+ S_d \right)
\ee
This is the diagram in the $\lambda$-representation, which generalizes the momentum representation valid
only for  constant fields. Unlike momenta, the spectrum of Dirac eigenvalues $\lambda$ may have various
spectral densities. In particular, there is a zero mode, corresponding to  zero mode in the original 4-dimensional
symmetric case. This describes the fermion production on various backgrounds, such as the exploding sphaleron.

Independent of what physical meaning  and spectrum of $\lambda$ is, the point is that one can perform multiplication of flavor matrices
and extract universal
function of $\lambda$, describing dependence of CP violation on the scale.

Using the standard form of the CKM matrix $\hat V$, in terms of known three angles and the CP-violating phase $\delta$, and also six known quark masses, 
one can perform the multiplication of these 8 flavor matrices and identify the lowest order  CP-violating term of the result.
Performing the multiplication in the combination above one finds a
complicated expression which does $not$ have $O(\delta)$ term,
so there is no lowest order CP violation. This agrees with a statements from earlier works
 that the leadng 4-th order diagram generates no operators.

Higher order diagrams however do have such contributions. For example,
if on top of 4 $W$ vertices with CKM matrices there are also two $Z$,
flavor trace looks as follows
\be
F_{ZZ}(\lambda) = 
\lambda^6 Tr\left(  \hat V S_d  \hat V^+ S_u  \hat V S_d Z S_d  \hat V^+ S_u Z S_u \right)
\ee
Now the flavor trace has the lowest order CP violation 
described by the following symmetric expression
$$  Im F_{ZZ}(\lambda) =  2  \lambda^6 {J (m_b^2 - m_d^2) (m_b^2 - m_s^2) (m_d^2 - m_s^2) (m_c^2 - m_t^2) (m_c^2 - m_u^2) (m_t^2 - m_u^2) \over \Pi_{f=1..6} (\lambda^2 +m_f^2)^2 } $$

 The numerator is the Jarlskog
combination of the CKM angles and differences of masses squared.
 As expected, the effect vanishes when the mass spectrum 
 of either u-type or d-type quarks gets degenerate.
 \begin{figure}[h!]
\begin{center}
\includegraphics[width=7.5cm]{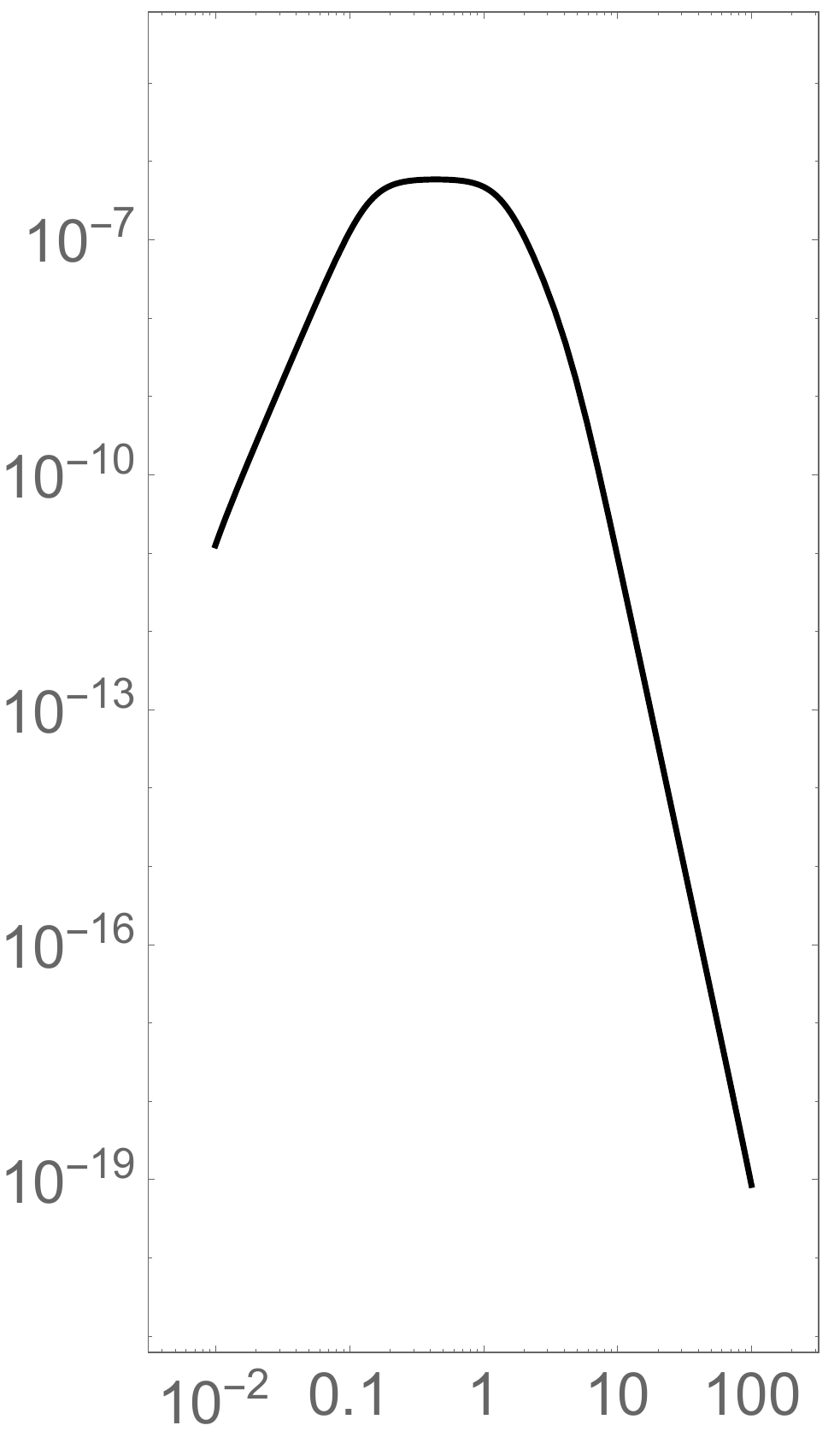}
\caption{The CP-violating part of the $W^4 Z^2$ diagram $ Im F_{ZZ}(\lambda)$ versus $\lambda$ (GeV).}
\label{fig_Fcp_lambda}
\end{center}
\end{figure}

A plot of this function\footnote{Quark masses are all taken as they are in our world, that is at Higgs field equal $v$. However,  in a ``hot spot" in which the sphaleron transition happens,
the Higgs expectation value is smaller than in the broken phase, $\phi< v$, by certain factor. One can take care of this by rescaling
all $\tilde\lambda$ by this $\phi/v$ factor. Since the function is dimensionless,
its values are preserved, and  the plot just moves horizontally as a whole by this factor.}
 is shown in Fig.\ref{fig_Fcp_lambda}. Because of the cancellation
between different quark flavors, it is very small at large $\lambda$, about $10^{-19}$ at the electroweak scale 
$\lambda\sim 100 \, GeV$ at the r.h.s. of the plot, which can be called ``the Jarlskog regime".
Yet at the scale near and below $\lambda =1\, GeV$ -- so to say ``the Smit regime" -- it is twelve orders of magnitude larger!

These calculations show that for sphalerons produced by cold electroweak scenario, with size at the scale $\rho\sim 1/80\, GeV$,
the CP odd effect is way too small to explain the baryon asymmetry. If some mechanism will be found, which can generate the sphaleron transitions with much large sizes, say $1\, GeV$,
the CP violation in the CKM matrix
would be sufficient! 

\subsection{ The CP violation in the background of exploding
sphalerons}

The  paper \cite{Kharzeev:2019rsy}
performed  evaluation of the CP violation in the background of exploding
sphaleron, which we will here follow. But first, a general discussion of why exploding
sphalerons are different from arbitrary background field discussed in the previous subsection.

One obvious point is that, as any classical object, sphaleron has strong fields $A\sim 1/g_{EW}$ and therefore in the expansion over the field operators there is no coupling constant.
It is however still true that one can have perturbative expansion, but with a parameter
being deviation of CKM matrix from the unit one. It is still true that CP violation requires at least
four CKM matrices. 

The nontrivial point made by \cite{Kharzeev:2019rsy} is the so called {\em ``topological stability"}:
the   Dirac operator  in the background of a sphaleron explosion still possesses
a  topological zero mode\footnote{Note that this scenario is different from earlier ones,
based on small $momenta$ $\vec p\approx 0$, which are $not$ topologically protected  from rescatterings.}, $\lambda=0$, surviving any gluon rescattering. 
For this mode the formulae from the previous section cannot be applied, and CP violation needs to be calculated in a different way. 

Not going into detail, we emphasize the physics. In the broken phase $T<T_{EW}$
six different quarks obtain six different masses, proportional to now nonzero Higgs VEV
$v(T)$. Suppose sphalerons produce first a particular quark, say $u$: in a CKM vertex
it can morph into a different ones, and then again. Combining amplitude and conjugated amplitudes one can single out interference diagrams with (say) four CKM  vertices. 
When quark propagates between those, it gets a phase proportional to
time and $m_Q^2/M_{KW}$, where $ m_Q$ is its mass, and $M_{KW}$
 ``thermal Klimov-Weldon" quark mass
 \be M_{KW}={g_s T \over \sqrt{6}} \sim 50 \, {\rm GeV }  \label{eqn_MKW}  
 \ee
induced by the real part of the forward scattering amplitude of a gluon on a quark.   
Different possible options with different intermediate flavors interfere with each other,
and produce CP violating imaginary part. This means that probability to produce
quarks and antiquarks are not the same, which leads to BAU.

Omitting details, let me go to the answer from  \cite{Kharzeev:2019rsy}:
for light quark ($u,d$) production the asymmetry is 
\be A_{CP}\sim J  {(m_b^2-m_s^2)(m_c^2-m_u^2) \rho^2 \over M_{KW}^2} \sim 10^{-9} \ee
where the sphaleron size $\rho$ stands for typical distance between vertices.  
It is also important that phase associated with $top$ quark is large and those terms were excluded, thus no top quark mass here. 

Is it enough for BAU observed? In a very crude way, very small sphaleron rate near
freezeout $\sim 10^{-16}$ is basically cancelled by very long time $T t \sim 10^{16}  $
in units of micro scale. Therefore, up to a couple of orders of magnitude, the CP 
asymmetry is the BAU ratio. SO, to my opinion, the scenario works. Needless to say,
there is and will be long process of its scrutiny. 

%
%

\subsection{Electroweak sphaleron explosion: other potential observables} 

{\bf Sounds and gravity waves} are naturally emitted at explosions, and sphaleron explosions are no exception. In a symmetric phase, $T>T_{EW}$, we have explicit  time-dependent solution
of the classical Yang-Mills equations \cite{Ostrovsky:2002cg,Shuryak:2002qz},
from which one can calculate the stress tensor components, and see that  the word ``explosion" is not a metaphor here. There is however a problem: in the symmetric phase, the sphaleron explosion is spherically symmetric. and therefore cannot radiate direct gravitational 
waves. However, the indirect gravitational waves can still be generated at this stage, 
through the process$$ {\rm sound+sound \rightarrow gravity \,\,wave}$$
  pointed out in~\cite{Kalaydzhyan:2014wca}. After the EWPT, at
 $T<T_{EW}$, the nonzero VEV breaks the symmetry and the sphalerons (and their explosions) are no longer spherically symmetric but elliptic. 
 With a nonzero and time-dependent quadrupole moment, they generate
 direct  gravitational radiation. The corresponding matrix elements 
 of the stress tensor were evaluated in \cite{Khazeev:2019rsy}. 

{\bf Magnetic clouds and their linkage}
In the broken phase, at  $T<T_{EW}$, the original third component of the  $SU(2)$ non-Abelian
gauge field  $A_\mu^3$  get mixed with electromagnetism, producing physical massive 
 $Z_\mu$ and a massless $a_\mu$. This mixing is controlled by the so called Weinberg angle
 $\theta_W$. Therefore sphaleron explosions in this phase creates electromagnetic
 magnetic clouds. Since there are no  electromagnetic
 magnetic  monopoles, they are not screened but survive till today, as so called 
 {\em intergalactic magnetic clouds}.
 
 The interesting feature of those is that chirality of fermions (electrons) produced by
 sphalerons can be directly transported to chirality of the magnetic field, known as
 ``magnetic linkage" or Abelian Chern-Simons number 
 \be 
\label{HEL}
\int d^3 x A B\sim B^2 \xi^4 \sim {\rm const }
\ee  
The configurations with nonzero (\ref{HEL}) are called  {\em  helical}. 
 
 Multiple sphaleron and anti-sphaleron transitions create large number of 
 left and right-linked objects, which would annihilate each other (as quarks and antiquarks do).
 But, if there is a ($CP$-violating) $asymmetry$ between sphaleron and anti-sphaleron transitions, there would be remaining linkage which will not be able to annihilate
 (as remaining quarks do). Therefore, potential observation of 
 helical misbalance in magnetic intergalactic fields today can be related to 
 BAU. 
 
 Of course, many magnetic phenomena may happen, between electroweak transition and now.
 One of them is  the so called {\em inverse cascade} effect, following from Maxwell equation in matter with chiral unbalance and current proportional to magnetic field (CME).  
Discussion of them goes well beyond these lectures.   
 


\section{QCD sphalerons}
\subsection{Sphaleron transitions at the initial stage of heavy ion collisions}
In these lectures we do not systematically discuss heavy ion collisions,
often referred to as ``the little Bangs". For
our current purposes it is enough to point out few similarities and important differences from 
cosmology, the Big Bang. Similarly, both are explosions of systems much larger than
the microscale of the matter involved, expanding slowly enough to warrant adiabatic
approximation and entropy conservation. In both cases, we are interesting in crossing
the phase transitions, electroweak or QCD ones, and phenomena associated with those.
It is assumed that most of excitations from the initial time has been relaxed 
to thermal equilibrium. 

Yet,
in both cases, there are certain long-wavelength acoustic modes, with their relaxation time being so long that
they do $not$ die out from the intitation (Bang) time.  Both in cosmology 
and heavy ion collisions those are visible in final matter distribution: as acoustic microwave
 background temperature fluctuations in cosmology and as azimuthal correlations in the Little Bang. Their physics is rather similar: but in Universe we speak about angular harmonics up to
 few thousands, and heavy ion collisions up to 7 or so. Both in principle
 tell us something about the ``initial state". For Big Bang it is amplitude and statistical
 properties of initial quantum oscillations of whatever fields there were at Big Bang
 (inflaton?), for Little Bang it is (much less mysterious) initial positions of the nucleons at
 the moment of the collision. 
 
Of course, in these lectures we are not interested in just initial distribution of matter
densities in space,  resulting in such acoustic modes excitation. Our interest is in the initial distribution in relevant {\em topological coordinates}, the Chern-Simons number,
and its relaxation to equilibrium via certain instanton/sphaleron transitions. 

High energy collisions of  nuclei start from highly excited  out-of-equilibrium initial stage
of mostly gluonic field. Unfortunately, it is not easy to make reliable ab-initio calculations
of it. All we experimentally know are the one-body density matrix, or
gluon density functions (PDF's) as a function of their momentum fraction $x$, $g(x)$.
Very little is known about two-gluon correlations or coherency of those gluons.
One popular view of it is {\em Color Glass Condesate} model, suggested by
 \cite{McLerran:1993ni} which views it as high occupancy 
classical gluonic field. If so, it is calculated via classical Yang-Mills equations with
certain sources. After collision, the field gets modified to (still 
out-of-equilibrium and classical) state of  glue  called GLASMA.
With system expansion and decreasing density, the occupation numbers 
eventually reach $O(1)$ and thermal state of the gluon plasma is established.

\begin{figure}[t]
\begin{center}
\includegraphics[width=10cm]{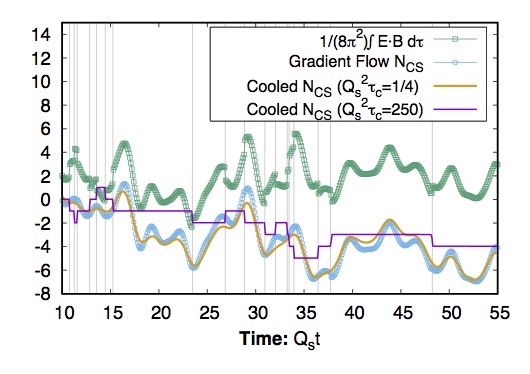}
\includegraphics[width=10cm]{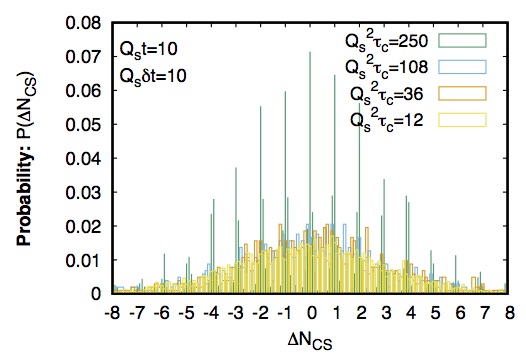}
\caption{Upper:Evolution of the Chern-Simons number for a 0.35
single non-equilibrium configuration. Different curves correspond to  different extraction procedures. Lower:
Histograms of the distribution over $N_{CS}$ at time $Q_s t=10$.}
\label{fig_mace}
\end{center}
\end{figure}

Studies of GLASMA are done by numerical real-time solution of classical Yang-Mills equation, 
technically similar to
simulations of out-of-equilibrium electroweak stage of the Big Bag we discussed above. 
In this section we will follow  work by Mace, Schlichting and  Venugopalan
\cite{Mace:2016svc} (in which you can also find references on the previous works on the subject).

During the evolution of GLASMA there is diffusive spread over the topological landscape,
so that measuring the Chern-Simons number as a function of time one observes a typical
diffusive behavior, see Fig.\ref{fig_mace} from \cite{Mace:2016svc}. ``Cooled" configurations, by
various method, descend to the bottom of the  topological landscape, that is toward zero energy and
integer values of $\Delta N_{CS}$, shown by (violet) rectangular curve in the upper figure. 

The lower figure
shows the histogram of the distribution over $\Delta N_{CS}$. Since in QCD there is no CP violation,
the curve is left-right symmetric, that is mean $<\Delta N_{CS}>=0$. 
The width of the distribution 
indicate the magnitude of the additional chiral imbalance in the typical events
created between the time $Q_s t$=10 during the time interval $Q_s \delta t$=10.
Taking say the characteristic parton momenta in GLASMA $Q_s\sim 2\, GeV$ one
finds that both the absolute time and time interval are about 1 $fm/c$. 
 Taking this histogram
as an example of that, we see than r.m.s. deviation corresponds to the diffusive motion adding 
$\Delta N_{CS}^{r.m.s.} \approx 3$ new transitions.  
For 3 light quark flavors, this corresponds to the added chiral charge of the configurations to be about
\be Q_5^{r.m.s.}=2N_f*\Delta N_{CS}^{r.m.s.}\approx 20 \ee
The rate of sphaleron transition decreases with time, so the first $fm/c$ is dominant in such 
calculations.

Note that we call diffusive component $\Delta N_{CS}$, not just  $ N_{CS}$, because this is the amount
of sphaleron transition generated in GLASMA, on top of what it is at the time  of the collision $N_{CS}(t=0)$.
This initial value is the subject of the next subsection. 

\subsection{Sphalerons from instant perturbations}

As we argue in most chapters of this book, the QCD vacuum contains virtual fields
which are topologically non-trivial. Given certain amount of energy, these virtual
fields may become real excitations observable experimentally. Before we 
discuss specific proposal for experiments, let us first discuss some solvable
 quantum-mechanical example  explaining the principle on which they are based.

\begin{figure}[t]
\includegraphics[width=10.cm]{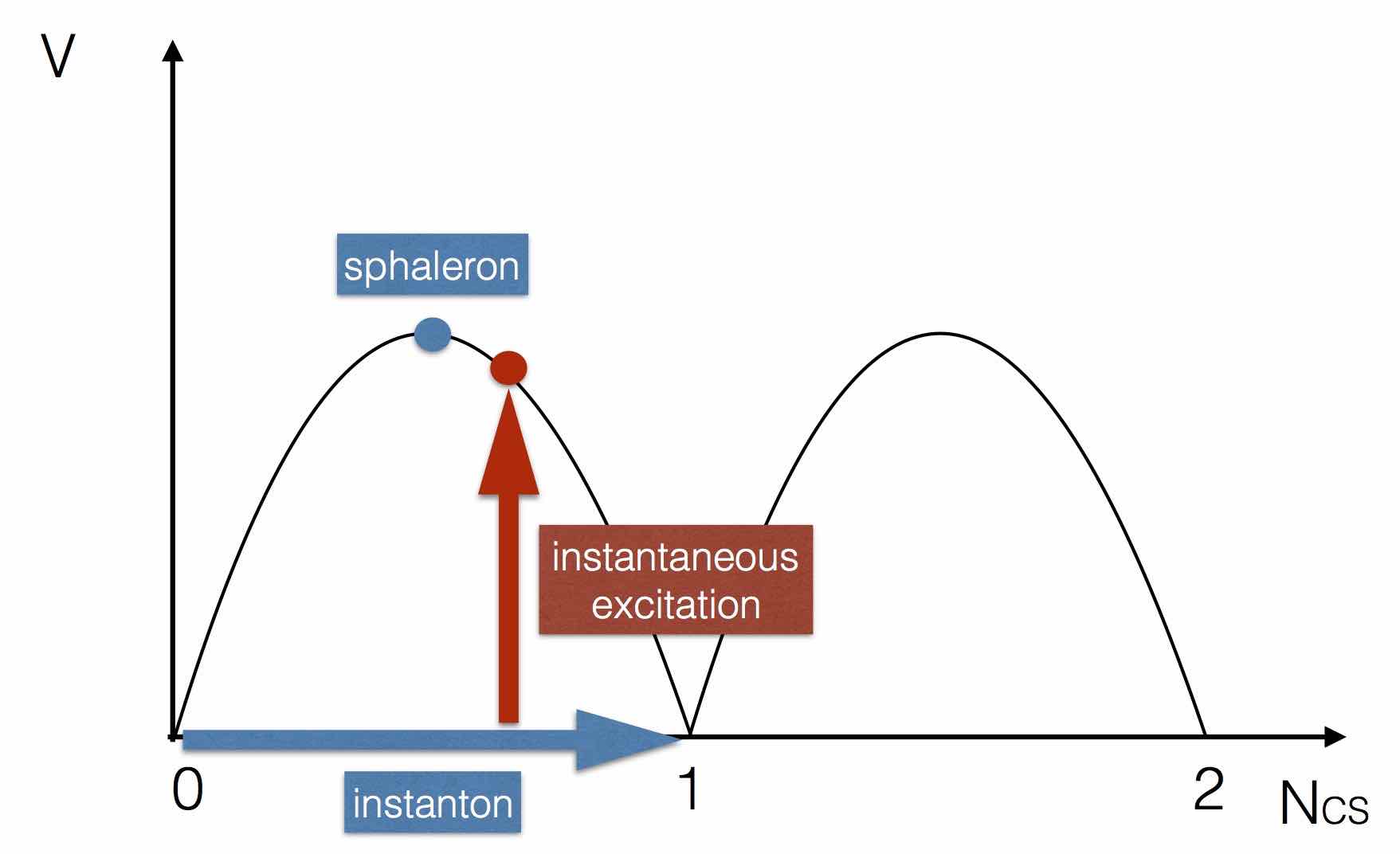}
\caption{Schematic representation of the potential as a function of Chern-Simons number. Horizontal blue arrow 
marked ``instanton" indicate the tunneling process in vacuum, while vertical red arrow shows
the direction of instantaneous excitation. } \label{fig_sketch_excitation}
\end{figure}

  We will show that 
basically  if one wants to make some virtual fields of the  vacuum real,
all one has to do is to clap the palms of one's hands strong enough!  Indeed, 
strong enough instantaneous perturbations  applied to a system at some coordinate
{\em under the barrier} will localize it in a state {\em near the top of the barrier}
at the same value of the coordinate. Schematic picture of such 
excitation is indicated in Fig.\ref{fig_sketch_excitation} by vertical red arrow.

In QCD setting we will study 
production of sphalrons in 
high energy collisions by
``instanton-induced processes" in the next chapter. Here we 
%
%
  only outline the argument by a
  quantum mechanical example from my paper \cite{Shuryak:2002an}.
  Its main idea is that near-instantaneous  perturbation
  does not leave time to move -- change any coordinates, 
  including the topological ones -- while the
  energy can be changed by the amount determined by energy-time uncetainty relation
  $\Delta E \sim 1/\Delta t$.  As we will see, under rapid perturbation the system jumps mostly
  {\em on the barrier},
  into the analogs of the ``sphaleron path" states we discuss
  in this chapter. 
  
\begin{figure}[b]
\includegraphics[width=5cm]{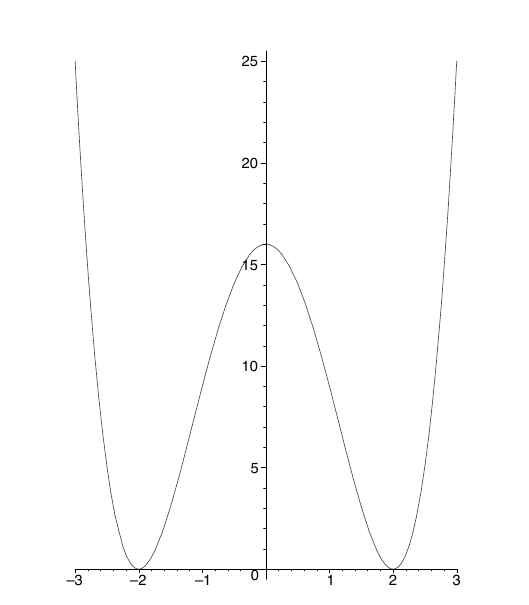} \includegraphics[width=5cm]{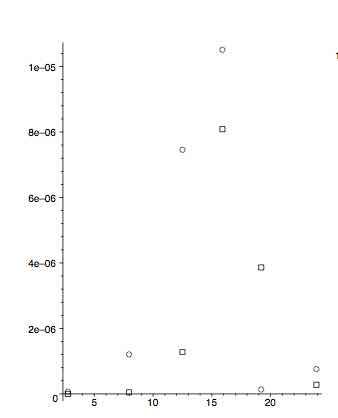}
\caption{(Left) the double well potential used. (Right)The excitation probability $P_n$ of the double-well system versus the excitation energy. Two sets of points are for two excitation functions  mentioned in the text.  } \label{fig_instantanous_pert}
\end{figure}

  Let us consider a double well potential (shown in Fig.\ref{fig_instantanous_pert})
  and
   try to detect a particle's presence  under the barrier,  near its middle $x=0$ point.
  To do so one may introduce an external perturbation  
   $\delta V\approx \delta(x) f(t)$ ocalized  near $x=0$  and 
   some time dependence $f(t)$. Standard solution is expanding $f(t)$ 
   into Fourier integral and observing that
    frequencies tuned to
the transition from the ground to the $n$-th level $\omega=E_n-E_0$
would create real transitions.
One can calculate the probability of the transition $P_n=<0|\delta V |n>|^2$.
The results are plotted versus $n$ as points in Fig.\ref{fig_instantanous_pert}
Note that the peak excitation energy corresponds to the ``sphaleron excitation",
the maximum of the potential, $V \approx 16$ in this example.

The reason for this is the following. While there can be enough energy to excite
the system to even higher states, the overlap matrix element to those higher states with
the ground wave function rapidly decreasing with excitation. Only near the maximum of the
potential the final spatial wave function is smooth enough, and thus the overlap with the
ground state wave function is large.

\subsection{QCD sphalerons in experiments}

As we already discussed above, the sphalerons were discovered in the context of electroweak theory,
but -- because there is no hope to produce an electroweak sphaleron experimentally --
it has been mostly discussed in relation to cosmological applications.
Now, returning to QCD, with its energy scale a factor 100 lower, one might think
that sphaleron production happens routinely in any hadronic or heavy ion collisions,
and is therefore studied throughout  in experiments. Yet, as we will now discuss,
only the former part of the previous sentence is true.


Three suggestions of how one can observe the QCD sphalerons experimentally were discussed in literature.
We will go over these ideas briefly here, as discussion of real experiments 
would take us too far from the main goals of this book.

Discussion of sphaleron excitation in high energy collisions has historically started in the setting of electroweak
theory.  If observed, it would be a spectacular demonstration of baryon and lepton number violation
in Standar Model: but, s we already 
mentioned several times, there are no prospects to do so
experimenally.  Still there were significant theoretical efforts made and important insights
were gained: for a good review see \cite{Mattis:1991bj}.

 Schrempp, Ringwald and collaborators applied this theory to lepton-hadron deep inelastic scattering,
 see \cite{Moch:1996bs} and subsequent works.  The idea was to
consider deep-inelastic $ep$ collisions with large momentum transfer $Q$, in which small-size instantons 
will  excite large mass $M\sim Q$ sphalerons. Unlike perturbative processes, resulting in a single
quark jet (or few jets, with radiated hard gluons), the sphaleron is  expectted to decay into a high multiplicity near-isotropic cluster of particles. The idea was to 
keep the scale in the weak coupling domain, so that the semiclassical calculation
be well controlled, say $M\sim 10\, GeV$. Unfortunately, in this regime the
cross section is too small, compared to various perturbative processes, and
the project eventually collapsed since it was not possible to tell the ``signal" from ``background".

 Another direction for experiments, which
 can be called  ``soft", 
 has been proposed by Zahed and myself  \cite{Shuryak:2003xz}.  Instead of hard gluons we propose to use
  soft Pomerons, focusing on the double-diffractive   $pp$ or $\gamma\gamma$ collisions. 
  The sphaleron is expected to be found among various gluonic clusters which
  ``jump out of the vacuum" at mid-rapidity. 
 Observation of the two scattered protons provide very constrained kinematics
and define the mass of the clusters. There are some old experiments   which observed 
few-GeV clusters of hadrons, with an isotropic decay: but theor quantum numbers
were never controlled. Our suggestion
was to look for 
 certain exclusive channels related to the anomaly, 
 which tells us that quark state produced must be of 
 the particular flavor-spin structure
\be  (\bar u_R u_L)(\bar d_R d_L)(\bar s_R s_L) + (L\leftrightarrow R) \ee
 which would confirm or reject
their topological origin. Similar argument was quite successful 
in the decays of $\eta_c$, which we will discuss in the next chapter.
 The scale of the
 sphaleron mass there should correspond to the average
 instanton size $\rho\sim .3 \, fm$, corresponding to the sphaleron mass of about 3 \, GeV
 (incidentally, close to the mass of  $\eta_c$.)
The proposal to do these experiments at RHIC STAR detector in the pp mode has been made,
but not yet done.

 The third approach was suggested by Kharzeev and collaborators \cite{Kharzeev:2007jp}.
Instead of looking for individual sphaleron transitions, they proposed to observe
total global chiral $L-R$ disbalance of the fireball 
 produced in heavy ion collision by fluctuating multiple sphaleron transitions. 
 Note that the chiral charge is CP odd quantity, and so its average value is of course zero in strong interactions.
 
 Its observation 
was proposed to do based on the so called Chiral Magnetic Effect (CME) 
\cite{Fukushima:2008xe}, according to which  the CP-odd chiral disbalance leads to an $electric$ current along
the applied $magnetic$ field\footnote{Note that the coefficient between the current and filed is in this case
T-even: so, unlike the usual Ohmic current this one is a $non-dissipative$ one. This observation
 will lead to multiple applications of the CME in condense matter physics and
perhaps even electronics.}

It is important that ambient matter should be
quark-gluon plasma, at $T>T_c$, in which chiral symmetry is unbroken and thus chiral disbalance
remains conserved. 
This suggestion has lead to significant experimental activity. 
We  of course cannot describe it here in detail: the effect is clearly seen, but possible backgrounds are
not yet completely understood. 

The experimental program continues, at both RHIC and LHC. Eventually we will learn the sphaleron rates,
both the ``primordial one", from nonzero topology in the vacuum wave function, as well as that
in GLASMA. Experimental check of the sphaleron theory in QCD will, no doubt, strengthen also our understanding
of electroweak sphalerons in the Big Bang.

\subsection{Diffractive production of sphalerons}
After the correspondence between the instanton-antiinstanton configurations
and sphalerons has been elucidated, let us return to high energy collisions, following 
Ref
\cite{Nowak:2000de}.

\begin{figure}[h]
\begin{center}
\includegraphics[width=8.cm]{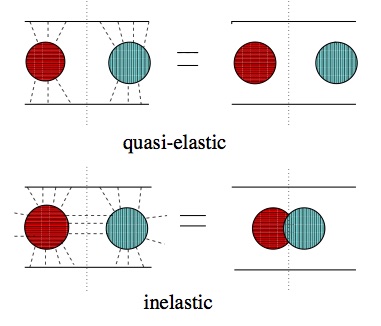}
\caption{Schematic representation of the amplitude
squared, with (without) gluon lines are shown in the left (right) side of the figure. The dotted vertical line is the uni- tarity cut. The upper panel illustrates the quasi-elastic (at the parton level) amplitudes where only color is exchanged. The lower panel depicts inelastic processes in which some gluons cross the unitarity cut, and some gluons are absorbed in the initial stage..}
\label{fig_elastic-inelastic}
\end{center}
\end{figure}

Calculations of the instanton-induced scattering proceeded gradually. First, one can consider a single gluon exchange
between the vacuum field of the instanton and the quarks: in this case a gluon is of course just a perturbative tail of an instanton, a weak dipole-shaped potential. Then came realization that any number of exchanged gluons are
summed up in the Wilson line, which can be easily analytically evaluated in the background field of the instanton.
The probability is the amplitude squared, so it can be represented by the upper picture in Fig.\ref{fig_elastic-inelastic},
in which it is assumed that the vertical line -- the unitarity cut -- is very much removed from the instanton,
so that its field at the cut can be ignored. The two parts of the plot are shown together, but they are in fact independent
matrix elements with a single soliton.

Next came discussion of inelastic processes, in which first a gluon or few were passing through the cut. 
Such situation is shown by the left lower part of Fig.\ref{fig_elastic-inelastic}. Summing those gluons up resulting in
a real breakthrough
in the realization that the whole process can be described
by a continuous semiclassical path, starting in the vacuum {\em under
 the barrier}, proceeding to a {\em turning point} and then to
a {\em real (Minkowskian) evolution}. 
Whatever way the system is driven, it emerges from under the barrier
 via what we will call  {\em ``a turning state''}, 
familiar from WKB semiclassical method in quantum mechanics.
 
 The turning states, released into the unitarity cut or
Minkowski world, are the states we have already discussed.
 From there starts the
 real time 
motion outside the barrier. 
Here the action is real and $|e^{iS}|=1$. That means  that
whatever happens  at this Minkowski stage has the probability 1 and cannot
 affect the total cross section of the process:
this part of the path is only needed for understanding of the
properties of the
final state. 

So, how we describe inelastic collisions and what is produced? 
Let me summaries the main lesson of
this section (and also of the section on instantaneous excitation in the previous chapter) in one sentence: if the particle under
the barrier is hit, it jumps into the lowest state {\em at
  the
barrier} with the same coordinate. The collision amplitude corresponds to the process 
described by ``streamline" configurations, describing classical solution with an external force along the 
topological valley, exciting it to the sphaleron path.



 \chapter{Chiral matter}  \label{sec_chiral_matter} 
 \index{chiral matter and chiral anomaly-induced effects}
For extensive discussion of the history of the chiral effects see e.g. \cite{Kharzeev:2013ffa}.
This terminology is rather recent, and we clearly need to start with its explanation.
What is meant by it is some form of matter in which the following two conditions hold:\\
(i)  a certain $imbalance$ between the occupation of the left- and right-handed 
fermions is created; \\
 (ii) the lifetime of the chirality $\tau_5$ is sufficiently long,
compared to the timescale of the phenomenon considered,
so that quantum effects induced by the  chiral anomaly
(discussed in the preceding chapter) can be observed.


In chapter on Instantons
we  discussed  breaking of the  chiral symmetry of QCD. We will see that
in the QCD vacuum $T=0$ and, more generally, for subcritical temperature $T<T_c$
the non-zero {\em quark condensate} violates the  chiral (left-right) symmetry
present in the Lagrangian. 
 In contrast to that, the QGP at $supercritical$ temperature $T>T_c$ keeps
 the  chiral symmetry
 unbroken, and  therefore, provided chiral imbalance is somehow created, it is our first example of a ``chiral matter".
 (Historically, it was also the first in which many effects to be discussed in this chapter
 were first considered.) Unfortunately, QGP is quite expensive -- one need to have a relativistic collider
to create it -- and so one may ask if there are other cheaper alternatives. 

To illustrate this distinction quantitatively, let me note that Dirac equation for a free massive fermion conserves the vector current -- the fermion number 
\be \partial_\mu \left(\bar\psi \gamma_\mu \psi\right) =0 \ee
but not the  axial current (the difference between the number  of the left and the right chirality components): 
\be
 \partial_\mu \left(\bar\psi \gamma_\mu \gamma_5 \psi \right) = 
 2 m  \left(\bar\psi \gamma_5 \psi \right) \ee
because the mass term connects the left and right components of the fields. 

In QGP one finds two light  quarks, called up and down or $u,d$. Their masses are
of the magnitude of few $MeV$, much smaller than any other 
relevant scales,  and since they play no role one can for simplicity consider them to be zero. 
If so, the r.h.s. of the equation above is also zero, and this means that axial charge
is also conserved. A simpler way to understand the situation is to state that the number of left and right-polarized quarks are conserved $independently$. This is {\em the chiral symmetry}.

Chiral electron quasiparticles can also exist in
 the so called {\em semimetals}.
The terminology which needs to be explained at this point is as follows. As 
the reader surely knows, the $metals$ have the chemical potential inside the
allowed zone of the electron state, and  thus there are Fermi spheres, with a non-zero 
surface area.
 $Insulators$ have the chemical potential located inside the forbidden energy zone, and
thus there is no  Fermi sphere of excitations with small energies.
The metals have free electrons to move, by external fields, and the insulators have not.

$Semimetals$ are in between, with the point-like touchings of the valence and conduction bands\footnote{Graphene  has linear
electron spectrum: but it is a 2-dimensional material, while the anomaly phenomenon
 we need exists in 1+3 dimensions. }.

The so called Dirac semimetals have the so called ``Dirac point", shared by ``left" and ``right" 
fermions, near which linear relativistic-like dispersion relation is valid. There exist also 
``Weyl semimetals",  for which two chiralities have two separate touching points. The 
quasiparticle modes near those
points have the following  Hamitonian 
\be H=\pm v_F \vec \sigma \cdot \vec k \ee
where $\vec \sigma$ being Pauli spin matrices. It is of the kind originally suggested by 
Weyl for uncharged massless fermions.
The first observation of CME in
condense matter setting \cite{Li:2014bha} in zirconium pentatelluride, $ZrTe_5$, a 
3d Dirac semimetal. We will discuss this experiment in section on Chiral Magnetic Effect.

\section{Electrodynamics in a CP violating matter}
Chirality -- the product of spin and momentum  $(\vec s \vec p)$-- for massless particles, fermions or 
bosons (e.g. photons) have two values, commonly refered to as ``left" and ``right" handed polarizations.
It is odd under P parity transformation $\vec x \rightarrow -\vec x$ since momentum changes sign
and spin does not. Thus matter in which there is left-right disbalance is P-odd (that is, not the same as
its mirror image).

The subject of this section is however going well beyond violation of P parity, to that of CP violating medium.
The physics of it was originally discussed in the context of the so called $axion$ dynamics,
which is still hypothetical and is well beyond the scope of this book. However it elucidated
many interesting new effects which existence of effective pseudoscalar field $\theta(x)$ 
brings with it. Our discussion of it -- in the framework of so called modified Maxwell-Chern-Simons electrodynamics 
 -- follows Ref.\cite{Kharzeev:2009fn}.

While the  Maxwell theory historically emerged, term by term, from certain ingenious experiments,
in modern textbooks all of them are derived from a single  {\em principle of gauge invariance}, requiring to
change all derivatives of the charged fields by their covariant form 
\be  \partial_\mu \rightarrow D_\mu=\partial_\mu-i e  A_\mu \ee
plus the statement that
the gauge action can only be given by the only dimension-4 gauge invariant operator
\be  L_{Maxwell}=-{1\over 4} F^{\mu\nu} F_{\mu\nu} \ee
The last point is required for consistency with expected symmetries, such as the $CP$ invariance.
The usual kinds of media modify Maxwellian theory only slightly, renormalizing the squares $(\vec E)^2,(\vec B)^2$
by coefficients known as electric and magnetic  permitivities $\epsilon$ and $\mu$. 

However, if the condition of
$CP$ invariance is not required, one have no reason not to add a term $(\vec E \vec B)$ of the same dimension.
We will use it in the form
\be L_{MCS}=-{1\over 4} F^{\mu\nu} F_{\mu\nu} -A_\mu J^\mu  - {c \over 4} \theta \tilde F^{\mu\nu} F_{\mu\nu} \ee
where $c$ is some coefficient while $\theta$ will be treated as time and space-dependent field. If 
one would add its kinetic and potential energies, this ``axion" field $\theta$ can be upgraded to a separate dynamical
entity: but we will not do so, and simply think of it as a matter-induced coefficient in a Lagrangian.

Now, if it is just a constant $\theta=const(t,\vec x)$, one finds that the last term does $not$ change the equations of motion because the term we added is in fact a full divergence 
\be  \tilde F^{\mu\nu} F_{\mu\nu} =\partial_\mu J_{CS}^\mu \ee
where the current is known as Chern-Simons current
\be  J_{CS}^\mu =\epsilon^{\mu\nu \rho\sigma} A_\nu F_{\rho\sigma} \ee
The 4-volume integral of the  full divergence  can be rewritten as 3-d integral of the current flux over
the volume boundaries -- ``large sphere" -- which unusally considered to be zero, provided
all fields vanish there.

If however $\theta$ is time and/or space dependent, the derivative can be passed to it by integration by parts,
so the last term in action can also be written as $+(c/4) (\partial_\mu \theta) J_{CS}^\mu $.
To make the lessons more familiar let us write the equations of motion in the non-relativistic form,
introducing the following notations for the vector and axial currents 
\be J_0=\rho, \vec J, M=\partial_0 \theta, \vec P=\vec \nabla \theta \ee
Here they are

\be \vec \nabla \cdot \vec E=\rho+c \vec P \cdot \vec B \ee
\be \vec \nabla \times \vec E+ {\partial \vec B \over \partial t}=0 \ee
\be \vec \nabla \cdot \vec B=0 \ee
\be  \vec \nabla \times \vec B -  {\partial \vec E \over \partial t}=\vec J+c(M \vec B - \vec P \cdot \vec E) \ee
One can see that both electric and magnetic fields get their sources -- the r.h.s. of the first and last eqns --
modified.  This leads to several new effects, some of them we will mention.

{\bf The Witten effect} appears for nonzero $\vec P$. For example consider a spherical ``defect", a region in which
$\theta$ vanishes, while it is nonzero outside. Such a defect can be a vortex or a magnetic monopole we will discuss
later on. Without new terms, those would support magnetic field only, without $\vec E$, but at  nonzero $\vec P$ 
the r.h.s. of the first equation sources the electric field as well. Thus a magnetic vortices or monopoles become
also electric. 

{\bf The electric charge separation} in external magnetic field also appears for nonzero $\vec P$: the 
r.h.s. of the first equation may be zero when the two terms cancel each other. 

If the $\theta$ is time dependent and $M\neq 0$ one finds more unusual effects. 

{\bf Chiral magnetic effect}
which we will discuss more in the nexts section is one of them: it is a vector current along the magnetic field
\be \vec J = - c M \vec B \ee
vanishing the r.h.s. of the last Maxwell-Chern-Simons equation. One can put $\vec E=0$
and take constant $\vec B$, which vanishes all other terms. 

\section{Chiral magnetic effect (CME) and the chiral anomaly} \label{sec_CME}
\index{chiral magnetic effect (CME)}
We start this section with some general discussion of space-time symmetries and
currents in the medium. It will explain the required conditions under which the CME $may$
exist.

The first expression we start with is the Ohmic current,
induced by the electric field
\be  \vec J=\sigma_{Ohm} \vec E \ee
If one watch this phenomenon in a mirror (which means performing the so called P-parity transformation $\vec x \rightarrow -\vec x$), both the l.h.s. and the r.h.s. vectors change sign,
so the coefficient $\sigma_{Ohm}$ remains unchanged. 

Imagine now that one flips the
sign of time  $t\rightarrow -t$, performing the so called $T$-parity transformation. The
current is related with velocity of the charge, and it changes sign. The electric field is
only related with $positions$ of the charges which created it: therefore it is unchanged.
The conclusion is that $\sigma_{Ohm}\rightarrow -\sigma_{Ohm}$, which is not
surprising since Ohmic current is dissipative, leading to increasing entropy
of the media, and thus it should be also dissipative in an imagined world in which time goes backward.

In superconductors the expression for a current, proposed by London, is
\be  \vec J=\sigma_{London} \vec A \ee
where the inducing field is the vector potential. If one recalls that $\vec E$ contains $\partial_t \vec A$, it be clear that $\vec A$ must be T-odd. Therefore, $\sigma_{London} $
should be the same in our world and in ``backward time" Universe\footnote{
According to Kharzeev, this argument has been made by V.I.Zakharov.
}. It follows that
it cannot lead to dissipation and the entropy growth. And indeed, supercurrents
are eternal and non-dissipative!

The CME is  the vector current along the $magnetic$ field 
\be  \vec J=\sigma_{CME} \vec B \ee 
Transformation in a mirror of $\vec J$ is sign changing, while  $\vec B$
remains unchanged. So, $\sigma_{CME}$ must be $P$-odd. So, one may think about
weak interaction effects, like with neutrinos, since they violate $P$ parity by
involving left-handed fermions only.

Under the $T$-parity transformation both $\vec J,\vec B $ change sign, as they are both
related with velocities of the charges. Thus $\sigma_{CME}$ must be $T$-even, and
therefore be non-dissipative(!), as the supercurrents are. 
This agrees with the idea that magnetic field, whicle exerting a force on a moving charge, does not make any work on it. 
So, if one can find the
conditions under which CME current can be produced, it will not dissipate. Unlike the
supercurrent, it does not seem to require coherence and low temperatures.

The {\em  chiral anomaly} not only lead to 
 existence of
CME in a chiral matter, but provides the universal coefficient of the effect. In its operator form the 
expression is 
\be \vec J= {e^2  \over 2\pi^2} \mu_5 \vec B  \label{eqn_CME}
\ee
where $\mu_5$ is the chemical potential for the chiral charge\footnote{
The first paper in which this formula was written was by Vilenkin. The setting was
 P-violating due to
left-handed fermions -- neutrinos -- of weakly interacting sector,
 near a rotating black hole. Existence of this current 
in equilibrium was questioned by C.N.Yang, and in the next paper  Vilenkin showed that
the equilibrium current is in fact zero. 
The resolution lies in the realization that the chiral matter is always metastable, not in equilibrium. }.

Kharzeev and collaborators first suggested to use QGP -- a good chiral matter -- and superstrong magnetic field created by two positive colliding ions, directed normal to the collision plane.
If vector CME current appears, it will create electric dipole along $\vec B$,
which would be observed by a charge dependence of the elliptic flow of secondaries.

However, since strong interactions
are CP-conserving, the chiral imbalance can only appear as a fluctuation. It means that we can only observe CP-even quadratic effect. Possible backgrounds of origins unrelated
to CME, or even to $\vec B$, can influence observations. Therefore, strickly speaking,
the observations of CME in heavy ion collisions are not yet completely clear: a special run
at RHIC with two nuclei, having the same number of nucleons but different charge $Z$,
are planned, to at least separate $\vec B$-dependent effects from others.

\begin{figure}[htbp]
\begin{center}
\includegraphics[width=8cm]{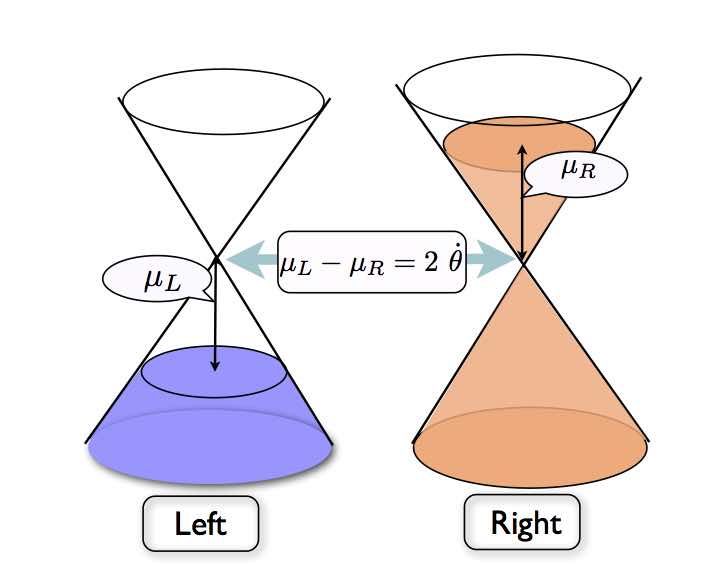}
\caption{Dirac cones of the left and right fermions. In the presence of the changing chiral charge there is an asymmetry between the Fermi energies of left and right fermions $\mu_L- \mu_R = 2\mu_5$ }
\label{fig_Dirac_cones}
\end{center}
\end{figure}

\index{chiral magnetic effect in semi-metals}
It is easier to explain the CME experiment with the semimetal.
In the absence of fields, there is  of course no chiral imbalance, $\mu_L=\mu_R$
and there are  equal number 
of left and right fermions. 
But 
when parallel electric and magnetic fields are applied, the  change in the chiral density $\rho_5$ appears, as illustrated in Fig.\ref{fig_Dirac_cones} from \cite{Kharzeev:2009fn}.
The time evolution of the chiral imbalance
can be written as 
\be {d\rho_5 \over dt}={e^2 \over 4 \pi^2} (\vec E \cdot \vec B) -{\rho_5\over \tau_5} \ee
where the first term in the r.h.s. again stems from  the chiral anomaly and the second is the chiral density relaxation.
The stationary condition is reached at late times then the l.h.s. is zero
because the gain and loss terms
in the r.h.s. cancel each other. So, one get
\be \rho_5(t \rightarrow \infty)= {e^2 \over 4 \pi^2} (\vec E \cdot \vec B) \tau_5 \ee
The chemical potential $\mu_5\sim (\vec E \cdot \vec B)  $ as well, and putting it into the
expression (\ref{eqn_CME}) one would obtain the current $\vec J\sim \vec B (\vec B \vec E)$.
This is indeed what is observed \cite{Li:2014bha} in zirconium pentatelluride, $ZrTe_5$,
and then in other materials. Potentially, the CME current may be used 
in electronics, providing non-dissipative currents at room temperatures. 

\section{Chiral vortical effect}
\index{chiral vortical effect}
Can a magnetic field $\vec B$ be substituted by another quantity, possessing a similar
$P,T$ parity, e.g. $vorticity$ which we define in relativistic notations by
 \be  \omega_\mu =({1 \over 2}) \epsilon^{\mu\nu\lambda\rho} u_\nu \partial_\lambda u_\rho  \ee  
with 4-velocity $u_\mu$. 
The chiral vortical effect (CVE)
introduced in \cite{Son:2009tf,Kharzeev:2009fn}
can be summarized by a relation
\be j_\mu =\sigma_{CVE}  \omega_\mu \ee
with a vector current propagating along the vorticity.
A similar relation -- but of course with a different kinetic coefficient -- can be
written for the entropy current $s_\mu$. Son and Surowka argued that entropy production
must be positive $\partial_\mu s_\mu>0 $, but Kharzeev then argued that it should in fact
be zero because of non-dissipative nature of the effect. These considerations
lead to a specific expression for the coefficient $\sigma_{CVE} $ in terms of matter EOS.

So far I am not aware of any specific applications of the CVE. In
ultrarelativistic  heavy ion collisions specifically, vorticity is not zero but very small
to be used for it.

\section{The chiral waves} 
The CME expression (\ref{eqn_CME}) has an analog: interchanging vector current to axial one, and axial chemical potential to the usual --  vector -- one $\mu$, one also get the following expression
\be \vec J_A = {e \over 2 \pi^2} \mu \vec B \ee
As argued in
\cite{Burnier:2011bf}, combining the two together one 
finds new oscillation mode called the {\em chiral magnetic wave}. Indeed, divergence of the currents can be
substituted by time derivatives of the corresponding densities, and two linear
equation combined produce one equation of the second order. 

So, a heavy ion collision starting with certain baryon number density and $\mu$,
leads to a quadrupole excitation in which the density and the chiral imbalance should oscillate
into each other.  In the previous chapter we had shown how the density oscillations -- the sound modes -- were observed. A search for the chiral magnetic wave is in progress.

In early Universe chiral waves may perhaps lead to (at least locally) chiral magnetic
fields: if it is true, it has a potential to contribute to Baryon asymmetry puzzle we discussed
before.

\chapter{Instanton-dyons }

\setcounter{footnote}{0}

\section{The Polyakov line and confinement}  
\subsection{Generalities}
Let me remind the setting. The finite temperature formulation of the periodic path integral defines the Euclidean time
$\tau$ on a circle with circumference 
$ \beta= \hbar/T $
A general mathematical construction allows closed loops around circles and toruses,
known as holonomies. In the gauge theory it generates a gauge invariant object known as
the Polyakov line
\be P= Pexp(i \oint A_\mu^a T^a dx_\mu ) \ee
where $T^a=t^a/2$ is the color generator in a particular representation of the color group,
namely the fundamental one. 

\begin{figure}[h!] 
   \includegraphics[width=7cm]{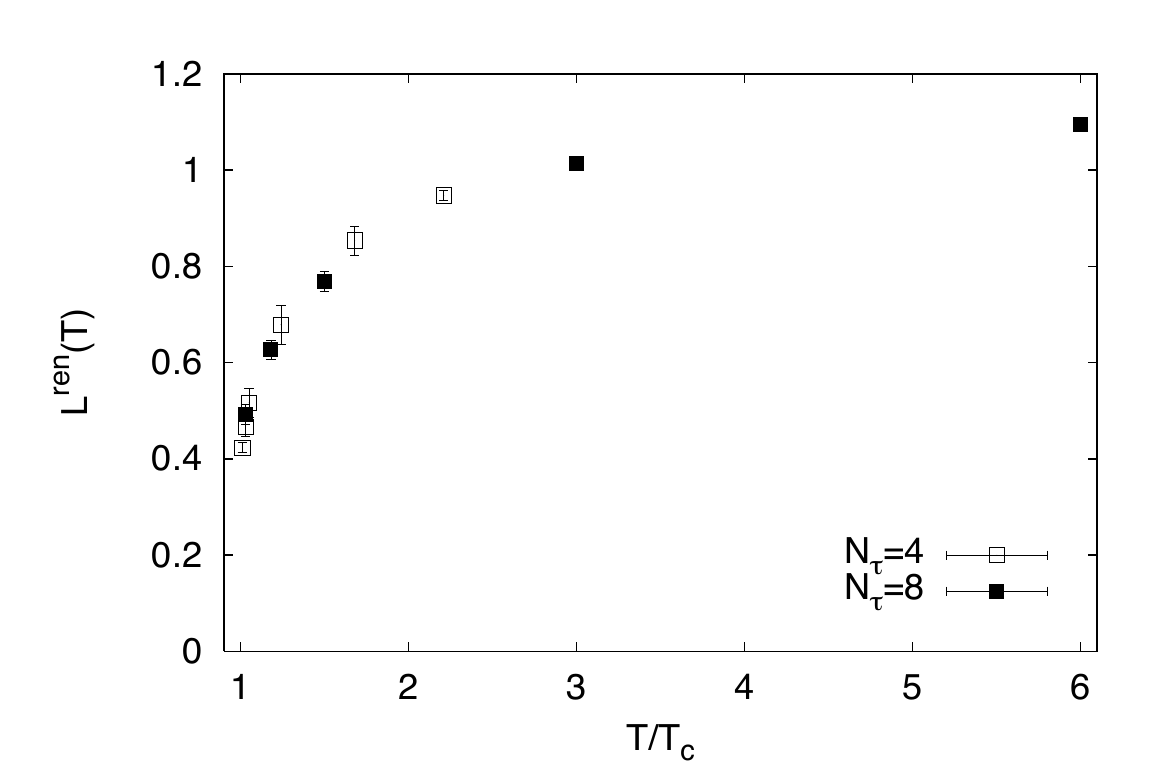} 
  \includegraphics[width=5cm]{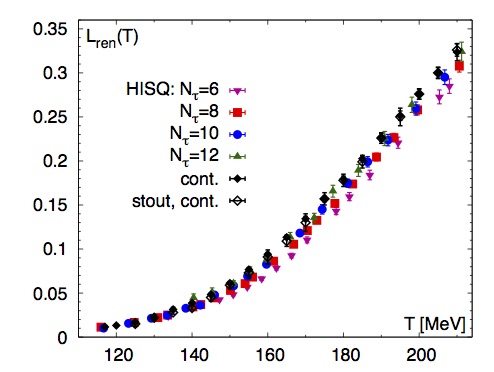} 
   \caption{The renormalized Polyakov line, in pure gauge SU(3) (left) and QCD with light quarks (right)
   }
   \label{fig_P}
\end{figure}

The temperature dependence of its VEV is plotted in Fig.\ref{fig_P}. The
left plot, from \cite{Kaczmarek:2002mc}, is for pure gauge theory. One can see that at high $T$
${1 \over N_c} <tr P>\rightarrow 1$: below we will call it ``trivial holonomy"
limit, because it corresponds to vanishing $A_4$. At $T<T_c$ the VEV is zero, which corresponds to
strict confinement. As one can see in the plot, there is a finite jump, from the value of about 0.4 to 0,
so the transition is of the first order.

The plot on the
   right side of Fig.\ref{fig_P}, from \cite{Bazavov:2013yv}, is for QCD with light quarks. In this case 
   the VEV is never strictly zero, but decreases to rather small values gradually. 

\subsection{The free energy of the static quark on the lattice}

The Polyakov line is gauge invariant and thus a physical quantity, related to the free energy of the static quark
\be <{1 \over N_c} tr P >= exp(-F_Q/T) \ee

\begin{figure}[h!] 
   \includegraphics[width=6cm]{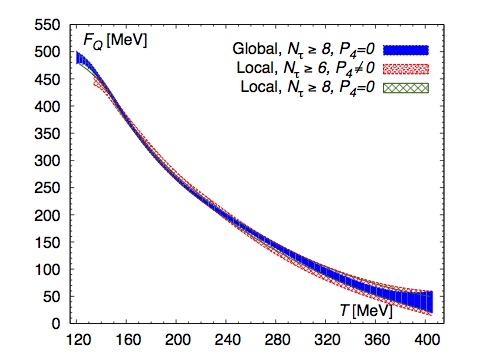} 
    \includegraphics[width=6cm]{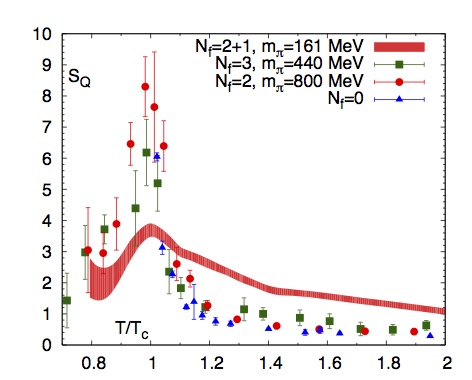} 
   \caption{Free energy and entropy
    of the static quark, extrapolated to physical QCD from different lattice measurements.
    The value of the quark mass is expressed via the pion mass: one can see that
    as it becomes smaller, the magnitude of the near-$T_c$ peak is reduced.}
   \label{fig_F_Q}
\end{figure}

Renormalized Polyakov loop was calculated in a wide range of temperatures 
by Bazavov et al \cite{Bazavov:2016uvm}. The resulting free energy of the static quark extrapolated to physical QCD
is shown in Fig.\ref{fig_F_Q}. 

In theories with light quark there is level-crossing transition between the heavy quark $\bar{Q}Q $ state and
the 4-quark meson-meson state $\bar{Q}q \bar{q}Q $. It is also of course ``dressed" with certain
vacuum polarization around the static meson.
The value of $F_Q$ at the lowest $T$, $\approx 500\, MeV$, corresponds to the effective free energy of the 
extra light quark. In vacuum, at $T=0$, it can be phenomenologically evaluated from the mass difference between
heavy-light B meson and the $b$ quark, $M_B-M_b$.

The so called Polyakov loop susceptibility is defined by
\be \chi= (V T^3) \left(< | P |^2>- < | P |>^2\right) \ee
and it is also calculated in \cite{Bazavov:2016uvm}.  Using the gradient 
flow method, one can study how it changes as quantum fluctuations
are reduced. 

\subsection{The color phases}
\begin{figure}
 \centering
\includegraphics[width=6.cm]{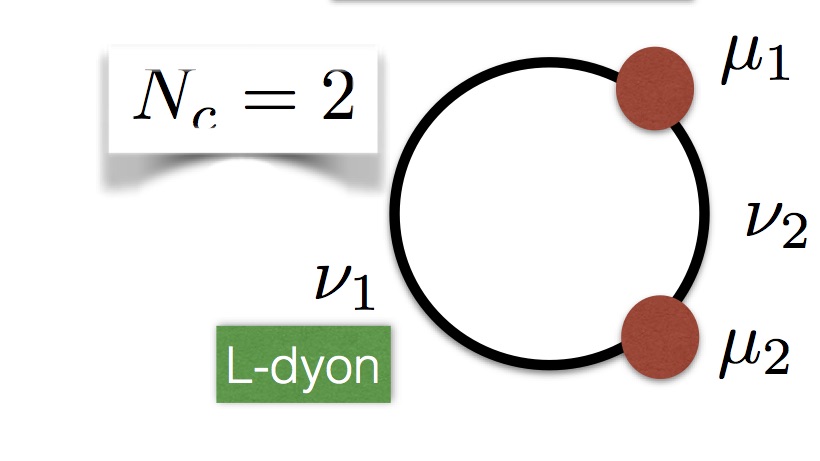}
\includegraphics[width=4.cm]{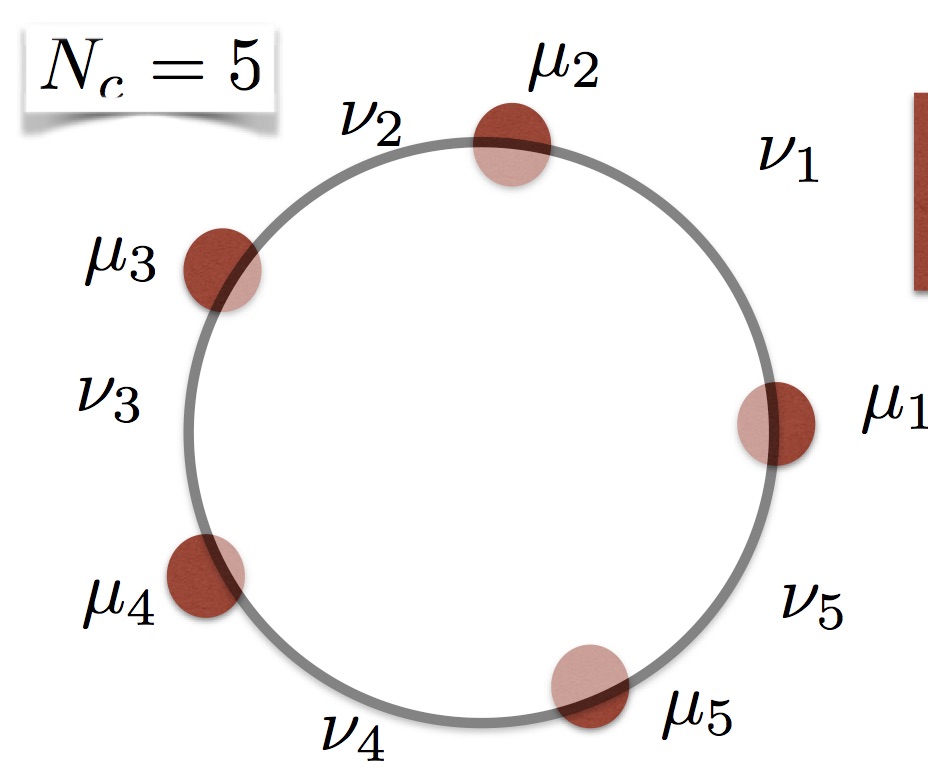}
\caption{ The holonomy circle and the definition of the parameters $\mu_i$ and $\nu_i$, for 2 and 5 colors.}
\label{fig_holonomy}
\end{figure}

Furthermore, it has temperature-dependent  VEV 
$< P(T) > $ which is a unitary matrix. Its eigenvalues are complex number with modulus 1,
so they can be written as phases, usually defined by
\be A_4= 2\pi T diag(\mu_1,\mu_2,...\mu_{N_c}) \ee
Assuming they are order in magnitude $\mu_1<mu_2...$ and introducing one more 
$\mu_{N_c+1}=1+\mu_1$ one can proceed to their differences 
\be \nu_m=\mu_{m+1}-\mu_m \ee
In Fig.\ref{fig_holonomy}(left)  we show locations of the holonomy eigenvalues for the simplest case of the 
 $SU(2)$ gauge group, for which most of the calculations is made. In this case $\mu_1=-\mu_2=\mu$, so there is only
 one parameter. 
 The VEV of the Polyakov line is 
 \be < {1\over 2} Tr P>=cos(\pi\nu) \ee
 At high $T$  $<P>\approx 1$, which means all $\mu_i\approx 0$. Only one $\nu_i\approx 1$
 However in the temperature interval
  $(2..1)T_c$ it changes to zero (or small value in QCD with quarks). Accounting for this phenomenon lead
 Pisarski  and collaborators to  ``semi-QGP"  paradigm \cite{Pisarski:2009zza} and construction of the so called PNJL model. 
  At $T<T_c$, in a confined phase, $<P>=0$ which means that $\nu=1/2$. 
  
%

For the SU(2) gauge group
 the selfdual ones are called $M$ with charges $(e,m)=(+,+)$ and $L$ with charges $(e,m)=(-,-)$, the anti-selfdual antidyons are called  
 $\bar{M}$, $(e,m)=(+,-)$ and  $\bar{L}$, $(e,m)=(-,+)$.

Generalization to $SU(N_c)$ follows the usual Cartan subalgebra of all diagonal (mutally commuting)
generators,
and generalize 2 dyons just described for the $SU(2)$ to $N_c$ of them, $N_c-1$ of M-type and one L-type.
We will briefly introduce notations to be used below, starting from the HIggs VEV
\be A_4(\infty)=(2\pi T) diag(\mu_1... \mu_{Nc}); \,\, \,\,\, \sum_i \mu_i=0\ee
where the latter condition follows from the required zero trace of $A_4$. Introducing $\mu_{Nc+1}=1+\mu_1$ and the differences
\be  \nu_i=\mu_{i+1}-\mu_{i}  ; \,\, \,\,\, \sum_i \nu_i=1 \ee
These differences will determined the masses of the corresponding charged
components and thus the core sizes, which are $\sim 1/(2\pi T \nu_i)$ for the i-th dyon.


It is also necessary to add the following. The so called Cartan subalgebra of all diagonal (mutally commuting)
generators for $SU(N_c)$ are made of $N_c-1$ matrices of the type
\be diag(1,-1, 0, ...0), diag(0, 1,-1, 0, ...0),... ,diag(0,0,...1,-1) \ee
The corresponding $N_c-1$ components of the gluon field remains massless, while the rest
of them get nonzero mass from the term in the action $[A_4, A_\mu]^2$. For the case
of two colors there is only one ``massless photon" $A_\mu^3$ and two ``massive gluons" $A_\mu^1,A_\mu^2$.
 For the physical case
of three colors there are two ``massless photons" and 6 ``massive gluons".
 
Since the instanton-dyons are basically SU(2) objects, they are made of the gauge fields with two
colors. If one asks which ones, those are defined by two ends of the  corresponding segment $\nu_i$
with which it is identified with, namely colors $i+1$ and $i$. Two subsequent dyon types, say identified 
with segments $\nu_i$ and $\nu_{i+1}$ have one color in common, namely $i+1$, with charges plus and minus,
respectively. The dyon types which are not subsequent on the circle have no common ``photon charges" and thus
cannot interact at large by Coulomb-like forces. 

\section{ Semiclassical instanton-dyons }
\subsection{The instanton-dyon field configuration}
In the chapter on monopoles it was many times stated that QCD-like theories
lack scalar fields. Some features depend heavily on that -- in particularly 
existence of chiral symmetries and physics related to them. But it also presents a number
of difficulties to the
physics of monopoles. 

We already mentioned that one possible way to proceed is to use the 4-th component
of the gauge field $A_4$ as adjoint scalar. Of course, this is not a Lorentz invariant choice --
but at nonzero $T$ the frame in which matter  is at rest is special anyway. More importantly,
if one attempts to analytically continue the theory to Minkowski formulation, 
 $A_4$ gets imaginary and the construction looses its meaning. 

It is perhaps worth repeating that unlike particle-monopoles discussed in the earlier chapter, those instanton-dyons are $not$ particles in the ordinary sense.  (The original name for instantons from \cite{Belavin:1975fg} for any objects of the kind has
been ``pseudo-particles".)  The distinction between the two types: ``Partcle" (or quasiparticle)  dyons are described in the usual Minkowski
world, their mass squared is  the sum of two positive terms, originating from $E^2+B^2$, both limited from below
by the so called Bogomolny bounds, proportional to their integer electric and magnetic charges $ n_e,n_m$, respectively.   

The ``pseudoparticle" dyons, like instantons, are selfdual (or antiselfdual) in Eiclidean formulation. In  Minkowski notations this means that
 \be \vec E=\pm i \vec B \ee 
and thus negative electric energy cancels the magnetic one to $E^2+B^2=0$. As it is well documented in the literature since 1970's, they semiclassically
describe vacuum transition between different gauge nonequivalent topological classical vacua, at zero classical energy.  


 Historically, their applications started ``from the top", starting from the finite-T instantons, or calorons,
generalized to the case of
a nonzero
holonomy, by \cite{Kraan:1998sn}. (More or less at the same time,
 Lee and Lu \cite{Lee:1998bb} derived it from certain brain construction related to AdS/CFT.)
 Only after plotting the action distribution it has been realized that instantons get
 split into $N_c$ independent clusters -- the instanton-monopoles or instanton-dyons.
 The technicality of the KvBLL solution is very interesting
 theoretically but rather involved:   we return to it a bit later.

If the instanton constituents are  very distant from each other, one can of course first study them
as separate solitons, and then try to do ``bottom up" superposition of them.  

Before we discuss the solutions, let us mention their quantum numbers. 
 For the simplest $SU(2)$ color group there are 4 types of dyons:
see Table \ref{tab_su2dyons}. 
The charges and the mass (in units of $8\pi^2/e^2 T$) for 4 SU(2) dyons cover all four possibilities for the electric and magnetic charges,

\begin{table}
\centering
\begin{tabular}{| r | c|c| c| c|} \hline
name & E & M & $S/S_{inst}$ \\ \hline
$M$         & + &+ &$\nu$ \\
$\bar{M}$ & + &- &$\nu$ \\
$L$          & - & - &$\bar \nu=1-\nu$ \\
$\bar{L}$ & - & + &$\bar \nu=1-\nu $ \\ \hline
\end{tabular} 
\caption{Electric charge, magnetic charge and action (in units of the instanton action) for four
types of the instanton-dyons of the $SU(2)$ gauge theory
} \label{tab_su2dyons}
\end{table}

The $M,\bar{M}$ dyons are the ``ordinary" BPS dyons, which in the spherical ``hedghog" gauge is
\be 
A^a_4=\mp n_a v \Phi(vr)  \\ \nonumber
A^a_i= \epsilon_{aij} n_j {1-R(vr) \over r}
\ee
where $n_a=r_a/r$ is the radial unit vector, the minus-plus is for self (antiself) dual solutions and the two functions are 
\be
  \Phi(vr)&=& [coth(vr)-{1\over vr}] \rightarrow \mp n_a v [[1-{1 \over vr}] +O(exp(-vr)) \\ \nonumber
R(vr)&=& vr/sinh(vr)
\ee
The corresponding magnetic field has the transverse and longitudinal stuctures, which contained derivatives of these functions, but can be rewritten without them as follows
\be 
B^a_i=(\delta_{ai}-n_an_i)[-v \Phi)vr) R(vr)/r]+n_an_i [R^2(vr)-1]/r^2
\ee
While the former one exponentially decreases with the distance, the longitudinal has $1/r^2$
behavior indicating the nonzero magnetic charge. It of course matches the electric charge.

Construction of the $L,\bar{L}$ dyons starts from the same expressions, in which the following substitution
\be v \rightarrow \bar{v}=2\pi T-v \ee
is made. But then, it has a ``wrong" asymptotics of the Higgs field $A_4$: this however can be  remedied
by the so called ``twist". It is gauge transformation with the {\em time-dependent}  matrix
\be U= exp \left[i 2\pi T x^4 (\tau^3/2)\right]\ee 
the time derivative of which subtracts the unwanted $2\pi T$ from the HIggs asymptotics.  
Note that the color matrics $\tau_3$ commutes with the diagonal (Abelian) part of the dyon field,
leaving it as before. But the ``core" of the L dyon is made of a charged fields with colors 1,2:
thus the core is time-dependent. So, the  $L,\bar{L}$ dyons are not static solutions, and thus they
do not exist in 3-d theory (and were not covered in the previous chapter). It can only be defined
in the finite-$T$ theory with the Matsubara time!

The actions of the dyons are $\nu_i(8\pi^2/e^2)$, so if all $selfdual$ ones are summed
using $\sum_i \nu_i=1$ , the result is the instanton action. This statement, so far demonstrated for a very distant (non-interacting) dyons,
should in fact be true in general because selfduality relates the total action of the PBS dyons to their
total charge. 

I was often asked\footnote{Once by Polyakov himself}, how is it possible to have 
separate objects with a non-integer topological charge $Q$. The answer is they are
independent in physical sense, not in mathematical one. 

As any monopoles, they
are interconnected by singular but invisible Dirac strings. 
This can be seen in two ways. One is explicit contruction starting with well-separated
monopoles, ``combed" into a gauge in which the color direction of ``Higgs" $<A_4>$
is some fixed direciton. SU(2) instanton is made of $L+M$ dyons, with a string 
connecting their centers. 

The explicit solution,  obtained for the single caloron in the  Kraan-van Baal paper \cite{Kraan:1998sn},  lead to this conclusion
after plotting  the action distribution.  The solution itself can be described
using the so called prepotential scalar function\footnote{In order to explain how the gauge potential $A_\mu$ was constructed one needs to understand Nahm construction, which I cannot describe better than done in Kraan-van Baal papers. Without that, the formulae 
for $A_\mu$ and $G_{\mu\nu}$ look like pure magic. I decided not to go into it, keeping only
definition of $\psi$ and the action distribution defined in terms of this function alone. 
}
\be \psi(r)=(1/2)tr  (A_N... A_1) -cos(2\pi r_4) \ee
where here and below the temperature and circumference of the Matsubara circle are 
temporarily put to $T=1/\beta=1$.
The matrices need to be multiplied in the order written,  right to left,
 in the order  corresponding to the magnitude of $\mu_i$ 
are ordered. 
The  the following  2$\times$ 2 matrices $A_m$ 
(not to be confused with the gauge potential $A_\mu$)
are defined by the product of two matrices
\be A_m= \left(  \begin{array}{cc}  1 &  (\vec y_m - \vec y_{m+1})/r_m \\
                                                      0 & r_{m+1}/r_m  \end{array}  \right)     
                  \left(  \begin{array}{cc} c_m &  s_m  \\ s_m & c_m   \end{array}    \right)    \ee
with $c_m=cosh(2\pi \nu_m r_m), s_m=sinh(2\pi \nu_m r_m)$. Here $r_m=|\vec r -\vec y_m|$
is the distance from the observation point to the m-th dyon center $\vec y_m$, and $\vec y_m - \vec y_{m+1}$ are vector distances between dyons. Note further that the first matrix is ``pure geometry" independent of holonomies, and the second is some hyperbolic rotation matrix by
the angle including m-th holonomy and distance to m-th dyon.

The action density can be expressed in terms of prepotential in a surprisingly simple way 
\be Tr G^2_{\mu\nu}= \partial^2 \partial^2 log(\psi) \ee
with two 4-dimensional Laplacians. Note that there are 4 derivatives because the matrixes
define  the ``prepotential" in terms of which the field potentials already have a derivative. Note also that
the matrices $A_m$ does not include time $\tau=r_4$, which is solely located in the cos function in the last term of $\psi$ (thus the
periodicity in the Matsubara box is explicit).

\begin{figure}[h]
\begin{center}
\includegraphics[width=10cm]{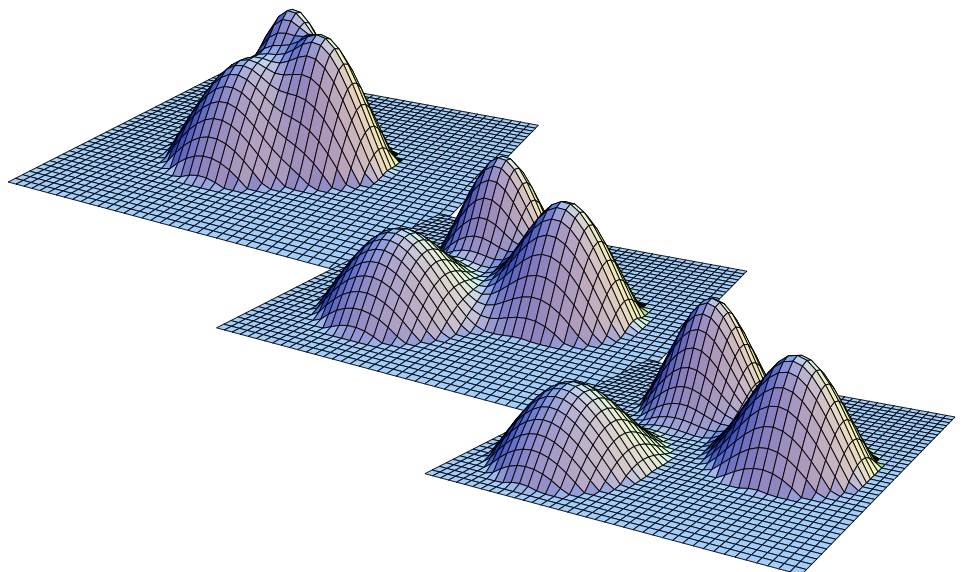}
\caption{Action densities for the SU(3) caloron at x0 = 0 in the plane defined by the
centers of the three constituents for 1/T = 1.5, 3 and 4 (increasing temperature from top
to bottom). We choose mass parameters $(\nu_1, \nu_2, \nu_3)$ = (0.4, 0.35, 0.25), implemented by
$(\mu_1,\mu_2, \mu_3)$ = (-17/60, -2/60, 19/60). The constituents are located at $\vec y_1$ = (-1 , 1 , 0), $\vec y_2$ = (0, 1,0) and $\vec y_3$ = (1,-1,0), in units of T . The profiles are given on equal logarithmic 
scales, cut-off at an action density below 1/e.}
\label{fig_3dyons}
\end{center}
\end{figure}

{\bf Exercise} Calculating $\psi$
for any holonomies and locations in Mathematica is very straightforward: Do it and use  the last expression for the action density. 
Do not look at the resulting huge expression but just
 plot  its distribution, e.g. for 
parameters corresponding to Fig. \ref{fig_3dyons}.

 The construction itself is based on ADHM multi-instanton construction and the so called Nahm 
 version of it, generalizing it to the monopoles. It is too technical to be presented here: see
 original paper \cite{Kraan:1998sn}. 

\section{ Instanton-dyon interactions  }
\subsection{Large distance Coulomb}
Since the objects we now study have electric and magnetic Coulomb fields at large distances,
$r\nu \ll 1 $, one might naively expect
that electric and magnetic Coulomb interaction are proportional to the usual products of the charges
\be V_{naive}(r)= { e_1 e_2 + m_1 m_2 \over r }\ee
Yet this naturally looking expression turned out to be $wrong$! 

Let me start with a counterexample showing that it $must$ be wrong. We know
that two self-dual solitons -- $M$ and $L$ in particular -- cannot classically interact at all,
since the total action is simply given by the total topological charge, which cannot depend
on the distance between them $r$.

And indeed, plugging solutions into the action and performing the large-distance
expamsion one finds another -- funny looking but correct -- expression
 \be V(r)\sim \left(   { - e_1 e_2 + m_1 m_2 \over r  } \right) \ee
with a minus sign in the electric term. Extra correction to the naive formulare above comes from the 
commutator of $A_4$ with other components of the gauge field.
So, in the L-M case, with charges from the Table above, there is no potential.

This expression for $\bar{M} M$ produces $double$ attraction, since their
magnetic charges are opposite and electric are the same. The $\bar{L} M$
case has the opposite, thus it generates a  $double$ repulsion. 

\subsection{The dyon-antidyon classical interaction}
\subsubsection{Combing the hedgehogs}

Superposition of the dyons at nonzero $A_4$ is  nontrivial 
since it should match not only in magnitude but also in its direction in the color space.
This is achieved by the following four-step procedure:\\
(i) ``combing", or going to a gauge in which the ``Higgs field" $A_4=v$ of a dyon at large distances is
the same in all directions and for all objects\\
(ii)  performing a time-dependent gauge transformation which removes  $v$ \\
(iii) superimposing the dyons in this gauge\\
(iv)  making one more  time-dependent gauge transformation, re-introducing $v$ back

%

The details of the combing.

The hedgehog ansatz is invariant with respect to gauge transformations of the
form
\be
U=e^{i\hat r\cdot \tau f(r)}\;.
\ee
However to superpose the dyon solutions we need to first have the "Higgs" go to
a constant value at infinity, which does not depend on the direction. This is
accomplished by doing gauge transformation which converts $\hat r\cdot \tau$
into one direction in the color space. Since the "Higgs" field is associated
with the zero component of the gauge field $A_0$, applying the time independent
gauge transformation simply rotates the color direction of the $A_0$ field.
Further, the selfdual and anti-selfdual sector have asymptotic values of the
"Higgs" $A_0$ differ by a sign (see \cite{Diakonov:2009jq}), so two matrices
have to be used to gauge comb the selfdual and the anti-selfdual dyon. The gauge
transformations are given by
\be\label{eqs:dgc}
&S_+=e^{-i\phi \tau_3/2}e^{i\theta \tau_2/2}e^{i\phi \tau_3/2}\\
&S_-=e^{-i\phi \tau_3/2}e^{-i(\pi-\theta) \tau_2/2}e^{i\phi \tau_3/2}\;.
\ee
We calculate the fields of a Dyon and anti-dyon in these fields. If the
selfduality (anti-selfdualtity) equation are satisfied then the solution to the
fields are given by
\be
&A_0^a=\mp v \Phi(vr) \hat r^a\;,\\
&A_i^a=\epsilon_{aij}n_j\frac{1-R(vr)}{r}\;,
\ee
where
\be
&\Phi(x)=\coth x-\frac{1}{x}\;,\\
&R(x)=\frac{x}{\sinh x}\;.
\ee
We work in cylindrical coordinates, in which this ansatz becomes
\be
&A_r=0\;,\\
&A_\theta=\frac{R(vr)-1}{r}\frac{\vec{\hat \phi}\cdot\vec \tau}{2}\;,\\
&A_\phi=\frac{1-R(vr)}{r}\frac{\vec{\hat \theta}\cdot\vec \tau}{2}\;,
\ee
The upper sign in all equation corresponds to selfdual solutions. This means
that if we want to superpose the selfdual dyon and it's anti-selfdual
counterpart, we first have to make sure that $A_0$ goes to a common constant
value. We accomplish this with the matrices (\ref{eqs:dgc}), namely the matrices
$S_{\pm}$ will take $\vec{\hat{ r}}\cdot\vec\tau\rightarrow \pm\tau^3$. The choice
of the matrices which accomplish this is, of course, not unique. We always have
a choice of residual $U(1)$ symmetry $U=\exp(i \varphi \tau^3/2)$.

Proceeding with the calculation we obtain that the holonomy of both the selfdual
and anti-selfdual dyon after this gauge transformation is
\be
A_0=v\frac{\tau^3}{2}\;,
\ee
and their spatial components are given as
\be
&A_r=0\;,\\
&A_\theta=\frac{R(vr)}{r}\frac{\vec{\hat \phi}\cdot\vec \tau}{2}\;,\\
&A_\phi=\frac{R(vr)}{r}\frac{\vec{\hat \rho}\cdot\vec
\tau}{2}-\frac{1}{r}\cot\frac{\theta}{2}\frac{\tau^3}{2};,
\ee
for the selfdual dyon and
\be
&A_r=0\;,\\
&A_\theta=\frac{R(vr)}{r}\frac{\vec{\hat \phi}\cdot\vec \tau}{2}\;,\\
&A_\phi=-\frac{R(vr)}{r}\frac{\vec{\hat \rho}\cdot\vec
\tau}{2}+\frac{1}{r}\tan\frac{\theta}{2}\frac{\tau^3}{2};,
\ee

However this gauge transformation, apart from possessing a singular Dirac string
in the 3rd color direction, has also multivalued color $1,2$. The reason for
this is multivaluednes of the gauge transformation $S_{+,-}$ for the lines
$\theta=\pi,0$ respectively. 

\begin{wrapfigure}{o}{8.cm}
\begin{minipage}{7.cm}
\centering
   \includegraphics[width=7cm]{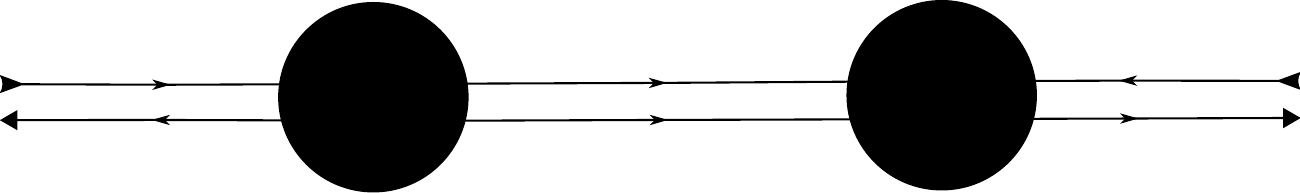} 
   \caption{$D\bar{D}$ with Dirac flux canceling outside, and adding
inside.}
   \label{fig:dstring}
   \end{minipage}
\end{wrapfigure}

We suggest a different gauge transformation in which the Dirac string splits in
$\pm z$ direction, but the fields remain single valued. What's more, since we
want to consider the superposition of dyon and anti-dyon, the dirac strings will
cancel everywhere except in between two objects like in Figure \ref{fig:dstring}

The gauge transformation we need is given by
\be  \nonumber
&U_+=e^{i\theta \;\tau_2/2}e^{i\phi \;\tau_3/2} \\ \nonumber
&U_-=e^{i(\pi-\theta)\; \tau_2/2}e^{i\phi \;\tau_3/2}
\ee
The gauge field then looks as follows 
\be
A_i=\left\{\begin{array}{l}
A_r=0\;,\\
A_\theta=\frac{R(vr)}{2 r}\tau_2\;\\
A_\phi=-\frac{R(vr)}{2r}\; \tau_1+\frac{1}{2 r}\cot\theta \;\tau_3
\end{array}
\right.
\ee
where $R(z)=z/\sinh(z)$.

However, superposing the two solutions is not as trivial as one may thing.
Surely the abelian components will superpose properly, and one may think that
the contribution of the core is negligible, as the core has exponentially small
influence. However the string singularity makes any contribution of the core
large, and one must proceed with caution.

We first examine a single monopole and compute it's magnetic field in spherical
coordinates. The metric is given by the line element
\be
ds^2=dr^2+r^2d\theta^2+r^2\sin^2\theta d\phi^2
\ee
The radial component of the magnetic field $B_r$ can then be calculated
as\footnote{The magnetic field $B$ is defined as
\[
B^i=\frac{1}{2\sqrt{g}}\epsilon^{ijk}F_{jk}
\]
where the $\sqrt{g}$ is put because $\epsilon^{ijk}$ is a tensor density.
}
\be
B^r=\frac{F_{\theta\phi}}{\sqrt{g}}\;.
\ee
where we calculate\footnote{the factors of $r\sin\theta$ and $r$ are inserted
because $A_\phi$ and $A_\theta$ are \emph{neither} contravariant nor covariant
components of the vector field $A_i$, but $A_\phi r\sin\theta$ and $A_\theta r$
are covariant components.}
\be
\vec F_{\theta\phi}=\partial_\theta (\vec A_\phi r\sin\theta)-\partial_\phi(\vec
A_\theta r)+\vec A_\theta\times\vec A_\phi r^2\sin\theta\;,
\ee

Taking into account just the abelian part we obtain exactly what we expect.
Namely the $\sin\theta$ term above turns $\cot\theta$ into $\cos\theta$ which,
upon differentiation, becomes $\sin\theta$. Then dividing by
$\sqrt{g}=r^2\sin\theta$ we obtain $B^{r,a}=\delta^{3,a}1/r^2$, where $a$ is the
color index. The commutator of the core of $A_\theta$ and $A_\phi$ will correct
the filed so that it is not divergent in the center. The more interesting
cancelation is one of the string and the core. In fact we will see that they
conspire to cancel the contribution of the string. The relevant term in $A_\phi$
is the core. Then 
\be
\partial_\theta(A_\phi r\sin\theta)=-\partial_\theta(R\sin\theta
\tau_1/2)+\dots=-R\cos\theta\;,
\ee
where dots indicate the abelian part which we argued contributes to the expected
field of the monopole. The commutator term becomes
\be
(\vec A_\theta\times\vec A_\phi r^2\sin\theta)\cdot\frac{\vec
\tau}{2}=R\cos\theta
\ee
which cancels the term in the derivative part of field strength tensor.

However, now we consider the superposition of the dyon and anti-dyon. We will
have then that the fields are that of dyon and the abelian part of the
anti-dyon, as the core contribution can be neglected. Therefore, apart from the
cancelation which we described before, there will be a term of the form
\be
-\partial_\theta R\pm \frac{r}{r_2}R\cot\theta_2=0\;.
\ee
where we assumed now that $R$ is a function of both $\theta$ and $r$. The sign
depends on whether the superposed field is a monopole or antimonopole, as well
as how it is oriented, i.e. whether the south poles of the
monopole-(anti)monopole system are facing eachother, or is the south pole of the
one facing the south pole of the other. We will see that the sign we want is
actually a relative minus sign between the two terms.

The equation above is just a simple, separable, differential equation. All that
we have to do is express $r_2,\theta_2$ in terms of the spherical coordinates
$r,\theta$. The relation is the following
\be
r_2=\sqrt{r^2+d^2+2rd\cos\theta}
\ee

The actual equation is
\be
\partial_\theta \ln R=\frac{r}{r_2}\tan\frac{\theta_2}{2}\;,
\ee
and the solution is
\be
-\frac{2(1-\xi_2+\xi^2)\cot\theta}{\xi
\xi_2}+\frac{2(\xi_2-\cos2\theta-1)}{\xi_2\sin\theta}-\frac{2(1+\xi)}{\xi}
E(\frac{\theta}{2},\frac{4\xi}{(1+\xi)^2})+\frac{2}{\xi(1+\xi)}F(\frac{\theta}{2
},\frac{4\xi}{(1+\xi)^2})
\ee


We start by writing the fields of in cylindrical coordinates. Then we may simply
superpose the two fields. The Dyon becomes
\be
&A_\rho=A_\theta\cos\theta=\frac{R(v r) z}{\rho^2+z^2}\;,\\
&A_\phi=\frac{R(vr)}{\sqrt{\rho^2+z^2}}\frac{\vec{\hat \rho}\cdot\vec
\tau}{2}-\frac{1}{r}\frac{\sqrt{\rho^2+z^2}+z}{\sqrt{\rho^2+z^2}-z}\frac{\tau^3}
{2};,\\
&A_z=-A_\theta\sin\theta=-\frac{R(vr)\rho}{\rho^2+z^2}\frac{\vec{\hat
\phi}\cdot\vec \tau}{2}\;,\\
\ee

%
%
%
%
%
%

\subsubsection{Following the gradient flow down the streamline.}
As it is well known, a ``combed" monopole or dyon must possess the Dirac string, a singular
gauge artifact propagating one unit of magnetic flux from infinity to the dyon center. 
By selecting appropriate gauge one can direct the Dirac string to have arbitrary direction.
Superimposing into a sum of two dyons with different directions of the Dirac string one gets  {\em non-equivalent}
configurations: the interference of singular and regular terms make the Dirac strings no longer invisible or pure gauge
artifact. (However, this is cured during the gradient flow process, as we will discuss below.)

\begin{figure}[t]
  \begin{center}
      \includegraphics[width=7cm]{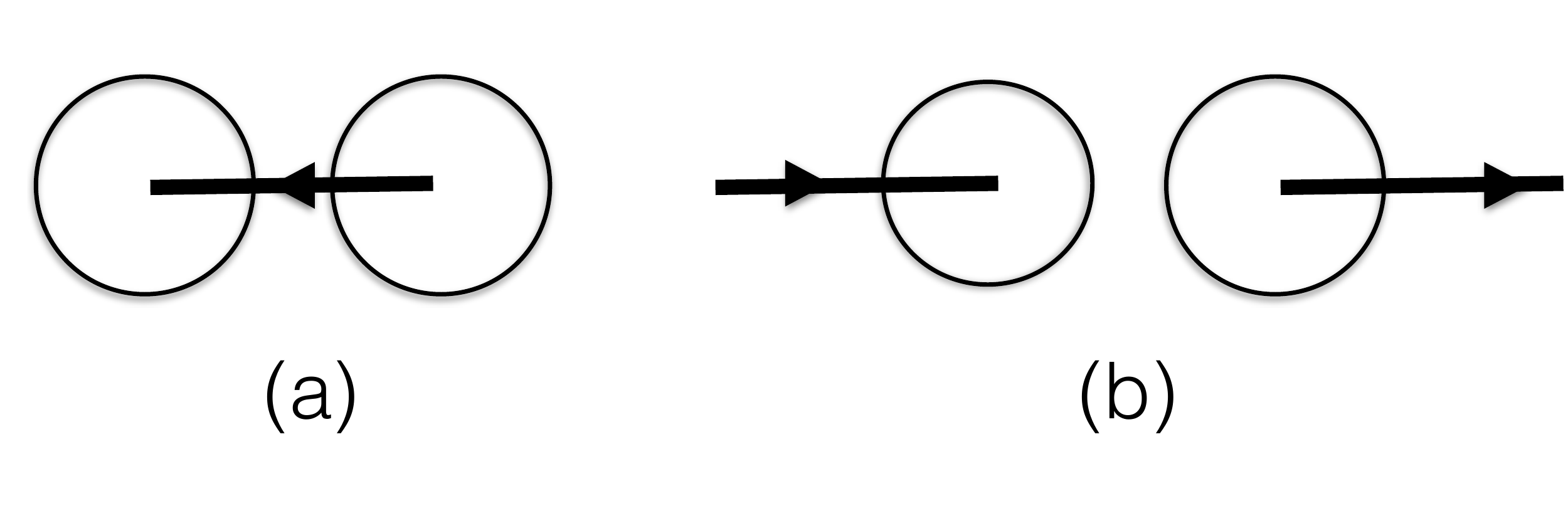}
   \caption{Two extreme positions for the Dirac strings, for  the $M\bar{M}$ pair.  
  \label{fig_Dstrings} }
  \end{center}
\end{figure}

 Two obvious extreme selections for the  Dirac strings are: (a) 
 a ``minimally connected dipole" when it goes along the  line
connecting two dyon centers;  and (b)  a ``maximally disconnected" pair,
in which the  Dirac strings go into the centers from two opposite directions, see Fig.\ref{fig_Dstrings}.
Under the gradient flow the former is supposed to reach magnetically trivial configuration,
while the latter must relax to a pure gauge Dirac-string-like state passing the flux through the system, from minus to plus infinity.
The former case seems to be simpler and more natural to use: but our experience has shown the opposite, 
that (b) generates smaller artifacts since the Dirac strings interfere less with the gradient flow changes between the two objects.
So we will use case (b) as our starting configurations below.

 An important role in what follows is a color current
 \be
	j_\mu^a=-\frac{\delta S}{\delta A_\mu^a}|_{A=A_\textrm{ansatz}}=(D_v^{ab}G_{v\mu}^b)|_{A=A_\textrm{ansatz}}\neq0.
\ee
The current vanishes for extrema (solutions of the YM equation, such as a single dyon), but 
it is nonzero for dyon-antidyon configurations which
we study.  It has the meaning of the force in the functional space showing the direction
towards a reduction of the action.  

The study of dyon-antidyon streamline configurations we will follow\footnote{We later learned that stable monopole-antimonopole configuration and the potential leading to it has also been
calculated in \cite{Shnir:2005te}.
} is due to  \cite{Larsen:2014yya}.
 The gradient flow is a process, in a computer time $\tau$,
 thus the current would be  the driving force. In the paper we follow in this section,
 a dyon-antidyon pair was put on the lattice\footnote{I was asked if it is possible to put it on the standard lattice with periodic boundary conditions. The answer is negative: the $M\bar{M}$ (and any other interesting dyon pairs) always have one of the charges -- electric or magnetic -- uncompensated. Furthermore, for arbitrary holonomy $\nu$, the pair also do not have an integer topological charge. So, the lattices we used had $no$ periodicity conditions at all. }. 

The gradient flow process was found to proceed via the following stages\footnote{Let me supply more colloquial names for these stages: running downhill to the stream, following the streamline, finding a lake, and then a waterfall.}: \\
 (i) {\em near initiation}: starting from relatively arbitrary ansatz one finds rapid disappearance of artifacts and convergence toward the streamline set\\
 (ii) following the {\em streamline itself}. The action decrease at this stage is small and steady. The dyons basically approach each other, with relatively small deformations:
 thus the concept of an interaction potential between them makes sense at this stage\\
 (iii) a {\em metastable state} at the streamline's end: the action remains constant, evolution is very slow and consists of internal deformation of the dyons rather than further approach\\
   (iv) {\em rapid collapse} into the perturbative fields plus some (pure gauge) remnants\\   
 
We will detail properties of these stages below, for now restricting  to  general comments.
One is  the existence of the stage (iii) which has not been anticipated on general grounds. Since all configurations corresponding to it
have the same action, one can perhaps lump all of them into a new class of states, corresponding to the same dyon-antidyon distance. 
Unlike the instanton-dyons themselves, such states have not yet been identified on the lattice.

Our other comment is the  action value even at the end of the streamline is not that far from the sum of the two dyon masses. In other words, the
classical interaction potential happens to be rather small numerically,   a welcoming feature for statistical mechanics simulations.  

It should also be noted that while the topological and magnetic charges of $M$ and $\bar{M}$ are
opposite, their electric charges are the same. Thus total value of it is 2, rather than zero.
That is why they cannot annihilate each other, as instanton and antiinstanton do.

Last but not least, we do observe the $universality$ of the streamline. As expected, independent on the 
initial dyon separation we found that gradient flow proceeds
through essentially the same set of configurations at stages (ii-iv). Thus one-parameter characterization of those is possible. A parameter 
we found most practical in this work is simply its $lifetime$ -- duration in our computer time $\tau$ needed for a particular configuration to reach a  final collapse.  
(Of course, for statistical mechanics applications one better map that into some collective coordinate, such as the dyon separation, whatever way it can be defined). 

\begin{figure}[h!]
\includegraphics[width=7cm]{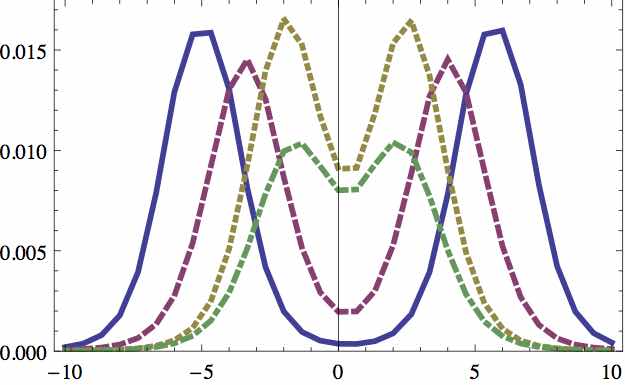}
\put(-245,145){\large{s(z)}}
\put(-15,-5){\large{z}}
\caption{Action density along the z axis in natural units for a separation $|r_M-r_{\bar{M}} |v=10$ between the centers of the 2 dyons. The configuration with the maximums furthest from each other is the start configuration. After 3000 iterations it has moved further towards the center. At 12000 iterations the configuration has reached the metastable configuration with a separation between the maximums of around 4. At 13700 the configuration has collapsed around halfway, and will continue to shrink until the action is 0. Times are as shown in Fig. \ref{Streamlines}. }
\label{Action_density}
\end{figure}

\begin{figure}[h!]
\includegraphics[width=7cm]{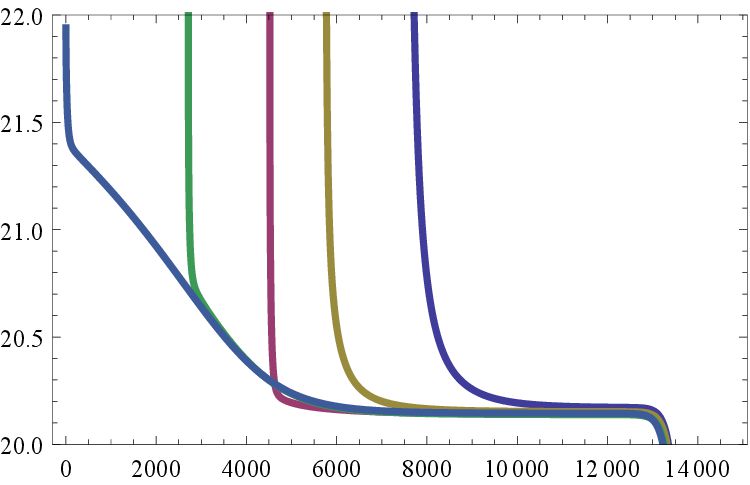}
\put(-245,130){\large{S}}
\put(-50,-10){\large{Iterations}}
\caption{Action for $v =1$ as a function of computer time (in units of iterations of all links) for a separation $|r_M-r_{\bar{M}} |v=0$, $2.5$, $5$, $7.5$, $10$ between the $M$ and $\bar{M}$ dyon from right to left in the graph. The action of two well separated dyons is 23.88.}
\label{Streamlines}
\end{figure}

We now show the results for a $M$ and $\bar{M}$ dyon separated by a distance (in natural units $1/v$) of the order 0 to 10 along the z-axis which is cooled using gradient flow. The action of an individual dyon on the lattice was found to be $11.94$, $5\%$ lower that the analytic value $4\pi$. This gives the action of $23.88$ for two well separated dyons. Any action lower than this therefore is ascribed to an attractive interaction between the dyons.

 
 \begin{figure}[b!]
\centering
\includegraphics[width=7cm]{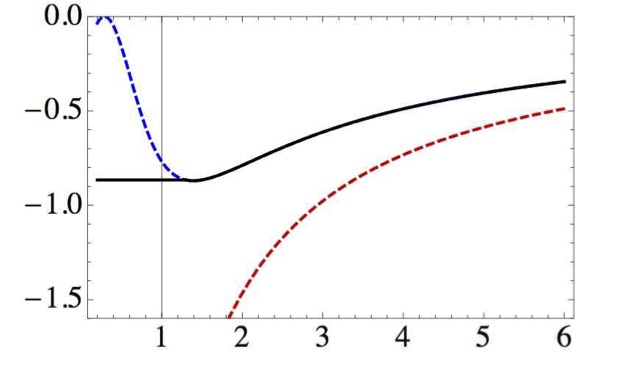}
\caption{Black solid line is the $\bar{M}M$ dimensionless potential (action), red dash line is the asymptotic Coulomb
term,  blue dashed line is a sketch of possible ``core" shape.}
\label{fig_dbard_pot}
\end{figure}

The simulations of the streamline for different configurations starts out with a slightly higher action around that of 2 individual dyons, this then converge to an almost stable configuration, the Streamline. 
A typical action history is shown in Fig. \ref{Streamlines}, for the initial separation values from 0 to 10. 
If the separation is bigger than 4, the two cores of the dyons are seen to move  towards each other and their action smoothy decreases.   We show four  important computer times  in Fig. \ref{Action_density}, where the action density is plotted along the z-axis (along the dyons separation) for at start configuration with a separation of 10 between the dyon and antidyon. 


\section{The partition function  in one loop}

\subsection{Electric screening}
   The basic physics of the electric screening can be explained most simply
   following the original derivation by one of us \cite{Shuryak:1977ut} in the Coulomb gauge. If some electrically charged object, with nonzero $A_4$, is immersed into
   a finite-tempreature QCD, gluons and quarks of the heat bath are scattered
   on it. The simplest diagram diagram comes from the quartic term in the gauge Lagrangian, $e^2 A_m^2 A_4^2$ which couples  the heat bath gluons directly to
  square of $A_4$, but there are also other diagrams contributing to the forward  
scattering amplitude. The result was the expression for the QCD Debye mass
\be M_D^2= e^2 T^2 (1+N_f/6) \label{eqn_Debye}\ee
In 1976, when QCD was only 3 years old, the main finding was its positivity,
which ensured $screening$ of a charge, as opposed to anticreening by vacuum loops:
thus the ``plasma" name. 

The next relevant paper is  \cite{Pisarski:1980md}  (PY)
who found in the one-loop action of the calorons (the finite-T instantons) the famous PY term $\sim \rho^2 M_D^2/e^2$, where $\rho$ is the instanton radius.  It also comes from
the forward scattering amplitude of the thermal plasma quanta on the $A_4^2$ of the instanton. This term is important, because it ensures that at $T>T_c$ (only in this
case there are thermal gluons!) the semiclassical expression provides finite instanton density.

Going forward to calorons at nonzero holonomy, the one-loop effective action
has been computed by Diakonov, Gromov, Petrov and Slizovskiy (DGPS) 
\cite{Diakonov:2004jn}. The caloron  is now a superposition of the M and L dyons,
separated by distance $r_{ML}$, and the basic expression from which the effect
came looks as follows
\be  \int d^3r ({1 \over r_L}-{1 \over r_M})^2=4\pi r_{ML}+...\ee
where $r_L,r_M$ are distances from the dyon centers. It thus creates 
an effect resembling linear confinement, with a force independent on the separation.
At zero holonomy this result matches the PY answer  because 
the relation between the instanton size and ML separation is
\be \rho^2={ r_{ML} \over \pi T} \ee

The bracket in this integral is nothing else than $A_4^2$.  One obvious consequence would be that if one would generalize the DGPS derivation to theory with fermions,
they will simply get extra factor $1+N_f/6$ as in (\ref{eqn_Debye}). Another one, worth
mentioning, is that for pairs $L\bar{L}$ or $M\bar{M}$ with the $same$ electric charge,
there will be plus in the integral above and thus the effect becomes repulsive. 

The electric screening effect ensures $LM$ ``binding" into finite-size instantons, 
into an object with a size $r_{ML} \sim e^2T/M_D^2$. (Note that the coupling $e$ cancels here,
it is because the nonperturbative fields are always $\sim 1/e$.) 

Although asymptoticlly at $N_f\rightarrow \infty$ this size is $O(1/N_f)$, the coefficient
1/6 in  (\ref{eqn_Debye}) makes it less important for ``interesting" $N_f=0..10$.
We will later see that the direct fermionic interaction discussed in the
 preceeding section binds $L\bar{L}$ pairs stronger than $LM$ interaction. 


Since we will be discussing charge-zero clusters consisting of all 4 dyons, let give an example
of the potential electric screening creates in this case. For simplicity we will only discuss $L\bar{L}$
at the same point, and $M,\bar{M}$ and $L\bar{L}$ to be on one line. The integral
\be  \int d^3r \left( {2 \over r}-{1 \over r_M} -{1 \over r_{\bar{M }} } \right)^2 \ee
leads to a potential for $M$ shown in Fig.\ref{fig_dbard_pot}. As one can see, like for DGPS case
it consists of linear segments, but is now deformed away from the companion dyon. (Note,
that it is not due to their Coulomb repulsion, which is also there but will be discussed in the next supsection.) 

%

\subsection{The one-loop measure, perturbative Coulomb corrections and the ``core"}
Another important result of the DGPS paper
\cite{Diakonov:2004jn}, see also the influential Diakonov's lectures 
\cite{Diakonov:2009jq}, was the derivation of the forces decreasing with the distance also coming from the one-loop gluonic determinant\footnote{More precisely,
what was calculated and presented below is the volume element of the ``moduli space"
of all multi-dyon configurations. The geometry, the metric tensor and the volume element of
space of all classical multi-monopole solutions has been studied in mathematical literature, 
especially by Attiya, Hitchin et al started from 1970's.
}. Combined with Manton's result for identical
dyons, it was generalized to the arbitrary number of selfdual $L,M$ dyons
The resulting volume element in space of collective variables is expressed elegantly  
$  \sqrt{g}=det[G] $ via a determinant of the so called Diakonov matrix $\hat G$, defined by  
 \be \hat G &=& \big[ \delta _{mn} \delta _{ij} ( 4\pi \nu_m-2\sum _{k\neq i}\frac{1}{T|x_{i,m}-x_{k,m}|} +2\sum _{k}\frac{1}{T|x_{i,m}-x_{k,p\neq m}|} ) \\
 & & 
 +2\delta_{mn}\frac{1}{T|x_{i,m}-x_{j,n}|}-2\delta_{m\neq n}\frac{1}{T|x_{i,m}-x_{j,n}|} \big], \nonumber
 \ee
Here $x_{i,m}$ denote the position of the i'th dyon of type m. This form is an interpolation of the exact metric between a $M$ and $L$ dyon,  true at any distance, with the metric of the two dyons of same type at large distances.
We  introduce a cutoff on the separation via $r \to \sqrt{r^2+cutoff ^2}$, such that for one pair of dyons of same type, the diagonal goes to 0 for $\nu=0.5$, instead of minus infinity.


Note that while the matrix itself has only 1/r terms, but after the effective one-loop action $log (det (G))$ is computed
one gets all powers of 1/r, involing nontrivial manybody interactions. 

 In subsequent works (see lectures \cite{Diakonov:2009jq} for summary and references) Diakonov had used this interaction in the $L,M$ sector to perform manybody
calculation. He argued that it can also leads to some integrable model,
which can be solved and even results in the holonomy potential leading to confinement.

We think those conclusions were a bit premature, because it somehow assumes that the selfdual $L,M$ sector and antiselfdual one
 $\bar{L},\bar{M}$ are invisible to each other and do not interact at all. Not only we do not see why this should be the case, we have found that dyon-antidyon interactions are
 important, and  in theory with many fermions even dominant. 
 
 Another issue is that for many configurations $det(G)<0$, which obviously makes no sense
 for a volume element. As some eigenvalue of $G$ approach zero, it means such
 configuration approach zero measure in the space of soluitons. What it means is that
 only those with $det(G)>0$ should be included in the partition function. This is relatively 
 easy to do in numerical studies, but was not taken care of in the original studies.

\section{ Fermionic zero modes  }
\subsection{How quark zero modes are shared between the dyons}

The instanton has unit topological charge $Q=1$ and, according to Attiyah-Singer index theorem,
has one fermionic zero mode. In the chapter about instantons we have discussed 't Hooft
effective Lagrangian arizing due to it.
Now we discuss non-zero holonomy environment at $T\neq 0, <P>\neq 1$, in which the instanton is split
into $N_c$ constituents. The question is how the zero mode is distributed among these constituents. 

As was determined by van Baal and collaborators, fermionic zero mode ``hops" from one type of dyon to the next
at certain critical values. 
Periodicity condition along the Matsubara circle can be defined with some arbitrary angles $\psi_f$ for quarks
with the flavor $f$. 
The resulting rule is: it belogs to the dyon corresponding to the segment of the
holonomy circle $\nu_i$ to which the periodicity phase belongs: $\mu_i<\psi_f<\mu_{i+1}$.

\begin{figure}[h]
\centering
\includegraphics[width=4.cm]{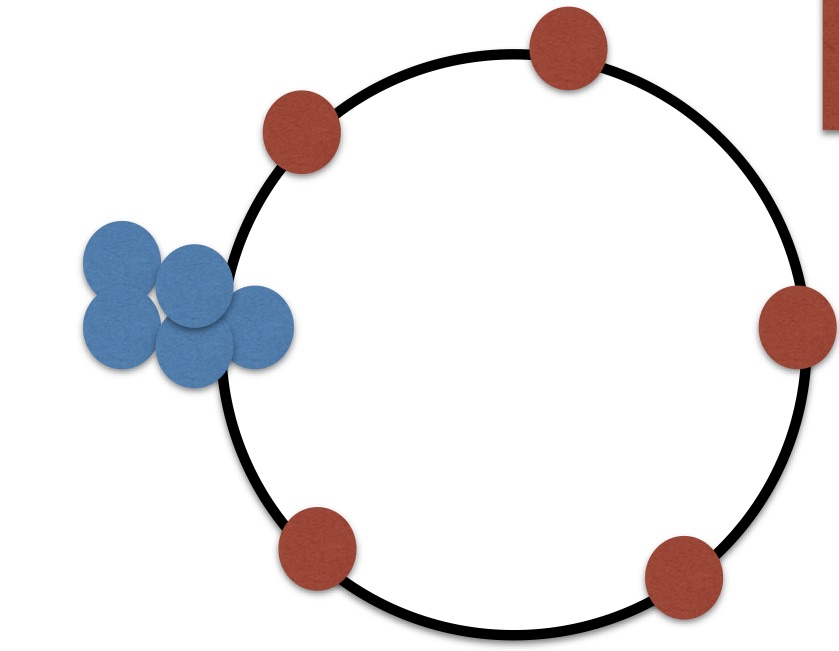}
\includegraphics[width=3.5cm]{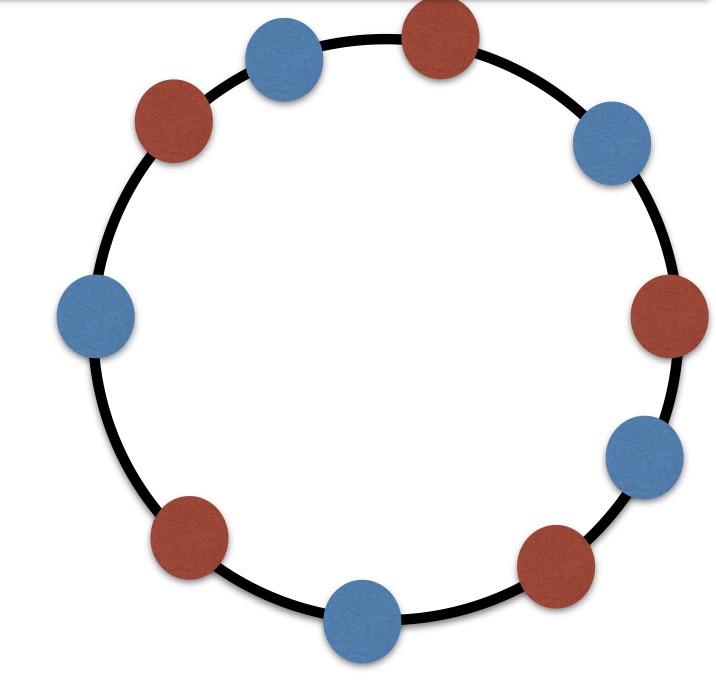}
\caption{ Schematic explanation of the difference between the usual QCD (left) and the $Z(N_c)$ QCD (right).}
\label{fig_circles}
\end{figure}

In physical QCD all quarks are fermions, and therefore $\psi_f=\pi$ for all $f$. This case is
schematically shown by blue dots in Fig.\ref{fig_circles}(left): all fermions fall on the same segment of the
circle, and therefore only one, of $N_c$ dyons, have zero modes and interact with quarks.

But one can introduce other arrangements of these phases. In particular, for $N_c=N_f$ the opposite extreme is 
the so called  $Z(N_c)$ QCD, proposed in \cite{Kouno:2013zr} , put them symmetrically around the circle, see Fig.\ref{fig_circles}(right).
In this case, the instanton-dyon framework becomes very symmetric: each dyon interact with ``its own" quark flavor.

\subsection{The zero mode for the fundamental fermion} 
We start with the Dirac equation
\be
\Dslash \psi=0\;,
\ee
and look for a normalizable solution within the hedgehog ansatz
\be
&A_i^a=\epsilon_{aji}\mathcal A\hat r^j\;,\\
&A_0^a=\mathcal H\hat r^a\;.
\ee

Since the Dirac operator is chiral, we may write the fermion in terms of upper and lower components $\psi_L$ and $\psi_R$. We do the calculation for the lower component, namely $\psi_R$. The Dirac equation then reads
\be\label{eq:dirac}
-(\sigma^\mu)_{\alpha\beta} (D_\mu)_{AB} (\psi_R)_\beta^B\;,
\ee
where we explicitly wrote the Dirac indices $\alpha,\beta$ and color indices $A,B$. Now we ansatz (see Shnir)
\be
\psi^{A}_\alpha=\alpha(r)\epsilon_{A\alpha}+\beta(r)[(\vec{\hat r}\cdot \vec \sigma)\epsilon]_{A\alpha}\;.
\ee

We may choose to consider the matrix 
\be
\eta_{A\alpha}=-\psi_{\beta}^A\epsilon_{\beta \alpha}
\ee
in which case 
\be
\eta=\alpha(r)\vec 1+\beta(r)\hat{ \vec{r}}\cdot \vec\sigma
\ee
The rule of acting with a color and the spin sigma matrices on this object is such that we multiply by a color matrix $\tau$ from the left, and if we multiply by a spin matrix $\sigma$, then we multiply from the right, and put a minus sign, i.e.
\be
\sigma \psi=\eta\epsilon \sigma^T=-\eta \sigma \epsilon\;.
\ee
If we wish to construct fermion density, we see that
\be
{\psi^*}_\alpha^A\psi_\alpha^A=\Tr(\eta^\dagger\eta)
\ee

We now plug the ansatz into \ref{eq:dirac} and obtain the following two equation
\be
&\alpha'(r)+\frac{\mathcal H+2\mathcal A}{2}\alpha+\frac{z}{\beta}\beta=0\;,\\
&\beta'(r)+\left(\frac{\mathcal H-2\mathcal A}{2} +\frac{2}{r}\right)\beta+\frac{z}{\beta}\alpha=0\;.
\ee
where we have assumed $\psi_R\propto e^{izt/\beta}$, i.e. that the Fermion has arbitrary periodicity condition in the imaginary time direction.

Let us look at the asymptotic behavior of dyons, i.e. when $\mathcal H(r\rightarrow \infty)=v$ and $\mathcal A(r\rightarrow\infty)=0$, then
\be
&\alpha'(r)+\frac{v}{2}\alpha(r)+\frac{z}{\beta}\beta=0\;,\\
&\beta'(r)+\frac{v}{2}\beta(r)-\frac{z}{\beta}\alpha=0\;.
\ee
This equation is easily solvable by taking the substitution $\alpha_\pm=\alpha\pm\beta$ we get
\be
\alpha_\pm=e^{-\left(\frac{v}{2}\pm \frac{z}{\beta}\right)}\;.
\ee

In order for the solution to be normalizable, we must have that both $\alpha_\pm$ vanish at infinity. This is only possible if $|z|<|v|\beta/2$

\begin{figure}[htbp] 
   \includegraphics[width=10cm]{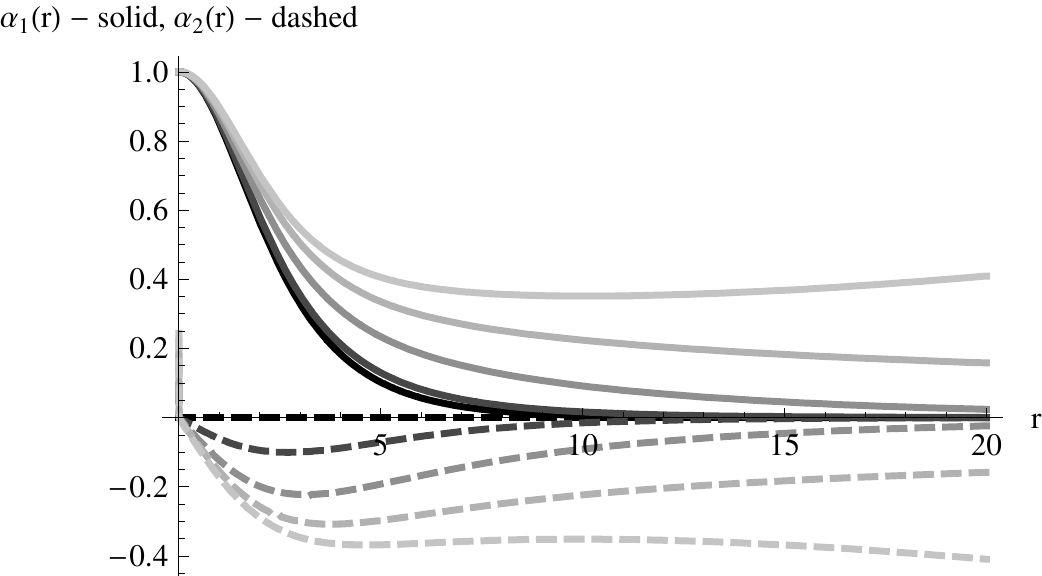} 
   \caption{Plot shows profile of zeromode components $\alpha_{1,2}$, for four different values of $z=0,0.2 v/\beta,0.4 v/\beta, 0.5 v/\beta, 0.55 v/\beta$. Note that the zero mode delocalizes at $z=0.5 v/\beta$}
   \label{fig:zeromodeprof}
\end{figure}

We now proceed to numerical solution of the differential equation. The plots in \ref{fig:zeromodeprof}

We may rewrite this equation in a two component form as
\be
\frac{d}{dr}\vec \alpha=-\vec M\vec \alpha\;,
\ee
where

We can write the formal solution as a path ordered exponent
\be
\vec\alpha=\mathcal P \exp\left(-\int_{r_0}^r \; \vec M\;dr\right)\vec \alpha(r_0)
\ee

The matrix $\vec M$ can be written as follows
\be
\vec M(r)=\left(\frac{\mathcal H}{2}+\frac{1}{r}\right)\vec 1 +\frac{z}{\beta}\sigma_1+\left(\mathcal A-\frac{1}{r}\right)\sigma_3
\ee

The factor proportional to the unit matrix commutes with everything else, and may be factored out. On the other hand

\[\mathcal A-1/r=-v/\sinh(v r)\sim e^{-vr}\]

is small and we may neglect it except at the origin. So for $r>> 1/v$ we may integrate the exponent, as it is proportional to $\sigma_3$ only, and then expand the result. Since 
\[
\int\left(\frac{\mathcal H}{2}+\frac{1}{r}\right)=\frac{1}{2}\ln\left[r\sinh(rv)\right]\;,
\]
we have that
\be
\vec\alpha(r)=\frac{e^{-\frac{v}{2}(r-r_0)}}{\sqrt{r/r_0}}
\times\left[\cosh\left(\frac{z(r-r_0)}{\beta}\right)\right.+\left.\sigma_1\sinh\left(\frac{z(r-r_0)}{\beta}\right)\right]\vec\alpha(r_0)
\ee

Now we will solve the equation exactly. To do this we separate the matrix $M(r)$ as
\be
M(r)=M_0(r)+M_1(r)\;,
\ee
where
\be
M_0(r)&=\left(\frac{\mathcal H}{2}+\frac{1}{r}\right)\vec 1=\;,\\
M_1(r)&=\frac{z}{\beta}\sigma_1+\left(\mathcal A-\frac{1}{r}\right)\sigma_3\\&=\frac{z}{\beta}\sigma_1-\frac{v}{\sinh(vr)}\sigma_3\;.
\ee

The solution can then be written as $\vec \alpha=\exp(-\int_{0}^rM_0(r)dr)\vec\chi$, or
\be
\vec\alpha=\frac{1}{\sqrt{r\sinh{r v}}}
\ee

with the differential equation for $\vec\chi$ reading
\be
\frac{d}{dr}\vec\chi=-M_1(r)\vec\chi\;,
\ee
i.e.
\be
&\chi_1'(r)=\frac{v}{\sinh(vr)}\chi_1(r)-\frac{z}{\beta}\chi_2(r)\;,\\
&\chi_2'(r)=-\frac{v}{\sinh(vr)}\chi_2(r)-\frac{z}{\beta}\chi_1(r)\;,
\ee
we may take a change of variables $\xi=r v$. Then the equation read
\be
&\chi_1'(\xi)=\frac{1}{\sinh(\xi)}\chi_1(r)-\varsigma\chi_2(r)\;,\\
&\chi_2'(\xi)=-\frac{1}{\sinh(\xi)}\chi_2(r)-\varsigma\chi_1(r)\;,
\ee
where we labeled $\varsigma=x/(v\beta)$
We now eliminate $\xi_2$, and obtain the second order differential equation
\be
-\frac{d^2}{d\xi^2}\chi_1-\frac{1}{2\cosh^2\frac{\xi}{2}}
\chi_1=-\varsigma^2\chi_1\;.
\ee

 a general solution with arbitrary constants $c_{1,2}$ is
\be
\chi_1(\xi)=c_1\left(-2\varsigma+\tanh\frac{\xi}{2}\right)e^{\varsigma
\xi}+c_2(2\varsigma+\tanh\frac{\xi}{2})e^{-\varsigma \xi}
\ee
. Using the first order equation we can write $\chi_2$ as
\be
\chi_2(\xi)=c_1\left(2\varsigma-\coth\frac{\xi}{2}\right)e^{\varsigma
\xi}+c_2(2\varsigma+\coth\frac{\xi}{2})e^{-\varsigma \xi}
\ee
The function $\chi_2(\xi)$ is divergent when $\xi\rightarrow 0$, except if $c_1=c_2$, in which case $\xi_2(0)=0$. Therefore $c_2=c_1$. The constant $c_1$ can be determined by overall normalization. The solution then becomes
\be \label{eq:chisol}
\chi_1(\xi)=2c_1\left(- 2\varsigma \sinh(\xi\varsigma)+\tanh\frac{\xi}{2}\cosh(\xi\varsigma)\right)\\
\chi_2(\xi)=2c_1\left(2\varsigma \cosh(\xi\varsigma)-\coth\frac{\xi}{2}\sinh(\xi\varsigma)\right) 
\ee
Finally we obtain
\be
\alpha_{1,2}=\frac{\sqrt{v}}{\sqrt{\xi\sinh\xi}}\chi_{1,2}\;.
\ee
$\sqrt{v}$ can be absorbed into constant $c_1$, and our final expression is
\be
\alpha_{1,2}=\frac{\chi_{1,2}}{\sqrt{\xi\sinh\xi}}\;.
\ee
with functions $\chi_{1,2}$ given by (\ref{eq:chisol}), $\xi=v r$, $\varsigma=z/(v\beta)$.
Note that the value of $\alpha_{1}(\xi\rightarrow 0)$ is given by
\be
c1(1-4\varsigma^2)\;,
\ee
and the solution is completely regular at $r=0$.

\subsubsection{ Elements of quark ``hopping matrix"}

We calculate the fermionic transition amplitude between the dyons, which is the
 operator $\Dslash$ with the proper background gauge field, expressed
 in the basis of the fermion zero modes  of all dyons and anti-dyons. The transition element will look like
\be
T_{D\bar D}=\int d^4x\; \psi^\dagger_{\bar D}\Dslash \psi_{D}\;,
\ee
where $\psi_{D,\bar D}$ are dyonic and anti-dyonic zero modes. We assume a superposition ansatz of dyons, and have that
\be
\Dslash=\partial\fslash-i {A \fslash}_D-i{A}\fslash _{\bar D}
\ee
then since $(\partial\fslash-i{A\fslash}_{D,\bar D})\psi_{D,\bar D}=0$
\be
T_{D\bar D}=\int d^4x\; \psi^\dagger_{\bar D}\partial\fslash \psi_{D}\;,
\ee

In our solution we use $\eta_{D,\bar D}$, and the transition element becomes
\be
-\Tr(\eta_{\bar D}^\dagger U_{\bar D}^\dagger \partial_\mu (U_D\eta_D) \sigma^\mu)
\ee
where $U_{D,\bar D}$ are gauge combing matrices for the dyon and anti-dyon respectively (remember that zero modes were found in a hedgehog gauge). Explicitly we use the form of the matrices $U_D=U_-(\theta,\phi)$ and $U_{\bar D}=U_-(\vartheta,\varphi)$, where $\theta,\phi$ and $\vartheta,\varphi$ are spherical angles in the center of the dyon and anti-dyon respectively.  These matrices  have the following properties
\be
&U_D \hat{\vec \theta}\cdot \vec\sigma U_D^\dagger=\sigma^1 &U_D \hat{\vec \phi}\cdot \vec\sigma U_D^\dagger=\sigma^2\\\nonumber
&U_D \hat{\vec r}\cdot \vec\sigma U_D^\dagger=\sigma^3\;,\\\nonumber\\
&U_{\bar D} \hat{\vec \vartheta}\cdot \vec\sigma U_{\bar D}^\dagger=-\sigma^1 &U_{\bar D} \hat{\vec \varphi}\cdot \vec\sigma U_{\bar D}^\dagger=\sigma^2\\\nonumber
&U_{\bar D} \hat{\vec s}\cdot \vec\sigma U_{\bar D}^\dagger=-\sigma^3\;, \ee
where $\hat {\vec r},\hat {\vec \theta},\hat {\vec \phi}$ and  $\hat {\vec s},\hat {\vec \vartheta},\hat {\vec \varphi}$ are spherical unit coordinate vectors around dyon and anti-dyon respectively.

We first consider the spatial index $\mu=i$. We may use the matrices $U_{D,\bar D}$ to transform $\eta_{D,\bar D}=\alpha_{D,\bar D}+\sigma\cdot \hat r_{D,\bar D} \beta_{D,\bar D}$ into an object $\zeta_{D,\bar D}=\alpha_{D,\bar D}\pm\beta_{D,\bar D}\sigma_3$, i.e. 
\[\zeta_{D,\bar D}=U_{D,\bar D}\eta_{D,\bar D}U_{D,\bar D}^\dagger\]

It can be shown that the transition element is given by the expression
\be
\Tr\left\{ U_{\bar D}U_{D}^\dagger \zeta_{\bar D}\partial_r\zeta_D\sigma_3\right.\\\left.+\zeta_{\bar D}^\dagger \zeta_D\left[\frac{1}{r} (\partial_\theta U_D)U^\dagger_D \sigma_1+\frac{1}{r\sin\theta} (\partial_\phi U_D)U^\dagger_D \sigma_2\right]\right\}
\ee
Since 
\be
&(\partial_\theta U_D)U_D=\frac{i\sigma_2}{2}\;,\\
&(\partial_\phi U_D)U_D=\frac{i}{2}(-\sigma_1\sin\theta+\sigma_3\cos\theta)\;,
\ee

After straightforward manipulations we obtain that the spatial part of the expression for the transition element is given by
\be
\Tr\left\{ \sigma_3U_{\bar D}U_{D}^\dagger \Big[\zeta_{\bar D}\partial_r\zeta_D\right.\\
\left.\left.+\frac{1}{r}\zeta_{\bar D}^\dagger \zeta_D\left(1-\frac{i\sigma_2}{2}\cot\theta\right)\right]\right\}
\label{eq:sptran}
\ee

The matrix $U_{\bar d}U^\dagger_d$ can be calculated
\be
&U_{\bar d}U^\dagger_{d}=\\\nonumber
&=\left(\sin{\frac{\vartheta+\theta}{2}}+i\sigma_2\cos{\frac{\vartheta+\theta}{2}}\right)\cos\frac{\Delta\phi}{2}\\\nonumber
&+\left(\sin{\frac{\vartheta-\theta}{2}}+i\sigma_2\cos{\frac{\vartheta-\theta}{2}}\right)i\sigma_3\sin\frac{\Delta\phi}{2}
\ee

We write this expression as
\be
U_{\bar d}U^\dagger_{d}=(a+i\sigma_2b)+(c+i\sigma_2 d)i\sigma_3\;.
\ee
where
\be
&a=\sin{\frac{\vartheta+\theta}{2}}\cos\frac{\Delta\phi}{2}\\
&b=\cos{\frac{\vartheta+\theta}{2}}\cos\frac{\Delta\phi}{2}\\
&c=\sin{\frac{\vartheta-\theta}{2}}\sin\frac{\Delta\phi}{2}\\
&d=\cos{\frac{\vartheta-\theta}{2}}\sin\frac{\Delta\phi}{2}
\ee

%

\subsection{Fermionic zero mode for a set of selfdual dyons}
As a special example we consider the so called instanton-antiinstanton molecule. Each of the instantons has one zero mode (for fundamental fermions),
and there is only one amplitude of fermion exchange $T_{IA}$ , so the 
contribution of zero modes to the fermionic determinant 
is in this case simply 
\be det \Dslash =- \left| T_{IA} \right|^2 \ee
which corresponds to the fact that instantons exchange one quark and one antiquark (per flavor).

At nonzero holonomy the instanton is described by $N_c$ dyons. However simple generalization of the determinant construction to
dyons would be wrong, as dyons posess only a fraction of the topological charge. 
 In the  instanton-antiinstanton setting there is still only one left and one right-handed fermionic zero modes. Those are however located in ``lumps"
near each dyon, with some holonomy-dependent coefficients normalized as below
\be \psi_{L,R}(x)  = \sum c^{L,R}_i(\mu) \psi_i(x); \,\,\,\,\,\sum |c^{L,R}_i|^2 =1\ee
where for $L$ and $R$ the sum runs only over dyons of particular self-duality. It is then obvious that the fermionic determinant is
\be |det \Dslash| = \left| <R|\Dslash |L> \right|^2 = \left| \sum_{i,j} c^{*,R}_i c^L_j < i | \Dslash |j> \right|^2 \ee
where $< i | \Dslash |j>$ is an amplitude of hopping between a dyon and antidyon. As usual, if the dyons do not strongly overlap in space-time,
such hopping amplitude is described by some coupling constants $a_i$, obtained by standard ``cutting the tail" procedure for modes, and the free fermionic propagator
\be  < i | \Dslash |j> = a_i^* a_j S( x_i-x_j)  \ee
Note that at nonzero holonomy fermions are massive, thus at zero $T=1/\beta$ $S\sim exp(-m_f r_{ij})/r_{ij}^3$ and at nonzero $T$ it depends differently on temporal and spatial
distance between the dyons. We also remind that different color fermions have different masses, although in the eigenframe of the holonomy the
mass matrix (and thus $S$ ) is diagonal in color. (Of course, fermions moving in the background color field of the gauge solitons do not conserve color, but they do when they move
in ``empty" space in between them.)

Explicit expressions for $c_i,a_i$ can be found from expression for zero modes worked out in ref.

It is now straightforward to generalize this to configurations which have Q instantons and Q antiinstantons, still with total topological charge zero: there are
Q left and Q right handed zero modes, and the fermionic determinant in the partition function 
can be approximated by the determinant of the ``hopping" matrix $T_{ij}$.S

\section{Instanton-dyons on the lattice are seen via their fermionic zero modes }
When we discuss topology on the lattice we only once mentioned the instanton-dyons, in the section on constrained Cooling, with the Polyakov line preserved \cite{Langfeld:2010nm}:
while the total topological charge of the lattice was always integer, the clusters observed
had smaller topological charge (and the same actions, as they were selfdual or antiselfdual).

Using fermionic method allows to get better understanding of these objects, since changing
the periodicity phases one can see $all$ types of the dyons separately. While ``cooling"
still distort the configurations, hunting for lowest (or even zero) Dirac eigenvalues
allows one to get the dyons {\em as they are} in the gauge ensemble. One of the early studies of the kind was done by Gattringer \cite{Gattringer:2002wh} who 
used quarks with modified periodicity phases as a tool to locate all kinds of instanton-dyons. 
Further studies along these lines have been continued by Mueller-Presussker, Ilgenfritz and collaborators,  see e.g.
\cite{Bornyakov:2015xao,Bornyakov:2016ccq}. 

\begin{figure}[h!]
\begin{center}
\includegraphics[width=7cm]{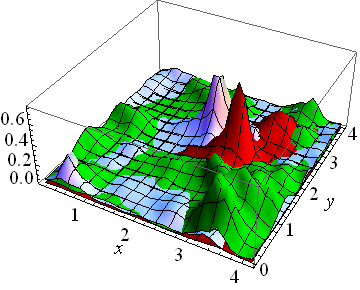}
\caption{Density $\rho (x,y)$ of the zero mode of conf. 2660 at $T= T_c$.  $\phi = \pi $(red), $\phi = \pi /3$(blue), $\phi = -\pi /3$(green). Peak height has been scaled to be similar to that of $\phi = \pi$.}
\label{2660_overlap}
\end{center}
\end{figure}

We however will jump to recent work by Larsen, Sharma and myself \cite{Larsen:2018crg}, which shows the underlying instanton-dyons in lattice  QCD with utmost clarity, due to application of the ``overlap" fermions possessing exact 
chiral symmetry and thus exact index theorems. Out of configurations at $T=1,1.08 T_c$ of QCD with realistic quark masses,
we selected those  which have $|Q_{top}| = 1$  thus with one exactly zero mode. By varying the periodicity phase, we can
identify location of all three type of dyons: see e.g. Fig.\ref{2660_overlap}
in which three dyons are see well separated. We also see configurations of strongly overlapping ones.

\begin{figure}[h!]
\begin{center}
\includegraphics[width=7cm]{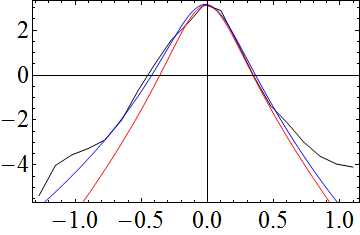}
\put(-170,100){$\log(\rho(x))$}
\put(-100,-5){$x$}
\caption{$\log(\rho(x))$ of the zero mode of conf. 2960 at $\phi = \pi $ (black) and the log of the analytic formula for Polyakov loop $P=0.4$ and $P=1$ though the maximum. $T=1.08T_c$. Red peak only has been scaled to fit in height, while blue peak uses the found normalization. The position of the other dyons are (blue) (0.13,0.1,0.0) and (0.1,-0.1,0.0) and (Red) (0.14,0.0,0.0) and (-0.14,0.0,0.0).}
\label{fig_2960pi_log}
\end{center}
\end{figure}

Not only we see that semiclassical formulae for zero modes well describe the lattics measurements when the dyons are far from each other, they also work well
in the case of partial or even complete overlap. In Fig.\ref{fig_2960pi_log} 
the profiles for single instanton is compared with that of overlapping dyons at the appropriate holonomy: the latter is
closer to the data. We have analyzed many cases, and in which the semiclassical expression
from Kraan and van Baal are in agreement with the data much better than expected.

\begin{figure}[h!]
\begin{center}
\includegraphics[width=7cm]{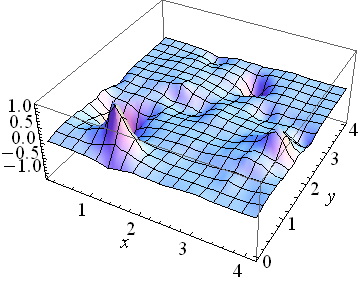}
\caption{Chiral density $\rho _5(x,y)$ of the first near-zero mode of conf. 2660 at $\phi = \pi $. $T=T_c$.}
\label{2660_pi_xy5sum_v1}
\end{center}
\end{figure}
Finally, the lowest non-zero modes show collectivized zero modes, see example in Fig.\ref{2660_pi_xy5sum_v1}. Furthermore, 
 near $T_c$ we can even see that one type of dyons show collectivized zero mode, and thus a nonzero chiral condensate, while in the same configuration the lowest nonzero modes of the other dyon type are still localized (with chiral symmetry unbroken).

\chapter{Instanton-dyon ensembles}

We will study 3 approaches to instanton-dyon ensembles: (i) the ``parametrically dilute" ones, (ii) then very dense,
in which case one can perhaps use the mean field methods, and (iii) statistical simulations which in principle
can work at any density. The last section of the chapter will be related with the so called ``flavor holonomies",
or complex chemical potentials, used as some diagnostic tool.

\section{Deformed QCD and  dilute ensembles with confinement}

\subsection{ Perturbative holonomy potential and deformed QCD}

In the previous chapter we had discussed holonomy field \be A_4=O(1/g)=const(x)\neq 0 \ee as some classical constant background.
At this level, it obviously does not lead to any energy, since the corresponding fields $G_{\mu\nu}=0$.

The next semiclassical approximation follows when one considers some quantum field $a_\mu=O(g^0) $
interacting with it. Since the holonomy is assumed to be diagonal in the color space, the commutator $[a_m, A_4]$ is non-zero for non-diagonal quantum gluons, which are ``higgsed" to become massive, like in Georgi-Gashow model.  
Quarks (fundamental fermions) interact with the holonomy via the $A_4 \bar q \gamma_4 q$ term in the 
Dirac Lagrangian, also getting a mass. Adjoint fermions, to be briefly discussed below, have color indices like those of the gauge fields\footnote{Similarly to gluons,   only the adjoint quark species corresponding to the non-diagonal color generators become massive. This is in contrast to the usual fundamental quarks, which are all affected by the Polyakov ``Higgsing".}  

All of this leads to certain positive free energy, first calculated in the (appendix of) the review by Gross, Pisarski and Yaffe
\cite{Gross:1980br}. In our current notation, for adjoint bosonic Majorana $N_a$ fermions  it has the form 
\be  V_{GPY}= (1\mp N_a )  {2 \pi^2T^4\over 3 } \sum_{i,j} (\mu_i -\mu_j)^2(\mu_i -\mu_j-1)^2 \label{GPY_adj} \ee
where minus sign stands for periodic and plus for antiperiodic boundary conditions.
In the SU(2) case it  simplifies to
\be V_{SU(2)}= (1\mp N_a ) {4 \pi^2 T^4\over 3 }\nu^2(1-\nu)^2\ee

In pure gauge theory, $N_a=0$, the potential is positive, and has minima at two symmetric points $\nu=0,\bar\nu=0$,
corresponding to the $trivial$ holonomy and its copy. In these limits one dyon is massless and another has
the action of the instanton.

Normal fermions are $anti-periodic$ on a circle\footnote{Recall that spinors get half rotation angle of the vectors,
so if the latter rotates by $2\pi$, spinor rotates by $\pi$.} , so the potential is the
 sum of two positive terms, as expected from the mass argument given above. 

However one may consider a theory in which the periodicity angle has any value one wishes\footnote{For example, it can be
interpreted as some external Abelian gauge field flux put through the circle in extra dimension.  
We will
return to this idea several time below, as it provides an excellent diagnostic tool to test our understanding of 
topological phenomena in gauge theories.}. 
Following \cite{Unsal:2007vu}, let us for now select  $bosonic$ spinor field with periodic boundary conditions.
This leads to the following observations: \\
(i) Standard finite-$T$ periodicity conditions are different for fermions and bosons, thus supersymmetry is violated
at $T\neq 0$. If however one forces  the same periodicity conditions, it is preserved.  
The $N_a=1$ case corresponds to $cal N$=1 supersymmetric theory, and 
correspondingly the potential vanishes as gluons and gluinoes contributions cancel each other.
 \\
(ii) at $N_a>1$ (no supersymmetry)
the sign is flipped and the minimum of the potential corresponds to $\nu=\bar\nu=1/2$ -- the confining value.
In the latter case -- one of the ``deformed QCD" versions -- there is confinement both at small and large
circle $\beta$, thus there is no deconfinement phase transition in this setting. 

Although we will not discuss them, let us mention another -- simpler -- versions of the ``deformed QCD",
also discussed in the literature. One may simply add to the QCD action some artificial potential,
depending on the Polyakov line, $V_{deform}(P)$, pushing the minimum, from the trivial $P=1,A_4=0$ 
point to the confining value. 

\subsection{The instanton-dyons in $N_a=1$ QCD= $\cal N$=1 SYM}
The pair gluon-gluino (adjoint Majorana fermion field) constitute the shortest 
supersymmetric multiplet, so the theory we are going to discuss in this section
is mostly known as $\cal N$=1 super-Yang-Mills. Before we turn to our main
subject --properties  of this theory compactified to $R^3\times S^1$  with a small circle
and periodic boundary conditions -- let me briefly desccribe what we know about this theory.
It is very much QCD-like, and if compactification
is ``thermal" (fermions are antiperiodic) it also has 
 deconfinement and chiral restoration phase transitions. According to lattice simulations
 \cite{Bergner:2015iva} those two transitions happen at about the same critical temperature $T_c$. 

The only difference with QCD is that the
number of zero modes for adjoint fermion is $2N_c$, so for $N_c=2$ 't Hooft effective vertex
 for an instanton has 4 zero modes. When chiral symmetry is broken, $<\lambda \lambda > \neq 0$,
 its $sign$ remains  undetermined: so unlike QCD this theory has remaining $Z_2$ symmetry, and two
equivalent vacua. As a result, this theory has domain walls, or kinks. 

From this point on we follow the work by Davis, Hollowood and Khose \cite{Davies:1999uw},  the first  
serious application of the
the instanton-dyons. The setting of the paper is supersymmetric Yang-Mills theory (SYM)  with a
single supersymmetry $\cal N$=1, defined on $R^3\times S^1$: it corresponds to $N_a=1$
and $periodic$ compactification, as described in the previous subsection.

The importance of this application is in the fact that it has resolved the so called {\em gluino condensate puzzle}.
Two methods to evaluate the value of the gluino condensate have two different answers, namely
\be < tr \lambda^2>_{WCI}= 16\pi^2 \Lambda_{PV}^3 \ee
 \be < tr \lambda^2>_{SCI}= 16\pi^2 \Lambda_{PV}^3 {2 \over [(N_c-1)! (3N_c-1)]^{1/N_c}} \ee
The abbreviations here stand for strong coupling instanton (SCI) and weak coupling instanton (WCI)
approaches. We will only review the former one.

Now back to the setting. Gluino $\lambda$ is the super-partner of a gluon, and $\cal N$=1 means that there is only
one type of the gluino. It is real adjoint field with spin 1/2, so there are two fermionic states. With 2 gluonic polarizations,
it completes the simplest SUSY multiplet. 
The selection of  periodicity condition $\alpha=0$ (periodic) for gluino preserves the supersymmetry,
which removes the GPY potential.

The circle $S^1$, if small in length $\beta \ll 1/\Lambda$, ensures weak coupling (like high-T). If the circle is large,
$\beta\rightarrow \infty$,
the theory is strongly coupled, like in the low-T QCD. The main difference between the two theories
is that -- unlike QCD -- in this setting for   the $\cal N$=1 SYM there are {\em neither deconfinement nor
chiral restoration phase transitions}, at any $\beta$!

Discussion in the previous subsection had prepared the reader to the conclusion that the holonomy can be
confining, at any $\beta$, since the main obstacle -- the GPY potential -- is in this case absent.
Now we need to understand why at small $\beta$ one may still have a broken chiral symmetry. 
Since it is a crucially important point, let us for the moment interrupt the discussion of instanton-dyons
and return to the discussion of the issue in historical order, starting with the instantons. 

Adjoint color  gluinoes, unlike the fundamentally charged quarks, have not one but $2N_c$ zero modes
per unit topological charge $Q$.
For the simplest gauge group we  discuss, $SU(2)$, it is 4 (instead of 2) fermionic sources in  the 't Hooft effective vertex 
 per topological charge. Therefore, unlike the $N_f=1$ QCD in which this vertex has the structure $\bar q q$, 
 in the $N_a=1$ theory it is instead  $\sim \lambda^4$, with 4 gluinoe lines. This is similar to 
$N_f=2$ QCD: but in this case we know that in this case the chiral symmetry is spontaneously broken 
only at sufficiently low $T<T_c$ (large $\beta$),   not at all $T$.

Here is an outline of the instanton-based calculation of the condensate.
The condensate has only 2 gluino fields: so in the
SCI calculation one did averaging of the $square$ of the condensate $< tr\lambda^2(x)  tr\lambda^2(y)>$
with a single instanton amplitude, and then argue that this function $f(x-y)$ cannot depend on the distance,
and thus is the same when $|x-y|\rightarrow \infty$ and one can apply the so called cluster decomposition
\be < tr\lambda^2(x)  tr\lambda^2(y)> \rightarrow < tr\lambda^2(x) ><  tr\lambda^2(y)> \ee
and get the SCI answer for the condensate mentioned above.

The alternative calculation \cite{Davies:1999uw,Hollowood:1999qn} was revolutionary in that they
had realized that in this setting
the semiclassical objects which needs to be used are not instantons but $instanton-dyons$. The reason for it
is that, even in weak coupling small circle setting, 
the holonomy is not  trivial  but the $confining$ one. For general $N_c$ $\mu_i$ 
have ``homogeneous distribution" on the circle
\be <A_4>=-({i \pi \over \beta }) diag \left( { N_c-1 \over N_c},{N_c-3 \over N_c},... -{ N_c-1 \over N_c}\right) \ee
For $N_c=2$ there are just two holonomy values, as usual. 

 With such holonomy setting and two colors, the fermionic zero modes are
 spread equally between $M$ and $L$ dyons. Since $M+L$=instanton, the total number of modes is still 4,
 thus it is 2 per dyon. 
 So, the situation is $not$ like in QCD with  $N_f=2$ and quartic vertex but rather like in the  $N_f=1$
 theory and the quadratic vertex! That is why there is no chiral restoration transition, and 
 the condensate $<tr \lambda^2>$  can be calculated directly from dilute gas of the dyons.  
 
 The result of the explicit calculation is\footnote{The original notations used in this paper are BPS monopole for M dyon and KK monopole for L dyon.}
 based on the dyon measure in the form
 \be dn_M=M_{PV}^3 e^{-S_M} d^3x ({g^2 S_M \over 2\pi})^{3/2} d\phi  ({g^2 S_M \over 2\pi v^2})^{1/2} {d\xi^2 \over 2g^2 S_M} \ee
 where 3-d $x$ is the center coordinate, $\phi\in[0,2\pi]$ is the 4-th collective coordinate corresponding to 
 dyon color rotation around the holonomy direction, and $\xi$ are Grassmanian fermionic coordinates
 for zero modes. The condensate is calculated in a standard diagram using zero modes
 \be <\lambda_\alpha(y) \lambda_\beta(y) >= \int dn \lambda_\alpha(y-x)  \lambda_\beta(y-x)   \ee
 and additional simplification at large distances 
 \be \lambda_\alpha\approx 8\pi S^\rho_\alpha (x)\xi_\rho, \,\,\,  S(x)={\gamma_\mu x^\mu \over 16\pi^2 |x|^2} \ee
 where S is the massless fermion propagator at zero Matsubara frequency (time integrated). 
 The result is 
 \be < tr \lambda^2>_{M}= < tr \lambda^2>_{L}=(16\pi^2) {M_{PV}^3 \over 2} exp(-{4\pi^2 \over g^2})=
 16\pi^2 \Lambda_{PV}^3 \ee
 and it agrees exactly with the WCI value but is different from the SCI one. 

The lesson: the vacuum of (bosonically compactified)  $\cal N$=1 SYM in weak coupling regime is a 
dilute gas of the $independent$ instanton-dyons, $not$ a gas of instantons.

\subsection{ QCD(adj)  with $N_a>1$ at very small circle: dilute molecular (or ``bion") ensembles} \label{QCDadj_2}

Study of the  QCD (adj) $N_a>1$ compactified to parametrically small circle 
 was due to Unsal 
\cite{Unsal:2007jx}.
The setting is the same as in the previous subsection, namely the periodicity is ``bosonic",
reversing the sign of the quark contribution to the GPY vacuum energy. As we had discussed previously,
for $N_a=1$ this cancels the GPY potential, but for $N_a>1$ the 
sign of the GPY potential is reversed, and its minimum corresponds to the confining holonomy.

The first point to focus on is the distinction between the $R^3$ space in the case of Polyakov's confinement and 
the $R^3\times S^1$ setting, with a small circle, we consider now. Both setting have time-independent monopoles
we call the $M$-type, but in the latter case there exist also the ``time-twisted", KK or $L$-type monopole as well.

The next question is what happens with chiral symmetries in such setting, when $N_a>1$?  
 In general the effective 't Hooft Lagrangian per dyon is $\sim \lambda^{2N_a}$.  For example, for  $N_a=2$ it is the
 4-fermion vertex similar to that of the NJL model. Since the effective NJL coupling -- the density of the dyons -- is exponentially small at weak coupling  setting (small circle), there is no spontaneously broken chiral symmetries.
 The density of the individual dyons are thus zero!

Note that this is the same phenomenon which we discussed in chapter on instantons in the usual QCD with massless quarks. 
  Since instantons  become effective vertices of the type $\sim \lambda^{2N_f}$, for $N_f>1$ 
  and since there is no chiral symmetry at high $T$, in QGP, the density of the individual instantons vanishes. 
  There remain however clusters with the topological charge zero, in particularly the instanton-antiinstanton
  ``molecules". The ensemble in this case
is a ``moleculecular gas", made of soliton-antisoliton pairs bound by quark exchanges \cite{Ilgenfritz:1988dh}.

In the confining setting on $R^3\times S^1$ for QCD(adj) 
there are dyon-antidyon pairs bound by fermion exchanges: Unsal \cite{Unsal:2007jx} call
these binary objects ``bions". Introducing deviation from confining holonomy as $\phi$ and magnetic holonomy\footnote{Note that in the Polyakov confinement subsection it was called $\chi$.} $\sigma$, one can write down amplitudes for all 4 types of dyons (of the SU(2) gauge group) 
$$ M=BPS\sim e^{-\phi +i\sigma-S_0/2} (\psi \psi)^{N_a} ; \,\,\,\,  \bar{M}=\bar{BPS}\sim e^{-\phi -i\sigma-S_0/2} (\bar \psi  \bar \psi )^{N_a}$$
$$ L=KK\sim e^{+\phi -i\sigma-S_0/2} (\psi \psi )^{N_a} ; \,\,\,\,  \bar{L}=\bar{KK}\sim e^{+\phi +i\sigma-S_0/2} (\bar \psi \bar \psi)^{N_a}  $$
where we use both Unsal' and our notations for the instanton-monopoles (instanton-dyons).
The instanton is LM pair, so in the combined amplitude -- the $LM$ product of individual amplitudes --
 all prefactors cancel out except
the instanton action $e^{-S_0}$, with  $2N_a N_c$- fermion operator.

Let us now, still following  \cite{Unsal:2007jx}, form all possible dyon-antidyon\footnote{Only for zero topological charge
objects one can correctly couple zero modes to each other, so that there remains no zero modes left.
}  pairs. 
The fermions are saturated between the pairs, for any $N_a$, so we do not write them anymore
(although they of course lead to extra factors in actual expression of the ``molec
$$  \bar{M} M =\bar{BPS}BPS\sim e^{-S_0 -2\phi}, \,\,\,\,\,\, \bar{L} L=\bar{KK}KK\sim e^{-S_0 +2\phi},$$
$$  \bar{L} M =\bar{KK}BPS\sim e^{-S_0 -2i\sigma}, \,\,\,\,\,\, \bar{M} L=\bar{BPS}KK\sim e^{-S_0 -2i\sigma}$$
The main idea of Unsal was to focus on the second raw,  the bions which are  $twice$ magnetically charged.
Since those have nonzero (but still exponentially small)  density and nonzero magnetic charge, 
they will screen the magnetic charge precisely as a single monopole does in the Polyakov confinement. 
So, these magnetically charged bions do enforce the confinement.

The next important point from this work is that the effective action due to all 4 types of the bions should lead,  if the $N_a=1$ case,  to
the effective action of the type
\be L_{eff} = ({1 \over 2}) \partial \phi^2+  ({1 \over 2}) \partial \sigma^2 + a e^{-S_0} (cos(2\sigma)-cosh(2\phi)) \ee
based on supersymmetry arguments. The first comment: the minimum of the potential requires $\sigma=\phi=0$, or it is the confining one. 
The second-- surprising-- observation follows from the expansion of this Lagrangian to $O(\sigma^2,\phi^2)$ terms:
 both holonomies, $\sigma, \phi$, should have in this theory {\em the same} 
screening masses! (As we will see below, in non-supersymmetric theories those masses are
always different, and have very different $T$-dependence.)
The question is, how can it be understood microscopically? Earlier in this chapter we 
discussed long distance classical binary interactions, and concluded that $\bar{M} M, \bar{L}L$ channels are
attractive while $\bar{M} L, \bar{L}M$ are repulsive. In the latter case the repulsion can be overcome by fermion exchanges, so the integral over the inter-dyon distance is converging both at small and at large distances.
However in the former case $both$ bosonic and fermionic interactions are attractive, the integral is 
thus converging only due to the ``core".  Why both integrals are the same? I cannot answer, it is one of the
miracles induced by supersymmetry.

%
%
 
\subsection{QCD(adj) with $N_a=2$ and periodic compactification on the lattice} 
In this subsection we continue to discuss the same setting, a theory on $R^3\times S^1$ with periodic compactification,
but add one more adjoint (Majorana) gluino. Of course, this theory is not supersymmetric.
For its general discussion see \cite{Myers:2007vc}.

However there were lattice studies of this theory \cite{Cossu:2009sq}, with variable circumference of the circle
$\beta$, called in this paper $L_c$. 

The lattice simulations had found {\em four distinct phases}, which we subsequently briefly describe:\\
(i) At large $L_c$ (low ``temperature") one finds the usual confining phase, with the 
Polyakov VEV $<P>=0$ and symmetric distribution of its eigenvalue, consistent with unbroken center symmetry.\\
(ii) As the $L_c$ gets shorter, one observes the deconfinement transition, in which $<P>\neq 0$ and its
eigenvalues distributed along one of the center elements, breaking the center symmetry.\\
(iii) As the $L_c$ gets even shorter, in some finite interval of  $L_c$ there exists $another$ deconfined phase,
in which eigenvalues distributed along a direction $opposite$ (making angle $\pi$)to that in the usual one (ii).
This phase was predicted by  \cite{Myers:2007vc}.
 \\
(iv) At very small $L_c$ (high ``temperature") the theory returns to center-symmetric confined phase.
This is consistent with the arguments made above, based on the GPY potential (\ref{GPY_adj}).
the authors called it ``re-confined" phase. 

\begin{figure}[h]
\begin{center}
\includegraphics[width=8cm]{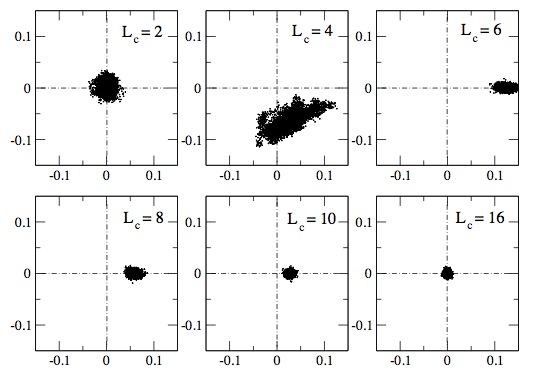}
\includegraphics[width=8cm]{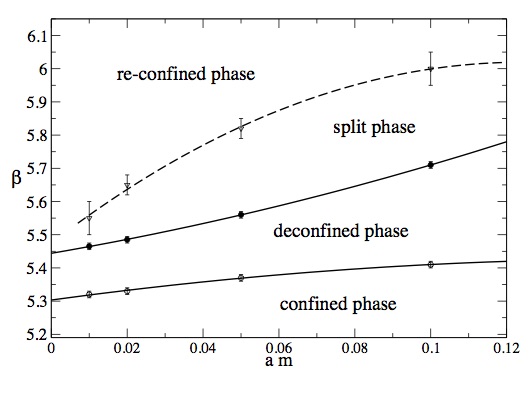}
\caption{(upper) Scatter plot of the Polyakov line eignevalues, as a function of the size
of the compactified dimension $L_c$: 4 subsequent phases are seen.
(lower)  The phase diagram on the plot quark mass $a m$ and lattice gauge coupling $\beta=6/g^2$.
Extrapolation of the phase boundaries to the chiral limit for split phase is not obvious.}
\label{fig_Na2}
\end{center}
\end{figure}

Fig.\ref{fig_Na2}(upper) from  \cite{Cossu:2009sq} shows Polyakov line eigenvalues as a function of the
circle circumference $L_c$. 
The most important consequence of this study is that it contradicts to  the conjectured "volume independence":
contrary to naive interpolation, 
two confined phases are $not$ smoothly connected. Since it is a lattice work, 
with a nonzero gluino masses and finite lattice spacings, one may ask
 to what extent their zero limits have been
reliably reached. Fig.\ref{fig_Na2}(lower)  shows, that in the $m\rightarrow 0$ limit the phase (iii) $may$
disappear, as such option is within the numerical accuracy, but the other deconfined phase (ii) seems definitely be there, also in the chiral limit. 

What about the fate of the chiral symmetry breaking, through all these transitions? Remarkably (but in agreement with other studies of the QCD(adj) with
thermal compactification) it was  observed in  \cite{Cossu:2009sq} that $<\bar \psi \psi>\neq 0$ in the whole region
of $L_c$ studied. So, massless fermions keep their nonzero ``constituent quark mass" throughout:
the two (or maybe one) deconfined phases are then a plasma of ``constituent quarks". 

The magnitude of the condensate howeever decreases, by about an order of magnitude from ``vacuum" (large $L_c$)
value, suggesting that chiral symmetry is going to be eventually restored, just at very small  $L_c$ not included
in these particular set of  simulations   \cite{Cossu:2009sq}. Eventually, the $N_a=2$ theory with small circle
should of course have zero gluino condesate, as was argued in the previous section.

\section{Dense dyon plasma in the mean field approximation} \label{sec_meanfield} 


The main idea of the mean field approximation is that a particle interact simultaneously with many\footnote{
Note that this condition is necessary but in general not sufficient. For example, the approximation is valid 
 in the perturbative Debye theory of plasma, e.g. for the monopole plasma discussed by Polyakov. It is not valid in strongly coupled plasmas, which may be strongly correlated liquids or even solids, producing mean fields
 very different from being space-independent.}. One may think that it can be used
in cases when the ensemble of the dyons is dense enough, producing 
strong  screening, which effectively reduces the
pair-wise correlations.  Anyway, using some average
 mean field is the simplest approach, in which one can get analytical evaluation of the observables.

In this section we will follow a series of papers by Liu, Zahed and myself using the mean field approximation.
The first paper of the series,  \cite{Liu:2015ufa},
had established the approximation in the technical sense.  Here there is no  place to
present technical details of these works, and we just summarize few important points and the results.

The main result is that
dense enough dyon ensemble does overcome the GPY potential and shift the
minimum of the free energy to the confining value, $\nu=1/2$ for the SU(2)
gauge theory considered.  The key expression for the partition function is
put into the form
\bea
{\rm ln} Z/V_3=-{ V_{ideal}}-\frac{1}{2}\int \frac{d^3p}{(2\pi)^3}
{\rm ln}\left|1-\frac {V^2(p)}{16}\frac{p^8 M_D^4}{(p^2+M_D^2)^4}\right|\nonumber\\
\label{1loop}
\eea
where ${V_{ideal}}$ describes free energy of the non-interaction dyons,  $V(p)$ is the Fourier transform of the
dyon-antidyon potential (classical or including one loop). The very presence of dyons, with electric charges, generates a Debye electric
  screening mass   $M_D=\sqrt{2n_D/T}$ related to the dyon density $n_D$. 
When this density is large, 
 $M$ is large. From expression of the partition function
 it follows that the effect of the potential (inside the logarithm) gets reduced (screened out),
which in principle justify the mean field method. Specific applicability limits of it
can be derived from requirement that the second term inside the log is $less$ than one\footnote{
Otherwise the argument is negative and logarithm gives an imaginary part, signaling appearance of an instability.}.

\begin{figure}[h]
\begin{center}
\includegraphics[width=8cm]{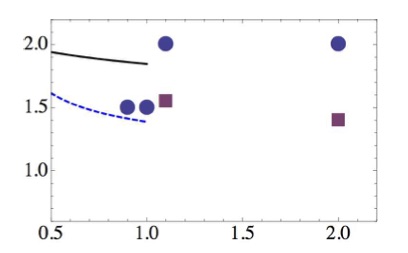}
\caption{The electric $M_E/T$ (dashed line) and magnetic $M_M /T$ (solid line) screening masses versus $T/T_c$. The points are the  lattice data for  $SU(2)$ gauge theory shown for comparison, (blue) circles are electric, (red) squares are magnetic.}
\label{fig_M_D_meanfield}
\end{center}
\end{figure}

Fig.\ref{fig_M_D_meanfield} illustrates one of the results of this work, the temperature dependence of the
electric and magnetic screening masses, in comparison to what has been derived from numerical simulations of the 
 $SU(2)$ gauge theory. Note that the electric screening mass\footnote{
 For clarity: while the calculation includes only dyons but not gluons, it does include the one-loop GPY potential.
 Its derivative over the holonomy value is the perturbative one-loop electric Debye mass, due to gluons. 
 } -- shown by the closed circles -- has a drop downward,
 as $T$ is reduced below the deconfinement transition, while the magnetic screening is expected to get larger
 than the electric one there.

\begin{figure}[htbp]
\begin{center}
\includegraphics[width=8cm]{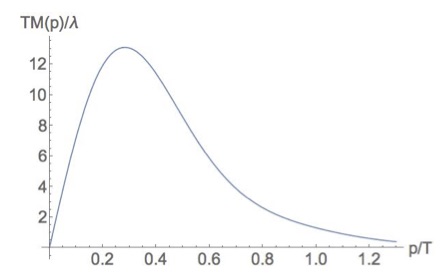}
\caption{The momentum dependent constituent quark mass $T M(p) /\lambda$ versus momentum in units of temperature $p/T$.}
\label{fig_M_of_p}
\end{center}
\end{figure}

The next work of the series 
\cite{Liu:2015jsa}
 applies MFA to the $N_c=2$ color theory with $N_f=2$ light quark flavors.
  At high density the minimum of the free energy still corresponds to the confining ensemble with $\nu=1/2$. 
The gap equation for the effective quark mass (proportional the quark condensate)  of \cite{Liu:2015jsa} the usual form
\be \int {d^3 p \over (2\pi)^3} {M_{eff}^2(p) \over p^2+M_{eff}^2(p)}= n_L
\ee
where the r.h.s. is the density of the dyon type possessing the fermion zero mode, namely the $L$-dyons.  The equation is actually for the parameter $\lambda$ in the effective mass 
$ M_{eff}(p)=\lambda p T(p)$, in which $T(p)$ being the Fourier transform of the  ``hopping matrix element"
calculated using the fermionic zero mode. 
Momentum dependence of $M(p)/\lambda$ is universal and is shown in Fig.\ref{fig_M_of_p}. 
In practice, the best way to
solve the gap equation is to calculate the momentum integral in its l.h.s. numerically, and then parameterize the dependence on parameter $\lambda$. 

A generalization of the mean field treatment to arbitrary number of colors and flavors  in \cite{Liu:2015jsa} shows that 
this gap equation has nonzero solutions for the quark condensate only  if 
\be N_f<2N_c \ee
So,  the critical number of flavors is $N_f=6$ for $N_c=3$.
The lattice simulation indeed show weakening of chiral symmetry violation effects with increasing $N_f$,
but specific results about 
on the end of chiral symmetry breaking are so far rather incomplete: for $N_c=3$ we know that
in the $N_f=4$ case the chiral symmetry is broken, the case $N_f=8$ is not yet completely decided and  $N_f=12$
seems to be already in the conformal window.

Another important generalization -- for quarks in the adjoint representation - is made in a separate paper \cite{Liu:2016mrk}.
The number of fermionic zero modes  increases, and they are more complicated.
In the symmetric dense phase both $M$ and $L$ dyons have two zero modes. 
But the actual difficulty is not some longer expressions but the fact that one of them has rather 
singular behavior -- gets delocalized -- exactly at the confining value of the holonomy, $\nu=1/2$.
Therefore, in the case of adjoint quarks the approach toward the confining phase  needs some special care.
In the case $N_c=2,N_a=1$ the deconfinement and chiral restoration happen at about the same temperature,
in agreement with lattice result we discussed above for this theory.

\section{Statistical simulations of the instanton-dyon ensembles} \label{sec_simulations}

\subsection{Holonomy potential and deconfinement in pure gauge theory}
The first direct simulation of the instanton-dyon ensemble with dynamical fermions
has been made by  \cite{Faccioli:2013ja}. The general setting
follows the example of the ``instanton liquid", it included the determinant'of the so called "hopping matrix", a part of the Dirac operator in the quasizero-mode
sector. It has been done for $SU(2)$ color group and the number of fermions
flavors $N_f=1,2,4$. Except in the last case, chiral symmetry breaking
has been clearly observed, for dense enough dyon ensemble.

The second one, by Larsen and myself  \cite{Larsen:2015vaa} uses direct numerical simulation of the instanton-dyon ensemble, both in the high-T dilute and low-T dense regime. The holonomy potential as a function
of all parameters of the model is determined and minimized.
 
\begin{figure}[t!]
  \begin{center}
  \includegraphics[width=8cm]{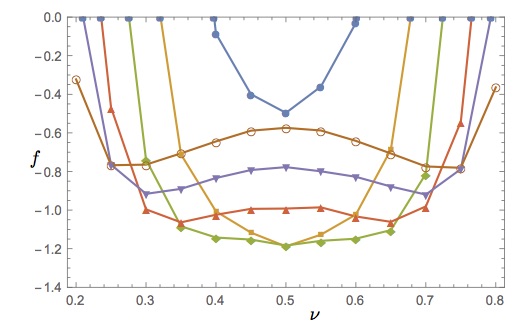}
    \includegraphics[width=8cm]{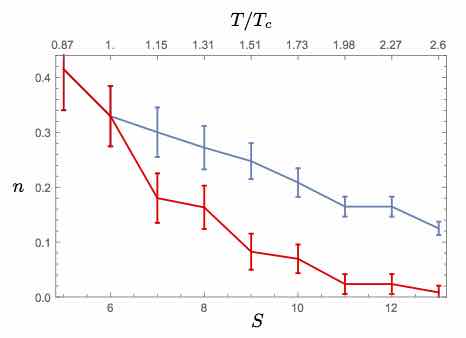}
%
   \caption{ (left) Free Energy density $f$ as a function of holonomy $\nu$ at $S=6$, $M_D=2$ and $N_M=N_L=16$. The different curves corresponds to different densities. $\bullet$ $n=0.53$, $\blacksquare$ $n=0.37$, $\blacklozenge$ $n=0.27$, $\blacktriangle$ $n=0.20$, $\blacktriangledown$ $n=0.15$, $\circ$ $n=0.12$.
   (right) Density $n$ (of an individual kind of dyons) as a function of action $S$ (lower scale) which is related to $T/T_c$ (upper scale) for M dyons(higher line) and L dyons (lower line). }
  \label{fig_potential}
  \end{center}
\end{figure}

 In Fig.\ref{fig_potential}(left) from this work we show the dependence of the 
total free energy on holonomy value, for different ensemble densities.
This important plot\footnote{Later analysis improved statistical accuracy
of the data points: we nevertheless show here the first plot in which the
confinement had came out of the simulations.
} had shown, for the first time in direct simulations,
that
at high density of the dyons their back reaction does generate confinement! 
Indeed, the minimum of the holonomy potentials shifts  to $\nu=1/2$,
 the confining value for SU(2) ($cos(\pi \nu)=0$).
 The self-consistent parameters of the
ensemble, minimizing the free energy, is  determined for each density. 

As the action parameter $S$ is growing, corresponding to growing temperature, the dyon-symmetric
phase goes into an asymmetric phase, in which the density of $M$ and $L$ dyons are not the same, see 
Fig.\ref{fig_potential}(right). The $l$ dyon has larger action due to time-dependent ``twist", and thus smaller
density.

\begin{figure}[h]
\begin{center}
\includegraphics[width=6cm]{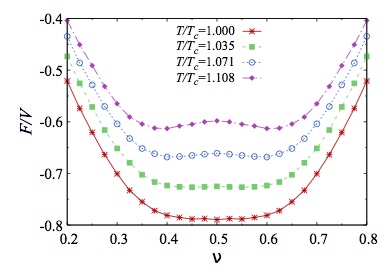}
\includegraphics[width=6cm]{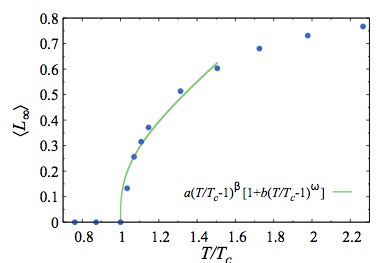}
\caption{(left) Free energy density of the dyon ensemble near Tc (S = 7); (right) Fit of the  average value of the Polyakov line to the expected 2-nd order critical point.}
\label{fig_P_near_Tc}
\end{center}
\end{figure}

The next simulations of the instanton-dyon ensemble for $SU(2)$ gauge group
has been done by Lopez-Ruiz, Y.~Jiang and J.~Liao \cite{Lopez-Ruiz:2016bjl}. 
In Fig.\ref{fig_P_near_Tc} from this work we show the shapes of the holonomy potential 
$V(\nu)$ near the critical point, and the fit to the average value of the Polyakov line,
fitted to the expected second-order behavior with indices of the 3D Ising model 
\be \beta \approx 0.3265, \,\,\,\,  \omega \approx 0.84 
\ee 
Within the statistical accuracy of the calculation, the expected second-order behavior is indeed observed.

\begin{figure}[h]
\begin{center}
\includegraphics[width=6cm]{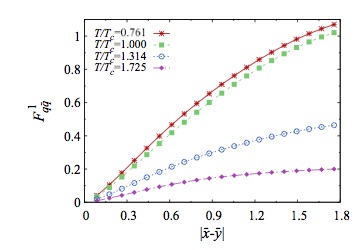}
\includegraphics[width=6cm]{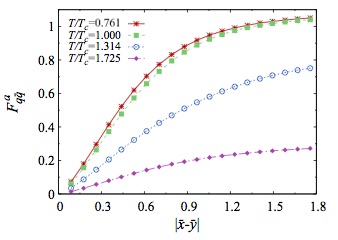}
\caption{The free singlet energy potential, for static charges in fundamental (left) and adjoint (right) color representations.}
\label{fig_F_of_r}
\end{center}
\end{figure}

The free energy potential between static quark and antiquark from  \cite{Lopez-Ruiz:2016bjl}
is shown in Fig.\ref{fig_F_of_r}. Note first, that the potential is nearly temperature
independent below $T_c$ (the first two sets), but rapidly decreases above $T_c$. 
While the fundamental quarks show nearly linear confining potential, the adjoint one shows
screening aboove certain distance. This is as expected for pure gauge theory,
without quarks\footnote{Note that the simulation includes dyons but not gluons, 
which however are also integer charged and can screen the adjoint charge. }. 

Another important issue addressed in  \cite{Lopez-Ruiz:2016bjl} is the area law for the
spatial Wilson loop. The corresponding measurements are reported, the existence of the spatial
tension is demonstrated, but its temperature dependence is not yet compared to
the avilable lattice data.

\subsection{Instanton-dyon ensemble and chiral symmetry breaking}

 The issue of chiral symmetry breaking using numerical simulations were addressed by \cite{Larsen:2015tso}.  Including the fermionic determinant 
in ``hopping" approximation we calculated the spectrum of the lowest Dirac eigenvalues.

Extracting the quark condensate is complicated, as usual, by finite-size effects.
Using two sizes of the system, with 64 and 128 dyons, we 
identify the finite-size effects in the eigenvalue distribution, and 
extrapolate to infinite size system.
The location of the chiral transition temperature is 
defined both by extrapolation of the quark condensate, from below, and
the so called ``gaps" in the Dirac spectra, from above.

\begin{figure}[h]
\centering
\includegraphics[width=7cm]{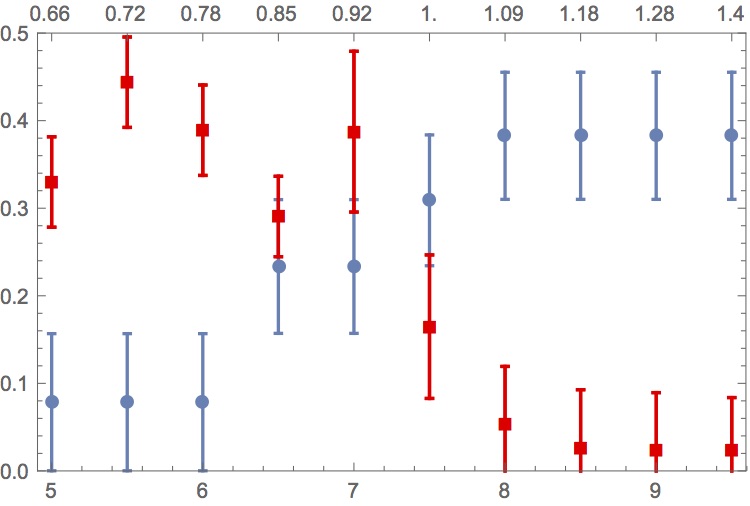}
\put(-100,-5){$S$}
\put(-100,150){$T/T_c$}
\put(-230,100){$\Sigma$}
\put(-230,20){$P$}
\caption{(Color online) The Polyakov loop $P$ (blue circles) and the chiral condensate $\Sigma$ (red squares) as a function of action $S=8\pi^2/g^2$ or temperature $T/T_c$. $\Sigma$ is scaled by 0.2.}
\label{fig_cond}
\end{figure}
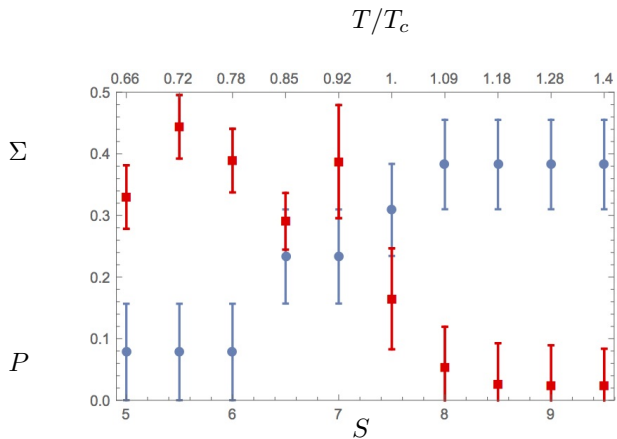

We do indeed observe, that for SU(2) gauge theory with 2 flavors of light fundamental quarks both the confinement-deconfinement transition
and chiral symmetry breaking,
 as the density of dyons goes up at lower temperature, see Fig.\ref{fig_cond}. 

 Determination of the transition point by vanishing of $<P>$ or $<\bar \psi \psi>$ 
 is difficult for technical reasons. 
  Since  both transitions appear to be in this case just a smooth crossover ones, it is by now well
 established procedure to define the transition points via maxima of corresponding susceptibilities.
  Those should correspond to {\em inflection points} (change of curvature) on the plots
  to be shown. Looking from this perspective at Fig. \ref{fig_cond},
  one would locate the  inflection points of both curves, for  $<P>$ or $<\bar \psi \psi>$ , at
  the same location, namely $S=7-7.5$. Thus, within the accuracy our simulations have,
  we conclude that both phase transitions happen at the same conditions.

\section{QCD with flavor-dependent periodicity phases}
For applications, such as heavy ion collisions, one needs to know properties of the QCD matter not only as a function
of the temperature $T$ but quark (baryon number, isospin etc) densities as well. Unfortunately,
Euclidean partition function at nonzero quark chemical potentials $\mu_f$ contains complex factor $e^{i \mu_i/T}$
which cannot be interpreted as probability: so standard Monte-Carlo simulation algorithms cannot be used. 

One can however introduced {\em imaginary chemical potentials}, proceed with calculations, and then 
extrapolate, in the $\mu^2$ plot, across zero. This is done by several lattice groups, but we will not
discuss those results here.

The reason is we are not really focus on the physical problems with chemical potential here, but 
on the use of the imaginary chemical potentials, or ``flavor holonomies" as they are also called, as some diagnostic tool. 
Observing how QCD ensembles react on their magnitude, on the lattice and in the models, we
hope to gain better understanding of the underlying mechanisms of the QCD phase transitions. 

\subsection{Imaginary chemical potentials and Roberge-Weiss transitions} \label{sec_RW}

The phase diagram of QCD-like theories with imaginary quark chemical potential has been discussed in
the fundamental paper by 
Roberge and Weiss \cite{Roberge:1986mm}. 
At imaginary chemical potential denoted by $\theta = i\mu/T$ appears in the periodicity condition over the
Matsubara circle
\be \psi(x,\beta)= e^{i\theta} \psi(x,0) \ee

As we already mentioned, for real $\theta$ 
the sign problem is absent and standard Monte Carlo algorithms can be applied to simulate lattice QCD.
QCD possesses rich phase structure at nonzero  $\theta$, details of which depend on the number of flavors $N_f$ and the quark mass (masses) $m_f$. Later in the section we will make it even reacher, by considering 
different  phases for different quark flavors $\theta_f$. 

Since $\theta= i \mu/T$  is an angle, it is obvious that the QCD partition function Z is a periodic function of it, with 
the period $2\pi$. However, as noted in \cite{Roberge:1986mm}, the period is actually smaller $2\pi / N_c$
and there is the so  called the Roberge-Weiss symmetry
\be Z(T,\theta)=Z(T,\theta+{2\pi \over N_c}) \ee
because of $N_c$ branches of the gluonic GPY potential. 

The main point is that the imaginary chemical potential $\theta$ simply shifts \be 2\pi\mu_j \rightarrow  2\pi\mu_j +\theta \ee in the quark term (\ref{eqn_GPY_q})
of the GPY one-loop effective potential.  In pure gauge theory at sufficiently large $T$
there is spontaneous breaking of the center symmetry and one of the $N_c$ branches is selected. For example,
$SU(3)$ theory has one real and two complex conjugated branches. Recall that in QCD the quark term is $not$
center symmetric, thus the free energy is tilted and the real branch is the preferred one: the mean $<P>$
 slowly moves as a function of $T$ along the real axes, as described above.

\begin{figure}[h!]
\begin{center}
\includegraphics[width=10.cm,angle=-1.]{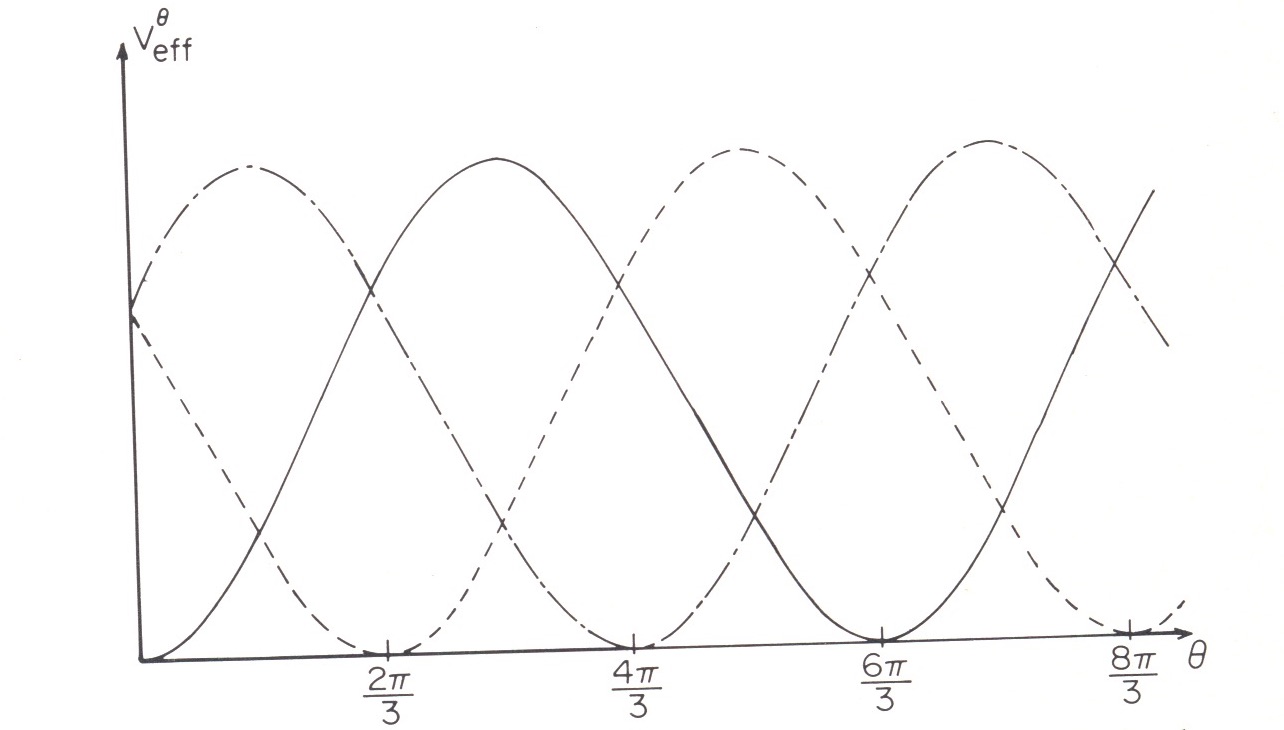}
\caption{Effective GPY potential as a function of the imaginary chemical potential $\theta$,
for 3 colors $N_c=3$}
\label{fig_RW_pot}
\end{center}
\end{figure}

If there is a nonzero $\theta$, it effectively rotates the quark part of the potential, and 
at certain values,  $\theta_k = (2k + 1)\pi/3$ in this theory, the free energies of different brunches cross. As a result,
there appear kinks -- the 1st order phase transitions -- at such values. These points are crossings of the
$N_c$ branches of the effective potential, as shown in Fig.\ref{fig_RW_pot}: at any $\theta$ the physical branch
is the lowest one.

Since these arguments are derived using the one-loop GPY potential, they  of course valid only at high $T$, or weak coupling. 
 In reality, the RW transition exists at $T > T_{RW}$, where $T_{RW}$ are  the critical endpoints 
of the Roberge-Weiss 1-st order transitions.


\begin{figure}[h!]
\begin{center}
\includegraphics[width=10.cm]{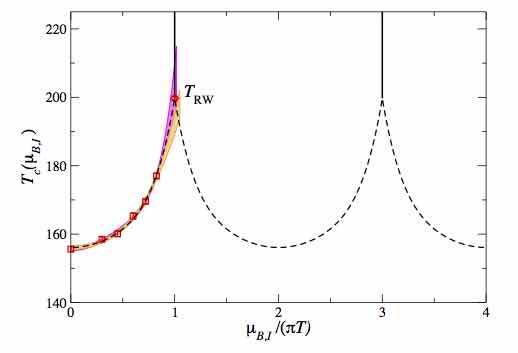}
\caption{Phase diagram of QCD in the presence of an imaginary baryon chemical potential obtained from numerical simulations on Nt = 8 lattices alone. Bands denote fits to polynomials in $\mu^2_B$: the orange (longer) band is obtained using terms up to order $\mu^4_B$ , the violet (shorter) one using up to $\mu^6_B$ terms.}
\label{fig_RW_diagram}
\end{center}
\end{figure}

At sufficiently low $T$ there is the confinement phase, with $<P>=0$, and the cusp disappears. First order transitions must end at some critical points. 
Where exactly it happens can be calculated on the lattice.
Recent lattice investigation \cite{Bonati:2016pwz}
has located it at
\be T_{RW}=1.34 (7) T_c=  208(5)\,\,\, MeV \ee
Does the pseudocritical line really get to the RW endpoint, as suggested by early studies on the subject?
Or two pseudocritical lines meat each other, and then go vertically to the critical point?
Fig.\ref{fig_RW_diagram} from \cite{Bonati:2016pwz}
illustrates current state of lattice answer to this question. Inside the accuracy of the
current data, the answer to this question seems to select the former option to this question.

\subsection{ $Z(N_c)$ QCD}
  The main idea of this deformation of $N_c=N_f$ QCD \cite{Kouno:2013zr}
  is a
   ``democratic" distribution of flavor holonomies, putting those in between 
   the subsequent holonomy phases $\mu_i$, see Fig.\ref{fig_circles} (right). 
   The framework in which it has been suggested is the PNJL model.
   
   In the theory of the instanton-dyons  $Z(N_c)$ QCD has a very simple meaning: 
   {\em  each dyon type get a zero mode of one quark flavor}.

The $Z(N_c)$ QCD has been studied  in the mean field framework \cite{Liu:2016yij},  
by statistical simulations \cite{Larsen:2016fvs} and also by lattice simulations \cite{Misumi:2015hfa}.
In the dilute limit it also has been studied by \cite{Cherman:2016hcd}.

The first two papers consider the $N_c=N_f=2$ version of the theory, while the last one focus on the
$N_c=N_f=3$. In the former case the set of phases are $\psi_f=0,\pi$, so one quark is a boson and one is a fermion. In the latter  $\psi_f=\pi/3,\pi,-\pi/3$.

\begin{figure}[h!]
\centering
\includegraphics[width=7.cm]{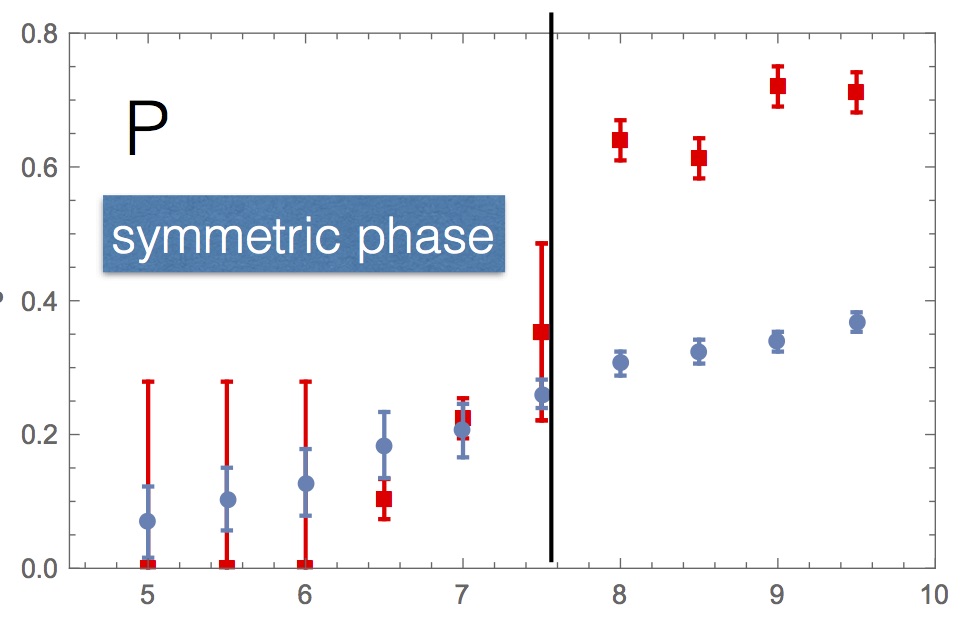}
\includegraphics[width=7.cm]{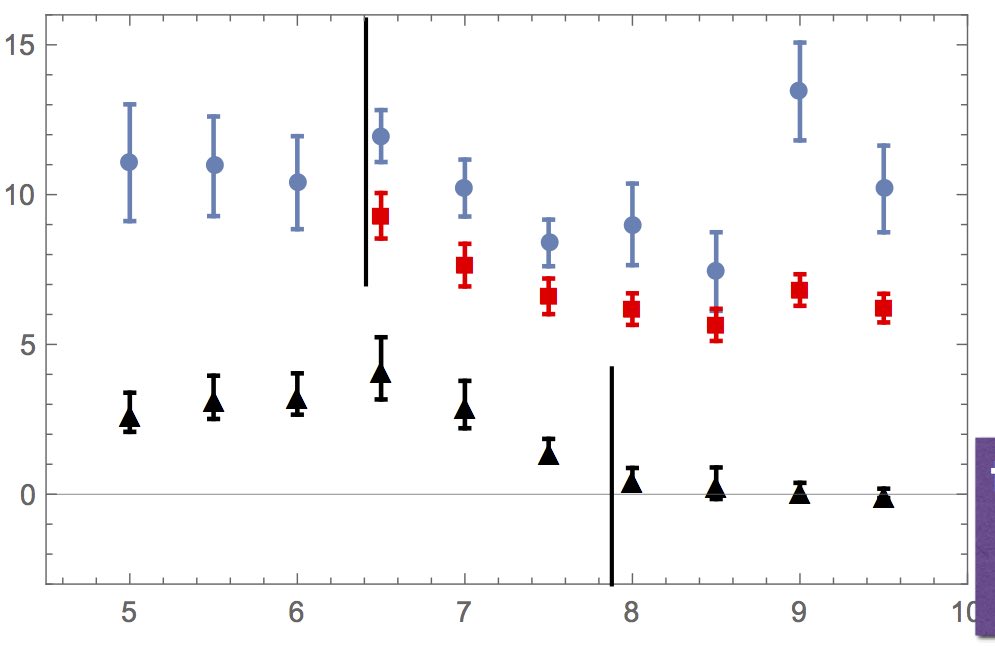}
\caption{(left) The mean Polyakov line P versus the density parameter $S$. Red squares
are for $Z_2 QCD$ while blue circles are for the usual QCD, both with $N_c=N_f=2$ .
(right) The quark condensate versus the density parameter $S$.
Black triangles correspond to the usual QCD: and they display chiral symmetry restoration.
Blue and red poins are for two flavor condensates of the $Z_2 QCD$: to the left of vertical
line there is a ``symmetric phase" in which both types of dyons and condensates are the same.
}
\label{fig_Z2}
\end{figure}

All  these works find deconfinement transition to strengthen significantly, compared to QCD 
with the same $N_c,N_f$ in which it is a very snooth crossover. While in \cite{Liu:2016yij}
the $<P>$ reaches zero smoothly, a la second order transition, the simulations \cite{Larsen:2016fvs}
and lattice \cite{Misumi:2015hfa} both see clear jump in its value indicated strong first order transition.
The red squares at Fig.\ref{fig_Z2}(left)  from \cite{Larsen:2016fvs} are comparing the behavior of the mean Polyakov line
in  $Z_2$ and ordinary QCD. The parameter $S$ used as measure of the dyon density
is the ``instanton action", related with the temperature by
\be S=({11N_c \over 3}- {2N_f \over 3})log({T \over \Lambda})
\ee
The dyons share it as $S_M=\nu S, S_L=\bar \nu S$.
So, larger $S$ at the r.h.s. of the figure correpond to high $T$ and thus to more dilute ensemble,
since densities contain $exp(-S_i)$.

All three studies see a non-zero chiral condensates in the studied region of densities:
perhaps no chiral restoration happens at all. The value fo the condensate are shown in 
 Fig.\ref{fig_Z2}(right)  from \cite{Larsen:2016fvs}.

 The simulation   \cite{Liu:2016yij} demonstrate that
the spectrum of the Dirac eigenvalues has a very specific ``triangular" shape, characteristic
of a single-flavor QCD. This explains why the $Z(N_c)$ QCD has much larger condensate
than ordinary QCD, at the same dyon density, and also why there is no tenedecy to restoration.
As expected, all works see different  condensates, $<\bar u u>\neq <\bar d d>$, but with
difference smaller than one could expect from the difference in the dyon density.

In \cite{Cherman:2016hcd} the authors study the periodic compactification to a small circle,
with $Z(N_c)$ QCD flavor holonomies. The main statement is that even in the 
limit of very small circle -- exponentially small dyon density -- in this setting the chiral symmetry remains spontaneously broken, but 
in a very specific way. There are however only $N_c-1$ massless pions, not $N_f^2-1$ as usual,
equal to the number of Cartan subalgebra generators.

\subsection{Roberge-Weiss transitions and instanton-dyons}
In subsection \ref{sec_RW} we introduced the Roberge-Weiss transitions and discussed
some lattice studies of them. 
Let us remind the reader, that imaginary chemical potential $\theta=i\mu/T$ is a phase, or flavor holonomy,
which effectively rotates quarks on the holonomy circle. 

Here we will discuss the role of $\theta\neq 0$ in the theory of the instanton-dyons.
As discussed in the previous chapter and also previous subsection, the fermionic zero modes 
``jump" from one kind of instanton-dyons to another, when the flavor and color holonomies coinside.
Therefore, instanton-dyons should play a very important role in the Roberge-Weiss transitions.

The first lattice study of this phenomenon has been performed in Ref\cite{Bornyakov:2016ccq}.
Its main result is demonstration of dramatic changes in the Dirac eigenvalue spectra 
as the lines of the Roberge-Weiss transitions are crossed. 

\begin{figure}[h]
\begin{center}
\includegraphics[width=6cm]{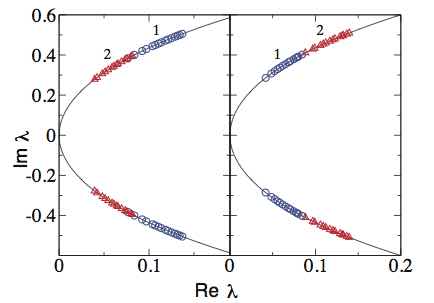}
\includegraphics[width=6cm]{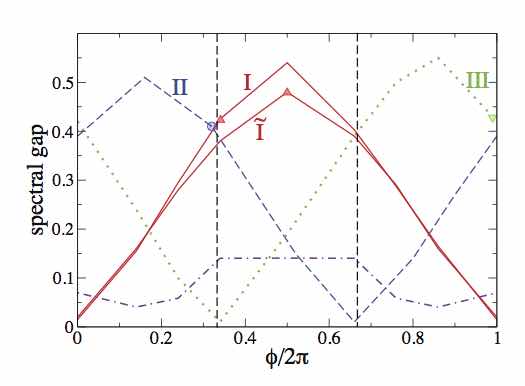}
\caption{
(left) Spectra of the overlap Dirac operator for two configurations. As the 
RW transition is crossed, the eigenvalues corresponding to different dyon types
exchange places.
(right) Spectral gap of the overlap Dirac operator as function of the periodicity angle $\phi$ for four thermal configurations (two red solid curves, one dashed blue curve and one green dotted curve) generated at $T = 1.35 T_c$. 
}
\label{fig_RW_gap}
\end{center}
\end{figure}

The authors use variable periodicity angle $\phi$ as a diagnostic tool,
allowing to monitor eigenvalues for each kind of dyons separately 
The left figure Fig.\ref{fig_RW_gap}
demonstrate eigenvalue spectra for two configurations on both sides of RW transition:
the dyon spectra are clearly interchanged. The right figure shows how
the spectral gap (for configurations above $T_c$) depends on the variable periodicity angle $\phi$.

All results are exactly as anticipated, based on the instanton-dyons mechanism of chiral symmetry breaking.

\chapter{The Poisson duality between the particle-monopole and the semiclassical (instanton) descriptions} \label{sec_Poisson}

We discussed a number of applications of the monopoles. In particular, the confinement is indeed a Bose-Einstein condensation of monopoles, at $T<T_c$.
There are multiple applications to physics of quark-gluon plasma and heavy ion collisions. Basically,  QGP is
a {\em ``dual plasma"} made of electrically charged quasiparticles, quarks and gluons, and magnetically charged monopoles leads to explanation of various observed phenomena. The key yo all of it is the notion that  monopoles can be treated as quasiparticles, and use them both in calculations  involving Euclidean (thermodynamics)  or Minkowski (kinetics) times
as needed.

However, the 't Hooft -Polyakov solution require an adjoint scalar (Higgs) with a non-zero VEV. This is the case  in the Georgi-Glashow model and in 
 other theories with an adjoint scalar field, notably in theories with extended supersymmetry ${\cal N} = 2,\,4$. Yet it is $not$ so in QCD-like theories without scalars, and thus
 one cannot use this solution. 
 
In the chapter devoted to instanton-dyons the   't Hooft -Polyakov solution will be used
with the time component of the gauge field $A_4$ as an adjoint scalar.
 The semiclassical theory built on them obviously can only be used in the Euclidean
 time formulation: an analytic continuation of   $A_4$ to Minkowski time include an imaginary
 field which makes no sense. So, the instanton-dyons cannot be used as quasiparticles.
  And yet, the presence of magnetic charge of the 
 instanton-dyons does suggest, that they should $somehow$ be related to particle-monopoles.
 
A gradual understanding of this statement began some time ago, but remained rather unnoticed by the larger community. One reason for that was the setting in which it was shown, which was based on extended supersymmetry. Only in these cases was one able to derive reliably {\em both} partition functions -- in terms of monopoles and instanton-dyons -- and show them to be equal \cite{Dorey:2000qc,Poppitz:2011wy,Poppitz:2012sw}. Furthermore, they were not summed up to an analytic answer, but shown instead to be related by the so-called ``Poisson duality."

\section{The rotator} 

Another classic example, which display features important for physics to be discussed
in this book, is a rotating object, which we will call the {\em rotator} or {\em the top}. What is special in this case
is that the coordinates describing its location are $angles$, which are always defined with some natural
periodicity conditions. Definition of the path integrals in such cases
require important additional features\footnote{ 
In the Feynman-Hibbs book  does not have its discussion, and contains only a 
comment that the authors cannot describe, say, an electron with spin $1/2$,
and that it was a ``serious limitation" of the approach.}. 

The key questions and solutions can be explained following Schulman 
\cite{Schulman:1968yv} using the simplest  
$SO(2)$ top,  a particle moving on a circle. Its location is defined by the angle $\alpha\in[0,2\pi]$ and its (initial) action contains only the kinetic term   
\be  S=\oint dt\, {\Lambda\over 2}\, \dot\alpha^2 \ee
with $\Lambda=m R^2$  the corresponding moment of inertia for rotation.

All possible paths are naturally split into topological homotopy classes, defined by their {\em winding  
 number}. The paths belonging to different classes cannot be continuously deformed to each
 other. Therefore a fundamental question arises: {\em How should one normalize those disjoint
 path integrals over classes of paths?} Clearly, there is no natural way to define 
 their relative normalization, or rather their relative $phase$. 
 
    Following Aharonov and Bohm  \cite{Aharonov:1959fk} one may provide a
    direct physical interpretation of this setting. Suppose our particle has an Abelian electric charge, and 
    certain device (existing in extra dimensions invisible to the rotator)  creates a nonzero
    magnetic field flux $\Phi\neq 0$  through the circle. Stokes theorem 
    relates it to the circulation of the gauge field 
    $$\oint d\alpha A_\alpha =\int \vec B d\vec S
   $$ 
 While $A_\mu(x)$ is gauge-dependent, its circulation (called holonomy)
is gauge invariant, since it is related to the field flux\footnote{For non-Abelian case there is no Stokes theorem, but gauge invariance of all closed paths is still true: it follows from direct
calculation of gauge transformation of path-ordered-exponents
.}. 

The extra phase is thus physical. Furthermore, it propagates into the energy spectra and
the partition function.  One can  write it in a Hamiltonian way,  as
the sum over states with the angular momentum $m$  at temperature $T$ 
\be   Z_1=\sum_{m=-\infty}^\infty \exp \bigg(-{m^2 \over 2\Lambda T}+i m \omega  \bigg)\,,
\ee   
 where $\omega$ is the holonomy phase, which is so far arbitrary.  
 
Although physical, the effect is invisible at the classical level. This can be seen from the
 inclusion of the additional term in the action $\sim (\omega/2\pi) \int d\tau  \dot \alpha$ 
 which would ``explain" the holonomy phase. This term in Lagrangian however is a full derivative, $ \dot \alpha$,  so the action depends on
 the endpoints of the paths only, and is insensitive to its smooth deformations. It therefore
 generates no contribution to classical equations of motion, thus failing to ``exert any force" on
 the particle in classical sense. In summary, an appearance of the  holonomy phase 
 is our first nontrivial quantum effect, not coming from the classical action.
 
 Now one can also use Lagrangian approach, looking for paths periodic in Euclidean time
 on the Matsubara circle. Classes of paths which make a different number $n$ of rotations around the original circle can be defined as ``straight" classical periodic paths
\be \alpha_n(\tau)=2\pi n {\tau \over \beta}\,, \ee 
plus small fluctuations around them. Carrying out a Gaussian integral over them leads to the following
partition function,
\be  Z_2=\sum_{n=-\infty}^\infty \sqrt{2\pi \Lambda T}\, \exp \bigg( -{T \Lambda \over 2}(2\pi n - \omega)^2 \bigg)\,. 
\ee
The key point here is that these quantum numbers,
$m$ used for $Z_1$ and $n$ for $Z_2$, are very different in nature.
The dependence on the temperature is different. 
Also, for $Z_1$ each term of the sum is periodic in $\omega$, while for $Z_2$, this property is also true, but recovered only after summation over $n$.
 
\begin{figure}[h!]
\begin{center}
\includegraphics[width=7cm]{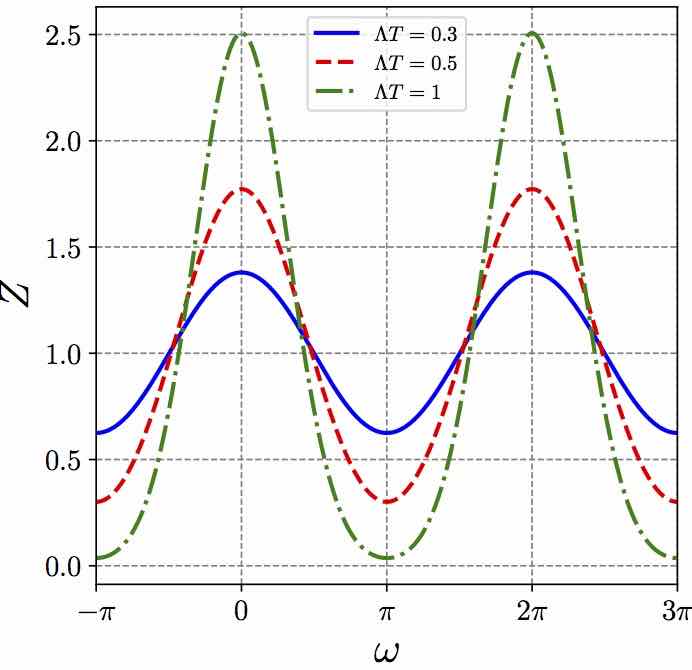}
\caption{The partition function $Z$ of the rotator as a function of the external
Aharonov-Bohm phase $\omega$ (two periods are shown to emphasize its periodicity). 
The (blue) solid, (red) dashed and (green) dash-dotted curves are for $\Lambda T=0.3,0.5,1.$}
\label{fig_rotator_Z}
\end{center}
\end{figure}

 In spite of such differences, both expressions are in fact the same! In this toy model, it is
 possible to do the sums numerically and plot the results. Furthermore, one can also derive the analytic expressions, expressible in terms of the elliptic theta function
of the third kind 
\be Z_1 = Z_2=  \theta_3\bigg(-{\omega \over 2},\, \exp\bigg(-{ 1 \over 2\Lambda T}\bigg) \bigg)\,, \ee
which is plotted in Fig. \ref{fig_rotator_Z} for few values of the temperature $T$.

 In order to prove that one may use   the Jacobi identity 
 $$\theta_3(z,t)=(-it)^{-1/2} e^{z^2/i\pi t} \theta_3(z/t,-1/t)$$
As emphasized by our recent work \cite{Ramamurti:2018evz}, one can observe that two statistical sums are related by the Poisson summation formula, in a form
\be \sum_{n=-\infty}^\infty f(\omega+n P) =\sum_{l=-\infty}^\infty {1 \over P} \tilde f \bigg({l\over P}\bigg)e^{i 2 \pi l \omega/P}\,, \label{eqn_Zygmund}
\ee 
where $f(x)$ is some function, $\tilde f$ is its Fourier transform, and $P$ is the period of both sums
as a function of the ``phase" $\omega$. In this particular example the function is Gaussian,
with Fourier transform being a periodic Gaussian: but we will later encounter examples of the Poisson duality 
with other functions as well.

 The generalization to path integrals defined on other groups can proceed similarly. 
  Schulman \cite{Schulman:1968yv} in particularly was interested in the rotation over the
   $SO(3)$ group, a manyfold with three Euler angles. Instead of infinitely many topological classes of paths, in this case there are
   two classes. The arbitrariness reduces to the relative sign between them: case plus
   leads to bosons and minus to fermions.  The interested reader should consult
   literature on path integrals over Lie groups: for our purposes the simplest $SO(2)$
   (circle) case would be sufficient. 

In summary: the rotator serves as an example of path integral on manifolds which  
have topologically distinct classes of paths. Lesson number one is that their
ambiguous relative normalization allows to recognize ``hidden quantum phase" $\omega$.
Lesson number two is that this is the simplest example in which the Hamiltonian and 
Lagrangian ways to calculate the statistical sum lead to differently looking, although ``Poisson dual",
results.  

\section{Monopoles versus instantons in extended supersymmetry}
The setting of these studies are in weak coupling $g\ll 1$ and   compactification to $R^3\times S^1$. In the ${\cal N}=4$ theory, the charge does not run and $g$ is simply an input parameter. In the  ${\cal N}=2$ theory, however, the coupling does run, and one needs to select the circumference of the circle $\beta$ to be small enough such that the corresponding frequencies $\sim 2\pi/\beta$ are large enough to ensure weak coupling. 
Compactification of one coordinate to the circle is needed to introduce ``holonomies,'' gauge invariant integrals over the circle
$ \oint dx_\mu A^\mu,\,  \oint dx_\mu C^\mu$ of the electric and magnetic potentials, respectively. Their values can have nonzero expectation values, which can be viewed as external parameters given by Aharonov-Bohm fluxes through the circle induced by fields in extra dimensions. These holonomies will play important role in what follows. Dorey et al. \cite{Dorey:2000dt} call these external parameters $\omega$ and $\sigma$, respectively.
Finally, in order to make the discussion simpler, one assumes the minimal non-Abelian color group SU($N_c$) with the number of colors $N_c=2$. This group has only one single diagonal generator $\tau^3$, breaking the color group SU(2)$\rightarrow$U(1).

The theories with extended supersymmetry ${\cal N} = 2,\,4$ have one and six adjoint scalar fields, respectively.  Recall that these two theories also have, respectively, 2 and 4 fermions, so that the balance between bosonic and fermionic degrees of freedom is perfect.  For simplicity, all vacuum expectation values (VEV) of the scalars, as well as both holonomies are assumed to be in this diagonal direction, so the scalar VEVs and  $\omega$ and $\sigma$ are single-valued parameters without indices. In the general  group SU($N_c$), the number of diagonal directions is the Abelian subgroup, and thus the number of parameters is $N_c-1$.

The particle-monopole mass is  
\be M=\bigg({4 \pi \over g^2}\bigg)\phi\,. \ee 
We will only discuss the ${\cal N}=4$ case, following  Dorey and collaborators \cite{Dorey:2000dt}. Six scalars and two holonomies can be combined to vacua parameterized by 8 scalars, extended by supersymmetry to 8 chiral supermultiplets. These 8 fermions have zero modes, describing their binding to monopoles. We will, however, not discuss any of those in detail.  

The SU(2) monopole has four collective coordinates, three of which are related with translational symmetry and location in space, while the fourth is rotation around the $\tau^3$ color direction, 
\be \hat \Omega={\rm exp}(i\alpha \hat \tau^3/2)\,. \ee
 Note that such rotation leaves unchanged the presumed VEVs of the Higgses and holonomies, as well as the Abelian $A^3_\mu\sim 1/r$ tails of the monopole solution. Nevertheless, these rotations are meaningful because they do rotate  the monopole core -- made up of non-Abelian $A^1_\mu,\,A^2_\mu$ fields -- nontrivially. It is this rotation in the angle $\alpha$ that makes the monopole problem similar to a quantum rotator.  As was explained by Julia and Zee \cite{Julia:1975ff}, the corresponding integer angular momentum is nothing but the electric charge of the rotating monopole, denoted by $q$. 

Now that we understand the monopoles and their rotated states, one can  define the partition function at certain temperature, which (anticipating the next sections) we will call $T\equiv1/\beta$,
\bea 
\label{Zmono}
Z_{mono}&&= \sum_{k=1}^\infty \sum_{q=-\infty}^\infty\bigg( {\beta \over g^2}\bigg)^8 
{k^{11/2} \over \beta^{3/2} M^{5/2} }  \nonumber\\
&&\times \exp\bigg( i k \sigma - i q \omega -\beta k M -{\beta \phi^2 q^2 \over 2 k M}\bigg)\nonumber\,,\\
\eea
where $k$ is the magnetic charge of the monopole. The derivation can be found in the original paper, and we only comment that the temperature in the exponent only appears twice, in the denominators of the mass and the rotation terms, as expected. The two other terms in the exponent, $\exp( i k \sigma - i q \omega)$, are the only places where holonomies appear, as the phases picked up by magnetic and electric charges over the circle.  

Now we derive an alternative 4d version of the theory, in which we will look at gauge field configurations in all coordinates including  the compactified  ``time coordinate" $\tau$. These objects are versions of instantons, split by a nonzero holonomy into instanton constituents.  Since these gauge field configurations need  to be periodic on the circle, and this condition can be satisfied by paths adding arbitrary number $n$ of rotations, their actions are
 \be 
 S_{mono}^n=\bigg({4 \pi \over g^2}\bigg)\bigg({\beta^2 |\phi|^2+| \omega - 2\pi n |^2} \bigg)^{\frac 12}\,,
 \ee 
including the contribution from the scalar VEV $\phi$, the electric holonomy $\omega$, and the winding number of the path $n$. In the absence of the holonomies, the first term would be $M/T$ as one would expect. 

The partition function then takes the form \cite{Dorey:2000dt}
\bea
&& Z_{inst}= \sum_{k=1}^\infty \sum_{n=-\infty}^\infty\bigg( {\beta \over g^2} \bigg)^9{k^6 \over (\beta M)^3} \nonumber\\
&&\times \exp\bigg( i k \sigma-\beta k M -{k M \over 2 \phi^2 \beta}(\omega - 2\pi n)^2 \bigg)\,,\nonumber\\
\label{Zinst}
\eea
where $M=(4\pi\phi/g^2)$, the BPS monopole mass without holonomies; thus the second term in the exponent is interpreted as just the Boltzmann factor. The ``temperature" appears in the unusual place in the last term (like for the rotator toy model). The actions of the instantons are large at high-$T$ (small circumference $\beta$); the semiclassical instanton theory works best
at high-$T$. 

The Poisson duality relation between these two partition functions, Eqs. (\ref{Zmono}) and (\ref{Zinst}), was originally pointed out by Dorey and collaborators \cite{Dorey:2000dt}. In this book, following \cite{Ramamurti:2018evz},
it was explained earlier  using the toy model of a {\em quantum rotator}. In fact the 
Poisson duality relation between two sums is in this case exactly the same.

\section{Monopole-instanton duality in QCD}

The authors of \cite{Ramamurti:2018evz} went further, performing the Poisson duality transformation over the semiclassical sum
over twisted instanton-dyons. 
 The resulting expression for the 
 semiclassical partition function is
\be 
Z_{inst}  =\sum_n e^{-\left(\frac{4\pi}{g_0^2}\right) | 2\pi n  - \omega |}
\label{ZINST}
\ee
It is periodic in the holonomy, as it should be. Note that, unlike in Eq. (\ref{Zinst}), it has a modulus rather than a square of the corresponding expression in the exponent. This is due to the fact that the sizes of $L_n$ and their masses are all defined by the same combination $ | 2\pi n -\omega |T$ and therefore the moment of inertia $\Lambda \sim 1/| 2\pi n \beta  - v |$.

Using the general Poisson relation, Eq. (\ref{eqn_Zygmund}),  the Fourier transform of the corresponding function appearing in the sum in Eq. (\ref{ZINST}) reads
\bea
F\left(e^{-A|x|}\right)\equiv && \int_{-\infty}^{\infty}  d x \, e^{i 2\pi\nu x -A|x|} \nonumber\\
= &&{2A \over A^2+(2\pi \nu)^2} \,,
 \eea
and therefore the monopole partition function is 
\be 
Z_{mono}\sim \sum_{q=-\infty}^\infty e^{i q \omega - S(q) }\,,
\ee
where 
\bea 
S(q)&&={\rm log}\bigg(\bigg({4\pi \over g_0^2}\bigg)^2 + q^2\bigg)\nonumber\\
 &&\approx  2{\rm log}\bigg({4\pi \over g_0^2}\bigg)+q^2 \bigg({g_0^2\over 4\pi}\bigg)^2+\ldots \,,
 \label{eqn_action_rotatingx} 
\eea
where the last equality is for $q \ll 4\pi/g_0^2$.

the resulting partition function can be interpreted as being generated by {\em moving and rotating monopoles}. The results are a bit surprising. First, the action of a monopole, although still formally large in weak coupling, is only a logarithm of the semiclassical parameter; these monopoles are therefore quite light. Second is the issue of monopole rotation. The very presence of an object that admits rotational states implies that the monopole core is not spherically symmetric.
The Poisson-rewritten partition function has demonstrated that the rotating monopoles are {\em not} the rigid rotators, because their action, Eq. (\ref{eqn_action_rotatingx}), depends on the angular momentum $q$ and is quadratic only for small values of $q$.
The slow (logarithmic) increase of the action with $q$ implies that the dyons are in fact shrinking with increased rotation. In the moment of inertia, this shrinkage is more important than the growth in the mass, as the size appears quadratically. As strange as it sounds, it reflects on the corresponding behavior of the instanton-dyons $L_n$ with the increasing $n$. 

Although such rotations are well known in principle as Julia-Zee dyons with {\em real} electric charge (unlike that of the instanton-dyons, which only exist in the Euclidean world) and studied in theories with extended supersymmetries, to our knowledge the existence of multiple rotational states of monopoles has not yet been explored in monopole-based phenomenology. In particular, one may wonder how the existence of multiple rotational states affects their Bose-condensation at $T<T_c$, the basic mechanism behind the deconfinement transition. The electric charges of the rotating monopoles  should, therefore, also contribute to the jet quenching parameter $\hat q$ and the viscosity, which was not yet included in literature.

\begin{figure}[h]
\begin{center}
\includegraphics[width=8cm]{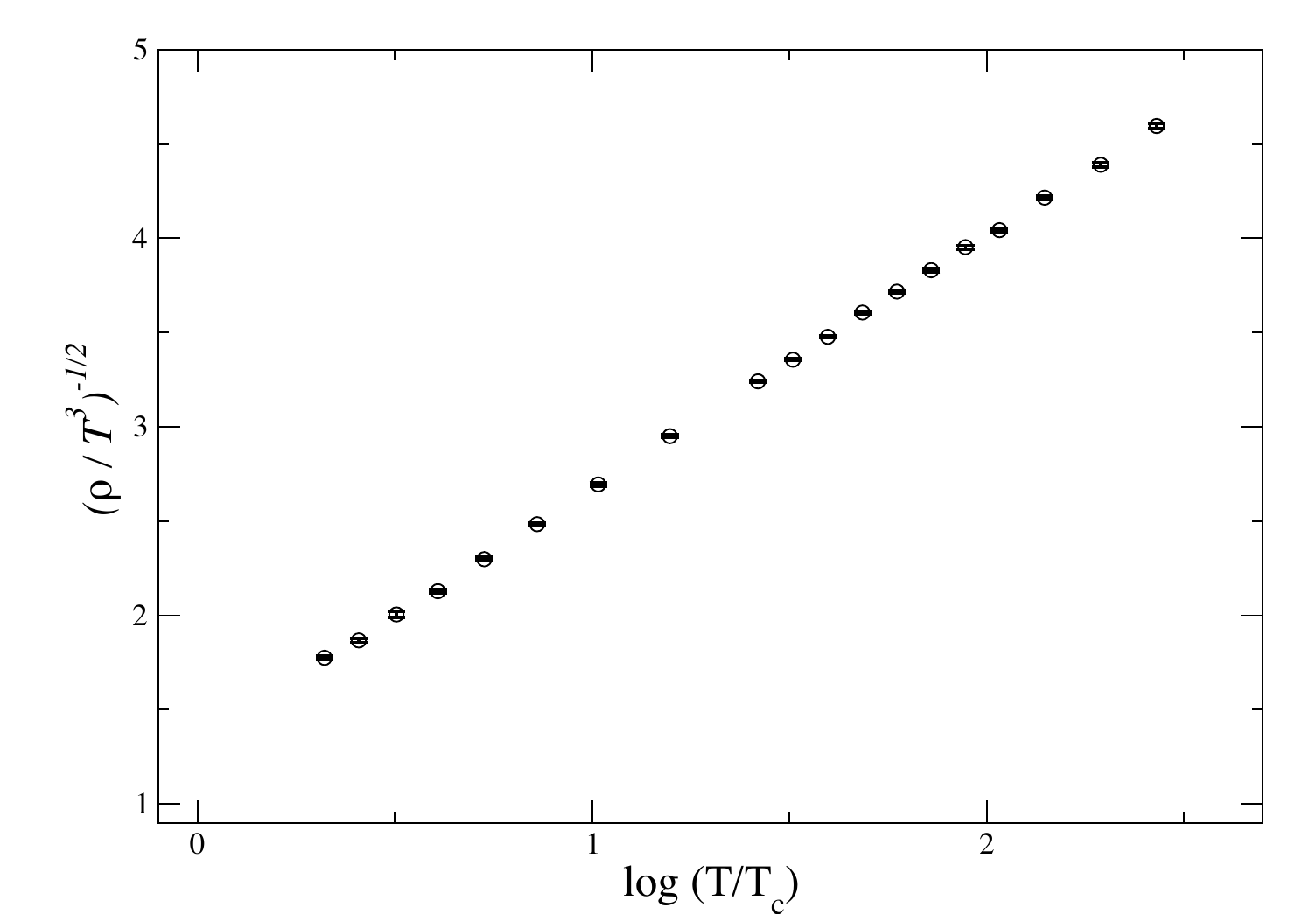}
\caption{The normalized monopole density in SU(2) gauge theory in power -1/2, $(\rho/T^3)^{-1/2}$ versus $log(T/T_c)$ shows an apparent linear dependence. 
}
\label{fig_mono_density_log}
\end{center}
\end{figure}

\section{Short summary}

Let me start with the main conclusions of this chapter: \\

{\bf I. Monopoles and instanton-dyons describe the same physics, and produce the same partition function.} They are just so to say Hamiltonian and Lagrangian (or Minkowskian and Euclidean) ways to describe it. One may call it two dual approaches. 
 Depending on the problem, one or the other should be used. 
Adding the contributions would be double counting, a sin in theoretical physics.

{\bf II. In QCD-like theories without adjoint scalars, monopoles are not classical objects.}

While it came with some surprise, the evidences for that where in front of us for a long time.
In particular, it has been demonstrated rather clearly by \cite{D'Alessandro:2007su}
that monopole density is $not$ decreasing as inverse power of $T$, but only as a power of its log: see Fig.\ref{fig_mono_density_log} from that paper. It is possible only when their action is $$S_{mono}\sim log(1/g^2)\sim log(log(T))$$ And yet, monopoles are Poisson dual to
instanton-dyons, whose densities are the inverse powers of $T$!

\chapter{The QCD flux tubes} \label{chap_strings} 
\section{History} 
  The story of QCD flux tubes started in 1960's,  prior to the discovery of QCD
, with two important $hints$.

 At the time experimental discoveries of multiple hadronic states were the main occupation
 of high energy physics, and discovery and tests of flavor $SU(3)$ symmetry was the main focus.
 It became obvious that mesons and baryons cannot be ``elementary particles", as they were expected to be earlier.  
 It was pointed out by \cite{Hagedorn:1965st}  that rapid growth of density of states  should lead to slow
 growth of temperature. In particularly the exponential one $\rho(m)\sim exp(m/T_H)$ 
 would lead to the ``ultimate temperature limit" of hadronic matter, 
 because the partition function
 \be Z=\int {d^3p \over (2\pi)^3}dm \rho(m) e^{(-\sqrt{p^2+m^2})/T} \sim \int  e^{\big[m({1\over T_H}-{1\over T})\big]} dm\ee
 become  divergent at
 \be T\rightarrow T_H \ee                  
   
  Another line of studies were related to hadron-hadron scattering. A phenomenological breakthrough was discovery that hadron seems to belong to certain {\em Regge trajectories},
  like quantum-mechanical bound states in some non-relativistic potentials. It means that
  there exist some formulae for angular momentum as a function of energy, producing
  energy levels when the values are integer. In relativistic notations such expression is
  written as  $l=\alpha(m^2)$. Furthermore, it was shown in many papers\footnote{
  For a recent review on Regge trajectories of light and heavy mesons see \cite{Sonnenschein:2014jwa}.} starting with \cite{Chew:1962eu} that the trajectories for mesons and baryons are
  approximately linear, and can be approximated by only two constants
  \be \alpha(t) \approx \alpha(0) + \alpha'(0) t \ee  
  the dimensionless ``intercept" $\alpha(0) $ and the ``slope" $\alpha'(0)$. The expression works not only for positive $t=m^2>0$ but also for $t<0$ in scattering. For high energies $s\gg |t|$ the cross sections
  have the following form
  \be {d\sigma(s,t) \over dt} \sim s^{\alpha(t)-1} \ee
in good agreement with the data. The largest $\alpha(t)$ belongs to the so called``leading" trajectory
called  the $Pomeron$, named after Pomeranchuck. Its  $\alpha(0)\approx 1.08$ is above one
and therefore all total cross sections grow with $s$, although in a rather slow pace.

Perhaps the most influential (and the most beautiful) paper of that period was \cite{Veneziano:1968yb} which constructed expression for the amplitude, based on linearity of the 
trajectories and possessing a marvelous $duality$ property\footnote{Perhaps the first time such notion was explicitly demonstrated, the same answer followed from two entirely different  
and seemingly unrelated derivations. Before this work the phenomenologists were inclined
to sum these contribution together, but Veneziano formula elucidated that it was a double counting.}
: it can be derived either
as sum of $s$-channel resonances or $t$-channel Regge exchanges.  

In due time it was realized that straight Regge trajectories and Veneziano amplitude indicated that the object under consideration 
is basically a {\em rotating string}. This development has lead to important historic point,
the birth\footnote{As a birthmark, proving the connection, note that modern string theorists
still call the string scale $\alpha'$.
} of the {\em string theory}. 

Of course, the QCD strings are not point-like but some complicated gluonic finite-size objects, with certain properties and structure we are going to discuss in this chapter. This was not tolerated by purists among string theorists, and a 
bifurcation happened, namely most of them proceeded to study theories of some idealized fundamental pointlike strings, not intimidated by the serious obstacles.  (One of them was that for that one needed to quit our 4-dimensional space-time and go into much larger number of dimensions, D=26.) Fortunately, many years later, with the advent of AdS/CFT duality, it became possible to re-unite the two theories together, describing the  QCD strings as $holograms$
of the pointlike fundamental strings in higher dimensions.  We will return to this point in section \ref{sec_string_holo}.

Dramatic events of 1970's included not only the discovery of QCD, but also experimental discoveries of heavy $c,b$ quarks and quarkonia states. It soon became apparent that
the potential needed to explain them was linear, $V(r)\sim r$, and thus the    QCD strings
got another name, the {\em confining flux tubes}. By the end of 1970's numerical studies of the non-Abelian fields
on the lattice have developed to the point that it was possible  \cite{Creutz:1980wj} to relate the 
Yang-Mills Lagrangian and asymptotic freedom to the string tension.

\section{The confining flux tubes on the lattice vs the ``dual Higgs" model}
Permanent
confinement of color-electric charges (or ``confinement", for short)
is the most famous non-perturbative feature of the gauge theories.
The Lagrangian of QCD-like gauge theories is similar to that of QED,
with massless photons substituted by massless gluons, and massive electrons 
by (very light, or even massless) quarks. 
There are multiple definition of the term itself, e.g. the statement that no object with a color
charge can appear in physical spectrum\footnote{There is a pending million dollar prize offered for a
mathematical proof that pure gauge theory has a finite mass gap. Physicists are already sure that it is the case, beyond any reasonable doubt. 
Billions of high energy collisions of hadrons and nuclei (we already briefly discussed
above) observed produced large number of secondaries, and none of them ever was 
a quark or a gluon. The formal limits on that are so small that there is no sense to even 
mention them.}.

There is perhaps no need to remind the reader the general setting of the lattice gauge theory,
or any technical details about it.  
Most physicists trust that the  limit of vanishing lattice spacing $a\rightarrow 0$ is taken
correctly, since, starting from the  pioneering work \cite{Creutz:1980wj}, it was many times
demonstrated  that the string tension (and other relevant quantities) do scale in correspondence with 
the correct renormalization group prescription, and are thus physical. 

What was observed on the lattice,for pure gauge theories, was that the electric flux from
a  color charge is not distributed radially outward, as in electrodynamics,
but instead, being expelled from the QCD vacuum,  is confined into a {\em flux tube} 
between the charges.
At large distances, the leading contribution to the static and heavy quark-antiquark
potential $V_0(r)$ in pure Yang-Mills theory is the famous linear potential
\be 
\label{0}
V_0(r) =\sigma_T r 
\ee 
with $\sigma_T$ as the fundamental string tension, the  energy per length.
Its numerical value (in QCD with physical quarks)  is 
\be \sigma_T\approx (420\, MeV)^2\approx 1\,\,\, GeV/fm \ee
also served\footnote{ Recall that
in QCD with light quarks
this behavior is only valid till some distance due to screening by light quarks
in the form of two heavy-light mesons. So now lattice units are usually set via
location of a point at which the potential times the distance take some prescribed value.
}
as the $definition$ of absolute units in any confining theory.

\begin{figure}[h!]
\begin{center}
\includegraphics[width=12cm]{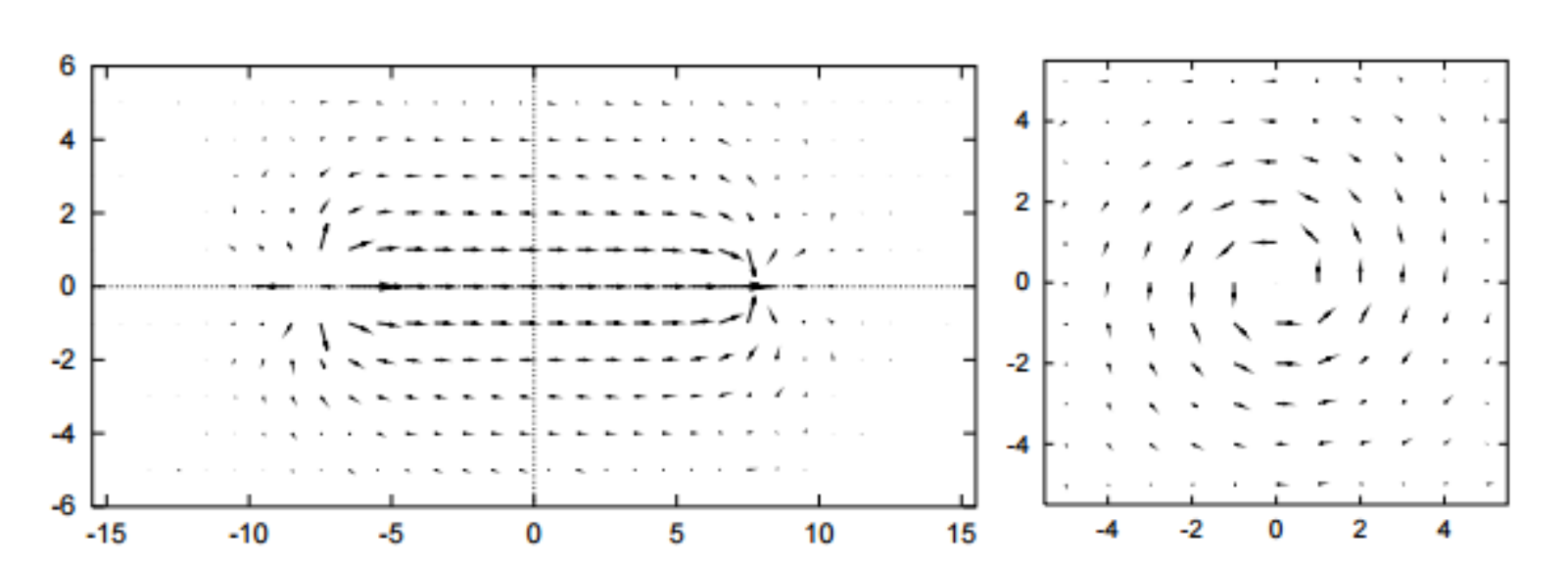}
\caption{Lattice data on distribution of the electric field strength (left) and the magnetic current (right), for two static quark-antiquark external sources.The profile of the electric field is shown by squares , where lines are just fits.}
\label{fig_conf_tube}
\end{center}
\end{figure}

Fig.\ref{fig_conf_tube} (displaying the result of lattice simulations summarized in the review \cite{Bali:1998de}) shows distribution of the electric field (left) 
along the flux tube, and magnetic current (right) in a transverse plane. So,  in  numerical
simulations of the gauge fields, not only the flux tube with
the longitudinal electric field is  clearly seen,
as well as the stabilizing ``coil" around it. Its physical origin we will discuss later in the chapter, in section   \ref{sec_tubes_mono}.

The ``dual superconductor" idea has been mentioned many times above, and need not to be repeated here. The specific relation between the QCD 
flux tubes, and their dual flux tubes in superconductors
has been pointed out in \cite{Nielsen:1973cs}.
 At this point one would like to test whether the duality relation between them is really quantitative. More specifically, let us test whether
 the shape of the confining flux tubes 
can indeed be described by the same generic effective model, the Ginzburg-Landau theory\footnote{Let me remind
that when Ginzburg-Landau paper was written, the physical nature of electric object which makes the condensate
was also unknown: they argues for the form of effective action on general grounds.} as the magnetic flux tubes in superconductors. 

There is no need to describe here the Ginzburg-Landau theory in detail: it is enough to say that its
expression for the effective free energy (analog of the action) includes Abelian QED gauge field and a charged scalar described by complex  field $\phi$, representing condensate of Cooper pairs\footnote{There was an instructive story about the charge of $\phi$. 
The GL paper was written well before the microscopic BCS theory of
superconductivity. Ginzburg initially put some ``effective charge" $e_{eff}$ but Landau objected, saying that if the charge be dependent on matter parameters, like temperature, it would spoil gauge invariance of electrodynamics, and so they put $e$, the electron charge. 
After BCS it became clear that the charge must indeed be fixed, but not to one but $2e$.}.

The key Maxwell equation
we will focus on is\footnote{A reminder: we are looking for a static solutions only, so time derivative of $\vec E$ in it is omitted.} 
$$ \vec \nabla \times \vec B=\vec j $$
It tells us that Abrikosov flux tube solution, with nonzero magnetic field $B$ inside the
tube and zero outside needs a ``coil" with current, confining the field inside. The current is
the gradient of the scalar's phase. In the dual case we discuss, one should substitute 
$$\vec B\rightarrow \vec E, \,\,\,\,  \vec j \rightarrow \vec j_{magnetic} $$
and $\phi$ representing the (magnetically charged) monopole condensate.

\begin{figure}[t!]
\begin{center}
\includegraphics[width=5.5cm]{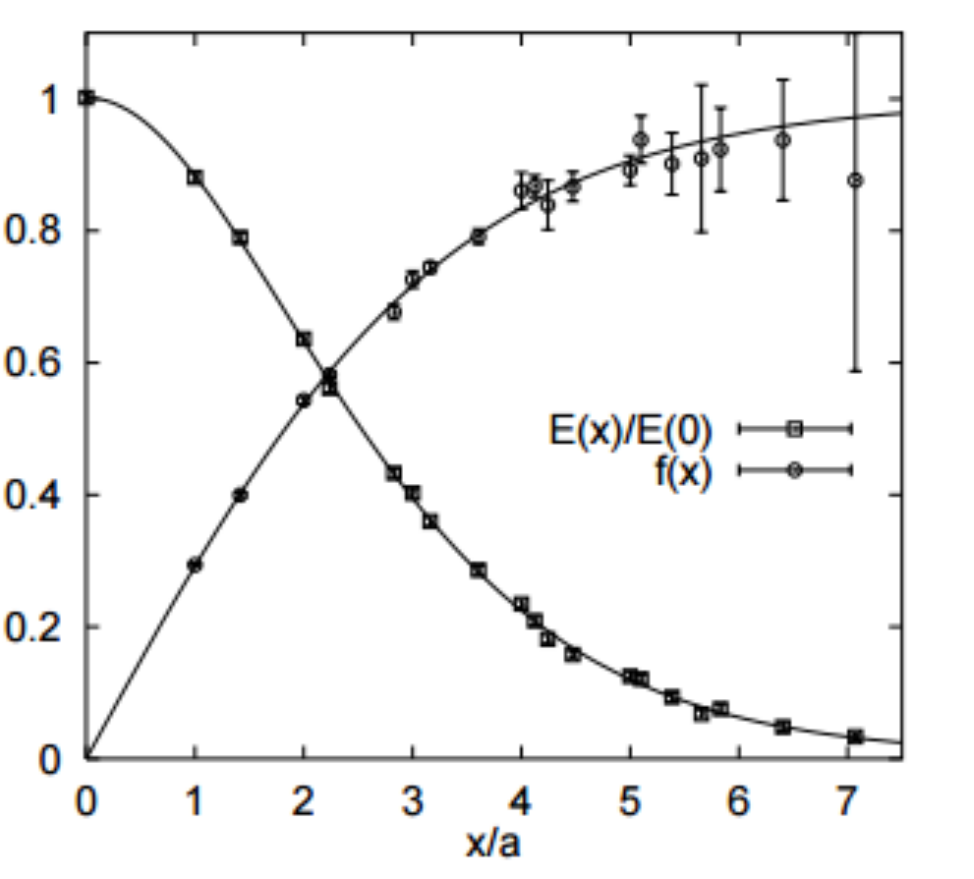}
\includegraphics[width=6cm]{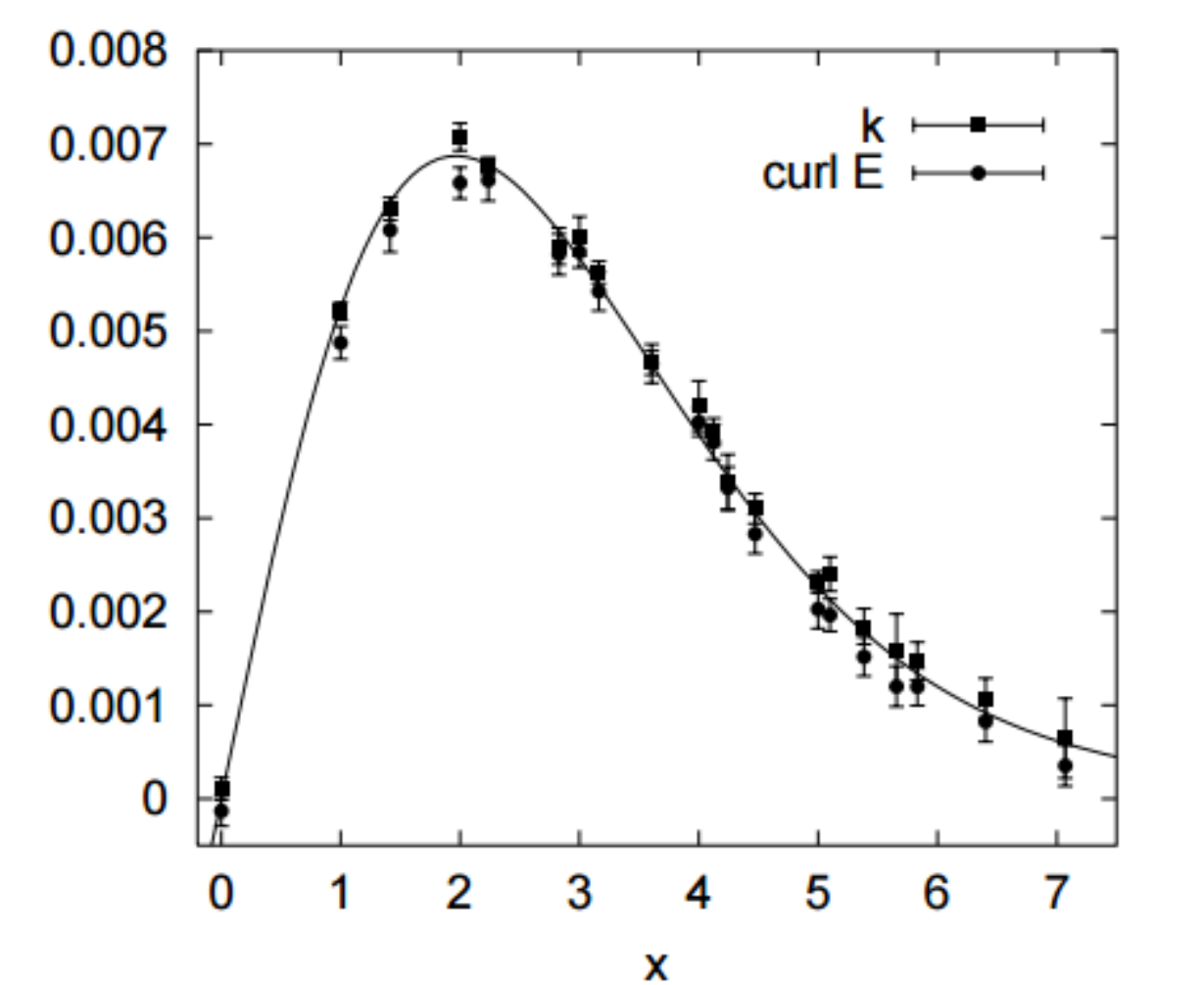}
\caption{(left) Transverse profile of the electric field and the condensate.  (right)
The transverse distribution of the magnetic current. The points in both are lattice data, and the lines are fits using Abrikosov's flux tube solution of the Ginzburg-Landau equations.
}
\label{fig_conf_tube2}
\end{center}
\end{figure}

 The curl of $E$ is shown  in Fig.\ref{fig_conf_tube2}(right) coincides well with the separately measured magnetic current $k$ from monopole motion. 
So, at least the second equation -- basically dual Maxwell equation -- is indeed satisfied. With some 
accuracy, also the first equation is satisfied, and $f(x)$ shown in Fig.\ref{fig_conf_tube2}(left)
is the radial profile of the condensate observed on the lattice. Note in particular, that like in the superconductors,  the  ``dual Higgs" scalar field vanishes at the
 center. The resulting parameters for two basic lengths, in ``physical units" obtained fixing lattice scale to
physical $\sigma$, are
\be \lambda=0.15\pm 0.02 \, fm, \,\,\,  \xi=0.251\pm 0.032 \,\,fm,   \kappa={\lambda \over \xi}=.59\pm 0.14 <{1 \over \sqrt{2}}
\ee
Recall that the so called  Ginzburg-Landau ratio of them 
 is $smaller$ than the critical value (shown at the end of the previous equation). 
 This implies  that the QCD vacuum (we live in) is the {\em dual superconductor of type I}
 \footnote{  
 Those who are not convinced by  not-too-impressive accuracy
of this numerical statement, may wander if there are more direct manifestation of it. We will
return to the issue of flux tube interaction in section on multi-string systems.}

Rather extensive calculations of the static potential, at large and also finite $r$,
were performed  in the framework of the ``dual Higgs model", see review \cite{Baker:1991bc}.
We will not reproduce here the results, but just provide few comments:

Comment 1: In this and subsequent works not only the static classical linear potential is described, but also the velocity-dependent 
relativistic corrections, ultimately rather successfully compared to phenomenological relativistic terms derived from the quarkonia spectra. Later Baker and collaborators had also calculated the Regge trajectories, see \cite{Baker:2002km}.
 
Comment 2: Generally speaking,   the ``dual Higgs models" constitute quite interesting examples of 
``an effective magnetic theory" approach. Not only the scalar fields in them are magnetically charged-- describing the BEC of monopoles, but the gauge field itself is also treated using
the dual potential rather than the usual $A_\mu$.  A specific form of the model is motivated 
and defined in
\cite{Baker:1991bc}.

As an example of further progress in lattice technology, let me mention \cite{Yanagihara:2018qqg}
in which the flux tube has been studied from the point of view of the underlying stress tensor. The QCD operator of the stress
tensor is of course well known, but for many years its direct evaluation has been blocked by very large statistical noise.
In the paper under consideration the authors used the {\em gradient flow} smoothening procedure, appended by
additional extrapolation back to zero value of the gradient flow time. 
The authors have demonstrated that the procedure
is consistent with lattice studies of the equation of state, for example their
$\langle T^{00}(T) \rangle$ agrees with the energy density, $\langle T^{11}(T) \rangle$ agrees with pressure, etc.

\begin{figure}[htbp]
\begin{center}
\includegraphics[width=14cm]{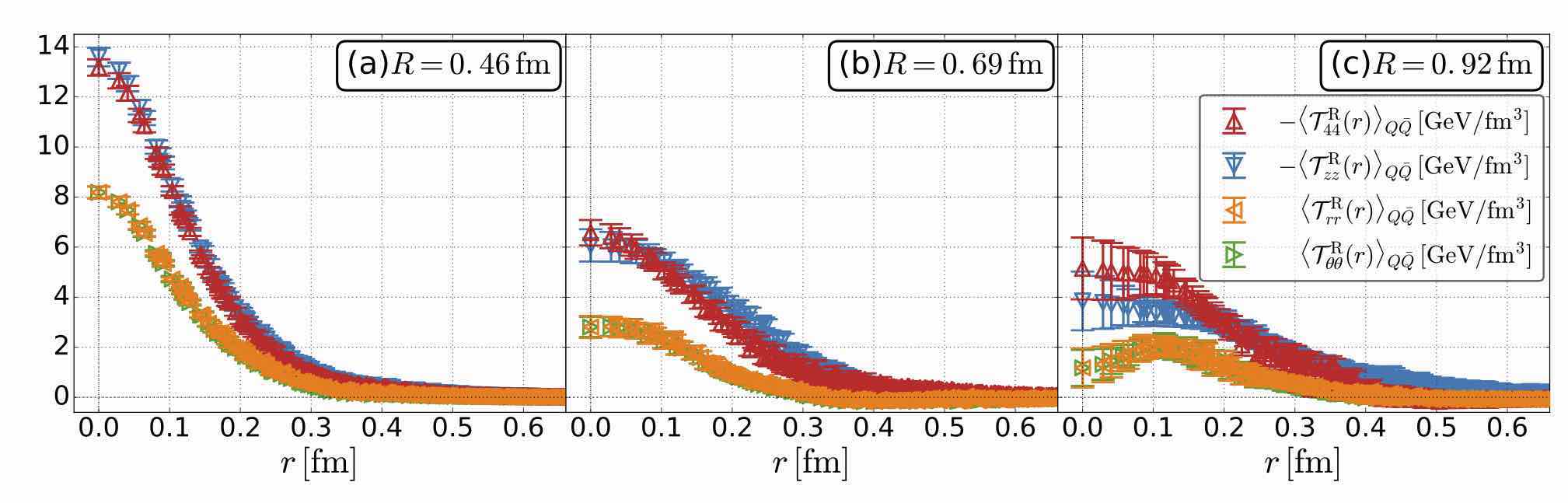}
\caption{Mid-plane distribution of various components of the stress tensor, in cylindrical coordinates.
Three pictures differ by the value of $R$, the distance between static quarks.  }
\label{fig_flux_stresstensor}
\end{center}
\end{figure}

Needless to say, in order to study the properties of the flux tube, one needs sufficient statistical precision  
to subtract these mean values, present everywhere. In Fig.
\ref{fig_flux_stresstensor} from this work one can see transverse distribution of the diagonal components of 
the stress tensor. Note that the sign of them, as shown, are selected in such a way, that in pure electric field
all four would be the same: this apparently this is $not$ the case. Yet two transverse pressures, the $rr$ and $\theta\theta$ components,
seem to be always the same. 

One other comment is that the energy density at the center of the flux tube, as read from figs.b and c, 
is about 6 $GeV/fm^3$. This is very large value. In particular,
it is 
about two orders of magnitude larger  (!) than the value suggested by the early
MIT bag model of 1970's, which tried to describe flux tube compressed by some
"Bag pressure". In
 those days the magnitude of non-perturbative effect were grossly
underestimated. 

\section{Regge trajectories and rotating strings} 

Static potentials are not the only place where one can infer existence of
the (fundamental) flux tubes. Another impressive confirmation comes from hadronic spectroscopy.

Quarkonia -- the non-relativistic bound state of heavy quark-antiquarks -- are indeed well described by 
the sum of Coulomb and linear potential. Hadrons made of light quarks also show a  very spectacular confirmation
to the idea, that
 mesons are basically quark-antiquarks  connected by a flux tube,  and (at least some) baryons can be approximated by similar quark-diquark systems, also connected by a flux tube. 
 
  \begin{figure}[h]
\begin{center}
\includegraphics[width=12cm]{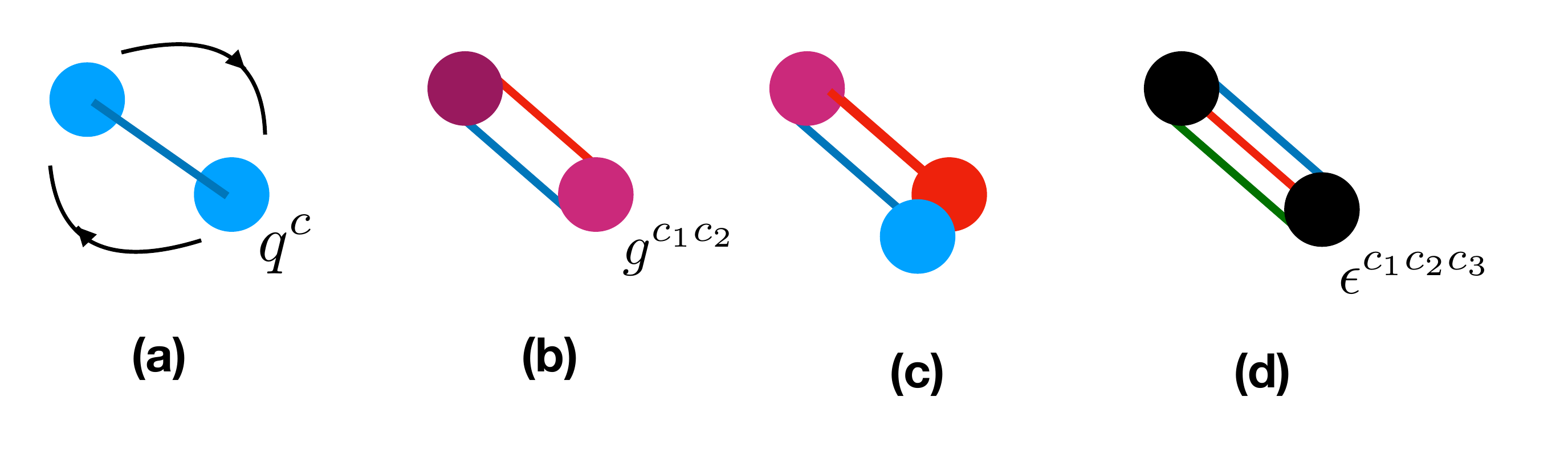}
\caption{Various types of Reggeons: (a) rotating $\bar q q$ pair, held by a string; (b) rotating gluon pair
(positive C-parity) held by two strings, (c) three gluons (negative C-parity) objects held by three strings, (d) rotating baryon junctions,  held by $N_c$ (=3) strings
 }
\label{fig_reggeon_types}
\end{center}
\end{figure}

\begin{figure}[h]
\begin{center}
\includegraphics[width=12cm]{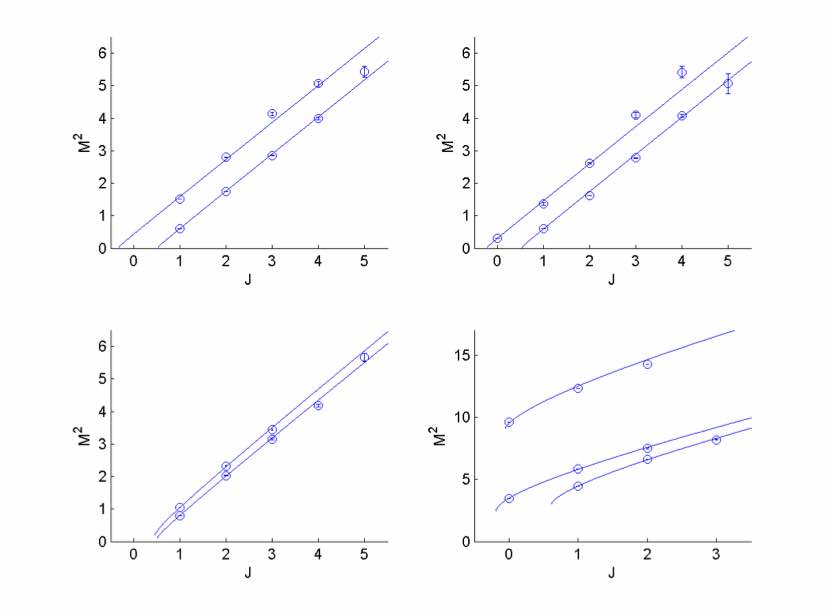}
\caption{
Nine Regge trajectories on inverted ($J,M^2$) Regge plot: top left $\pi,\rho$ (pseudoscalars and vectors with isospin $I=1$), top right $\eta,\omega$ (pseudoscalars and vectors with isospin $I=0$), bottom left $K^*,\phi$ (vectors with one or two strange quarks), bottom right $D,D^*_s,J/\Psi$ (with one or two charmed quarks). }
\label{fig_meson_Regge}
\end{center}
\end{figure}

 Quantization of such system with a string predicts that such hadrons should appear in form of Regge trajectories. As it was noticed in 1960's,  excited states of light mesons and baryons 
 are indeed located on near-linear trajectories, in the total angular momentum $J$ - squared mass $M^2$. Let me not present historic Chu-Frautchi plot, but go into relatively recent 
 Fig.\ref{fig_meson_Regge} from \cite{Sonnenschein:2014jwa}, in which there are many states and also
 lines indicating the model we will be discussing
 \footnote{For completeness, let me mention
 that slope is universal $\alpha'=.884\, GeV^2$, and effective quark masses (those including the chiral symmetry breaking, not the ones in QCD Lagrangian) are $60, 220, 1500\, MeV$,
 for light, strange and charm quark, respectively. 
 }. For future reference let me note that all vector trajectories at $M^2\rightarrow 0$ go to
 $\alpha(0)\approx 0.5$.
 
\begin{figure}[h!]
\begin{center}
\includegraphics[width=6cm]{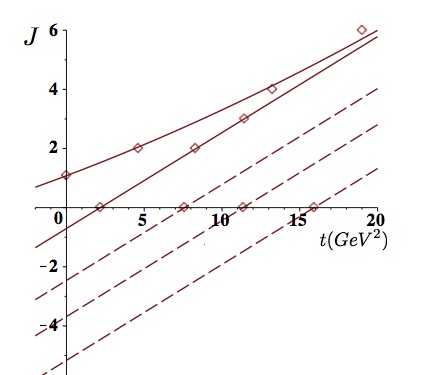}
\includegraphics[width=5.5cm]{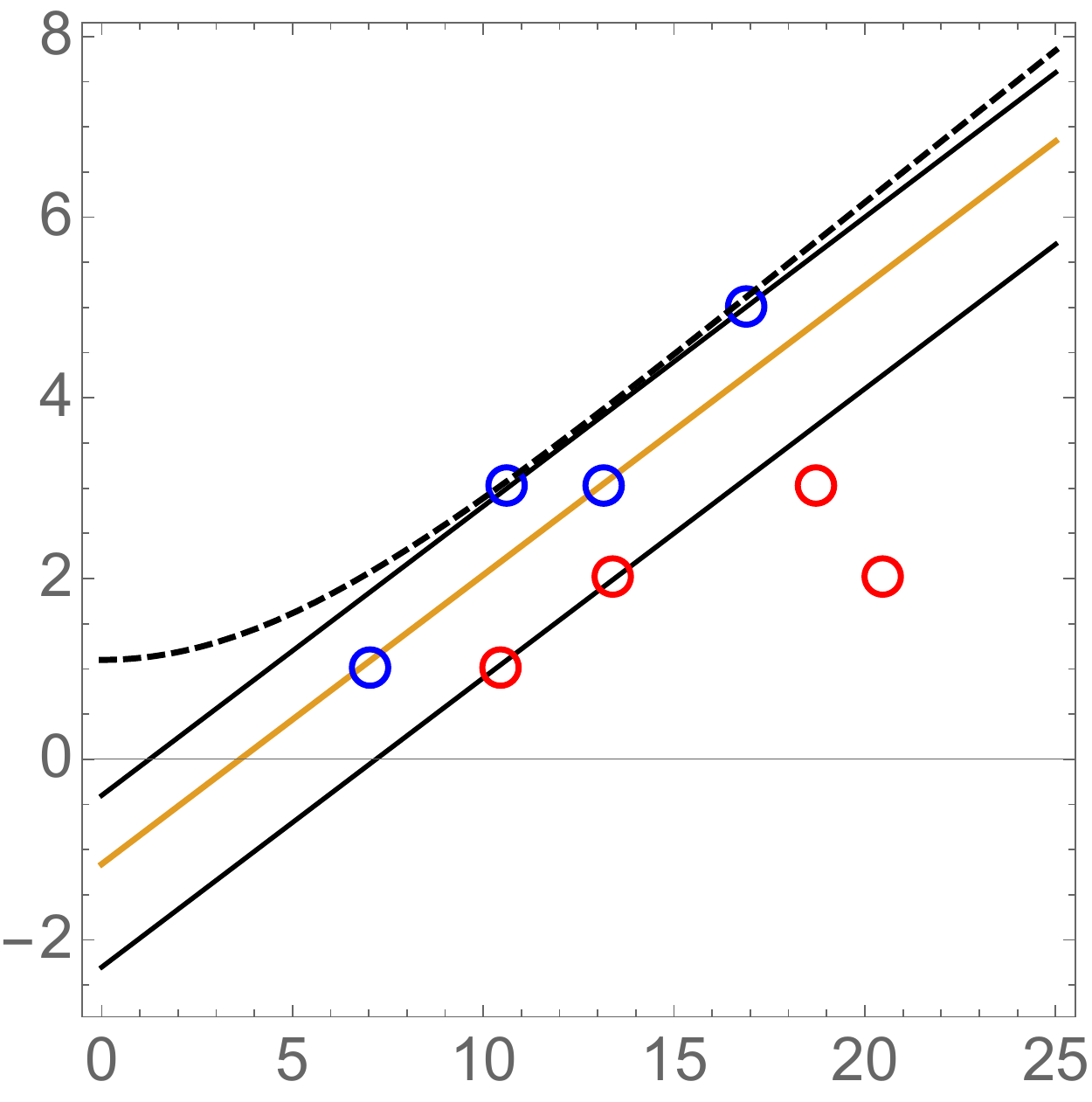}
\caption{Glueballs with positive charge parity $ C=1$ (left) and negative $C=-1$  (right)
 on Regge plots, 
 their angular momentum $J$ versus their squared mass $M^2 ({\rm GeV})^2$.
  The two upper (blue) points and lines are for the negative $spatial$ parity $P=-1$
  gluballs, and the lower (red) ones are for the $P=+1$. 
The lines are the (hypothetical) Regge trajectories.}
\label{fig_glueballs}
\end{center}
\end{figure}

Furthermore, even  the hadrons $without$ quarks -- the glueballs -- can be described
in terms of rotating $closed$ strings, forming another set of   Regge trajectories.  The
lattice data on spectrosopy of pure gauge theories have, in my view, produced rather
signifiant support to this statement. Since it is much less known, and will be needed in 
connection to Pomerons we will discuss later in this chapter, let us see them, following
\cite{Kharzeev:2017azf}. The corresponding plot for  the masses of  glueball with positive charge $C$ parity, taken
from the lattice study \cite{Meyer:2004gx}, is shown in 
Fig.~\ref{fig_glueballs}. 

{\em A comment}: Naive approach to rotating closed string states suggests that the string tension
should be doubled, (two strings rather than one), or that the slope of
C-even glueballs Regge trajectories should be a $half$ of 
 that in mesons . Yet, as seen from
these plots, the mesonic $\alpha'=0.88\, GeV^{-2}$ and the glueball slope
(calculated from J=0 and J=3) is  $\alpha'_{C=1}=0.36\, GeV^{-2}$, so
this ratio is 2.4 rather than 2.  What it implies is that the two strings must not be independent but interacting with each other.  
For $C=-1$ plot one again finds that in three pairs the slope is the same,  $\alpha'_{C=-1}=0.33\, GeV^{-2}$, not far from the other glueballs. Does it mean that,like baryons, the 3-gluon states are
not $Y$-shaped, but 1+2 gluon type? Perhaps.
To my knowledge, nobody had worked
it  out. Also I am not aware of any calculations of the odderon slope. 
  
  QCD has one more mysterious colored gluonic object, the {\em baryonic junction}. In case of $N_c$ 
  colors it connects together  $N_c$ flux tubes with all colors. Its algebraic structure is 
  antisymmetric $\epsilon^{c_1,c_2...c_{N_c}}$. 
  
  One place where it can appear 
  is in the so called Y-shape baryons, with three strings joint at the junction at the center. 
  Rotation of it should then lead to Regge trajectories with slope 1/3 of the usual, as there are three strings
  involved.
  However, Regge trajectories and other theoretical studies had convincingly shown that
  at least nucleon-like baryons are $not$ of this type, having quark-diquark structure instead
  and standard single slope, the same as mesons have.
  
  Another use\footnote{This paragraph is based on \cite{Kharzeev:1996sq} and recent
  private communication from D.Kharzeev, who is working on so far unpublished paper on this issue.
  } of baryonic junctions appeared in hadronic (e.g. $pp$) collisions in which ``stopped baryons" are observed far from the beam rapidity, for example near the center of mass energy
 (where colliders have best detection capabilities). Effective Regge diagram for this process 
 must include ``the junction Reggeon" (nobody proposed any name for it so far) with (unknown to me)
 intercept and the slope close to $1/3$ of the usual one. I am not aware of any resonance or state attributed to this trajectory.

In summary: confining (fundamental) flux tubes have been seen and studied on the lattice, and they also have strong support via hadronic spectroscopy and reactions. Their properties are in agreement with predictions based on the dual Higgs models. 

\section{Flux tubes and finite temperatures: the role of monopoles} \label{sec_tubes_mono}

In this section we extend our discussion of the flux tubes to QCD {\em at finite temperatures}.
A general expectation -- based on analogy to superconductors -- is that they exist in the confining phase $T<T_c$ and disappear
above it. However, as we will see shortly, the situation turned out to be posessing
unexpected and rather peculiar features, not present in the case of  superconductors.

At finite temperatures the natural quantity to calculate, for the observed flux tubes
between static charges, is the {\em free energy}. It can be written as
 \be F(r,T) =V(r,T)- T S(r,T), \,\,\, S(r,T)={\partial F(r,T)  \over \partial T}   \ee 
where $S(r,T)$ is the $entropy$ associated with the pair of static quarks. 
Since it can be calculated from the free energy itself, as indicated in the r.h.s., one can 
subtract it and plot also the {\em potential energy} $V(r)$. The derivatives over $r$ of both potentials -- the force -- is what we call the {\em string tension}.

\begin{figure}[h]
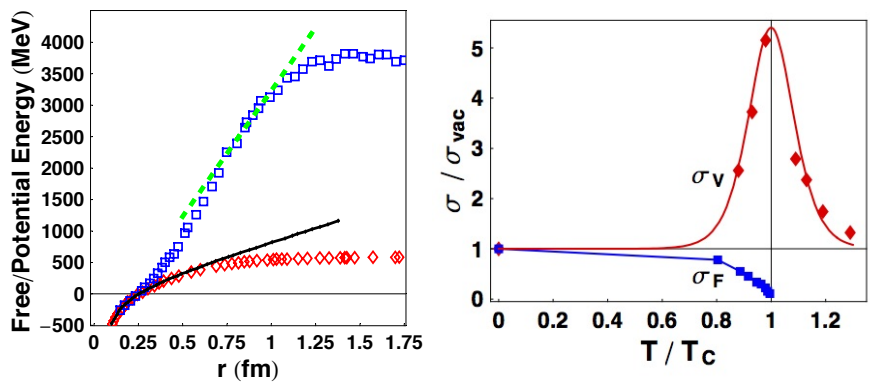

\begin{center}
\includegraphics[width=5.5cm]{Splitting}
\includegraphics[width=6cm]{tension_F_U}
\caption{Left: Free (red rhombs) energy $F(r)$ and potential (blue squares) energy $V(r)$,
at $T_c$, compared to the zero temperature potential (black line). 
Right: Effective string tension for the free 
 and the internal energy. 
}
\label{fig_tension_F_U2}
\end{center}
\end{figure}

The lattice calculations have shown that in certain range of $r$ the tension is
constant (the potential is approximately linear in $r$). We do not show this but proceed
directly to the temperature dependence of the two resulting tensions, 
shown in Fig.\ref{fig_tension_F_U2}(left) (based on lattice calculations 
by the Bielefeld-BNL group, see  \cite{Kaczmarek:2005gi} and earlier works mentioned there). 

The tension of the free energy shows the expected behavior: $\sigma_F(T)$ vanishes as
$T\rightarrow T_c$.  But the  tension of the potential energy $\sigma_V(T)$
shows drastically different behavior, with large $maximum$ at $T_c$, and non-zero value above it. This unexpected behavior was hidden in $\sigma_F(T)$, studied in many previous works, because in it a large energy and a large entropy cancel each other. 

The explanation to this effect has been proposed by \cite{Liao:2008vj}, which we will here
follow. But before we follow this particular explanation, related with monopoles, let us make some general comments:

Comment 1: A large entropy implies exponentially large $exp\big[S(T,r)\big]$ number of states, associated with static quark pair. Furthermore, the nonzero tension --derivative over $r$ -- mean that such states are not concentrated at the string's end, but are also
distributed along the string. 
What can the physical origin of those states be?

Comment 2: A nonzero tension $\sigma_V(r)$ at $T>T_c$ implies the existence of flux tubes 
$above$ $T_c$. 

Comment 3: 
The $free$ energy, by its nature, 
corresponds to physical conditions of complete thermal equilibrium,
which can only be reached at long time. If the time is limited -- for example 
if color dipole is only created for a finite time, or in other situations with 
moving (non-static) charges, there would be deviations from equilibrium,
and therefore cancellations between energy and entropy may be only partial.

Comment 4: Therefore, the effective potentials for quarkonia, which have non-relativistic but 
still non-static heavy quarks, should be somewhat intermediate between $F(r)$ and $V(r)$.
 \cite{Liao:2008vj} discussed a setting in which heavy quark and antiquark slowly
  move away from each other with some velocity $v$, and argue that the entropy production can
  be calculated using Landau-Zener theory of level crossings, used originally for 
  description of bi-atomic molecules.  This theory describes how the resulting population of both crossed levels depends on $v$. 
 
 Now let us proceed to the dynamical explanation proposed by \cite{Liao:2008vj}. 
 Its main point is that the ``dual superconductor" picture is not sufficient:
 one should also recognize the existence of $uncondenced$ (or ``normal")
 component\footnote{Note that in the BCS superconductors
 there are no ``uncondences" Cooper pairs.}
  of the monopole density. 
 
 Liao and myself  argued  \cite{Liao:2008vj}  that electric flux tubes can  be  (mechanically) stable even without the ``dual superconductor", or BEC of monopoles. Indeed
 presence of magnetic flux tubes in various plasmas are well known:
\footnote{
In fact, in a good telescope one can 
directly see hundreds of them in solar corona! Bunches of flux tubes compose the black spots on the Sun made  famous by observations by Galileo, who used them
to discover solar rotation. The Pope of the time interpreted black spots on the Sun as allegory
criticizing him personally, and initiated trials ended in Galileo home-arrests.
}.  The difference between such flux tubes, at $T>T_c$, and  the one due to 
``dual superconductor" at  $T<T_c$  is that in the latter case the ``coil" includes
non-dissipative supercurrent, making them permanently stable, while the latter ones
have Ohmic losses and are therefore metastable.
%
%

Using elliptic coordinates,  \cite{Liao:2008vj} had derived a solution for the electric field
in the monopole plasma, even for finite distance between the quark charges, reproducing
the potential from Coulomb-like behavior at small distances to long flux tubes at large. 
We will not give any details here, and only note that because at high $T$ the monopole
density drops rapidly at $T>1.5T_c$, and thus the metastable flux tubes do not exist there.

\begin{figure}[htbp]
\begin{center}
\includegraphics[width=9cm]{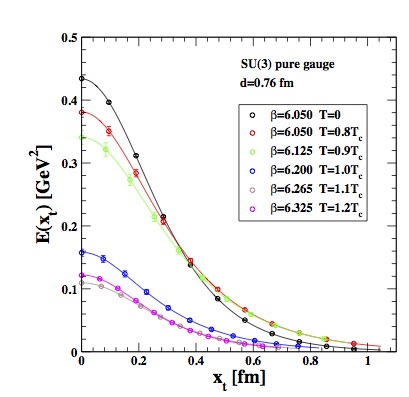}
\caption{ The longitudinal electric field as a function of $transverse$ coordinate is measured,
for a number of temperatures, for pure gauge $SU(3)$ theory, from  \protect\cite{Cea:2017bsa}.}
\label{fig_profile_T2}
\end{center}
\end{figure}

Finally, let me add some comments about more recent lattice study of
the flux tubes at finite temperature  \cite{Cea:2017bsa}. Using certain smoothening procedure,
the shape of the longitudinal electric field as a function of $transverse$ coordinate is measured,
for a number of temperatures, for pure gauge $SU(3)$ theory and for QCD with realistic quark masses. In Fig.\ref{fig_profile_T2} we show some of their results. Note that this theory has
the first order transition, seen as a jump in the field strength. And yet, the overall
flux tube shape persists, in the pictures of the electric field, approximately till $T=1.5 T_c$.
These observations support the idea that the flux tubes do exists in the QGP, in spite of
absence of the ``the dual superconductor" there.  

\section{ Effective string theory (EST) versus precise lattice data}  \label{sec_EST}
This section starts with a pedagogical introduction, introducing classical string solutions and explaining elements of string quantization, and then jumps to a brief review of the
current status of EST, in connection with empirical and lattice data.

A particle  moving in $D$ space-time dimensions can be described by its path  $X^\mu(\tau)$
with proper time $\tau$ and $\mu=1..D$. Similarly, a 
string moving in $D$ space-time dimensions is described by coordinates $X^\mu(\tau,\sigma) $ with two $internal$ coordinates, time-like $\tau$ and space-like $\sigma$. The simplest geometrical Nambu-Goto action is simply the area of the corresponding ``membrane" times the tension
\be S_{NG}=-\sigma_T \int d\tau d\sigma \sqrt{-h}, \,\,\,\, 
h_{\alpha\beta}=\eta_{\mu\nu} \partial_\alpha X^\mu \partial_\beta X^\nu \ee
where $\eta_{\mu\nu} $ is called the external metric, assumed for now  just flat Minkowski metric,
$h_\alpha,\beta, \alpha,beta=1,2$ is the $internal$ metric of worldsheet. 
The area element include the  2*2 determinant of it, $h=det(h_{\alpha\beta})$. We will also use a
dot for derivative in $\tau$ and a prime for derivative in $\sigma$, e.g. $\dot X^\mu,X'^\mu$.
If the endpoints are massive, the following ``particle" term is added to the action
\be S_{ends}=-m\int d\tau \sqrt{-\dot X^2} \ee
(For simplicity we assume both masses to be the same $m$.)

Classical equation of motion for the string is 
\be \partial_\alpha \big(\sqrt{-h} h^{\alpha\beta} \partial_\beta X^\mu \big) =0 \ee
and the boundary conditions should be
\be \sigma_T \sqrt{-h} \partial^\sigma X^\mu \pm m \partial_\tau \big({\dot X^\mu \over \sqrt{-\dot X^2}} \big) =0\ee
It may look complicated, but we will not study complicated string dynamics. The most straightforward rotating string configuration is given by
\be X^0=\tau, X^1=\sigma cos(\omega \tau), X^2=\sigma sin(\omega \tau) \ee
solves the string equation of motion. The boundary condition takes the form
\be \sigma_T \sqrt{1-\omega^2 l^2}={m \omega^2 l \over \sqrt{1-\omega^2 l^2} }\ee
which has the obvious meaning, $l$ is half of string length. Using standard Noether procedure to calculate 
the energy and angular momentum, and substituting this solution into them, one gets
\be E={2m \over  \sqrt{1-\omega^2 l^2}}+\sigma_T\int_{-l}^l {d\sigma \over \sqrt{1-\omega^2 \sigma^2}}\ee
  \be J={2m \omega l^2  \over  \sqrt{1-\omega^2 l^2}}+\sigma_T \omega   
\int_{-l}^l {d\sigma \sigma^2 \over \sqrt{1-\omega^2 \sigma^2}}\ee
Using simpler notation $v\equiv \omega l$, the velocity of the string ends,
 and performing the integrals one gets the following rather intuitive results 
\be E={2m \over \sqrt{1-v^2}} +2\sigma_T l {arcsin(v) \over v} \ee
\be J={2m v l\over  \sqrt{1-v^2}} +\sigma_T l^2 \big( {arcsin(v)-v\sqrt{1-v^2} \over v^2 }\big)
\ee
the length $2l$
can be substituted from the boundary condition $\sigma_T l=m v^2/(1-v^2) $, and 
one has  the resulting Regge trajectory, in a parametric form.  
How well it describes
the mesons one can see in \cite{Sonnenschein:2014jwa}, let us just mention 
the case of light quarks, small $m$ and $v$ close to 1. Expanding the function
in the r.h.s. one has then the linear Regge trajectory with corrections
\be J=\alpha' E^2 \big( 1- {8\sqrt{\pi} \over 3} ({m \over E })^{3/2}  +...\big) \ee
where we also use
the standard relation between the string tension $\sigma_T$ and the Regge slope
is 
\be \alpha'\equiv {1 \over 2\pi \sigma_T} \ee

$Quantization$ of the problem must include not only quantum motion of masses, but also 
that of the string, and it is not  simple at all\footnote{
String quantization is a complicated topic going well beyond this course. Unless one deals with the so called critical dimension of space-time $D=26$
for bosonic string, certain anomalies appear. Their cancellation is possible via
complicated addition to string Lagrangian. To my knowledge, it is not important issue
for stationary string, but appear e.g. for rotating one, dealed with by Sonnenshein and collaborators.
For stringy Pomeron solution to be discussed below it is not yet resolved, to my knowledge.
}.  People obviously did first
the case in which string ends are fixed. For the Nambu-Goto action one can 
solve this problem and obtain string energy including quantum  string vibrations \cite{Arvis:1983fp} 
\be E(r)=\sigma_T r \sqrt{ 1-{\pi \over 6} {1 \over \sigma_T r^2} } \ee
which appends the classical linear potential by a quantum factor close to one at large $r$,
but generating certain expansion in powers of ${1/\sigma_T r^2}$.

For Regge trajectories  transition from classical 
to quantum results are more involved in general, but for
  massless endpoints it  can be done by
the following $additive$ substitution
$$ J=\alpha' E^2 \,\,\,\, \rightarrow  J+n-a=\alpha' E^2 $$
where $n$ is the quantum number for radial excitations and $a$ is the ``quantum addition to the intercept"
\footnote{In \cite{Hellerman:2013kba} it was shown, using very general assumptions,
that for massless endpoints $a=1$.}. Note how elegantly two quantum numbers for orbital and radial excitations -- $J$ and $n$ -- appear together. In a Chew-Frautschi plot $J(M^2)$ various integer values of $n$ generate ``daughter" trajectories, which are simply shifted downward from the ``parents"
by one or more units. We will not discuss them, but just mention that those predict correctly
certain observed mesons and baryons as well. 
  
As a parting comment, let us note that masses at the string ends can be viewed holographically, as just extra piece 
of a string, reaching in the 5-th dimension to the ``flavor brane". For a review on holography-inspired 
stringy hadrons see \cite{Sonnenschein:2016pim} .

Now we have completed the pedagogical part introducing stringy potentials and Regge trajectories in their simplest form. 
Now we will address much more difficult questions related with real-life QCD
strings. Those are complicated extended objects, and one has no general reasons
to assume that they simply follow Nambu-Goto geometric action. Twisting of a string
may cause extra energy: therefore higher order terms may appear in the effective string action. Let us briefly summarize what is known about them at this time.

Long strings are described uniquely by the 
 expanded form of the Nambu-Goto action
%
\be 
\label{1}
S=-\sigma_T \int_M d^2x \,(1+\partial_\alpha X^i \partial^\alpha X_i) 
\ee
The integration is over the world-volume  of the string $M$ with embedded coordinates $X^i$
in D-dimensions. The first contribution is the area of the world-sheet, and the second contribution
captures the fluctuations of the world-sheet in leading order in the derivatives.

Since the QCD string is extended and therefore not fundamental, its description in terms
of an action is  ``effective" in the generic sense, organized in increasing derivative contributions
each with new coefficients.  These contributions are generically split into bulk $M$ and boundary 
$\partial M$ terms. The former add pairs of derivatives to the Polyakov action. The first of such 
contribution in the gauge (fixed as in (\ref{1})) was proposed by Polyakov~\cite{Polyakov:1986cs}

\be 
\label{2}
+\frac 1\kappa \int d^2x\, \left(\dot{\dot {X}}^\mu\dot{\dot{X}}_\mu +2\dot{X}^{\prime \mu}\dot{X}^\prime_\mu+
X^{\prime\prime\mu}X^{\prime\prime}_\mu\right)
\ee
which is seen to be conformal with 
the dimensionless extrinsic curvature.
Higher derivative contributions are restricted by
Lorentz (rotational in Euclidean time) symmetry. The boundary contributions are also restricted 
by symmetry. The leading contribution is a constant $\mu$, plus higher derivatives. We will only
consider the so-called $b_2$ contribution with specifically
\be 
\label{3}
S_b=\int_{\partial M} d^2x \left (\mu +b_2\, (\partial_0\partial_1 X^i)^2\right)
\ee
All the terms in  (\ref{1}-\ref{3})  contribute to the static potential (\ref{0}). The first contribution stems from the string
vibrations as described in the quadratic term (\ref{1}),\be
\label{4}
\sigma_Tr\left(1+\frac {V_0}{\sigma_Tr^2}\right)
\ee
It is Luscher universal term with $V_0=-\pi/12$ in 4-dimensions.  Using string dualities,
\cite{Luscher:2004ib} have shown that also the next  two terms are $universal$ 
(true for any string action)
\be
\label{5}
\sigma_Tr\left(1+\frac {V_0}{\sigma_Tr^2}-\frac 18 \left(\frac{V_0}{\sigma_Tr^2}\right)^2\right)
\ee
Note that these contributions are the two terms of quantum string contributions re-summed  \cite{Arvis:1983fp} mentioned above, so they also
follow from the Nambu-Goto action, as of course they should. But the next terms can be modified.
For further discussion of the static $Q\bar Q$ potential stemming from the EST we refer to \cite{Aharony:2010db}.  

Summarizing: quantum and boundary corrections to the potential,
at large $r$ to order $1/r^4$, has the form   
\be
\label{VB1}
V(r)\approx \sigma_T r -\mu-\frac {\pi D_\perp }{24 r}-\frac{\pi^2}{2\sigma r^3}\left(\frac{D_\perp}{24}\right)^2 +{\tilde b_2 \over r^4} +...
\ee
 The third and fourth contributions in (\ref{VB1}) 
are Luscher and Luscher-Weisz universal terms in arbitrary dimensions, both reproduced by expanding
Arvis potential; see \cite{Petrov:2014jya} for a related discussion of the role of Luscher terms in the Pomeron structure. The last contribution is induced by the derivative-dependent string boundary contribution (\ref{3}). 

Comment 0: even if the string ends are constant in
external space, they still may depend on the  2-dimensional  coordinates on the worldvolume of a vibrating string.

Comment 1: The number of transverse dimensions $D_\perp=D-2$ is 2, if string vibrations
occur in the usual $D=4$ space-time. However, in holography $D=5, D_\perp=3$. An extra
vibration can be physically viewed as radial string excitation, as we will discuss in section
\ref{sec_string_holo}.

Comment 2: The $\mu$ term receives both perturbative and non-perturbative  contributions.
The former are UV sensitive and in dimensional regularization renormalize to zero,
as we  assume throughout. The latter are not accounted for  in the conformal  Nambu-Goto string, 
but arise from the extrinsic curvature term (\ref{3}) in the form~\cite{Hidaka:2009xh,Qian:2014jna,footnote}.

\be
\label{VB2}
\frac{D_\perp}4\sqrt{\sigma\kappa}\rightarrow \mu
\ee
Note that this contribution amounts to a negative boundary mass term in 
(\ref{3}), and vanishes for $D=4$ spacetime-dimensions. It 
 is finite for $D_\perp> 2$ in the holographic AdS/QCD approach.
 
 We will not discuss the extensive holographic studies of the EST  and related  potential~\cite{Aharony:2010db}, 
but proceed to lattice simulations of the heavy-quark potential. These studies have now reached 
a high degree of precision, shedding  light on the relevance and limitation of the string 
description.  In a recent investigation by Brandt~\cite{Brandt:2017yzw} considerable accuracy was
obtained for the potential at zero temperature and for pure gauge SU(2) and SU(3) theories.
As can be seen from Fig.~3 in~\cite{Brandt:2017yzw}, the inter-quark potential is described to
an accuracy of one-per-mille, clearly showing that both  Luscher's universal terms, $1/r, 1/r^3$
are correctly reproduced  by the numerical simulations. Indeed, for  $r/r_0>1.5$ (or $r>0.75\,$  fm 
for Sommer's parameter $r_0=0.5\,$ fm) these two contributions  describe the potential
extremely well.

Expanding further to order  $1/r^5$, or keeping the complete square root in  Arvis potential, would not improve the agreement with the lattice potential, since the measured 
potential turns up and opposite to the expansion. Brandt  lattice simulations~\cite{Brandt:2017yzw}
have convincingly demonstrated that the next correction is  of order $1/r^4$ with the opposite sign. 
The extracted contribution fixes the $b_2$ coefficient   in (\ref{VB1}) as

\be 
\tilde b_2= - {\pi^3 D_\perp \over 60 } b_2 
 \ee
with the numerically fitted values

\bea
b_2^{SU(2)} \sigma_T^{3/2} = && -0.0257 (3)(38)(17)(3) \nonumber\\
b_2^{SU(3)} \sigma_T^{3/2}=&&-0.0187 (2)(13)(4) (2) 
\eea
(for the details and explanation regarding  the procedure and meaning of the errors we refer to~\cite{Brandt:2017yzw}).
Note that the overall contribution of this term to the potential is positive.

In summary: according to modern lattice studies, at $r\approx r_0=0.5\,$ fm the static potential 
contains a wiggle, visible however only with a good magnifying glass since its relative magnitude is  $10^{-3}$. 
Above this point EST describes the potential accurately, with 4 terms of the expansion defined.

Applications of QCD strings in general (and EST in particular) include not only (i) the
static potential (the Wilson loop), but, via certain duality transformation, also
two more important applications:\\
 \hspace{1cm} (ii) the correlator of two Polyakov lines at finite temperatures;\\
 \hspace{1cm} (iii)  the ``stringy Pomeron", or the tube-like stringy instanton describing amplitude of the elastic hadron-hadron scattering at high energies. \\

Basically, all three applications stem from description of a rectangular
piece of stringy membrane.  
Therefore any progress in understanding of (i) thus induces some progress in (ii) and (iii) as well. We will briefly describe those for (iii), following
\cite{Kharzeev:2017azf}, in the next section.

\section{The stringy  Pomeron} \label{sec_Pome}

Let us start this section with a general motivation, explaining why 
I decided go into the subject 
of hadron scattering and Pomerons, in spite of its apparent complexity.

Basically, there are two motivations. One is that hadronic scattering amplitudes
depend on properties of QCD strings $exponentially$. At large impact parameter $b$
(exceeding the r.m.s. hadronic sizes),
the stringy configuration produced must be $virtual$,  instanton-like, so that
the amplitude has a tunneling form  $\sim exp(-S_{cl})\sim exp(- b^2)$. 
It is indeed confirmed by the experimental data at large $b$. Furthermore, the experimentally observed 
peripheral collisions reach\footnote{At very small scattering angles or $t$, electromagnetic Coulomb
forces dominate the strong interactions.} $b$ as large as $2\, fm$,  which means there are string lengths $larger$
than what we study in quarkonia or Regions. The second motivation
is that, put in the exponent, even relatively small quantum corrections 
can be seen more clearly. 

One of the specific important issues of the field
is whether one would be able to locate a transition between the perturbative regime at
small $b$  and ``stringy one" at large $b$. We have shown above that in static potentials
the transition is now detected: similar study is badly needed as a funciton of $t$ or $b$.

The Pomeron can be defined as the  $non-positive$ $t$ (zero or negative near-zero mass squared) 
object located at the leading (highest $\alpha(t)$ ) Regge trajectory.
It has the vacuum quantum number, which means nothing\footnote{The Regge trajectory with e.g. 
the pion has isospin, and thus can be studied via isospin transfer reactions like $p n \rightarrow np$.} is transfered from one beam to another.
The universal behavior of all hadronic elastic amplitudes at large $s$ $\sim s^{\alpha(t)-1}$ 
In Fig.\ref{fig_glueballs} we already presented current data on the glueball spectroscopy,
and located this trajectory, containing $J=2,4,6$ lowest mass states. Phenomenologically, the
Pomeron {\em intercept}  $\alpha(t=0)\approx 1.08$\footnote{ 
High sensitivity to the Pomeron parameters can be illustrated by the fact
that that this small deviation from 1
is the reason why all cross section slowly grow with $s$. This ``small effect" is in fact responsible for
about doubled $NN$ cross section, between the collision energies used in 1960's and today. 
}. 

The Regge calculus, with Reggeon exchange diagrams, have been created in phenomenologically in 1960's, mostly by Pomeranchuck, Gribov and Veneziano.
With the development of pQCD it has been derived from resummation of ladder diagrams,
descibing multiple production of gluons. The so called BFKL Pomeron \cite{Kuraev:1976ge}
collects collinear logarithms and produce power of $s$ known as the BFKL Pomeron intersept
\be \alpha^{\scriptscriptstyle \bf BFKL}=1+\frac{g^2 N_c}{\pi^2}\,{\rm ln}\,2  
\ee
The Pomeron slope, as dimensional quantity, cannot of course occur in pQCD.

There is ongoing debate about both the experimental observation of the Odderon,
the C-odd exchange which would make a difference between the $pp$ and $\bar{p}p$
elastic amplitudes, and the pQCD predictions for its trajectory. It is supposed to be
calculated from the bound state of three Reggeized gluons: some studies put its intercept  at $\alpha(0)<1$, some exactly at $\alpha(0)=1$. 

Let us now introduce the {\em semi-classical  stringy}  Pomeron. It originates from 
 the  paper \cite{Basar:2012jb} and thus will be called the BKYZ Pomeron. Using the instanton method and stringy Lagrangian, these authors had
 calculated the forward scattering amplitude between two small dipoles relativistically moving relative to each other.
 
   In order to explain the stringy Pomeron, let me first take a detour and consider 
   related classic problem {\em of the
 $e+e-$ pair production in constant electric field}. It is widely known as the Schwinger process,
 as he solved it in detail in 1950's. However we will not discuss neither the Schwinger paper, nor
 even earlier Heisenberg-Euler paper, but much earlier semiclassical work \cite{Sauter:1931zz}
 from 1931 (well before anyone else).  
 
 The EOM of a charge relativistically moving in constant electric field is a classic
 problem which everybody had encounter in E/M classes. Writing it in a form
 \be {dp\over dt}={d \over dt}\big( {v \over \sqrt{1-v^2} } \big)={eE\over m}\equiv a \ee
 one finds the solution
 \be v(t)={a t \over \sqrt{1-a^2 t^2}},\,\,\,\, x(t)={1\over a}\big( \sqrt{1+a^2t^2}-1\big) \ee
 (check small and large time limits).
 
 Transformation into Euclidean time $\tau=it$ of the trajectory yields 
 \be x_E(\tau)={1\over a} \big( \sqrt{1-a^2\tau^2}-1 \big) 
 \ee
 and between $\tau=-1/a$ and $\tau=1/a$ it describes the Euclidean path in shape of the semicircle. This should not surprise us: in Euclidean world time is no different from
 other coordinates, and electric field $G_{01}$ is no different from the magnetic ones, so in the 0-1 plane
 the paths are circles, like they are in all other planes.
 
 The physical meaning of the semicircle is as follows:
 it describes tunneling through the ``mass gap" in the spectrum of states: there
 are no states
 between $E=-\sqrt{p^2+m^2}$ and $E=\sqrt{p^2+m^2}$ with real momentum $p$,
 but on the  Euclidean path we found the momentum is imaginary.  
 
 Calculating the action 
 $ S=\int (-m ds-eEx dt) $ one gets the Euclidean version of it for the semicircle $S_E=\pi m^2/2eE$. The semiclassical probability (square of the amplitude) of
 the pair production is then
 \be P\sim e^{-2S_E}\sim e^{- \pi m^2/eE} \ee
 
  In QCD problem we want to solve there are two colliding dipoles, $each$ having
  a flux tube in between two charges. If $each$ of them produces
  $\bar q q$ pair,  one notice that quarks  are under constant tension force, so the problem is
  analogous to that just considered. The probability 
  to create two quark pair (and split each dipole into $two$) 
  would corresponds to trajectories of 
  quarks making $two$ circles, on the worldsheet of each dipole. 
  
\begin{figure}[t]
  \begin{center}
  \includegraphics[width=8cm]{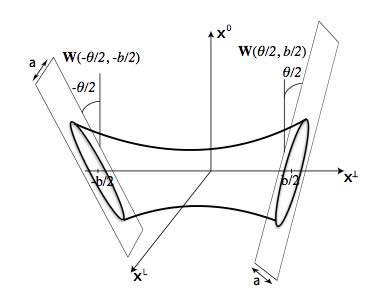}
  \caption{The dipole-dipole scattering due to closed string exchange. The impact parameter  ${\bf b}$ is the transverse distance between two colliding dipoles. Reclined
  by angles $\pm \theta/2$ dipole paths will become colliding after the result is
  transferred to Minkowski kinematics via $\theta \rightarrow iy$ where $y$ is relative rapidity.}
  \label{fig_tube}
  \end{center}
\end{figure}

  Now, imagine that, instead of production of massive quarks, we would like to think of purely stringy process,
  in which two ``circular holes" on the worldsheets get connected by some ``tube-like" configuration
  connecting the worldsheets of two dipoles. The setting is sketched in Fig.\ref{fig_tube}. 
  Such  stringy object with minimal Euclidean action (area times the tension) is the ``stringy instanton"
  describing tunneling between two colliding dipole worldsheets. The semiclassical probability of 
  it to happen in the forward scattering amplitude will give us the ``stringy Pomeron".
 
 Some introduction to the formal setting of BKYZ paper is perhaps needed. 
 The starting expression is
 \be { i \over 2s} T(\theta,q) =\int d^2b e^{i(\vec{q}_\perp\vec{b})} \langle (W(-\theta/2,-b/2)-1)  (W(\theta/2,b/2)-1)\rangle \ee
 where $q_\perp$ is the momentum transfer, $b$ is impact parameter and $W$ is the Wilson loop for a dipole 
 $$ W(\theta,b)={1 \over N_c} Tr \big( Pexp (ig\oint dx A) \big) $$
 For clarity, the authors start in perturbation theory and calculate this amplitude due to two-gluon
 exchange $$T(\theta,b)\approx {N_c^2-1 \over  32\pi^2 N_c^2} \big({g a \over b}\big)^4cotan^2(\theta) $$
 with $a$ being the dipole size. Minkowski analytic continuation is done via $\theta \rightarrow iy$ where $y$ is relative rapidity of the dipoles. Note that the scattering profile (its $b$-dependence) is power-like and can be obtained just by dimensional argument: pQCD, lacking
any dimensionful quantities, cannot give anything else. 
 
 The effective string theory has a parameter $\sigma_T b^2$, which, as we will see,
 will appear in the action and then in the scattering profile.
 
 The classical solution itself is obtained with simplified Polyakov form of the action
 $$ S={\sigma_T \over 2} \int_0^T d\tau \int_0^1 d\sigma ( \dot{X}^\mu \dot{X}_\mu +X'^\mu X'_\mu) $$
 The length of the tube is obviously the impact parameter between the two dipoles (protons) $b$, assumed to be large. The
 circumference of the tube is $\beta=2\pi b /\chi$ where the quantity in denominator represents the collision energy $\chi=log(s/s_0)$, 
 $s=(p_1+p_2)^2$ is the Mandelstam invariant related to the collision energy. $\chi$ can also be viewed as 
 the rapidity difference between the two colliding beams. 
 
 The product $b\beta$
 is the tube area, which,  times the string tension $\sigma_T$, gives the  action of stringy instanton (presumed large for 
 semiclassical setting to be valid). 
 
One can 
map the two problems -- static potential and the Pomeron -- to each other, as
  discussed in the Appendix of~\cite{Shuryak:2017phz}. It is done  via some duality relation, by exchanging time and space. One  also needs to
add another mirror image of a potential, to match the boundary conditions, which explains appearance of factor 2 below. 
In this case the two partition functions of the string and its excitations become identical. 
The explicit transformation is 
\be 
2b \leftrightarrow {\hbar \over T}, \,\,\, \,\, \beta \leftrightarrow 2r
\ee 
Assuming the correspondence between the potential and the Pomeron is exact
we can map the potential (\ref{VB1}) onto
the Pomeron scattering  amplitude in in impact parameter space  as
\be
\label{VB3}
{\cal A}{(\beta, b)}\approx 2is\,{\bf K}\approx 2is \,e^{-{ S}(\beta,b)} \ee
with 
\be
\label{SPOT}
{ S(\beta,b)}=\sigma_T\beta b-2\mu b-\frac{\pi D_\perp}{6} \frac{b}{\beta}
-\frac{8\pi^2}{\sigma}\frac{b}{\beta^3}\left(\frac{D_\perp}{24}\right)^2 
-{ 2^5 b \tilde b_2\over \beta^4}
\ee

with $\sigma=\sigma_T/2$ and $2\pi\sigma_T=1/\alpha^\prime_{\bf R}$. 
Now, we can recall the parameters of the ``tube" and set
 $\beta=2\pi b/\chi, \chi={\rm ln}(s/s_0)$
  
 Following this substitution, one observes that the leading and subleading  terms
have very different roles and  energy dependence. The leading two contributions 
 \be
 \label{VB4}
e^{\chi\frac{D_\perp}{12}-\frac{b^2}{4\chi\alpha_P^\prime}}
\ee
give the  Pomeron form of the amplitude, with the intercept value $$\alpha(0)-1=\Delta=\frac{D_\perp}{12}$$
 The Gaussian dependence on $b$ is consequence of the famous ``Gribov diffusion", derived originally
 in perturbative setting, due to random emission of gluons in ladder diagrams. In fact strings also
follow the same
``diffusive law"
\be 
b^2 \sim \chi ={\rm ln}\left({s\over s_0}\right)
\label{eqn_diff}
 \ee  
which exists equally for perturbative gluons and strings.

Furthermore, one should recognize that the stringy Pomeron approach exists in two versions, the flat space  and the holographic
ones. In the former case the space has two flat transverse directions $D_\perp=2$,
while in the latter the string  also propagates in the third  and curved dimension. Since Gribov diffusion
also takes place along this coordinate, identified with the ``scale" of the incoming dipoles,
the expressions we will use are a bit modified from the standard expressions.
One such effect, derived for the  the  BKYZ Pomeron  is the modification of the Pomeron intercept due to extra dimension
\be 
{D_\perp \over 12} \rightarrow {D_\perp \over 12}\left(1-{3(D_\perp -1)^2 \over 2D_\perp \sqrt{\lambda} }\right) 
\label{eqn_Delta_P}
\ee
 Here $D_\perp=3$ and $\lambda=g^2N_c$ is the 't Hooft coupling, assumed to be  large. 
 In the range of $\lambda=20-40$,  (\ref{eqn_Delta_P})  is in the range 0.14-0.18. For the numerical analyses to
 follow, we will use for the Pomeron intercept the value $\alpha_{\bf P}(0)-1=\Delta_{\bf P}=0.18$.
 (This  happens to be  not far from the flat space value of $\frac 16=0.166$.) 

%
 
    Experimentally, the Pomeron scattering amplitude exibits both a real and imaginary part. 
The real part can in fact be measured at two locations:  \\(i) at  small $t\approx 0$, by observing the
interference with the electromagnetic scattering induced by a photon exchange; (ii)   at the location  of 
the diffractive node $t_{node}$ where the imaginary part vanishes and the subleading real part gets visible.  For the interference measurement,
the results are expressed in terms of the so called $\rho$ parameter 
\be
 \rho={{\rm Re}({\cal A}(s,t=0)) \over {\rm Im}({\cal A}(s,t=0)) }
\ee
  The  TOTEM data~\cite{Antchev:2016vpy} give
  \be 
\rho(\sqrt{s}=8\, {\rm TeV})=0.12\pm 0.03, \,\,\,\,  
\rho(\sqrt{s}=13\, {\rm TeV})=0.098\pm 0.01
\ee
The textbook 
description of the Regge  scattering amplitudes relates the $\rho-$parameter with the signature factor 
%
%
which for small $t$ is captured by the phase factor $e^{i \pi \Delta_{\bf P}}$. It is small
if $ \Delta_{\bf P}$ is small, in agreement with data.

A real part of the amplitude may appear in Reggeon calculus because    
the Pomeron can be exchange both is $s$ and cross-channel $u$. In
 the Euclidean calculation of the scattering amplitude one needs to include two contributions, with both the
Euclidean angle $\theta$ as well as $\pi+\theta$, representing the  $u$ channel.
The resulting amplitude, after analytic continuation to Minkowski space and  the Fourier-transforming from the impact parameter, to momentum transfer $\sqrt{-t}$, has the form
$s^{\alpha(t)}+u^{\alpha(t)}$.
    
\begin{figure}
\begin{center}
\includegraphics[width=6cm]{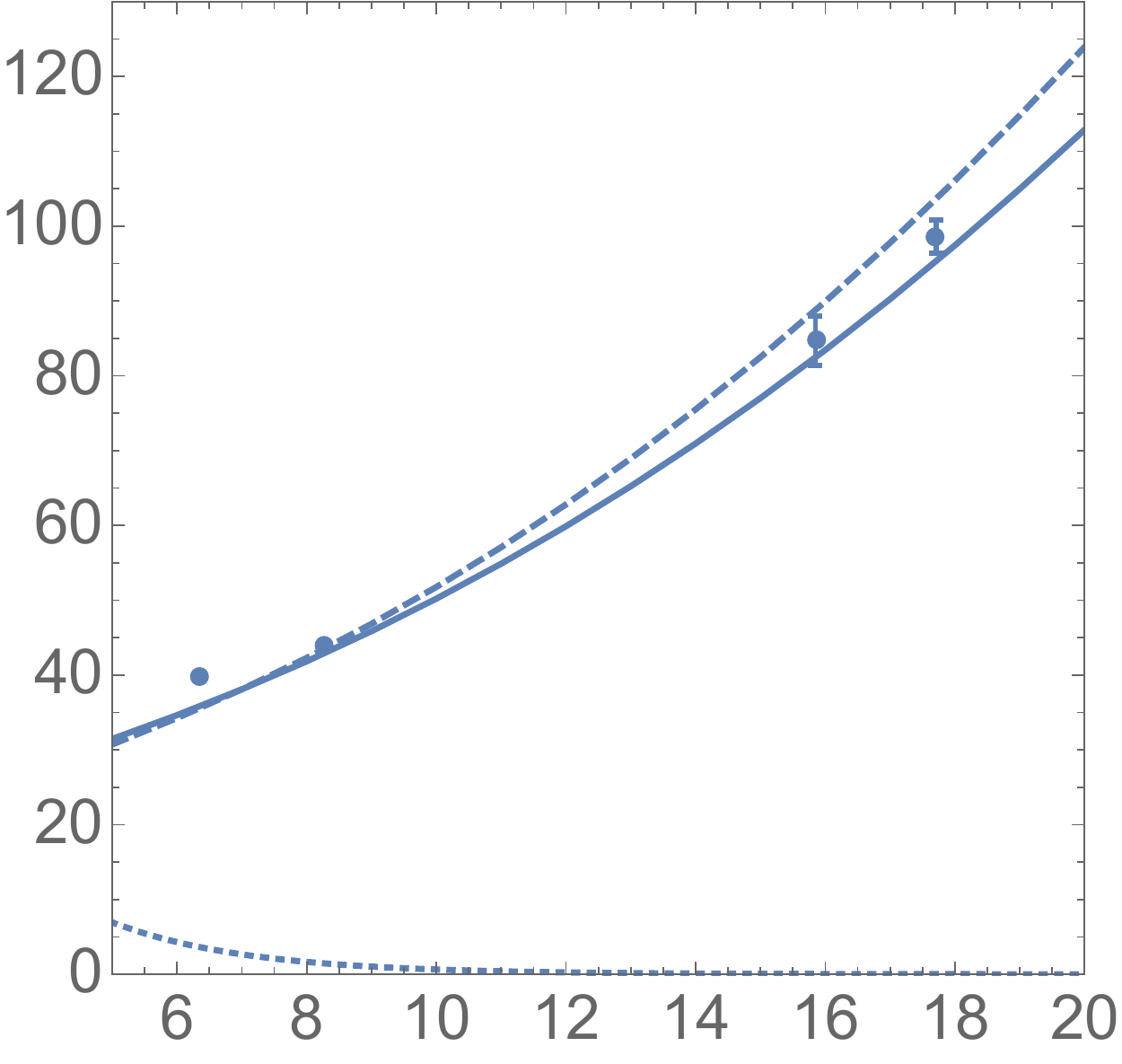}
\includegraphics[width=6cm]{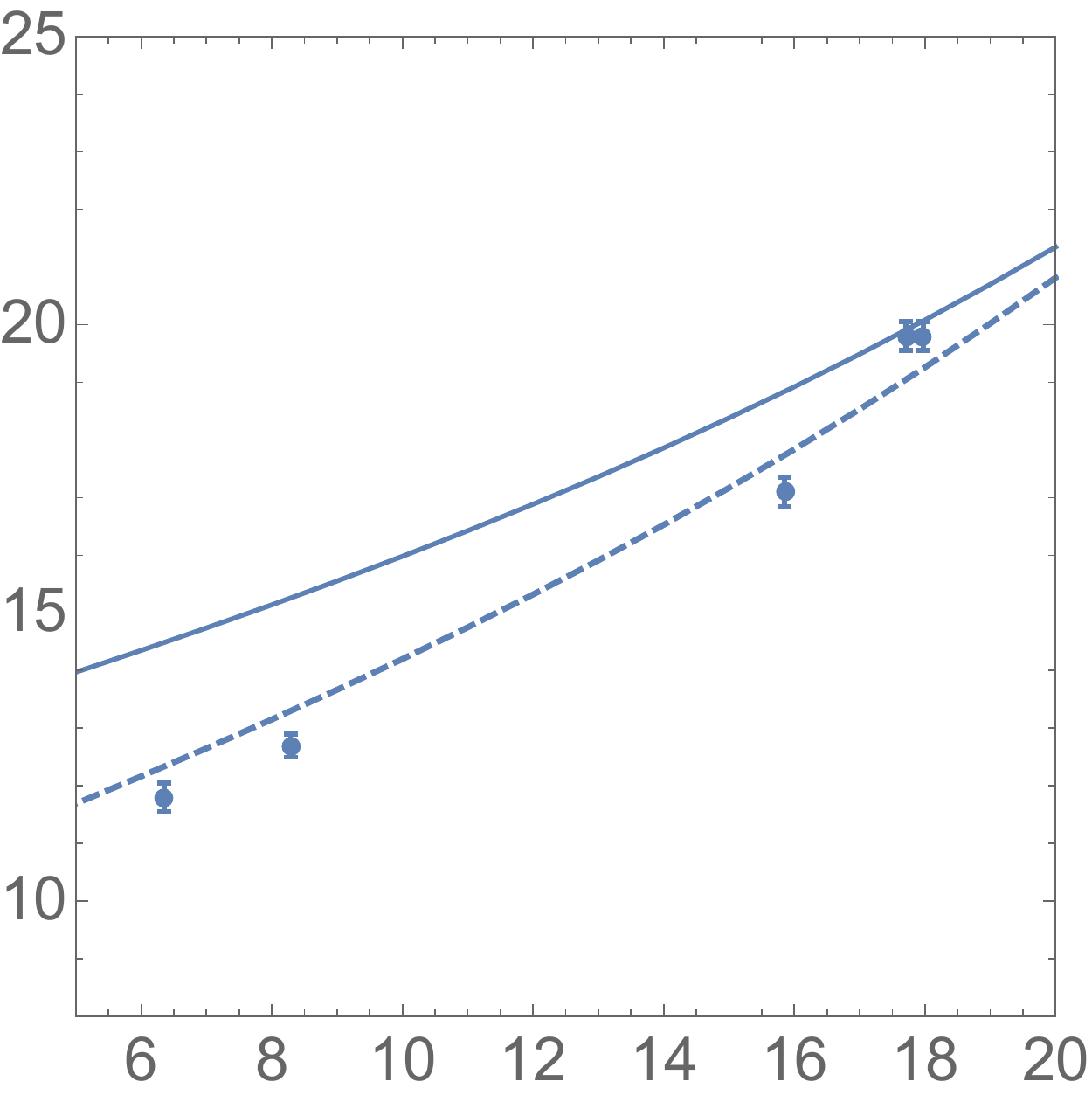}
\caption{
The upper plot shows $\sigma_{\rm tot}$ in $mb$ versus the log of the collision energy $\chi={\rm ln}(s/s_0)$. 
The left-side (low energy) data points at $\sqrt{s}=27,63\, {\rm GeV}$  are from the old ISR measurements, and the three right-side points, for 
$\sqrt{s}=2.76, 7, 8 \, {\rm TeV}$ are from the TOTEM measurements. The dotted line in the lower plot indicates the contribution of the Reggeons other than the Pomeron (from the PDG fit.)
The right plot shows the elastic slope  $B$ (${\rm GeV}^{-2}$). The curves are for model profiles discussed in our paper}.
\label{fig_pheno_from_profile6}
\end{center}
\end{figure}

The main information we have about the profile of the scattering amplitude
can be summarized as follows. One observable is the total cross section
with few datapoints shown in 
the upper plot of Fig.~\ref{fig_pheno_from_profile6}. 
Another important parameter is the so called  slope of the elastic scattering amplitude $B$ 
\be
B(t=0)=\left(-\frac{d{\rm ln}\sigma_e}{d|t|}\right)_0=\frac 12 \langle b^2\rangle 
\ee
The corresponding data are shown  in 
the right plot of Fig.~\ref{fig_pheno_from_profile6}. Both 
 grow with the collision energy due to the effective growth
of the proton size induced by Gribov diffusion process. 


%

Our last comment is that there is a serious problem with the description via Pomerons of the $pp$ collisions at LHC energies: at small enough $b<b_{\rm bd}$  the protons basically are black discs, with probability of scattering very close to unity. 
 Obviously any
structure in the amplitude inside the black disc is unobservable. One can model it with
multi-Pomeron expressions and unitarization of the amplitude, but inherently
there is no accuracy at small $b$. The only information remaining is the large-$b$ or small-$t$
slope we discussed above. 
To go around this difficulty one can hope to get data on $\gamma p$ collisions from the future Electron-Ion Collider: the photon coupling to Pomeron is small and no multi-Pomeron processes will be needed.

\section{Interaction of QCD strings: lattice, AdS/QCD and experiments} \label{sec_string_holo}

In the ``dual superconductor" approach the electric flux tubes 
 are treated as dual to Abrikosov's solution for magnetic flux tubes  in semiconductors. 
Depending on the ratio of the two lengths of the problem, associated with the gauge field and
``Higgs" field  masses, their interaction at large distances can be attractive or
repulsive. For superconductors this generates type-I and type-II superconductors.
We already discussed above  that QCD vacuum is of the type-I, which means that QCD strings 
should attract each other at large distances. 

 Obviously, one can  arrange lattice configuration with four static charges and two strings, and  study their mutual interaction. And indeed, for pure gauge theories their mutual attraction has been confirmed.
However, the situation is different for pure gauge theories and QCD with the light quarks. 
In the former case the lightest hadron is the scalar glueball, with a mass of about $m_{0^{++}}\approx 1.5\, GeV$.
Therefore  the interactions can only be very short-range $\sim exp(-m_{0^{++}} r)/r $.

In the real-world QCD  the lightest mesons are pions, of mass $m_\pi=0.138 \, GeV$ and its scalar chiral partner $\sigma$ meson with a mass of about $m_\sigma\approx 0.5 \, GeV$,
and therefore much longer-range string-string interactions are possible.
The pion is isovector, and cannot be emitted by pure glue state, so we are left with sigma\footnote{Analogy to $NN$ nuclear forces suggest that one needs to include 
the isoscalar vector $\omega$ meson as well, as its repulsive force nearly cancel the
attractive sigma term. In the application discuss below we have all kind of string pairs,
string-string and string-antistrings, with equal probability, so one may think in this case
the sign-changing vector exchange averages out to zero. Yet in case of a Pomeron or glueball
Reggeons, or the baryon junction Reggeon, we have strong-antistring or three strings, respectively. Perhaps in this case the omega meson exchanges need to be included. 
}.

\begin{figure}[htbp]
\begin{center}
\includegraphics[width=10cm]{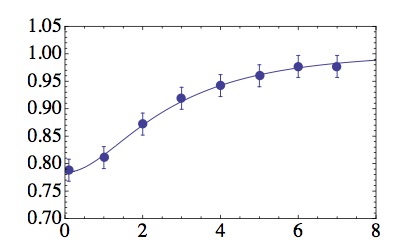}
\caption{The normalized chiral condensate perturbed by a flux tube, as a funciton of the
coordinate transverse to the tube. The lattice data are from \protect\cite{Iritani:2013rla}, the
curve is a fit with the  sigma meson 2d propagator described in the text.}
\label{fig_iritani_fit}
\end{center}
\end{figure}

In order to understand how QCD vacuum is modified around a string, one can
perform lattice studies, measuring VEVs of various operators around it.
For sigma meson the operator is the isoscalar scalar quark density $\langle\sigma(x)\rangle=\langle\bar q q(x)\rangle$. In Fig.\ref{fig_iritani_fit}
the (normalized to vacuum) value of this VEV is shown as a function
of transverse distance from the string (in lattice units), from \cite{Iritani:2013rla}.
 Note that all deviations from 1 -- the vacuum value of the quark condensate -- is
 indeed small, as it is expected to be suppressed by $1/N_c^2$. Even at the string center the suppression effect is only about 1/5 or so. 

The curve at this plot is from \cite{Kalaydzhyan:2014zqa}, it is a fit to the lattice data by the expression
\be {\langle\bar q q(x_\perp) W\rangle \over \langle W \rangle}=1-C K_0(m_\sigma \tilde x_\perp) \ee
where the regulated transverse distance is defined by $\tilde x_\perp^2=x_\perp^2+s_{string}^2$. The $K_0$ is Bessel function, corresponding
to massive scalar propagator in $d-1=2$ spatial dimensions normal to the string.
The fit parameters used are $C=.26, s_{string}=.176\, fm, m_\sigma=600\, MeV$. 
In the string-string interaction the following dimensionless parameter enters
\be g_N\sigma_T={ \langle\sigma\rangle^2 C^2 \over 4 \sigma_T}\ll 1 \ee
which is of the order of few percents. Thus one finds string ensembles subject to
scalar effective description\footnote{
Another theoretical approach in which string interactions can be studied is based on 
 $holographic$ models, originating from
the AdS/CFT correspondence,   of the conformal $\cal N$=4 supersymmetric gluodynamics
to string theory in $AdS_5 \times S^5$  10-dimensional space-time. We put some elementary introduction to it in Appendix. 
Since these models start with string theories with 10-d superstrings, the  strings are
 natural point-like objects ``in the bulk". Their ends lead naturally to fundamental charges on the
 boundary -- that is, in the 4-d manifold where the gauge theory (and ourselves) are located. 
While the string shape in
curved space is not so simple, its total energy (static potential) is $V(r)\sim 1/r$. Indeed,
it is obvious by dimension, because conformal theories lack any dimensional parameters.
The original  AdS/CFT correspondence has been generalized to some ``bottom-up" holographic models, collectively known as  
AdS/QCD: for a review see  \cite{Gursoy:2007cb}. }.

The final topic in this section, devoted to flux tube interactions, is its experimental aspect.
In fact producing a single  string is hardly possible:  the color field flux needs to be returned.  
We have already mentioned that because the Pomeron can be viewed 
either as an exchange of a closed string or production of two strings connecting the
colliding hadrons: thus the minimal number of produced strings is $two$.
But there are occasions in which many more strings are produced\footnote{
For example, when a proton flies through a diameter of a heavy nucleus the mean number of 
protons it interacts with at LHC energies is about  $ n_A \sigma_{NN} (2R_A) \approx16 $ If so, one needs to deal with at (minimum) 32 strings in such
``central $pA$ events". This number  of course returns to two strings or a single Pomeron for very peripheral  $pA$ collisions.
The open question is: at which impact parameters one has to describe the system as a set of strings, and at which all strings get ``collectivized" into a common QGP fireball? }.

  Completing the subject of QCD string-string interactions, let  us briefly discuss
  what such interactions imply for multi-string systems. 
Collective interaction of an ensemble of strings were studied by
\cite{Kalaydzhyan:2014zqa} using the sigma exchange in 1+3 d space time,
and in holographic setting later. In both of them, the strings were assumed to be
stretched in the same longitudinal direction, as it is the case in not-too-early time
in high energy collisions. We call configurations with many parallel strings a ``spaghetti" state.

\begin{figure}[h]
\begin{center}
\includegraphics[width=10cm]{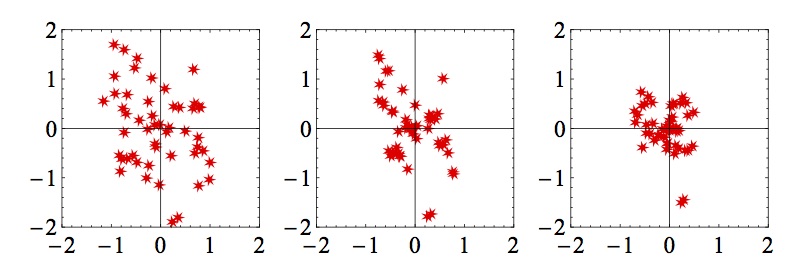}
\caption{Snapshots of multi-string configuration evolution in time, at $t=0.1,0.5, 1. \, fm/c$. The string locations (in fm) in the transverse plane are shown by stars, they are assumed to be
all strenched in the same longitudinal direction (not shown). }
\label{fig_string_collapse}
\end{center}
\end{figure}

So, classical string dynamics is restricted to motion 
the transverse d-1=2 or 3 space, in which strings are just points. The simulations
are Molecular dynamics (MD), or simply solving classical equation of motion of
the strings. In Fig.\ref{fig_string_collapse} we show an example of snapshots at subsequent time: as one can see, the central part of the multi-string system
undergoes clustering resembling the gravitational collapse. For each configuration
one can calculate the value of the quark condensate, modified according to collective influence of all strings: typically it rapidly develop regions in which such suppression
is about complete. What it means is that there is chiral symmetry restoration
at the center of the system, or that multi-string systems rapidly create a QGP fireball.

\section{String balls}
    Historically, studies of the {\em self-interacting} string balls started in the framework
of fundamental string theory in
   critical dimensions (26 for bosonic strings). The theoretical questions discussed were related to the understanding of the transition from the free strings, via {\em string balls}, to black holes.


     The main question relevant for this  transition is very simple. 
As we discussed in the previous section, the interaction between QCD strings
is weak, and the same weak coupling regime is believed to hold for fundamental
strings\footnote{We remind that 
   massless modes of closed strings include gravitons; therefore, it is a candidate for the theory of quantum gravity.}. So, for a short string the selfinteraction is negligible.       
A very large string (or in fact any large object), if described by gravity which grows with mass more than any other interaction, is subject to gravitational collapse.
Therefore, gravity becomes a dominant force, and  sufficiently massive strings should  be black holes of the classical gravity. These two limits are obvious.
The main idea is that {\em at some intermediate mass range} the gravity and other
forces can be balanced, producing stable gravitationally bound object\footnote{
For example stars and (gaseous) planets exist due to balance between thermal
pressure and gravity.}.
 
 Let us start  with free strings. A ``random walk" process,   of $M/M_s$ steps, where $M_s\sim 1/\sqrt{\alpha'}$
  is the typical mass of a straight string segment. If so, the string entropy  scales as the number of segments
  \be S_{ball}\sim M/M_s
  \ee
 The Schwarzschild radius of a black hole in $d$ spatial dimensions is
  \be  R_{BH}\sim \left( M\right) ^{1 \over (d-2)} \label{R_BH} \ee
   and the Bekenstein entropy
   \be S_{BH}\sim {Area }\sim M^{d-1\over d-2} \ee
 Thus, the equality $S_{ball}= S_{BH}$ can  only be reached
   at some special critical mass $M_c$.  When this happens, the Hawking temperature of the black hole
   is  exactly the string Hagedorn value $T_H$ and the radius is at the string scale.
   So, at least at such  value of the mass, a near-critical string ball
   can be identified -- at least thermodynamically -- with a black hole.

   However, in order to understand how exactly this state is reached, one should first address the
   following puzzle.  Considering a free string ball (described by the Polyakov's near-critical  random walk), one would estimate its radius to be
   \be {R_{ball,r.w.} \over l_s} \sim \sqrt{ M }
   \ee
for any dimension $d$.  This  answer does not fit the Schwarzschild radius $ R_{BH}$ given above (\ref{R_BH}).

The important element  missing is the self-interaction of the string ball: perhaps, Susskind was the first who pointed it out.
 A more quantitative study  \cite{Horowitz:1997jc} had used the mean field approach,  and then  \cite{Damour:1999aw}
completed the argument, by using the correction to the ball's mass due to the self-interaction. Their reasoning can be nicely summarized by
 the following schematic expression for the entropy of a self-interacting  string ball of radius $R$ and mass $M$,
\be
S(M,R)\sim M\left(1-{1\over R^2}\right) \left(1-{R^2 \over M^2}\right)\left(1+{ g^2 M \over R^{d-2}}\right)
\ee
where all numerical constants are for brevity suppressed and all dimensional quantities are in string units
given by its tension.
The coupling
$g$ in the last bracket is the string self-coupling constant to be much discussed below.
 For a very weak coupling, the last term in the last bracket can
be ignored and the entropy maximum will be given by the first two terms; this brings us back to the random walk string ball. However, even
for a very small $g$, the importance
of the last term depends not on $g$ but on $g^2M$. So, very massive balls can be influenced by a very weak
 gravity (what, indeed, happens with planets and stars).
If the last term is large compared to 1, the self-interacting string balls become much smaller in size and eventually fit the Schwarzschild radius.

Let us now switch back to QCD strings. In the preceding section their long-range interaction has been ascribed to $\sigma$ meson exchanges. We also
have demonstrated there that a sufficiently dense multi-string states can collapse.
(In this case, into a QGP fireball.)
Now we wander if this attractive force can be balanced by entropy, leading to some
stable configurations at some intermediate parameters of the problem.

The numerical model we use to study the
string balls with self-interaction. While we discuss the details of the setting below in this
section, let us emphasize on the onset its main physics prerequisites, namely, that
the ball surface should be approximately near the Hagedorn temperature, making the
string fluctuate widely outward. The string-string interaction established in the previous section is in the vacuum,
$T=0$, while the string balls we are discussing are expected to be produced
at $T\approx T_c$. Therefore, the effective $\sigma$ meson mass
is expected to be reduced, in fact to zero in QCD with strictly massless quarks.

  Following a bit Wilson's strong coupling expansion,
we place the strings on links of a $(d=3)$-dimensional lattice.  Strings are assumed to be in contact with a
heat bath, and a partition function includes all possible string configurations.

Intersections of the strings are not included because
of the repulsive interaction at small distances. Even for the Abelian fields, which
add up simply as vectors,
the action is quadratic in fields (no commutators), and intersections are energetically not favorable.
An exception (in the lattice geometry) is the case of exactly oppositely directed fluxes,
when a part of the string should basically disappear.
We had not included this complication believing that the total entropy and
energy of the string ball will not be affected much.

Instead of using boxes (with or without periodic boundary conditions) as is customary in the lattice gauge
theory and many other statistical  applications, we opted for an infinite space (no box).  Instead the temperature $T$ is {\em space
dependent}. We think it better corresponds to the experimental situation. Furthermore, the
string ball surface is automatically near criticality and thus strongly fluctuating;
this aspect will be important for our application of initial deformations below.

The ``physical units" in gluodynamics, as in lattice tradition, are set by putting
the string tension to its value in the real world:
$ \sigma_T = (0.42\, GeV)^2 $. Numerical lattice simulations
have shown that gluodynamics with $N_c>2$ has a first-order
deconfinement phase transition, with $T_c/\sqrt{\sigma_T}$ very weakly dependent
on $N_c$  (for review, see, e.g., Refs.~\cite{Teper:2009uf}).  Numerically, the critical temperature of the gluodynamics is $T_c\approx 270\, \mathrm{MeV}$.

It has been further shown that the effective string tension of the $free$ energy $\sigma_{F}(T)$ decreases with $T$; a point
where it vanishes is known as the Hagedorn point. Since this point is above $T_c$, some
attempts have been made \cite{Bringoltz:2005xx} to get closer to it by ``superheating" the
hadronic phase, yet some amount of extrapolation is still needed. The resulting value was found to be
\be {T_H \over T_c}= 1.11    \label{eqn_T_H} \ee

The nature of the lattice model we use is very different from that of the lattice gauge theory (LGT).
First of all, we do not want to study quantum strings and generate two-dimensional surfaces in the Matsubara $R^d S^1$
space, restricting ourselves to the thermodynamics of strings in $d$ spatial dimensions.

The lattice spacing $a$ in LGT is a technical cutoff, which at the end of the calculation is expected to
be extrapolated to zero, reaching the so-called continuum limit. In our case $a$ is a physical parameter
characterizing QCD strings: its value is selected from the requirement that it determines the
correct density of states.
Since we postulate that the string can go to any of $2d-1$ directions from each point
(going backward on itself is prohibited), we have $(2d-1)^{L/a}$ possible strings of length $L$.
Our partition function is given by
\be Z\sim \int d L \exp \left[ {L \over a} \ln( 2d-1)- {\sigma_T L \over T} \right]\,,
\ee
and hence the Hagedorn divergence happens at
\be T_H={\sigma_T a \over  \ln( 2d-1)}. \ee
Setting $T_H=0.30\, GeV$, according to the lattice data mentioned above and the string tension,
we fix the three-dimensional spacing to be
\be a_3= 2.73 \, GeV^{-1} \approx 0.54\, fm. \ee
It is, therefore, a much more coarse lattice, compared to the ones usually used in LGT.

If no external charges are involved, the excitations are closed strings.
At low $T$ one may expect to excite only the smallest ones. With the ``no self-crossing" rule we apply,
that would be an elementary plaquette with four links. Its mass,
\be E_{plaquette}=4\sigma_T a\approx 1.9 \, GeV \,,\ee
is amusingly in the ballpark of the lowest glueball masses of QCD.
(For completeness, the lowest ``meson" is one link or mass 0.5 $GeV$, and the lowest ``baryon" is three links -- $1.5\, GeV$ of string energy --  plus that of the ``baryon junction".)

At temperatures below and not  close to $T_H$, one finds extremely dilute $\mathcal{O}(e^{-10})$
gas of glueballs, or straight initial strings we put in. Only close to $T_H$ do multiple string states get excited;
the strings rapidly grow and start occupying a larger and larger fraction of the available space.

Before we show the results of the simulation, let us discuss the opposite ``dense" limit of our model.
We do not allow strings to overlap; the minimal distance between them is one link length, or again about $0.5 \, fm$.
Is it large enough for the string to be considered well separated? We think so, as it is about three times the string radius

%

The most compact (volume-filling or Hamiltonian) string wrapping visits each site of the lattice. If the string is closed, then the number of occupied links is the same as the number of occupied sites. Since in $d=3$ each site is shared among eight neighboring cubes, there is effectively only one occupied link per unit cube, and this wrapping produces the maximal energy density,
\be {\epsilon_{max} \over T_c^4} = {\sigma_T a \over a^3 T_c^4} \approx 4.4 \ee
(we normalized it to a power of $T_c$, the highest temperature of the hadronic phase).
It is instructive to compare it to the energy density of the gluonic plasma, for which we use the free Stefan-Boltzmann
value
\be   {\epsilon_{gluons} \over T^4}=(N_c^2-1) {\pi^2 \over 15}\approx
5.26 \ee
and conclude that our model's maximal energy density  is comparable to the physical maximal energy density
of the mixed phase we would like to study.

The algorithm consists of a sequence of updates for the each string segment, such that the configuration gradually approaches equilibrium. The spatial distribution
over all three coordinates is close to the Gaussian one, as is exemplified in the upper figure. Yet it is not
just a Gaussian ensemble of random points, as the points constitute extended objects - strings.

 \begin{figure}[t]
  \begin{center}
  \includegraphics[width=6cm]{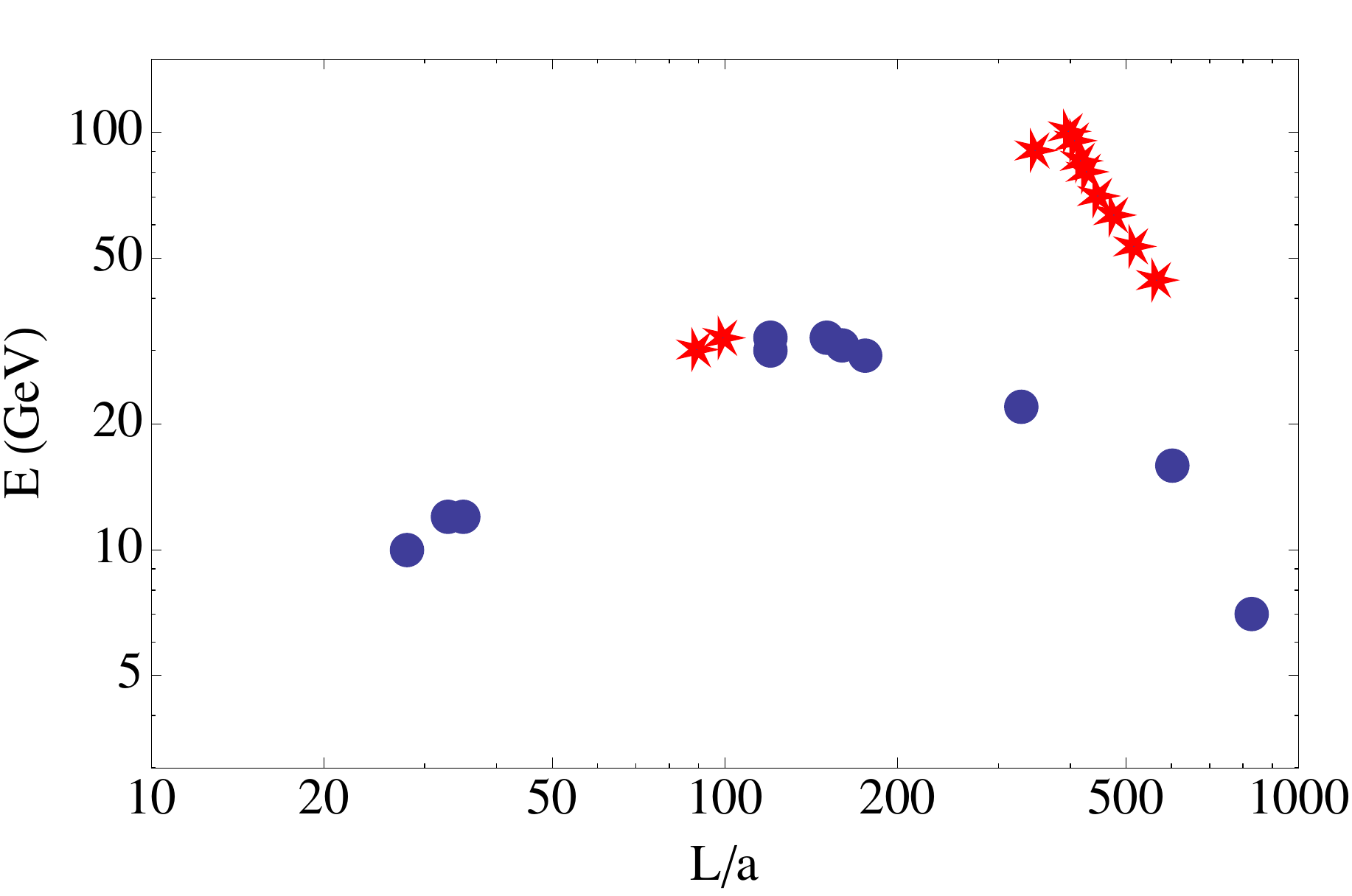}
      \includegraphics[width=6cm]{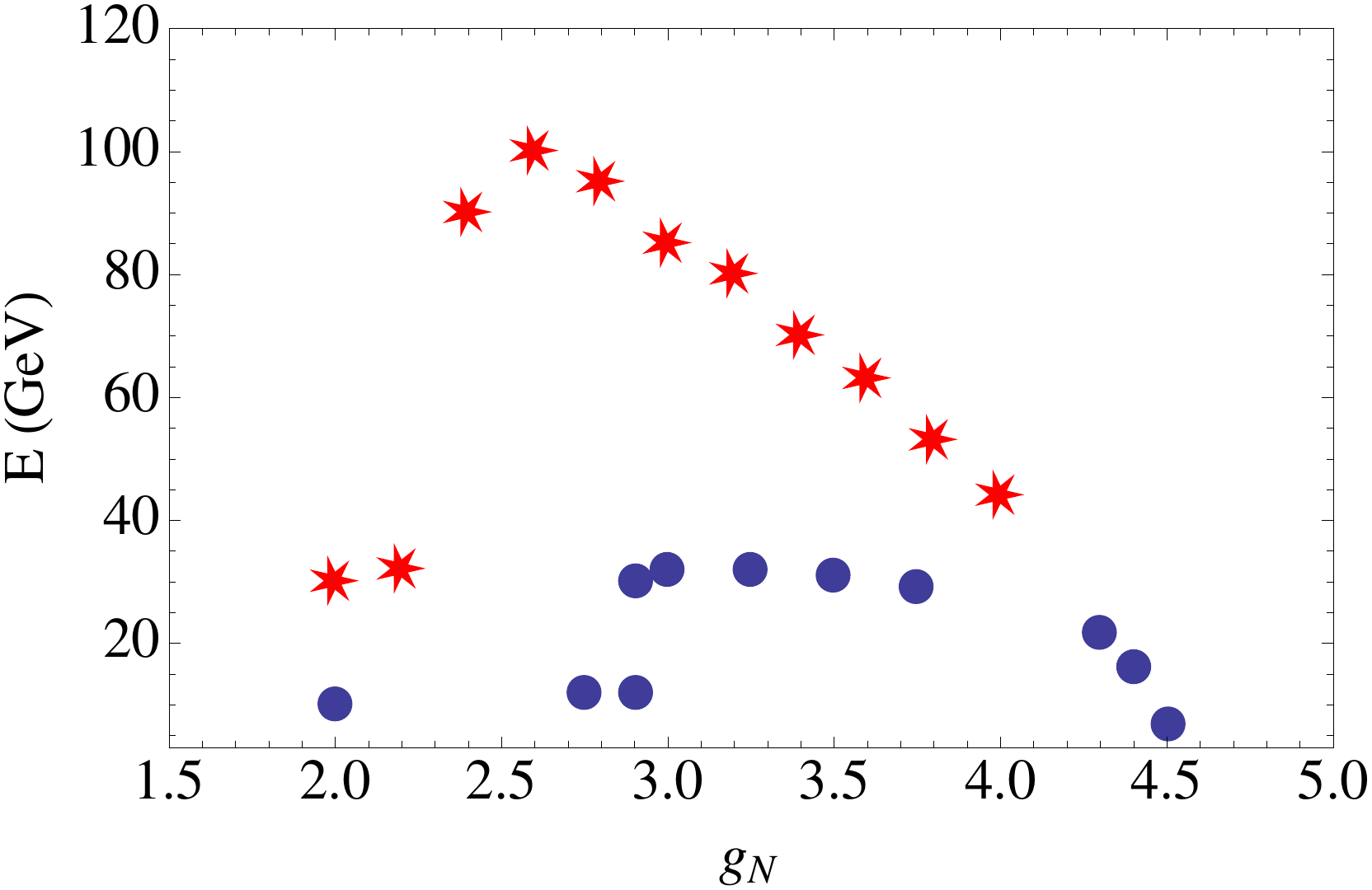}
  \caption{ Left plot: the mean energy of the cluster $E(g_N)\, [GeV]$ vs the mean length of the string $L(g_N)/a$.
   Lower plot: the mean energy of the cluster $E(g_N)\, [GeV]$ vs the ``Newton coupling" $g_N\, [GeV^{-2}]$.
Points show the results of the simulations in setting  $T_0=1\, GeV$ and size of the ball $s_T=1.5a, 2a$,
for circles and stars, respectively.
}
  \label{fig_runs0}
  \end{center}
\end{figure}

In Fig.~\ref{fig_runs0} (left figure), we show the calculated
relation between the average string length $L$ and its energy $E$.
Each point is a run of about $10^4$ iterations of the entire string updates after equilibration.
 While at small coupling
$E$ and $L$ are simply proportional to each other, like for noninteracting strings described above,
this behavior changes abruptly.  As the negative self-interaction energy
become important, the total energy $E$ of the ball becomes $decreasing$ with the string length  $L$.
 In Fig.~\ref{fig_runs0} (right figure), we show more details of this behavior: this plot demonstrates how total energy $E$
depends on the coupling value $g_N$.
We find a jump  at the critical coupling (for this setting) $g_N^{c1}$,
which in a simulation looks like a first-order transition, with double-maxima distributions in the energy and length.
As is seen from the figure, the precise value of the coupling somewhat depends on the system size.
At this coupling the jump in energy is always about a factor 3, and the jump in string length (or entropy)
is even larger.

In this way we observe a new regime for our system, which we will call
the ``entropy-rich self-balanced string balls". For a given fixed mass $M$, we thus find
 that string balls may belong to two very distinct
classes: (i) small near-random balls and (ii) large ones in which the string can be very long
but balances its tension by a comparable collective attraction. Discovery of this second regime
 is the main result of this paper.

  \begin{figure}[t]
  \begin{center}
  \includegraphics[width=6cm]{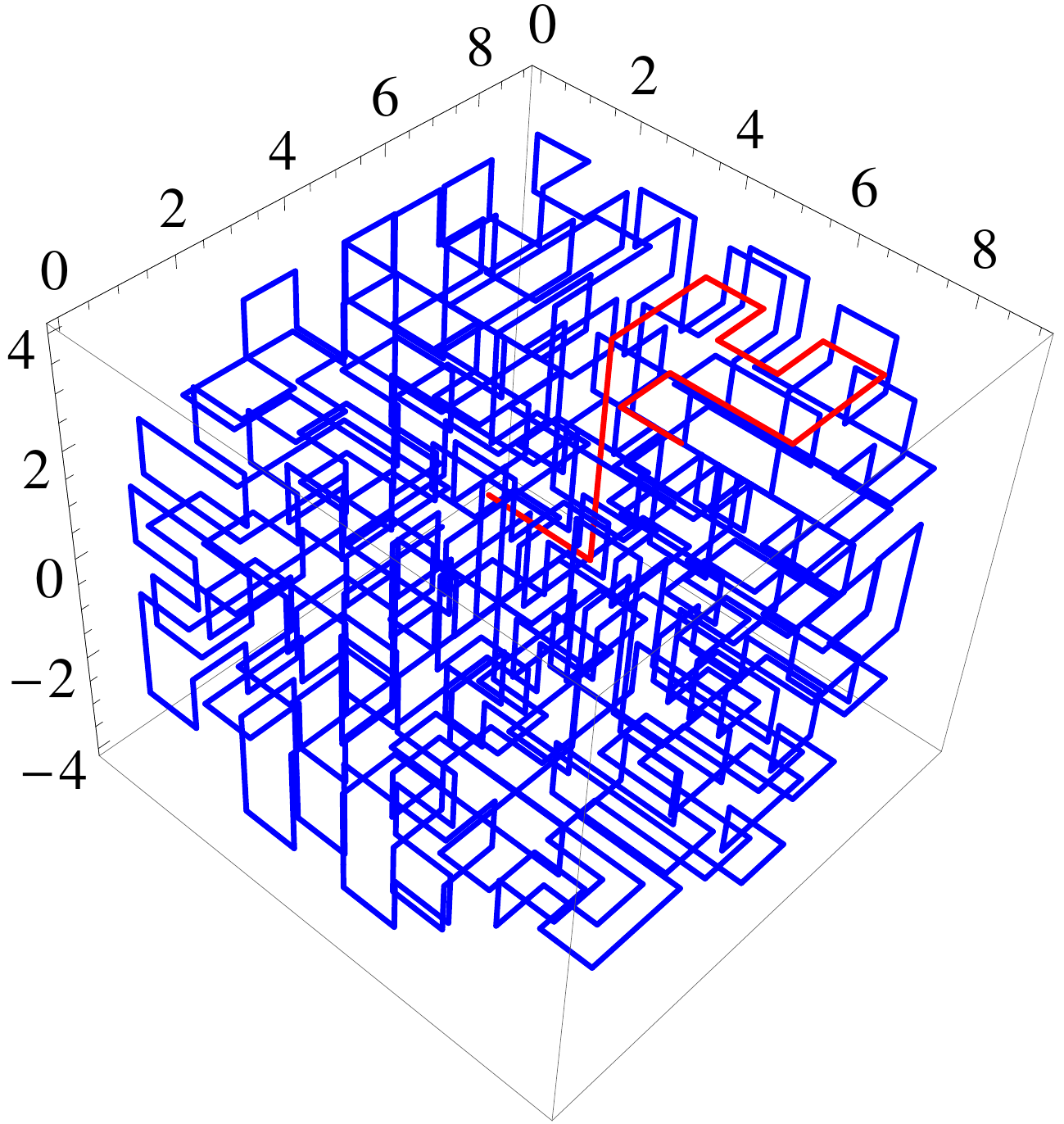}
 \caption{(Color online)
  A typical configuration in the entropy-rich self-balanced string balls ensemble. Simulation
  parameters: $T_0=1\, GeV,\, s_T=1.5a,\, g_N=4.4\, GeV^{-2}$. }
  \label{fig_31_config}
  \end{center}
\end{figure}

An example of a corresponding configuration is shown in Fig.~\ref{fig_31_config}. Note that,
in spite of a very large string length $L/a\sim 700$, the total energy is only $E\approx 17\, GeV$,
as a result of the balancing between the mass and self-interaction.
Note furthermore that that configurations are very asymmetric:
  one string is excited much more than the other, since the longer string has many more states
  than the shorter one. The same feature has been noticed on the lattice as well: typically, one very long string
  forms a large cluster, dominating over a few small clusters.
  Note further that nearly all space inside the ball with $T>T_H$ is occupied.
High entropy corresponds to a (astronomically) large number of shapes
this string may have.

 Finally,
there exists the second critical coupling, which is found to be
$g_N^{c2}\approx 4.5 \, GeV^{-2}$, above which
balancing the energy becomes impossible and simulations show immediate
collapse of the system, in which the energy quickly falls to large negative values, clearly
of no physical meaning.


Finally, admitting that  the  QCD string balls still remain a theoretical  
construction,
let us discuss whether the QCD string balls can still be produced in experiments.

We already discussed, in the preceeding section that ``tube" geometry of the surface naturally leads to a periodic coordinate and thermal description:
 the  circumference of the
tube is identified with the Matsubara time $\tau=1/T$ , inverse to the
 effective string temperature. 
 At certain values of the
 impact parameter $b$ this temperature
corresponds to the Hagedorn value; the effective tension of the string decreases, and
its high excitations become possible. As a result, as one can expect (and, indeed, sees it directly in the
observed elastic scattering profile), the scattering amplitude for such $b$ exceeds the value
interpolated by a Pomeron string expression from large $b$. 

There are two explanations proposed in literature for this rapid increase of the
scattering profile at certain $b$. The ``mainstream" one is that Pomeron amplitude
becomes too large and needs ``uitarization", or shadowing by certain multi-Pomeron amplitudes. It is also possible that
the Hagedorn transition suggested
 in Ref.~\cite{Shuryak:2013sra} is at play, so that it is the mixed phase
with long strings. 

Whatever is the interpretation, the way to experimentally proceed is to
study {\em double diffraction}, or Pomeron-Pomeron collisions. As discussed
in Refs for a long time, we already seen production of scalar and tensor glueballs.
Studying in detail the created system with few-GeV mass is a way to go.
 
\chapter{Holographic Gauge-Gravity duality} \label{sec_holo}
In the QCD-related setting, the first example of the duality is that between the {\em weak-coupling}
description in UV, in which the fields are quarks and gluons, and the {\em chiral effective Lagrangians} describing IR properties in terms of light mesons.

More attention in 
 these lectures was devoted to {\em electric-magnetic duality}, also related with the RG
 evolution from the UV to IR momentum scales. It goes from the ``electric" description in terms of 
quarks and gluons, to ``magnetic" one in terms of monopoles and dyons. 

The dualities we are going to discuss in this and the next chapters are a bit different.
They are ``holographic", the UV and IR ends are in this case connected by
an extra coordinate (called $z$ or $u=1/z$). The Lagrangians and equations of motion are
written for fields living ``in the bulk", between the two limits. The RG flow becomes 
just their dynamics in extra dimensions.  After the problem in the bulk is solved, the
physical predictions are extracted ``holographically", by projecting via certain
procedures of the results ``on the boundary", where the QCD-like gauge theories are
located. 

As this projection goes, pointlike fundamental objects in the bulk become somehow ``blurred"
on the UV boundary.
A single point source in a bulk generates  an instanton on the boundary, of some finite size $\rho$ equal to the distance to the source. A pointlike fundamental string
in the bulk corresponds to finite-width QCD string. Any object moving into IR
appears growing in size on the boundary: this is how one describe ``fireball explosions" of high energy collisions. 

\section{D-branes}
D-branes are topological solitons of the string theory: but since we are not expecting/ requiring
its knowledge we will not discuss their structure in-depth. (The reader  interested in their
historical  origins and structure should consult e.g. the TASI lecture of Polchinski \cite{hep-th/0611050} or the original papers mentioned there.)
Our aim in this chapter is to explain the geometric properties of these objects, and  their connections
to our main object of interest, the gauge theories.  

There is no place here to present string theory, but we need to start with it, to some elementary extent.
Strings are extended object with one spatial extension moving in time. Their dynamics is described by $D=d+1$ dimensional $external$ coordinates $X^\mu, \mu=0..d$ of some  manifold $M$, all being the functions of two
internal coordinates $\sigma_a,a=0,1$, also sometimes called $\tau,\sigma$, with time-like and space-like signatures respectively. 
The resulting 2-d surface is a world history of the string propagation.
Obviously the string action should be invariant under re-parameterization of both sets of coordinates. Classic example is the Nambu-Goto
action 
\be S= {1 \over 2\pi \alpha' } \int_M d^2\sigma \partial_a X^\mu  \partial_a X_\mu   \ee
where the (dimensional)  coefficient for historical reasons  is written like this, with  the constant $\alpha'$ has prime
(derivative) because it has been first introduced as the ``slope of the Regge trajectories".
For QCD strings it defined their tension, and also for fundamental string theory its value defines the basic  scale of the theory. The bosonic strings are consistent\footnote{Anomalies appearing during string quantization cancels in these dimensions. We will not consider string quantization: 	but we discussed some of its consequences in chapter on the QCD strings.} in $d=26$ while superstrings (with fermions) can be formulated in $d=10$ or 11.

The branes, or more properly  $D_p$  branes, can be seen as a generalization of the concept of a string, to an object with $p$ spatial coordinates.
The letter $D$ in the
 name came from the Dirichlet boundary condition $X^\mu=const$ on the manifold boundary
$\partial M$ ( as opposed to Neumann boundary condition for the derivative at the boundary). 
Brane world history is $p+1$ dimensional manifold, as they can move in time.
At $p=1$, the $D_1$ brane is a string-like object like the fundamental string: it has the same classical action, except a coefficient, the string tension,
is different. 

For reasons soon to become obvious, a choice most popular for applications  is the $D_3$ branes.
Including time, this one has  $p+1=4$  internal coordinates. 
 Since this number is less than the space-time  of the superstring
 theory, D=10, there are 6  extra dimensions in which such branes can move and interact.

As for any soliton, symmetries induce the zero modes of the branes,
such as simple shifts of the object as a whole. When collective coordinates
corresponding to shifts (and other zero modes) are gently modulated (coordinate dependence have small gradients), these near-symmetries generate massless Goldstone modes.  One can classify all of them and calculate the corresponding effective Lagrangian. It turns out that
spin of these modes can be 2,1,0. Therefore interactions induced by their exchange
between branes appear to be gravity-like\footnote{We will not discuss ``theory of everything"
based on superstrings, aiming to find a theory unifying gravity with the Standard Model.
}, vector-like and scalar-like.

\section{Brane perturbations induce effective gauge theories}
Suppose one finds a $p$-dimensional membrane solution for the D-branes, with $p+1<d$. Without knowing its internal structure one can in general determine its lowest excitations. The issue is no different from any other membrane, e.g. that of a drum. One may start from a flat membrane with p+1 coordinates inside it, $x^\mu, \mu=0..p$, and then deform it  to 
\be y^\mu=x^\mu + A^\mu(x) \ee
The displacements $A_\mu$ are of two kinds: \\ (i) one with $\mu>p+1$ $external$ to the brane, \\ (ii) and another with $\mu=0..p$
 $internal$ to the brane.\\ From the point of view of a p-dimensional observer {\em living on the brane} 
the first type corresponds to some scalar fields, while the second are vector fields, as the index direction can be recognized.

The corresponding effective action of the deformed membrane is known as the Born-Infeld action. It can be determined from general
principles, if the action is proportional to the area 
times some undetermined coefficient, the tension $T_p$.  
\be S_p=-T_p \int d^{p+1}x e^{-\phi}\sqrt{det \left(G_{ab} +B_{ab}+2\pi\alpha' F_{ab}\right) }
 \ee
 The differential geometry tells us that the square root of the determinant of the internal metric $G_{ab},a,b=0..p$ is needed
 in order to get the invariant volume element, and the dilaton $\phi$
 appear from this determinant as well. 
 
 The term $F_{ab}$, central to our discussion, is the (Abelian) field strength corresponding to vector displacement
 field $A_a$ (internal part). If the gauge field coefficient is considered small, the square root can be expanded to second order, resulting in the $( F_{ab})^2$ Lagrangian 
 for the effective $U(1)$ theory, or ``induced  electrodynamics". One can understand its appearance as follows. Suppose 2 dimensions $X^1,X^2$ do
 belong to the brane and there is
 one nonzero field component $F_{12}$, so that one can set the displacement to be $A_2=X^1F_{12}$. What it means is certain linear stretching of a brane along 2-direction
 by amount proportional to $X^1$, such that the volume element will increase, now containing a factor $dX^1 \sqrt{1+ (2\pi\alpha' F_{12})^2}$.
 This is precisely a correction which a nonzero $F_{12}=-F_{21}$ in the Born-Infeld  determinant will produce in such setting. (The factor $2\pi \alpha'$  
 is just a matter of field normalization.)

   The term $B_{ab}$ is related with the magnetic charge density of the brane: let me not specify its role and structure at this point,
   as it is not that important for what follows.

 Let us now make a more complicated example with $N$ p-branes placed together, at the same place. Open strings
 then obtain indices at both ends, $i,j=1..N$. The displacements can be described by the matrix-valued
 vector fields $A^\mu_{ij}$. The central consequence of this construction is that the corresponding  Born-Infeld action
 will then contain the correct non-Abelian field strength $(F^{\mu\nu})^2$, with the commutator term as prescribed
by Yang and Mills. This means that the vibrations of $N$ p-branes are described by some effective non-Abelian gauge theory.

This is the reason why {\em the string theory can be  related with the physical gauge theories.} 
 How important this observation is depends on the 
opinion, which are vary vastly across the theory community. 

The $maximalists$ in the
 string theory community think that 
the $U(1)\times SU(2)\times SU(3) $ gauge groups of the Standard Model are effective theories in this very sense, that we do live on the $D_3$
brane, and  
 hope that one day we will be able to find out how exactly the compactification of unwanted 6  extra  dimensions will explain ``everything",
 or even discover these extra dimensions experimentally.
 
The $minimalists$ (to which most people outside the string community like myself belong) take this fact as a mathematical statement, providing
very useful theoretical tools to model the problem one wants to solve. 

   Whatever people's hopes may be, the  fact remains that we have no good ideas about compactification scale of the extra dimensions. Experimentally,
  it can be from something like TeV to the inverse Plank mass or $1/10^{19} GeV$. Thus, even if we do live on a brane,  we may never be able to find that out, given obvious experimental limitations.  

   But nothing stops us from using the theoretical models which originated from this
   construction, as a mathematical tool. Indeed,
as we will see below,  one can invent certain brane constructions whose vibration spectrum
is capable of reproducing the QCD  spectrum of hadrons. Furthermore, one can
calculate effective chiral Lagrangians, thermal and even   transport properties of strongly coupled Quark-Gluon Plasma\footnote{There are of course also applications of the gauge-string correspondence outside QCD,
in the strongly-coupled systems in condense matter theory, which we of course will not discuss.}.

\section{Brane constructions}
\subsection{A stack of $D_3$ branes}
The main construction leading to AdS/CFT is a stack of $N_c$
``coincident $D_3$ branes", that is located at the same spatial point in
6 external coordinates. A brane is a location at which string can end.
The (nonexcited) string has a mass equal  to its length times the tension. 
So, if the length is zero, we have $N_c \times N_c$ massless vector fields:
thus this construction has just enough vector fields to generate $SU(N_c)$ 
effective gauge theory.

As another example one can consider  ``non-coincident $D_3$ branes",
separated in external coordinates by some distances $R_{ij}, i,j=1..N_c$.  
The strings between them would have masses proportional to these distances. 
In this way one can generate models with ``Higgsing", like Georgy-Glashow
model of weak interactions.

Let us just point out that 
the extra coordinate can be compactified to a circle. 
This will generate strings with the lengths proportional to $N_c$ lengths of the segments.
This model will split gluons into massless and massive ones, modelling the adjoint Higgsing
happening when the Polyakov line has a nonzero VEV. Recall that 
precisely in this setting we came to understand instanton-dyons,
or specifically 
why in this case the instanton (a complete $D_1$
wrapping of the whole circle) can
be seen as consisting out of $N_c$ massive segments, the $N_c$ types of
the (selfdual) instanton-dyons. We will not discuss this idea in full: see the original paper \cite{Lee:1998bb}. 

\subsection{The Seiberg-Witten curve from the branes}
One more beautiful brane construction \cite{hep-th/9703166}, explains how the celebrated curve for $\cal N$=2 SYM and related theories can be interpreted (and calculated) 
directly from the equilibrium (minimal action) brane shape. 

\begin{figure}
\includegraphics[width=10cm]{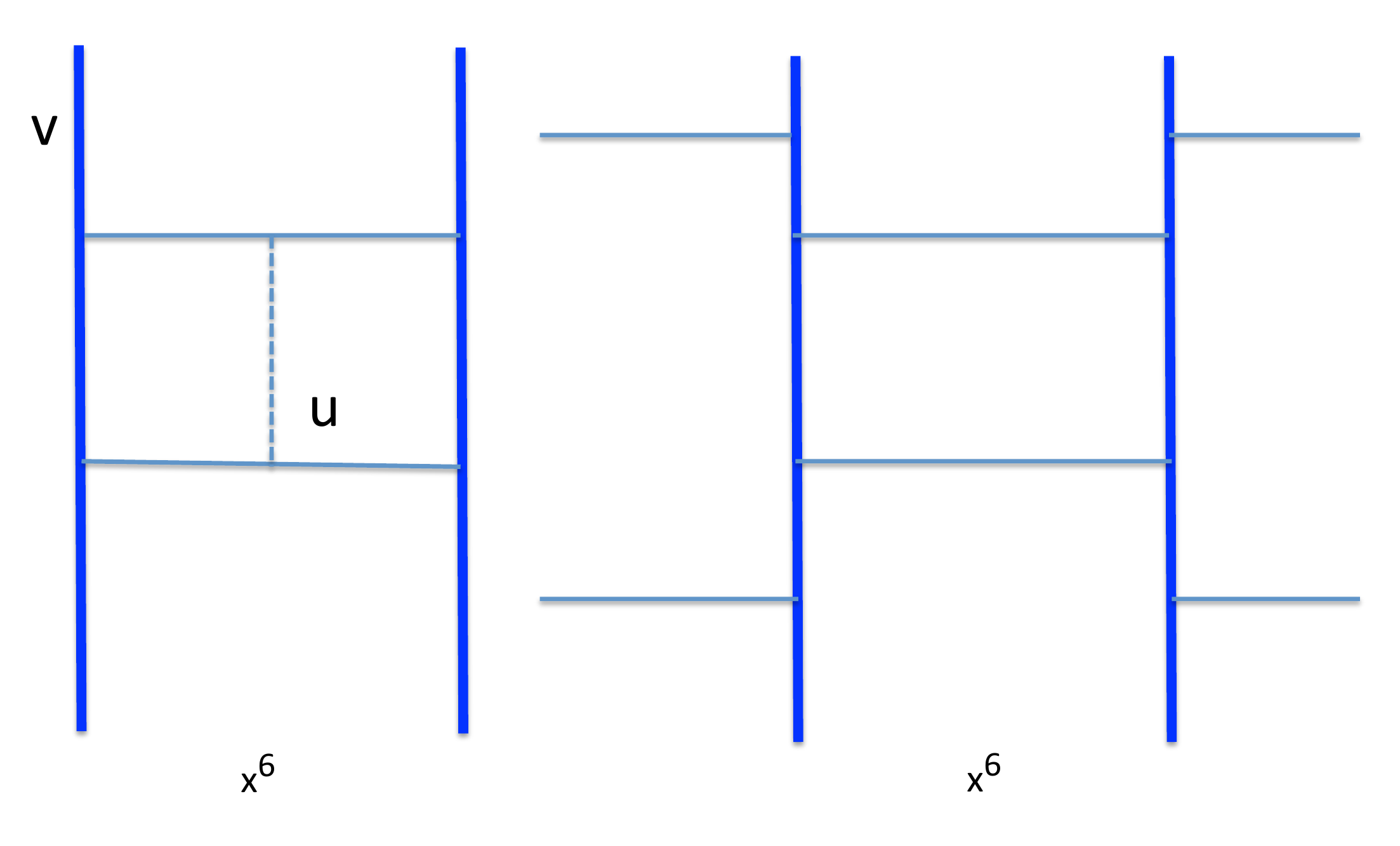}
\caption{\label{fig_W_brane_constr} 
The vertical thick lines are 5-branes, this  
direction corresponds to complexified coordinate $v$. The horizontal
lines (along $x^6$) are 4-branes, we show two of them which
corresponds to SU(2) color. Left picture has adjoint matter and
the forces on the 5-branes are unbalanced: thus the branes
are logariphmically bent at large $v$. Right picture has 4 external
lines depicting 4 flavors of fundamentally charged  matter: those
balanced the forces on 5-branes and make its position at large $v$
balanced, which thus produces a conformal theory with fixed coupling. Thin dashed line on the left connecting 4-branes will
correspond to adjoint matter particle (or monopole): 
its length is associated with the Higgs VEV $a$ (or dual $a_D$).
}
\end{figure}

It includes  5-branes extended in coordinates (0..5), located at
$x^7=x^8=x^9=0$ and $some$
 $x^6$ values to be called $\alpha_i$.
Like in a previous subsection, those are connected by some
lower dimensional branes, now of the dimension 4\footnote{
Type IIB string theory has odd-dimensional branes which
naturally generate gauge fields/theories, and so far we dealt with
them only. Dimension 4 branes lead to multi-index forms and
to get a gauge field out of it is done by another procedure
related with nonzero holonomy for those forms: see the paper.
}.  They have infinite extension in coordinates (0,1,2,3) and finite in $x^6$, namely
between the positions of the $D_5$ $\alpha_i<x^6<\alpha_{i+1}$.
It is depicted in the Fig. \ref{fig_W_brane_constr} in which $x^6$ is the horizontal direction  and the complexified coordinate $v=x^4+ix^5$ runs vertically.  (One may think
of $x^6$ to be also complexified
together with the 11-th coordinate $x^{10}$ into 
\be t=exp[-(x^6+i x^{10})/R] \ee
where $R$ is the radius of a circle to which the latter can be
compactified. (This variable is then single-valued.)

The horizontal branes would pull the vertical ones,  and if the
tensions are not balanced (as in the left picture) they would displace
the vertical branse.
 Furthermore, they would create certain
``dimples" on the $D_5$. At large distance (in $|v|$) from such
dimple the curve $x^6(v)$ must satisfy the Laplace eqn
\be \nabla^2 x^6(v)=0 \ee
 which in 2 dimensions (1 complex $v$) produce solution
 \be x^6\sim log(v)  \ee 
which is not constant even at large $v$. Adding or subtracting those
change the coefficient of the $log(v)$: this is a change of a
perturbative beta function. 

If tensions are balanced (like the right picture with 4 flavors of fundamental matter), a conformal
theory emerges. (In fact this paper lead to may new conformal models.) The positions of those horizontal branes in $v$
correspond to masses of all those flavors.

The 4-d world and also the 4-d gauge theory appears in the world volume of horizontal
6-branes: and their  mutual  HIggsing depends in the 
(vertical) distance in $v$ (see the dashed line). 
and the log correspond to perturbative running of that coupling.

Lower dimensional ones
(4-branes) can be considered as becoming narrow tubes 
made of 5-branes. 
Furthermore,  simple plots in the figure are for rigid strings, represented by the straight lines. In fact they are flexible and will bend to get to a
minimal energy shape. (Its zero Laplacian is obtained, as usual,
by complexification and then using function depending on $z$ but not 
its conjugate $z^*$.) 

Let us imagine the shape of the 5-branes is given by some equation.  Since there are two of them, it should be quadratic
\be A(v)t^2+B(v) t + C(v)=0 \ee
the power of $v$ in the coefficient is $k$, the number of horizontal
branes (e.g. 2 in our left figure but 6 on the right one). 
Zero $t$ corresponds to infinite $x^6$: so roots of $C(v)$
are right-going branes. A root of $A(v)$ is $t=\infty$ or $x^6=-\infty$,
so those count lines going to the left. 

Suppose we don't want any flavors of fundamental matter: then $A,C$
can be constants, rescaled to the equation
\be  t^2+B(v)t +1=0\ee   
 which can be further rescaled by $\tilde{t}=t+B/2$ into
 a form 
 \be \tilde{t}^2={1\over 4} B(v)^2 -1 \ee
with $B$ being degree-k polynomial, which is a known curve for SU(k) SYM. Adding fundamental matter is then making
$A,C$ non-constant.  

In summary:
this brane construction leads to explicit $derivation$
of  the Seiberg-Witten curve,
while they had to guess it in their original paper. 


\section{Brane interactions  AdS}
Closed strings form certain
massless bosonic excitations, described by 
fields in the d=10 dimensional external string space. Those  are:
 the graviton $g_{\mu\nu}$, dilaton $\phi$, antisymmetric tensor $B^{\mu\nu}$ and vector field $A_\mu$. The corresponding classical (tree level) effective action 
is
\be S={1 \over 2 \kappa_0^2 } \int d^{10}x e^{-2\phi} \left( R+4(\nabla \phi )^2  -{1\over 12} H_{\mu\nu\kappa} H^{\mu\nu\kappa} \right)  
- {c \over 4} e^{-\phi} F_{\mu\nu} F^{\mu\nu} 
\ee
The constants $\kappa_0, c$ as well as
loop corrections of higher  orders in $\alpha'$ can in principle be calculated. Here $R$ is the scalar curvature defined via the metric tensor (Einstein-Hilbert action of general relativity) and $H,F$ are the field strength tensors for $B,A$ fields.
Supersymmetric extension of this action makes full classical 10-d supergravity (SUGRA). 

Any objects existing in the theory interact at large distances via an exchange of those massless fields.
D-branes are such objects, and in fact they do have certain mass and charge densities (in respect to $\phi,A,B$ fields).

Let us now be more specific and discuss the fields induced by the branes around them in the bulk.  As they have no extension
in some coordinates (e.g. just mentioned $x_4...x_9$ coordinates for  $D_3$ brane),
their fields are that of a point objects: in general relativity this naturally implies
that they are black hole-like in such coordinates. As they are extended
in other coordinates, the proper name of such objects is {\em black branes}.
To find out their properties in classical (no loops or quantum fluctuations)
approximation, one has to solve
 coupled Einstein-Maxwell-Scalar eqns, looking for static spherically symmetric
 solutions.
This is no more difficult than to find the Schwartzschild solution for the usual
black hole, which we remind has a metric tensor in spherical coordinates
\be ds^2=g_{\mu\nu}dx^\mu dx^\nu= -(1-r_h/r)dt^2+{dr^2\over (1-r_h/r)}+r^2d\Omega^2\ee 
and the horizon radius (in fulll units) is $r_h=2G_NM/c^2$,
containing the mass $M$ and Newton constant $G_N$.  The horizon -- zero of $g_{00}$ -- is the ``event horizon": a distant observer
cannot see beyond it\footnote{Note that both metric elements  $g_{00},g_{rr}$ change
sign at the horizon, so in a way the time and space exchange places. 
It indicates that the Schwartzschild-like metrics cannot be used beyond horizon. Some other versions of metric however can do it: see GR textbooks.}. Similar expressions, with appropriate powers of the distance
can be derived for p-branes.

Note that at large distance $r\ll r_h$ deviations from flat metrics is a small correction -- there the 
nonrelativistic potential description of Newtonian gravity is possible.
Similarly far from a brane  the fields are just given by Newton+Coulomb+scalar
formulae in corresponding dimensions, and it is not hard to figure
out what are their mutual interactions.
It is important that these densities are related in a nontrivial way.
Like the BPS magnetic monopoles we had studied above, the  D-branes are BPS objects. It means attractive
forces (Newtonian due to gravity as well as to the dilaton exchange) and repulsive Coulomb forces due to vector exchanges cancel each other, so that two parallel $D$-branes do not interact. (The D and anti-D are not BPS.) As a result, a ``brane engineer" may consider a number of parallel branes to be put
at some random points: and they will stay there.
Large ($N_c\rightarrow\infty$) number of branes put into 
the same point combine their mass and thus create strong gravity field, justifying
the use of classical  Einstein/Maxwell eqns. Open string states, which
keep their ends on some branes i and j lead to effective gauge theory
on the brane with the $U(N_c)$ group. 

(The BPS objects still have quantum interaction and can form bound states. In particular electric particles (quarks,gluons) as well as monopoles and dyons can form positronium-like
series of Coulombic bound states. Explicit examples are e.g. well studied in the context of $\cal{N}$=2 SYM, the Seiberg-Witten theory.
Generalization of all of that to multidimensional SUGRA had also been studied: in this case of course gravity is added to  Coulombic electric/magnetic
forces as well as (massless) Yukawa ones.  Tuning the parameters one may make gravity to become even dominant:
the corresponding bound states have been called ``galaxies" in such limits. The reader interested in this developments can consult e.g.
the lectures by ????? )

When the black hole has a nonzero vector charge, the Gauss' law insists on constant
flux through sphere
of any radius,
thus there are nonzero fields at large $r$ and Schwartzschild solution gets modified
into a ``charged black hole". For asymptotically flat spaces there is the famous
statement, much emphasized by Wheeler:  there are ``no scalar hairs"
of black holes, as there is no Gauss theorem to protect them.
This happens to be not  true in general, and for spaces which are
AdS-like
thus scalars should $not$ be left over\footnote{In fact there is a whole recent direction
based on solution with a ``scalar atmosphere" around black branes, providing
gravity dual to superconductors on the boundary.}. 

 The solution for $D_3$ brane happens to be the so called
 $extremal$ (6-dimensional) charged black hole, which
has the lowest possible mass for a given charge. When the mass
decreases to the extreme value, the horizon shrinks to nothing,
and thus it is especially simple.
Spherical symmetry in 6d allows one to separate the 5 angles
(making the 5-dim sphere $S_5$) from the radius $r$ (in 6-dimensions
orthogonal to ``our world" space $x_1,x-2,x-3$)
and write the resulting 10-d metrics  as follows
 \begin{eqnarray} ds^2={-dt^2+dx_1^2+dx_2^2+dx_3^2 \over \sqrt{1+L^4/r^4}}+ 
\sqrt{1+L^4/r^4}(dr^2+r^2d\Omega_5^2) \label{eqn_10dBH}
 \end{eqnarray}
 Again, at large $r$ all corrections are small and we have
 the asymptotically flat space there.

\section{AdS/CFT correspondence}
AdS/CFT correspondence \cite{Maldacena:1997re} is a $duality$ between specific
gauge theory known as $\cal N$=4 super-Yang-Mills theory (SYM)
in 4 dimensions and the (10-dimensional) superstring theory in
a specific setting. 

Step one toward the $AdS_5$ space starts from  the black hole metric (\ref{eqn_10dBH})
which can be in certain region be simplified. 
In fact we will only need the metric 
  in the ``near-horizon region",  at $r<<L$, when 1 in both roots can be ignored.
  If so,  in the last term two $r^2$ cancels out
and the 5-dimensional sphere element gets constant coefficient and thus
gets decoupled from 5 other coordinates. 
Thus quantum numbers or motion in $S^5$ becomes kind of internal
quantum numbers like flavor in QCD, and it will  be 
mostly ignored from now on. What is left
is very simple 5-dimensional metric known as  Anti-de-Sitter metric. 
Using 
a new coordinate $z=L^2/r$ we get it into the  ``standard $AdS_5$ form'' used below:
 \begin{eqnarray}  ds^2={-dt^2+dx_1^2+dx_2^2+dx_3^2+dz^2 \over z^2}   \label{eqn_AdS5metric}
\end{eqnarray}     
Note that $z$ counts the distance from ``the AdS boundary'' $z=0$.
This metric has no scale and is $not$ asymptotically flat even at the boundary.
Performing dilatation on these 5 coordinates we find that the metric remains
invariant: in fact one can do any conformal transformation.
It is this metric which
is AdS in the AdS/CFT correspondence, and string theory
in this background is ``holographically gravity dual" to some conformal gauge theory at 
the boundary.

So far  nothing
 unusual happened: all formulae came straight from string and general 
relativity textbooks. A truly remarkable
theoretical discovery is the so called
``holography'': the exact duality (one-to-one
correspondence) between the 5-dim ``bulk'' effective theory in $AdS_5$
to 4-dim ``boundary'' ($r\rightarrow\infty$)  gauge theory. 
There is a dictionary, relating any observable in the gauge
theory to another one in string theory: the duality implies that all
answers are the same in both formulations. We will see below how it
works ``by examples".

The last step, which makes it  useful, is the Maldacena relations between
the gauge coupling, the AdS radius $L$ and the string tension
$\alpha'$ (which comes from the total mass of the brane set):
 \begin{eqnarray} L^4=g^2 N_c (\alpha')^2=\lambda(\alpha')^2  
\end{eqnarray} 
It tells us that  large gauge 
coupling $\lambda>>1$ corresponds to large
AdS radius (in string units) and one can use
classical (rather than quantum) gravity. At the same time the string 
and gravity couplings $g_s\sim g^2$ may remain small:
 so one may do perturbative
calculations in the bulk! 

At this point many readers are probably very confused
 by new 5-th dimension of space. One possible approach is
to think of it as 
just a mathematical trick, somewhat analogous to more familiar
introduction of the complex variables\footnote{
Suppose an Experimentalist
 measured some complicated cross section which is approximately
a sum of Breit-Wigner resonances. His friend Phenomenologist
may be able to write the answer as an analytic function with certain
pole singularities in the complex energy plane,
which will help for fitting and for evaluating integrals.
 Even better, their other friend
Theorist  cleverly  developed a ``bulk theory'',
deriving the pole positions from some
 interaction laws on the complex plane.  }.
 However, there is a perfectly 
  physical meaning of the
5-th coordinate.
One hint is provided by the fact that distance along it
$\int_a^b dl=\int_a^b dz/z=log(b/a)$ is the logarithm of the ratio.
Thus its meaning is the ``scale'',  the
argument of the renormalization group. If one takes a bulk
object and move it into larger $z$, its hologram at the boundary
(z=0) grows in size: this direction thus corresponds
to the infrared direction. 
The running coupling constant
would thus be a $z$-dependent field
called ``dilaton''.   Indeed, there are  theories with
gravity dual, in which this field (related to the coupling) 
does ``run" in $z$:
 unfortunately, known examples do not
(yet?) include QCD! In spite of that, there are efforts to built its gravity
dual ``bottom-up'', introducing weak coupling at ultraviolet
(small z) \cite{Shuryak:2007uq}
and confinement in infrared (large z) \cite{Karch:2006zz,Gursoy:2007cb}
 by certain modification of the
rules. These approaches -- known as AdS/QCD-- we would not discuss
in this review, except briefly in the subsection on bulk viscosity.

\begin{figure}
\centering
\includegraphics[width=4cm]{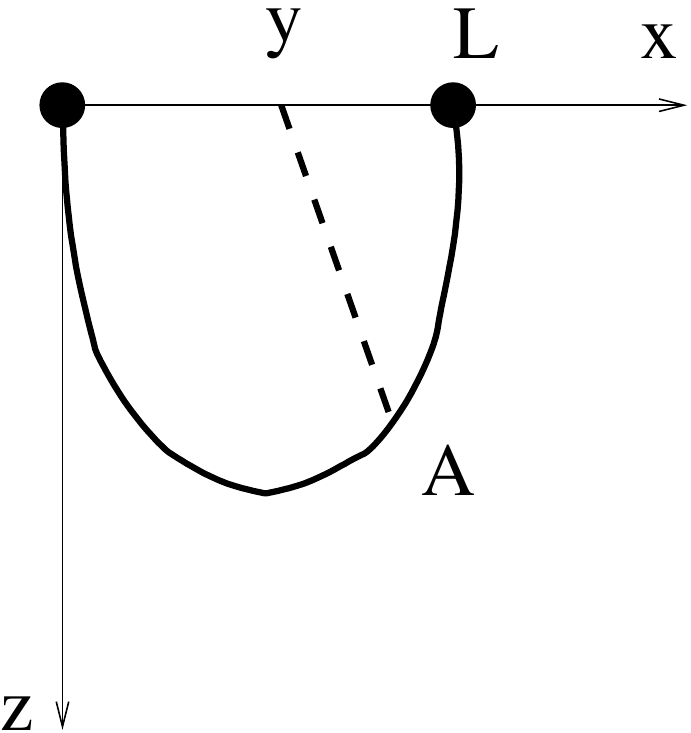}
\caption{\label{fig_M_dipole} 
Setting of the Maldacena dipole: two charges
at the boundary (black dots) are connected by the string
 (shown by solid curve) pending under
gravity toward the AdS center $z\rightarrow\infty$. Classical graviton  propagator
(the dashed line) should be used
 to calculate the ``hologram'' -- the stress tensor at
the observation point $y$. The string is the gravity source;  
the point $A$ has to be integrated
over.
}
\end{figure}

The simplified $AdS_5$ metric above is very simple, it apparently has no scale and has a conformal symmetry matching that of the  effective
boundary theory which is $\cal N$=4 SYM. That observation was the basis for Maldacena AdS/CFT correspondence. 
 Let me  start with 
 our first  example of the ``AdS/CFT at work'', related with
 the 
strong-coupling version of the Coulomb law calculated in
 \cite{MALDA2_REY}. 
The setting-- to be called ``the Maldecena dipole''
-- is shown in Fig.\ref{eos-shuryak-fig_dipole}(a), includes two static
charges (heavy fundamental quarks) separated by the distance $R$.

At weak coupling -- the usual QED -- we think of one
 charge creating the electric potential in which the other is
 placed, leading to the usual Coulomb law which in our notation is
 \be V(L)= -{g^2\over 4\pi}{1\over  L}  \ee

In the $\cal N$=4 theory at $weak$ coupling the only difference is that one can
exchange massless scalars on top of gluons. It is always attractive, and for
two heavy quarks leads to cancellation of the force, with doubling for
quark-antiquark (we discuss now). The QED coupling $g^2$ changes to
't Hooft coupling $\lambda=g^2N_c$ proportional to   the number of colors $N_c$.

Now we turn to the  AdS/CFT for the $\cal N$=4 theory at $strong$ coupling 
\cite{Maldacena:1998im} .
The electric flux in the bulk forms a 
singular object -- the fundamental or $F$ string (shown by the solid curve in Fig.\ref{eos-shuryak-fig_dipole}(a)) --
which bends from the boundary $z=0$ due to gravity force into the 5-th dimension,
like in the famous catenary (chain) problem\footnote{Another --
more  Einsteinian --way to explain it is to note that this is simply 
the shortest
string possible: it is not  straight because the space is
curved.
It is the same reason why  the shortest path from New York to London
does not look straight on the map.
}.  
The calculation thus follows from Nambu-Goto action for the string, whose general form
is
\be S={1\over 2\pi\alpha'} \int d\sigma d\tau \sqrt{det G_{MN} \partial_\alpha X^M \partial_\beta X^N  } 
\ee
where 2 coordinates $\sigma,\tau$ parameterize the string world 
line $X^M(\tau,\sigma)$, 
where
 $M,N$ are space-time indices in the whole space (10dim reduced to 5d in AdS/CFT). 
$G_{MN}$ is the space metric and $det$ stands for 2*2 matrix with all $\alpha,\beta$.
In the $AdS_5$ metric we need the components $-G_{00}=G_{11}=G_{55}=1/z^2$,
and we can think of $\sigma,\tau$ as our coordinates $x,t$: the string is then
described by only one function $z(t,x)$ and its action is reduced to
\be S\sim \int dt dx {1\over z^2}\sqrt{1+ (\partial z/\partial x)^2- (\partial z/\partial t)^2}\ee 
We will use this action for ``falling strings" below, and now proceed to further
simplifications for static string, for which there is no time derivative
and the function is $z(x)$. Maldacena uses  $u(x)=1/z(x)$ and thus the Lagrangian
becomes $L=\sqrt{(u_{,x})^2+u^4}$ with comma meaning the x- derivative.
 One more simplification comes from the fact
that $x$ does not appear in it: thus an ``energy" is conserved
\be H= p\dot q -L={\partial L \over \partial u_{,x}} u_{,x}-L=E=const\ee 
which reduces the EOM from second order eqn to just $(u_{,x})^2=u^4(u^4/E^2-1)$ which 
can finally be directly integrated
 to \be x(u)=\int_{u_m}^u {du' \over (u')^2 \sqrt{(u')^4/E^2-1)} }\ee
 The minimum position of the string $u_m$ is related to $E$ by the relation
 following from this formula at
$x=L/2$. Plugging the solution back into action and removing
divergence (which is independent of $L$)
  gives finally the total string energy, which is
the celebrated {\em new Coulomb law at strong coupling}
\be \label{eqn_new_Coulomb}
V(L)= -{4\pi^2  \over \Gamma(1/4)^4 }{\sqrt{\lambda} \over  L}
\label{coulomb}
\ee
The power of distance $1/L$ is in fact the only one possible by dimension, as
the theory is conformal and has no scales of its own. What is 
remarkable is the (now famous) $\sqrt{\lambda}$ appearing instead of $\lambda$ in
the weak coupling. 
(The numerical coefficient in the first bracket is 0.228,
to be compared to the result from a diagrammatic
re-summation below.)

What is the reason for this modification? For pedagogical reasons
let me start with two ``naive but reasonable guesses", both to be
shown to be wrong later in this section:\\ (i) One idea is that
 strongly coupled vacuum acts like some
kind of a space-independent dielectric constant, $\epsilon \sim 1/\sqrt{\lambda}$
which is reducing the effect of the Coulomb  field, similarly at all points.\\
(ii) Perhaps such dielectric constant has nonlinear effects, and thus is
not the same at different points: but the fields created by static dipole
are still just the electric field $\vec E$.

   As we get glimpse of some first results from AdS/CFT  we see that they are quite different from 
   those in weak coupling.  One would like
    to understand them better, both from the bulk (gravity) side
as well as  from the
   {\em gauge theory} side.
   
   But before we discuss this problem, we perhaps need to recall how this problem is
   solved for the familiar Coulomb forces in weak coupling.
   Since we consider a static (time-independent)  problem, one can
rotate time $t$ into its Euclidean version $\tau=i t$, which  
simplifies calculation of the perturbative diagrams.  A static charge generates a Wilson line operator
\be  W=Pexp(ig\int d\tau A_0)\ee
and the negative charge produce a conjugated line: the two are separated by the distance $L$.
The exponents can be expanded in powers of the coupling $g$: the lowest order is the first one.
The correlator of the two $A$ fields is the propagator.  Its endpoints resign at the two lines: thus a double
integral running over both lines:
\be V(L)(T_{max}) \sim  -g^2\int_0^{T_{max}} {d\tau_1 d\tau_2 \over L^2+
  (\tau_1-\tau_2)^2}\sim -{g^2 T_{max}\over L} \ee
 where  large regulator time $T_{max}$ is introduced. 
 The denominator -- the square of the Euclidean distance between the
gluon emission and absorption points-- comes from the  Euclidean
propagator in flat 3+1 space-time. Note that the propagation time of a virtual exchange gluon is of the order of the distance $(\tau_1-\tau_2)\sim L$: this cancels one power of L and 
thus one gets the usual  Coulomb potential $1/L$.
   
   It turns out the first is relatively easy. For example, both in the total energy
   and energy density we get $\sqrt{\lambda}$ because this factor
   is in front of the Nambu-Goto Lagrangian (in proper units). The reason
   the field decays as $y^7$ is extremely natural:
    in the $AdS_5$ space-time the field of a static object (a single time integral over the 5-d
    propagator, analogous to Coulomb 
   $1/r$ in flat 3d) has that very power of distance. In fact it is
   \be P_s={15\over{4\pi}} \frac{z^2}{ (z^2+r^2 )^{7\over 2}}  \ee
   with $z$ being the 5-th coordinate of the source and $r$ the 3-distance
   between it and the observation point. Thus it is likely to be a bulk gravity/scalr exchanges
   rather than the vector fields. (Note
   that gravity/scalar forces have no cancellations
   between opposite charges, unlike the vector ones.)
   
    In order to understand the same results from the gauge side we will
    need a bit of pedagogical
   introduction: the resolution will be given by
   the idea  of {\em short color correlation time}
  by Zahed and myself \cite{Shuryak:2003ja}.
 In QCD, with its running coupling, 
 higher-order effects modify the zeroth order Coulomb field/potential.

 Higher order diagrams include self-coupling of gluons/scalars
and multiple interactions with the charges. A
famous simplification  proposed by 't Hooft is the large number of colors limit
in which only $planar$ diagrams should be considered. People suggested that
as $g$ grows those diagrams are becoming ``fishnets'' with smaller
and smaller holes, converging to a ``membrane'' or string worldline:
but although this idea
was fueling decades of studies trying to cast gauge theory into
stringy form it have not strictly speaking succeeded.
It may still be true: just nobody was smart enough to sum up 
all planar diagrams\footnote{Well, AdS/CFT is kind of a solution,
actually, but it is doing it indirectly.}.

  If one does not want to give up on re-summation idea, one may
consider a subset of those  -- the $ladders$ --
which can be  summed up.
Semenoff and Zarembo \cite{Semenoff:2002kk} have done that: let us look what have they found.
The first point is that in order that each rung of the ladder contributes a factor $N_c$,
 emission time ordering should be strictly
enforced, on each charge; let us call
these time moments $s_1>s_2>s_3...$ and $t_1>t_2>t_3...$.
Ladder diagrams must connect $s_1$ to $t_1$, etc, otherwise it is nonplanar
and subleading diagram.
 Thus the main difference from
the Abelian theory comes from the dynamics of the color vector.
The (re-summed) Bethe-Salpeter kernel $\Gamma(s,t)$, describing
the evolution from time zero to times $s,t$ at two lines,
satisfies the following integral equation
\be \label{eqn_BS}
\Gamma({\cal S},{\cal T})=1+{\lambda \over 4\pi^2}\int_0^{\cal S} ds
\int_0^{\cal T} dt{1\over
    (s-t)^2+L^2} \Gamma(s,t) \ee
If this eqn is solved, one gets  re-summation of all the
ladder diagram. The kernel obviously
satisfies the boundary condition $\Gamma ({\cal S},0) =\Gamma(0,{\cal T})=1$.
If the equation is solved, the ladder-generated potential is
\be
V_{\rm lad}(L) 
=-\lim_{T\to{+\infty}}{\frac 1{\cal T} \Gamma\, ({\cal T},{\cal T})}\,\,,
\label{0a}
\ee
In weak coupling $\Gamma\approx 1$  and the integral on the rhs is
easily taken, resulting in the usual Coulomb law. 
For solving it at any coupling, it is convenient
to switch to the differential equation

\be
\frac{\partial^2\Gamma}{\partial {\cal S}\,\partial {\cal T}} =
\,\frac{\lambda/4\pi^2}{({\cal S-T})^2+L^2}
\Gamma ({\cal S,T})\,\,\,.
\label{1a}
\ee
and change variables to
$x=({\cal S-T})/L$ and $y=({\cal S+T})/L$ through
\be
\Gamma (x,y) =\sum_{m}\,{\bf C}_m \gamma_m (x)\,e^{\omega_m y/2}
\label{2a}
\ee
with the corresponding boundary condition $\Gamma (x,|x|)=1$. The
dependence of the kernel $\Gamma$ on the relative times $x$ follows
from the differential equation

\be
\left(-\frac{d^2}{dx^2} -
\frac{\lambda/4\pi^2}{x^2+1} \right)
\,\gamma_m (x) = -\frac {\omega_m^2}{4}\,\gamma^m (x)
\label{3a}
\ee
For large $\lambda$ the dominant part of the potential in (\ref{3a})
is from {\it small} relative times $x$ resulting into a harmonic
equation~\cite{Semenoff:2002kk}

At large 
times ${\cal T}$,  the kernel is dominated by the lowest harmonic mode. For large times ${\cal S\approx T}$ that is small $x$ and large 
$y$ 

\be
\Gamma (x,y)\approx {\bf C}_0\,e^{-\sqrt{\lambda}\,x^2/4\pi}\,
e^{\sqrt{\lambda}\,y/2\pi}\,\,.
\label{5a}
\ee
From (\ref{0a}) it follows that
in the strong coupling limit the ladder generated potential
is \be V_{\rm lad}(L)= -\frac{\sqrt{\lambda}/\pi}L \ee which 
has {\em the same parametric form}  as the one derived from the
AdS/CFT correspondence (\ref{eqn_new_Coulomb}) except for the
overall coefficient. Note that the difference
is not so large,  since $1/\pi=0.318$ is larger than the exact value  
0.228 by about 1/3. 
Why did it happened that the potential is reduced relative to 
the Coulomb law by $1/\sqrt{\lambda}$? It is because the relative time
between gluon emissions is no longer $\sim L$, as in the Abelian case,
but reduced to parametrically small time of relative color coherence  $\tau_c\sim 1/L\lambda^{1/2}$. Thus  
we learned an important lesson: in the strong coupling regime even the
static charges communicate with each other via high frequency
gluons and scalars,  propagating
(in Euclidean formulation!) with a  super-luminal velocity
$v\approx \lambda^{1/2}\gg 1$. 

This idea --although it 
 seemed to be too bizarre to be true --
 as we will see below to explain some of the AdS/CFT results.
Klebanov, Maldacena and Thorn \cite{Klebanov:2006jj}
have pointed out that  the reason the stress tensor around the
dipole is  different from perturbative one by a factor
$\sim  (L/r)/\sqrt{\lambda}$ is actually explained by  
 limited   relative emission time by color coherence time.
 I am sure possible usage of this idea does not ends here.
 
Before we leave the subject of Maldacena dipole, one more interesting
question is what happen if one of the charges makes a small accelerated motion
near its original position. Will there be a radiation? In a somewhat different
setting than used above, the answer was provided by Mikhailov \cite{Mikhailov:2003er}. He studied perturbations of the string and found that 
the radiated energy is described by familiar classical Li\'enard formula
\begin{eqnarray}\label{Lienard}
\Delta E= A
\int_{-\infty}^{\infty} {\ddot{\vec{x}}^{\;2}-
\left.[\dot{\vec{x}}\times \ddot{\vec{x}}]\right.^2\over
(\,1-\dot{\vec{x}}^{\;2}\,)^{\scriptstyle 3\atop  }}dt
\end{eqnarray}
in which QED weak coupling constant $A={2\over 3}e^2$
is substituted by CFT strong coupling $A={\sqrt{\lambda}\over 2\pi}$.

\section{Holography at work}
   For pedagogical reasons, let me split this idea into two parts.

The first one is the ``usual" bulk-boundary relation. In general, let us recall our setting: there is the time $X^0$, $p$ ``internal" and $d-p$ ``external"   coordinates.
 Imagine some creatures which live inside the internal coordinates and are not aware of the external ones. We of course do live on the surface of the Earth, with p=2,
 but we can see up and down and are aware of stars above, so at this stage one should imagine e.g. some blind insects which cannot jump or dig in, with only these two
 coordinates imprinted in their branes.    Suppose these creatures are intelligent enough to develop their theory of the world, obviously $p+1$ dimensional, 
 describing what happens in it. Say consequences of Earthquakes or fallen meteorites, while originated externally, create 
  displacements of the surface which can be described by certain scalar and vector fields as we described above.  Perhaps our creatures can send surface vibrations
  themselves, and communicate by those as we use sounds. Perhaps they can even travel and understand that Earth is topologically a sphere and even develop a theory about
  existence of the radial ``extra dimension",  unifying their previously separate scalar and vector fields into common vector displacement fields in 3+1 dimensions. 
 
 (It may be that our 4 dimension is a brane in a larger dimensions, and events we observe like our Big Bang have in fact simple external causes. 
 There are many fascinating papers about that, but this text
 is not a place to follow this, at this point,  purely speculative possibility.)

 \begin{figure}[t]
   \includegraphics[height=3.cm]{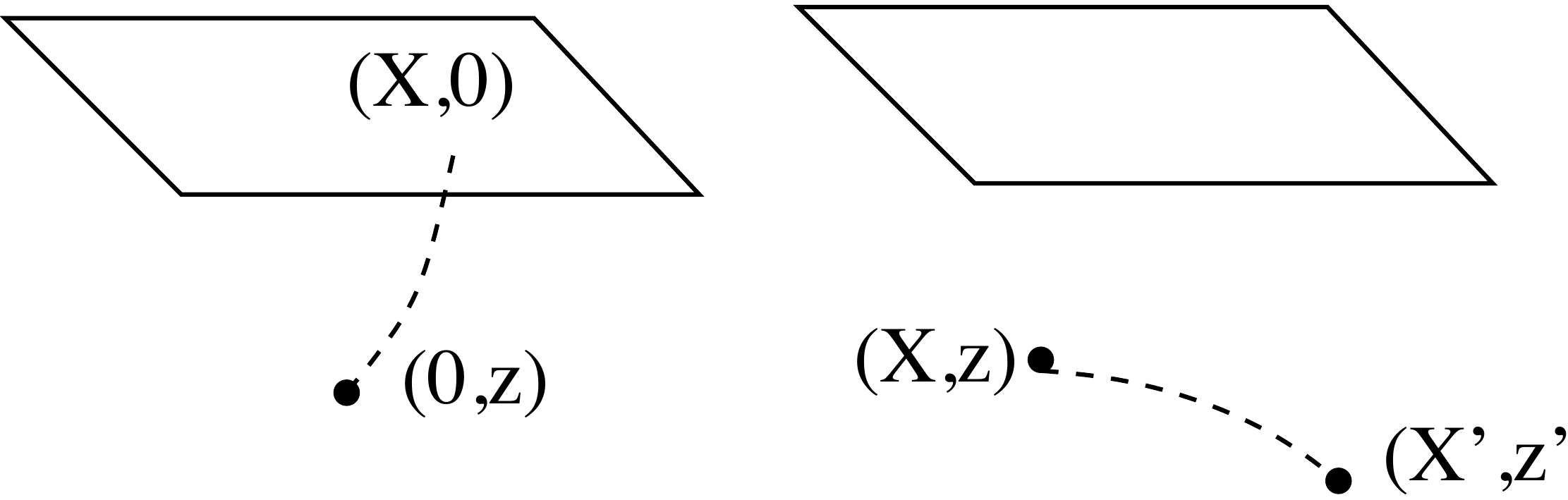} 
\caption{ (a) Setting of the bulk point event (black dots) at distance $z$ from the boundary
at the boundary, or (b) two such events (black dots) 
\label{fig_ads_inst}
}
\end{figure}

 The second statement is truly unusual, it deals with the ``holography" between the near-horizon
 region of the bulk and  the observers at the boundary of the black holes. 
It historically started with the discussions of the information paradox 
of classical gravity,
 which states that information can fall into a black hole and get lost for the outside observers\footnote{
 Loss of information, which is the same as entropy growth, is of course rather standard
 phenomenon in statistical mechanics. Everyone knows that if you throw a book into a bonfire,
 the information contained in it is lost. Same may happen if you drop a book into a black hole.
 However, from
quantum mechanical point of view 
  the pure state should remain pure state, so  the information should be somehow recoverable.
  We will not discuss huge literature on this ``information paradox".
 }.
 
 Hawking famously found that black holes should emit thermal radiation, with
 $T$ given by their total mass $M$. Bekenstein argued that since 
 in thermodynamics $dM=T dS$ one can evaluate the {\em universal
 amount of entropy} produced when you increase the black hole mass by $dM$
 \footnote{So, dropping your Ph.D. thesis weighting 1kg, 
 or an equivalent 1 kg stone, produce the same amount of entropy!}.
  The black hole entropy scales as its area, so it is expected to be
 stored  in the vicinity of the horizon. 
 These developments lead to an idea that some $holographic$\footnote{A hologram keeps all 3-d information in a 2d plane.} relation should exist between the dynamics/history observed far from black hole in $d$ dimensional space
 and its ``compactified form" (by strong gravity) in the $d-1$ dimensional region
  near horizon. 
    
 The holographic AdS/CFT duality is in fact a precise correspondence between the two, namely that  $  certain$ boundary and near-horizon theories
 can be {\em dual} to each other,  in the sense of complete/exact equivalence.
  The boundary effective theory is not only applicable to describe something like
 long-wavelength sounds, as is always the case, but it is in fact an exact copy of the bulk quantum theory.  
 AdS/CFT correspondence is believed to be one of such examples: many test have been performed and they all passed well, yet we don't have a formal proof of it.

 \subsection{A hologram of a point space-time source: an $instanton$!}
 Instead of discussing deep issues related to holography, let me demonstrate few examples of how one use it in applications.
Our first example would be
an image of a point space-time event -- sometimes called the $D_{-1}$ 
brane because its word history has dimension $p+1=0$-- seen on the boundary, see Fig.\ref{fig_ads_inst}(a). Since a point cannot produce 
vector or tensor field (it lacks directions for the indices)
 the only light bulk field it can emit are the two scalars, the axion and the dilaton. It thus immediately follows that the hologrphic image would be lacking any
energy and momentum, since only $g_{\mu\nu}$ can induce the stress tensor $T_{\mu\nu}$, but there would be non-zero values of the operators $G^2$ and $G\tilde G$. If the object
has the same couplings to both, the images be also equal (perhaps up to a sign).

In order to calculate this simplest diagram one needs  bulk-to-boundary propagators in $AdS_5$
which we discuss in Appendix. In the usual coordinates in which the metric is
\be ds^2 = { dX^\mu dX_\mu+ dz^2 \over z^2} \ee
with $\mu=0..3$ and convolution done with the usual Minkowski metric, with minus one for the zeroth component.
 Feynman (the analytically continued  Euclidean) propagator is expected to be just the function of the invariant distance $d(X,X')$ between the two points,
which in this space is
\be cosh(d(X,X')) =1+{ z^2+z'^2+\sum_{i=0}^p(X^p-X'^p)^2 \over z z' } \ee
in fact the propagator is just a power of the combination in the r.h.s. above, namely
\be   D(X,X') \sim  {z^4 z'^4 \over  \left[ X^\mu X_\mu +z^2+z'^2 \right]^4 } \ee
The ``holography" is a projection from the (d+1 dimensional) bulk to (d-dimensional) boundary.
It is
 done in two steps: (i) one should take one point to the boundary, $z'\rightarrow 0$; (ii) the power of $z'$ of the amplitude corresponding
to the dimensionality of the boundary operator in question should be ``amputated". 

{\bf Example}: Both operators $G^2$ and $G\tilde G$ have mass dimension 4, thus one should
amputate exactly the 4 powers of $z'$ in the propagator.
Thus we recover the distributions of these quantities at the boundary
\be  G^2(X),G\tilde G (X)\sim  {z^4  \over  \left[ X^\mu X_\mu +z^2 \right]^4 }
\ee
So, it now becomes clear that the object found on the boundary, as {\em an image of the bulk point event}, is nothing else but the
{\em gauge field instanton}! The 5-th coordinate of the bulk point $z$ is nothing else but the instanton radius $\rho$, confirming the idea that it is a ``scale" variable.

Note further that a bulk point even cannot be couple to gravitons, as a point lacks needed indices to make a tensor field. In the example at hand, it implies that gauge instantons 
must have {\em zero stress tensor} $T^{\mu\nu}=0$ at any point, which is also 
true and nontrivial\footnote{Note that the fact that instantons describe classical tunneling paths only
require zero stress at endpoints, not locally everywhere.}.  

Relations between the gauge theory instantons in $\cal{N}$=4 SYM and the $AdS_5$ space has been
noticed independently and simultaneously with the AdS/CFT correspondence, while studying the semiclassical (weak coupling) instantons in 
\cite{Dorey:1999pd}, see also a review  \cite{Dorey:2002ik}. Basically, the instanton moduli space includes  the $AdS_5$ space, if $\rho=z$ is the 5-th coordinate. For example, the volume element of the $AdS_5$ metric (\ref{eqn_AdS5metric}) is
\be \sqrt{det(g)}={1 \over \rho^5} \ee
which is a factor well familiar from the semiclassical instanton measure.  What this agreement tells us is that instantons of the $\cal{N}$=4 SYM
populate the internal $AdS_5$ space $homogeneously$.

The same logic then implies that the {\em interaction between  
instanton-antiinstanton pair}  is given by the diagram Fig.\ref{fig_ads_inst} (b)  in which one should use the same  bulk-to-bulk scalar propagator in $AdS_5$, which is
the function of ``geodesic distance" between two points in the bulk.
One indeed obtains this nontrivial result, now in much simpler
(and ``geometrically motivated")  way that we did in the instanton chapter.

Let us complete this example with one important comment: the ``holography" just described cannot give us boundary gauge field $A_\mu$ of the instanton
(or in fact anything having a color index): we only get colorless operators    $G^2(X),G\tilde G (X)$ and a statement about $T_{\mu\nu}=0$
which are coupled to bulk massless fields.  

\begin{figure}
\centering
\includegraphics[width=4cm]{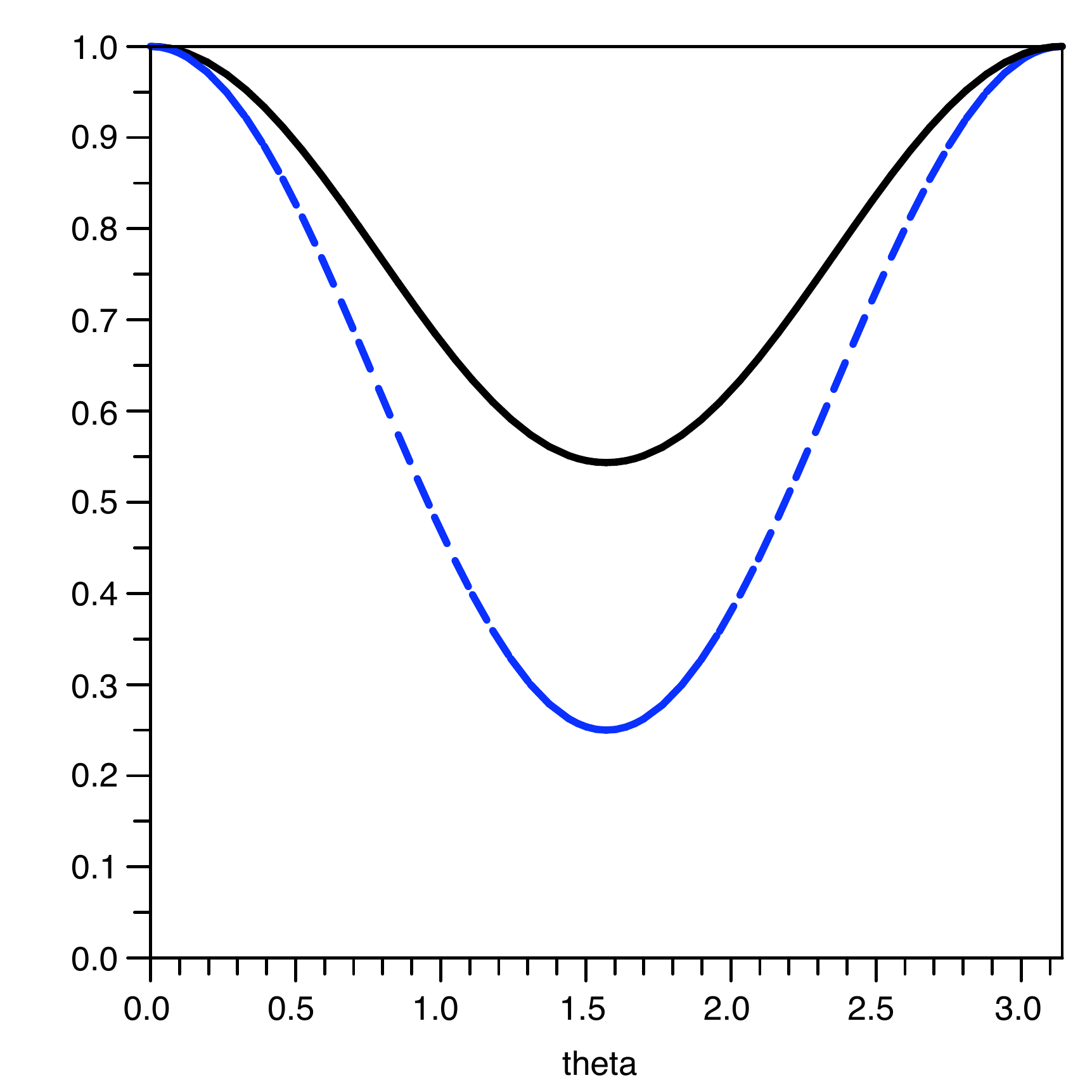}
\caption{ Angular distribution of the  far field energy versus the polar angle
 ($cos(\theta)=y_1/|y|)$.
 Solid black line is the AdS.CFT result,
  compared to the perturbative dipole energy $(3cos^2\theta+1)/4$ 
(the dashed blue line), both normalized at zero angle.
\label{eos-shuryak-fig_dipole}
}
\end{figure}

\subsection{A hologram of the Maldacena dipole}

In the leading order in weak coupling, the only field is electric, and the field of a dipole   is just a linear superposition of
the two Coulomb fields of the charges
   \be E_m(y) =({g\over 4\pi})\left({y_m-(L/2)e_m \over
 |y_m-(L/2)e_m|^3}-{y_m+(L/2)e_m \over |y_m+(L/2)e_m|^3}\right)
 \label{eqn:dipole_weak} \nonumber \ee
Because of the cancellation between these two terms,
 at large distances from the dipole
the field decays as $E\sim L/y^3$. The corresponding  
 energy density (and other components of the
stress tensor) are thus of the order  $T_{00}\sim E^2\sim g^2 L^2/y^6$.

 
 In the strong coupling, one has to calculate the hologram of the ``pending string" construction explained above.
 The stress tensor  of matter  $<T_{\mu\nu}(y)>$ at any  point $y$ on the boundary comes from the gravitational field of that string, calculated in the bulk 
 and then approximated near the boundary.
 Lin and myself
\cite{Lin:2007pv} have found it:
the solution is rather technical to be presented here in full, and even the
  resulting stress tensor expressions are rather  long. Let me just comment that there are some general
  requirements which should be satisfied, used to  verify explicitly that no mistake in the calculation was made.
   The stress tensor should  be traceless  $T^\mu_\mu=0$ because the boundary theory is conformal, and it should
have zero covariant divergence $T_{\mu\nu}^{;\nu}$ because there are no sources away from the charges themselves.
 Let me just illustrate one point, using 
 the leading large-distance term, at the 
$y>>L$.

\be\label{eq:ff}
T_{00}=\sqrt{\lambda} L^3
\left(\frac{C_1y_1^2+C_2 y^2}{|y|^9}\right)f(\theta)
\ee
where $C_1,C_2$ are numerical constants whose values can be looked up in the paper
and $f(\theta)$ is the angular distribution. It is shown by solid line in  Fig.\ref{eos-shuryak-fig_dipole}(b), to be compared to
  that in  weak coupling (the dashed line). 

\subsection{A hologram of a particle falling in the AdS bulk}

The  example in this subsection can be put  in a setting quite different in scope from 
AdS/CFT correspondence. 
 Now we ask the reader to imagine, that instead of using it as a mathematical tool 
 (similar in spirit to extending functions to complex plane),  
 we for a moment assume it to be the true picture of the physical Universe.
 Specifically, imagine that we
 indeed live on the $D_3$ brane, where all gauge fields of the Standard Model are defined,
but there are more dimensions of space and gravity act ``in bulk". 

What would  be our perception (the hologram) of a {\em bulk particle} -- so to say a bulk meteorite --flying by?   Of course any particle\footnote{The example before
was a point even in the bulk or $D_{-1}$ brane. A particle has a path, it is a $D_0$ brane, no space extension but with time dependent path. } must fly on
some geodesic, as a stone thrown into a nonzero gravity field. For example, at $t<0$ it can fly toward the boundary in $z$ direction, then stop
at $t=0$ at some distance from it, and then fall back to the
AdS center. Since ``stone" has a nonzero stress tensor $T^{stone}_{\mu\nu}=m\dot{X^\mu} \dot{X^\nu}$, it can be coupled to the bulk metric perturbations $\delta g_{\mu\nu}$
and, via the gravitational propagator,  produce a holographic   stress tensor on the boundary.
The hologram can again be obtained by findding the bulk $\delta g_{\mu\nu}(z,X)$, expanding it at small $z$ and reading the coefficient of the 4-th power of $z$, as we did above. The result 
is  a spherically symmetric  implosion and then explosion.

However, this time we will use a different method, using this occasion to learn a bit more about 
geometry of the $AdS_5$ space.  We will follow  \cite{hep-th/0611005}
which uses 
 the so called global coordinates. Recall that e.g. a $D$-dimensional sphere of radius $R$ can be
described by its imbedding into the $D+1$ space, in which it is defined by the equation $(X^1)^2+(X^D)^2=R^2$. 
Similarly, the $AdS_5$ space can be described by the equation
\be -(X^{-1})^2- (X^0)^2+(X^1)^2+(X^2)^2+(X^3)^2+(X^4)^2=-L^2 \ee
in the 6-dimensional space $X^{-1}..X^4$. (Note, that there are $two$ negative signs!)
The relation to the coordinates, the usual ones for hyperboloids, is
\be X^{-1}=\sqrt{L^2+\rho^2}cos({\tau \over L}), \,\, X^{0}=\sqrt{L^2+\rho^2}sin({\tau \over L}) ,\,\, X^{i}=\rho \Omega^i \ee
The last term contains the coordinates of the 3d unit sphere, with  standard line element
\be d\Omega^2=d\chi^2 + sin^2\chi (d\theta^2+sin^2\theta d\phi^2)
 \ee

The point is there exist another set, known as Poincare coordinates, defined by (i=1..3)
\be X^{-1}={z \over 2}(1+{L^2+\vec x^2-t^2 \over z^2 }) ,\, \, X^0=L {t \over z},  \\ \nonumber 
X^i=L {x^i \over z}, \,\,\, X^4={z \over 2}(-1+{L^2-\vec x^2+t^2 \over z^2 }) 
\ee 
One can eliminate global coordinates and get the relation between these two sets, $\tau,\rho,\chi,\theta,\phi$ and $t,\vec{x}$ (left as an exercise).

Understanding of the construction can be started from the boundary, the large $\rho \gg L$ limit in which all $X^A\sim \rho$.   Here $\tau$ just  runs a circle and $\Omega$ is a 3-d sphere with constant spatial curvature. This world is known as ``Einstein's static Universe.
The Poincare coordinates at large $z$ are just Minkowskian.  
The boundary relation is relatively simpler to do, it is
\be {t \over L}={sin(\tau/L) \over cos(\tau/L) +cos(\chi)},  {\vec{x} \over L}={sin(\chi) \over cos(\tau/L) +cos(\chi)} \Omega^i\ee
This is a map between the curved Einsteinian and flat Minkowskian 4-d Universes.

Let us now switch to the relation between the differentials of both set of coordinates, and work out their correspondence and the metrics.
After going through it one finds the following metric tensor
\be ds^2=-dt^2+(dx^i)^2= W^2(-d\tau^2+L^2 d\Omega^2)\ee
where the so called {\em conformal factor} is
\be W^2={ 1 \over (cos(\tau/L) +cos(\chi)  )^2}={t^2 \over L^2}+{1 \over 4} \left(1+{r^2\over L^2}-{t^2\over L^2}\right)^2 \label{eqn_conf_factor}\ee

The next step is upgrading the AdS to the so called Global AdS-Schwartzschield Black Hole,  GAdSBH, a charged 5-d black hole
which (we will need below anyway). Its line element  corresponds to the metric
\be  ds^2=-f d\tau^2 + {d\rho^2 \over f} 
+\rho^2 d\Omega^2\,\, f=1-{\rho_0^2 \over \rho^2}+ {\rho^2 \over L^2}
\ee
where $\tau,\rho,\Omega$ have the same meaning as ``standard" coordinates above.
This metric is a solution to the Einstein equation
with a particular cosmological term
\be R_{ab}+{4 \over L^2} g_{ab}=0 \ee
and $\rho_0,L$ are two parameters related to the mass and charge of the black hole. Here, for reference, are such relations
including the Bekenstein entropy and the Hawking temperature
\be M={3\pi^2 \rho_0^2 \over \kappa^2}, \,\, S={4\pi^3 \rho_h^3 \over \kappa^2}, \,\, T={\rho_h \over \pi L^2}(1+{L^2 \over 2 \rho_h^2}) \ee 
where the horizon radius, the upper root of $g_{00}=f=0$, is 
\be \rho_h/L=\sqrt{\sqrt{(1+4\rho_0^2/L^2}-1)/2} \ee
$\kappa$ is the 5-d Newton constant, which is related to $L$ and the brane number $N$ via
\be  {L^3 \over \kappa^2}=({N \over 2\pi})^2\ee
Note that if $N\gg 1$ is large, than the space parameter $L$ is large compared to the 5d Plank scale and thus gravity
is classical. 

This black hole is to play a role of a ``meteorite" flying in the bulk away from our ``brane world". 
In principle, one can consider its mass, and thus $\rho_h$ to be  small parameters, and expand in it to the 1st order: but it is not really
needed.
In the way we just described it,
it is a static BH sitting at the origin $\rho=0$: but this is just because standard coordinates used describe everything ``from the BH point of view".
``Our coordinates", the same Poincare ones as defined above, are related with the standard ones by the $same$ coordinate transformation.
Since those are moving (time-dependent) relative to the BH, in such coordinates the BH does fly.

  The elegant solution we describe calculates $T_{\mu\nu}$ using {\em static} black hole metric in global AdS, in standard coordinates,
  and then just transform this tensor in the Poincare coordinates.
The result (written in the usual Minkowski spherical coordinates  $t,r,\theta,\phi$) takes relatively simple form  
\be T_{\mu\nu} \sim \left( \begin{array}{cccc}   
 3L^4+4 t^2 r^2/W^2 &  -2 t r (L^2+t^2+r^2)/W^2 &  0 &  0\\
  -2 t r (L^2+t^2+r^2)/W^2 &  L^4+4 t^2 r^2/W^2 &  0 &  0\\
   0 &  0 &  r^2 L^4 &  0\\    0 &  0 &  0 &  r^2 L^4 sin^2(\theta)\\
\end{array} \right) \nonumber
\ee
where $L$ is a parameter and  $W$ is the  conformal factor (\ref{eqn_conf_factor})
of metric transformation given above. 
It is instructive to
check that this stress tensor is  traceless and conserved, \be T_{\mu\nu}^{;\nu}=0 \ee so there is no source for it in the Minkowski world.
It is so-to-say spherical tsunami\footnote{Note however, that this phenomenon has nothing to do with hydrodynamics,
as this tensor $cannot$ be written in the form of a moving fluid.}, an implosion at $t<0$ and explosion at $t>0$, coming and going to infinity at $t=\pm \infty$.

\subsection{
 ``Holographic $e^+e^-$ collisions" show no signs of jets!}
Here we present results of calculation \cite{Lin:2006rf,Lin:2007fa} modelling what happens
when the two ends of the open string, ``charges" located on the boundary, move away from each other with certain velocities\footnote{If $v=0$ it is static Maldacena dipole. If $v$
is small, it complements modified Coulomb law by the modified Ampere law. 
These case are discussed in the paper but in this subsection we only consider ultrarelativistic $v$.} $\pm v$. 

This setting corresponds to a process in which some neutral object (highly virtual photon or $Z$ boson) decays into quark-antiquark pair. In QCD, due to asymptotic freedom, at large total mass $M\gg \Lambda_{QCD}$, the corresponding coupling $g^2(M)$ is weak, and
quarks produce a sequence of rather distant gluons along their way. This is indeed
what is observed as two-jet events experimentally.

The question we now address is: what happens in strong coupling, when $g^2(M)$ is large and
gluons interact strongly, between quarks and themselves? In particular, would there be any visible jets along the direction of motion of the charges?

\begin{figure}[h!]
\begin{center}
\includegraphics[width=8cm]{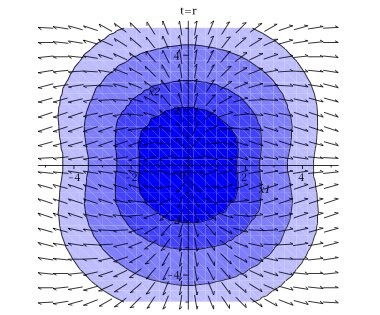}
\caption{Contour plot of momentum density $T^{0m}$. The direction of the momentum density is indicated by normalized arrows. }
\label{fig_stress_from_jets}
\end{center}
\end{figure}

In order to answer it one needs to solve first equation of motion for the string, simultaneously
falling into the AdS space due to gravity and expansing between the receeding charges. 
Then, one needs to calculate the effective stress tensor source, the falling string provides.
And finally, using propagators in the AdS space, one can calculate the induced stress tensor
perturbation on the boundary. All these steps are highly technical, and details should be looked at in the original papers.  In Fig.\ref{fig_stress_from_jets} we show as an example the momentum density distribution on the boundary (charges move relativistically in the horizontal direction). The answer to the question raised above is now clear: at strong coupling there are no jets. There is some kind of collective explosion. It however cannot be a hydrodynamical
flow of some ideal fluid: we see this because stress tensor is very anisotropic even in the rest frame of each element. 

\section{Thermal AdS/CFT and strongly coupled plasma}
 Let me briefly remind few more facts
from the Black Hole toolbox. Studies of how
quantum field theory can sit in a  background of 
a classical black hole metrics have resulted in two major
discoveries: the {\em Hawking radiation}  and the {\em Bekenshtein entropy},
related to horizon radius and area, respectively. Hawking radiation
makes black holes in asymptotically flat space unstable: it 
and heats the Universe till the black hole disappears. Putting black hole
into a finite box also does not help: it is generically thermodynamically unstable
and gets smaller and hotter till it finally burns out. Only in appropriate curved spaces
black branes can be in thermal equilibrium with their ``Universe",
 filled with radiation at  some finite temperature $T$. As shown by Witten,
all one has to do to get its metric is to consider non-extreme (excited extreme)
black brane solution, which has a horizon.

 Let me now make a logical jump 
and  instead of just giving the metric for finite-T solution let me first
provide an ``intuitive picture'' for non-experts,
explaining
the finite-temperature Witten's settings
in which most\footnote{The exception is heavy quark diffusion
  constant  calculated by Casalderrey and Teaney\cite{Casalderrey-Solana:2006rq} which
 needs more complicated
settings, with a
 Kruskal metric connecting a World to an Anti-world through
the black hole.
}  pertinent calculations are done., 
The 
3-dimensional space boundary z=0 
is flat (Minkowskian) and corresponds
to ``our world'', where the gauge theory lives. In the bulk 
 there is a black hole metric with horizon
at $z=z_h$. The b.h. center
is  located at $z=\infty$, but all $z>z_h$ are in fact irrelevant
as they are not observable from the boundary. Studies of finite-T 
conformal plasma by AdS/CFT famously started exactly by 
evaluation of the  Bekenstein entropy \cite{Gubser:1998nz}, 
$S=A/4$ via calculating the horizon  $area$  $A$. 

Now comes the promised ``intuitive picture'':
 this setting can be seen as a {\em swimming pool}, with
 the gauge theory (and us,  to be referred below as ``distant
 observers'')  living on its surface, at zero depth $z=0$, enjoying
the desired temperature $T$. In order to achieve that, the
 pool's bottom 
looks $infinitely$ hot for observers which are sitting at some fixed
$z$ close to its coordinate $z_h$: thus diving to such
depth is not recommended.
Strong gravity takes care and stabilizes
 this setting thermodynamically: recall that time units, as well as
those of energy and temperature are subject to ``warping''
with $g_{00}$ component of the metric, which vanishes at  $z_h$.

 When astronomers found evidences for black holes
 and accretion into those, the physics of black hole became a regular
 part of physics since lots of problem have to be solved.
  Here important step forward was the so called {\em membrane paradigm}
  developed by many people and best formulated
by Thorne and collaborators \cite{Thorne:1986iy}, known also under the name of
``stretched horizons". Its main idea is to imagine that there is a
physical membrane at some small distance $\epsilon$ away from
the horizon, and that it has properties  exactly such that all
the eqns (Maxwell's, Einstein's
etc) would have the same solutions outside it as $without$ a membrane but
with a continuation through horizon.
For example, a charged black hole would have a membrane
with a nonzero charge density, to terminate the electric field lines.
The fact that Poynting vector at the membrane must be pointed
inward means some (time-odd!) relation between $\vec E$ and $\vec B$:
this is achieved by giving the membrane finite $conductivity$, which in turn
leads to ohmic losses, heat and entropy generation in it. Furthermore,
as shown by Damour back in 1980's, displacements of the membrane
and relations for gravitational analog of the Poynting vector
leads to nonzero $viscosity$ of the membrane,
and its effective low frequency theory take the form
of Navier-Stokes hydrodynamics. 
For a bit more modern derivations of the effective action
and field theory point of view look at Parikh nd Wilczek \cite{Parikh:1997ma}.
As we will see below in this chapter,
all of those ideas have resurfaced now in AdS context, generating
new energy of young string theorists who now pushed the ``hydrodynamics
of the horizon memebrane" well beyond the
Navier-Stokes to a regular construction of systematic derivative expansions
to any needed order.


\chapter{Holographic QCD}
As we emphasized it few times already, there is $no$ known holographic dual for QCD.
And yet, people tried to build it,  so-to-say, by
``bottom-up" approach. It does not have the same status as AdS/CFT correspondence discussed above, but
 obviously a set of made-up models,  which howeve can be rather productive and instructive. 
We start with two crude ideas, to be refined later.

\section{Witten and Sakai-Sugimoto models}

The way from the holographic dual of the $\cal N$=4 (SYM) theory to the usual Yang-Mills gluodynamics 
(YM) was suggesting in \cite{Witten:1998zw}. He emphasized that this step included
the high-temperature limit of YM theory, implying breaking of sypersymmetry, yet preserving holographic correspondence. 

In short, the idea is to start with $D_4$ branes (rather than $D_3$ ones used for ADS/CFT). The ``unwanted" direction of space $x^4$ is compactified on a circle with 
circumference $\beta_4\equiv 2\pi/M_{KK}$, the $M_{KK}$ being the so called Kaluza-Kline
mass. The metric of the Witten background is
\be ds^2=\big( {u \over R}\big)^{3/2} \big( dx^\mu dx_\mu +f(u) dx_4^2 \big) 
+ \big( {R\over u}\big)^{3/2} {du^2 \over f(u)} +R^{3/2} u^{1/2} d\Omega_4^2 \ee
where $f(u)=1-u_0^3/u^3$ is the warping factor\footnote{It is completely analogous
to Schwarzschild solution of the usual black hole, except the power of distance in
the Newtonian potential is not $1/r$, as in 3 dimensions, but $1/r^3$ in 5 dimensions. 
} depending on the holographic coordinate $u$,
defined for $u>u_0$. 
There is also a dilaton field and 4-form Ramon-Ramon field in this background, which
we would not go into.
Here $\mu=0,1,2,3$ are Minkowski coordinates, and the extra spatial coordinate is $x_4$. 
The key feature here is that the radius of $x_4$ vanishes at $u_0$: the geometry thus has
the so called ``cigar" form with its tip at $u_0$. 

The Euclidean form of the finite-$T$ theory corresponds to the time, $\tau=i x_0$, also compactified 
to the Matsubara time, with circumference $\beta=1/T$. Thus there are two competing
phases of this theory: a cigar in one ciclic variable and ``tube" in another. The lowest
free energy corresponds to the winning phase. This phase transition Witten identified
with the deconfinement phase transition in thermal YM theory.  The deconfinement
temperature scale is therefore related with $M_{KK}$, defined above. 

Generalization of this model to nonzero $\theta$ angle can be done by turning on nonzero 
1-form potential $C_1$ along the $x_4$ circle, Aharonov-Bohm style, so that there is a nontrivial flux 
of the field $F_2=dC_1$ through the cigar. 

In order to get a QCD-like model, one needs to complement Witten's construction for YM theory
by inclusion of the light quarks. This can be done by incorporating some $N_f$ copies of extra branes
posessing fundamental fermions. Since the number of flavors $N_f$ is assumed to be small
compared to $N_c$, they are treated in ``probe approximation", that is neglecting their role
in overall metric formation.

\begin{figure}[htbp]
\begin{center}
\includegraphics[width=8cm]{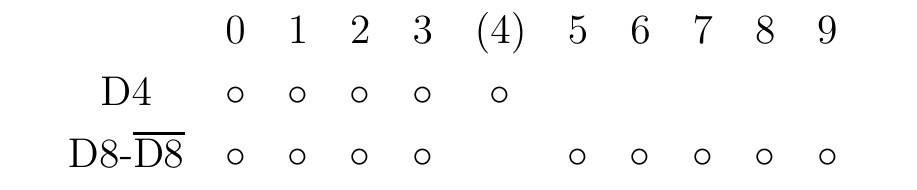} \\ \vskip 0.2cm
\includegraphics[width=10cm]{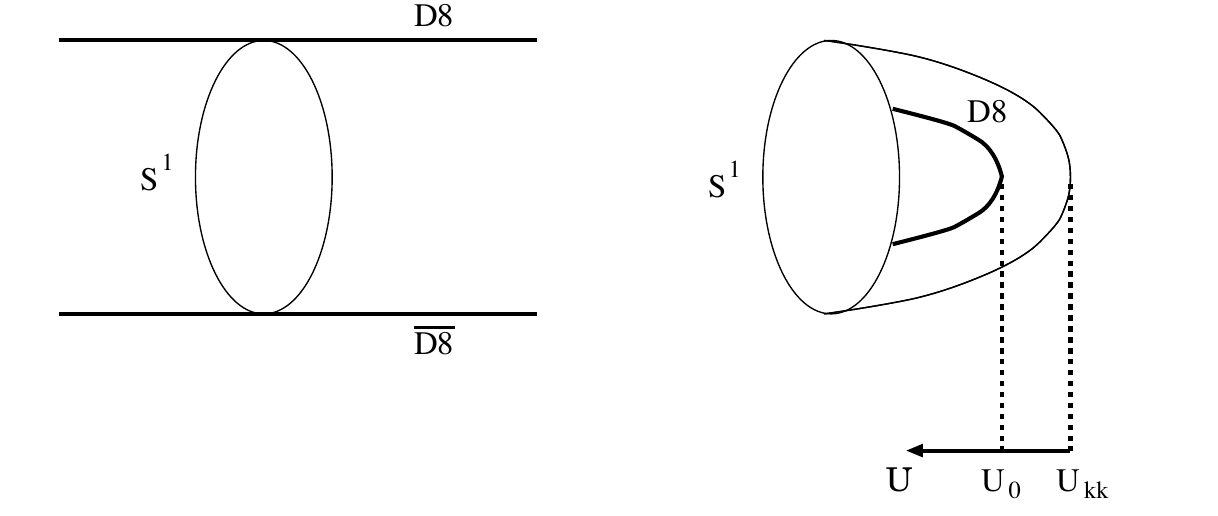}
\caption{Upper: extensions of the coordinates of $D_4$ and $D_8$ branes.
Lower, two phases, with chiral symmetry and with its breaking.}
\label{fig_D8_branes}
\end{center}
\end{figure}

The model we will now discuss is Sakai-Sugimoto model  \cite{Sakai:2004cn}, based on $N_f$ sets of $D_8$ and $\bar D_8$ branes located at the opposite sides of the Witten's cigar. In upper part of Fig.\ref{fig_D8_branes} from this paper we show extensions of the $D4$ and $D8$ branes.
The left lower sketch shows how they are connected, with the circle being in $x_4$ coordinate
which $D8$ lack. If the geometry has two idependent $D8$ branes, corresponding to left and right quarks, the chiral symmetry is unbroken. It gets however broken if the geometry is as shown in lower right corner, in which there is a single $D8$ brane turning back after it reached 
some point $U_0$. 

In the WSS geometry thus constructed one can move further, studying small perturbation of
this brane construction. The displacement of the  $D8$ is described by 9-dimensional gauge field $A_M$. Those with $M=0,1,2,3$ would be Lorentz 4-vector, while others scalars. The
action of perturbation has a general form 
\be S=-T \int d^9x e^{-\phi} \sqrt{ - det\big( g_{MN}+2\pi \alpha' F_{mn} \big) }\ee  
where $T$ is tension, $F_{MN}$ is the field strength corresponding to $A_M$. Assuming the 
field is small, one can expand the square roots in powers of the field. In quadratic order in $F_{MN}$ it describes masses
of (pseudo)vector and (pseudo)scalar mesons.  We would not go into details, but just
comment that mass ratios between those can be calculated from the model, and their
relation to experiment is reasonable. Let me emphasize that in fact we have much more
then predictions of the meson masses here: we have a complete form of {\em effective
chiral Lagrangian}, a dream of Weinberg, Leutwyler etc, which in QCD is so hard to build! Note further, that the relation between vector and scalar mesons fulfill also another dream, 
that of {\em hidden gauge symmetry} related with the flavor $SU(N_f)$ group!

Since in these lectures we devoted a lot of attention to chiral anomaly, in its relation to
topology and also to chiral effects in general, let us explain how \cite{Sakai:2004cn}
had approached these issue holographically. They added the following topological action term
\be  S_{topo}= {N_c \over 24\pi^2} \int_{D8}  \omega_5(A) 
\ee
where $N_c$ comes from integral over $F_4$ charge, of the  the RR 4-form. The Chern-Simons form in 5 dimensions $x_0,x_1,x_2,x_3,z$  is
\be  \omega_5(A) =Tr\big( A F^2-{1\over 2} A^3 F +{1 \over 10} A^5 \big) \ee
Recall that the CS form is not by itself gauge invariant, but its change 
is, because\footnote{This expression is analogous to $d\omega_3=Tr(F^2)$ 
relation we used relating 3d to 4d gauge topology. 
}  $$d\omega_5=Tr(F^3)$$ 
Therefore gauge invariant action, reproducing QCD chiral anomaly, can be defined as the difference between the boundary values
at two ends of $D8$, the left minus the right-handed one. 


We end our discussion of Witten-Sakai-Sugimoto settings by
 introducing the $
\theta$ dependence of the vacuum energy and calculate it.
 For details we refer to
  \cite{Bigazzi:2019eks}
  and only note that always $E_{vac}\sim cos(\theta)$, as in the dilute instanton gas.
  
  To make this connection more clearly, we would like to mention the topological susceptibilties
  obtained in these two settings.
  In N=4 or SYM theory they have, at finite temperature setting
  \be \chi_{SYM}(T)={15 \over 128} \pi^{3/2} \sqrt{N_c} T^4  exp\big(-{8\pi^2\over g_{YM}^2}\big)  \ee
  and the exponent indeed corresponds to the instanton action. The preexponent
  however has strange square root of $N_c$ and no coupling:  both features 
I cannot explain in simple terms.  
  
  In WYM (Witten-Sakai-Sugimoto) high T phase the topological .susceptibility (deduced from the $\theta$- dependence of the vacuum energy $\sim cos(\theta)$ )
  the result is 
  \be \chi_{WYM}(T)={3285\pi^{3/2} \over 42} \big({4\pi \over 3}\big)^4 \sqrt{N_c \over \lambda} T^4 exp\big(-{8\pi^2\over g_{YM}^2}\big) \ee
  In this case
 $N_c$ and $\lambda\sim g^2_{YM} N_c$ dependence is such that they cancel  dependence on $N_c$ and only the YM coupling remains. 
 
 In summary: holographic model of the theta vacua/axions produce predictions for the topological susceptibilities which looks very much like
  the semiclassical instanton density in exponent, but with very different pre-exponential
  factors. This is hardly surprising by itself, since in weak and strong coupling
  the fluctuating fields (making the bosonic and fermionic determinants) are quite different. 
  Yet some ``microscopic" derivation of expressions given above would be much appreciated.

\section{Instantons as baryonic solitons}  
Let me start admitting that we had not covered in this book the {\em skyrmions}
picturing baryons as heavy classical solitons made of the pion field. This approach 
is natural in the limit $N_c\rightarrow \infty$ (in which baryon mass is large, $O(N_c)$)
and the chiral limit $m_q\rightarrow 0$ (in which pions are massless). The so called Weiss-Zumino-Witten term in the chiral Lagrangian  provides the topological definition of the
baryon number which therefore is conserved. 

Skyrmions naturally predict large-distance forces between nucleons to be
pion exchanges. However we know from nuclear physics that phenomenology suggest
nuclear forces to be described rather in terms of scalar $\sigma$ and vector $\rho,\omega$  
mesons, producing attractive and repulsive parts of the forces. Note furthermore,
that attraction and repulsion obviously are well tuned , nearly cancelling  each other
in sum. 

The WSS model just described provide a very curious realization of ``mesonic soliton" idea,
in the setting explained in the previous section. As shown by
\cite{Hata:2007mb}, the classical solution to  4-d Euclidean  YM equations, used at
the lectures at the top of the course describing tunneling in the topological (Chern-Simons) landscape, 
can now be interpreted in 4d spatial dimentions of the WSS model as a soliton.

Its ``scalar part" corresponds to the Skyrmion pion  cloud, while the
vector field $A_M$ made $U(N_f)$ vector mesons, $\rho,\omega$-mesons,
represent the repulsive baryon core,
missing from the original Skyrme model. 
 Since instanton is a solution to the nonlinear YM equations,
it means that interactions between all types of mesons are taken into account consistently. 

The excited states of instanton-baryons, with say other meson attached, was used as
models for pentaquark hadrons. 
 
The study of high density of instantons, turned to be  a 4-d crystalline lattice, has been first constructed in
\cite{Shuryak:1989cx} ,  it was the first example of ``time crystal" later emphasized by Wilczek.
The same crystalline lattice in the holography setting was later reused as a model for 
dense baryonic matter. If the density is high, instantons can be further split into its
constituents with fractional topological charge, see  \cite{Rho:2009ym}.

\section{Confining holographic models with ``walls" in the inferared}
In spite of strong coupling regime, the AdS/CFT setting is scale invariant,
with Coulomb-like forces between charges. (Only the coefficients of 1/r changes
at strong coupling, as we already discussed.) 
In QCD-like holographic models one needs to somhow {\em generate confinement}.

The first crude idea we  discuss aims at it
 was introduced in \cite{Polchinski:2001tt} who 
 suggested to simply cut off the most infrared part of the $AdS_5$ space,
  by ``brute force".   Indeed,
if there is no space for the bending string to fall into, it would reach ``the bottom of space" and thus generate a linear -- confining -- potential\footnote{The reader should be warned that
this idea cannot be really true, as it also limit $D_1$ magnetic strings, and thus confines the magnetic monopoles. We are not expected to do so in QCD! But let us for some time ignore this unwanted feature, to be repaired later in the section ???}.

A generic idea can be formulated as follows: hadrons would correspond to normalizable modes
of the AdS space, modified by some confinement-related cut-off.  Let us show how it works, following Ref \cite{Karch:2006pv}.
Suppose the metric takes the form 
\be ds^2=exp(2A(z))\left(dz^2+\eta_{\mu\nu} dx^\mu dx^\nu \right)
\ee
where $\eta_{\mu\nu}=diag(-1,1,1,1)$ is the flat Minkowski metric. Another function, the dilaton profile, is in the 
action
\be S=\int d^5 x\sqrt{g}e^{-\Phi(x)} \cal{L}\ee
The model has a general minimal Lagrangian containing the scalar field $X$ and left/right vector fields $A_L,A_R$,
both matrices in terms of some flavor group $SU(N_f)$. 
\be {\cal{L}}=|DX|^2+3 X^2 -{1 \over 2 g^2_5}(F_L^2  +F_R^2) \ee
where $D_\mu X=\partial_\mu X-iA_{L\mu}X+i A_{R\mu}X$ and $F_{L,R}$ are the fields strength made of  $A_L,A_R$.
We will be using the gauge $A_z=0$.
Vector/axial fields are $V,A=(A_L\pm A_R)/2$ as usual. 
They satisfy the eigenvalue equation ($B(z)=\Phi(z)-A(z)$)
\be \partial_z e^{-B}  \partial_z  v_n(z) +m_4^2(n) e^{-B} v_n(z)=0\ee
where we already look for a 4-d plain wave solution and substituted the 4-momentum squared by the mass squared.
Standard substitution $v_n=e^{B/2}\psi_n$ transfers it to Schreodinger-like form
\be \psi''_n+\left({(B')^2 \over 4}-{B''\over 2}\right) \psi_n=0 \ee

 It was argued that the best way to cut the space off is by a 
Gaussian factor in the metric $B\sim z^2$ at large $z$.
One motivation \cite{Karch:2006pv}   is that bulk  fields generate in this case
quite perfect Regge trajectories, as we will see shortly. Another motivation \cite{Shuryak:2006yx}  is that the instanton size distribution in QCD
(unlike in the CFT) does show a Gaussian cutoff, and we now know
that the instanton size $\rho$ is in holography just the 5-th coordinate
of its pointlike source.

 For the preferred choice of the functions
 \be B=z^2+log(z), {(B')^2 \over 4}-{B''\over 2}=z^2+{3\over 4 z^2} \ee
there is the following analytical solution
\be m_n^2=4(n+1), v_n=z^2 L^1_n(z^2) \sqrt{{2 n\! \over (1+n)\!}} \ee
where $L^m_n$ are associated Laguerre polynomial. Note the equidistant spectrum of the vector meson
masses $m^2$, which is phenomenologically correct.
Furthermore, introducing spin $S$ fields,as 5-d symmetric tensors of rank $S$   \cite{Karch:2006pv}  
one see that the equation of motion is modified to 
 \be  B=z^2+(2S-1)ln(z) \ee
 and the spectrum to 
 \be   m_{n,S}^2=4(n+S) \ee
 which is the (phenomenologically correct) linear Regge trajectory.
 
 The model can  include explicit and spontaneous chiral symmetry breaking  \cite{Erlich:2005qh}.
by
 Including the scalar $X$ with the boundary conditions $X=M z/2+\Sigma z^3/2+... z\rightarrow 0$,
 where $M$ is the (flavor matrix) of quark masses and $\Sigma$ is the quark condensate, Karch et al
 also got reasonable parameters of the scalar and axial mesons.
 Schafer \cite{Schafer:2007qy} have demonstrated 
 that the whole Euclidean
 vector and axial correlators  $V(\tau),A(\tau)$ 
 built out of these modes  reproduce experimental data within 10-20\% 
accuracy, at any Euclidean time $\tau$. 
(These results are extremely similar
to what was obtained by Schafer and myself in much more sophisticated
 instanton model \cite{Schafer:2000rv}. The remaining
 10\% deviations, at least at small distances,
are clearly just neglected
 perturbative correction $1+\alpha_s/\pi+...$.)

We will however argue that vector/axial channels are
in fact $ exceptions $ rather than the rule,
and for generic operators/fields one cannot hope to use
 this simplest model.
 Coupling of the boundary sources to
  fields propagating in the (modified-AdS) bulk does indeed fit to the general philosophy of holography. Yet I do
  not see any justification to why all those spin-S bulk fields may be massless in the 5-dimensional sense. 
  String theory can only supply the massless states up to gravitons ($S$=2): and even for strict AdS/CFT
  a generic bulk fields have
  nonzero (and large at large spins and  strong coupling) anomalous dimensions. In the form of the 5-d mass
  those would appear in the effective Schreodinger potential as follows  
  \be V(z)=z^2+2(S-1)+{S^2+M_5^2-1/4 \over z^2}
 \label{eqn-pot} \ee
(Only the vector fields can be exempt, on the basis of vector current conservation.) 

 We follow notations of KKSS \cite{Karch:2006pv}, except that
we have added extra
 5-d mass term.  
  Standard substitution
 $\phi=e^{B/2}\psi$ transform this into a Schreodinger-like eqn without first order
 derivatives
 \be  -\psi'' +V(z)\psi=m_4^2\psi \ee
 with \be V(z)=z^2+2(S-1)+{S^2+M_5^2-1/4 \over z^2}
 \label{eqn-pot} \ee
 KKSS needed only bound state wave functions,
 given in terms of  Laguerre polynomials.   Quadratic IR potential was tuned  by
\cite{Karch:2006pv} to reproduce nice Regge trajectories : for absent bulk mass $M_5=0$
they are linear and  $m_4^2=  4(n+S)$. 

   The absolute units we will use also follow from
    KKSS notations,  can be fixed
 by calculate some physical mass.
 Rho meson is an example of the $protected$ state, associated with
 conserved vector current: it has
  with $M_5=0$ and $S=1$: a solution without
 nodes (n=0) gives $m_\rho^2=4$. Using it as an input we fix our unit of length 
 as \be  length\, unit =2/m_\rho =  0.51 fm \ee
  The expected position of the  domain wall at large $N_c$ is
 \be z_{dw} \approx 0.4 fm =.777 (length\, unit) \ee 
 Thus
 {\em most of the wave function is in fact located in the strongly coupled
 domain} and modifications due to weakly coupled domain
 at $z<z_{dw}$ turns out to be very small, as far as the 4-d spectroscopy goes.
 However hard processes will be sensitive to this small tail of the wave function,
and their relative normalization is crucial for what follows.

\begin{figure}
\centering
\vskip -.4cm
 \includegraphics[width=6.cm]{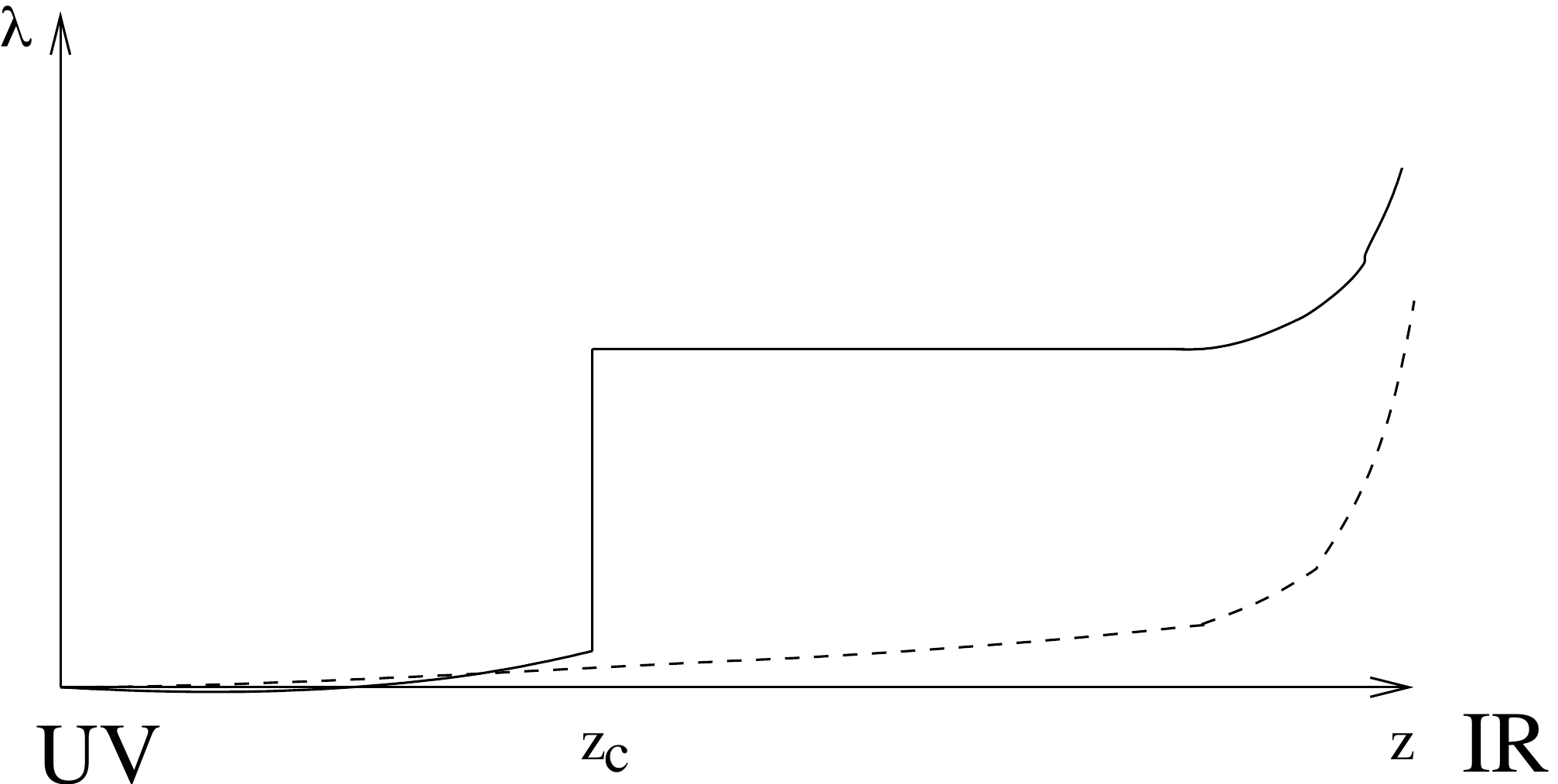}
  \caption{Schematic dependence of the t'Hooft coupling $\lambda$
on the holographic 5-th coordinate $z$. 
  } \label{fig_jump}
\end{figure}

\section{The ``domain wall"  in the ultraviolet?}

The second crude idea try to deal with an unpleasant fact: the QCD coupling runs into
weak in the UV, and so the AdS/CFT-like description cannot possibly be valid there.
I have proposed \cite{Shuryak:2006yx} to include this feature by introducing the so called
{\em UV wall} beyond which the space is still AdS but all 5-d masses 
of the fields return to their non-interacting dimensions. 

Where the UV wall should be placed? What is its physical nature?
The guiding idea is again related with the instantons.
We had discussed large-$N_c$ limit of the instanton ensemble, and have
concluded that their size distribution becomes a delta function, at some limiting instanton size $\rho_*$
 
 What can be calculated in this model? Example include hard exclusive processes,
 and transition from strong coupling at large distances to weak coupling at small one.
 A jump in power should be seen, as a consequence of the change from large anomalous dimensions
 in strong coupling to ``naive" dimensions for non-inetracting fields.
 
 specific example The simplest object of the kind is the pion formfactor.
Unfortunately experiment has not yet seen how transition
to old perturbative prediction takes place,
as it is very hard to measure it at large $Q$. 
All we know is that the observed $Q^2F_\pi(Q)$  at $Q\sim 1-2\, GeV$
is about twice large than the asymptotic value,
so some decrease and then leveling at new level should happen.
As far as I know, nobody have seen it on the lattice either. The 
 magnitude of the pion formfactor at $Q\sim 1-2\, GeV$ induced by
 instantons has been calculated 
in ref. \cite{Faccioli:2003ve}, but  
unfortunately 
the approximation made
are not good at large enough $Q$,
so the transition to the perturbative
regime has not been seen in this approach as well. 
More or less similar situation is with the nucleon formfactor
and many other exclusive reactions.

Another hard reaction involving the pion is
 pion diffractive dissociation into two jets 
\be \pi \rightarrow  jet(k_1)+ jet(k_2)\ee
 first
theoretically discussed by Frankfurt et al \cite{Frankfurt:1993it}
and studied experimentally by
the Fermilab experiment E791. It seems to be showing  
 a  transition from the non-perturbative to the perturbative
regime we are looking for.

 \section{Improved holographic QCD}

 ``Realistic bottom-up" holographic models with the dilaton/axion potentials
 were gradually developed, and in this section we will 
follow the approach of the Kiritsis group
  \cite{Gursoy:2007er,Gursoy:2007cb}, for review see \cite{Gursoy:2010fj}.   
  They have developed  a systematic
approach 
toward building a gravity dual model to QCD. The fields in the action 
include the (tensor) metric, (scalar) dilaton and (pseudoscalar) axion  in the common
Lagrangian. One more scalar field, the ``tachion", is added later, to model the chiral symmetry breaking and
the quark condensate. 

Let us start with the pure YM theory: for its holographic dual one uses
generic form of the (5d) action, with the bulk and boundary terms
\be S=-M_p^3 N_c^2 \int d^5x \sqrt{g} \big[ R -{4\over 3} (\partial \Phi)^2+V(\Phi)\big]
+ 2 M_p^3 N_c^2 \int_{\partial M} d^4x \sqrt{h}K \ee
Here $R$ and $K$ are bulk and boundary curvature, the factors in front
correspond to 5-d Newton constant $G_5=1/(16\pi  M_p^3 N_c^2)$ small in the large $N_c$ 
limit.  The exponent of the dilaton $exp(\Phi)=\lambda$ is identified with running t'Hooft
coupling $N_c g^2_{YM}$. 

What needs to be carefully chosen is the potential $V(\Phi)$, as it incorporate the beta function of the theory. These details can be found in the original works.
  A clever trick allows to incorporate
the beta function (which demands first order differential eqn)
with Lagrangian formalism which demands second order ones.

\begin{figure}[htbp]
\begin{center}
\includegraphics[width=6cm]{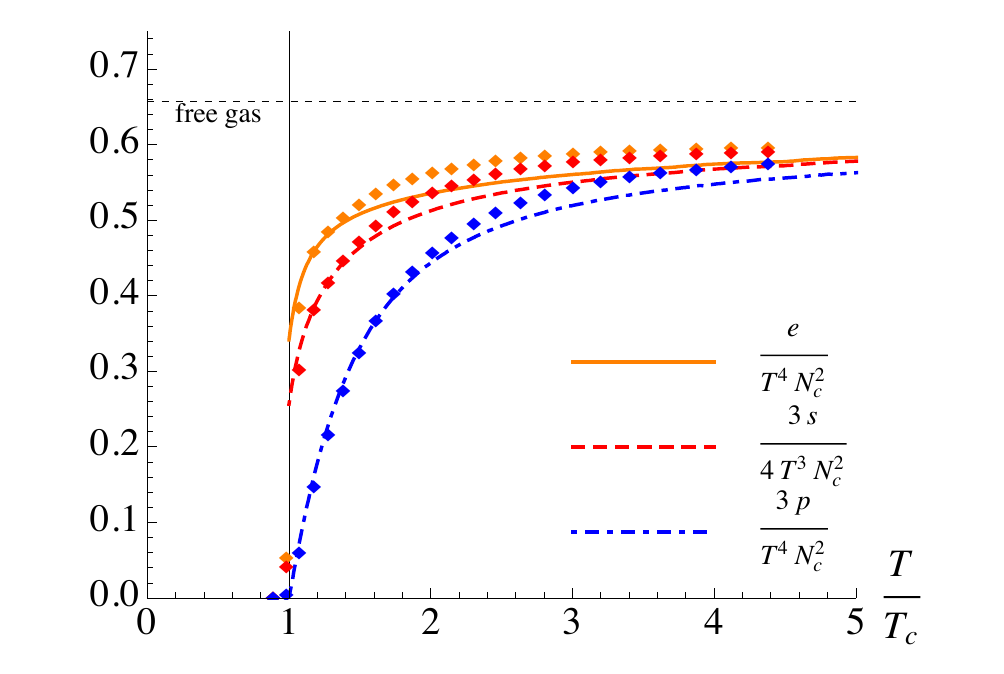}
\includegraphics[width=6cm]{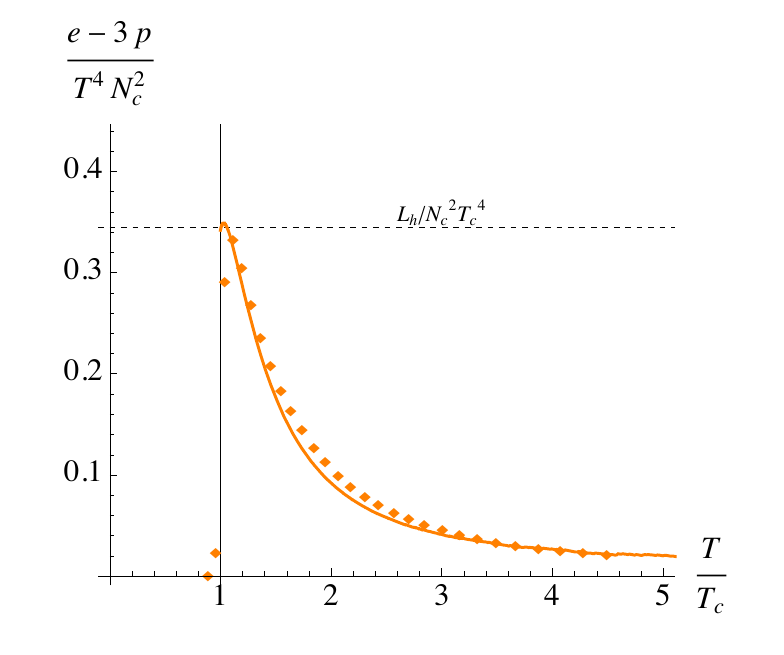}
\caption{Temperature dependence of the dimesionless combinations, normalized so that
they all have large-$T$ limit $\pi^2/15$ (dahsed horizontal line). The second plot is the so called trace anomaly combination, the measure of the non-conformal behavior.}
\label{fig_holo_gluo}
\end{center}
\end{figure}

There are basically two types of the solutions, with qualitatively different metric behavior,
without or with the black hole. Aiming at thermal physics of quark-gluon plasma, we
start with Euclidean time which, with slight abuse of notations, we still call $t$. 
In the former case the following simple ansatz is used (m=1,2,3):
\be  ds^2=b_0^2(r) \big(dt^2+dr^2 +dx_m dx^m \big) \ee
In the latter 
\be  ds^2=b^2(r) \big[ f(r)dt^2+{dr^2 \over f(r)} +dx_m dx^m \big] \ee
the warping factor $f(r)$ has zero at horizon location $f(r_h)=0$ and only $r<r_h$
are considered. These two geometries correspond to low-T and high-T phases,
as calculation of their free energy shows. The obtained temperature dependence of the thermodynamical
quantities -- energy and entropy densities $e,s$ and pressure $p$ --  are compared in Fig.\ref{fig_holo_gluo} with the  YM lattice data (dots). In spite of the model assuming large $N_c$ limit and the lattice data at physical $N_c=3$, the agreement is rather good.

The phase transition is of the first order, and, as e.g. is the case for elementary Van-der-Waals case, there is the third
solution (here called ``the small black hole") which is unstable and correspond to free energy maximum between the two minima. 

We will not go into discussion of kinetic quantities, the diffusion constant, the viscosities ,
the jet quenching parameter etc -- which can also be calculated in this model.

The next extension of the model is adding the pseudoscalar $axion$ field $a$: this allows to
fix negative parity glueballs and also to calculate the topological susceptibility: but we will not
discuss it. Finally, one should add quarks, to construct really holographic model for QCD.
The topic has its history which we do not go to, but just jump to the paper \cite{Jarvinen:2011qe} in which quark-generating $D_4$-anti $D_4$ branes are not considered in the ``probe approximation"\footnote{
We remind the reader that in real-world QGP quark-antiquark degrees of freedom are twice
more important than those of gluons.}.  In this work 
 the so called {\em Veneziano limit} is taken
in which  the ratio of the number of flavors to the number of colors
\be x_{Veneziano}\equiv {N_f \over N_c} \ee 
 is kept fixed in the $N_c,N_f\rightarrow \infty$ limit. Its value is from zero to the maximal value $x_{max}=11/2$ at which the asymptotic freedom is lost. 
 
 The model action is supplied with the fermionic part
 \be S_f=-x_{Veneziano} N_p^3 N_c^2 \int d^5x V(\lambda,\tau) \sqrt{det(g_{\mu\nu}+h(\lambda)\partial_\mu \tau \partial_\nu \tau^+ )} \ee
containing the ``tachion" field $\tau$. 

One interesting aspect of this model is existence of critical value $x_c$, approximately near 4, separating ``conformal window" at $x_{max}> x>x_c$, in which the system flows in IR into the zero of the beta function, and the more common regime at $x<x_c$ in which the tachion 
value at the boundary is related to the nonzero quark condensate. 
The solution to EOM is numerical and highly technical:
details can be found in the original papers, so let me only present some results.

\begin{figure}[htbp]
\begin{center}
\includegraphics[width=8cm]{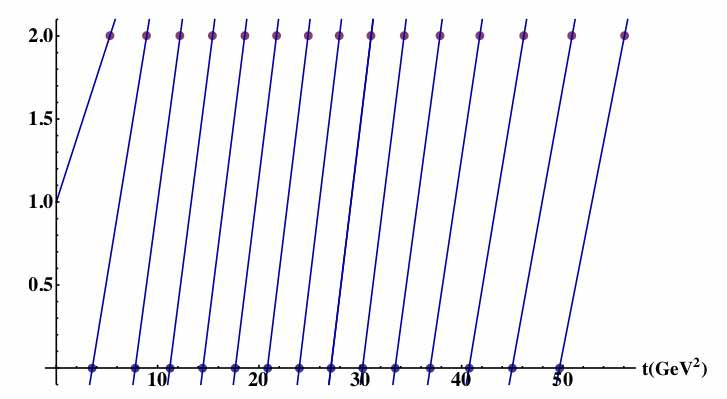}
\caption{The $2^+$ and $0^+$ glueball masses on the Regge plot $J-M^2$,
calculated from improved holographic QCD in the Veneziano limit
at $x_{Veneziano}=1$ by \protect\cite{Iatrakis:2015rga}. }
\label{fig_glueball_masses}
\end{center}
\end{figure}

After the background field configuration is established, one can study 
quantum deviations from it. Quadratic terms, as usual, give the
excitation spectrum of the fields involved.  For example, excitations
of the graviton and dilaton fields give the spectrum of tensor and scalar glueballs .
Connecting those one can obtain the lower parts of multiple ``daughters" Regge trajectories, as shown in Fig. \ref{fig_glueball_masses}. 

Meson scalar gives its own set of excitations. More precisely, the dilaton and tachion excitations
(scalar glueballs  and scalar mesons) 
 get mixed together. I will discuss this mixing, following \cite{Iatrakis:2015rga}, because it was important for understanding
interaction strength of the QCD strings, and (perhaps more importantly) it is the part of 
AdS/QCD I was involved myself. 

\begin{figure}[htbp]
\begin{center}
\includegraphics[width=8cm]{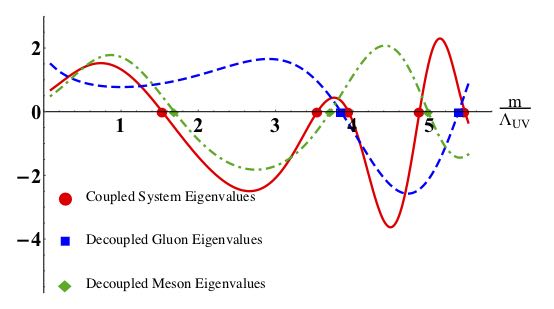} 
\caption{
The determinant of the UV boundary value of two linearly independent solutions
 of the scalar fluctuation equations, versus the mass parameter in $\Lambda_{UV}$ units.
The red solid line with 5 zeroes indicate values of the lowest scalars from full Hamiltonian.
Blue dashed lines and squares indicate unmixed glueball masses, and green dash-dotted curve
and green diamonds  indicate unmixed meson masses. The lowest state -- sigma meson $f_0(500)$ -- shifts little, but the second meson is close to the first glueball and their mixing is more serious. In fact these two states collide at $x_{Veneziano}\approx 1.5$, but
including mixing one observe classic ``avoided crossings" of the levels.
}
\label{fig_scalar_mixing}
\end{center}
\end{figure}

As a brief historic introduction, let me  just state that in hadronic spectroscopy one can hardly find 
more confused subject than scalar mesons. Experimentally, there are more resonances than
the simple $\bar q q$ mesons may exist. Clearly there should be scalar glueballs and 4-quark
mesons as well, all mixed up in some proportions. AdS/QCD is at least a self-consistent model,
providing a well-defined method to calculate these mixings.   
How it works is explained in Fig.\ref{fig_scalar_mixing}. The strings  

Completing this section, let me note that expanding the action to terms quadratic in 
all fields deviations from the background  is not the only thing one can do. Nothing 
(except technical complexity) can stop anyone to expand the action further, e.g. to 
the third order, defining the interaction constants between all these states. 

The question is not only the magnitude of the coupling, but the index structure of the 
corresponding vertices as well. For example, as shown by 
\cite{Anderson:2014jia}, the index structure of  triple vertex of two tensor fields and $pseudoscalar$ $0^-$ production can be determined uniquely
\be V_{h h o^-}\sim \epsilon_{\alpha\beta\gamma\delta} q_1^\alpha  q_2^\beta h^{\gamma\sigma}  h^{\delta\sigma} 
\ee  
Indeed, the $2^+$ Pomeron is identified with symmetric tensor $h^{\mu\nu}$, which cannot be directly paired with an
antisymmetric epsilon symbol. Two momenta of the Pomerons, $q_1,q_2$  need to be also included
for obvious reasons. If transverse part of those are parallel, $\vec q_1^\perp \sim \vec q_2^\perp$, the vertex vanishes. In general, it is predicted to be proportional to $sin(\phi_{12})$
(the azimuthal angle between  $\vec q_1^\perp ,  \vec q_2^\perp$), and the data
is in agreement with this prediction.

Unfortunately, the index structure of  triple vertex of three tensor fields $V_{hh2^+}$ is not unique, it
may be  written in several possible forms. The holography, however, suggests
 the correct one: if all object are holographically related to gravitons, one can try to mimic
  the Einstein-Hilbert action, expanded to the third order in $h^{\mu\mu}$ perturbation
  \footnote{ Other evidences that the Pomeron should be accompanied by a tensor polarization
 were given by \cite{Ewerz:2016onn}.  }
  .
This idea by \cite{Iatrakis:2016rvj} was compared to the experimentally observed
 Pomeron-Pomeron-Tensors  (PPT) vertex. It was shown that the Einstein-Hilbert
 vertex indeed works for tensor glueball $f_2(2300)$, but not for ordinary tensor mesons like $f_2(1200)$.

\section{QCD strings and  multi-string ``spaghetti"} 
\begin{figure}[h!]
\begin{center}
\includegraphics[width=7cm]{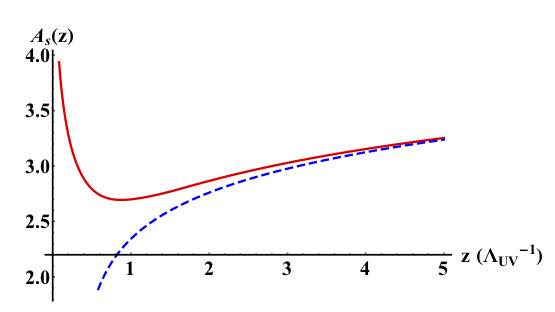}
\caption{The red solid lin shows effective potential for a string in the bulk, as a function of holographic potential $z$. The minimum corresponds to location of a string in equilibrium.}
\label{fig_holo_string}
\end{center}
\end{figure}

In the holographic QCD there are bulk strings, moving in a potential created by nontrivial profile of the gravitational metrics as well as the profile of the dilaton. The combined potential is shown in Fig.\ref{fig_holo_string}
from \cite{Iatrakis:2015rga}. The bulk string is pointlike, but since it is at some distance
from the boundary, its hologram has finite size, the radius of the QCD string. The oscillations
of the bulk string around the minimum of the potential are perceived at the boundary as the ``breathing mode" of the QCD string. 

The string interaction with gravity and dilaton mean that they can interact via exchanges
of tensor and scalar glueballs. (Recall their masses displayed in Fig.\ref{fig_glueball_masses}
in the preceding section.) The problem is, both are rather heavy, and the resulting interaction range 
$V\sim exp\big(-M_{glueball} r\big)$ is very short. This is where the glueball-meson mixing 
becomes relevant:
it allows the string interact via sigma meson ($f_0(500)$ in particle data book now) exchange.
In chapter on QCD strings we discussed it and found that this idea fits the lattice data
rather well. 

\begin{figure}[h!]
\begin{center}
\includegraphics[width=8cm]{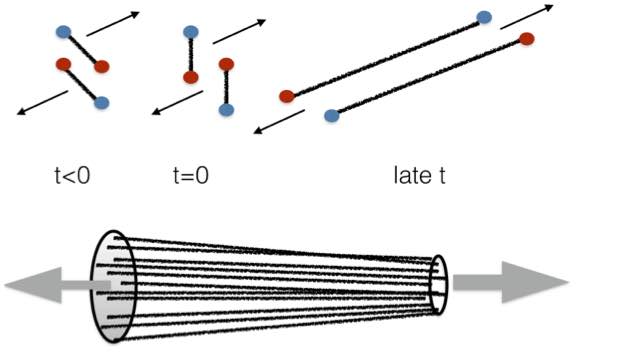}
\caption{Upper part display production of two strings in a Pomeron, a result of color exchange of the incoming color dipoles. The lower part is a sketch explaining the ``spaghetti" multi-string configuration.}
\label{fig_spaghetti}
\end{center}
\end{figure}

Multi-string configurations are produced in high energy collisions. Recall that we discussed
that Pomeron exchange leads to production of (at least) two strings. The collisions of the
proton with $Pb$ nucleus at LHC, depending on impact parameter, includes a range of string number from two to about 30 at central collision. As the string ends -- the through-going quarks -- fly along the beam directions, one finds string in ``spaghetti" situation, see Fig.\ref{fig_spaghetti}

\begin{figure}[h!]
\begin{center}
\includegraphics[width=4.9cm]{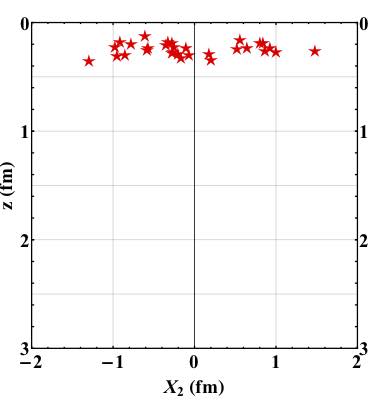}
\includegraphics[width=5.cm]{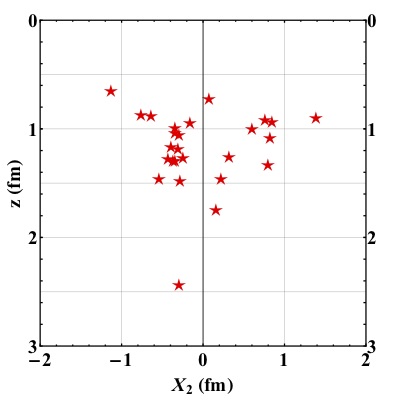}
\caption{Early and late time locations of 30 interacting parallel strings, in a plane transverse 
coordinate $x_2$ - holographic coordinate $z$. The set is falling into IR under its common weight.}
\label{fig_multistring}
\end{center}
\end{figure}

In the transverse ``plane" (which is in holography 3-dimensional, $\vec x_\perp,z$ ) 
these strings has no extension and are shown as just points.  Weak attraction which
string have to each other only become important when the density of strings is sufficiently large. 
The study of collective multi-string collapse has been studied by  \cite{Iatrakis:2015rga}, using
their EOM in the transverse plane. Two snapshots, at early and late time of some particular configuration, are shown in Fig.\ref{fig_multistring}. One can see that they become closer to each other, as a result of attraction. Also, they all fall into curved $z$ direction. On the boundary
this motion is seen as coalescing into some common ``fireball". 

\chapter{Summary} \label{chap_summary}
At the end of these lectures, let me summarize again the main content
of this course. 

\section{Semiclassical theory}

The semiclassical theory, in quantum mechanics and QFTs, is the main technical  
method on which the applications were based. It always implies existence of some 
parameter due to which the action of certain classical solutions are considered large
$S\gg \hbar$. These solutions are generally extrema of the path integrals, in imaginary (Euclidean) time. 

The contributions of those extrema to the  path integral can be then calculated by writing the
path (in QM) or field configuration (in QFT) as classical+quantum fluctuation $\phi=\phi_{cl}+\delta \phi$  substituting
it to the action and reading from it the {\em Green functions} (propagators) -- inverse to
the operator of order $O(\delta \phi^2)$ -- and $vertices$ of higher orders in $\delta \phi$.
One can then use standard Feynman diagrams to calculate the effects of these fluctuations, to the desired order. 

The most simple example of classical Euclidean paths are $fluctons$, for which this procedure
is most straightforward. They are dependent on the point $x_0$ (field configuration) for which the density matrix is evaluated. Semiclassical theory require the corresponding action be large
$S(x_0)\gg \hbar$.

We also discussed quantum-mechanical $instantons$, the tunneling events in quantum mechanics, and {\em monopoles, instantons and instanton-dyons} in QFTs. 
In all those case large action was due to presumed 
small coupling appearing in denominator $S\sim 1/g^2 \gg 1$. 

In all those cases there are {\em bosonic zero modes} induced by symmetries of the solutions,
like displacements, scale change, rotations etc. Zero modes correspond to non-Gaussian
integrals: those over {\em collective coordinates} were defined. A general new feature
is appearance of the  delta function ensuring that quantum fluctuations are orthogonal to 
all zero modes. The Jacobian generated by this  delta function leads to new Feynman diagrams, not coming from the Lagrangian. 

While the amplitudes for {\em quantum-mechanical instantons} has been calculated for a number of models up to 3 loops, for the QFTs and {\em gauge field instantons} the semiclassical theory 
has only been carried to the one-loop (determinant) order. 

(In one case, the static magnetic balls known as $sphalerons$, there is a mode corresponding to instability of these solutions. The classical solution corresponding to sphaleron decay,
in Minkowski space-time, has also been discussed.)

After (i) single-soliton semiclassical amplitudes are established, the next steps are\\
(ii) calculation of the soliton interactions; \\(iii) formulation of the partition function of their ensemble;\\
(iv) analytic or numerical integration over all collective coordinates.\\

 The interaction can be of $classical$ or $quantum$ origin. 
In general, extrema of integrals are connected by certain ``lines", streamlines  or ``thimbles"
\footnote{In many cases, for the usual integrals as well as the QM and QFT settings, it may require  $complexification$ of the paths/configurations.
The notion of ``thimbles"  explains and generalizes the famous ``Stokes phenomena" known for the ordinary integrals
and special functions, and it is related to the phase transitions in the partition functions in QFT and statistical mechanics settings. All of them can be seen as thimble reconnections, happening at some values of the parameters.  
},
being solutions of the gradient flow equation containing the force $\partial S/\partial \phi$ . We met three examples of those: for 
instanton-antiinstanton pairs in QM and gauge theory, and for instanton-dyon-antidyon pairs.  
In the dyon-dyon case (when classical interaction is absent) we also studied the
quantum one-loop interaction, generated by the determinant in two-soliton background. 

 Significant role in our discussion has been played by the {\em fermionic zero modes}.
 For monopoles those are just bound states of fermions, for QFT instantons those 
 lead to 't Hooft effective Lagrangian, in QCD with $2N_f$ fermion legs and 
 in electroweak sector with 9 quark and 3 lepton left-handed legs. Fermion-induced
 forces between instantons can be considered as diagrams with this effective Lagrangian.  

%

The semiclassical theory defines amplitudes in a form of $transseries$, complementing
perturbative series by terms of the order of $$e^{-{const  \over g^2}}(instanton\, series\, in\, g)$$ as well as
terms with exponent and powers of $log(1/g^2)$ coming from the ``streamlines".
 In QM applications certain {\em  resurgence relations} are known, relating series in
 powers of $g$ for all these terms. In QFT case such relations are not (yet?) known.

\section{Magnetic monopoles and the near-$T_c$ QCD matter as a dual plasma}

Perhaps the most discussed  3d soliton is the {\em magnetic monopole}. While electrodynamical
monopoles were never found, in non-Abelian gauge theories they can exist. The particular classical solution we discussed is the t' Hooft-Polyakov monopole. Recall that the setting was the
Georgi-Glashow model possessing the gauge field and the {\em adjoint scalar}. 

Since  adjoint scalars naturally appear in {\em extended supersymmetric theories $\cal N$=2,4}
the monopoles were extensively studied in that setting. Recall that supersymmetry forbid
appearance of nonzero vacuum energy, therefore vacua with different VEVs of the scalars are
all degenerate and form the so called ``moduli space". Seiberg and Witten has found that in the $\cal N$=2 theory
at weak coupling monopoles are heavier than ``electric" $W^{\pm}$ particles, but at certain
points the coupling goes to infinity and monopole (or dyon) mass to zero. The renormalization group flow connects those two limits smoothly. So, the same theory appear as weakly coupled non-Abelian theory with asymptotic freedom in one end of the moduli space, and as weakly coupled dual (magnetic) electrodynamics with monopoles.  The Dirac condition -- product of electric times the magnetic coupling must be an integer -- is preserved everywhere.

Fermions (gluino's and quarks) have bound state with the monopoles. If SUSY is unbroken,
those make certain supermultiplets. These statement has been explicitly verified
semiclasiscally. 
The $\cal N$=4 theory was the most important case: it was found to be {\em electric-magnetic self-dual}. This fact, without any perturbative digramms calculated, explains why it
must have $zero$ beta function: it should be the same for g and 1/g. 
(Quarks in QCD also should be able to be bound to monopoles: these issues are however not yet studied in 
any details.)

The monopoles in pure gauge and QCD-like theories were 
studied extensively by lattice numerical simulations. We know that their density increases toward $T_c$,and at $T<T_c$  they form Bose-Einstein condensate. 

We discussed phenomenology of the hadronic matter
near the QCD phase transition, its thermo and kinetic properties, related to heavy ion collision experiments at RHIC and LHC
colliders.
%
%
 While the equation of state predicted by the lattice was well confirmed by hydro explosions, the subsequent discoveries yield rather
 unexpected values  of the kinetic coefficients,
 such as entropy-density-to-sheer-viscosity $s/\eta$ ratio, heavy quark diffusion coefficient $D/T$
 and jet quenching parameter $\hat q/T^3$. We argue further that all of them have
 rather peculiar $T$-dependence, indicating extremely small mean free path
 of the matter constituents, especially between $T_c$ and $2T_c$. This transition is
 the deconfinement transition.
 
 Rescattering between constituents is proportional to their densities. Yet the (dimensionless) densities of
 quarks and gluons $n_{q,g}/T^3$ rapidly decrease as temperature goes down to critical value $T_c$, due to confinement. The only density
 which is $peaked$ at $T_c$ is that of the magnetic monopoles $n_m$. This argument, and more detailed 
 calculations we discussed in that chapter, indicate that significant density of the particle-monopoles
 is the most probable  cause of the unusual sQGP kinetics observed experimentally. 

In short, QGP near $T_c$ seems to be a ``dual plasma", containing a density of monopoles comparable to that
of ``electric" objects, quarks and gluons. A number of theoretical methods were used subsequently to
study such dual plasmas, from classical molecular dynamics, to quantum cross section and kinetic theory,
to quantum path integral monte carlo (PIMC). Those reproduce correlation functions of monopoles
observed on the lattice in detail.

\section{Instanton-dyons, deconfinement and chiral restoration phase transitions}

Another version of non-perturbative theory at $T\sim T_c$ is based on semiclassical objects,
the {\em instanton-dyons}.

  
 Incorporation of {\em nonzero VEV of the Polyakov line} -- called holonomy --
 lead to a shift  from studies of instantons to studies of their constituents -- {\em instanton-dyons}
 or instanton-monopoles.
 Recent papers on their ensembles,  done by a variety of methods, we discussed in the previous chapter.
  lead to very significant advances. 
  
  Unlike instantons,
  these objects have three different set of charges, {\em topological, magnetic and electric},
  therefore back-reacting on the holonomy potential. 
%
%
The calculations showed that at sufficiently large density of the dyons, the  minimum of the free energy {\em shifts
to  confining value of the holonomy}, at which the mean value of the Polyakov line 
vanishes, $<P>=0$. This creates the so called symmetric phase, in which all types of the dyons obtain equal actions
and densities.

Unlike instantons, the dyons posess magnetic charges, and thus their ensemble
generates {\em the magnetic screening mass}. Recall that perturbative polarization
tensor does not generate it \cite{Shuryak:1977ut}: but, according to lattice 
data, in the near-$T_c$ region it even surpasses the electric mass. While we have not discuss it above, let me just mention that it clearly indicates a transition from
electric (QGP) to magnetic plasma, as the coupling grows with decreasing temperature.


Further remarkable findings were obtained using a very sensitive tool, deforming QCD-like theories with
{\em nonzero flavor holonomies} $\theta_f$, also known as imaginary chemical potentials,
 or modified periodicity phases.
 The so called Roberge-Weiss transitions, originally postulated at high $T$ only on perturbative grounds,
  are no confirmed to be related with the ``hopping" of quark zero modes from one type of dyon to the next.
A ``democratic" distribution of flavor holonomies, corresponding to the so called $Z(N_c)$ QCD, 
were shown to modify the phase transition in a very dramatic way.
Deconfinement transition is strengthened, to the strong first order transition, while
chiral symmetry restoration transition is weakened to become non-existing at all, at any temperature.

These finding complement another kind of ``deformed QCD", in which quarks are substituted
by adjoint gluinoes periodically compactified on Matsubara circle. The theory with $one$ such gluino,
  $\cal{N}$=1 SYM  ($N_a=1$)  preserves both confinement and chiral symmetry breaking all the
  way to very small circle (very high $``T"$), eliminating phase transitions.   The theory with $two$ such gluinoes
  shows even more spectacular behavior, with 3 phase transitions in the Polyakov line and 4 subsequent phases (confined-deconfined-
  another deconfined - reconfined), all of them in the chirally broken phase!  

One should stress, that flavor holonomies constitute very ``soft" modification of the theory,
in a periodicity condition on the Matsubara circle. All gluons and perturbative RG parameters 
of the theory remain unchanged -- and yet all phase transitions are changed dramatically.
It would be impossible to explain it using theory of instantons alone: the number of
their fermionic zero modes is defined solely by the topological charge, and are insensitive to these holonomies.
While those changes appear natural in the framework of the instanton-dyon theory,
so far no other known explanation of them exists. It is a significant challenge 
now to any other model of the 
deconfinement and chiral transitions to explain these phenomena. 
Lut us even speculate that we are now perhaps approaching ``the point of no return", 
at which the mechanism of the QCD phase transitions we discussed in this book is going to be finalized.

We also discuss lattice studies based on the so called ``fermion method" to study underlying topology.
It is unmistakedly found to be the instanton-dyons. The semiclassical model
can be extensively checked on the lattice: so far we see unexpectedly high accuracy of its
predictions, at least for the lowest Dirac eigenstates.

\section{The ``Poisson duality" between the monopole and the instanton-dyon  descriptions}

Let us start with formulation of the physics behind this duality. For a description of a thermal system of any particles one can adopt two well known strategies, which produce two
different forms of the partition function. 
(One may call them Hamiltonian and Lagrangian approaches.)
Needless to say, they both describe the same system
and thus must be identical.  While to show its equivalence explicitly is often difficult, as they are
simple in two opposite limits, of low and high $T$, it was in fact possible in certain highly symmetric examples, to be discussed in this section.

 The standard {\em approach 1} is to find all the
states of the system and perform the usual statistical sum $Tr[exp(-\hat H/T)]$. 
It is working best at low
temperatures, where only some lowest states needs to be included.

The   {\em approach 2} is to go to Euclidean formulation and evaluating
the partition sum using the path integral over the  paths periodic in the Matsubara time. 
Those can be classified by their winding number (also known as BEC cluster number).
This approach works best at high $T$ or small Matsubara circumference $\beta=1/T$,
in which case the paths with zero winding number dominate.   

In order to see how these two approaches work for monopoles, one needs 
a setting in which both the monopoles and instanton-dyons are well defined
semiclassical objects. Convenient settings thus include 
theories with extended supersymmetry, because those have scalar fields and 't Hooft-Polyakov monopoles. Both of these theories attracted a lot of theorist's attention
in the 1990's: some of that will be useful for our current goal, which can be formulated as
{\em getting some understanding on inter-relation between the physics of particle-monopoles and 
instanton-dyons}. 

In section \ref{QCDadj_2} we had discussed QCD with two (or more) adjoint gluinos. By adding appropriate
number of scalars, those theories can be upgraded to theories with extended supersymmetry. Specifically,
adding one complex scalar $a$ to the $N_a=2$ theory one gets  $\cal N$=2 SYM, and by adding 6 scalars 
to the $N_a=4$ theory one gets  $\cal N$=4 SYM. 
This has been done in literature, first for  $\cal N$=4 SYM by  N.Dorey and collaborators  \cite{Dorey:2000dt,Dorey:2000qc,Chen:2010yr}, and then also for  $\cal N$=2 SYM
by Poppitz and Unsal
 \cite{Poppitz:2011wy}. 

What is common to all those papers is that they start by compactifying one of the 
dimensions to a circle $S^1$ of a circumference $\beta$. In contrast to the thermal theory,
however, the gluino fields are assumed to be $periodic$ on this circle, and therefore 
the supersymmetry is $not$ going to be broken. We will still call this direction the 0-th one.
Furthermore, the compactification allows one to define two holonomies, the Polyakov loop
with the $A_0$ field and  the magnetic field with a dual magnetic potential. 
Dorey et al call those $\omega$ and $\sigma$, respectively, and their VEVs can be considered 
two main parameters of the settings, together with $\beta$. In order to make the
discussion simpler, one assumes the minimal number of colors $N_c=2$,
in which there is only one diagonal generator $\tau^3$, breaking $SU(2)\rightarrow u(1)$
so these VEV's   $\omega$ and $\sigma$ are just parameters.

On top of that, the $\cal N$=4 theory is discussed\footnote{The Dorey et al also discussed 
two brane constructions corresponding to two Poisson-dual formulations, which we would not discuss here. 
} in the so called Coulomb branch,
which means that, on top of 
the holonomies $\omega$ and $\sigma$,  one -- or more of 6 available -- 
scalars is also assumed to have a
nonzero VEV, called $\phi$, for the same color-diagonal component.   
This of course leads to monopoles with the action %
%
\be S_{m}=({4 \pi \over g^2})\sqrt{\beta^2 |\phi|^2+| \omega - 2\pi n |^2} \ee 
including the contribution from the scalar VEV $\phi$,  electric holonomy $\omega$ and
the winding number of the path in the $S^1$ circle $n$. The $n=0$ term is what
we called above the $M$-type instanton-dyon, the $n=-1$ the $L$ dyon,
and higher $n$ correspond to the paths with stronger time-dependent twists.
 We would not derive the partition function\footnote{We simplify it here
a bit, compared to
the original paper, by putting one more external parameter of the setting, the CP-odd $\theta $ angle, to zero.}   but just present the resulting expression
\be Z_{inst}= \sum_{k=1}^\infty \sum_{n=-\infty}^\infty\big( {\beta \over g^2} \big)^9{k^6 \over (\beta M)^3}
exp\big[ i k \sigma-\beta k M -{k M \over 2 \phi^2 \beta}(\omega - 2\pi n)^2 \big]
\ee 
where $M=(4\pi\phi/g^2)$, the BPS monopole mass without holonomies,
and thus the second term in exponent is interpreted as just the Boltzmann factor.
The index $k$ is the magnetic charge of the configuration. Note that it appears in the exponent
times the magnetic holonomy $\sigma$, and since $k$ is an integer, the expression is periodic in 
$\sigma$ with the $2\pi$ period, as it should. The second index $n$ is the winding number of the path in the $\beta$ circle.  Note that all values of $n$  need to be included in the sum, because $Z$  should be periodic 
in the electric holonomy $\omega$ as well as in $\sigma$.
Finally note that the last term of the exponent has unusual position of $\beta$ (or ``temperature" in the numerator): the sum over $n$ therefore converges better at 
small $\beta$ (high $T$) limit. 

Now we switch to other description, which is better convergent in the opposite case, of
large $\beta$ and  low $T$. It operates with states of motion of monopoles in its 4 collective coordinates. Three of those are locations of the monopole and are included in trivial way.
The fourth collective coordinate is the angle $\alpha$ of monopole color rotation which preserves the
holonomy, $\Omega=exp(i \alpha \tau^3)$ defined on $another$ circle $S^1$. Therefore,
the problem includes a ``quantum rotator". As was explained by Julia and Zee ,   
the corresponding integer angular momentum $q$ is nothing but the electric charge
of the rotating monopole. The partition function looks in this approach as follows
\be Z_{mono}= \sum_{k=1}^\infty \sum_{q=-\infty}^\infty\big( {\beta \over g^2}\big)^8 
{k^{11/2} \over \beta^{3/2} M^{5/2} } exp( i k \sigma - i q \omega -\beta k M -{\beta \phi^2 q^2 \over 2 k M})
\ee
Now both holonomies appear with appropriate integers, so the periodicity in $\sigma,\omega$ is as required.
The last term in exponent is due to the kinetic energy of the rotation, with angular momentum squared and
the rest being nothing else but the momentum of inertia of the monopole (in denominator).
Note also that $\beta$ (or temperature) is in this term in the usual thermal position,
so that the sum in $Z_{mono}$ is more suppressed at higher $\beta$ or lower temperature.

We have copied  those expressions from the original work in order to demonstrate the key statement of this section, pointed out by Dorey and called 
the {\em Poisson duality}: the two seemingly different 
expressions, addressing seemingly different motions in two different circles,
do in fact lead to the $same$ partition function:
$$Z_{inst}=Z_{mono}$$ Indeed, performing the sum over $q$ in the latter
expression -- the discrete Fourier transform of a Gaussian -- one gets the so called
{\em periodic Gaussian} given by the sum over $n$ in the former expression.

In fact, both expressions generate the same function. Let us understand why it is so. The first exponents
(outside of the sums) are the same for obvious reasons: the masses of the particle-monopole and instanton-dyon
are the same, as so are their magnetic charge -- thus $e^{i\sigma}$. To understand what is happening
with the sums, let us simplify the issue to the simplest problem possible, in this case a particle
moving in a circle. Thermal Euclidean time theory is thus defined on two circles $S^1\times S^1$.

The low-$T$ theory starts with defining the spectrum of excitations. As usual, since there is no dependence on the 
position on the circle, the angular momentum $l$ is conserved (commutes with the Hamiltonian) and the
excited states are numerated by it. The spectrum of a rotator is $E_l = l^2/2mR^2$ where $m$ is the particle mass
and $R$ is the circle radius. Furthermore, if there is a magnetic flux $\Phi$ through the circle, so that our particle
 gets the Aharonov-Bohm phase, the spectrum shifts to $E_l = (l-e\Phi)^2/2mR^2$. The partition function is
 $Z=\sum_l exp(-E_l/T)$. 
 
 The dual description at high $T$ describe paths of the particle in terms of how many times it is ``winding"
 around the circle. At high $T$ the thermal circle is very short, so most of the periodic paths would be 
 approximately time-independent. But as the thermal circle gets longer, there appear paths in which a particle
 rotates by additional $2\pi n_w$ times around the spatial circle. It is not difficult to calculate the 
 action for such paths and get another representation for $Z$, better converging at high $T$.
 Since both descriptions correspond to the same quantum mechanical problem,
 one should not be surprised that both of them give the same $Z$. 

The next step was to apply this Poisson duality to QCD, calculating semiclassical sum with all winding numbers into
the monopole-like identical sum. What was found is that in the QCD case the monopole action is
$S\sim log(1/g^2)\sim log(log(T))$, and thus the density 
$$ exp(-S)\sim {1 \over log(T)^2}$$ 
This explains long-known lattice data on the monopole density. It also tells us that in pure gauge and QCD-like
theories the monopoles are $not$ classical objects(!)

\section{The QCD vacuum and correlation functions: instantons}


I was asked many times: why the topic of gauge topology needs to be studied,
in view of the fact that lattice gauge theories include automatically all of these objects anyway
? 
Why do we need any dedicated studies of these degrees of freedom and
 effective models, in view of the fact that  lattice gauge theory simulations do reproduce the hadronic spectrum,
 equation of state of thermal hadronic matter, and many other observables?  
 
 The  answer is that degrees of freedom associated with topological objects are
 much more important than millions other degrees of freedom simulated by modern supercomputers.
 The first answer is that the topological solitons have zero modes, which translate them into effective
 multi-fermion operators. Without any idea about topology and understanding
 of the corresponding the index theorems, and these zero modes themselves, it would be 
 probably next to impossible to figure out where these interaction between quarks come from.
 
 It is very important, that topology-induced  interaction between quarks are in fact the strongest
 forces shaping the observed hadronic spectrum. It is this interaction which makes pions
 $\pi$ near-massless 
 and $\sigma$ quite light, $$ m_\pi^2\approx 0, \,\,\,\,\, m_\sigma^2\sim 0.2\, GeV^2$$
  while $\eta '$ and spin-0 isospin-1 meson we called $\delta$
 are nearly as heavy as the nucleon, $$m_{\eta'}^2\sim m_\delta^2\sim 1 \, \,GeV^2$$.  
 We systematically studied the QCD point-to-point correlation functions in the corresponding chapter,
 and had shown how their splitting develops as a function of the distance between the operators,
 ``probing" the QCD vacuum, and have seen that they are indeed stem from the 
 topology-induced effective 't Hooft Lagrangian. 
 
 Moreover, the  distances at which such non-perturbative
 phenomena turns out to be rather small, indicated that these topologically nontrivial gauge fields are
 rather strong. They are even more important for spin-zero gluonic operators, dominating
 the perturbative effects (and thus limiting any pQCD applications) for momentum transfer $Q^2 < 10 \, GeV^2$.
 
 We have shown above that the gauge topology is key to understanding of the $SU(N_f)$ and $U(1)_a$ chiral symmetries. Lattice practitioners know well the Casher-Banks relation , relating the density of the Dirac eigenvalues
 near zero to the quark condensate. Much less widely known is the notion of the {\em zero modes zone} (ZMZ),
 a thin layer of Dirac eigenvalues made out of collectivized zero modes. All textbooks 
 repeat the standard notion of the light quark masses $m_u,m_d,m_s$, as being much smaller compared to some strong coupling scale
 $\Lambda_{QCD}$ and thus justifying a Taylor expansion in their powers. 
 In reality, many masses used on the lattice are actually comparable to the ZMZ width.
 
 As a result, lattice practitioners  continue to be surprised by large deviations from 
 linear chiral perturbation theory, for quark masses they routinely use. The reason for that is an observation --
 quite elementary in terms of instanton ensembles and 
 going back to early 1980's -- that the hopping matrix elements for quark jumping between topological solitons, and thus 
 the width of the ZMZ, are of the order of only 20-30$\, MeV$, an order of magnitude smaller than   $\Lambda_{QCD}$.
 
 The reason the ZMZ so small  width is also of the topological origin. No matter how small or large are perturbations
of the solitons, their topology remains integer-valued and their zero modes remains $unperturbed$, at
zero Dirac eigenvalues. The only effect which does perturb them is a presence of another topological anti-soliton
nearby.  
 
 Finally, about some practical effects of topology. 
Modern supercomputers simulate path integrals QCD-like gauge theories with millions of variables:
and yet the results -- such as correlation functions, hadronic masses and other properties -- are still subject
 of large fluctuations, from configuration to configuration. Also some observables -- e.g. mean topological charge
 $<Q>$ -- may show nonzero values, in contradiction to CP invariance of the theory.
 The reason for both phenomena is the fact, that local update algorithms used are notoriously inefficient
 in updating the topology of the configurations. This can in principle be improved by new algorithms
 able to identify/create/destroy the topological objects and appropriately update their collective coordinates.
 Furthermore, even the largest simulated volumes contain
 only $O(10)$ topological solitons, not a very large number.

\section{Outlook}

Finally, let us at least enumerate some important issues which, for various reasons, was $not$ discussed above.

One of the goals of the book was to help to
bridge an existing gap between  two communities, which can be loosely called -- following hep-th and hep-ph arXiv
denomination -- as the $phenomenologists$
(lattice, hadron spectra, hadron properties, quark gluon plasma) and hard $theorists$ (supersymmetric theories, string theory, holographic dualities). Unfortunately, the communication between them is weak.

As we emphasized, repeatedly from the Introduction,
topological solitons are not just some cute exotic objects from a mathematical zoo. 
Their ensembles have various phases, many of which are (or may be in future) important
for very practical applications. The vortices in type-II superconductors need to be in a crystalline form,
to produce good industrial magnets. The density of monopoles (or instanton-dyons) should be  
 sufficient to generate color confinement and chiral symmetry breaking. We also discussed evidences
 that an interplay between electrically and magnetically charged quasiparticles are at the hard of
 unusual kinetics of quark-gluon plasma. Sphaleron transitions provide chiral imbalance
 in matter, driving a non-dissipative current due to chiral magnetic effect, now extending
 into condense matter applications, and perhaps even into electronics.
 The instanton-induced multi-fermion interactions generate
 the so called color superconductivity, expected to dominate thermodynamics and kinetics of the dense matter
 deep inside the neutron stars.
 
On the theory side, interest to electric-magnetic duality has been superseded, at the end of 1990's, 
by the holographic dualities of the   AdS/CFT type. Originally found for conformal $\cal{N}$=4 SYM,
it has been extended, by softly broken supersymmetry, to $\cal{N}$=2 and so on. The Seiberg-Witten
elliptic curve has been reformulated in terms of certain brane constructions. While 
the top-down holography extension
to $\cal{N}$=1 and to non-supersymmetric ($\cal{N}$=0) theories  meets with problems, multiple down-top
models were developed, commonly known as AdS/QCD. Those describe rather well many aspects
of hadronic spectroscopy and QGP phenomenology. 


 \section*{Acknowledgements}
The progress reported
 would not be possible without contributions by Pierre van Baal and
 Mitya Diakonov,  who are no longer with us but whose legacy  is quite visible in these lectures.
 I am indebted to many colleagues and all my collaborators, for explaining to me 
 many issues discussed in the lectures. Special thanks go
  to  my long-time collaborators at Stony Brook, Jac Verbaarschot and Ismail Zahed,
  and to my former students Pietro Faccioli, Jinfeng Liao, Shu Lin, Rasmus Larsen and Adith Ramamurti.

\appendix
\begin{appendix}
\chapter{Notations, units}

\section{Some abbreviations used}
{\bf AdS} \ = \ anti-de Sitter spacetime \\
{\bf CFL=CSC3} \ = \ Color-Flavor Locked phase, or Color superconducting phase
with 3 flavors\\
{\bf CSC2} \ = \ Color superconducting phase with 2 flavors\\
{\bf CFT} \ = \ conformal field theory \\
{\bf DIS} \ = \ Deep inelastic scattering \\
{\bf EoS} \ = \ Equation of state \\
{\bf $N_c$ and $N_f$} are numbers of quark colors and flavors\\
{\bf NJL} \ = \ Nambu-Jona Lasinio model \\
{\bf MFA} and {\bf  RPA} are the Mean Field and Random Phase Approximations\\
{\bf OPE } \ = \ Operator product expansion \\
{\bf QCD } \ = \ Quantum Chromodynamics, pQCD is its perturbative version \\
{\bf QED} \ = \ Quantum Electrodynamics \\
{\bf QGP } \ = \ Quark-Gluon PLasma \\
{\bf RG} \ = \ Renormalization Group\\
{\bf RILM } and {\bf IILM } are Random Instanton Liquid Model and Interacting Instanton Liquid Model\\
{\bf SUSY} \ = \ Supersymmetric\\
{\bf SYM}= \ Supersymmetric\ Yang-Mills theory\
{\bf UV} and {\bf IR } are ultraviolet and infrared limits, meaning the limits of
large and small momentum scales \\
{\bf VEV } \ = \ Vacuum expectation value \\
{\bf ZMZ}\ = \ Zero mode zone \\

\section{Units} 
We use standard ``natural units'' of high energy/nuclear physics
in which the speed of light and the Plank constant $\hbar=c=1$.
Thus length and time has the same dimension, the inverse of momentum
and energy. Transition between units occurs by a convenient
substitution
of 1 according to
$$ 
1= 0.19732\, fm\, \cdot GeV\, 
$$
and then cancellation of femto-meters ($fm=10^{-15}\, m$, also known as $fermis$)
or GeV ($10^9 \, eV$ or Giga-electron-volts) as needed.

Discussion of the temperature $T$ uses also Boltzmann constant $k_B=1$,
so it is measured in  energy units (e.g. $GeV$). 

\section{Space-time and other indices, standard matrices}
We follow standard physics convention that an index appearing twice
on one side of the equation is a dummy variable with the summation
implied, e.g. $a_m b_m\equiv \sum_m a_m b_m$. 

We use Latin letters $a,b...$ to count color generators, 1-8 or
 1-$N_c^2-1$,
 and $i,j..$
to count colors, 1-3 or 1-$N_c$. 
We use letters $l,m,n$ also to count spatial vectors 1-3.

Greek letters are generally used for space-time. Standard Minkowski
metrics $g^M_{\mu\nu}=diag(1,-1,-1,-1)$ is implied in sums,
$$ a_\mu b_\mu\equiv \sum_{\mu\nu}g_{\mu\nu} a_\mu b_\nu$$  

Transition to Euclidean time is done with 
$$ x_0^{M}=-ix_4^E\qquad x_m^{M}=x_m^E$$
 The Euclidean metrics is 
just $g^E_{\mu\nu}=\delta_{\mu\nu}=diag(1,1,1,1)$.

Pauli matrices $\tau^m_{ij}$, all indices 1..3, are twice the
generators
of the SU(2) rotations. They satisfy the basic relation
$$ \tau^a \tau^b =\delta^{ab}+i\epsilon^{abc} \tau^c $$  

Color SU(3) generators are half of the Gell-Mann matrices  $T^a=t^a/2$,
a=1..8. We also use a notation $\lambda^a$ for the same set
of matrices, and also use those for SU(3) flavor.
 Their product can be written in a for similar to that for Pauli matrices 
$$ t^a t^b ={2\over 3}\delta^{ab}+t^c (d^{abc}+if^{abc})  $$ 
where $d,f$ are some standard numerical tensors of the SU(3) group.

\section{ Angular momentum in four dimensions and t'Hooft $\eta$ symbol}
\label{app_eta}

Angular momentum algebra related with 3-dimensional rotation group $O(3)$
is assumed to be familiar from quantum mechanics textbooks.
Just reminding,  classically one defines the angular momentum as $\vec x \times \vec p$, so its quantum version is
$$\hat L_i=\epsilon_{ijk} x_j  \big( -i {\partial \over \partial x_k} \big) $$ 
Its square $\hat L_i^2$ enters Laplace-Bertrami operator of the Laplacian,
producing the so called centrifugal potential   $$V_l={\hat L^2 \over r^2}={l(l+1) \over r^2}$$
 
In 4 dimensions  the group of rotations $O(4)$  has 6 generators. There exist a standard formalism with
total angular momentum and different angular functions for each dimension in math literature, but we would not
follow it here. Instead we will view these 6 generators as {\em two pairs} of $O(3)$ or $SU(2)$ algebras,
much more familiar to us. Often these two are called left and right-handed sets of angular momenta.

In taking this road, one has 
to face  a certain dilemma:, one can either  keep (i) the standard normalization of the angular momentum
; or (ii)  the standard form of the  $SU(2)$ 
commutation relation. Following t' Hooft,  we use the latter choice,  defining  these {\em two pairs} of angular momenta operators by
\be L_1^a=-{i \over 2} \eta_{a\mu\nu} x^\mu {\partial \over \partial x_\nu} \ee
\be L_2^a=-{i \over 2} \bar\eta_{a\mu\nu} x^\mu {\partial \over \partial x_\nu} \ee
which do not have standard normalization (due to extra 1/2) but commute in the standard way, namely
$$\big[ L^a_p L^b_q \big]=i \delta_{pq}\epsilon_{abc} L^c_p $$
Note that $\delta_{pq}$ indicate that two sets are mutually commuting. 
Note extra $1/2$ in definition of $L_{1,2}$ compared to standard quantum mechanics in 3d:
it leads to extra factor 4 in angular part of the Laplacian. With it
the radial equation adds the centrifugal potential of the form   $+2(L_1^2+L_2^2)/r^2$ .

These notations imply that some simple object look a bit unusual. For example the 4-vector 
is not simply $l=1$ object, as it is in 3d, but a tensor made of left and right spinors
  $(l_1,l_2)=(1/2,1/2)$ representation
of $O(4)$. The corresponding centrifugal term in the vector channel obtains the coefficient
of the centrifugal potential $2l_1(l_1+1)+2l_2(l_2+1)=3$.  

(One more reason physicists are using such notations is that the $O(4)$ group is obviously a close relative of the 
Lorentz group, of 1+3 Minkowski space-time. One of the $SU(2)$ is rotations, and another 
(modified) is that of the boosts. )

 Before defining 't Hooft symbol $ \eta_{a\mu\nu} $, we  elevate 3 Pauli matrices to  4 quaternions
\be
    \tau_\mu^{\pm} = (\vec\tau,\mp i),
\ee
where $\tau^a\tau^b=\delta^{ab}+i\epsilon^{abc}\tau^c$ and
\be
 \tau_\mu^+\tau_\nu^- &=& \delta_{\mu\nu}+i\eta_{a\mu\nu}\tau^a ,\\
 \tau_\mu^-\tau_\nu^+ &=& \delta_{\mu\nu}+i\bar\eta_{a\mu\nu}\tau^a ,
\ee
with the $\eta$-symbols given by
\be
 \eta_{a\mu\nu} &=& \epsilon_{a\mu\nu} + \delta_{a\mu}\delta_{\nu 4}
                       - \delta_{a\nu}\delta_{\mu 4}, \\
 \bar\eta_{a\mu\nu} &=& \epsilon_{a\mu\nu} - \delta_{a\mu}\delta_{\nu 4}
                       + \delta_{a\nu}\delta_{\mu 4} .
\ee
The $\eta$-symbols are (anti) self-dual in the vector indices 
\be
 \eta_{a\mu\nu} = \frac 12 \epsilon_{\mu\nu\alpha\beta}
    \eta_{a\alpha\beta}, \hspace{1cm}
 \bar\eta_{a\mu\nu} = -\frac 12 \epsilon_{\mu\nu\alpha\beta}
    \bar\eta_{a\alpha\beta} \hspace{1cm} 
 \eta_{a\mu\nu} = -\eta_{a\nu\mu} . 
\ee
We have the following useful relations for contractions involving
$\eta$ symbols
\be
 \eta_{a\mu\nu}\eta_{b\mu\nu} &=& 4\delta_{ab},\\ 
 \eta_{a\mu\nu}\eta_{a\mu\rho} &=& 3\delta_{\nu\rho}, \\
 \eta_{a\mu\nu}\eta_{a\mu\nu} &=& 12, \\
 \eta_{a\mu\nu}\eta_{a\rho\lambda} &=&
       \delta_{\mu\rho}\delta_{\nu\lambda}
      -\delta_{\mu\lambda}\delta_{\nu\rho}
      +\epsilon_{\mu\nu\rho\lambda},\\
 \eta_{a\mu\nu}\eta_{b\mu\rho} &=& \delta_{ab}\delta_{\nu\rho} 
      + \epsilon_{abc}\eta_{c\nu\rho}, \\
 \eta_{a\mu\nu}\bar\eta_{b\mu\nu} &=& 0.
\ee
The same relations hold for $\bar\eta_{a\mu\nu}$, except for
\be
\bar \eta_{a\mu\nu}\bar\eta_{a\rho\lambda} =
       \delta_{\mu\rho}\delta_{\nu\lambda}
      -\delta_{\mu\lambda}\delta_{\nu\rho}
      -\epsilon_{\mu\nu\rho\lambda} .
\ee
Some additional relations are
\be
 \epsilon_{abc}\eta_{b\mu\nu}\eta_{c\rho\lambda} &=&
 \delta_{\mu\rho}\eta_{a\nu\lambda} -
 \delta_{\mu\lambda}\eta_{a\nu\rho} +
 \delta_{\nu\lambda}\eta_{a\mu\rho} -
 \delta_{\nu\rho}\eta_{a\mu\lambda} ,\\
 \epsilon_{\lambda\mu\nu\sigma}\eta_{a\rho\sigma} &=&
  \delta_{\rho\lambda}\eta_{a\mu\nu} +
  \delta_{\rho\nu}\eta_{a\lambda\mu} +
  \delta_{\rho\mu}\eta_{a\nu\lambda}.
\ee

\chapter{Conventions for fields in Euclidean vs Minkowskian space-time}
\section{The gauge fields}
The QED/QCD gauge part of the Lagrangians
is
$$ S=-{1\over 4} \int d^4x (G_{\mu\nu}^a)^2$$ 
where the QCD field
$$ G_{\mu\nu}^a=\partial_\mu A^a_\nu-\partial_\nu A^a_\mu+gf^{abc}
 A^b_\mu 
 A^c_\nu$$ 
where $f^{abc}$ are structure constant of the $SU(N_c)$ Lee algebra
(For SU(2) it is $\epsilon_{abc}$). Alternative form of the 
gauge fields is a matrix notations, in which the generator are
 included together with fields $A_\mu=\equiv A^a_\mu T^a$:
then the
second
 term
is the commutator.

In pQCD one uses the so called {\em perturbative definition} in which the
coupling
constant is explicitly written in the non-linear terms, while the
kinetic terms are free of it. We will use the non-perturbative definition, used
in lattice and instanton studies, which is obtained by an inclusion of g
into
$\tilde A=gA$ and $\tilde G=gG$ so that $g$ no longer appears in front of the nonlinear
terms,
but is placed instead in front of the action $S={-1\over 4 g^2}\int
d^x \tilde G^2$. 

The transition to Euclidean time is done by
$$ A_0^M=iA_4^E \qquad A_m^M=-A_m^E$$ 
Note the minus sign on spatial components, different from what happens
with coordinates themselves. This is done in order
not to modify covariant derivatives, so that both E and M read as
$$  iD_\mu=i\partial_\mu+{g\over 2}A^a_\mu t^a $$ 

\section{Fermionic path integrals}
Introduction of fermion fields into the path integral needs
special definitions appropriate for
Grassmannian (anti-commuting) variables.
Those have been defined in a classic work  \cite{Berezin:1974du}
and is  discussed in any modern textbook on QFTs.
Additional subtleties appear with its transformation
into the Euclidean space-time, in which $\bar q$ and $q$ must be
treated as
independent variables.

    I have to explain in what sense the integral over $\psi$
should be defined. It is a fermionic variable, not just the ordinary field.
They are called Grassman variables, or anti-commuting ones
$\chi_1\chi_2=-\chi_2\chi_1$. In particular,
for such variables  one has $\chi^2=0$, which represents the Pauli principle.
These variables have funny rules for the integrals. They are given essentially
by two basic integrals
\be \int d\chi =0; \hspace{2cm} \int \chi d\chi=1 \ee
  One can then derive the following formula
\be \int exp(-\chi^*_k M_{k,l}\chi_l) d\chi^*_1d\chi_1\cdot\cdot\cdot
d\chi^*_Nd\chi_N= det M \ee which can be proved in the 
eigenvector basis by decomposition of the exponent.
Note, that the ordinary Gaussian integral with N variables  of such
type
is equal to 
\be \int exp(-\sum_{k,l}\phi_k M_{k,l}\phi_l/2) d\phi_1d\cdot\cdot\cdot
d\phi_N= {(2\pi)^{N/2}\over \sqrt{det M}} \ee
and the determinant stands in the denominator.
So, if one of the eigenvalues is zero, the fermionic integral vanishes while
in the corresponding bosonic one  diverges.

Note also in passing, that one fermionic
integral can compensate two bosonic ones, if the
eigenvalue spectra happen to be equal.
This comment is important for supersymmetric theories, in which
fermionic and bosonic integrals do indeed
 compensate each other, in the vacuum energy
and many other observables.

Fortunately, all of the above can be bypassed because a general
integration
over the fermions in QCD can be made in a simple form because those
appear in the action only $linearly$.
The schematic  master formula for it is 
\be \label{eqn_Grassman_int}
\int D\bar q Dq e^{\bar q M q}=det(M)\ee
where M is in general a matrix in all fermionic indices and also 
possibly a differential  operator\footnote{
To prove it imagine that the operator is diagonalized, the exponent
is expanded and the ``Pauli principle'' in the form $q^2=0$ holds.}.
In QCD $M=iD_\mu\gamma_\mu$ is the Dirac operator, with the color 
matrix in covariant derivative obviously in fundamental representation.
  The final comment, as the operator $i \hat D$ is Hermitian,  its
eigenvalues $\lambda$ are all real. However, one may still ask how it
happens that at non-zero mass (entering with $i$ in the Euclidean formulation)
$$ \prod_f det[ i\hat D(A_\mu(x)) +im_f] $$
and non-positive $\lambda$, the ratio of the determinants happen to
be positive (otherwise one cannot use probability language). 
We noticed that the former Dirac operator is hermitian, therefore its 
eigenvalues $\lambda$ are real.   Due to chiral symmetry, they go in pairs and
 therefore
we have always  $$\prod_{\lambda > 0}(\lambda+im)(-\lambda+im)=
 -\prod_{\lambda > 0}(m^2+\lambda^2)$$, so this factor   is 
real too, and even  has definite sign and can be with proper definition 
made positive. This is important, as we want to prescribe to it the meaning of 
the probability of occurrence of the corresponding configuration in the vacuum.

\section{Quark fields}
We denote quarks fields as $\psi$ or $q$, usually omitting but
implying
their spinor index $\alpha=1..4$, color $i=1..N_c$ and flavor
$f=1..N_f$. The QED/QCD Lagrangian is
$$ S=\int d^4x \bar q (i\gamma_\mu D_\mu -m)q $$ 

The transition to Euclidean time is done by
$$ q^E=q^M,\,\, \bar q^M=-i\bar q^E,$$
and the gamma-matrices change by $$ \gamma_4^E=\gamma_0^M, \,
\gamma_m^E=-i\gamma_m^M,$$
We remind that the anti-commutators are
$$   \{ \gamma^M_\mu, \gamma^M_\nu\}=2g_{\mu\nu} \qquad
 \{ \gamma^E_\mu, \gamma^E_\nu\}=2\delta_{\mu\nu} $$
Another often used notation is a slash or a hat, indicating a
convolution of a 4-vector with the gamma matrices.
For example, the relations just described can be written
using these notations as
$$\hat a \hat b+\hat b\hat a=(ab)$$
where $a_\mu,b_\mu$ are any 4-vectors.

Finally, the Euclidean fermionic action looks like
$$ S^E=-iS^M=\int d^4x \bar q(-i\gamma_\mu D_\mu -im) q $$

Let me explicitly mention the
definitions used here
$$\psi_e=\psi_M, \bar \psi_M=-i\bar \psi_E, \gamma^0_E=\gamma^0_M,
\gamma^m_E=-i\gamma^m_E$$
We get $$S^f_E=\int d^4x\bar\psi_E (i\hat D +im) \psi$$
thus the complete partition function of QCD is

$$ Z=\int DA_\mu(x)D\bar\psi D\psi exp(-S_E-S^f_E)$$

\chapter{Perturbative QCD}
%

\section{Renormalization group and asymptotic freedom} \label{sec_as_freedom}
 A systematic approach using continuous renormalization group (RG) has been developed originally  in \cite{GellMann:1954fq} and describes the renormalization of the
charge as a function of (momentum) scale. They introduced the so called beta function
 defined as the charge $derivative$ over the scale magnitude
at which it is defined
\be{ \partial g \over \partial \log\mu}\equiv \beta(g) \ee
 If the beta function is known, this eqn. can be integrated 
\be \log({\mu \over \mu_0})=\int_{g_0}^g {dg'\over
\beta(g')}\ee

\begin{figure}[h!]
\centering
\includegraphics[width=10cm]{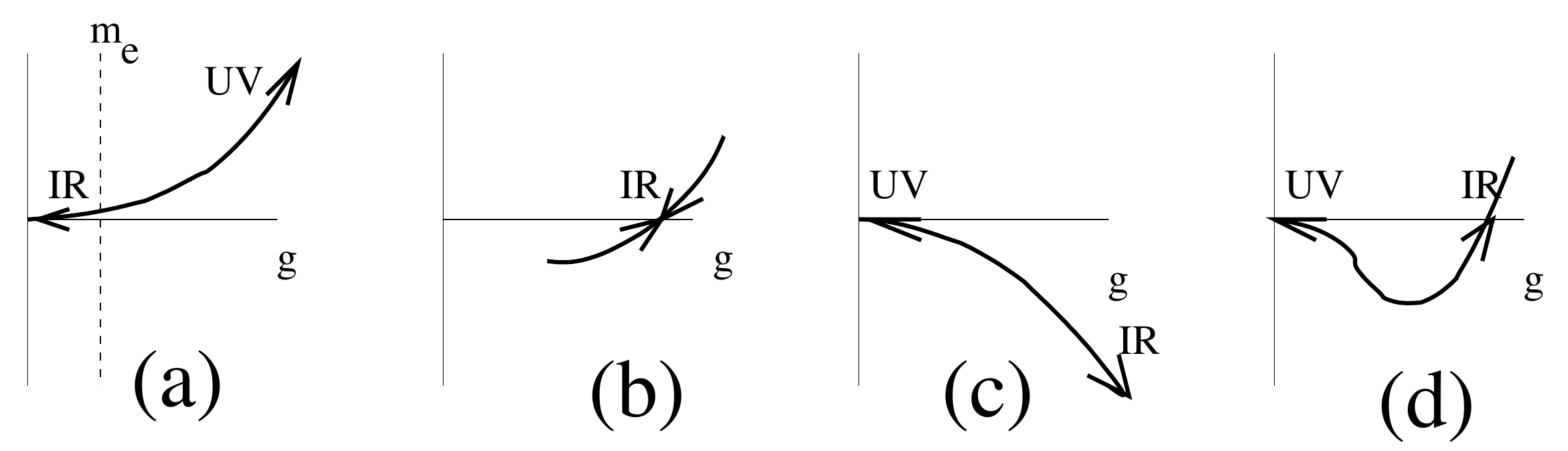}
\caption{\label{fig_betafunctions} 
Schematic behavior of the beta functions in the 4 examples considered:
(a) QED, (b) second order phase transition, (c) QCD and (d) QCD with 
sufficiently large number of light flavors. IR and UV with arrows
indicate direction of the charge motion at small and large momenta, respectively.
}
\end{figure}

For small coupling values the beta function can be determined perturbatively,
and its standard
definition includes the so called first and second beta function
coefficients 
\be \beta(g)=-b {g^3\over 16\pi^2}-b' {g^5\over (16\pi^2)^2}+... \ee
which are computed from one and two-loop diagrams.
Let us consider a number
of  examples with different beta functions, see Fig.\ref{fig_betafunctions}.

{\bf Example 0.}\\
A very special class of QFTs are 
$conformal$ (CFT). In those theories $\beta(g)=0$ and the coupling
is independent of the scale. The famous example in 4d is 
$\cal N$=4 supersymmetric
 gluodynamics. It has 4 kinds of gluinos and 6 scalars,
and their contributions to the beta function cancel that of the gluons,
order by order. This theory was studied in 1990's and was historically central for a discovery of 
 the so called AdS/CFT duality. \\


{\bf Example 1}\\  Quantum electrodynamics (QED) is the first QFT studied since 1950's.
 In one loop the so called
 vacuum polarization diagram leads to the following answer
\be\beta(e)={e^3 \over 12\pi^2}\ee
The solution for the charge-vs-scale is therefore
\be e^2(\mu)={e^2({\mu_0}) \over 1- (e^2({\mu_0})/6\pi^2)log(\mu/\mu_0)}\ee
and when $\mu$ grows (the charge is measured deeper inside the electron, at 
distances $\sim 1/\mu$) it appears
larger. Eventually one finds zero of the denominator,
the so called Landau pole, where the formula tells us the charge is infinite.
Before taking this answer at its face value one should be reminded that
the formula has applicability limitations, and already
when the coupling becomes of the order 1  we cannot any more trust any perturbative
results. What actually happens in QED at such small distances remains unclear.
 QED might be the first QFT studied, but by no means it
  is a well-defined theory: it lacks any non-perturbative definition. Therefore, 
  one cannot answer,
 or even
in principle formulate how to answer, many questions.\\

{\bf Example 2.}\\ The second order phase transitions or scalar theories in 3 dimensions. In this case the beta function changes sign and has a zero at $g^*$, the ``fixed point"",
so when the coupling reaches this value its ``running" stops. 
 Assuming that close to its zero it can be
approximated by a linear function
$\beta(g)=a(g-g^*)$, one finds power-like dependence near the fixed point:
\be g(\mu)-g^*=(g(\mu_0)-g^*)(\mu/\mu_0)^a \ee\\

{\bf Example 3.}\\
 QCD-like gauge theories have the so called ultraviolet fixed point, near which the coupling is small.
\be
\beta(g)=-b {g^3\over 16\pi^2}-b' {g^5\over (16\pi^2)^2}+... \ee
with $positive$ \be b=(11/3)N_c-(2/3)N_f . \ee
where $N_c,N_f$ are the number of colors and quark flavors, both can be taken to be 3 in QCD.
Positivity of the first coefficient of RG beta function behavior is also known as  ``asymptotic freedom".
Its first published version (for $N_c=2,N_f=0$) was obtained in \cite{Khriplovich:1969aa},
using Coulomb gauge quantization.
  In 1973  \cite{Gross:1973ju,Politzer:1973fx} 
have not only obtained this result in covariant gauges, but, more
importantly,  they related this feature with the experimentally observed (in DIS at SLAC) 
 weakly interacting point-like quarks,  suggesting that the fundamental theory of strong interactions,
  Quantum Chromodynamics, describe it.

Integrating the beta function, one has the explicit asymptotic freedom formula
\be g^2(\mu)={g^2_{\mu0} \over 1+ b(g^2_{\mu0}/8\pi^2)  log(\mu/\mu0)}\ee
and at large $\mu$ (small distances) the charge goes to $zero$.
There is less and less color charge deep inside the quarks.

The expression can be written as
\be g^2(L) = { 8\pi^2 \over [(11/3)N_c-(2/3)N_f] log[1/L \Lambda_{QCD}]}\ee
where the constant $g(a)$ is traded for a dimensional one,
$\Lambda_{QCD}$, giving
to the strong interaction theory its natural scale.
\\
 
{\bf Example 4.}\\
The QCD-like theories close to the $b=0$ line have both infrared and ultraviolet
fixed points \cite{Banks:1981nn}. The expression for the second (two-loop)
coefficient is 
\be b'=(34/3)N^2_c-(13/3) N_c N_f + (N_f/N_c) \ee
and so when $b=0$ (for example, in a theory with $N_c=3$ colors and $N_f=33/2$ flavors) $b'$ is
$negative$.
Putting expressions for $b,b'$ to the definition of  the beta function, one finds that
it possesses a zero (fixed point)  at 
\be {g^2_*\over 16\pi^2}=|b/b'| \ee
Note that it is small if $b/b'$ is small, in this case the
 charge is $always$ small. Therefore hadrons
cannot exist and the correlation functions have power-like, rather than exponential, decrease with the distance. 

Studies of this ``conformal regime", at zero and nonzero temperatures,
require to handle many quark species and therefore they
only recently became available to lattice practitioners, due to advances in computing.
There evidences that QCD with  $N_c=3$ colors and $N_f=12$ flavors is already in this regime.
We  will not discuss this interesting subject any more, since it is  too close to
the cutting edge of supercomputer ability, at the time of this writing. 
\\

 For supersymmetric theories it has been suggested that rather than calculate beta function
 in vacuum, one can rather find it from calculation of the instanton amplitude. We 
have discussed the so called NSVZ beta function in section \ref{sec_NSVZ}.

\section{Gross-Pisarski-Yaffe one-loop free energy for nonzero holonomy} \label{sec_holonomy_potential}
The notations used in their paper \cite{Gross:1980br} for the 4-th component of the gauge potential on the Matsubara circle, of circumference $\beta$, are 
\be A_4= {2\pi \over \beta} diag({q \over 2 }) \ee
so their variables $q^i,i=1...N_c$ are related to the phase fractions we use just by $\mu_i=q^i/2$.
Note further that  they use  $[q]_+=[q]_{mod2}-1, [q]_-=[q+1]_{mod2}-1$ so that $[q]_\pm\in [-1,1]$. 

The results for the fundamental and adjoint color fermionic determinants come from generic sums
\be ln det_\pm [ (\partial_\mu + q \xi_\mu)^2 ]=2 Re \int {d^3k V_3 \over (2\pi)^3}ln(1\mp e^{-\beta |k|+i k q}) 
\ee
$$=-{\pi^2 V_3 \over \beta^3}\left({1 \over 45} - {1 \over 24} (1-[q]_\pm^2)^2\right)$$
where $\pm$ are for periodic and antiperiodic boundary conditions ,
Using that for 
periodic adjoints (gluons and periodically compactified gluinoes) one then get
\be ln det \Dslash =-{\pi^2 V_3\over \beta^3} 
\left( {N_c^2- 1\over 45} - {1 \over 6} tr[ (ln{P \over i \pi}) (1-ln{P \over i \pi}) ]^2 \right) \label{eqn_GPY_g} \ee
while for quarks -- antiperiodic fundamental charges --
it is
\be ln det (-D^2) =-{2\pi^2 V_3\over \beta^3} \left( {N_c \over 45} - {1 \over 24} tr[ 1-(ln{P \over i \pi})^2 ]^2 \right)
\label{eqn_GPY_q}
\ee
This last expression should be rotated by complex chemical potential angle $\theta$ if it is non-zero,
as we discussed in section devoted to Roberge-Weiss symmetry.

\chapter{Instanton-related formulae}
\label{app_basics}
\section{Instanton gauge potential}
\label{app_inst_pot}  

  We use the following conventions for Euclidean gauge fields:
The gauge potential is $A_\mu = A_\mu^a\frac{\lambda^a}{2}$, 
where the $SU(N)$ generators are normalized according to ${\rm tr}
[\lambda^a,\lambda^b] = 2\delta^{ab}$. The covariant derivative is 
given by $D_\mu = \partial_\mu + A_\mu$ and the field strength tensor 
is
\be
 F_{\mu\nu} = [D_\mu,D_\nu] = \partial_\mu A_\nu - \partial_\nu A_\mu
         +[A_\mu,A_\nu] .
\ee
In our conventions, the coupling constant is absorbed into the 
gauge fields. Standard perturbative notation corresponds to
the replacement $A_\mu\to gA_\mu$. The single instanton solution 
in regular gauge is given by
\be 
 A_\mu^a = \frac{2 \eta_{a\mu\nu} x_\nu}{x^2+\rho^2},
\ee
and the corresponding field strength is
\be
 G_{\mu\nu}^a  &=& \frac{4 \eta_{a\mu\nu}\rho^2}
                          {(x^2+\rho^2)^2}, \\
 (G_{\mu\nu}^a)^2 &=& \frac{192\rho^4}{(x^2+\rho^2)^4}.
\ee
The gauge potential and field strength in singular gauge are
\be 
\label{A_sing}
 A_\mu^a &=&  \frac{2 \bar\eta_{a\mu\nu}x_\nu\rho^2}
      {x^2(x^2+\rho^2)},\\
\label{G_sing}
 G_{\mu\nu}^a &=& -\frac{4\rho^2}{(x^2+\rho^2)^2} 
  \left( \bar\eta_{a\mu\nu} -2\bar\eta_{a\mu\alpha}
    \frac{x_\alpha x_\nu}{x^2} -2\bar\eta_{a\alpha\nu}
    \frac{x_\mu x_\alpha}{x^2} \right)\; .
\ee
Finally, an $n$-instanton solution in singular gauge is given by
\be
\label{A_n_inst}
 A_\mu^a &=&  \bar\eta_{a\mu\nu} \partial_\nu\ln\Pi(x) ,\\
 \Pi(x)  &=& 1+\sum_{i=1}^n \frac{\rho_i^2}{(x-z_i)^2} .
\ee
Note that all instantons have the same color orientation. For
a construction that gives the most general $n$-instanton
solution \cite{Atiyah:1978ri}.

\section{Fermion zero modes and overlap integrals}
\label{app_zm}

  In singular gauge, the zero mode wave function $iD\!\!\!\!/\,\phi_0 =0$
is given by 
\be
 \phi_{a\nu} = \frac{1}{2\sqrt{2}\pi\rho} \sqrt{\Pi}
\left[ \partial\!\!\!/ \left(\frac\Phi\Pi \right)
\right]_{\nu\mu} U_{ab}\epsilon_{\nu b} ,
\ee
where $\Phi=\Pi-1$. For the single instanton solution, we get
\be
\phi_{a\nu} (x) = \frac{\rho}{\pi}\frac{1}{(x^2+\rho^2)^{3/2}}
 \left( \frac{1-\gamma_5}{2}\right) \frac{x\!\!\! /}{\sqrt{x^2}}
  U_{ab}\epsilon_{\nu b}  .
\ee
The instanton-instanton zero mode density matrices are
\be
\phi_I(x)_{i\alpha}\phi_J^\dagger(y)_{j\beta} &=& 
\frac{1}{8} \varphi_I(x)\varphi_J(y) 
\left( x\!\!\! / \gamma_\mu\gamma_\nu y\!\!\! / 
\frac{1-\gamma_5}{2} \right)_{ij}
\otimes \left( U_I \tau_\mu^-\tau_\nu^+ U_J \right)_{\alpha\beta}, \\
\phi_I(x)_{i\alpha}\phi^\dagger_A(y)_{j\beta} &=& 
-\frac{i}{2} \varphi_I(x)\varphi_A(y) 
\left( x\!\!\! / \gamma_\mu y\!\!\! / 
\frac{1-\gamma_5}{2} \right)_{ij}
\otimes \left( U_I \tau_\mu^- U_A^\dagger \right)_{\alpha\beta}, \\
\phi_A(x)_{i\alpha}\phi^\dagger_I(y)_{j\beta} &=& 
\frac{i}{2} \varphi_A(x)\varphi_I(y) 
\left( x\!\!\! / \gamma_\mu y\!\!\! / 
\frac{1+\gamma_5}{2} \right)_{ij}
\otimes \left( U_A \tau_\mu^+ U_I^\dagger \right)_{\alpha\beta} ,
\ee
with
\be
\varphi (x)= \frac{\rho}{\pi}\frac{1}{\sqrt{x^2}(x^2+\rho^2)^{3/2}}.
\ee
The overlap matrix element is given by
\be
 T_{AI} &=& \int d^4x\, \phi_A^\dagger (x-z_A)
 iD\!\!\!\! / \phi_I(x-z_I) \nonumber \\
        &=& r_\mu\,{\rm Tr}(U_I\tau_\mu^- U^\dagger_A)\,
            \frac{1}{2\pi^2 r}\frac{d}{dr} M(r) ,
\ee
with
\be
 M(r) = \frac{1}{r}\, \int\limits_0^\infty dp\, p^2|\varphi(p)|^2 J_1(pr) .
\ee 
The Fourier transform of zero mode profile is given by
\be
 \varphi(p) = \pi\rho^2 \left.\frac{d}{dx}\left(I_0(x)K_0(x)-
 I_1(x)K_1(x)\right)\right|_{x=\frac{p\rho}{2}}.
\ee

\section{Group integration and Fierz transformations}
\label{app_group}

    In order to perform averages over the color group, we need
the following integrands over the invariant $SU(3)$ measure

$$ \int dU\, U_{ij}U^\dagger_{kl} = \frac{1}{N_c}
 \delta_{jk}\delta_{li} , $$
$$ \int dU\, U_{ij}U^\dagger_{kl}U_{mn}U^\dagger_{op} = \frac{1}{N_c^2}
 \delta_{jk}\delta_{li}\delta_{no}\delta_{mp}
 + \frac{1}{4(N_c^2-1)}(\lambda^a)_{kj} (\lambda^b)_{il} 
        (\lambda^a)_{on} (\lambda^b)_{mp}
$$

The so called Fierz transformations are, in general,  identities
allowing to rewrite multi-fermion operators in different forms.
For four-fermion operators written as product of two ``neutral" brackets (meaning all indices, Dirac, color and flavor, convoluted inside a bracket) transformed into a sum of different brackets. Since a given quark can be convoluted with 3 others, there are 3 possible ``channels" (corresponding to three Mandelstam kinematical variables $s,t,u$.) Their generic form is  
$$ \sum_A \Gamma_{ij}^A  \Gamma_{kl}^A = \sum_B  C_B \Gamma_{ik}^B  \Gamma_{jl}^B $$
 where indices $i,j,k,l$ (no sums) are quark variables, and indices $A,B$ number all possible matrices, with $C_B$ being some coefficients of the Fierz identity.   

Since we mostly deal with QCD quarks with three colors, let us start with Fierz relations for 
$SU(3)$ spinor generators $$t^a={1\over 2}\lambda^a, \,\,\,\,a=1..8$$ where $\lambda^a$ are eight traceless
Gell-Mann matrices. If $\Gamma^a=t^a$ and $i,j,k,l=1,2,3$ color indices, the corresponding relations follow from the identity 
  \be t^a_{ik} t^a_{jl}={1\over 2} \big( \delta_{il} \delta _{kj} -{1 \over 3 }\delta_{ik} \delta _{jl} \big) \label{eqn_color_Fierz} \ee
In original papers and textbooks  validity of this and other identities is usually proven by doing multiple convolutions, e.g. with
$\delta_{ik}\delta_{jl}$ or $\delta_{ij}\delta_{kl}$ plus some additional considerations which 
force us to think too much. Nowadays one can check them in seconds, directly using Mathematica. Indeed, the l.h.s. and r.h.s. are 4-index tables, with just $3^4=81$ components,
so there is no problem to calculate them $all$ for both sides. One
indeed  finds that all 81 of them $are$ the same, namely
(in Mathematica table notations with i,j,k,l order) 
$$ \{\{\{\{1/3, 0, 0\}, \{0, -(1/6), 0\}, \{0, 0, -(1/6)\}\},
 \{\{0, 0, 0\}, \{1/2, 0,  0\}, \{0, 0, 0\}\},\{\{0, 0, 0\}, \{0, 0, 0\}, \{1/2, 0, 0\}\}\}, $$
 $$    \{\{\{0, 1/2,  0\}, \{0, 0, 0\}, \{0, 0, 0\}\}, \{\{-(1/6), 0, 0\}, \{0, 1/3, 0\}, \{0, 
    0, -(1/6)\}\}, \{\{0, 0, 0\}, \{0, 0, 0\}, \{0, 1/2, 0\}\}\}, $$
$$     \{\{\{0, 0, 1/2\}, \{0, 0, 0\}, \{0, 0, 0\}\}, \{\{0, 0, 0\}, \{0, 0, 1/2\}, \{0, 0, 
    0\}\}, \{\{-(1/6), 0, 0\}, \{0, -(1/6), 0\}, \{0, 0, 1/3\}\}\}\}
$$ 
  The somewhat simplified form of such relations are often used, in which matrices are mentioned without explicit indices. For example, the relation given above can be rewritten
using such notations  as 
  \be \big(\hat 1  \bigotimes \hat 1\big)_{direct}=
\big( 2 t^a  \bigotimes t^a+ {1 \over 3} \hat 1  \bigotimes \hat 1 \big)_{cross}   \ee
 Another relation of the type is
 \be   \big( t^a  \bigotimes t^a \big)_{direct}=
\big( -{1\over 3} t^a  \bigotimes t^a+ {4 \over 3} \hat 1  \bigotimes \hat 1 \big)_{cross}  
  \ee
and using both one can transfer any expression contaning color convolution from one channel to the next. 

Let me add that so far no assumptions about open 4 indices were made : but sometimes 
their symmetry is known,  simplifying relations further. For example, if the cross channel combines two quarks (diquark) in the baryon, we know that in this case only $antisymmetric$ color indices are allowed. Then in the r.h.s. out of 9 matrices
$t^a,\hat 1$ only $three$ ones, which are antisymmetric $t^a_{anti} =t^2,t^4,t^7$ 
produce nonzero contributions.  

Quark isospin matrices are related to $SU(2)$ generators, Pauli matrices $\sigma^a,a=1,2,3$
divided by two. Of course, those coinside with the first three Gell-Mann generators. Here are
corresponding relation for Pauli matrices 
\be  \big(\hat 1  \bigotimes \hat 1\big)_{direct}= \big({1\over 2} \hat 1  \bigotimes \hat 1+{1\over 2}  \sigma^a  \bigotimes \sigma^a   \big)_{cross}   \ee
\be   \big( \sigma^a  \bigotimes \sigma^a \big)_{direct}= \big( {3 \over 2}  \hat 1  \bigotimes \hat 1-{1 \over 2} \sigma^a  \bigotimes \sigma^a   \big)_{cross} \ee 
  In diquark channel with isospin zero, and $antisymmetric$ indices in cross channel,
  there remains only one anstisymmetric Pauli matrix $\sigma^2$,
  so r.h.s. in this case is simply proportional to $(ud -du)$ flavor combination. It is 
  of course the same
  as the wave function for two spins 1/2 added into total spin zero.  
  

Similarly, one can define the matrix of Fierz transformation for all 16 Dirac 
matrices. Defining symbolically five products of quark bilinars, 
$$S=1\bigotimes 1,\, V=\gamma_\mu \bigotimes \gamma_\mu, \,A=\gamma_5\gamma_\mu \bigotimes \gamma_5 \gamma_\mu, \,T=\sigma_{\mu\nu} \bigotimes \sigma_{\mu\nu}, \,P=\gamma_5 \bigotimes \gamma_5 $$ one can perform Fierz transformation from direct channel ($s$) to cross channel ($u$) (coupling antiquark to another quark) and find the following  matrix of Fierz 
transformations 
\be \left( \begin{array}{c} S \\ V \\ T \\ A \\ P  \end{array}
\right) _{direct}= \left( \begin{array}{c c c c c} 
1/4 & 1/4 & -1/8 & -1/4 & 1/4 \\
1 & -1/2 & 0 & -1/2 & -1 \\
-3 & 0 & -1/2 & 0 & -3 \\
-1 & -1/2 & 0 & -1/2 & 1 \\
1/4 & -1/4 & -1/8 & 1/4 & 1/4  \end{array} \right)  \left( \begin{array}{c} S \\ V \\ T \\ A \\ P  \end{array}
\right)_{cross}
\ee 

\chapter{Some special theories}
\section{Gauge theory with the exceptional group $G_2$}
Interest to this particular theory is related with the question of whether
the $Z_N$ symmetry of the $SU(N)$ gauge theories is or is not related to confinement.
(The answer is, {\em not at all}.)

The construction starts with the $SO(7)$ group,  which has 21 generators and rank 3.
As for any SO group, its $7\times 7$ real matrices satisfy
\be det\Omega=1,\,\,\, \Omega^{-1}=\Omega^T \ee
$G_2$ is its subgroup which additionally satisfy 7 more relations
\be T_{abc}=T_{def} \Omega_{da} \Omega_{eb} \Omega_{fc} \ee
where $T$ is an antisymmetric tensor such that
\be T_{123} =T_{176} =T_{145} =T_{257} =T_{246} =T_{347} =T_{365} =1 \ee
This leaves us 14 generators
listed e.g. in \cite{Cossu:2007dk}. The rank of $G_2$ is 2, so, like in SU(3), a nonzero Polyakov
line leaves 2 massless U(1)'s to make the Abelian monopole charges. In fact SU(3) is the largest subgroup of the
 $G_2$, with its 8 generators. The rest can be constructed via 6 SU(2)
 subgroups: and such construction is explicitly used in lattice
updates.
 
 Yet the center of this SU(3) does not commute with extra generators of SU(2)'s, leaving $G_2$ without a nontrivial center (only the unit matrix commutes
 with all generators). Therefore there is no ``spontaneous breaking of the center
 symmetry" in the deconfined phase! 

\section{ $\cal N$=2 SYM and SQCD, and their Seiberg-Witten solution } \label{sec_SW}
The $\cal N$=1 and corresponding SUSY QCD theories will be discussed only   
in one chapter, on instanton-dyons: this is how the puzzle of quark condensate 
was resolved. In general, these theories have properties similar to ordinary QCD.

Therefore we just jump to  $\cal N$=2 and then  $\cal N$=4 theories, in which theoretical progress
was more impressive\footnote{This fits general mathematical expectations: the more symmetry
the problem has, the
easier it is to solve. Unfortunately, real world around us is not that symmetric as one would like.}.

\subsection{The field content and RG flows}

Let us start with the {\bf field content} of those theories. The  $\cal N$=2 gluodynamics or super-Yang-Mills (SYM) 
theory has
gluons (spin 1), two real gluinoes $\lambda,\chi$ (spin 1/2), and a complex scalar (spin 0) which we will
call $a$. Each of them has two degrees of freedom, thus 4 bosonic and 4 fermionic ones.

The $\cal N$=2 QCD is a theory with additional matter supermultiplets of structure $\psi_f,\phi_f$ with spin 1/2 and 0,
respectively. We will call $N_f$ the number of Dirac quarks, as in QCD, or $2N_f$ Majorana ones.

Two different Higgsing possible, defining ``branches" of these theories. If $<a>\neq 0, <\phi>=0$ 
Higgsing is like in Georgi-Glashow model, with massless photon/photino multiplet: this branch is
called the ``Coulomb branch". However if both $<a>\neq 0, <\phi>\neq 0$, HIggsing is like in 
the Weinberg-Salam model, with all gluons massive: this is called the ``Higgs branch".

The {\bf coupling renormalization} in these theories is done only via the
 one-loop beta function, with  the coefficient 
\be b=4-N_f \ee
while two-loop and higher coefficients vanish. The explanation for that was given in
 the instanton chapter, see NSVZ beta function.
 
 Start with $N_f=0$ or $\cal N$=2 SYM.
 
At the opposite end, at  $N_f=4$ QCD,  one finds zero beta function and is thus a conformal theory:
we will not discuss it.

\subsection{The moduli }

\begin{figure}[htbp]
\begin{center}
\includegraphics[width=10cm]{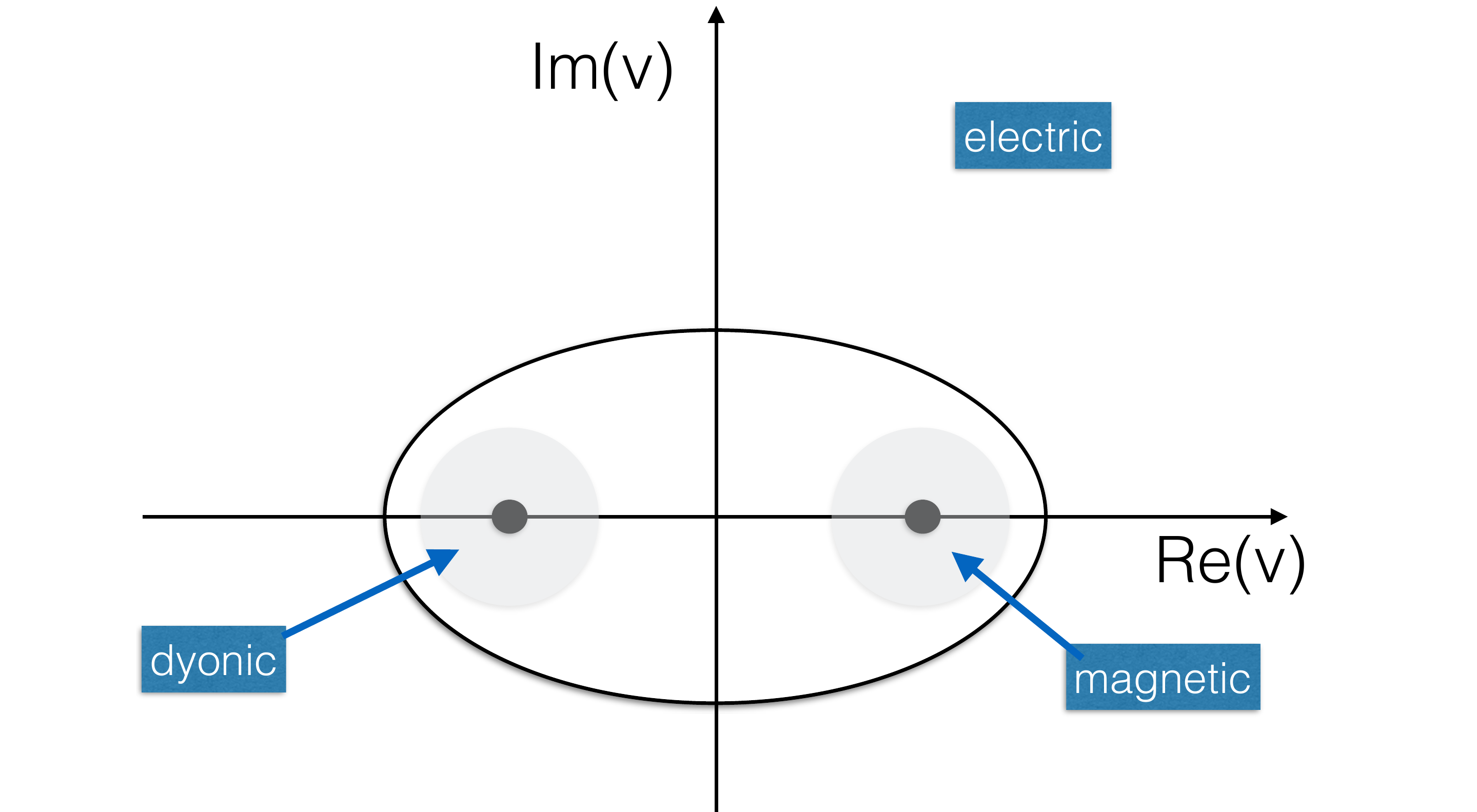}
\caption{The map of the moduli space according to Seiberg-Witten solution.}
\label{fig_SW_moduli2}
\end{center}
\end{figure}

 Supersymmetry also require that for $any$ value of $v$ the vacuum energy remains zero: thus
there is a whole {\em manyfold of non-equivalent vacua}, known as {\em moduli space}, labeled by a complex number $v$.
All properties of the system are expressed as derivatives of one fundamental holomorphic function 
$F(u)$, in particularly
the effective charge and the theta angle are combined into a variable $\tau$ which is given by its second derivative
\be \tau(u)= {\theta \over 2\pi}+ {4\pi i \over e^2(u)}={\partial^2 F(u) \over \partial u^2} \ee
We will return to the exact form of this function later, in connection with its instanton-based description.

The map of the moduli space is sckematically given in Fig.\ref{fig_SW_moduli2}
There are three distinct patches on the $v$ plane:\\
(i) at large values of $v\rightarrow \infty$ there is a ``perturbative patch"
, in which the coupling $e^2(v)/4\pi \ll 1$ is weak. It is dominated by electric
particles -- gluons, gluinoes and higgses -- with small masses $O(ev)$, which determine the beta function.
Monopoles have large masses 
$4\pi v/e^2(v)$ there, and can be treated semiclassically.\\
(ii) a ``magnetic patch" around the $v=\Lambda$ point, in which the coupling  is infinitely strong $e^2\rightarrow \infty$, the monopole mass
goes to zero as well as the magnetic charge $g\sim 1/e \rightarrow 0$.  \\
(iii) a ``dyonic patch" around the  $v=-\Lambda$ point, in which a dyon (particle with electric and magnetic charges both being 1) gets massless.

\subsection{Singularities for $\cal N$=2 QCD }
 Let us now focus on the next case,  $N_f=3$. Here are the limiting cases discussed by
 Seiberg-Witten. Suppose first the quark/squark triplet has large mass $m$. This means that
 one needs large value of the VEV $<a>$ to cancel it: the structure of the   $<a>$ plane is as follows:
 two original singular points, inherited from  $N_f=0$ or $\cal N$=2 SYM, plus a triply degenerate
 singularity at large  $<a>$. As $m_f \rightarrow 0$, one by one, one may reason what happens
 (see details in the second  Seiberg-Witten paper). When the process is complete, they came up with
 the following unusual structure: a 4-degenerate zero where monopoles ($n_m=1,n_e=0$) get massless,
 plus another single zero where some particle with  $n_m=2,n_e=1$  gets massless. The former 
 quartet of monopoles includes 3 monopole+occupied zero mode by one quark plus perhaps the old unoccuplied
 monopole (the singularity existing before). The latter  $n_m=2,n_e=1$  state is some kind
 of di-monopole bound state bound by quarks: its possible nature has been discussed at the end
 of the monopole chapter.
 
 \section{$\cal N$=4 super-Yang-Mills theory}
 As the name suggests, it has 4 supersymmetries, and thus fermions $\chi_i$ have 
 (``flavor" or R-charge) index $i=1..4$, on top of assumed adjoint color index. This makes 8 fermionic degrees of freedom, compensated by 2 polarizations of the gauge field and 6 complex scalars $\phi_{ij}$ in corresponding representation  
 of the R-charge. The usual cancellations between bosonic and fermionic loops follow. 
 
 This remarkable theory has been called ``a harmonic oscillator of the 21-st
 century"\footnote{This saying is attributed to D.Gross, but I do not know if it is indeed true.
 }. 
 As follows from NSVZ beta function, derived from instanton above, in this case
 bosonic and fermionic parts of beta function cancel {\em to all loops}, and thus it is a super-conformal theory. The charge keeps its value, independent of the scale.

It is due to this remarkable high symmetry of this theory
(and to J.Maldacena, of course) that we now have AdS/CFT
holographic correspondence.

We already discussed another derivation of this fact from electric-magnetic duality, in the monopole chapter. The monopole dressed by all types of gluinoes 
make spin-1, spin1/2 and spin-0 objects, which together form  precisely a 
 $\cal N$=4 supermultiplet. So, its magnetic theory, with charge $1/g$, is
 the same as the electric theory, with charge $g$. Beta functions must be the same for electric and magnetic theory, up to a sign:
  the only possible solution is that it is identically zero!

\chapter{AdS/CFT correspondence}
\section{Black holes and branes}
The original Schwartzschild solution for the usual
black hole in asymptotically flat 3+1 dimensions has a metric tensor (in spherical coordinates)
\be ds^2=g_{\mu\nu}dx^\mu dx^\nu= -(1-r_h/r)dt^2+{dr^2\over (1-r_h/r)}+r^2d\Omega^2\ee 
and the horizon radius (in fulll units) is $r_h=2G_NM/c^2$,
containing the mass $M$, the Newton constant $G_N$ and speed of light $c$.

generalization to 10 dimensions for a stack of $N$ $D_3$ branes

adding charge

\section{Colors and the brane stack,  the road to AdS/CFT}
the 10-d solution
 \begin{eqnarray} ds^2={-dt^2+dx_1^2+dx_2^2+dx_3^2 \over \sqrt{1+L^4/r^4}}+ 
\sqrt{1+L^4/r^4}(dr^2+r^2d\Omega_5^2) \label{eqn_10dBH_2}
 \end{eqnarray}

$AdS_5$ limit:  the ``near-horizon region",  at $r<<L$, when 1 in both roots can be ignored
and the metric splits into noninteracting $AdS_5\times S_5$. 
Using 
a new coordinate $z=L^2/r$ we get it into the  ``standard $AdS_5$ form'' used below:
 \begin{eqnarray}  ds^2={-dt^2+dx_1^2+dx_2^2+dx_3^2+dz^2 \over z^2}   \label{eqn_AdS5metric}
\end{eqnarray}     

he Maldacena relations between
the gauge coupling, the AdS radius $L$ and the string tension
$\alpha'$ (which comes from the total mass of the brane set):
 \begin{eqnarray} L^4=g^2 N_c (\alpha')^2=\lambda(\alpha')^2  
\end{eqnarray}

$AdS_5$ in 6d Global coordinates, standard and Poincare coordinates

 the $AdS_5$ space can be described by the equation
\be -(X^{-1})^2- (X^0)^2+(X^1)^2+(X^2)^2+(X^3)^2+(X^4)^2=-L^2 \ee
in the 6-dimensional space $X^{-1}..X^4$.
``Standard" coordinates
\be X^{-1}=\sqrt{L^2+\rho^2}cos({\tau \over L}), \,\, X^{0}=\sqrt{L^2+\rho^2}sin({\tau \over L}) ,\,\, X^{i}=\rho \Omega^i \ee
The last term contains the coordinates of the 3d unit sphere, with the standard line element
\be d\Omega^2=d\chi^2 + sin^2\chi (d\theta^2+sin^2\theta d\phi^2)
 \ee
Poincare coordinates, defined by (i=1..3)
\be X^{-1}={z \over 2}(1+{L^2+\vec x^2-t^2 \over z^2 }) ,\, \, X^0=L {t \over z},  \\ \nonumber 
X^i=L {x^i \over z}, \,\,\, X^4={z \over 2}(-1+{L^2-\vec x^2+t^2 \over z^2 }) 
\ee 
Relation between the two sets
\be {z\over L}=\left[\sqrt{1+{\rho^2 \over L^2}}cos({\tau \over L})  + {\rho \over L} cos\chi \right]^{-1},\,\,\,
t=z \sqrt{1+{\rho^2 \over L^2}}sin({\tau \over L}), \, \vec{x}=z{\rho \over L}sin\chi\vec{\Omega} 
 \ee
the metric tensor in both sets are related
\be ds^2=-dt^2+(dx^i)^2= W^2(-d\tau^2+L^2 d\Omega^2)\ee
where the so called {\em conformal factor} is
\be W^2={ 1 \over (cos(\tau/L) +cos(\chi)  )^2}={t^2 \over L^2}+{1 \over 4} \left(1+{r^2\over L^2}-{t^2\over L^2}\right)^2 \label{eqn_conf_factor}\ee

\section{Propagators in $AdS_5$} 

The Laplace-Bertrami operator is in general
\be \Delta= {1 \over \sqrt{g}} \partial_\mu \sqrt{g} g^{\mu\nu}  \partial_\nu \ee
and the equation for massive propagator is obtained by simply adding the mass term
\be \Delta D(x,x') - m^2 D(x,x') = {1 \over \sqrt{g}} \delta(x-x') \ee
where the mass can be negative but not the r.h.s. of 
\be \nu^2=m^2+d^2/4 >0 \ee

The main object one needs for holography is Eucledean bulk-to-boundary propagators: those were
constructed in \cite{Witten:1998qj} in the following way. The propagator is a scalar field of a point charge, placed somewhere
in the space under consideration. Clearly the simplest case is when we put this charge at the $origin$ of the space,
$z=\infty$. In this case no dependence on $x^i,i=1..d$ is expected by symmetry, and so one can keep only the $z$ part of the Laplacian. Considering the massless case for simplicity, we only have one term in the equation
\be  {d \over dz} z^{-d+1}  {d \over dz}D(z)=0  \ee
and the only solution which vanishes at z=0 is 
\be  D(z)\sim z^d  \ee 
It is singular at $z=\infty$, as of course it should, as there is a source there. (One can show it is exactly the right delta function.) 

The next step is to use conformal invariance and move the source to a place we actually want it to be, namely at the boundary.
This is done by the conformal transformation
\be   x^i \rightarrow  { x^i \over z^2+\sum_{j=1}^d (x^j)^2} , \,\,\,\,  z \rightarrow  { z \over z^2+\sum_{j=1}^d (x^j)^2} \ee
which moves the source to the point $z=0,x^i=0$ and changes the solution to
\be D\sim \left[  { z \over z^2+\sum_{j=1}^d (x^j)^2} \ \right]^d \ee
All what remains is to shift the point to arbitrary location, by $x^i\rightarrow x^i-x'^i$ and finding the normalization constant.

For more complete discussion of propagators we suggest  \cite{Danielsson:1998wt}: we will not copy its beginning with a concise description
of what vacuum one should select, in AdS curved space: yet it is worth reading.

The basic geometry: the distance $d(X,X')$ between the two points
is defined as the line element integrated along the geodesic.
In this space it is
\be cosh(d(X,X')) =1+{ z^2+z'^2+(X^0-X'0)^2... (X^3-X'^3)^2 \over z z' } \ee
(see that it makes sense along each of the axes). 

$AdS_5$ corresponds to spatial dimension $d=2m=4$ and the solution for massive bulk-to-bulk Euclidean propagator is
\be D={1\over 8\pi^2 sinh(u)}{d \over du}{e^{\nu u} \over sinh(u)}
\ee
where we use $u=d(X,X')$ defined above. Note that the function is more complicated than just the power of the distance:
that was a large-distance tail, which is the only one needed for bulk-to-boundary applications. 

Classical applications sometimes need retarded bulk propagators, which are not obtainable from
the Euclidean version by an analytic contiuation. Those are only nonzero for real timelike distance $v$ 
\be cos(v)=1- {t^2-r^2-(z-z')^2 \over 2 z z'}\,\,\, 
\ee
and we only give a simple answer  for integer $\nu$
\be D= - {cos(\nu v) \over 2\pi sin(v)}\ee
\section{Non-zero temperatures in holography}
black branes with a horizon: nonzero T

Global AdS-Schwartzschield Black Hole,  GAdSBH, is a charged 5-d black hole.
the Einstein equation
with a particular cosmological term
\be R_{ab}+{4 \over L^2} g_{ab}=0 \ee
  the metric
\be  ds^2=-f d\tau^2 + {d\rho^2 \over f} 
+\rho^2 d\Omega^2\,\, f=1-{\rho_0^2 \over \rho^2}+ {\rho^2 \over L^2}
\ee
where $\tau,\rho,\Omega$ have the same meaning as in ``standard" AdS coordinates above.

Parameters $\rho_0,L$ are two parameters related to the mass 
\be   M={3\pi^2 \rho_0^2 \over \kappa^2}, \,\, \ee
and charge of the black hole. 

the Hawking temperature
\be  T={\rho_h \over \pi L^2}(1+{L^2 \over 2 \rho_h^2}) \ee
 the Bekenstein entropy 
\be S={4\pi^3 \rho_h^3 \over \kappa^2}, \,\, \ee 
where the horizon radius, the upper root of $g_{00}=f=0$, is 
\be \rho_h/L=\sqrt{\sqrt{(1+4\rho_0^2/L^2}-1)/2} \ee
$\kappa$ is the 5-d Newton constant, which is related to $L$ and the brane number $N$ via
\be  {L^3 \over \kappa^2}=({N \over 2\pi})^2\ee
Note that if $N\gg 1$ is large, than the space parameter $L$ is large compared to the 5d Plank scale and thus gravity
is classical.


\end{appendix}


\bibliography{allbib,hi,hi2,hi3,hi4,hi5,hi6}{}

\begin{thebibliography}{}

\bibitem[, ]{}


\bibitem[foo, ]{footnote}


\bibitem[MAL, ]{MALDA2_REY}


\bibitem[Cas, ]{Casalderrey-Solana:2006rq}


\bibitem[Adamczyk et~al., 2017]{STAR:2017ckg}
Adamczyk, L. et~al. (2017).
\newblock {Global $\Lambda$ hyperon polarization in nuclear collisions:
  evidence for the most vortical fluid}.
\newblock {\em Nature}, 548:62--65.

\bibitem[Aharonov and Bohm, 1959]{Aharonov:1959fk}
Aharonov, Y. and Bohm, D. (1959).
\newblock {Significance of electromagnetic potentials in the quantum theory}.
\newblock {\em Phys. Rev.}, 115:485--491.
\newblock [,95(1959)].

\bibitem[Aharony and Klinghoffer, 2010]{Aharony:2010db}
Aharony, O. and Klinghoffer, N. (2010).
\newblock {Corrections to Nambu-Goto energy levels from the effective string
  action}.
\newblock {\em JHEP}, 12:058.
\newblock 1008.2648.

\bibitem[Alday et~al., 2010]{Alday:2009aq}
Alday, L.~F., Gaiotto, D., and Tachikawa, Y. (2010).
\newblock {Liouville Correlation Functions from Four-dimensional Gauge
  Theories}.
\newblock {\em Lett. Math. Phys.}, 91:167--197.
\newblock 0906.3219.

\bibitem[Aleinikov and Shuryak, 1987]{Aleinikov:1987wx}
Aleinikov, A.~A. and Shuryak, E.~V. (1987).
\newblock {Instantons in quantum mechanics, two-loop effects .}
\newblock {\em Yad. Fiz.}, 46:122--129.

\bibitem[Alexandru et~al., 2016]{Alexandru:2015xva}
Alexandru, A., Basar, G., and Bedaque, P. (2016).
\newblock {Monte Carlo algorithm for simulating fermions on Lefschetz
  thimbles}.
\newblock {\em Phys. Rev.}, D93(1):014504.
\newblock 1510.03258.

\bibitem[Alford et~al., 1998]{Alford:1997zt}
Alford, M.~G., Rajagopal, K., and Wilczek, F. (1998).
\newblock {QCD at finite baryon density: Nucleon droplets and color
  superconductivity}.
\newblock {\em Phys. Lett.}, B422:247--256.
\newblock hep-ph/9711395.

\bibitem[Anderson et~al., 2014]{Anderson:2014jia}
Anderson, N., Domokos, S.~K., Harvey, J.~A., and Mann, N. (2014).
\newblock {Central production of $\eta$ and $\eta?$ via double Pomeron exchange
  in the Sakai-Sugimoto model}.
\newblock {\em Phys. Rev.}, D90(8):086010.

\bibitem[Andrei and Gross, 1978]{Andrei:1978xg}
Andrei, N. and Gross, D.~J. (1978).
\newblock {The Effect of Instantons on the Short Distance Structure of Hadronic
  Currents}.
\newblock {\em Phys. Rev.}, D18:468.

\bibitem[Antchev et~al., 2016]{Antchev:2016vpy}
Antchev, G. et~al. (2016).
\newblock {Measurement of elastic pp scattering at $\sqrt{s}=8 \, TeV$ in the
  Coulomb-nuclear interference region: determination of the $\mathbf {\rho }$
  -parameter and the total cross-section}.
\newblock {\em Eur. Phys. J.}, C76(12):661.
\newblock 1610.00603.

\bibitem[Arnold and McLerran, 1987]{Arnold:1987mh}
Arnold, P.~B. and McLerran, L.~D. (1987).
\newblock {Sphalerons, Small Fluctuations and Baryon Number Violation in
  Electroweak Theory}.
\newblock {\em Phys. Rev.}, D36:581.

\bibitem[Arnold et~al., 1997]{Arnold:1996dy}
Arnold, P.~B., Son, D., and Yaffe, L.~G. (1997).
\newblock {The Hot baryon violation rate is O (alpha-w**5 T**4)}.
\newblock {\em Phys. Rev.}, D55:6264--6273.
\newblock hep-ph/9609481.

\bibitem[Arvis, 1983]{Arvis:1983fp}
Arvis, J.~F. (1983).
\newblock {The Exact $q \bar{q}$ Potential in Nambu String Theory}.
\newblock {\em Phys. Lett.}, 127B:106--108.

\bibitem[Athenodorou et~al., 2018]{Athenodorou:2018jwu}
Athenodorou, A., Boucaud, P., De~Soto, F., Rodriguez-Quintero, J., and
  Zafeiropoulos, S. (2018).
\newblock {Instanton liquid properties from lattice QCD}.
\newblock {\em JHEP}, 02:140.
\newblock 1801.10155.

\bibitem[Atiyah et~al., 1978]{Atiyah:1978ri}
Atiyah, M.~F., Hitchin, N.~J., Drinfeld, V.~G., and Manin, {\relax Yu}.~I.
  (1978).
\newblock {Construction of Instantons}.
\newblock {\em Phys. Lett.}, A65:185--187.
\newblock [,133(1978)].

\bibitem[Baier et~al., 1996]{Baier:1996vi}
Baier, R., Dokshitzer, Y.~L., Mueller, A.~H., Peigne, S., and Schiff, D.
  (1996).
\newblock {The Landau-Pomeranchuk-Migdal effect in QED}.
\newblock {\em Nucl. Phys.}, B478:577--597.

\bibitem[Baker et~al., 1991]{Baker:1991bc}
Baker, M., Ball, J.~S., and Zachariasen, F. (1991).
\newblock {Dual QCD: A Review}.
\newblock {\em Phys. Rept.}, 209:73--127.

\bibitem[Baker and Steinke, 2002]{Baker:2002km}
Baker, M. and Steinke, R. (2002).
\newblock {Semiclassical quantization of effective string theory and Regge
  trajectories}.
\newblock {\em Phys. Rev.}, D65:094042.
\newblock hep-th/0201169.

\bibitem[Bali, 1998]{Bali:1998de}
Bali, G.~S. (1998).
\newblock {The Mechanism of quark confinement}.
\newblock In {\em {Quark confinement and the hadron spectrum III. Proceedings,
  3rd International Conference, Newport News, USA, June 7-12, 1998}}, pages
  17--36.
\newblock hep-ph/9809351.

\bibitem[Balitsky and Yung, 1986]{Balitsky:1986qn}
Balitsky, I.~I. and Yung, A.~V. (1986).
\newblock {Collective - Coordinate Method for Quasizero Modes}.
\newblock {\em Phys. Lett.}, B168:113--119.
\newblock [,273(1986)].

\bibitem[Banks and Casher, 1980]{Banks:1979yr}
Banks, T. and Casher, A. (1980).
\newblock {Chiral Symmetry Breaking in Confining Theories}.
\newblock {\em Nucl. Phys.}, B169:103--125.

\bibitem[Banks and Zaks, 1982]{Banks:1981nn}
Banks, T. and Zaks, A. (1982).
\newblock {On the Phase Structure of Vector-Like Gauge Theories with Massless
  Fermions}.
\newblock {\em Nucl. Phys.}, B196:189--204.

\bibitem[Basar et~al., 2012]{Basar:2012jb}
Basar, G., Kharzeev, D.~E., Yee, H.-U., and Zahed, I. (2012).
\newblock {Holographic Pomeron and the Schwinger Mechanism}.
\newblock {\em Phys. Rev.}, D85:105005.
\newblock 1202.0831.

\bibitem[Bazavov et~al., 2016]{Bazavov:2016uvm}
Bazavov, A., Brambilla, N., Ding, H.~T., Petreczky, P., Schadler, H.~P., Vairo,
  A., and Weber, J.~H. (2016).
\newblock {Polyakov loop in 2+1 flavor QCD from low to high temperatures}.
\newblock {\em Phys. Rev.}, D93(11):114502.
\newblock 1603.06637.

\bibitem[Bazavov and Petreczky, 2013]{Bazavov:2013yv}
Bazavov, A. and Petreczky, P. (2013).
\newblock {Polyakov loop in 2+1 flavor QCD}.
\newblock {\em Phys. Rev.}, D87(9):094505.
\newblock 1301.3943.

\bibitem[Behtash et~al., 2016]{Behtash:2015zha}
Behtash, A., Dunne, G.~V., Schafer, T., Sulejmanpasic, T., and Unsal, M.
  (2016).
\newblock {Complexified path integrals, exact saddles and supersymmetry}.
\newblock {\em Phys. Rev. Lett.}, 116(1):011601.
\newblock 1510.00978.

\bibitem[Behtash et~al., 2015]{Behtash:2015kna}
Behtash, A., Sulejmanpasic, T., Schafer, T., and Unsal, M. (2015).
\newblock {Hidden topological angles and Lefschetz thimbles}.
\newblock {\em Phys. Rev. Lett.}, 115(4):041601.
\newblock 1502.06624.

\bibitem[Belavin et~al., 1975]{Belavin:1975fg}
Belavin, A.~A., Polyakov, A.~M., Schwartz, A.~S., and Tyupkin, {\relax Yu}.~S.
  (1975).
\newblock {Pseudoparticle Solutions of the Yang-Mills Equations}.
\newblock {\em Phys. Lett.}, B59:85--87.
\newblock [,350(1975)].

\bibitem[Berezin, 1975]{Berezin:1974du}
Berezin, F.~A. (1975).
\newblock {General Concept of Quantization}.
\newblock {\em Commun. Math. Phys.}, 40:153--174.

\bibitem[Bergner et~al., 2014]{Bergner:2015iva}
Bergner, G., Giudice, P., Munster, G., Piemonte, S., and Sandbrink, D. (2014).
\newblock {First studies of the phase diagram of N=1 supersymmetric Yang-Mills
  theory}.
\newblock {\em PoS}, LATTICE2014:262.
\newblock 1501.02746.

\bibitem[Bigazzi et~al., 2019]{Bigazzi:2019eks}
Bigazzi, F., Caddeo, A., Cotrone, A.~L., Di~Vecchia, P., and Marzolla, A.
  (2019).
\newblock {The Holographic QCD Axion}.

\bibitem[Bjorken, 2000]{Bjorken:2000ni}
Bjorken, J.~D. (2000).
\newblock {Intersections 2000: What's new in hadron physics}.
\newblock {\em AIP Conf. Proc.}, 549(1):211--229.
\newblock hep-ph/0008048.

\bibitem[Bonati et~al., 2012]{Bonati:2011jv}
Bonati, C., Cossu, G., D'Elia, M., and Di~Giacomo, A. (2012).
\newblock {The disorder parameter of dual superconductivity in QCD revisited}.
\newblock {\em Phys. Rev.}, D85:065001.
\newblock 1111.1541.

\bibitem[Bonati et~al., 2016a]{Bonati:2015vqz}
Bonati, C., D'Elia, M., Mariti, M., Martinelli, G., Mesiti, M., Negro, F.,
  Sanfilippo, F., and Villadoro, G. (2016a).
\newblock {Axion phenomenology and $\theta$-dependence from $N_f = 2+1$ lattice
  QCD}.
\newblock {\em JHEP}, 03:155.
\newblock 1512.06746.

\bibitem[Bonati et~al., 2016b]{Bonati:2016pwz}
Bonati, C., D'Elia, M., Mariti, M., Mesiti, M., Negro, F., and Sanfilippo, F.
  (2016b).
\newblock {Roberge-Weiss endpoint at the physical point of $N_f = 2+1$ QCD}.
\newblock {\em Phys. Rev.}, D93(7):074504.
\newblock 1602.01426.

\bibitem[Bonati et~al., 2016c]{Bonati:2016tvi}
Bonati, C., D'Elia, M., Rossi, P., and Vicari, E. (2016c).
\newblock {$\theta$ dependence of 4D $SU(N)$ gauge theories in the large-$N$
  limit}.
\newblock {\em Phys. Rev.}, D94(8):085017.
\newblock 1607.06360.

\bibitem[Bornyakov et~al., 2017]{Bornyakov:2016ccq}
Bornyakov, V.~G., Boyda, D.~L., Goy, V.~A., Ilgenfritz, E.~M., Martemyanov,
  B.~V., Molochkov, A.~V., Nakamura, A., Nikolaev, A.~A., and Zakharov, V.~I.
  (2017).
\newblock {Dyons and Roberge - Weiss transition in lattice QCD}.
\newblock {\em EPJ Web Conf.}, 137:03002.
\newblock 1611.07789.

\bibitem[Bornyakov et~al., 2016]{Bornyakov:2015xao}
Bornyakov, V.~G., Ilgenfritz, E.~M., Martemyanov, B.~V., and Muller-Preussker,
  M. (2016).
\newblock {Dyons near the transition temperature in lattice QCD}.
\newblock {\em Phys. Rev.}, D93(7):074508.
\newblock 1512.03217.

\bibitem[Borsanyi et~al., 2016]{Borsanyi:2016ksw}
Borsanyi, S. et~al. (2016).
\newblock {Calculation of the axion mass based on high-temperature lattice
  quantum chromodynamics}.
\newblock {\em Nature}, 539(7627):69--71.
\newblock 1606.07494.

\bibitem[Boulware et~al., 1976]{Boulware:1976tv}
Boulware, D.~G., Brown, L.~S., Cahn, R.~N., Ellis, S.~D., and Lee, C.-k.
  (1976).
\newblock {Scattering on Magnetic Charge}.
\newblock {\em Phys. Rev.}, D14:2708.

\bibitem[Brandt, 2017]{Brandt:2017yzw}
Brandt, B.~B. (2017).
\newblock {Spectrum of the open QCD flux tube and its effective string
  description I: 3d static potential in SU(N = 2, 3)}.
\newblock {\em JHEP}, 07:008.
\newblock 1705.03828.

\bibitem[Braun, 2006]{Braun:2006hn}
Braun, V.~M. (2006).
\newblock {Nucleons on the light-cone: Theory and phenomenology of baryon
  distribution amplitudes}.
\newblock In {\em {Continuous advances in QCD. Proceedings, 7th Workshop, QCD
  2006, Minneapolis, USA, May 11-14, 2006}}, pages 42--57.

\bibitem[Brauner et~al., 2012]{Brauner:2012gu}
Brauner, T., Taanila, O., Tranberg, A., and Vuorinen, A. (2012).
\newblock {Computing the temperature dependence of effective CP violation in
  the standard model}.
\newblock {\em JHEP}, 11:076.
\newblock 1208.5609.

\bibitem[Bringoltz and Teper, 2006]{Bringoltz:2005xx}
Bringoltz, B. and Teper, M. (2006).
\newblock {In search of a Hagedorn transition in SU(N) lattice gauge theories
  at large-N}.
\newblock {\em Phys. Rev.}, D73:014517.
\newblock hep-lat/0508021.

\bibitem[Brodsky and de~Teramond, 2006]{Brodsky:2006uqa}
Brodsky, S.~J. and de~Teramond, G.~F. (2006).
\newblock {Hadronic spectra and light-front wavefunctions in holographic QCD}.
\newblock {\em Phys. Rev. Lett.}, 96:201601.

\bibitem[Brodsky and Lepage, 1989]{Brodsky:1989pv}
Brodsky, S.~J. and Lepage, G.~P. (1989).
\newblock {Exclusive Processes in Quantum Chromodynamics}.
\newblock {\em Adv. Ser. Direct. High Energy Phys.}, 5:93--240.

\bibitem[Brown et~al., 1978]{Brown:1977eb}
Brown, L.~S., Carlitz, R.~D., Creamer, D.~B., and Lee, C.-k. (1978).
\newblock {Propagation Functions in Pseudoparticle Fields}.
\newblock {\em Phys. Rev.}, D17:1583.
\newblock [,168(1977)].

\bibitem[Brown and Creamer, 1978]{Brown:1978yj}
Brown, L.~S. and Creamer, D.~B. (1978).
\newblock {VACUUM POLARIZATION ABOUT INSTANTONS}.
\newblock {\em Phys. Rev.}, D18:3695.

\bibitem[Burnier et~al., 2011]{Burnier:2011bf}
Burnier, Y., Kharzeev, D.~E., Liao, J., and Yee, H.-U. (2011).
\newblock {Chiral magnetic wave at finite baryon density and the electric
  quadrupole moment of quark-gluon plasma in heavy ion collisions}.
\newblock {\em Phys. Rev. Lett.}, 107:052303.
\newblock 1103.1307.

\bibitem[Burnier et~al., 2006]{Burnier:2005hp}
Burnier, Y., Laine, M., and Shaposhnikov, M. (2006).
\newblock {Baryon and lepton number violation rates across the electroweak
  crossover}.
\newblock {\em JCAP}, 0602:007.
\newblock hep-ph/0511246.

\bibitem[Callan et~al., 1979]{Callan:1978bm}
Callan, Jr., C.~G., Dashen, R.~F., and Gross, D.~J. (1979).
\newblock {A Theory of Hadronic Structure}.
\newblock {\em Phys. Rev.}, D19:1826.

\bibitem[Callan et~al., 1978]{Callan:1978ye}
Callan, Jr., C.~G., Dashen, R.~F., Gross, D.~J., Wilczek, F., and Zee, A.
  (1978).
\newblock {The Effect of Instantons on the Heavy Quark Potential}.
\newblock {\em Phys. Rev.}, D18:4684.

\bibitem[Carlitz and Creamer, 1979]{Carlitz:1978yj}
Carlitz, R.~D. and Creamer, D.~B. (1979).
\newblock {Light Quarks and Instantons}.
\newblock {\em Annals Phys.}, 118:429.

\bibitem[Carson et~al., 1990]{Carson:1990jm}
Carson, L., Li, X., McLerran, L.~D., and Wang, R.-T. (1990).
\newblock {Exact Computation of the Small Fluctuation Determinant Around a
  Sphaleron}.
\newblock {\em Phys. Rev.}, D42:2127--2143.

\bibitem[Cea et~al., 2018]{Cea:2017bsa}
Cea, P., Cosmai, L., Cuteri, F., and Papa, A. (2018).
\newblock {QCD flux tubes across the deconfinement phase transition}.
\newblock {\em EPJ Web Conf.}, 175:12006.
\newblock 1710.01963.

\bibitem[Chen et~al., 2010]{Chen:2010yr}
Chen, H.-Y., Dorey, N., and Petunin, K. (2010).
\newblock {Wall Crossing and Instantons in Compactified Gauge Theory}.
\newblock {\em JHEP}, 06:024.
\newblock 1004.0703.

\bibitem[Cherman et~al., 2016]{Cherman:2016hcd}
Cherman, A., Schaefer, T., and Unsal, M. (2016).
\newblock {Chiral Lagrangian from Duality and Monopole Operators in
  Compactified QCD}.
\newblock {\em Phys. Rev. Lett.}, 117(8):081601.
\newblock 1604.06108.

\bibitem[Chern and Simons, 1974]{Chern:1974ft}
Chern, S.-S. and Simons, J. (1974).
\newblock {Characteristic forms and geometric invariants}.
\newblock {\em Annals Math.}, 99:48--69.

\bibitem[Chernyak et~al., 1989]{Chernyak:1987nv}
Chernyak, V.~L., Ogloblin, A.~A., and Zhitnitsky, I.~R. (1989).
\newblock {Calculation of Exclusive Processes With Baryons}.
\newblock {\em Z. Phys.}, C42:583.
\newblock [Sov. J. Nucl. Phys.48,889(1988)].

\bibitem[Chernyak and Zhitnitsky, 1984]{Chernyak:1983ej}
Chernyak, V.~L. and Zhitnitsky, A.~R. (1984).
\newblock {Asymptotic Behavior of Exclusive Processes in QCD}.
\newblock {\em Phys. Rept.}, 112:173.

\bibitem[Chew and Frautschi, 1962]{Chew:1962eu}
Chew, G.~F. and Frautschi, S.~C. (1962).
\newblock {Regge Trajectories and the Principle of Maximum Strength for Strong
  Interactions}.
\newblock {\em Phys. Rev. Lett.}, 8:41--44.

\bibitem[Chu et~al., 1994]{Chu:1994vi}
Chu, M.~C., Grandy, J.~M., Huang, S., and Negele, J.~W. (1994).
\newblock {Evidence for the role of instantons in hadron structure from lattice
  QCD}.
\newblock {\em Phys. Rev.}, D49:6039--6050.
\newblock hep-lat/9312071.

\bibitem[Corrigan et~al., 1979]{Corrigan:1979di}
Corrigan, E., Goddard, P., Osborn, H., and Templeton, S. (1979).
\newblock {Zeta Function Regularization and Multi - Instanton Determinants}.
\newblock {\em Nucl. Phys.}, B159:469--496.

\bibitem[Cossu and D'Elia, 2009]{Cossu:2009sq}
Cossu, G. and D'Elia, M. (2009).
\newblock {Finite size phase transitions in QCD with adjoint fermions}.
\newblock {\em JHEP}, 07:048.
\newblock 0904.1353.

\bibitem[Cossu et~al., 2007]{Cossu:2007dk}
Cossu, G., D'Elia, M., Di~Giacomo, A., Lucini, B., and Pica, C. (2007).
\newblock {G(2) gauge theory at finite temperature}.
\newblock {\em JHEP}, 10:100.

\bibitem[Creutz, 1980]{Creutz:1980wj}
Creutz, M. (1980).
\newblock {Asymptotic Freedom Scales}.
\newblock {\em Phys. Rev. Lett.}, 45:313.

\bibitem[Crichigno et~al., 2010]{Crichigno:2010ky}
Crichigno, M.~P., Flambaum, V.~V., Kuchiev, M.~{\relax Yu}., and Shuryak, E.
  (2010).
\newblock {The W-Z-Top Bags}.
\newblock {\em Phys. Rev.}, D82:073018.
\newblock 1006.0645.

\bibitem[D'Alessandro and D'Elia, 2008]{D'Alessandro:2007su}
D'Alessandro, A. and D'Elia, M. (2008).
\newblock {Magnetic monopoles in the high temperature phase of Yang-Mills
  theories}.
\newblock {\em Nucl. Phys.}, B799:241--254.
\newblock 0711.1266.

\bibitem[D'Alessandro et~al., 2010]{D'Alessandro:2010xg}
D'Alessandro, A., D'Elia, M., and Shuryak, E.~V. (2010).
\newblock {Thermal Monopole Condensation and Confinement in finite temperature
  Yang-Mills Theories}.
\newblock {\em Phys. Rev.}, D81:094501.
\newblock 1002.4161.

\bibitem[Damour and Veneziano, 2000]{Damour:1999aw}
Damour, T. and Veneziano, G. (2000).
\newblock {Selfgravitating fundamental strings and black holes}.
\newblock {\em Nucl. Phys.}, B568:93--119.
\newblock hep-th/9907030.

\bibitem[Danielsson et~al., 1999]{Danielsson:1998wt}
Danielsson, U.~H., Keski-Vakkuri, E., and Kruczenski, M. (1999).
\newblock {Vacua, propagators, and holographic probes in AdS / CFT}.
\newblock {\em JHEP}, 01:002.

\bibitem[Dashen et~al., 1974]{Dashen:1974ck}
Dashen, R.~F., Hasslacher, B., and Neveu, A. (1974).
\newblock {Nonperturbative Methods and Extended Hadron Models in Field Theory.
  3. Four-Dimensional Nonabelian Models}.
\newblock {\em Phys. Rev.}, D10:4138.

\bibitem[Davies et~al., 1999]{Davies:1999uw}
Davies, N.~M., Hollowood, T.~J., Khoze, V.~V., and Mattis, M.~P. (1999).
\newblock {Gluino condensate and magnetic monopoles in supersymmetric
  gluodynamics}.
\newblock {\em Nucl. Phys.}, B559:123--142.
\newblock hep-th/9905015.

\bibitem[de~Forcrand et~al., 1997]{deForcrand:1997esx}
de~Forcrand, P., Garcia~Perez, M., and Stamatescu, I.-O. (1997).
\newblock {Topology of the SU(2) vacuum: A Lattice study using improved
  cooling}.
\newblock {\em Nucl. Phys.}, B499:409--449.
\newblock hep-lat/9701012.

\bibitem[de~Forcrand and Liu, 1993]{deForcrand:1992ww}
de~Forcrand, P. and Liu, K.-F. (1993).
\newblock {Glueball wave functions}.
\newblock {\em Nucl. Phys. Proc. Suppl.}, 30:521--524.

\bibitem[Diakonov, 2009]{Diakonov:2009jq}
Diakonov, D. (2009).
\newblock {Topology and confinement}.
\newblock {\em Nucl. Phys. Proc. Suppl.}, 195:5--45.
\newblock 0906.2456.

\bibitem[Diakonov et~al., 2004]{Diakonov:2004jn}
Diakonov, D., Gromov, N., Petrov, V., and Slizovskiy, S. (2004).
\newblock {Quantum weights of dyons and of instantons with nontrivial
  holonomy}.
\newblock {\em Phys. Rev.}, D70:036003.
\newblock hep-th/0404042.

\bibitem[Diakonov and Petrov, 1984]{Diakonov:1983hh}
Diakonov, D. and Petrov, V.~{\relax Yu}. (1984).
\newblock {Instanton Based Vacuum from Feynman Variational Principle}.
\newblock {\em Nucl. Phys.}, B245:259--292.

\bibitem[Diakonov and Petrov, 1986]{Diakonov:1985eg}
Diakonov, D. and Petrov, V.~{\relax Yu}. (1986).
\newblock {A Theory of Light Quarks in the Instanton Vacuum}.
\newblock {\em Nucl. Phys.}, B272:457--489.

\bibitem[Dine and Kusenko, 2003]{Dine:2003ax}
Dine, M. and Kusenko, A. (2003).
\newblock {The Origin of the matter - antimatter asymmetry}.
\newblock {\em Rev. Mod. Phys.}, 76:1.
\newblock hep-ph/0303065.

\bibitem[Dirac, 1931]{Dirac:1931kp}
Dirac, P. A.~M. (1931).
\newblock {Quantized Singularities in the Electromagnetic Field}.
\newblock {\em Proc. Roy. Soc. Lond.}, A133:60--72.
\newblock [,278(1931)].

\bibitem[D'Onofrio et~al., 2014]{DOnofrio:2014rug}
D'Onofrio, M., Rummukainen, K., and Tranberg, A. (2014).
\newblock {Sphaleron Rate in the Minimal Standard Model}.
\newblock {\em Phys. Rev. Lett.}, 113(14):141602.

\bibitem[Dorey, 2001]{Dorey:2000dt}
Dorey, N. (2001).
\newblock {Instantons, compactification and S-duality in N=4 SUSY Yang-Mills
  theory. 1.}
\newblock {\em JHEP}, 04:008.
\newblock hep-th/0010115.

\bibitem[Dorey et~al., 2002]{Dorey:2002ik}
Dorey, N., Hollowood, T.~J., Khoze, V.~V., and Mattis, M.~P. (2002).
\newblock {The Calculus of many instantons}.
\newblock {\em Phys. Rept.}, 371:231--459.

\bibitem[Dorey et~al., 1999]{Dorey:1999pd}
Dorey, N., Hollowood, T.~J., Khoze, V.~V., Mattis, M.~P., and Vandoren, S.
  (1999).
\newblock {Multi-instanton calculus and the AdS / CFT correspondence in N=4
  superconformal field theory}.
\newblock {\em Nucl. Phys.}, B552:88--168.

\bibitem[Dorey and Parnachev, 2001]{Dorey:2000qc}
Dorey, N. and Parnachev, A. (2001).
\newblock {Instantons, compactification and S duality in N=4 SUSY Yang-Mills
  theory. 2.}
\newblock {\em JHEP}, 08:059.
\newblock hep-th/0011202.

\bibitem[Dorokhov and Kochelev, 1993]{Dorokhov:1993fc}
Dorokhov, A.~E. and Kochelev, N.~I. (1993).
\newblock {Instanton induced asymmetric quark configurations in the nucleon and
  parton sum rules}.
\newblock {\em Phys. Lett.}, B304:167--175.

\bibitem[Dunne and Unsal, 2014]{Dunne:2013ada}
Dunne, G.~V. and Unsal, M. (2014).
\newblock {Generating nonperturbative physics from perturbation theory}.
\newblock {\em Phys. Rev.}, D89(4):041701.
\newblock 1306.4405.

\bibitem[Dyson, 1952]{Dyson:1952tj}
Dyson, F.~J. (1952).
\newblock {Divergence of perturbation theory in quantum electrodynamics}.
\newblock {\em Phys. Rev.}, 85:631--632.

\bibitem[Erlich et~al., 2005]{Erlich:2005qh}
Erlich, J., Katz, E., Son, D.~T., and Stephanov, M.~A. (2005).
\newblock {QCD and a holographic model of hadrons}.
\newblock {\em Phys. Rev. Lett.}, 95:261602.

\bibitem[Escobar-Ruiz et~al., 2015]{Escobar-Ruiz:2015nsa}
Escobar-Ruiz, M.~A., Shuryak, E., and Turbiner, A.~V. (2015).
\newblock {Three-loop Correction to the Instanton Density. I. The Quartic
  Double Well Potential}.
\newblock {\em Phys. Rev.}, D92(2):025046.
\newblock [Erratum: Phys. Rev.D92,no.8,089902(2015)].

\bibitem[Escobar-Ruiz et~al., 2016]{Escobar-Ruiz:2016aqv}
Escobar-Ruiz, M.~A., Shuryak, E., and Turbiner, A.~V. (2016).
\newblock {Quantum and thermal fluctuations in quantum mechanics and field
  theories from a new version of semiclassical theory}.
\newblock {\em Phys. Rev.}, D93(10):105039.
\newblock 1601.03964.

\bibitem[Escobar-Ruiz et~al., 2017]{Escobar-Ruiz:2017uhx}
Escobar-Ruiz, M.~A., Shuryak, E., and Turbiner, A.~V. (2017).
\newblock {Fluctuations in quantum mechanics and field theories from a new
  version of semiclassical theory. II}.
\newblock {\em Phys. Rev.}, D96(4):045005.
\newblock 1705.06159.

\bibitem[Ewerz et~al., 2016]{Ewerz:2016onn}
Ewerz, C., Lebiedowicz, P., Nachtmann, O., and Szczurek, A. (2016).
\newblock {Helicity in proton-proton elastic scattering and the spin structure
  of the pomeron}.
\newblock {\em Phys. Lett.}, B763:382--387.

\bibitem[Faccioli, 2011]{Faccioli:2011eg}
Faccioli, P. (2011).
\newblock {Investigating Biological Matter with Theoretical Nuclear Physics
  Methods}.
\newblock {\em J. Phys. Conf. Ser.}, 336:012030.
\newblock 1108.5074.

\bibitem[Faccioli and DeGrand, 2003]{Faccioli:2003qz}
Faccioli, P. and DeGrand, T.~A. (2003).
\newblock {Evidence for instanton induced dynamics, from lattice QCD}.
\newblock {\em Phys. Rev. Lett.}, 91:182001.

\bibitem[Faccioli et~al., 2004]{Faccioli:2003ve}
Faccioli, P., Schwenk, A., and Shuryak, E.~V. (2004).
\newblock {Instanton contribution to the pion and proton electromagnetic
  form-factors at Q**2 approximately greater than 1-GeV**2}.
\newblock {\em Fizika}, B13:193--200.

\bibitem[Faccioli and Shuryak, 2013]{Faccioli:2013ja}
Faccioli, P. and Shuryak, E. (2013).
\newblock {QCD topology at finite temperature: Statistical mechanics of
  self-dual dyons}.
\newblock {\em Phys. Rev.}, D87(7):074009.
\newblock 1301.2523.

\bibitem[Feynman, 1972]{Feynman_SM}
Feynman, R. (1972).
\newblock {\em {Statistical Mechanics: A set of lecture}s}.
\newblock W.A.Benjamin Inc., Reading, M.A.

\bibitem[Feynman and Hibbs, 1965]{FH_65}
Feynman, R. and Hibbs, H. (1965).
\newblock {\em {Quantum Mechanics and Path Integrals}}.
\newblock Mcgraw-Hill, New York.

\bibitem[Flambaum and Shuryak, 2010]{Flambaum:2010fp}
Flambaum, V.~V. and Shuryak, E. (2010).
\newblock {Possible Role of the WZ-Top-Quark Bags in Baryogenesis}.
\newblock {\em Phys. Rev.}, D82:073019.
\newblock 1006.0249.

\bibitem[Frankfurt et~al., 1993]{Frankfurt:1993it}
Frankfurt, L., Miller, G.~A., and Strikman, M. (1993).
\newblock {Coherent nuclear diffractive production of mini - jets: Illuminating
  color transparency}.
\newblock {\em Phys. Lett.}, B304:1--7.

\bibitem[Friess et~al., 2007]{hep-th/0611005}
Friess, J.~J., Gubser, S.~S., Michalogiorgakis, G., and Pufu, S.~S. (2007).
\newblock {Expanding plasmas and quasinormal modes of anti-de Sitter black
  holes}.
\newblock {\em JHEP}, 04:080.

\bibitem[Fukushima et~al., 2008]{Fukushima:2008xe}
Fukushima, K., Kharzeev, D.~E., and Warringa, H.~J. (2008).
\newblock {The Chiral Magnetic Effect}.
\newblock {\em Phys. Rev.}, D78:074033.
\newblock 0808.3382.

\bibitem[Gamow, 1928]{Gamow}
Gamow, G. (1928).
\newblock {Zur Quantentheorie des Atomkernes}.
\newblock {\em Zeitschrift fur Physik}, 51:204.

\bibitem[Garcia-Bellido et~al., 2004]{GarciaBellido:2003wd}
Garcia-Bellido, J., Garcia-Perez, M., and Gonzalez-Arroyo, A. (2004).
\newblock {Chern-Simons production during preheating in hybrid inflation
  models}.
\newblock {\em Phys. Rev.}, D69:023504.
\newblock hep-ph/0304285.

\bibitem[Garcia-Recio and Salcedo, 2009]{GarciaRecio:2009zp}
Garcia-Recio, C. and Salcedo, L.~L. (2009).
\newblock {CP violation in the effective action of the Standard Model}.
\newblock {\em JHEP}, 07:015.
\newblock 0903.5494.

\bibitem[Gattringer, 2003]{Gattringer:2002wh}
Gattringer, C. (2003).
\newblock {Calorons, instantons and constituent monopoles in SU(3) lattice
  gauge theory}.
\newblock {\em Phys. Rev.}, D67:034507.
\newblock hep-lat/0210001.

\bibitem[Gauntlett, 1994]{Gauntlett:1993sh}
Gauntlett, J.~P. (1994).
\newblock {Low-energy dynamics of N=2 supersymmetric monopoles}.
\newblock {\em Nucl. Phys.}, B411:443--460.

\bibitem[Gauntlett and Harvey, 1996]{Gauntlett:1995fu}
Gauntlett, J.~P. and Harvey, J.~A. (1996).
\newblock {S duality and the dyon spectrum in N=2 superYang-Mills theory}.
\newblock {\em Nucl. Phys.}, B463:287--314.
\newblock hep-th/9508156.

\bibitem[Geesaman and Reimer, 2019]{Geesaman:2018ixo}
Geesaman, D.~F. and Reimer, P.~E. (2019).
\newblock {The sea of quarks and antiquarks in the nucleon}.
\newblock {\em Rept. Prog. Phys.}, 82(4):046301.

\bibitem[Gell-Mann and Low, 1954]{GellMann:1954fq}
Gell-Mann, M. and Low, F.~E. (1954).
\newblock {Quantum electrodynamics at small distances}.
\newblock {\em Phys. Rev.}, 95:1300--1312.

\bibitem[Gross et~al., 1981]{Gross:1980br}
Gross, D.~J., Pisarski, R.~D., and Yaffe, L.~G. (1981).
\newblock {QCD and Instantons at Finite Temperature}.
\newblock {\em Rev. Mod. Phys.}, 53:43.

\bibitem[Gross and Wilczek, 1973]{Gross:1973ju}
Gross, D.~J. and Wilczek, F. (1973).
\newblock {Asymptotically Free Gauge Theories - I}.
\newblock {\em Phys. Rev.}, D8:3633--3652.

\bibitem[Gubser et~al., 1998]{Gubser:1998nz}
Gubser, S.~S., Klebanov, I.~R., and Tseytlin, A.~A. (1998).
\newblock {Coupling constant dependence in the thermodynamics of N=4
  supersymmetric Yang-Mills theory}.
\newblock {\em Nucl. Phys.}, B534:202--222.

\bibitem[Gurney and Condon, 1928]{CG_28}
Gurney, R. and Condon, E. (1928).
\newblock {Quantum Mechanics and Radioactive Disintegration}.
\newblock {\em Nature}, 122:439.

\bibitem[Gursoy and Kiritsis, 2008]{Gursoy:2007cb}
Gursoy, U. and Kiritsis, E. (2008).
\newblock {Exploring improved holographic theories for QCD: Part I}.
\newblock {\em JHEP}, 02:032.
\newblock 0707.1324.

\bibitem[Gursoy et~al., 2011]{Gursoy:2010fj}
Gursoy, U., Kiritsis, E., Mazzanti, L., Michalogiorgakis, G., and Nitti, F.
  (2011).
\newblock {Improved Holographic QCD}.
\newblock {\em Lect. Notes Phys.}, 828:79--146.

\bibitem[Gursoy et~al., 2008]{Gursoy:2007er}
Gursoy, U., Kiritsis, E., and Nitti, F. (2008).
\newblock {Exploring improved holographic theories for QCD: Part II}.
\newblock {\em JHEP}, 02:019.

\bibitem[Hagedorn, 1965]{Hagedorn:1965st}
Hagedorn, R. (1965).
\newblock {Statistical thermodynamics of strong interactions at high-energies}.
\newblock {\em Nuovo Cim. Suppl.}, 3:147--186.

\bibitem[Harvey, 1996]{Harvey:1996ur}
Harvey, J.~A. (1996).
\newblock {Magnetic monopoles, duality and supersymmetry}.
\newblock In {\em {Fields, strings and duality. Proceedings, Summer School,
  Theoretical Advanced Study Institute in Elementary Particle Physics, TASI'96,
  Boulder, USA, June 2-28, 1996}}.
\newblock hep-th/9603086.

\bibitem[Hasenfratz, 2000]{Hasenfratz:1999ng}
Hasenfratz, A. (2000).
\newblock {Spatial correlation of the topological charge in pure SU(3) gauge
  theory and in QCD}.
\newblock {\em Phys. Lett.}, B476:188--192.
\newblock hep-lat/9912053.

\bibitem[Hata et~al., 2007]{Hata:2007mb}
Hata, H., Sakai, T., Sugimoto, S., and Yamato, S. (2007).
\newblock {Baryons from instantons in holographic QCD}.
\newblock {\em Prog. Theor. Phys.}, 117:1157.

\bibitem[Hellerman and Swanson, 2015]{Hellerman:2013kba}
Hellerman, S. and Swanson, I. (2015).
\newblock {String Theory of the Regge Intercept}.
\newblock {\em Phys. Rev. Lett.}, 114(11):111601.
\newblock 1312.0999.

\bibitem[Hernandez et~al., 2009]{Hernandez:2008db}
Hernandez, A., Konstandin, T., and Schmidt, M.~G. (2009).
\newblock {Sizable CP Violation in the Bosonized Standard Model}.
\newblock {\em Nucl. Phys.}, B812:290--300.
\newblock 0810.4092.

\bibitem[Hidaka and Pisarski, 2009]{Hidaka:2009xh}
Hidaka, Y. and Pisarski, R.~D. (2009).
\newblock {Zero Point Energy of Renormalized Wilson Loops}.
\newblock {\em Phys. Rev.}, D80:074504.
\newblock 0907.4609.

\bibitem[Hollowood et~al., 2000]{Hollowood:1999qn}
Hollowood, T.~J., Khoze, V.~V., Lee, W.-J., and Mattis, M.~P. (2000).
\newblock {Breakdown of cluster decomposition in instanton calculations of the
  gluino condensate}.
\newblock {\em Nucl. Phys.}, B570:241--266.
\newblock hep-th/9904116.

\bibitem[Horowitz and Polchinski, 1998]{Horowitz:1997jc}
Horowitz, G.~T. and Polchinski, J. (1998).
\newblock {Selfgravitating fundamental strings}.
\newblock {\em Phys. Rev.}, D57:2557--2563.
\newblock hep-th/9707170.

\bibitem[Iatrakis et~al., 2015]{Iatrakis:2015rga}
Iatrakis, I., Ramamurti, A., and Shuryak, E. (2015).
\newblock {Collective String Interactions in AdS/QCD and High-Multiplicity pA
  Collisions}.
\newblock {\em Phys. Rev.}, D92(1):014011.
\newblock 1503.04759.

\bibitem[Iatrakis et~al., 2016]{Iatrakis:2016rvj}
Iatrakis, I., Ramamurti, A., and Shuryak, E. (2016).
\newblock {Pomeron Interactions from the Einstein-Hilbert Action}.
\newblock {\em Phys. Rev.}, D94(4):045005.

\bibitem[Ilgenfritz et~al., 1986]{Ilgenfritz:1985dz}
Ilgenfritz, E.-M., Laursen, M.~L., Schierholz, G., Muller-Preussker, M., and
  Schiller, H. (1986).
\newblock {First Evidence for the Existence of Instantons in the Quantized
  SU(2) Lattice Vacuum}.
\newblock {\em Nucl. Phys.}, B268:693.

\bibitem[Ilgenfritz and Muller-Preussker, 1981]{Ilgenfritz:1980bm}
Ilgenfritz, E.-M. and Muller-Preussker, M. (1981).
\newblock {Interacting Instantons, 1/$N$ Expansion and the Gluon Condensate}.
\newblock {\em Phys. Lett.}, 99B:128.
\newblock [,87(1980)].

\bibitem[Ilgenfritz and Shuryak, 1989]{Ilgenfritz:1988dh}
Ilgenfritz, E.-M. and Shuryak, E.~V. (1989).
\newblock {Chiral Symmetry Restoration at Finite Temperature in the Instanton
  Liquid}.
\newblock {\em Nucl. Phys.}, B319:511--520.

\bibitem[Ilgenfritz and Shuryak, 1994]{Ilgenfritz:1994nt}
Ilgenfritz, E.-M. and Shuryak, E.~V. (1994).
\newblock {Quark induced correlations between instantons drive the chiral phase
  transition}.
\newblock {\em Phys. Lett.}, B325:263--266.
\newblock hep-ph/9401285.

\bibitem[Iritani et~al., 2014]{Iritani:2013rla}
Iritani, T., Cossu, G., and Hashimoto, S. (2014).
\newblock {Analysis of topological structure of the QCD vacuum with
  overlap-Dirac operator eigenmode}.
\newblock {\em PoS}, LATTICE2013:376.
\newblock 1311.0218.

\bibitem[Isgur and Wise, 1989]{Isgur:1989vq}
Isgur, N. and Wise, M.~B. (1989).
\newblock {Weak Decays of Heavy Mesons in the Static Quark Approximation}.
\newblock {\em Phys. Lett.}, B232:113--117.

\bibitem[Jackiw and Rebbi, 1976]{Jackiw:1975fn}
Jackiw, R. and Rebbi, C. (1976).
\newblock {Solitons with Fermion Number 1/2}.
\newblock {\em Phys. Rev.}, D13:3398--3409.

\bibitem[Jarvinen and Kiritsis, 2012]{Jarvinen:2011qe}
Jarvinen, M. and Kiritsis, E. (2012).
\newblock {Holographic Models for QCD in the Veneziano Limit}.
\newblock {\em JHEP}, 03:002.

\bibitem[Jia and Vary, 2019]{Jia:2018ary}
Jia, S. and Vary, J.~P. (2019).
\newblock {Basis light front quantization for the charged light mesons with
  color singlet Nambu–Jona-Lasinio interactions}.
\newblock {\em Phys. Rev.}, C99(3):035206.

\bibitem[Julia and Zee, 1975]{Julia:1975ff}
Julia, B. and Zee, A. (1975).
\newblock {Poles with Both Magnetic and Electric Charges in Nonabelian Gauge
  Theory}.
\newblock {\em Phys. Rev.}, D11:2227--2232.

\bibitem[Kaczmarek et~al., 2002]{Kaczmarek:2002mc}
Kaczmarek, O., Karsch, F., Petreczky, P., and Zantow, F. (2002).
\newblock {Heavy quark anti-quark free energy and the renormalized Polyakov
  loop}.
\newblock {\em Phys. Lett.}, B543:41--47.
\newblock hep-lat/0207002.

\bibitem[Kaczmarek and Zantow, 2005]{Kaczmarek:2005gi}
Kaczmarek, O. and Zantow, F. (2005).
\newblock {Static quark anti-quark interactions at zero and finite temperature
  QCD. II. Quark anti-quark internal energy and entropy}.
\newblock hep-lat/0506019.

\bibitem[Kalaydzhyan and Shuryak, 2014]{Kalaydzhyan:2014zqa}
Kalaydzhyan, T. and Shuryak, E. (2014).
\newblock {Collective interaction of QCD strings and early stages of high
  multiplicity pA collisions}.
\newblock {\em Phys. Rev.}, C90(1):014901.
\newblock 1404.1888.

\bibitem[Kapusta et~al., 2019]{Kapusta:2019ktm}
Kapusta, J.~I., Rrapaj, E., and Rudaz, S. (2019).
\newblock {Is Hyperon Polarization in Relativistic Heavy Ion Collisions
  Connected to Axial U(1) Symmetry Breaking at High Temperature?}

\bibitem[Karch et~al., 2006a]{Karch:2006pv}
Karch, A., Katz, E., Son, D.~T., and Stephanov, M.~A. (2006a).
\newblock {Linear confinement and AdS/QCD}.
\newblock {\em Phys. Rev.}, D74:015005.

\bibitem[Karch et~al., 2006b]{Karch:2006zz}
Karch, A., Son, D.~T., Katz, E., and Stephanov, M.~A. (2006b).
\newblock {Linear confinement and AdS/QCD}.
\newblock In {\em {Continuous advances in QCD. Proceedings, 7th Workshop, QCD
  2006, Minneapolis, USA, May 11-14, 2006}}, pages 96--102.

\bibitem[Kazama et~al., 1977]{Kazama:1976fm}
Kazama, Y., Yang, C.~N., and Goldhaber, A.~S. (1977).
\newblock {Scattering of a Dirac Particle with Charge Ze by a Fixed Magnetic
  Monopole}.
\newblock {\em Phys. Rev.}, D15:2287--2299.

\bibitem[Kharzeev, 1996]{Kharzeev:1996sq}
Kharzeev, D. (1996).
\newblock {Can gluons trace baryon number?}
\newblock {\em Phys. Lett.}, B378:238--246.
\newblock nucl-th/9602027.

\bibitem[Kharzeev et~al., 2018]{Kharzeev:2017azf}
Kharzeev, D., Shuryak, E., and Zahed, I. (2018).
\newblock {Higher order string effects and the properties of the Pomeron}.
\newblock {\em Phys. Rev.}, D97(1):016008.
\newblock 1709.04007.

\bibitem[Kharzeev et~al., 2019]{Kharzeev:2019rsy}
Kharzeev, D., Shuryak, E., and Zahed, I. (2019).
\newblock {Baryogenesis and helical magnetogenesis from the electroweak
  transition of the minimal Standard Model}.

\bibitem[Kharzeev, 2010]{Kharzeev:2009fn}
Kharzeev, D.~E. (2010).
\newblock {Topologically induced local P and CP violation in QCD x QED}.
\newblock {\em Annals Phys.}, 325:205--218.
\newblock 0911.3715.

\bibitem[Kharzeev, 2014]{Kharzeev:2013ffa}
Kharzeev, D.~E. (2014).
\newblock {The Chiral Magnetic Effect and Anomaly-Induced Transport}.
\newblock {\em Prog. Part. Nucl. Phys.}, 75:133--151.
\newblock 1312.3348.

\bibitem[Kharzeev and Levin, 2017]{Kharzeev:2017qzs}
Kharzeev, D.~E. and Levin, E.~M. (2017).
\newblock {Deep inelastic scattering as a probe of entanglement}.
\newblock {\em Phys. Rev.}, D95(11):114008.

\bibitem[Kharzeev et~al., 2008]{Kharzeev:2007jp}
Kharzeev, D.~E., McLerran, L.~D., and Warringa, H.~J. (2008).
\newblock {The Effects of topological charge change in heavy ion collisions:
  'Event by event P and CP violation'}.
\newblock {\em Nucl. Phys.}, A803:227--253.
\newblock 0711.0950.

\bibitem[Khriplovich, 1969]{Khriplovich:1969aa}
Khriplovich, I.~B. (1969).
\newblock {Green's functions in theories with non-abelian gauge group.}
\newblock {\em Sov. J. Nucl. Phys.}, 10:235--242.
\newblock [Yad. Fiz.10,409(1969)].

\bibitem[Klebanov et~al., 2006]{Klebanov:2006jj}
Klebanov, I.~R., Maldacena, J.~M., and Thorn, III, C.~B. (2006).
\newblock {Dynamics of flux tubes in large N gauge theories}.
\newblock {\em JHEP}, 04:024.

\bibitem[Klinkhamer and Manton, 1984]{Klinkhamer:1984di}
Klinkhamer, F.~R. and Manton, N.~S. (1984).
\newblock {A Saddle Point Solution in the Weinberg-Salam Theory}.
\newblock {\em Phys. Rev.}, D30:2212.

\bibitem[Kochelev, 1998]{Kochelev:1996pv}
Kochelev, N.~I. (1998).
\newblock {Anomalous quark chromomagnetic moment induced by instantons}.
\newblock {\em Phys. Lett.}, B426:149--153.

\bibitem[Koma et~al., 2003]{Koma:2003gq}
Koma, Y., Koma, M., Ilgenfritz, E.-M., Suzuki, T., and Polikarpov, M.~I.
  (2003).
\newblock {Duality of gauge field singularities and the structure of the flux
  tube in Abelian projected SU(2) gauge theory and the dual Abelian Higgs
  model}.
\newblock {\em Phys. Rev.}, D68:094018.
\newblock hep-lat/0302006.

\bibitem[Kouno et~al., 2013]{Kouno:2013zr}
Kouno, H., Makiyama, T., Sasaki, T., Sakai, Y., and Yahiro, M. (2013).
\newblock {Confinement and $\mathbb{Z}_{3}$ symmetry in three-flavor QCD}.
\newblock {\em J. Phys.}, G40:095003.
\newblock 1301.4013.

\bibitem[Kozcaz et~al., 2016]{Kozcaz:2016wvy}
Kozcaz, C., Sulejmanpasic, T., Tanizaki, Y., and Unsal, M. (2016).
\newblock {Cheshire Cat resurgence, Self-resurgence and Quasi-Exact Solvable
  Systems}.
\newblock 1609.06198.

\bibitem[Kraan and van Baal, 1998]{Kraan:1998sn}
Kraan, T.~C. and van Baal, P. (1998).
\newblock {Monopole constituents inside SU(n) calorons}.
\newblock {\em Phys. Lett.}, B435:389--395.
\newblock hep-th/9806034.

\bibitem[Kuraev et~al., 1976]{Kuraev:1976ge}
Kuraev, E.~A., Lipatov, L.~N., and Fadin, V.~S. (1976).
\newblock {Multi - Reggeon Processes in the Yang-Mills Theory}.
\newblock {\em Sov. Phys. JETP}, 44:443--450.
\newblock [Zh. Eksp. Teor. Fiz.71,840(1976)].

\bibitem[Langfeld and Ilgenfritz, 2011]{Langfeld:2010nm}
Langfeld, K. and Ilgenfritz, E.-M. (2011).
\newblock {Confinement from semiclassical gluon fields in SU(2) gauge theory}.
\newblock {\em Nucl. Phys.}, B848:33--61.
\newblock 1012.1214.

\bibitem[Larsen and Shuryak, 2015]{Larsen:2015vaa}
Larsen, R. and Shuryak, E. (2015).
\newblock {Interacting ensemble of the instanton-dyons and the deconfinement
  phase transition in the SU(2) gauge theory}.
\newblock {\em Phys. Rev.}, D92(9):094022.
\newblock 1504.03341.

\bibitem[Larsen and Shuryak, 2016a]{Larsen:2014yya}
Larsen, R. and Shuryak, E. (2016a).
\newblock {Classical interactions of the instanton-dyons with antidyons}.
\newblock {\em Nucl. Phys.}, A950:110--128.
\newblock 1408.6563.

\bibitem[Larsen and Shuryak, 2016b]{Larsen:2015tso}
Larsen, R. and Shuryak, E. (2016b).
\newblock {Instanton-dyon Ensemble with two Dynamical Quarks: the Chiral
  Symmetry Breaking}.
\newblock {\em Phys. Rev.}, D93(5):054029.
\newblock 1511.02237.

\bibitem[Larsen and Shuryak, 2016c]{Larsen:2016fvs}
Larsen, R. and Shuryak, E. (2016c).
\newblock {Instanton-dyon ensembles with quarks with modified boundary
  conditions}.
\newblock {\em Phys. Rev.}, D94(9):094009.
\newblock 1605.07474.

\bibitem[Larsen et~al., 2018]{Larsen:2018crg}
Larsen, R.~N., Sharma, S., and Shuryak, E. (2018).
\newblock {The topological objects near the chiral crossover transition in
  QCD}.

\bibitem[Laursen and Schierholz, 1988]{Laursen:1987eb}
Laursen, M.~L. and Schierholz, G. (1988).
\newblock {Evidence for Monopoles in the Quantized SU(2) Lattice Vacuum: A
  Study at Finite Temperature}.
\newblock {\em Z. Phys.}, C38:501.

\bibitem[Lee and Lu, 1998]{Lee:1998bb}
Lee, K.-M. and Lu, C.-h. (1998).
\newblock {SU(2) calorons and magnetic monopoles}.
\newblock {\em Phys. Rev.}, D58:025011.
\newblock hep-th/9802108.

\bibitem[Levine and Yaffe, 1979]{Levine:1978ge}
Levine, H. and Yaffe, L.~G. (1979).
\newblock {HIGHER ORDER INSTANTON EFFECTS}.
\newblock {\em Phys. Rev.}, D19:1225.

\bibitem[Li et~al., 2016]{Li:2014bha}
Li, Q., Kharzeev, D.~E., Zhang, C., Huang, Y., Pletikosic, I., Fedorov, A.~V.,
  Zhong, R.~D., Schneeloch, J.~A., Gu, G.~D., and Valla, T. (2016).
\newblock {Observation of the chiral magnetic effect in ZrTe5}.
\newblock {\em Nature Phys.}, 12:550--554.
\newblock 1412.6543.

\bibitem[Liao and Shuryak, 2007]{Liao:2006ry}
Liao, J. and Shuryak, E. (2007).
\newblock {Strongly coupled plasma with electric and magnetic charges}.
\newblock {\em Phys. Rev.}, C75:054907.
\newblock hep-ph/0611131.

\bibitem[Liao and Shuryak, 2008a]{Liao:2007mj}
Liao, J. and Shuryak, E. (2008a).
\newblock {Electric Flux Tube in Magnetic Plasma}.
\newblock {\em Phys. Rev.}, C77:064905.
\newblock 0706.4465.

\bibitem[Liao and Shuryak, 2008b]{Liao:2008jg}
Liao, J. and Shuryak, E. (2008b).
\newblock {Magnetic Component of Quark-Gluon Plasma is also a Liquid!}
\newblock {\em Phys. Rev. Lett.}, 101:162302.
\newblock 0804.0255.

\bibitem[Liao and Shuryak, 2009]{Liao:2008dk}
Liao, J. and Shuryak, E. (2009).
\newblock {Angular Dependence of Jet Quenching Indicates Its Strong Enhancement
  Near the QCD Phase Transition}.
\newblock {\em Phys. Rev. Lett.}, 102:202302.
\newblock 0810.4116.

\bibitem[Liao and Shuryak, 2010]{Liao:2008vj}
Liao, J. and Shuryak, E. (2010).
\newblock {Static $\bar Q Q$ Potentials and the Magnetic Component of QCD
  Plasma near $T_c$}.
\newblock {\em Phys. Rev.}, D82:094007.
\newblock 0804.4890.

\bibitem[Liao and Shuryak, 2012]{Liao:2012tw}
Liao, J. and Shuryak, E. (2012).
\newblock {Effect of Light Fermions on the Confinement Transition in QCD-like
  Theories}.
\newblock {\em Phys. Rev. Lett.}, 109:152001.
\newblock 1206.3989.

\bibitem[Lin and Shuryak, 2007]{Lin:2007pv}
Lin, S. and Shuryak, E. (2007).
\newblock {Stress tensor of static dipoles in strongly coupled N = 4 gauge
  theory}.
\newblock {\em Phys. Rev.}, D76:085014.

\bibitem[Lin and Shuryak, 2008a]{Lin:2007fa}
Lin, S. and Shuryak, E. (2008a).
\newblock {Toward the AdS/CFT gravity dual for high energy collisions. 2. The
  stress tensor on the boundary}.
\newblock {\em Phys. Rev.}, D77:085014.

\bibitem[Lin and Shuryak, 2008b]{Lin:2006rf}
Lin, S. and Shuryak, E. (2008b).
\newblock {Toward the AdS/CFT gravity dual for High Energy Collisions:
  I.Falling into the AdS}.
\newblock {\em Phys. Rev.}, D77:085013.
\newblock hep-ph/0610168.

\bibitem[Liu et~al., 2015a]{Liu:2015ufa}
Liu, Y., Shuryak, E., and Zahed, I. (2015a).
\newblock {Confining dyon-antidyon Coulomb liquid model. I.}
\newblock {\em Phys. Rev.}, D92(8):085006.
\newblock 1503.03058.

\bibitem[Liu et~al., 2015b]{Liu:2015jsa}
Liu, Y., Shuryak, E., and Zahed, I. (2015b).
\newblock {Light quarks in the screened dyon-antidyon Coulomb liquid model.
  II.}
\newblock {\em Phys. Rev.}, D92(8):085007.
\newblock 1503.09148.

\bibitem[Liu et~al., 2016a]{Liu:2016mrk}
Liu, Y., Shuryak, E., and Zahed, I. (2016a).
\newblock {Light Adjoint Quarks in the Instanton-Dyon Liquid Model IV}.
\newblock {\em Phys. Rev.}, D94(10):105012.
\newblock 1605.07584.

\bibitem[Liu et~al., 2016b]{Liu:2016yij}
Liu, Y., Shuryak, E., and Zahed, I. (2016b).
\newblock {The Instanton-Dyon Liquid Model V: Twisted Light Quarks}.
\newblock {\em Phys. Rev.}, D94(10):105013.
\newblock 1606.02996.

\bibitem[Lopez-Ruiz et~al., 2018]{Lopez-Ruiz:2016bjl}
Lopez-Ruiz, M.~A., Jiang, Y., and Liao, J. (2018).
\newblock {Confinement, Holonomy and Correlated Instanton-Dyon Ensemble I:
  SU(2) Yang-Mills Theory}.
\newblock {\em Phys. Rev.}, D97(5):054026.
\newblock 1611.02539.

\bibitem[Lucini and Teper, 2001]{Lucini:2001ej}
Lucini, B. and Teper, M. (2001).
\newblock {SU(N) gauge theories in four-dimensions: Exploring the approach to N
  = infinity}.
\newblock {\em JHEP}, 06:050.

\bibitem[Luscher and Weisz, 2004]{Luscher:2004ib}
Luscher, M. and Weisz, P. (2004).
\newblock {String excitation energies in SU(N) gauge theories beyond the
  free-string approximation}.
\newblock {\em JHEP}, 07:014.
\newblock hep-th/0406205.

\bibitem[Mace et~al., 2016]{Mace:2016svc}
Mace, M., Schlichting, S., and Venugopalan, R. (2016).
\newblock {Off-equilibrium sphaleron transitions in the Glasma}.
\newblock {\em Phys. Rev.}, D93(7):074036.
\newblock 1601.07342.

\bibitem[Maldacena, 1998]{Maldacena:1998im}
Maldacena, J.~M. (1998).
\newblock {Wilson loops in large N field theories}.
\newblock {\em Phys. Rev. Lett.}, 80:4859--4862.

\bibitem[Maldacena, 1999]{Maldacena:1997re}
Maldacena, J.~M. (1999).
\newblock {The Large N limit of superconformal field theories and
  supergravity}.
\newblock {\em Int. J. Theor. Phys.}, 38:1113--1133.
\newblock [Adv. Theor. Math. Phys.2,231(1998)].

\bibitem[Mandelstam and Leontowitsch, 1928]{tunneling2}
Mandelstam, L. and Leontowitsch, M. (1928).
\newblock {Zur Theorie der Schroedingerschen Gleichung}.
\newblock {\em Zeitschrift fur Physik}, 47 (1-2):131--136.

\bibitem[Mandelstam, 1976]{Mandelstam:1974pi}
Mandelstam, S. (1976).
\newblock {Vortices and Quark Confinement in Nonabelian Gauge Theories}.
\newblock {\em Phys. Rept.}, 23:245--249.

\bibitem[Marshakov et~al., 2009]{Marshakov:2009kj}
Marshakov, A., Mironov, A., and Morozov, A. (2009).
\newblock {Zamolodchikov asymptotic formula and instanton expansion in N=2 SUSY
  N(f) = 2N(c) QCD}.
\newblock {\em JHEP}, 11:048.
\newblock 0909.3338.

\bibitem[Mattis, 1992]{Mattis:1991bj}
Mattis, M.~P. (1992).
\newblock {The Riddle of high-energy baryon number violation}.
\newblock {\em Phys. Rept.}, 214:159--221.

\bibitem[McLerran, 1989]{McLerran:1988ja}
McLerran, L.~D. (1989).
\newblock {Anomalies, Sphalerons and Baryon Number Violation in Electroweak
  Theory}.
\newblock {\em Acta Phys. Polon.}, B20:249--286.

\bibitem[McLerran and Venugopalan, 1994]{McLerran:1993ni}
McLerran, L.~D. and Venugopalan, R. (1994).
\newblock {Computing quark and gluon distribution functions for very large
  nuclei}.
\newblock {\em Phys. Rev.}, D49:2233--2241.
\newblock hep-ph/9309289.

\bibitem[Meggiolaro:1997dy, 1998]{Meggiolaro:1997dy}
Meggiolaro:1997dy, E. (1998).
\newblock {High-energy quark quark scattering and the eikonal approximation}.
\newblock {\em Nucl. Phys. Proc. Suppl.}, 64:191--196.

\bibitem[Meyer, 2004]{Meyer:2004gx}
Meyer, H.~B. (2004).
\newblock {\em {Glueball regge trajectories}}.
\newblock PhD thesis, Oxford U.
\newblock hep-lat/0508002.

\bibitem[Mikhailov, 2003]{Mikhailov:2003er}
Mikhailov, A. (2003).
\newblock {Nonlinear waves in AdS / CFT correspondence}.

\bibitem[Milton, 2006]{Milton:2006cp}
Milton, K.~A. (2006).
\newblock {Theoretical and experimental status of magnetic monopoles}.
\newblock {\em Rept. Prog. Phys.}, 69:1637--1712.
\newblock hep-ex/0602040.

\bibitem[Misumi et~al., 2016]{Misumi:2015hfa}
Misumi, T., Iritani, T., and Itou, E. (2016).
\newblock {Finite-temperature phase transition of $N_{f}=3$ QCD with exact
  center symmetry}.
\newblock {\em PoS}, LATTICE2015:152.
\newblock 1510.07227.

\bibitem[Miura et~al., 2012]{Miura:2011mc}
Miura, K., Lombardo, M.~P., and Pallante, E. (2012).
\newblock {Chiral phase transition at finite temperature and conformal dynamics
  in large Nf QCD}.
\newblock {\em Phys. Lett.}, B710:676--682.
\newblock 1110.3152.

\bibitem[Moch et~al., 1997]{Moch:1996bs}
Moch, S., Ringwald, A., and Schrempp, F. (1997).
\newblock {Instantons in deep inelastic scattering: The Simplest process}.
\newblock {\em Nucl. Phys.}, B507:134--156.
\newblock hep-ph/9609445.

\bibitem[Moore, 1996]{Moore:1995jv}
Moore, G.~D. (1996).
\newblock {Fermion determinant and the sphaleron bound}.
\newblock {\em Phys. Rev.}, D53:5906--5917.
\newblock hep-ph/9508405.

\bibitem[Musakhanov, 2018]{Musakhanov:2018sdu}
Musakhanov, M. (2018).
\newblock {Gluons, Heavy and Light Quarks in the QCD Vacuum}.
\newblock In {\em {Proceedings, 6th International Conference on New Frontiers
  in Physics (ICNFP 2017): Crete, Greece, August 17-29, 2017}}.

\bibitem[Musakhanov and Egamberdiev, 2018]{Musakhanov:2017erp}
Musakhanov, M. and Egamberdiev, O. (2018).
\newblock {Dynamical gluon mass in the instanton vacuum model}.
\newblock {\em Phys. Lett.}, B779:206--209.

\bibitem[Myers and Ogilvie, 2008]{Myers:2007vc}
Myers, J.~C. and Ogilvie, M.~C. (2008).
\newblock {New phases of SU(3) and SU(4) at finite temperature}.
\newblock {\em Phys. Rev.}, D77:125030.
\newblock 0707.1869.

\bibitem[Nachtmann, 1997]{Nachtmann:1996kt}
Nachtmann, O. (1997).
\newblock {High-energy collisions and nonperturbative QCD}.
\newblock {\em Lect. Notes Phys.}, 479:49--138.
\newblock [Lect. Notes Phys.496,1(1997)].

\bibitem[Nambu, 1974]{Nambu:1974zg}
Nambu, Y. (1974).
\newblock {Strings, Monopoles and Gauge Fields}.
\newblock {\em Phys. Rev.}, D10:4262.
\newblock [,310(1974)].

\bibitem[Nambu and Jona-Lasinio, 1961]{Nambu:1961tp}
Nambu, Y. and Jona-Lasinio, G. (1961).
\newblock {Dynamical Model of Elementary Particles Based on an Analogy with
  Superconductivity. 1.}
\newblock {\em Phys. Rev.}, 122:345--358.
\newblock [,127(1961)].

\bibitem[Nekrasov, 2003]{Nekrasov:2002qd}
Nekrasov, N.~A. (2003).
\newblock {Seiberg-Witten prepotential from instanton counting}.
\newblock {\em Adv. Theor. Math. Phys.}, 7(5):831--864.
\newblock hep-th/0206161.

\bibitem[Neuberger, 1980]{Neuberger:1980as}
Neuberger, H. (1980).
\newblock {INSTANTONS AS A BRIDGEHEAD AT N = infinity}.
\newblock {\em Phys. Lett.}, 94B:199--202.

\bibitem[Nielsen and Olesen, 1973]{Nielsen:1973cs}
Nielsen, H.~B. and Olesen, P. (1973).
\newblock {Vortex Line Models for Dual Strings}.
\newblock {\em Nucl. Phys.}, B61:45--61.
\newblock [,302(1973)].

\bibitem[Novikov et~al., 1981]{Novikov:1981xi}
Novikov, V.~A., Shifman, M.~A., Vainshtein, A.~I., and Zakharov, V.~I. (1981).
\newblock {Are All Hadrons Alike? {DESY}-check = Moscow Inst. Theor. Exp. Phys.
  Gkae - Itef-81-048 (81,rec.jun.) 32 $P$ and Nucl. Phys. B191 (1981) 301-369
  and Moscow Inst. Theor. Exp. Phys. Gkae - Itef-81-042 (81,rec.apr.) 70 $P$.
  (104907)}.
\newblock {\em Nucl. Phys.}, B191:301--369.

\bibitem[Novikov et~al., 1986]{Novikov:1985rd}
Novikov, V.~A., Shifman, M.~A., Vainshtein, A.~I., and Zakharov, V.~I. (1986).
\newblock {Beta Function in Supersymmetric Gauge Theories: Instantons Versus
  Traditional Approach}.
\newblock {\em Phys. Lett.}, 166B:329--333.
\newblock [Yad. Fiz.43,459(1986)].

\bibitem[Nowak et~al., 2001]{Nowak:2000de}
Nowak, M.~A., Shuryak, E.~V., and Zahed, I. (2001).
\newblock {Instanton induced inelastic collisions in QCD}.
\newblock {\em Phys. Rev.}, D64:034008.
\newblock hep-ph/0012232.

\bibitem[Nussinov and Lampert, 2002]{Nussinov:1999sx}
Nussinov, S. and Lampert, M.~A. (2002).
\newblock {QCD inequalities}.
\newblock {\em Phys. Rept.}, 362:193--301.

\bibitem[Ostrovsky et~al., 2002]{Ostrovsky:2002cg}
Ostrovsky, D.~M., Carter, G.~W., and Shuryak, E.~V. (2002).
\newblock {Forced tunneling and turning state explosion in pure Yang-Mills
  theory}.
\newblock {\em Phys. Rev.}, D66:036004.
\newblock hep-ph/0204224.

\bibitem[Parikh and Wilczek, 1998]{Parikh:1997ma}
Parikh, M. and Wilczek, F. (1998).
\newblock {An Action for black hole membranes}.
\newblock {\em Phys. Rev.}, D58:064011.

\bibitem[Petrov and Ryutin, 2015]{Petrov:2014jya}
Petrov, V.~A. and Ryutin, R.~A. (2015).
\newblock {High-energy scattering versus static QCD strings}.
\newblock {\em Mod. Phys. Lett.}, A30(18):1550081.
\newblock 1409.8425.

\bibitem[Pisarski, 2009]{Pisarski:2009zza}
Pisarski, R.~D. (2009).
\newblock {Towards a theory of the semi-Quark Gluon Plasma}.
\newblock {\em Nucl. Phys. Proc. Suppl.}, 195:157--198.

\bibitem[Pisarski and Yaffe, 1980]{Pisarski:1980md}
Pisarski, R.~D. and Yaffe, L.~G. (1980).
\newblock {THE DENSITY OF INSTANTONS AT FINITE TEMPERATURE}.
\newblock {\em Phys. Lett.}, 97B:110--112.

\bibitem[Polchinski and Strassler, 2002]{Polchinski:2001tt}
Polchinski, J. and Strassler, M.~J. (2002).
\newblock {Hard scattering and gauge / string duality}.
\newblock {\em Phys. Rev. Lett.}, 88:031601.

\bibitem[Politzer, 1973]{Politzer:1973fx}
Politzer, H.~D. (1973).
\newblock {Reliable Perturbative Results for Strong Interactions?}
\newblock {\em Phys. Rev. Lett.}, 30:1346--1349.
\newblock [,274(1973)].

\bibitem[Polyakov, 1974]{Polyakov:1974ek}
Polyakov, A.~M. (1974).
\newblock {Particle Spectrum in the Quantum Field Theory}.
\newblock {\em JETP Lett.}, 20:194--195.

\bibitem[Polyakov, 1977]{Polyakov:1976fu}
Polyakov, A.~M. (1977).
\newblock {Quark Confinement and Topology of Gauge Groups}.
\newblock {\em Nucl. Phys.}, B120:429--458.

\bibitem[Polyakov, 1986]{Polyakov:1986cs}
Polyakov, A.~M. (1986).
\newblock {Fine Structure of Strings}.
\newblock {\em Nucl. Phys.}, B268:406--412.

\bibitem[Poppitz et~al., 2012]{Poppitz:2012sw}
Poppitz, E., Schafer, T., and Unsal, M. (2012).
\newblock {Continuity, Deconfinement, and (Super) Yang-Mills Theory}.
\newblock {\em JHEP}, 10:115.
\newblock 1205.0290.

\bibitem[Poppitz and Unsal, 2011]{Poppitz:2011wy}
Poppitz, E. and Unsal, M. (2011).
\newblock {Seiberg-Witten and 'Polyakov-like' magnetic bion confinements are
  continuously connected}.
\newblock {\em JHEP}, 07:082.
\newblock 1105.3969.

\bibitem[Qian and Zahed, 2015]{Qian:2014jna}
Qian, Y. and Zahed, I. (2015).
\newblock {A Stringy (Holographic) Pomeron with Extrinsic Curvature}.
\newblock {\em Phys. Rev.}, D92(8):085012.
\newblock 1410.1092.

\bibitem[Ramamurti and Shuryak, 2017]{Ramamurti:2017fdn}
Ramamurti, A. and Shuryak, E. (2017).
\newblock {Effective Model of QCD Magnetic Monopoles From Numerical Study of
  One- and Two-Component Coulomb Quantum Bose Gases}.
\newblock {\em Phys. Rev.}, D95(7):076019.
\newblock 1702.07723.

\bibitem[Ramamurti and Shuryak, 2018a]{Ramamurti:2018hdh}
Ramamurti, A. and Shuryak, E. (2018a).
\newblock {Chiral symmetry breaking and monopoles in gauge theories}.
\newblock 1801.06922.

\bibitem[Ramamurti and Shuryak, 2018b]{Ramamurti:2017zjn}
Ramamurti, A. and Shuryak, E. (2018b).
\newblock {Role of QCD monopoles in jet quenching}.
\newblock {\em Phys. Rev.}, D97(1):016010.
\newblock 1708.04254.

\bibitem[Ramamurti et~al., 2018]{Ramamurti:2018evz}
Ramamurti, A., Shuryak, E., and Zahed, I. (2018).
\newblock {Are there monopoles in the quark-gluon plasma?}
\newblock 1802.10509.

\bibitem[Randall et~al., 1999]{Randall:1998ra}
Randall, L., Rattazzi, R., and Shuryak, E.~V. (1999).
\newblock {Implication of exact SUSY gauge couplings for QCD}.
\newblock {\em Phys. Rev.}, D59:035005.
\newblock hep-ph/9803258.

\bibitem[Rapp et~al., 1998]{Rapp:1997zu}
Rapp, R., Schafer, T., Shuryak, E.~V., and Velkovsky, M. (1998).
\newblock {Diquark Bose condensates in high density matter and instantons}.
\newblock {\em Phys. Rev. Lett.}, 81:53--56.
\newblock hep-ph/9711396.

\bibitem[Rapp et~al., 2000]{Rapp:1999qa}
Rapp, R., Schafer, T., Shuryak, E.~V., and Velkovsky, M. (2000).
\newblock {High density QCD and instantons}.
\newblock {\em Annals Phys.}, 280:35--99.

\bibitem[Ratti and Shuryak, 2009]{Ratti:2008jz}
Ratti, C. and Shuryak, E. (2009).
\newblock {The Role of monopoles in a Gluon Plasma}.
\newblock {\em Phys. Rev.}, D80:034004.
\newblock 0811.4174.

\bibitem[Rho et~al., 2010]{Rho:2009ym}
Rho, M., Sin, S.-J., and Zahed, I. (2010).
\newblock {Dense QCD: A Holographic Dyonic Salt}.
\newblock {\em Phys. Lett.}, B689:23--27.

\bibitem[Roberge and Weiss, 1986]{Roberge:1986mm}
Roberge, A. and Weiss, N. (1986).
\newblock {Gauge Theories With Imaginary Chemical Potential and the Phases of
  {QCD}}.
\newblock {\em Nucl. Phys.}, B275:734--745.

\bibitem[Rodrigues and Wotzasek, 2006]{hep-th/0611050}
Rodrigues, D.~C. and Wotzasek, C. (2006).
\newblock {3D and 4D noncommutative electromagnetic duality and the role of the
  slowly varying fields limit}.
\newblock {\em PoS}, IC2006:048.

\bibitem[Sakai and Sugimoto, 2005]{Sakai:2004cn}
Sakai, T. and Sugimoto, S. (2005).
\newblock {Low energy hadron physics in holographic QCD}.
\newblock {\em Prog. Theor. Phys.}, 113:843--882.

\bibitem[Sauter, 1931]{Sauter:1931zz}
Sauter, F. (1931).
\newblock {Uber das Verhalten eines Elektrons im homogenen elektrischen Feld
  nach der relativistischen Theorie Diracs}.
\newblock {\em Z. Phys.}, 69:742--764.

\bibitem[Schafer, 2002]{Schafer:2002af}
Schafer, T. (2002).
\newblock {Instantons in QCD with many colors}.
\newblock {\em Phys. Rev.}, D66:076009.
\newblock hep-ph/0206062.

\bibitem[Schafer, 2008]{Schafer:2007qy}
Schafer, T. (2008).
\newblock {Euclidean correlation functions in a holographic model of QCD}.
\newblock {\em Phys. Rev.}, D77:126010.

\bibitem[Schafer and Shuryak, 1995]{Schafer:1994fd}
Schafer, T. and Shuryak, E.~V. (1995).
\newblock {Glueballs and instantons}.
\newblock {\em Phys. Rev. Lett.}, 75:1707--1710.
\newblock hep-ph/9410372.

\bibitem[Schafer and Shuryak, 1996]{Schafer:1995pz}
Schafer, T. and Shuryak, E.~V. (1996).
\newblock {The Interacting instanton liquid in QCD at zero and finite
  temperature}.
\newblock {\em Phys. Rev.}, D53:6522--6542.
\newblock hep-ph/9509337.

\bibitem[Schafer and Shuryak, 1998]{Schafer:1996wv}
Schafer, T. and Shuryak, E.~V. (1998).
\newblock {Instantons in QCD}.
\newblock {\em Rev. Mod. Phys.}, 70:323--426.
\newblock hep-ph/9610451.

\bibitem[Schafer and Shuryak, 2001a]{Schafer:2000rv}
Schafer, T. and Shuryak, E.~V. (2001a).
\newblock {Implications of the ALEPH tau lepton decay data for perturbative and
  nonperturbative QCD}.
\newblock {\em Phys. Rev. Lett.}, 86:3973--3976.

\bibitem[Schafer and Shuryak, 2001b]{Schafer:2000et}
Schafer, T. and Shuryak, E.~V. (2001b).
\newblock {Phases of QCD at high baryon density}.
\newblock {\em Lect. Notes Phys.}, 578:203--217.
\newblock [,203(2000)].

\bibitem[Schafer et~al., 1994]{Schafer:1993ra}
Schafer, T., Shuryak, E.~V., and Verbaarschot, J. J.~M. (1994).
\newblock {Baryonic correlators in the random instanton vacuum}.
\newblock {\em Nucl. Phys.}, B412:143--168.
\newblock hep-ph/9306220.

\bibitem[Schulman, 1968]{Schulman:1968yv}
Schulman, L. (1968).
\newblock {A Path integral for spin}.
\newblock {\em Phys. Rev.}, 176:1558--1569.

\bibitem[Schwinger et~al., 1976]{Schwinger:1976fr}
Schwinger, J.~S., Milton, K.~A., Tsai, W.-y., DeRaad, Jr., L.~L., and Clark,
  D.~C. (1976).
\newblock {Nonrelativistic Dyon-Dyon Scattering}.
\newblock {\em Annals Phys.}, 101:451.

\bibitem[Seiberg and Witten, 1994a]{Seiberg:1994rs}
Seiberg, N. and Witten, E. (1994a).
\newblock {Electric - magnetic duality, monopole condensation, and confinement
  in N=2 supersymmetric Yang-Mills theory}.
\newblock {\em Nucl. Phys.}, B426:19--52.
\newblock [Erratum: Nucl. Phys.B430,485(1994)].

\bibitem[Seiberg and Witten, 1994b]{Seiberg:1994aj}
Seiberg, N. and Witten, E. (1994b).
\newblock {Monopoles, duality and chiral symmetry breaking in N=2
  supersymmetric QCD}.
\newblock {\em Nucl. Phys.}, B431:484--550.
\newblock hep-th/9408099.

\bibitem[Semenoff and Zarembo, 2002]{Semenoff:2002kk}
Semenoff, G.~W. and Zarembo, K. (2002).
\newblock {Wilson loops in SYM theory: From weak to strong coupling}.
\newblock {\em Nucl. Phys. Proc. Suppl.}, 108:106--112.
\newblock [,106(2002)].

\bibitem[Sen, 1994]{Sen:1994yi}
Sen, A. (1994).
\newblock {Dyon - monopole bound states, selfdual harmonic forms on the multi -
  monopole moduli space, and SL(2,Z) invariance in string theory}.
\newblock {\em Phys. Lett.}, B329:217--221.
\newblock hep-th/9402032.

\bibitem[Shifman et~al., 1979]{Shifman:1978bx}
Shifman, M.~A., Vainshtein, A.~I., and Zakharov, V.~I. (1979).
\newblock {QCD and Resonance Physics. Theoretical Foundations}.
\newblock {\em Nucl. Phys.}, B147:385--447.

\bibitem[Shnir, 2005a]{Shnir:2005te}
Shnir, Y. (2005a).
\newblock {Electromagnetic interaction in the system of multimonopoles and
  vortex rings}.
\newblock {\em Phys. Rev.}, D72:055016.

\bibitem[Shnir, 2005b]{Balian:2005joa}
Shnir, Y.~M. (2005b).
\newblock {\em {Magnetic Monopoles}}.
\newblock Text and Monographs in Physics. Springer, Berlin/Heidelberg.

\bibitem[Shuryak, 2003]{Shuryak:2002an}
Shuryak, E. (2003).
\newblock {How quantum mechanics of the Yang-Mills fields may help us
  understand the RHIC puzzles}.
\newblock {\em Nucl. Phys.}, A715:289--298.
\newblock hep-ph/0205031.

\bibitem[Shuryak, 2006]{Shuryak:2006yx}
Shuryak, E. (2006).
\newblock {Building a 'holographic dual' to QCD in the AdS(5): Instantons and
  confinement}.

\bibitem[Shuryak, 2007]{Shuryak:2007uq}
Shuryak, E. (2007).
\newblock {A 'Domain Wall' Scenario for the AdS/QCD}.

\bibitem[Shuryak, 2019]{Shuryak:2019zhv}
Shuryak, E. (2019).
\newblock {Light-front wave functions of mesons, baryons and pentaquarks, with
  topology-induced local 4-quark interaction}.

\bibitem[Shuryak and Zahed, 2003a]{Shuryak:2002qz}
Shuryak, E. and Zahed, I. (2003a).
\newblock {Prompt quark production by exploding sphalerons}.
\newblock {\em Phys. Rev.}, D67:014006.
\newblock hep-ph/0206022.

\bibitem[Shuryak and Zahed, 2003b]{Shuryak:2003xz}
Shuryak, E. and Zahed, I. (2003b).
\newblock {Semiclassical double pomeron production of glueballs and eta-prime}.
\newblock {\em Phys. Rev.}, D68:034001.
\newblock hep-ph/0302231.

\bibitem[Shuryak and Zahed, 2004a]{Shuryak:2003ja}
Shuryak, E. and Zahed, I. (2004a).
\newblock {Understanding the strong coupling limit of the N=4 supersymmetric
  Yang-Mills at finite temperature}.
\newblock {\em Phys. Rev.}, D69:046005.

\bibitem[Shuryak and Zahed, 2014]{Shuryak:2013sra}
Shuryak, E. and Zahed, I. (2014).
\newblock {New regimes of the stringy (holographic) Pomeron and
  high-multiplicity $pp$ and $pA$ collisions}.
\newblock {\em Phys. Rev.}, D89(9):094001.
\newblock 1311.0836.

\bibitem[Shuryak and Zahed, 2016]{Shuryak:2016ipj}
Shuryak, E. and Zahed, I. (2016).
\newblock {CP violation during the electroweak sphaleron transitions}.
\newblock 1610.05144.

\bibitem[Shuryak and Zahed, 2018]{Shuryak:2017phz}
Shuryak, E. and Zahed, I. (2018).
\newblock {Regimes of the Pomeron and its Intrinsic Entropy}.
\newblock {\em Annals Phys.}, 396:1--17.
\newblock 1707.01885.

\bibitem[Shuryak, 1978]{Shuryak:1977ut}
Shuryak, E.~V. (1978).
\newblock {Theory of Hadronic Plasma}.
\newblock {\em Sov. Phys. JETP}, 47:212--219.
\newblock [Zh. Eksp. Teor. Fiz.74,408(1978)].

\bibitem[Shuryak, 1982a]{Shuryak:1981fza}
Shuryak, E.~V. (1982a).
\newblock {Hadrons Containing a Heavy Quark and QCD Sum Rules}.
\newblock {\em Nucl. Phys.}, B198:83--101.

\bibitem[Shuryak, 1982b]{Shuryak:1981ff}
Shuryak, E.~V. (1982b).
\newblock {The Role of Instantons in Quantum Chromodynamics. 1. Physical
  Vacuum}.
\newblock {\em Nucl. Phys.}, B203:93.

\bibitem[Shuryak, 1983]{Shuryak:1982qx}
Shuryak, E.~V. (1983).
\newblock {Pseudoscalar Mesons and Instantons}.
\newblock {\em Nucl. Phys.}, B214:237--252.

\bibitem[Shuryak, 1988]{Shuryak:1987tr}
Shuryak, E.~V. (1988).
\newblock {Toward the Quantitative Theory of the 'Instanton Liquid' 4.
  Tunneling in the Double Well Potential}.
\newblock {\em Nucl. Phys.}, B302:621--644.

\bibitem[Shuryak, 1989]{Shuryak:1988rf}
Shuryak, E.~V. (1989).
\newblock {Instantons in {QCD}. 1. Properties of the 'Instanton Liquid'}.
\newblock {\em Nucl. Phys.}, B319:521--540.

\bibitem[Shuryak, 1993]{Shuryak:1993kg}
Shuryak, E.~V. (1993).
\newblock {Correlation functions in the QCD vacuum}.
\newblock {\em Rev. Mod. Phys.}, 65:1--46.

\bibitem[Shuryak, 1995]{Shuryak:1995pv}
Shuryak, E.~V. (1995).
\newblock {Instanton size distribution: Repulsion or the infrared fixed point?}
\newblock {\em Phys. Rev.}, D52:5370--5373.
\newblock hep-ph/9503467.

\bibitem[Shuryak, 1999]{Shuryak:1999fe}
Shuryak, E.~V. (1999).
\newblock {Probing the boundary of the nonperturbative QCD by small size
  instantons}.
\newblock hep-ph/9909458.

\bibitem[Shuryak, 2002]{Shuryak:2001me}
Shuryak, E.~V. (2002).
\newblock {The Azimuthal asymmetry at large p(t) seem to be too large for a
  `jet quenching'}.
\newblock {\em Phys. Rev.}, C66:027902.

\bibitem[Shuryak and Velkovsky, 1994]{Shuryak:1994ay}
Shuryak, E.~V. and Velkovsky, M. (1994).
\newblock {The Instanton density at finite temperatures}.
\newblock {\em Phys. Rev.}, D50:3323--3327.

\bibitem[Shuryak and Verbaarschot, 1990]{Shuryak:1989cx}
Shuryak, E.~V. and Verbaarschot, J. J.~M. (1990).
\newblock {Chiral Symmetry Breaking and Correlations in the Instanton Liquid}.
\newblock {\em Nucl. Phys.}, B341:1--26.

\bibitem[Shuryak and Verbaarschot, 1993a]{Shuryak:1992ke}
Shuryak, E.~V. and Verbaarschot, J. J.~M. (1993a).
\newblock {Mesonic correlation functions in the random instanton vacuum}.
\newblock {\em Nucl. Phys.}, B410:55--89.
\newblock hep-ph/9302239.

\bibitem[Shuryak and Verbaarschot, 1993b]{Shuryak:1992jz}
Shuryak, E.~V. and Verbaarschot, J. J.~M. (1993b).
\newblock {Quark propagation in the random instanton vacuum}.
\newblock {\em Nucl. Phys.}, B410:37--54.
\newblock hep-ph/9302238.

\bibitem[Shuryak and Zahed, 2000]{Shuryak:2000df}
Shuryak, E.~V. and Zahed, I. (2000).
\newblock {Instanton induced effects in QCD high-energy scattering}.
\newblock {\em Phys. Rev.}, D62:085014.
\newblock hep-ph/0005152.

\bibitem[Shuryak and Zahed, 2004b]{Shuryak:2003rb}
Shuryak, E.~V. and Zahed, I. (2004b).
\newblock {Understanding the nonperturbative deep inelastic scattering:
  Instanton induced inelastic dipole dipole cross-section}.
\newblock {\em Phys. Rev.}, D69:014011.
\newblock hep-ph/0307103.

\bibitem[Son and Surowka, 2009]{Son:2009tf}
Son, D.~T. and Surowka, P. (2009).
\newblock {Hydrodynamics with Triangle Anomalies}.
\newblock {\em Phys. Rev. Lett.}, 103:191601.

\bibitem[Sonnenschein, 2017]{Sonnenschein:2016pim}
Sonnenschein, J. (2017).
\newblock {Holography Inspired Stringy Hadrons}.
\newblock {\em Prog. Part. Nucl. Phys.}, 92:1--49.
\newblock 1602.00704.

\bibitem[Sonnenschein and Weissman, 2014]{Sonnenschein:2014jwa}
Sonnenschein, J. and Weissman, D. (2014).
\newblock {Rotating strings confronting PDG mesons}.
\newblock {\em JHEP}, 08:013.
\newblock 1402.5603.

\bibitem[Suzuki et~al., 2009]{Suzuki:2009xy}
Suzuki, T., Hasegawa, M., Ishiguro, K., Koma, Y., and Sekido, T. (2009).
\newblock {Gauge invariance of color confinement due to the dual Meissner
  effect caused by Abelian monopoles}.
\newblock {\em Phys. Rev.}, D80:054504.
\newblock 0907.0583.

\bibitem['t~Hooft, 1974]{tHooft:1974kcl}
't~Hooft, G. (1974).
\newblock {Magnetic Monopoles in Unified Gauge Theories}.
\newblock {\em Nucl. Phys.}, B79:276--284.

\bibitem['t~Hooft, 1976]{tHooft:1976snw}
't~Hooft, G. (1976).
\newblock {Computation of the Quantum Effects Due to a Four-Dimensional
  Pseudoparticle}.
\newblock {\em Phys. Rev.}, D14:3432--3450.
\newblock [,70(1976)].

\bibitem['t~Hooft, 1978a]{'tHooft:1977hy}
't~Hooft, G. (1978a).
\newblock {On the Phase Transition Towards Permanent Quark Confinement}.
\newblock {\em Nucl. Phys.}, B138:1--25.

\bibitem['t~Hooft, 1978b]{tHooft:1977nqb}
't~Hooft, G. (1978b).
\newblock {On the Phase Transition Towards Permanent Quark Confinement}.
\newblock {\em Nucl. Phys.}, B138:1--25.

\bibitem[Teper, 1986]{Teper:1985ek}
Teper, M. (1986).
\newblock {The Topological Susceptibility in SU(2) Lattice Gauge Theory: An
  Exploratory Study}.
\newblock {\em Phys. Lett.}, B171:86--94.

\bibitem[Teper, 2009]{Teper:2009uf}
Teper, M. (2009).
\newblock {Large N and confining flux tubes as strings - a view from the
  lattice}.
\newblock {\em Acta Phys. Polon.}, B40:3249--3320.
\newblock 0912.3339.

\bibitem[Teper, 1980]{Teper:1979tq}
Teper, M.~J. (1980).
\newblock {Instantons and the 1/$N$ Expansion}.
\newblock {\em Z. Phys.}, C5:233.

\bibitem[Thorne et~al., 1986]{Thorne:1986iy}
Thorne, K.~S., Price, R.~H., and Macdonald, D.~A., editors (1986).
\newblock {\em {BLACK HOLES: THE MEMBRANE PARADIGM}}.

\bibitem[Tranberg and Smit, 2003]{Tranberg:2003gi}
Tranberg, A. and Smit, J. (2003).
\newblock {Baryon asymmetry from electroweak tachyonic preheating}.
\newblock {\em JHEP}, 11:016.
\newblock hep-ph/0310342.

\bibitem[Turbiner, 2010]{Turbiner:2009ns}
Turbiner, A.~V. (2010).
\newblock {Double Well Potential: Perturbation Theory, Tunneling, WKB (beyond
  instantons)}.
\newblock {\em Int. J. Mod. Phys.}, A25(02n03):647--658.
\newblock 0907.4485.

\bibitem[Turbiner, 2016]{Turbiner:2016aum}
Turbiner, A.~V. (2016).
\newblock {One-dimensional quasi-exactly solvable Schroedinger equations}.
\newblock {\em Phys. Rept.}, 642:1--71.
\newblock 1603.02992.

\bibitem[Unsal, 2008]{Unsal:2007vu}
Unsal, M. (2008).
\newblock {Abelian duality, confinement, and chiral symmetry breaking in
  QCD(adj)}.
\newblock {\em Phys. Rev. Lett.}, 100:032005.
\newblock 0708.1772.

\bibitem[Unsal, 2009]{Unsal:2007jx}
Unsal, M. (2009).
\newblock {Magnetic bion condensation: A New mechanism of confinement and mass
  gap in four dimensions}.
\newblock {\em Phys. Rev.}, D80:065001.
\newblock 0709.3269.

\bibitem[Vainshtein et~al., 1982]{Vainshtein:1981wh}
Vainshtein, A.~I., Zakharov, V.~I., Novikov, V.~A., and Shifman, M.~A. (1982).
\newblock {ABC's of Instantons}.
\newblock {\em Sov. Phys. Usp.}, 25:195.
\newblock [,201(1981)].

\bibitem[Veneziano, 1968]{Veneziano:1968yb}
Veneziano, G. (1968).
\newblock {Construction of a crossing - symmetric, Regge behaved amplitude for
  linearly rising trajectories}.
\newblock {\em Nuovo Cim.}, A57:190--197.

\bibitem[Verbaarschot, 1991]{Verbaarschot:1991sq}
Verbaarschot, J. J.~M. (1991).
\newblock {Streamlines and conformal invariance in Yang-Mills theories}.
\newblock {\em Nucl. Phys.}, B362:33--53.
\newblock [Erratum: Nucl. Phys.B386,236(1992)].

\bibitem[Weinberg, 1994]{Weinberg:1993sg}
Weinberg, E.~J. (1994).
\newblock {Monopole vector spherical harmonics}.
\newblock {\em Phys. Rev.}, D49:1086--1092.

\bibitem[Weingarten, 1983]{Weingarten:1983uj}
Weingarten, D. (1983).
\newblock {Mass Inequalities for QCD}.
\newblock {\em Phys. Rev. Lett.}, 51:1830.

\bibitem[Wilson, 1974]{Wilson:1974sk}
Wilson, K.~G. (1974).
\newblock {Confinement of Quarks}.
\newblock {\em Phys. Rev.}, D10:2445--2459.
\newblock [,319(1974)].

\bibitem[Witten, 1979]{Witten:1978bc}
Witten, E. (1979).
\newblock {Instantons, the Quark Model, and the 1/n Expansion}.
\newblock {\em Nucl. Phys.}, B149:285--320.

\bibitem[Witten, 1997]{hep-th/9703166}
Witten, E. (1997).
\newblock {Solutions of four-dimensional field theories via M theory}.
\newblock {\em Nucl. Phys.}, B500:3--42.
\newblock [,452(1997)].

\bibitem[Witten, 1998a]{Witten:1998qj}
Witten, E. (1998a).
\newblock {Anti-de Sitter space and holography}.
\newblock {\em Adv. Theor. Math. Phys.}, 2:253--291.

\bibitem[Witten, 1998b]{Witten:1998zw}
Witten, E. (1998b).
\newblock {Anti-de Sitter space, thermal phase transition, and confinement in
  gauge theories}.
\newblock {\em Adv. Theor. Math. Phys.}, 2:505--532.
\newblock [,89(1998)].

\bibitem[Witten, 2010]{Witten:2010zr}
Witten, E. (2010).
\newblock {A New Look At The Path Integral Of Quantum Mechanics}.
\newblock 1009.6032.

\bibitem[Wohler and Shuryak, 1994]{Wohler:1994pg}
Wohler, C.~F. and Shuryak, E.~V. (1994).
\newblock {Two loop correction to the instanton density for the double well
  potential}.
\newblock {\em Phys. Lett.}, B333:467--470.
\newblock hep-ph/9402287.

\bibitem[Xu et~al., 2016]{Xu:2015bbz}
Xu, J., Liao, J., and Gyulassy, M. (2016).
\newblock {Bridging Soft-Hard Transport Properties of Quark-Gluon Plasmas with
  CUJET3.0}.
\newblock {\em JHEP}, 02:169.
\newblock 1508.00552.

\bibitem[Yakhshiev et~al., 2018]{Yakhshiev:2018juj}
Yakhshiev, U., Kim, H.-C., and Hiyama, E. (2018).
\newblock {Instanton effects on charmonium states}.

\bibitem[Yanagihara et~al., 2018]{Yanagihara:2018qqg}
Yanagihara, R., Iritani, T., Kitazawa, M., Asakawa, M., and Hatsuda, T. (2018).
\newblock {Distribution of Stress Tensor around Static Quark--Anti-Quark from
  Yang-Mills Gradient Flow}.
\newblock 1803.05656.

\bibitem[Yang and Mills, 1954]{Yang:1954ek}
Yang, C.-N. and Mills, R.~L. (1954).
\newblock {Conservation of Isotopic Spin and Isotopic Gauge Invariance}.
\newblock {\em Phys. Rev.}, 96:191--195.
\newblock [,150(1954)].

\bibitem[Zarembo, 1996]{Zarembo:1995am}
Zarembo, K. (1996).
\newblock {Monopole determinant in Yang-Mills theory at finite temperature}.
\newblock {\em Nucl. Phys.}, B463:73--98.
\newblock hep-th/9510031.

\bibitem[Zetocha and Schafer, 2003]{Zetocha:2002as}
Zetocha, V. and Schafer, T. (2003).
\newblock {Instanton contribution to scalar charmonium and glueball decays}.
\newblock {\em Phys. Rev.}, D67:114003.
\newblock hep-ph/0212125.

\bibitem[Zinn-Justin and Jentschura, 2004]{ZinnJustin:2004cg}
Zinn-Justin, J. and Jentschura, U.~D. (2004).
\newblock {Multi-instantons and exact results II: Specific cases, higher-order
  effects, and numerical calculations}.
\newblock {\em Annals Phys.}, 313:269--325.
\newblock quant-ph/0501137.

\end{thebibliography}
\bibliographystyle{apalike}
\end{document}